\pdfoutput=1
\documentclass[acmsmall,screen]{acmart}
\usepackage{listings}
\usepackage{pgfplots}
\usepackage{pgfplotstable}
\usepackage{microtype}
\usepackage{mathtools} 

\usepackage{todonotes}

\usepackage[all,british]{foreign}

\lstdefinelanguage{SaC}[]{}{%
    language=C,
    morekeywords={inline,return,for,shape,dim,with,fold,modarray,genarray},
    otherkeywords={?,:,->},
    comment=[l][\color{green!50!brown}]{//},
    morecomment=[s][\color{blue!70!gray}]{int[}{]},
    morecomment=[s][\color{blue!70!gray}]{double[}{]}
}
\newcommand*{\xlbrace}{[\mskip-5mu[}
\newcommand*{\xrbrace}{]\mskip-5mu]}

\newcommand*{\ldblbrace}{\{\mskip-5mu\{}
\newcommand*{\rdblbrace}{\}\mskip-5mu\}}

\usepackage{bm}
\usepackage{agda}
\usepackage{newunicodechar}

\usepackage{mathpartir}
\usepackage{varwidth}
\usepackage{microtype} 

\newcommand\codeblock[1]{%
  {\begin{varwidth}{0.9\textwidth}#1\end{varwidth}}
}

\newunicodechar{₀}{\ensuremath{_0}}
\newunicodechar{₁}{\ensuremath{_1}}
\newunicodechar{₂}{\ensuremath{_2}}
\newunicodechar{₃}{\ensuremath{_3}}
\newunicodechar{₄}{\ensuremath{_4}}
\newunicodechar{₅}{\ensuremath{_5}}
\newunicodechar{₆}{\ensuremath{_6}}
\newunicodechar{₇}{\ensuremath{_7}}
\newunicodechar{₈}{\ensuremath{_8}}
\newunicodechar{₉}{\ensuremath{_9}}
\newunicodechar{ι}{\ensuremath{\iota}}
\newunicodechar{ₗ}{\ensuremath{_l}}
\newunicodechar{ₐ}{\ensuremath{_a}}
\newunicodechar{ₖ}{\ensuremath{_k}}
\newunicodechar{ᵢ}{\ensuremath{_i}}
\newunicodechar{ⱼ}{\ensuremath{_j}}
\newunicodechar{ᵥ}{\ensuremath{_v}}
\newunicodechar{ₕ}{\ensuremath{_h}}
\newunicodechar{ₚ}{\ensuremath{_p}}
\newunicodechar{ₛ}{\ensuremath{_s}}
\newunicodechar{ₒ}{\ensuremath{_o}}

\newunicodechar{ᵣ}{\ensuremath{_r}}
\newunicodechar{ʳ}{\ensuremath{^r}}
\newunicodechar{ˡ}{\ensuremath{^l}}
\newunicodechar{ℕ}{\ensuremath{\mathbb{N}}}
\newunicodechar{ℝ}{\ensuremath{\mathbb{R}}}
\newunicodechar{∀}{\ensuremath{\forall}}
\newunicodechar{∃}{\ensuremath{\exists}}
\newunicodechar{≡}{\ensuremath{\equiv}}
\newunicodechar{≈}{\ensuremath{\approx}}
\newunicodechar{≟}{\ensuremath{\stackrel{?}{=}}}
\newunicodechar{∎}{\ensuremath{\blacksquare}}
\newunicodechar{⊛}{\ensuremath{\circledast}}
\newunicodechar{⊗}{\ensuremath{\otimes}}
\newunicodechar{⊕}{\ensuremath{\oplus}}
\newunicodechar{⊝}{\ensuremath{\ominus}}
\newunicodechar{≤}{\ensuremath{\le}}
\newunicodechar{φ}{\ensuremath{\phi}}
\newunicodechar{ψ}{\ensuremath{\psi}}
\newunicodechar{ε}{\ensuremath{\epsilon}}
\newunicodechar{δ}{\ensuremath{\delta}}
\newunicodechar{λ}{\ensuremath{\lambda}}
\newunicodechar{σ}{\ensuremath{\sigma}}
\newunicodechar{ρ}{\ensuremath{\rho}}
\newunicodechar{∂}{\ensuremath{\partial}}

\newunicodechar{Γ}{\ensuremath{\Gamma}}
\newunicodechar{Δ}{\ensuremath{\Delta}}
\newunicodechar{Ψ}{\ensuremath{\Phi}}
\newunicodechar{Ξ}{\ensuremath{\Xi}}

\newunicodechar{′}{{'}}
\newunicodechar{∷}{\ensuremath{\dblcolon}}
\newunicodechar{↔}{\ensuremath{\leftrightarrow}}
\newunicodechar{↦}{\ensuremath{\mapsto}}
\newunicodechar{∘}{\ensuremath{\circ}}
\newunicodechar{⁻}{\ensuremath{^{-}}}
\newunicodechar{∙}{\ensuremath{\boldsymbol{\cdot}}}
\newunicodechar{▹}{\ensuremath{\triangleright}}
\newunicodechar{⋯}{\ensuremath{\cdots}}
\newunicodechar{∇}{\ensuremath{\nabla}}
\newunicodechar{Σ}{\ensuremath{\Sigma}}
\newunicodechar{∈}{\ensuremath{\in}}
\newunicodechar{⟦}{\ensuremath{\llbracket}}
\newunicodechar{⟧}{\ensuremath{\rrbracket}}
\newunicodechar{⦃}{\ensuremath{\ldblbrace}}
\newunicodechar{⦄}{\ensuremath{\rdblbrace}}
\newunicodechar{⌊}{\ensuremath{\lfloor}}
\newunicodechar{⌋}{\ensuremath{\rfloor}}

\newunicodechar{⇒}{\ensuremath{\Rightarrow}}
\newunicodechar{⟪}{\ensuremath{\xlbrace}}
\newunicodechar{⟫}{\ensuremath{\xrbrace}}
\newunicodechar{⊠}{\ensuremath{\boxtimes}}
\newunicodechar{⊞}{\ensuremath{\boxplus}}
\newunicodechar{∋}{\ensuremath{\ni}}
\newunicodechar{∶}{\ensuremath{\bm{:}}}
\newunicodechar{↦}{\ensuremath{\mapsto}}
\newunicodechar{⊤}{\ensuremath{\top}}

\newcommand{\AD}[1]{\AgdaDatatype{#1}}
\newcommand{\AC}[1]{\AgdaInductiveConstructor{#1}}
\newcommand{\AF}[1]{\AgdaFunction{#1}}
\newcommand{\AB}[1]{\AgdaBound{#1}}

\renewcommand{\AgdaCommentFontStyle}[1]{\textrm{#1}}
\renewcommand{\AgdaFontStyle}[1]{\textrm{#1}}
\renewcommand{\AgdaKeywordFontStyle}[1]{\textrm{#1}}

\pgfplotstableset{col sep=comma}

\title{Correctness is Demanding, Performance is Frustrating}

\author{Artjoms {\v{S}}inkarovs}
\email{A.Sinkarovs@soton.ac.uk}
\orcid{0000-0003-3292-2985}
\affiliation{%
  \institution{University of Southampton}
  \streetaddress{University of Southampton, University Road}
  \city{Southampton}
  \country{UK}
  \postcode{SO17 1BJ}
}
\author{Thomas Koopman}
\email{Thomas.Koopman@ru.nl}
\orcid{0009-0001-1031-7226}
\affiliation{%
  \institution{Radboud University}
  \streetaddress{Houtlaan 4}
  \city{Nijmegen}
  \country{Netherlands}
  \postcode{6525 XZ}
}
\author{Sven-Bodo Scholz}
\email{SvenBodo.Scholz@ru.nl}
\orcid{0000-0002-8663-1043}
\affiliation{%
  \institution{Radboud University}
  \streetaddress{Houtlaan 4}
  \city{Nijmegen}
  \country{Netherlands}
  \postcode{6525 XZ}
}

\pgfkeys{/pgf/number format/.cd,fixed,precision=0}


\begin{document}

\begin{abstract}
In this paper we demonstrate a technique for developing high performance applications
with strong correctness guarantees.  We use a theorem prover to derive a high-level
specification of the application that includes correctness invariants of our choice.
After that, within the same theorem prover, we implement an extraction of the
specified application into a high-performance language of our choice.  Concretely,
we are using Agda to specify a framework for automatic differentiation (reverse mode)
that is focused on index-safe tensors.  This framework comes
with an optimiser for tensor expressions and the ability to translate these
expressions into SaC and C.  We specify a canonical convolutional neural network
within the proposed framework, compute the derivatives needed for the training
phase and then demonstrate that the generated code matches the performance of hand-written
code when running on a multi-core machine.
\end{abstract}

\maketitle

\section{Introduction\label{sec:intro}}

The year is 2024, and we still have to make a formidable choice between
correctness and performance for all the programming projects that we start.
Low level programming languages such as C or Fortran make it possible to leverage
intricate hardware features for the price of poor analysability and
correctness guarantees.  Dependently-typed systems such as Lean or Agda make it possible
to describe arbitrary invariants within the given program, yet they can
rarely generate high-performance code.

The idealist inside of us is exasperated, because there should be a perfect
solution that caters for both cases.  However, the practitioner within us
proposes a different perspective.
Instead of using a single language, we may use two languages
in cooperation.  Specifically, we envision an alliance between a
proof assistant and some high-performance language of our choice.

To explore this idea, we investigate the concrete problem of automatic differentiation (AD) which
is often found in machine learning applications.  This is a convenient case study
as it comes with the following challenges.  From the correctness perspective,
it is crucially important to track the shapes and ranks of the tensors,
guaranteeing the absence of out-of-bound indexing.
This is a very common source of errors that can be incredibly difficult
to find.   Secondly, we have to compute derivatives of the given tensor expressions,
preserve safe indexing guarantees while we do so, and we have to be able to translate
the computed expressions into some high-performance language.
As machine learning applications
are known to be numerically intensive problems, our performance challenge lies
in running the program as fast as we can on the chosen hardware architecture.

We follow~\cite{cnn-array} which demonstrates that it is possible to implement
one of the canonical convolutional neural network (CNN) in the array
language SaC~\cite{sac1, sac2}, obtaining good sequential and parallel performance
that is competitive with TensorFlow~\cite{ad-tf} and PyTorch~\cite{ad-pytorch}.
Focussing on correctness, we propose
a theory of rank-polymorphic arrays~\cite{rank-poly} in Agda~\cite{agda-2-6-3}.
Within this framework, we encode the CNN from~\cite{cnn-array} and lift it 
into an embedded DSL.  We implement AD
(reverse mode) and domain-specific optimisations for expressions in that DSL.
Finally, we implement an extraction 
into SaC (functional array language) and C (low-level imperative language).

As a result, we demonstrate an approach where the entire specification,
optimiser, AD and code generation are available to us within a proof
assistant of our choice.  We can prove facts about all the
stages of the pipeline and easily adjust them to our liking.
We argue that such a liberating
approach is feasible in practice, at least for the times of dialectic
of correctness and performance.

The contributions of this paper are as follows:
\begin{enumerate}
  \item a rank-polymorphic array theory in Agda;
  \item an implementation of the CNN from~\cite{cnn-array} in Agda;
  \item an embedded DSL in Agda which supports AD (reverse mode);
  \item an extraction mechanism for generating SaC or C code from the DSL; and
  \item an experimental evaluation of the generated codes.
\end{enumerate}

This paper is written in literate Agda, which guarantees that all the code
snippets have been type-checked.

\section{Background\label{sec:background}}

Automatic differentiation has been around for many decades~\cite{early-ad1, early-ad2},
so it is well-understood at a conceptual level.  However,
a number of questions related to bringing AD into the context of
programming languages remain open.  Recent successes in machine learning
have spurred further interest in AD which has led to several new developments.
For the context of this paper, we focus on recent work that contributes to 
the perspective of balancing correctness guarantees and performance.
Our selection here is by no means exhaustive, for
a broader scope we refer the reader to~\cite{autodiff-survey}.

There has been a number of programming-language-oriented approaches that explain
how to add AD to a programming language of choice. Examples of these include
Futhark~\cite{futhark/sc22ad}, Haskell~\cite{ad-haskell}, and
Jax~\cite{ad-jax}. Furthermore, a number of machine learning
frameworks that incorporate AD have been proposed in recent years: TensorFlow~\cite{ad-tf},
PyTorch~\cite{ad-pytorch}, MXNet~\cite{ad-mxnet} and many more.
While in particular the dedicated frameworks tend to find widespread 
acceptance by practitioners, both, correctness and performance leave
two open questions: (i) how is it possible to
ensure that the AD algorithm is implemented correctly? (ii) if the
language or the framework do not perform as expected, what are the
options to solve this?  Unfortunately, for many cases the answer to
both questions is unsatisfying.  Most of the languages/frameworks do not
come with formal correctness guarantees, so one has to trust the
implementers of these tools.  One can run tests as well to gain trust 
in the implementation but that is far from a 
formal guarantee.  If one relies
on the AD provided by a chosen language/framework, and the generated code does not
perform well, one has to modify the language/framework, as these solutions
are tightly integrated with the tools. The problem here is that most of of these tools
have very large and sophisticated implementations typically comprising
of hundreds of thousands of lines of code.  Furthermore, these systems
often rely on sophisticated 
combinations of sub-systems that need to be fine-tuned to the executing hardware.

Another line of work studies high-level principles of AD using
category theory~\cite{ad-theor1, ad-theor2, ad-theor3}.
While this indeed comes with great correctness guarantees due to
some naturality principles, it is not always clear how to implement
this in a way that leads to efficiently executable specifications.  Also, the
entire treatment of index-safe tensors is unclear.

In~\cite{ad-elliott} the author proposes to view AD problem using
the language of cartesian categories.  It has been shown that
this approach can be used in practice by implementing the proposed
technique in Haskell.  Type classes are a vehicle to restrict expressions
that are differentiable.  There is a Haskell plugin that translates
expressions that are instances of the mentioned type classes into
categorical primitives, AD is performed on these and the code is reflected
back to Haskell.  This is a nice approach that makes it easy
to verify the correctness of the algorithm.  However, the treatment
of tensors and general extractability remains a little unclear.
While it is briefly mentioned that representable functors
are supported, it is unclear whether this is sufficient to
represent rank-polymorphic arrays with strict bound checks.
Also, correctness guarantees are inevitably restricted by the
Haskell type system, so we are likely to find invariants that
are inexpressible in that setup.

\section{Array Theory\label{sec:array-theory}}

\begin{code}[hide]%
\>[0]\AgdaKeyword{open}\AgdaSpace{}%
\AgdaKeyword{import}\AgdaSpace{}%
\AgdaModule{Relation.Binary.PropositionalEquality}\<%
\\
\>[0]\AgdaKeyword{open}\AgdaSpace{}%
\AgdaKeyword{import}\AgdaSpace{}%
\AgdaModule{Relation.Nullary}\<%
\\
\>[0]\AgdaKeyword{open}\AgdaSpace{}%
\AgdaKeyword{import}\AgdaSpace{}%
\AgdaModule{Data.List}\AgdaSpace{}%
\AgdaKeyword{using}\AgdaSpace{}%
\AgdaSymbol{(}\AgdaDatatype{List}\AgdaSymbol{;}\AgdaSpace{}%
\AgdaInductiveConstructor{[]}\AgdaSymbol{;}\AgdaSpace{}%
\AgdaOperator{\AgdaInductiveConstructor{\AgdaUnderscore{}∷\AgdaUnderscore{}}}\AgdaSymbol{)}\<%
\\
\>[0]\AgdaKeyword{open}\AgdaSpace{}%
\AgdaKeyword{import}\AgdaSpace{}%
\AgdaModule{Data.Empty}\<%
\\
\>[0]\AgdaKeyword{open}\AgdaSpace{}%
\AgdaKeyword{import}\AgdaSpace{}%
\AgdaModule{Function}\<%
\\
\\[\AgdaEmptyExtraSkip]%
\>[0]\AgdaKeyword{module}\AgdaSpace{}%
\AgdaModule{\AgdaUnderscore{}}\AgdaSpace{}%
\AgdaKeyword{where}\<%
\\
\>[0]\AgdaKeyword{module}\AgdaSpace{}%
\AgdaModule{Array}\AgdaSpace{}%
\AgdaKeyword{where}\<%
\\
\>[0][@{}l@{\AgdaIndent{0}}]%
\>[2]\AgdaKeyword{open}\AgdaSpace{}%
\AgdaKeyword{import}\AgdaSpace{}%
\AgdaModule{Data.Nat}\AgdaSpace{}%
\AgdaKeyword{using}\AgdaSpace{}%
\AgdaSymbol{(}\AgdaInductiveConstructor{zero}\AgdaSymbol{;}\AgdaSpace{}%
\AgdaInductiveConstructor{suc}\AgdaSymbol{;}\AgdaSpace{}%
\AgdaDatatype{ℕ}\AgdaSymbol{;}\AgdaSpace{}%
\AgdaOperator{\AgdaPrimitive{\AgdaUnderscore{}+\AgdaUnderscore{}}}\AgdaSymbol{;}\AgdaSpace{}%
\AgdaOperator{\AgdaPrimitive{\AgdaUnderscore{}*\AgdaUnderscore{}}}\AgdaSymbol{;}\AgdaSpace{}%
\AgdaOperator{\AgdaDatatype{\AgdaUnderscore{}≤\AgdaUnderscore{}}}\AgdaSymbol{;}\AgdaSpace{}%
\AgdaInductiveConstructor{s≤s}\AgdaSymbol{;}\AgdaSpace{}%
\AgdaInductiveConstructor{z≤n}\AgdaSymbol{;}\AgdaSpace{}%
\AgdaOperator{\AgdaFunction{\AgdaUnderscore{}<\AgdaUnderscore{}}}\AgdaSymbol{)}\<%
\\
\>[2]\AgdaKeyword{open}\AgdaSpace{}%
\AgdaKeyword{import}\AgdaSpace{}%
\AgdaModule{Data.Nat.Properties}\AgdaSpace{}%
\AgdaKeyword{using}\AgdaSpace{}%
\AgdaSymbol{(}\AgdaFunction{+-mono-≤}\AgdaSymbol{;}\AgdaSpace{}%
\AgdaFunction{≤-step}\AgdaSymbol{;}\AgdaSpace{}%
\AgdaFunction{≤-pred}\AgdaSymbol{;}\AgdaSpace{}%
\AgdaOperator{\AgdaFunction{\AgdaUnderscore{}≟\AgdaUnderscore{}}}\AgdaSymbol{;}\AgdaSpace{}%
\AgdaFunction{+-comm}\AgdaSymbol{;}\AgdaSpace{}%
\AgdaFunction{+-suc}\AgdaSymbol{)}\<%
\\
\>[2]\AgdaKeyword{open}\AgdaSpace{}%
\AgdaKeyword{import}\AgdaSpace{}%
\AgdaModule{Data.Fin}\AgdaSpace{}%
\AgdaKeyword{using}\AgdaSpace{}%
\AgdaSymbol{(}\AgdaInductiveConstructor{zero}\AgdaSymbol{;}\AgdaSpace{}%
\AgdaInductiveConstructor{suc}\AgdaSymbol{;}\AgdaSpace{}%
\AgdaDatatype{Fin}\AgdaSymbol{;}\AgdaSpace{}%
\AgdaFunction{combine}\AgdaSymbol{;}\AgdaSpace{}%
\AgdaFunction{remQuot}\AgdaSymbol{;}\AgdaSpace{}%
\AgdaFunction{fromℕ<}\AgdaSymbol{;}\AgdaSpace{}%
\AgdaFunction{inject+}\AgdaSymbol{;}\AgdaSpace{}%
\AgdaFunction{splitAt}\AgdaSymbol{)}\<%
\\
\>[2]\AgdaKeyword{open}\AgdaSpace{}%
\AgdaKeyword{import}\AgdaSpace{}%
\AgdaModule{Data.Fin.Properties}\AgdaSpace{}%
\AgdaKeyword{using}\AgdaSpace{}%
\AgdaSymbol{(}\AgdaFunction{suc-injective}\AgdaSymbol{;}\AgdaSpace{}%
\AgdaFunction{toℕ<n}\AgdaSymbol{;}\AgdaSpace{}%
\AgdaFunction{splitAt-inject+}\AgdaSymbol{)}\<%
\\
\>[2]\AgdaComment{--open\ import\ Fin2\ using\ (Fin;\ \#\AgdaUnderscore{};\ combine;\ remQuot;\ zerof;\ sucf;\ \AgdaUnderscore{}⊕\AgdaUnderscore{};\ \AgdaUnderscore{}⊝\AgdaUnderscore{})}\<%
\\
\>[2]\AgdaKeyword{open}\AgdaSpace{}%
\AgdaKeyword{import}\AgdaSpace{}%
\AgdaModule{Data.Sum}\AgdaSpace{}%
\AgdaKeyword{using}\AgdaSpace{}%
\AgdaSymbol{(}\AgdaOperator{\AgdaDatatype{\AgdaUnderscore{}⊎\AgdaUnderscore{}}}\AgdaSymbol{;}\AgdaSpace{}%
\AgdaInductiveConstructor{inj₁}\AgdaSymbol{;}\AgdaSpace{}%
\AgdaInductiveConstructor{inj₂}\AgdaSymbol{)}\<%
\\
\>[2]\AgdaKeyword{open}\AgdaSpace{}%
\AgdaKeyword{import}\AgdaSpace{}%
\AgdaModule{Data.Product}\AgdaSpace{}%
\AgdaKeyword{using}\AgdaSpace{}%
\AgdaSymbol{(}\AgdaFunction{∃}\AgdaSymbol{;}\AgdaSpace{}%
\AgdaOperator{\AgdaInductiveConstructor{\AgdaUnderscore{},\AgdaUnderscore{}}}\AgdaSymbol{)}\<%
\end{code}

The central data structures of our case study are multi-dimensional
arrays.  This section is dedicated to defining a minimalist array theory
in Agda which is well-suited for a specification of CNNs.

We assume that the reader is sufficiently familiar with Agda's syntax.
For gentle introductions we refer to one of the tutorials that are freely available
online\footnote{See \url{https://agda.readthedocs.io/en/v2.6.3/getting-started/tutorial-list.html}.}.

The conciseness of the specification
in~\cite{cnn-array} relies on rank-polymorphism, which is the ability to operate
on arrays of arbitrary rank.  We capture this feature in in our array theory.
The central consideration when working with dependent types is how to represent data.
Some encodings are better suited for reasoning, others are more efficient
at runtime.  Due to our two-language setup, our choice of representation is
driven by proof considerations only.
This enables us to represent arrays as functions from indices to values.

An absence of out-of-bound errors requires that all array indices fall within
the shapes of the arrays that they are selecting from.
The shape of an array describes the extent of each of its axes.  We represent shapes
as binary trees of natural numbers using the data type \AD{S}.
Leaves of the shape tree are constructed with \AC{ι} which takes one
argument.  The \AC{\_⊗\_} constructor makes a tree of two sub-trees.
Note that underscores in \AC{\_⊗\_} specify the position where arguments
go, therefore \AC{⊗} is an infix binary operation.

Array positions (indices) are given by the dependent type \AD{P} which
is indexed by shapes.  A position within an array of shape \AB{s} has exactly the
same tree structure as \AB{s}, but the leaves are natural numbers that
are bounded by the corresponding shape elements.

Arrays are given by the data type \AF{Ar} which is indexed by a shape
and an element type.  The formal definitions of \AF{S}, \AF{P} and \AF{Ar} are
as follows:
\begin{mathpar}
\codeblock{\begin{code}%
\>[2]\AgdaKeyword{data}\AgdaSpace{}%
\AgdaDatatype{S}\AgdaSpace{}%
\AgdaSymbol{:}\AgdaSpace{}%
\AgdaPrimitive{Set}\AgdaSpace{}%
\AgdaKeyword{where}\<%
\\
\>[2][@{}l@{\AgdaIndent{0}}]%
\>[4]\AgdaInductiveConstructor{ι}%
\>[9]\AgdaSymbol{:}\AgdaSpace{}%
\AgdaDatatype{ℕ}\AgdaSpace{}%
\AgdaSymbol{→}\AgdaSpace{}%
\AgdaDatatype{S}\<%
\\
\>[4]\AgdaOperator{\AgdaInductiveConstructor{\AgdaUnderscore{}⊗\AgdaUnderscore{}}}%
\>[9]\AgdaSymbol{:}\AgdaSpace{}%
\AgdaDatatype{S}\AgdaSpace{}%
\AgdaSymbol{→}\AgdaSpace{}%
\AgdaDatatype{S}\AgdaSpace{}%
\AgdaSymbol{→}\AgdaSpace{}%
\AgdaDatatype{S}\<%
\end{code}
\begin{code}[hide]%
\>[2]\AgdaKeyword{variable}\<%
\\
\>[2][@{}l@{\AgdaIndent{0}}]%
\>[4]\AgdaGeneralizable{m}\AgdaSpace{}%
\AgdaGeneralizable{n}\AgdaSpace{}%
\AgdaGeneralizable{k}\AgdaSpace{}%
\AgdaSymbol{:}\AgdaSpace{}%
\AgdaDatatype{ℕ}\<%
\\
\>[4]\AgdaGeneralizable{s}\AgdaSpace{}%
\AgdaGeneralizable{p}\AgdaSpace{}%
\AgdaGeneralizable{q}\AgdaSpace{}%
\AgdaGeneralizable{r}\AgdaSpace{}%
\AgdaGeneralizable{u}\AgdaSpace{}%
\AgdaGeneralizable{w}\AgdaSpace{}%
\AgdaSymbol{:}\AgdaSpace{}%
\AgdaDatatype{S}\<%
\\
\>[4]\AgdaGeneralizable{X}\AgdaSpace{}%
\AgdaGeneralizable{Y}\AgdaSpace{}%
\AgdaGeneralizable{Z}\AgdaSpace{}%
\AgdaSymbol{:}\AgdaSpace{}%
\AgdaPrimitive{Set}\<%
\end{code}}
\and
\codeblock{\begin{code}%
\>[2]\AgdaKeyword{data}\AgdaSpace{}%
\AgdaDatatype{P}\AgdaSpace{}%
\AgdaSymbol{:}\AgdaSpace{}%
\AgdaDatatype{S}\AgdaSpace{}%
\AgdaSymbol{→}\AgdaSpace{}%
\AgdaPrimitive{Set}\AgdaSpace{}%
\AgdaKeyword{where}\<%
\\
\>[2][@{}l@{\AgdaIndent{0}}]%
\>[4]\AgdaInductiveConstructor{ι}%
\>[9]\AgdaSymbol{:}\AgdaSpace{}%
\AgdaDatatype{Fin}\AgdaSpace{}%
\AgdaGeneralizable{n}\AgdaSpace{}%
\AgdaSymbol{→}\AgdaSpace{}%
\AgdaDatatype{P}\AgdaSpace{}%
\AgdaSymbol{(}\AgdaInductiveConstructor{ι}\AgdaSpace{}%
\AgdaGeneralizable{n}\AgdaSymbol{)}\<%
\\
\>[4]\AgdaOperator{\AgdaInductiveConstructor{\AgdaUnderscore{}⊗\AgdaUnderscore{}}}%
\>[9]\AgdaSymbol{:}\AgdaSpace{}%
\AgdaDatatype{P}\AgdaSpace{}%
\AgdaGeneralizable{s}\AgdaSpace{}%
\AgdaSymbol{→}\AgdaSpace{}%
\AgdaDatatype{P}\AgdaSpace{}%
\AgdaGeneralizable{p}\AgdaSpace{}%
\AgdaSymbol{→}\AgdaSpace{}%
\AgdaDatatype{P}\AgdaSpace{}%
\AgdaSymbol{(}\AgdaGeneralizable{s}\AgdaSpace{}%
\AgdaOperator{\AgdaInductiveConstructor{⊗}}\AgdaSpace{}%
\AgdaGeneralizable{p}\AgdaSymbol{)}\<%
\end{code}}
\and
\codeblock{\begin{code}%
\>[2]\AgdaFunction{Ar}\AgdaSpace{}%
\AgdaSymbol{:}\AgdaSpace{}%
\AgdaDatatype{S}\AgdaSpace{}%
\AgdaSymbol{→}\AgdaSpace{}%
\AgdaPrimitive{Set}\AgdaSpace{}%
\AgdaSymbol{→}\AgdaSpace{}%
\AgdaPrimitive{Set}\<%
\\
\>[2]\AgdaFunction{Ar}\AgdaSpace{}%
\AgdaBound{s}\AgdaSpace{}%
\AgdaBound{X}\AgdaSpace{}%
\AgdaSymbol{=}\AgdaSpace{}%
\AgdaDatatype{P}\AgdaSpace{}%
\AgdaBound{s}\AgdaSpace{}%
\AgdaSymbol{→}\AgdaSpace{}%
\AgdaBound{X}\<%
\end{code}}
\end{mathpar}
As arrays are functions, selections are function applications and
array construction is a function definition (\eg{} via $\lambda$-abstraction).

\paragraph{Array Combinators} It is helpful to invest a little time
in defining array combinators.  First, we can observe that \AD{Ar} of
a fixed shape is an applicative functor~\cite{applicative}, so we can trivially derive:
\AF{K}\ \AB{x} to produce a constant array; \AF{map}\ \AB{f}\ \AB{a}
to apply \AB{f} to all the elements of \AB{a}; and \AF{zipWith}\ \AB{f}
\ \AB{a}\ \AB{b} to point-wise apply the binary operation 
\AB{f} to \AB{a} and \AB{b}.
\begin{mathpar}
\codeblock{\begin{code}%
\>[2]\AgdaFunction{K}\AgdaSpace{}%
\AgdaSymbol{:}\AgdaSpace{}%
\AgdaGeneralizable{X}\AgdaSpace{}%
\AgdaSymbol{→}\AgdaSpace{}%
\AgdaFunction{Ar}\AgdaSpace{}%
\AgdaGeneralizable{s}\AgdaSpace{}%
\AgdaGeneralizable{X}\<%
\\
\>[2]\AgdaFunction{K}\AgdaSpace{}%
\AgdaBound{x}\AgdaSpace{}%
\AgdaBound{i}\AgdaSpace{}%
\AgdaSymbol{=}\AgdaSpace{}%
\AgdaBound{x}\<%
\end{code}}
\and
\codeblock{\begin{code}%
\>[2]\AgdaFunction{map}\AgdaSpace{}%
\AgdaSymbol{:}\AgdaSpace{}%
\AgdaSymbol{(}\AgdaGeneralizable{X}\AgdaSpace{}%
\AgdaSymbol{→}\AgdaSpace{}%
\AgdaGeneralizable{Y}\AgdaSymbol{)}\AgdaSpace{}%
\AgdaSymbol{→}\AgdaSpace{}%
\AgdaFunction{Ar}\AgdaSpace{}%
\AgdaGeneralizable{s}\AgdaSpace{}%
\AgdaGeneralizable{X}\AgdaSpace{}%
\AgdaSymbol{→}\AgdaSpace{}%
\AgdaFunction{Ar}\AgdaSpace{}%
\AgdaGeneralizable{s}\AgdaSpace{}%
\AgdaGeneralizable{Y}\<%
\\
\>[2]\AgdaFunction{map}\AgdaSpace{}%
\AgdaBound{f}\AgdaSpace{}%
\AgdaBound{a}\AgdaSpace{}%
\AgdaBound{i}\AgdaSpace{}%
\AgdaSymbol{=}\AgdaSpace{}%
\AgdaBound{f}\AgdaSpace{}%
\AgdaSymbol{(}\AgdaBound{a}\AgdaSpace{}%
\AgdaBound{i}\AgdaSymbol{)}\<%
\end{code}}
\and
\codeblock{\begin{code}%
\>[2]\AgdaFunction{zipWith}\AgdaSpace{}%
\AgdaSymbol{:}\AgdaSpace{}%
\AgdaSymbol{(}\AgdaGeneralizable{X}\AgdaSpace{}%
\AgdaSymbol{→}\AgdaSpace{}%
\AgdaGeneralizable{Y}\AgdaSpace{}%
\AgdaSymbol{→}\AgdaSpace{}%
\AgdaGeneralizable{Z}\AgdaSymbol{)}\AgdaSpace{}%
\AgdaSymbol{→}\AgdaSpace{}%
\AgdaFunction{Ar}\AgdaSpace{}%
\AgdaGeneralizable{s}\AgdaSpace{}%
\AgdaGeneralizable{X}\AgdaSpace{}%
\AgdaSymbol{→}\AgdaSpace{}%
\AgdaFunction{Ar}\AgdaSpace{}%
\AgdaGeneralizable{s}\AgdaSpace{}%
\AgdaGeneralizable{Y}\AgdaSpace{}%
\AgdaSymbol{→}\AgdaSpace{}%
\AgdaFunction{Ar}\AgdaSpace{}%
\AgdaGeneralizable{s}\AgdaSpace{}%
\AgdaGeneralizable{Z}\<%
\\
\>[2]\AgdaFunction{zipWith}\AgdaSpace{}%
\AgdaBound{f}\AgdaSpace{}%
\AgdaBound{a}\AgdaSpace{}%
\AgdaBound{b}\AgdaSpace{}%
\AgdaBound{i}\AgdaSpace{}%
\AgdaSymbol{=}\AgdaSpace{}%
\AgdaBound{f}\AgdaSpace{}%
\AgdaSymbol{(}\AgdaBound{a}\AgdaSpace{}%
\AgdaBound{i}\AgdaSymbol{)}\AgdaSpace{}%
\AgdaSymbol{(}\AgdaBound{b}\AgdaSpace{}%
\AgdaBound{i}\AgdaSymbol{)}\<%
\end{code}}
\end{mathpar}

Arrays are homogeneously nested, \ie{} the shapes of all the sub-arrays
have to be the same.  Therefore, we can switch between the array of a product
shape and the nested array (array of arrays).  This operation is very similar
to currying except it happens at the level of shapes.  The combinators that
achieve this are named \AF{nest} and \AF{unnest} and their definitions are:
\begin{mathpar}
\codeblock{\begin{code}%
\>[2]\AgdaFunction{nest}\AgdaSpace{}%
\AgdaSymbol{:}\AgdaSpace{}%
\AgdaFunction{Ar}\AgdaSpace{}%
\AgdaSymbol{(}\AgdaGeneralizable{s}\AgdaSpace{}%
\AgdaOperator{\AgdaInductiveConstructor{⊗}}\AgdaSpace{}%
\AgdaGeneralizable{p}\AgdaSymbol{)}\AgdaSpace{}%
\AgdaGeneralizable{X}\AgdaSpace{}%
\AgdaSymbol{→}\AgdaSpace{}%
\AgdaFunction{Ar}\AgdaSpace{}%
\AgdaGeneralizable{s}\AgdaSpace{}%
\AgdaSymbol{(}\AgdaFunction{Ar}\AgdaSpace{}%
\AgdaGeneralizable{p}\AgdaSpace{}%
\AgdaGeneralizable{X}\AgdaSymbol{)}\<%
\\
\>[2]\AgdaFunction{nest}\AgdaSpace{}%
\AgdaBound{a}\AgdaSpace{}%
\AgdaBound{i}\AgdaSpace{}%
\AgdaBound{j}\AgdaSpace{}%
\AgdaSymbol{=}\AgdaSpace{}%
\AgdaBound{a}\AgdaSpace{}%
\AgdaSymbol{(}\AgdaBound{i}\AgdaSpace{}%
\AgdaOperator{\AgdaInductiveConstructor{⊗}}\AgdaSpace{}%
\AgdaBound{j}\AgdaSymbol{)}\<%
\end{code}}
\and
\codeblock{\begin{code} %
\>[2]\AgdaFunction{unnest}\AgdaSpace{}%
\AgdaSymbol{:}\AgdaSpace{}%
\AgdaFunction{Ar}\AgdaSpace{}%
\AgdaGeneralizable{s}\AgdaSpace{}%
\AgdaSymbol{(}\AgdaFunction{Ar}\AgdaSpace{}%
\AgdaGeneralizable{p}\AgdaSpace{}%
\AgdaGeneralizable{X}\AgdaSymbol{)}\AgdaSpace{}%
\AgdaSymbol{→}\AgdaSpace{}%
\AgdaFunction{Ar}\AgdaSpace{}%
\AgdaSymbol{(}\AgdaGeneralizable{s}\AgdaSpace{}%
\AgdaOperator{\AgdaInductiveConstructor{⊗}}\AgdaSpace{}%
\AgdaGeneralizable{p}\AgdaSymbol{)}\AgdaSpace{}%
\AgdaGeneralizable{X}\<%
\\
\>[2]\AgdaFunction{unnest}\AgdaSpace{}%
\AgdaBound{a}\AgdaSpace{}%
\AgdaSymbol{(}\AgdaBound{i}\AgdaSpace{}%
\AgdaOperator{\AgdaInductiveConstructor{⊗}}\AgdaSpace{}%
\AgdaBound{j}\AgdaSymbol{)}\AgdaSpace{}%
\AgdaSymbol{=}\AgdaSpace{}%
\AgdaBound{a}\AgdaSpace{}%
\AgdaBound{i}\AgdaSpace{}%
\AgdaBound{j}\<%
\end{code}}
\end{mathpar}

\paragraph{Reduction} We implement reduction of the binary operations
over arrays in two steps.  Firstly, we define 1-d reductions  that
we call \AD{sum₁} which is very similar to right fold on lists.
The arrays of higher ranks iterate \AF{sum₁} bottom-up.  The definition
of the primitives are as follows:
\begin{mathpar}
\codeblock{\begin{code}%
\>[2]\AgdaFunction{ιsuc}\AgdaSpace{}%
\AgdaSymbol{:}\AgdaSpace{}%
\AgdaDatatype{P}\AgdaSpace{}%
\AgdaSymbol{(}\AgdaInductiveConstructor{ι}\AgdaSpace{}%
\AgdaGeneralizable{n}\AgdaSymbol{)}\AgdaSpace{}%
\AgdaSymbol{→}\AgdaSpace{}%
\AgdaDatatype{P}\AgdaSpace{}%
\AgdaSymbol{(}\AgdaInductiveConstructor{ι}\AgdaSpace{}%
\AgdaSymbol{(}\AgdaInductiveConstructor{suc}\AgdaSpace{}%
\AgdaGeneralizable{n}\AgdaSymbol{))}\<%
\\
\>[2]\AgdaFunction{ιsuc}\AgdaSpace{}%
\AgdaSymbol{(}\AgdaInductiveConstructor{ι}\AgdaSpace{}%
\AgdaBound{i}\AgdaSymbol{)}\AgdaSpace{}%
\AgdaSymbol{=}\AgdaSpace{}%
\AgdaInductiveConstructor{ι}\AgdaSpace{}%
\AgdaSymbol{(}\AgdaInductiveConstructor{suc}\AgdaSpace{}%
\AgdaBound{i}\AgdaSymbol{)}\<%
\end{code}}
\and
\codeblock{\begin{code}%
\>[2]\AgdaFunction{sum₁}\AgdaSpace{}%
\AgdaSymbol{:}\AgdaSpace{}%
\AgdaSymbol{(}\AgdaGeneralizable{X}\AgdaSpace{}%
\AgdaSymbol{→}\AgdaSpace{}%
\AgdaGeneralizable{X}\AgdaSpace{}%
\AgdaSymbol{→}\AgdaSpace{}%
\AgdaGeneralizable{X}\AgdaSymbol{)}\AgdaSpace{}%
\AgdaSymbol{→}\AgdaSpace{}%
\AgdaGeneralizable{X}\AgdaSpace{}%
\AgdaSymbol{→}\AgdaSpace{}%
\AgdaFunction{Ar}\AgdaSpace{}%
\AgdaSymbol{(}\AgdaInductiveConstructor{ι}\AgdaSpace{}%
\AgdaGeneralizable{n}\AgdaSymbol{)}\AgdaSpace{}%
\AgdaGeneralizable{X}\AgdaSpace{}%
\AgdaSymbol{→}\AgdaSpace{}%
\AgdaGeneralizable{X}\<%
\\
\>[2]\AgdaFunction{sum₁}\AgdaSpace{}%
\AgdaSymbol{\{}\AgdaArgument{n}\AgdaSpace{}%
\AgdaSymbol{=}\AgdaSpace{}%
\AgdaInductiveConstructor{zero}\AgdaSymbol{\}}%
\>[20]\AgdaBound{f}\AgdaSpace{}%
\AgdaBound{ε}\AgdaSpace{}%
\AgdaBound{a}\AgdaSpace{}%
\AgdaSymbol{=}\AgdaSpace{}%
\AgdaBound{ε}\<%
\\
\>[2]\AgdaFunction{sum₁}\AgdaSpace{}%
\AgdaSymbol{\{}\AgdaArgument{n}\AgdaSpace{}%
\AgdaSymbol{=}\AgdaSpace{}%
\AgdaInductiveConstructor{suc}\AgdaSpace{}%
\AgdaBound{n}\AgdaSymbol{\}}%
\>[20]\AgdaBound{f}\AgdaSpace{}%
\AgdaBound{ε}\AgdaSpace{}%
\AgdaBound{a}\AgdaSpace{}%
\AgdaSymbol{=}\AgdaSpace{}%
\AgdaBound{f}\AgdaSpace{}%
\AgdaSymbol{(}\AgdaBound{a}\AgdaSpace{}%
\AgdaSymbol{(}\AgdaInductiveConstructor{ι}\AgdaSpace{}%
\AgdaInductiveConstructor{zero}\AgdaSymbol{))}\AgdaSpace{}%
\AgdaSymbol{(}\AgdaFunction{sum₁}\AgdaSpace{}%
\AgdaBound{f}\AgdaSpace{}%
\AgdaBound{ε}\AgdaSpace{}%
\AgdaSymbol{(}\AgdaBound{a}\AgdaSpace{}%
\AgdaOperator{\AgdaFunction{∘}}\AgdaSpace{}%
\AgdaFunction{ιsuc}\AgdaSymbol{))}\<%
\end{code}}
\and
\codeblock{\begin{code}%
\>[2]\AgdaFunction{sum}\AgdaSpace{}%
\AgdaSymbol{:}\AgdaSpace{}%
\AgdaSymbol{(}\AgdaGeneralizable{X}\AgdaSpace{}%
\AgdaSymbol{→}\AgdaSpace{}%
\AgdaGeneralizable{X}\AgdaSpace{}%
\AgdaSymbol{→}\AgdaSpace{}%
\AgdaGeneralizable{X}\AgdaSymbol{)}\AgdaSpace{}%
\AgdaSymbol{→}\AgdaSpace{}%
\AgdaGeneralizable{X}\AgdaSpace{}%
\AgdaSymbol{→}\AgdaSpace{}%
\AgdaFunction{Ar}\AgdaSpace{}%
\AgdaGeneralizable{s}\AgdaSpace{}%
\AgdaGeneralizable{X}\AgdaSpace{}%
\AgdaSymbol{→}\AgdaSpace{}%
\AgdaGeneralizable{X}\<%
\\
\>[2]\AgdaFunction{sum}\AgdaSpace{}%
\AgdaSymbol{\{}\AgdaArgument{s}\AgdaSpace{}%
\AgdaSymbol{=}\AgdaSpace{}%
\AgdaInductiveConstructor{ι}\AgdaSpace{}%
\AgdaBound{n}\AgdaSymbol{\}}%
\>[19]\AgdaBound{f}\AgdaSpace{}%
\AgdaBound{ε}\AgdaSpace{}%
\AgdaBound{a}\AgdaSpace{}%
\AgdaSymbol{=}\AgdaSpace{}%
\AgdaFunction{sum₁}\AgdaSpace{}%
\AgdaBound{f}\AgdaSpace{}%
\AgdaBound{ε}\AgdaSpace{}%
\AgdaBound{a}\<%
\\
\>[2]\AgdaFunction{sum}\AgdaSpace{}%
\AgdaSymbol{\{}\AgdaArgument{s}\AgdaSpace{}%
\AgdaSymbol{=}\AgdaSpace{}%
\AgdaBound{s}\AgdaSpace{}%
\AgdaOperator{\AgdaInductiveConstructor{⊗}}\AgdaSpace{}%
\AgdaBound{p}\AgdaSymbol{\}}%
\>[19]\AgdaBound{f}\AgdaSpace{}%
\AgdaBound{ε}\AgdaSpace{}%
\AgdaBound{a}\AgdaSpace{}%
\AgdaSymbol{=}\AgdaSpace{}%
\AgdaFunction{sum}\AgdaSpace{}%
\AgdaBound{f}\AgdaSpace{}%
\AgdaBound{ε}\AgdaSpace{}%
\AgdaOperator{\AgdaFunction{\$}}\AgdaSpace{}%
\AgdaFunction{map}\AgdaSpace{}%
\AgdaSymbol{(}\AgdaFunction{sum}\AgdaSpace{}%
\AgdaBound{f}\AgdaSpace{}%
\AgdaBound{ε}\AgdaSymbol{)}\AgdaSpace{}%
\AgdaSymbol{(}\AgdaFunction{nest}\AgdaSpace{}%
\AgdaBound{a}\AgdaSymbol{)}\<%
\end{code}}
\end{mathpar}

Note that our reduction forces the types of the arguments of the binary
operation to be the same, which is different from the usual foldr definition.
While we do not need this functionality for our example,
it is worth noting that the standard behaviour can be recovered\footnote{
We recover regular fold behaviour by running \AD{sum} over function composition:
\begin{code}%
\>[2]\AgdaFunction{sum′}\AgdaSpace{}%
\AgdaSymbol{:}\AgdaSpace{}%
\AgdaSymbol{(}\AgdaGeneralizable{X}\AgdaSpace{}%
\AgdaSymbol{→}\AgdaSpace{}%
\AgdaGeneralizable{Y}\AgdaSpace{}%
\AgdaSymbol{→}\AgdaSpace{}%
\AgdaGeneralizable{Y}\AgdaSymbol{)}\AgdaSpace{}%
\AgdaSymbol{→}\AgdaSpace{}%
\AgdaGeneralizable{Y}\AgdaSpace{}%
\AgdaSymbol{→}\AgdaSpace{}%
\AgdaFunction{Ar}\AgdaSpace{}%
\AgdaGeneralizable{s}\AgdaSpace{}%
\AgdaGeneralizable{X}\AgdaSpace{}%
\AgdaSymbol{→}\AgdaSpace{}%
\AgdaGeneralizable{Y}\<%
\\
\>[2]\AgdaFunction{sum′}\AgdaSpace{}%
\AgdaBound{f}\AgdaSpace{}%
\AgdaBound{ε}\AgdaSpace{}%
\AgdaBound{a}\AgdaSpace{}%
\AgdaSymbol{=}\AgdaSpace{}%
\AgdaFunction{sum}\AgdaSpace{}%
\AgdaOperator{\AgdaFunction{\AgdaUnderscore{}∘′\AgdaUnderscore{}}}\AgdaSpace{}%
\AgdaFunction{id}\AgdaSpace{}%
\AgdaSymbol{(}\AgdaFunction{map}\AgdaSpace{}%
\AgdaBound{f}\AgdaSpace{}%
\AgdaBound{a}\AgdaSymbol{)}\AgdaSpace{}%
\AgdaBound{ε}\<%
\end{code}
} through reduction of function composition.

\paragraph{Reshaping}
One common operation on arrays is element-preserving change of shape.  We call
such an operation \AF{reshape}.  It is clear that array elements can be preserved only in
cases when the number of elements in the original array and the reshaped one
is the same.  We define an inductive relation \AF{Reshape} that relates
only those shapes that preserve the number of array elements.  
\begin{code}[hide]%
\>[2]\AgdaKeyword{infixr}\AgdaSpace{}%
\AgdaNumber{5}\AgdaSpace{}%
\AgdaOperator{\AgdaInductiveConstructor{\AgdaUnderscore{}∙\AgdaUnderscore{}}}\<%
\\
\>[2]\AgdaComment{--infixl\ 10\ \AgdaUnderscore{},\AgdaUnderscore{}}\<%
\end{code}
\begin{mathpar}
\codeblock{\begin{code}%
\>[2]\AgdaKeyword{data}\AgdaSpace{}%
\AgdaDatatype{Reshape}\AgdaSpace{}%
\AgdaSymbol{:}\AgdaSpace{}%
\AgdaDatatype{S}\AgdaSpace{}%
\AgdaSymbol{→}\AgdaSpace{}%
\AgdaDatatype{S}\AgdaSpace{}%
\AgdaSymbol{→}\AgdaSpace{}%
\AgdaPrimitive{Set}\AgdaSpace{}%
\AgdaKeyword{where}\<%
\\
\>[2][@{}l@{\AgdaIndent{0}}]%
\>[4]\AgdaInductiveConstructor{eq}%
\>[12]\AgdaSymbol{:}\AgdaSpace{}%
\AgdaDatatype{Reshape}\AgdaSpace{}%
\AgdaGeneralizable{s}\AgdaSpace{}%
\AgdaGeneralizable{s}\<%
\\
\>[4]\AgdaOperator{\AgdaInductiveConstructor{\AgdaUnderscore{}∙\AgdaUnderscore{}}}%
\>[12]\AgdaSymbol{:}\AgdaSpace{}%
\AgdaDatatype{Reshape}\AgdaSpace{}%
\AgdaGeneralizable{p}\AgdaSpace{}%
\AgdaGeneralizable{q}\AgdaSpace{}%
\AgdaSymbol{→}\AgdaSpace{}%
\AgdaDatatype{Reshape}\AgdaSpace{}%
\AgdaGeneralizable{s}\AgdaSpace{}%
\AgdaGeneralizable{p}\AgdaSpace{}%
\AgdaSymbol{→}\AgdaSpace{}%
\AgdaDatatype{Reshape}\AgdaSpace{}%
\AgdaGeneralizable{s}\AgdaSpace{}%
\AgdaGeneralizable{q}\<%
\\
\>[4]\AgdaOperator{\AgdaInductiveConstructor{\AgdaUnderscore{},\AgdaUnderscore{}}}%
\>[12]\AgdaSymbol{:}\AgdaSpace{}%
\AgdaDatatype{Reshape}\AgdaSpace{}%
\AgdaGeneralizable{s}\AgdaSpace{}%
\AgdaGeneralizable{p}\AgdaSpace{}%
\AgdaSymbol{→}\AgdaSpace{}%
\AgdaDatatype{Reshape}\AgdaSpace{}%
\AgdaGeneralizable{q}\AgdaSpace{}%
\AgdaGeneralizable{r}\AgdaSpace{}%
\AgdaSymbol{→}\AgdaSpace{}%
\AgdaDatatype{Reshape}\AgdaSpace{}%
\AgdaSymbol{(}\AgdaGeneralizable{s}\AgdaSpace{}%
\AgdaOperator{\AgdaInductiveConstructor{⊗}}\AgdaSpace{}%
\AgdaGeneralizable{q}\AgdaSymbol{)}\AgdaSpace{}%
\AgdaSymbol{(}\AgdaGeneralizable{p}\AgdaSpace{}%
\AgdaOperator{\AgdaInductiveConstructor{⊗}}\AgdaSpace{}%
\AgdaGeneralizable{r}\AgdaSymbol{)}\<%
\\
\>[4]\AgdaInductiveConstructor{split}%
\>[12]\AgdaSymbol{:}\AgdaSpace{}%
\AgdaDatatype{Reshape}\AgdaSpace{}%
\AgdaSymbol{(}\AgdaInductiveConstructor{ι}\AgdaSpace{}%
\AgdaSymbol{(}\AgdaGeneralizable{m}\AgdaSpace{}%
\AgdaOperator{\AgdaPrimitive{*}}\AgdaSpace{}%
\AgdaGeneralizable{n}\AgdaSymbol{))}\AgdaSpace{}%
\AgdaSymbol{(}\AgdaInductiveConstructor{ι}\AgdaSpace{}%
\AgdaGeneralizable{m}\AgdaSpace{}%
\AgdaOperator{\AgdaInductiveConstructor{⊗}}\AgdaSpace{}%
\AgdaInductiveConstructor{ι}\AgdaSpace{}%
\AgdaGeneralizable{n}\AgdaSymbol{)}\<%
\\
\>[4]\AgdaInductiveConstructor{flat}%
\>[12]\AgdaSymbol{:}\AgdaSpace{}%
\AgdaDatatype{Reshape}\AgdaSpace{}%
\AgdaSymbol{(}\AgdaInductiveConstructor{ι}\AgdaSpace{}%
\AgdaGeneralizable{m}\AgdaSpace{}%
\AgdaOperator{\AgdaInductiveConstructor{⊗}}\AgdaSpace{}%
\AgdaInductiveConstructor{ι}\AgdaSpace{}%
\AgdaGeneralizable{n}\AgdaSymbol{)}\AgdaSpace{}%
\AgdaSymbol{(}\AgdaInductiveConstructor{ι}\AgdaSpace{}%
\AgdaSymbol{(}\AgdaGeneralizable{m}\AgdaSpace{}%
\AgdaOperator{\AgdaPrimitive{*}}\AgdaSpace{}%
\AgdaGeneralizable{n}\AgdaSymbol{))}\<%
\\
\>[4]\AgdaInductiveConstructor{swap}%
\>[12]\AgdaSymbol{:}\AgdaSpace{}%
\AgdaDatatype{Reshape}\AgdaSpace{}%
\AgdaSymbol{(}\AgdaGeneralizable{s}\AgdaSpace{}%
\AgdaOperator{\AgdaInductiveConstructor{⊗}}\AgdaSpace{}%
\AgdaGeneralizable{p}\AgdaSymbol{)}\AgdaSpace{}%
\AgdaSymbol{(}\AgdaGeneralizable{p}\AgdaSpace{}%
\AgdaOperator{\AgdaInductiveConstructor{⊗}}\AgdaSpace{}%
\AgdaGeneralizable{s}\AgdaSymbol{)}\<%
\\
\>[4]\AgdaInductiveConstructor{assocl}%
\>[12]\AgdaSymbol{:}\AgdaSpace{}%
\AgdaDatatype{Reshape}\AgdaSpace{}%
\AgdaSymbol{(}\AgdaGeneralizable{s}\AgdaSpace{}%
\AgdaOperator{\AgdaInductiveConstructor{⊗}}\AgdaSpace{}%
\AgdaSymbol{(}\AgdaGeneralizable{p}\AgdaSpace{}%
\AgdaOperator{\AgdaInductiveConstructor{⊗}}\AgdaSpace{}%
\AgdaGeneralizable{q}\AgdaSymbol{))}\AgdaSpace{}%
\AgdaSymbol{((}\AgdaGeneralizable{s}\AgdaSpace{}%
\AgdaOperator{\AgdaInductiveConstructor{⊗}}\AgdaSpace{}%
\AgdaGeneralizable{p}\AgdaSymbol{)}\AgdaSpace{}%
\AgdaOperator{\AgdaInductiveConstructor{⊗}}\AgdaSpace{}%
\AgdaGeneralizable{q}\AgdaSymbol{)}\<%
\\
\>[4]\AgdaInductiveConstructor{assocr}%
\>[12]\AgdaSymbol{:}\AgdaSpace{}%
\AgdaDatatype{Reshape}\AgdaSpace{}%
\AgdaSymbol{((}\AgdaGeneralizable{s}\AgdaSpace{}%
\AgdaOperator{\AgdaInductiveConstructor{⊗}}\AgdaSpace{}%
\AgdaGeneralizable{p}\AgdaSymbol{)}\AgdaSpace{}%
\AgdaOperator{\AgdaInductiveConstructor{⊗}}\AgdaSpace{}%
\AgdaGeneralizable{q}\AgdaSymbol{)}\AgdaSpace{}%
\AgdaSymbol{(}\AgdaGeneralizable{s}\AgdaSpace{}%
\AgdaOperator{\AgdaInductiveConstructor{⊗}}\AgdaSpace{}%
\AgdaSymbol{(}\AgdaGeneralizable{p}\AgdaSpace{}%
\AgdaOperator{\AgdaInductiveConstructor{⊗}}\AgdaSpace{}%
\AgdaGeneralizable{q}\AgdaSymbol{))}\<%
\end{code}}
\end{mathpar}
Any expression $r$ of
the type (\AF{Reshape} \AB{s} \AB{p}) comes with a straight-forward action on
indices that we denote \AF{\_⟨\_⟩} (its definition is omitted).
Such a (contravariant) action translates
the index within the shape \AB{p} into the index within the shape \AB{s}.
Given this translation, we can easily define \AF{reshape} as shown below.
\AF{Reshape} is constructed such that if $s$ and $p$ are related, then 
$p$ and $s$ are related too.  This fact is given by the function \AF{rev}
(its definition is omitted) and it immediately implies that all the
actions on indices as well as array \AF{reshape}s are invertible.

Note that two shapes can be related by \AF{Reshape} in more than
one way, which results in different array reshapes.  
For example, consider \AF{Reshape} (\AC{ι} 5 \AC{⊗} \AC{ι} 4) (\AC{ι} 5 \AC{⊗} \AC{ι} 4)
given by \AC{swap} or through (\AC{split} \AC{∙} \AC{flat}).  While the former transposes 
the array elements, the latter does not.
\begin{mathpar}
\codeblock{\begin{code}%
\>[2]\AgdaOperator{\AgdaFunction{\AgdaUnderscore{}⟨\AgdaUnderscore{}⟩}}\AgdaSpace{}%
\AgdaSymbol{:}\AgdaSpace{}%
\AgdaDatatype{P}\AgdaSpace{}%
\AgdaGeneralizable{p}\AgdaSpace{}%
\AgdaSymbol{→}\AgdaSpace{}%
\AgdaDatatype{Reshape}\AgdaSpace{}%
\AgdaGeneralizable{s}\AgdaSpace{}%
\AgdaGeneralizable{p}\AgdaSpace{}%
\AgdaSymbol{→}\AgdaSpace{}%
\AgdaDatatype{P}\AgdaSpace{}%
\AgdaGeneralizable{s}\<%
\end{code}}
\and
\codeblock{\begin{code}%
\>[2]\AgdaFunction{reshape}\AgdaSpace{}%
\AgdaSymbol{:}\AgdaSpace{}%
\AgdaDatatype{Reshape}\AgdaSpace{}%
\AgdaGeneralizable{s}\AgdaSpace{}%
\AgdaGeneralizable{p}\AgdaSpace{}%
\AgdaSymbol{→}\AgdaSpace{}%
\AgdaFunction{Ar}\AgdaSpace{}%
\AgdaGeneralizable{s}\AgdaSpace{}%
\AgdaGeneralizable{X}\AgdaSpace{}%
\AgdaSymbol{→}\AgdaSpace{}%
\AgdaFunction{Ar}\AgdaSpace{}%
\AgdaGeneralizable{p}\AgdaSpace{}%
\AgdaGeneralizable{X}\<%
\\
\>[2]\AgdaFunction{reshape}\AgdaSpace{}%
\AgdaBound{r}\AgdaSpace{}%
\AgdaBound{a}\AgdaSpace{}%
\AgdaSymbol{=}\AgdaSpace{}%
\AgdaSymbol{λ}\AgdaSpace{}%
\AgdaBound{ix}\AgdaSpace{}%
\AgdaSymbol{→}\AgdaSpace{}%
\AgdaBound{a}\AgdaSpace{}%
\AgdaSymbol{(}\AgdaBound{ix}\AgdaSpace{}%
\AgdaOperator{\AgdaFunction{⟨}}\AgdaSpace{}%
\AgdaBound{r}\AgdaSpace{}%
\AgdaOperator{\AgdaFunction{⟩}}\AgdaSymbol{)}\<%
\end{code}}
\and
\codeblock{\begin{code}%
\>[2]\AgdaFunction{rev}\AgdaSpace{}%
\AgdaSymbol{:}\AgdaSpace{}%
\AgdaDatatype{Reshape}\AgdaSpace{}%
\AgdaGeneralizable{s}\AgdaSpace{}%
\AgdaGeneralizable{p}\AgdaSpace{}%
\AgdaSymbol{→}\AgdaSpace{}%
\AgdaDatatype{Reshape}\AgdaSpace{}%
\AgdaGeneralizable{p}\AgdaSpace{}%
\AgdaGeneralizable{s}\<%
\end{code}}
\end{mathpar}
From the perspective of category theory, if \AF{S} is an object then \AF{Reshape}
is a Hom set, where \AC{eq} is identity and \AC{\_∙\_} is a composition with
the expected properties.  In the language of containers~\cite{containers}, \AF{Ar} is
a container and \AF{Reshape} is an inductive subset of cartesian container morphisms.

\begin{code}[hide]%
\>[2]\AgdaBound{i}\AgdaSpace{}%
\AgdaOperator{\AgdaFunction{⟨}}\AgdaSpace{}%
\AgdaInductiveConstructor{eq}\AgdaSpace{}%
\AgdaOperator{\AgdaFunction{⟩}}\AgdaSpace{}%
\AgdaSymbol{=}\AgdaSpace{}%
\AgdaBound{i}\<%
\\
\>[2]\AgdaSymbol{(}\AgdaBound{i}\AgdaSpace{}%
\AgdaOperator{\AgdaInductiveConstructor{⊗}}\AgdaSpace{}%
\AgdaBound{j}\AgdaSymbol{)}\AgdaSpace{}%
\AgdaOperator{\AgdaFunction{⟨}}\AgdaSpace{}%
\AgdaBound{r}\AgdaSpace{}%
\AgdaOperator{\AgdaInductiveConstructor{,}}\AgdaSpace{}%
\AgdaBound{r₁}\AgdaSpace{}%
\AgdaOperator{\AgdaFunction{⟩}}\AgdaSpace{}%
\AgdaSymbol{=}\AgdaSpace{}%
\AgdaSymbol{(}\AgdaBound{i}\AgdaSpace{}%
\AgdaOperator{\AgdaFunction{⟨}}\AgdaSpace{}%
\AgdaBound{r}\AgdaSpace{}%
\AgdaOperator{\AgdaFunction{⟩}}\AgdaSymbol{)}\AgdaSpace{}%
\AgdaOperator{\AgdaInductiveConstructor{⊗}}\AgdaSpace{}%
\AgdaSymbol{(}\AgdaBound{j}\AgdaSpace{}%
\AgdaOperator{\AgdaFunction{⟨}}\AgdaSpace{}%
\AgdaBound{r₁}\AgdaSpace{}%
\AgdaOperator{\AgdaFunction{⟩}}\AgdaSymbol{)}\<%
\\
\>[2]\AgdaBound{i}\AgdaSpace{}%
\AgdaOperator{\AgdaFunction{⟨}}\AgdaSpace{}%
\AgdaBound{r}\AgdaSpace{}%
\AgdaOperator{\AgdaInductiveConstructor{∙}}\AgdaSpace{}%
\AgdaBound{r₁}\AgdaSpace{}%
\AgdaOperator{\AgdaFunction{⟩}}\AgdaSpace{}%
\AgdaSymbol{=}\AgdaSpace{}%
\AgdaBound{i}\AgdaSpace{}%
\AgdaOperator{\AgdaFunction{⟨}}\AgdaSpace{}%
\AgdaBound{r}\AgdaSpace{}%
\AgdaOperator{\AgdaFunction{⟩}}\AgdaSpace{}%
\AgdaOperator{\AgdaFunction{⟨}}\AgdaSpace{}%
\AgdaBound{r₁}\AgdaSpace{}%
\AgdaOperator{\AgdaFunction{⟩}}\<%
\\
\>[2]\AgdaSymbol{(}\AgdaInductiveConstructor{ι}\AgdaSpace{}%
\AgdaBound{i}\AgdaSpace{}%
\AgdaOperator{\AgdaInductiveConstructor{⊗}}\AgdaSpace{}%
\AgdaInductiveConstructor{ι}\AgdaSpace{}%
\AgdaBound{j}\AgdaSymbol{)}\AgdaSpace{}%
\AgdaOperator{\AgdaFunction{⟨}}\AgdaSpace{}%
\AgdaInductiveConstructor{split}\AgdaSpace{}%
\AgdaOperator{\AgdaFunction{⟩}}\AgdaSpace{}%
\AgdaSymbol{=}\AgdaSpace{}%
\AgdaInductiveConstructor{ι}\AgdaSpace{}%
\AgdaSymbol{(}\AgdaFunction{combine}\AgdaSpace{}%
\AgdaBound{i}\AgdaSpace{}%
\AgdaBound{j}\AgdaSymbol{)}\<%
\\
\>[2]\AgdaInductiveConstructor{ι}\AgdaSpace{}%
\AgdaBound{i}\AgdaSpace{}%
\AgdaOperator{\AgdaFunction{⟨}}\AgdaSpace{}%
\AgdaInductiveConstructor{flat}\AgdaSpace{}%
\AgdaOperator{\AgdaFunction{⟩}}\AgdaSpace{}%
\AgdaSymbol{=}\AgdaSpace{}%
\AgdaKeyword{let}\AgdaSpace{}%
\AgdaBound{a}\AgdaSpace{}%
\AgdaOperator{\AgdaInductiveConstructor{,}}\AgdaSpace{}%
\AgdaBound{b}\AgdaSpace{}%
\AgdaSymbol{=}\AgdaSpace{}%
\AgdaFunction{remQuot}\AgdaSpace{}%
\AgdaSymbol{\AgdaUnderscore{}}\AgdaSpace{}%
\AgdaBound{i}\AgdaSpace{}%
\AgdaKeyword{in}\AgdaSpace{}%
\AgdaInductiveConstructor{ι}\AgdaSpace{}%
\AgdaBound{a}\AgdaSpace{}%
\AgdaOperator{\AgdaInductiveConstructor{⊗}}\AgdaSpace{}%
\AgdaInductiveConstructor{ι}\AgdaSpace{}%
\AgdaBound{b}\<%
\\
\>[2]\AgdaSymbol{(}\AgdaBound{i}\AgdaSpace{}%
\AgdaOperator{\AgdaInductiveConstructor{⊗}}\AgdaSpace{}%
\AgdaBound{j}\AgdaSymbol{)}\AgdaSpace{}%
\AgdaOperator{\AgdaFunction{⟨}}\AgdaSpace{}%
\AgdaInductiveConstructor{swap}\AgdaSpace{}%
\AgdaOperator{\AgdaFunction{⟩}}\AgdaSpace{}%
\AgdaSymbol{=}\AgdaSpace{}%
\AgdaBound{j}\AgdaSpace{}%
\AgdaOperator{\AgdaInductiveConstructor{⊗}}\AgdaSpace{}%
\AgdaBound{i}\<%
\\
\>[2]\AgdaSymbol{((}\AgdaBound{i}\AgdaSpace{}%
\AgdaOperator{\AgdaInductiveConstructor{⊗}}\AgdaSpace{}%
\AgdaBound{j}\AgdaSymbol{)}\AgdaSpace{}%
\AgdaOperator{\AgdaInductiveConstructor{⊗}}\AgdaSpace{}%
\AgdaBound{k}\AgdaSymbol{)}\AgdaSpace{}%
\AgdaOperator{\AgdaFunction{⟨}}\AgdaSpace{}%
\AgdaInductiveConstructor{assocl}\AgdaSpace{}%
\AgdaOperator{\AgdaFunction{⟩}}\AgdaSpace{}%
\AgdaSymbol{=}\AgdaSpace{}%
\AgdaBound{i}\AgdaSpace{}%
\AgdaOperator{\AgdaInductiveConstructor{⊗}}\AgdaSpace{}%
\AgdaSymbol{(}\AgdaBound{j}\AgdaSpace{}%
\AgdaOperator{\AgdaInductiveConstructor{⊗}}\AgdaSpace{}%
\AgdaBound{k}\AgdaSymbol{)}\<%
\\
\>[2]\AgdaSymbol{(}\AgdaBound{i}\AgdaSpace{}%
\AgdaOperator{\AgdaInductiveConstructor{⊗}}\AgdaSpace{}%
\AgdaSymbol{(}\AgdaBound{j}\AgdaSpace{}%
\AgdaOperator{\AgdaInductiveConstructor{⊗}}\AgdaSpace{}%
\AgdaBound{k}\AgdaSymbol{))}\AgdaSpace{}%
\AgdaOperator{\AgdaFunction{⟨}}\AgdaSpace{}%
\AgdaInductiveConstructor{assocr}\AgdaSpace{}%
\AgdaOperator{\AgdaFunction{⟩}}\AgdaSpace{}%
\AgdaSymbol{=}\AgdaSpace{}%
\AgdaSymbol{(}\AgdaBound{i}\AgdaSpace{}%
\AgdaOperator{\AgdaInductiveConstructor{⊗}}\AgdaSpace{}%
\AgdaBound{j}\AgdaSymbol{)}\AgdaSpace{}%
\AgdaOperator{\AgdaInductiveConstructor{⊗}}\AgdaSpace{}%
\AgdaBound{k}\<%
\\
\>[0]\<%
\\
\>[0]\<%
\\
\>[2]\AgdaFunction{rev}\AgdaSpace{}%
\AgdaInductiveConstructor{eq}\AgdaSpace{}%
\AgdaSymbol{=}\AgdaSpace{}%
\AgdaInductiveConstructor{eq}\<%
\\
\>[2]\AgdaFunction{rev}\AgdaSpace{}%
\AgdaSymbol{(}\AgdaBound{r}\AgdaSpace{}%
\AgdaOperator{\AgdaInductiveConstructor{,}}\AgdaSpace{}%
\AgdaBound{r₁}\AgdaSymbol{)}\AgdaSpace{}%
\AgdaSymbol{=}\AgdaSpace{}%
\AgdaFunction{rev}\AgdaSpace{}%
\AgdaBound{r}\AgdaSpace{}%
\AgdaOperator{\AgdaInductiveConstructor{,}}\AgdaSpace{}%
\AgdaFunction{rev}\AgdaSpace{}%
\AgdaBound{r₁}\<%
\\
\>[2]\AgdaFunction{rev}\AgdaSpace{}%
\AgdaSymbol{(}\AgdaBound{r}\AgdaSpace{}%
\AgdaOperator{\AgdaInductiveConstructor{∙}}\AgdaSpace{}%
\AgdaBound{r₁}\AgdaSymbol{)}\AgdaSpace{}%
\AgdaSymbol{=}\AgdaSpace{}%
\AgdaFunction{rev}\AgdaSpace{}%
\AgdaBound{r₁}\AgdaSpace{}%
\AgdaOperator{\AgdaInductiveConstructor{∙}}\AgdaSpace{}%
\AgdaFunction{rev}\AgdaSpace{}%
\AgdaBound{r}\<%
\\
\>[2]\AgdaFunction{rev}\AgdaSpace{}%
\AgdaInductiveConstructor{split}\AgdaSpace{}%
\AgdaSymbol{=}\AgdaSpace{}%
\AgdaInductiveConstructor{flat}\<%
\\
\>[2]\AgdaFunction{rev}\AgdaSpace{}%
\AgdaInductiveConstructor{flat}\AgdaSpace{}%
\AgdaSymbol{=}\AgdaSpace{}%
\AgdaInductiveConstructor{split}\<%
\\
\>[2]\AgdaFunction{rev}\AgdaSpace{}%
\AgdaInductiveConstructor{swap}\AgdaSpace{}%
\AgdaSymbol{=}\AgdaSpace{}%
\AgdaInductiveConstructor{swap}\<%
\\
\>[2]\AgdaFunction{rev}\AgdaSpace{}%
\AgdaInductiveConstructor{assocl}\AgdaSpace{}%
\AgdaSymbol{=}\AgdaSpace{}%
\AgdaInductiveConstructor{assocr}\<%
\\
\>[2]\AgdaFunction{rev}\AgdaSpace{}%
\AgdaInductiveConstructor{assocr}\AgdaSpace{}%
\AgdaSymbol{=}\AgdaSpace{}%
\AgdaInductiveConstructor{assocl}\<%
\end{code}

\section{CNN Building Blocks\label{sec:cnn}}

With the array theory from the previous section we can define the actual primitives 
that are required for our case study.

\subsection{One-dimensional convolution}
We start with plus and minus operations for 1-d indices which will
be used in the definition of convolution:
\begin{code}[hide]%
\>[2]\AgdaFunction{inject-left}\AgdaSpace{}%
\AgdaSymbol{:}\AgdaSpace{}%
\AgdaDatatype{Fin}\AgdaSpace{}%
\AgdaSymbol{(}\AgdaInductiveConstructor{suc}\AgdaSpace{}%
\AgdaGeneralizable{m}\AgdaSymbol{)}\AgdaSpace{}%
\AgdaSymbol{→}\AgdaSpace{}%
\AgdaDatatype{Fin}\AgdaSpace{}%
\AgdaSymbol{(}\AgdaInductiveConstructor{suc}\AgdaSpace{}%
\AgdaSymbol{(}\AgdaGeneralizable{n}\AgdaSpace{}%
\AgdaOperator{\AgdaPrimitive{+}}\AgdaSpace{}%
\AgdaGeneralizable{m}\AgdaSymbol{))}\<%
\\
\>[2]\AgdaFunction{inject-left}\AgdaSpace{}%
\AgdaSymbol{\{}\AgdaBound{m}\AgdaSymbol{\}}\AgdaSpace{}%
\AgdaSymbol{\{}\AgdaBound{n}\AgdaSymbol{\}}\AgdaSpace{}%
\AgdaBound{i}\AgdaSpace{}%
\AgdaKeyword{rewrite}\AgdaSpace{}%
\AgdaFunction{+-comm}\AgdaSpace{}%
\AgdaBound{n}\AgdaSpace{}%
\AgdaBound{m}%
\>[44]\AgdaSymbol{=}\AgdaSpace{}%
\AgdaFunction{inject+}\AgdaSpace{}%
\AgdaSymbol{\AgdaUnderscore{}}\AgdaSpace{}%
\AgdaBound{i}\<%
\\
\>[0]\<%
\\
\>[2]\AgdaFunction{split-inj₁}\AgdaSpace{}%
\AgdaSymbol{:}\AgdaSpace{}%
\AgdaSymbol{(}\AgdaBound{i}\AgdaSpace{}%
\AgdaSymbol{:}\AgdaSpace{}%
\AgdaDatatype{Fin}\AgdaSpace{}%
\AgdaSymbol{(}\AgdaGeneralizable{m}\AgdaSpace{}%
\AgdaOperator{\AgdaPrimitive{+}}\AgdaSpace{}%
\AgdaGeneralizable{n}\AgdaSymbol{))}\AgdaSpace{}%
\AgdaSymbol{(}\AgdaBound{k}\AgdaSpace{}%
\AgdaSymbol{:}\AgdaSpace{}%
\AgdaDatatype{Fin}\AgdaSpace{}%
\AgdaGeneralizable{m}\AgdaSymbol{)}\AgdaSpace{}%
\AgdaSymbol{→}\AgdaSpace{}%
\AgdaFunction{splitAt}\AgdaSpace{}%
\AgdaGeneralizable{m}\AgdaSpace{}%
\AgdaBound{i}\AgdaSpace{}%
\AgdaOperator{\AgdaDatatype{≡}}\AgdaSpace{}%
\AgdaInductiveConstructor{inj₁}\AgdaSpace{}%
\AgdaBound{k}\AgdaSpace{}%
\AgdaSymbol{→}\AgdaSpace{}%
\AgdaFunction{inject+}\AgdaSpace{}%
\AgdaSymbol{\AgdaUnderscore{}}\AgdaSpace{}%
\AgdaBound{k}\AgdaSpace{}%
\AgdaOperator{\AgdaDatatype{≡}}\AgdaSpace{}%
\AgdaBound{i}\<%
\\
\>[2]\AgdaFunction{split-inj₁}\AgdaSpace{}%
\AgdaSymbol{\{}\AgdaInductiveConstructor{suc}\AgdaSpace{}%
\AgdaBound{m}\AgdaSymbol{\}}\AgdaSpace{}%
\AgdaInductiveConstructor{zero}\AgdaSpace{}%
\AgdaDottedPattern{\AgdaSymbol{.}}\AgdaDottedPattern{\AgdaInductiveConstructor{zero}}\AgdaSpace{}%
\AgdaInductiveConstructor{refl}\AgdaSpace{}%
\AgdaSymbol{=}\AgdaSpace{}%
\AgdaInductiveConstructor{refl}\<%
\\
\>[2]\AgdaFunction{split-inj₁}\AgdaSpace{}%
\AgdaSymbol{\{}\AgdaInductiveConstructor{suc}\AgdaSpace{}%
\AgdaBound{m}\AgdaSymbol{\}}\AgdaSpace{}%
\AgdaSymbol{(}\AgdaInductiveConstructor{suc}\AgdaSpace{}%
\AgdaBound{i}\AgdaSymbol{)}\AgdaSpace{}%
\AgdaInductiveConstructor{zero}\AgdaSpace{}%
\AgdaBound{p}\AgdaSpace{}%
\AgdaKeyword{with}\AgdaSpace{}%
\AgdaFunction{splitAt}\AgdaSpace{}%
\AgdaBound{m}\AgdaSpace{}%
\AgdaBound{i}\AgdaSpace{}%
\AgdaSymbol{|}\AgdaSpace{}%
\AgdaFunction{inspect}\AgdaSpace{}%
\AgdaSymbol{(}\AgdaFunction{splitAt}\AgdaSpace{}%
\AgdaBound{m}\AgdaSymbol{)}\AgdaSpace{}%
\AgdaBound{i}\<%
\\
\>[2]\AgdaFunction{split-inj₁}\AgdaSpace{}%
\AgdaSymbol{\{}\AgdaInductiveConstructor{suc}\AgdaSpace{}%
\AgdaBound{m}\AgdaSymbol{\}}\AgdaSpace{}%
\AgdaSymbol{(}\AgdaInductiveConstructor{suc}\AgdaSpace{}%
\AgdaBound{i}\AgdaSymbol{)}\AgdaSpace{}%
\AgdaInductiveConstructor{zero}\AgdaSpace{}%
\AgdaSymbol{()}\AgdaSpace{}%
\AgdaSymbol{|}\AgdaSpace{}%
\AgdaInductiveConstructor{inj₁}\AgdaSpace{}%
\AgdaBound{x}\AgdaSpace{}%
\AgdaSymbol{|}\AgdaSpace{}%
\AgdaOperator{\AgdaInductiveConstructor{[}}\AgdaSpace{}%
\AgdaBound{r}\AgdaSpace{}%
\AgdaOperator{\AgdaInductiveConstructor{]}}\<%
\\
\>[2]\AgdaFunction{split-inj₁}\AgdaSpace{}%
\AgdaSymbol{\{}\AgdaInductiveConstructor{suc}\AgdaSpace{}%
\AgdaBound{m}\AgdaSymbol{\}}\AgdaSpace{}%
\AgdaSymbol{(}\AgdaInductiveConstructor{suc}\AgdaSpace{}%
\AgdaBound{i}\AgdaSymbol{)}\AgdaSpace{}%
\AgdaInductiveConstructor{zero}\AgdaSpace{}%
\AgdaSymbol{()}\AgdaSpace{}%
\AgdaSymbol{|}\AgdaSpace{}%
\AgdaInductiveConstructor{inj₂}\AgdaSpace{}%
\AgdaBound{y}\AgdaSpace{}%
\AgdaSymbol{|}\AgdaSpace{}%
\AgdaOperator{\AgdaInductiveConstructor{[}}\AgdaSpace{}%
\AgdaBound{r}\AgdaSpace{}%
\AgdaOperator{\AgdaInductiveConstructor{]}}\<%
\\
\>[2]\AgdaFunction{split-inj₁}\AgdaSpace{}%
\AgdaSymbol{\{}\AgdaInductiveConstructor{suc}\AgdaSpace{}%
\AgdaBound{m}\AgdaSymbol{\}}\AgdaSpace{}%
\AgdaSymbol{(}\AgdaInductiveConstructor{suc}\AgdaSpace{}%
\AgdaBound{i}\AgdaSymbol{)}\AgdaSpace{}%
\AgdaSymbol{(}\AgdaInductiveConstructor{suc}\AgdaSpace{}%
\AgdaBound{k}\AgdaSymbol{)}\AgdaSpace{}%
\AgdaBound{p}\AgdaSpace{}%
\AgdaKeyword{with}\AgdaSpace{}%
\AgdaFunction{splitAt}\AgdaSpace{}%
\AgdaBound{m}\AgdaSpace{}%
\AgdaBound{i}\AgdaSpace{}%
\AgdaSymbol{|}\AgdaSpace{}%
\AgdaFunction{inspect}\AgdaSpace{}%
\AgdaSymbol{(}\AgdaFunction{splitAt}\AgdaSpace{}%
\AgdaBound{m}\AgdaSymbol{)}\AgdaSpace{}%
\AgdaBound{i}\<%
\\
\>[2]\AgdaFunction{split-inj₁}\AgdaSpace{}%
\AgdaSymbol{\{}\AgdaInductiveConstructor{suc}\AgdaSpace{}%
\AgdaBound{m}\AgdaSymbol{\}}\AgdaSpace{}%
\AgdaSymbol{(}\AgdaInductiveConstructor{suc}\AgdaSpace{}%
\AgdaBound{i}\AgdaSymbol{)}\AgdaSpace{}%
\AgdaSymbol{(}\AgdaInductiveConstructor{suc}\AgdaSpace{}%
\AgdaDottedPattern{\AgdaSymbol{.}}\AgdaDottedPattern{\AgdaBound{x}}\AgdaSymbol{)}\AgdaSpace{}%
\AgdaInductiveConstructor{refl}\AgdaSpace{}%
\AgdaSymbol{|}\AgdaSpace{}%
\AgdaInductiveConstructor{inj₁}\AgdaSpace{}%
\AgdaBound{x}\AgdaSpace{}%
\AgdaSymbol{|}\AgdaSpace{}%
\AgdaOperator{\AgdaInductiveConstructor{[}}\AgdaSpace{}%
\AgdaBound{r}\AgdaSpace{}%
\AgdaOperator{\AgdaInductiveConstructor{]}}\AgdaSpace{}%
\AgdaSymbol{=}\AgdaSpace{}%
\AgdaFunction{cong}\AgdaSpace{}%
\AgdaInductiveConstructor{suc}\AgdaSpace{}%
\AgdaSymbol{(}\AgdaFunction{split-inj₁}\AgdaSpace{}%
\AgdaBound{i}\AgdaSpace{}%
\AgdaBound{x}\AgdaSpace{}%
\AgdaBound{r}\AgdaSymbol{)}\<%
\\
\>[0]\<%
\\
\>[2]\AgdaFunction{inj₁₂}\AgdaSpace{}%
\AgdaSymbol{:}\AgdaSpace{}%
\AgdaSymbol{\{}\AgdaBound{A}\AgdaSpace{}%
\AgdaBound{B}\AgdaSpace{}%
\AgdaSymbol{:}\AgdaSpace{}%
\AgdaPrimitive{Set}\AgdaSymbol{\}\{}\AgdaBound{x}\AgdaSpace{}%
\AgdaSymbol{:}\AgdaSpace{}%
\AgdaBound{A}\AgdaSymbol{\}\{}\AgdaBound{y}\AgdaSpace{}%
\AgdaSymbol{:}\AgdaSpace{}%
\AgdaBound{B}\AgdaSymbol{\}}\AgdaSpace{}%
\AgdaSymbol{→}\AgdaSpace{}%
\AgdaInductiveConstructor{inj₁}\AgdaSpace{}%
\AgdaBound{x}\AgdaSpace{}%
\AgdaOperator{\AgdaDatatype{≡}}\AgdaSpace{}%
\AgdaInductiveConstructor{inj₂}\AgdaSpace{}%
\AgdaBound{y}\AgdaSpace{}%
\AgdaSymbol{→}\AgdaSpace{}%
\AgdaDatatype{⊥}\<%
\\
\>[2]\AgdaFunction{inj₁₂}\AgdaSpace{}%
\AgdaSymbol{()}\<%
\end{code}
\begin{mathpar}
\codeblock{\begin{code}%
\>[2]\AgdaOperator{\AgdaFunction{\AgdaUnderscore{}⊕\AgdaUnderscore{}}}\AgdaSpace{}%
\AgdaSymbol{:}\AgdaSpace{}%
\AgdaDatatype{Fin}\AgdaSpace{}%
\AgdaGeneralizable{m}\AgdaSpace{}%
\AgdaSymbol{→}\AgdaSpace{}%
\AgdaDatatype{Fin}\AgdaSpace{}%
\AgdaSymbol{(}\AgdaNumber{1}\AgdaSpace{}%
\AgdaOperator{\AgdaPrimitive{+}}\AgdaSpace{}%
\AgdaGeneralizable{n}\AgdaSymbol{)}\AgdaSpace{}%
\AgdaSymbol{→}\AgdaSpace{}%
\AgdaDatatype{Fin}\AgdaSpace{}%
\AgdaSymbol{(}\AgdaGeneralizable{m}\AgdaSpace{}%
\AgdaOperator{\AgdaPrimitive{+}}\AgdaSpace{}%
\AgdaGeneralizable{n}\AgdaSymbol{)}\<%
\\
\>[2]\AgdaInductiveConstructor{zero}%
\>[9]\AgdaOperator{\AgdaFunction{⊕}}\AgdaSpace{}%
\AgdaBound{j}\AgdaSpace{}%
\AgdaSymbol{=}\AgdaSpace{}%
\AgdaFunction{inject-left}\AgdaSpace{}%
\AgdaBound{j}\<%
\\
\>[2]\AgdaInductiveConstructor{suc}\AgdaSpace{}%
\AgdaBound{i}%
\>[9]\AgdaOperator{\AgdaFunction{⊕}}\AgdaSpace{}%
\AgdaBound{j}\AgdaSpace{}%
\AgdaSymbol{=}\AgdaSpace{}%
\AgdaInductiveConstructor{suc}\AgdaSpace{}%
\AgdaSymbol{(}\AgdaBound{i}\AgdaSpace{}%
\AgdaOperator{\AgdaFunction{⊕}}\AgdaSpace{}%
\AgdaBound{j}\AgdaSymbol{)}\<%
\end{code}}
\and
\codeblock{\begin{code}%
\>[2]\AgdaOperator{\AgdaFunction{\AgdaUnderscore{}⊝\AgdaUnderscore{}}}%
\>[785I]\AgdaSymbol{:}\AgdaSpace{}%
\AgdaSymbol{(}\AgdaBound{i}\AgdaSpace{}%
\AgdaSymbol{:}\AgdaSpace{}%
\AgdaDatatype{Fin}\AgdaSpace{}%
\AgdaSymbol{(}\AgdaGeneralizable{m}\AgdaSpace{}%
\AgdaOperator{\AgdaPrimitive{+}}\AgdaSpace{}%
\AgdaGeneralizable{n}\AgdaSymbol{))}\AgdaSpace{}%
\AgdaSymbol{(}\AgdaBound{j}\AgdaSpace{}%
\AgdaSymbol{:}\AgdaSpace{}%
\AgdaDatatype{Fin}\AgdaSpace{}%
\AgdaGeneralizable{m}\AgdaSymbol{)}\<%
\\
\>[.][@{}l@{}]\<[785I]%
\>[6]\AgdaSymbol{→}\AgdaSpace{}%
\AgdaRecord{Dec}\AgdaSpace{}%
\AgdaSymbol{(}\AgdaFunction{∃}\AgdaSpace{}%
\AgdaSymbol{λ}\AgdaSpace{}%
\AgdaBound{k}\AgdaSpace{}%
\AgdaSymbol{→}\AgdaSpace{}%
\AgdaBound{j}\AgdaSpace{}%
\AgdaOperator{\AgdaFunction{⊕}}\AgdaSpace{}%
\AgdaBound{k}\AgdaSpace{}%
\AgdaOperator{\AgdaDatatype{≡}}\AgdaSpace{}%
\AgdaBound{i}\AgdaSymbol{)}\<%
\end{code}}
\end{mathpar}
\begin{code}[hide]%
\>[2]\AgdaOperator{\AgdaFunction{\AgdaUnderscore{}⊝\AgdaUnderscore{}}}\AgdaSpace{}%
\AgdaSymbol{\{}\AgdaInductiveConstructor{suc}\AgdaSpace{}%
\AgdaBound{m}\AgdaSymbol{\}}\AgdaSpace{}%
\AgdaSymbol{\{}\AgdaBound{n}\AgdaSymbol{\}}\AgdaSpace{}%
\AgdaBound{i}\AgdaSpace{}%
\AgdaInductiveConstructor{zero}\AgdaSpace{}%
\AgdaKeyword{rewrite}\AgdaSpace{}%
\AgdaFunction{+-comm}\AgdaSpace{}%
\AgdaBound{m}\AgdaSpace{}%
\AgdaBound{n}\AgdaSpace{}%
\AgdaKeyword{with}\AgdaSpace{}%
\AgdaFunction{splitAt}\AgdaSpace{}%
\AgdaSymbol{(}\AgdaInductiveConstructor{suc}\AgdaSpace{}%
\AgdaBound{n}\AgdaSymbol{)}\AgdaSpace{}%
\AgdaBound{i}\AgdaSpace{}%
\AgdaSymbol{|}\AgdaSpace{}%
\AgdaFunction{inspect}\AgdaSpace{}%
\AgdaSymbol{(}\AgdaFunction{splitAt}\AgdaSpace{}%
\AgdaSymbol{(}\AgdaInductiveConstructor{suc}\AgdaSpace{}%
\AgdaBound{n}\AgdaSymbol{))}\AgdaSpace{}%
\AgdaBound{i}\<%
\\
\>[2]\AgdaSymbol{...}\AgdaSpace{}%
\AgdaSymbol{|}\AgdaSpace{}%
\AgdaInductiveConstructor{inj₁}\AgdaSpace{}%
\AgdaBound{k}\AgdaSpace{}%
\AgdaSymbol{|}\AgdaSpace{}%
\AgdaOperator{\AgdaInductiveConstructor{[}}\AgdaSpace{}%
\AgdaBound{r}\AgdaSpace{}%
\AgdaOperator{\AgdaInductiveConstructor{]}}\AgdaSpace{}%
\AgdaSymbol{=}\AgdaSpace{}%
\AgdaInductiveConstructor{yes}\AgdaSpace{}%
\AgdaSymbol{(}\AgdaBound{k}\AgdaSpace{}%
\AgdaOperator{\AgdaInductiveConstructor{,}}\AgdaSpace{}%
\AgdaFunction{split-inj₁}\AgdaSpace{}%
\AgdaBound{i}\AgdaSpace{}%
\AgdaBound{k}\AgdaSpace{}%
\AgdaBound{r}\AgdaSymbol{)}\<%
\\
\>[2]\AgdaSymbol{...}\AgdaSpace{}%
\AgdaSymbol{|}\AgdaSpace{}%
\AgdaInductiveConstructor{inj₂}\AgdaSpace{}%
\AgdaBound{k}\AgdaSpace{}%
\AgdaSymbol{|}\AgdaSpace{}%
\AgdaOperator{\AgdaInductiveConstructor{[}}\AgdaSpace{}%
\AgdaBound{r}\AgdaSpace{}%
\AgdaOperator{\AgdaInductiveConstructor{]}}\AgdaSpace{}%
\AgdaSymbol{=}\AgdaSpace{}%
\AgdaInductiveConstructor{no}\AgdaSpace{}%
\AgdaFunction{reason}\<%
\\
\>[2][@{}l@{\AgdaIndent{0}}]%
\>[4]\AgdaKeyword{where}\<%
\\
\>[4][@{}l@{\AgdaIndent{0}}]%
\>[6]\AgdaFunction{reason}\AgdaSpace{}%
\AgdaSymbol{:}\AgdaSpace{}%
\AgdaSymbol{\AgdaUnderscore{}}\<%
\\
\>[6]\AgdaFunction{reason}\AgdaSpace{}%
\AgdaSymbol{(}\AgdaBound{k}\AgdaSpace{}%
\AgdaOperator{\AgdaInductiveConstructor{,}}\AgdaSpace{}%
\AgdaInductiveConstructor{refl}\AgdaSymbol{)}\AgdaSpace{}%
\AgdaKeyword{rewrite}\AgdaSpace{}%
\AgdaFunction{splitAt-inject+}\AgdaSpace{}%
\AgdaSymbol{(}\AgdaInductiveConstructor{suc}\AgdaSpace{}%
\AgdaBound{n}\AgdaSymbol{)}\AgdaSpace{}%
\AgdaBound{m}\AgdaSpace{}%
\AgdaBound{k}\AgdaSpace{}%
\AgdaSymbol{=}\AgdaSpace{}%
\AgdaFunction{inj₁₂}\AgdaSpace{}%
\AgdaBound{r}\<%
\\
\>[2]\AgdaInductiveConstructor{zero}\AgdaSpace{}%
\AgdaOperator{\AgdaFunction{⊝}}\AgdaSpace{}%
\AgdaInductiveConstructor{suc}\AgdaSpace{}%
\AgdaBound{j}\AgdaSpace{}%
\AgdaSymbol{=}\AgdaSpace{}%
\AgdaInductiveConstructor{no}\AgdaSpace{}%
\AgdaSymbol{λ}\AgdaSpace{}%
\AgdaSymbol{\{}\AgdaSpace{}%
\AgdaSymbol{(}\AgdaBound{k}\AgdaSpace{}%
\AgdaOperator{\AgdaInductiveConstructor{,}}\AgdaSpace{}%
\AgdaSymbol{())}\AgdaSpace{}%
\AgdaSymbol{\}}\<%
\\
\>[2]\AgdaInductiveConstructor{suc}\AgdaSpace{}%
\AgdaBound{i}\AgdaSpace{}%
\AgdaOperator{\AgdaFunction{⊝}}\AgdaSpace{}%
\AgdaInductiveConstructor{suc}\AgdaSpace{}%
\AgdaBound{j}\AgdaSpace{}%
\AgdaKeyword{with}\AgdaSpace{}%
\AgdaBound{i}\AgdaSpace{}%
\AgdaOperator{\AgdaFunction{⊝}}\AgdaSpace{}%
\AgdaBound{j}\<%
\\
\>[2]\AgdaSymbol{...}\AgdaSpace{}%
\AgdaSymbol{|}\AgdaSpace{}%
\AgdaInductiveConstructor{yes}\AgdaSpace{}%
\AgdaSymbol{(}\AgdaBound{k}\AgdaSpace{}%
\AgdaOperator{\AgdaInductiveConstructor{,}}\AgdaSpace{}%
\AgdaBound{p}\AgdaSymbol{)}\AgdaSpace{}%
\AgdaSymbol{=}\AgdaSpace{}%
\AgdaInductiveConstructor{yes}\AgdaSpace{}%
\AgdaSymbol{(}\AgdaBound{k}\AgdaSpace{}%
\AgdaOperator{\AgdaInductiveConstructor{,}}\AgdaSpace{}%
\AgdaFunction{cong}\AgdaSpace{}%
\AgdaInductiveConstructor{suc}\AgdaSpace{}%
\AgdaBound{p}\AgdaSymbol{)}\<%
\\
\>[2]\AgdaSymbol{...}\AgdaSpace{}%
\AgdaSymbol{|}\AgdaSpace{}%
\AgdaInductiveConstructor{no}\AgdaSpace{}%
\AgdaBound{¬p}\AgdaSpace{}%
\AgdaSymbol{=}\AgdaSpace{}%
\AgdaInductiveConstructor{no}\AgdaSpace{}%
\AgdaSymbol{λ}\AgdaSpace{}%
\AgdaSymbol{\{}\AgdaSpace{}%
\AgdaSymbol{(}\AgdaBound{k}\AgdaSpace{}%
\AgdaOperator{\AgdaInductiveConstructor{,}}\AgdaSpace{}%
\AgdaBound{p}\AgdaSymbol{)}\AgdaSpace{}%
\AgdaSymbol{→}\AgdaSpace{}%
\AgdaBound{¬p}\AgdaSpace{}%
\AgdaSymbol{(}\AgdaBound{k}\AgdaSpace{}%
\AgdaOperator{\AgdaInductiveConstructor{,}}\AgdaSpace{}%
\AgdaFunction{suc-injective}\AgdaSpace{}%
\AgdaBound{p}\AgdaSymbol{)}\AgdaSpace{}%
\AgdaSymbol{\}}\<%
\\
\\[\AgdaEmptyExtraSkip]%
\>[2]\AgdaFunction{inject-left-zero}\AgdaSpace{}%
\AgdaSymbol{:}\AgdaSpace{}%
\AgdaFunction{inject-left}\AgdaSpace{}%
\AgdaSymbol{\{}\AgdaGeneralizable{m}\AgdaSymbol{\}}\AgdaSpace{}%
\AgdaSymbol{\{}\AgdaGeneralizable{n}\AgdaSymbol{\}}\AgdaSpace{}%
\AgdaInductiveConstructor{zero}\AgdaSpace{}%
\AgdaOperator{\AgdaDatatype{≡}}\AgdaSpace{}%
\AgdaInductiveConstructor{zero}\<%
\\
\>[2]\AgdaFunction{inject-left-zero}\AgdaSpace{}%
\AgdaSymbol{\{}\AgdaBound{m}\AgdaSymbol{\}}\AgdaSpace{}%
\AgdaSymbol{\{}\AgdaBound{n}\AgdaSymbol{\}}\AgdaSpace{}%
\AgdaKeyword{rewrite}\AgdaSpace{}%
\AgdaFunction{+-comm}\AgdaSpace{}%
\AgdaBound{n}\AgdaSpace{}%
\AgdaBound{m}\AgdaSpace{}%
\AgdaSymbol{=}\AgdaSpace{}%
\AgdaInductiveConstructor{refl}\<%
\\
\\[\AgdaEmptyExtraSkip]%
\>[2]\AgdaFunction{suc-not-zero}\AgdaSpace{}%
\AgdaSymbol{:}\AgdaSpace{}%
\AgdaSymbol{\{}\AgdaBound{i}\AgdaSpace{}%
\AgdaSymbol{:}\AgdaSpace{}%
\AgdaDatatype{Fin}\AgdaSpace{}%
\AgdaGeneralizable{m}\AgdaSymbol{\}}\AgdaSpace{}%
\AgdaSymbol{→}\AgdaSpace{}%
\AgdaOperator{\AgdaDatatype{\AgdaUnderscore{}≡\AgdaUnderscore{}}}\AgdaSpace{}%
\AgdaSymbol{\{}\AgdaArgument{A}\AgdaSpace{}%
\AgdaSymbol{=}\AgdaSpace{}%
\AgdaDatatype{Fin}\AgdaSpace{}%
\AgdaSymbol{(}\AgdaInductiveConstructor{suc}\AgdaSpace{}%
\AgdaGeneralizable{m}\AgdaSymbol{)\}}\AgdaSpace{}%
\AgdaSymbol{(}\AgdaInductiveConstructor{suc}\AgdaSpace{}%
\AgdaBound{i}\AgdaSymbol{)}\AgdaSpace{}%
\AgdaInductiveConstructor{zero}\AgdaSpace{}%
\AgdaSymbol{→}\AgdaSpace{}%
\AgdaDatatype{⊥}\<%
\\
\>[2]\AgdaFunction{suc-not-zero}\AgdaSpace{}%
\AgdaSymbol{()}\<%
\\
\\[\AgdaEmptyExtraSkip]%
\>[2]\AgdaFunction{inject-left-suc}\AgdaSpace{}%
\AgdaSymbol{:}\AgdaSpace{}%
\AgdaSymbol{∀}\AgdaSpace{}%
\AgdaSymbol{(}\AgdaBound{i}\AgdaSpace{}%
\AgdaSymbol{:}\AgdaSpace{}%
\AgdaDatatype{Fin}\AgdaSpace{}%
\AgdaGeneralizable{m}\AgdaSymbol{)}\AgdaSpace{}%
\AgdaSymbol{→}\AgdaSpace{}%
\AgdaFunction{inject-left}\AgdaSpace{}%
\AgdaSymbol{\{}\AgdaGeneralizable{m}\AgdaSymbol{\}}\AgdaSpace{}%
\AgdaSymbol{\{}\AgdaGeneralizable{n}\AgdaSymbol{\}}\AgdaSpace{}%
\AgdaSymbol{(}\AgdaInductiveConstructor{suc}\AgdaSpace{}%
\AgdaBound{i}\AgdaSymbol{)}\AgdaSpace{}%
\AgdaOperator{\AgdaDatatype{≡}}\AgdaSpace{}%
\AgdaInductiveConstructor{zero}\AgdaSpace{}%
\AgdaSymbol{→}\AgdaSpace{}%
\AgdaDatatype{⊥}\<%
\\
\>[2]\AgdaFunction{inject-left-suc}\AgdaSpace{}%
\AgdaSymbol{\{}\AgdaBound{m}\AgdaSymbol{\}}\AgdaSpace{}%
\AgdaSymbol{\{}\AgdaBound{n}\AgdaSymbol{\}}\AgdaSpace{}%
\AgdaBound{i}\AgdaSpace{}%
\AgdaBound{p}\AgdaSpace{}%
\AgdaKeyword{rewrite}\AgdaSpace{}%
\AgdaFunction{+-comm}\AgdaSpace{}%
\AgdaBound{n}\AgdaSpace{}%
\AgdaBound{m}\AgdaSpace{}%
\AgdaSymbol{=}\AgdaSpace{}%
\AgdaFunction{suc-not-zero}\AgdaSpace{}%
\AgdaBound{p}\<%
\\
\\[\AgdaEmptyExtraSkip]%
\>[2]\AgdaFunction{zero-suc-⊥}\AgdaSpace{}%
\AgdaSymbol{:}\AgdaSpace{}%
\AgdaSymbol{∀}\AgdaSpace{}%
\AgdaSymbol{\{}\AgdaBound{i}\AgdaSpace{}%
\AgdaSymbol{:}\AgdaSpace{}%
\AgdaDatatype{Fin}\AgdaSpace{}%
\AgdaGeneralizable{n}\AgdaSymbol{\}}\AgdaSpace{}%
\AgdaSymbol{→}\AgdaSpace{}%
\AgdaOperator{\AgdaDatatype{\AgdaUnderscore{}≡\AgdaUnderscore{}}}\AgdaSpace{}%
\AgdaSymbol{\{}\AgdaArgument{A}\AgdaSpace{}%
\AgdaSymbol{=}\AgdaSpace{}%
\AgdaDatatype{Fin}\AgdaSpace{}%
\AgdaSymbol{(}\AgdaInductiveConstructor{suc}\AgdaSpace{}%
\AgdaGeneralizable{n}\AgdaSymbol{)\}}\AgdaSpace{}%
\AgdaInductiveConstructor{zero}\AgdaSpace{}%
\AgdaSymbol{(}\AgdaInductiveConstructor{suc}\AgdaSpace{}%
\AgdaBound{i}\AgdaSymbol{)}\AgdaSpace{}%
\AgdaSymbol{→}\AgdaSpace{}%
\AgdaDatatype{⊥}\<%
\\
\>[2]\AgdaFunction{zero-suc-⊥}\AgdaSpace{}%
\AgdaSymbol{()}\<%
\\
\\[\AgdaEmptyExtraSkip]%
\>[2]\AgdaComment{--\ TODO:\ this\ is\ annoying\ to\ do\ inductively\ on\ Fin,\ it\ is\ easier\ to}\<%
\\
\>[2]\AgdaComment{--\ \ \ \ \ \ \ implement\ this\ via\ Fin\ n\ =\ Σ\ ℕ\ (\AgdaUnderscore{}<\ n)\ representation}\<%
\\
\>[2]\AgdaComment{--\ minusx\ :\ (i\ :\ Fin\ (m\ +\ n))\ →\ (j\ :\ Fin\ (suc\ n))\ →\ Dec\ (∃\ λ\ k\ →\ k\ ⊕\ j\ ≡\ i)}\<%
\\
\>[2]\AgdaComment{--\ minusx\ \{zero\}\ i\ zero\ =\ no\ λ\ \{\ (()\ ,\ \AgdaUnderscore{})\ \}}\<%
\\
\>[2]\AgdaComment{--\ minusx\ \{suc\ m\}\ \{n\}\ zero\ zero\ =\ yes\ (zero\ ,\ inject-left-zero\ \{n\}\ \{m\})}\<%
\\
\>[2]\AgdaComment{--\ minusx\ \{suc\ m\}\ \{n\}\ (suc\ i)\ zero\ with\ minusx\ \{m\}\ i\ zero}\<%
\\
\>[2]\AgdaComment{--\ ...\ |\ yes\ (j\ ,\ p)\ =\ yes\ (suc\ j\ ,\ cong\ suc\ p)}\<%
\\
\>[2]\AgdaComment{--\ ...\ |\ no\ ¬p\ =\ no\ λ\ \{\ (zero\ ,\ p)\ →\ let\ rr\ =\ trans\ (sym\ \$\ inject-left-zero\ \{n\}\ \{m\})\ p\ }\<%
\\
\>[2]\AgdaComment{--\ \ \ \ \ \ \ \ \ \ \ \ \ \ \ \ \ \ \ \ \ \ \ \ \ \ \ \ \ \ \ \ \ \ \ in\ zero-suc-⊥\ rr}\<%
\\
\>[2]\AgdaComment{--\ \ \ \ \ \ \ \ \ \ \ \ \ \ \ \ \ \ \ \ ;\ (suc\ j\ ,\ p)\ →\ ¬p\ (j\ ,\ suc-injective\ p)\ \}}\<%
\\
\\[\AgdaEmptyExtraSkip]%
\>[2]\AgdaComment{--\ minusx\ \{zero\}\ i\ (suc\ j)\ =\ no\ λ\ \{\ (()\ ,\ p)\ \}}\<%
\\
\>[2]\AgdaComment{--\ minusx\ \{suc\ m\}\ zero\ (suc\ j)\ =\ no\ λ\ \{\ (zero\ ,\ p)\ →\ inject-left-suc\ j\ p}\<%
\\
\>[2]\AgdaComment{--\ \ \ \ \ \ \ \ \ \ \ \ \ \ \ \ \ \ \ \ \ \ \ \ \ \ \ \ \ \ \ \ \ \ \ \ ;\ (suc\ k\ ,\ ())\ \}}\<%
\\
\>[2]\AgdaComment{--\ minusx\ \{suc\ m\}\ \{suc\ n\}\ (suc\ i)\ (suc\ j)\ =\ ?\ }\<%
\end{code}
While the definitions look very innocent, their types carry non-trivial
information.  Consider \AF{\_⊕\_} which is addition of bounded $i$ and $j$.
However, the type says:
\begin{mathpar}
  \inferrule*
    {i < m \and j < 1 + n}
    {i+j < m + n}
\end{mathpar}
This looks a little surprising, but this indeed holds for natural numbers.
A reader may convince herself by considering the maximum value that $i$ and $j$
can possibly take.  The \AF{\_⊕\_} have partial inverses making it possible
to define left and right subtraction.  We consider left subtraction \AF{\_⊝\_}.
Its type says that there exists a decision procedure for finding $k$ of type
\AF{Fin} (1 + \AB{n}) together with the proof that $k$ is an inverse.
In some sense \AF{Dec} is similar to \AF{Maybe} type, except it forces one
to prove why the value does not exist as opposed to just returning \AC{nothing}.
This happens to be very useful, as it is really easy to introduce off-by-one
errors otherwise.

With the above definitions we can define convolution for 1-dimensional
cases.  A side note for mathematically inclined readers.  We use the term
\emph{convolution} in the way it is used in machine learning.  Technically,
we compute a cross-correlation, because the array of weights is not flipped.
However, in practice this is not a problem, as we assume that weights are
stored flipped in memory.

We define a handy shortcut \AF{Vec} and \AF{Ix} which are \AF{Ar} and \AF{P}
for 1-dimensional cases.\begin{mathpar}
\codeblock{\begin{code}%
\>[2]\AgdaFunction{Vec}\AgdaSpace{}%
\AgdaSymbol{:}\AgdaSpace{}%
\AgdaDatatype{ℕ}\AgdaSpace{}%
\AgdaSymbol{→}\AgdaSpace{}%
\AgdaPrimitive{Set}\AgdaSpace{}%
\AgdaSymbol{→}\AgdaSpace{}%
\AgdaPrimitive{Set}\<%
\\
\>[2]\AgdaFunction{Vec}\AgdaSpace{}%
\AgdaBound{m}\AgdaSpace{}%
\AgdaBound{X}\AgdaSpace{}%
\AgdaSymbol{=}\AgdaSpace{}%
\AgdaFunction{Ar}\AgdaSpace{}%
\AgdaSymbol{(}\AgdaInductiveConstructor{ι}\AgdaSpace{}%
\AgdaBound{m}\AgdaSymbol{)}\AgdaSpace{}%
\AgdaBound{X}\<%
\end{code}}
\and
\codeblock{\begin{code}%
\>[2]\AgdaFunction{Ix}\AgdaSpace{}%
\AgdaSymbol{:}\AgdaSpace{}%
\AgdaDatatype{ℕ}\AgdaSpace{}%
\AgdaSymbol{→}\AgdaSpace{}%
\AgdaPrimitive{Set}\<%
\\
\>[2]\AgdaFunction{Ix}\AgdaSpace{}%
\AgdaBound{m}\AgdaSpace{}%
\AgdaSymbol{=}\AgdaSpace{}%
\AgdaDatatype{P}\AgdaSpace{}%
\AgdaSymbol{(}\AgdaInductiveConstructor{ι}\AgdaSpace{}%
\AgdaBound{m}\AgdaSymbol{)}\<%
\end{code}}
\end{mathpar}
We introduce the \AF{slide₁} primitive that selects a $(1+n)$-element vector
from the $(m+n)$-element vector starting at the offset $i$.  Then,
following~\cite{cnn-array}, we compute $m$-element array of slides
and then sum it up.
\begin{mathpar}
\codeblock{\begin{code}%
\>[2]\AgdaFunction{slide₁}\AgdaSpace{}%
\AgdaSymbol{:}\AgdaSpace{}%
\AgdaFunction{Ix}\AgdaSpace{}%
\AgdaGeneralizable{m}\AgdaSpace{}%
\AgdaSymbol{→}\AgdaSpace{}%
\AgdaFunction{Vec}\AgdaSpace{}%
\AgdaSymbol{(}\AgdaGeneralizable{m}\AgdaSpace{}%
\AgdaOperator{\AgdaPrimitive{+}}\AgdaSpace{}%
\AgdaGeneralizable{n}\AgdaSymbol{)}\AgdaSpace{}%
\AgdaGeneralizable{X}\AgdaSpace{}%
\AgdaSymbol{→}\AgdaSpace{}%
\AgdaFunction{Vec}\AgdaSpace{}%
\AgdaSymbol{(}\AgdaNumber{1}\AgdaSpace{}%
\AgdaOperator{\AgdaPrimitive{+}}\AgdaSpace{}%
\AgdaGeneralizable{n}\AgdaSymbol{)}\AgdaSpace{}%
\AgdaGeneralizable{X}\<%
\\
\>[2]\AgdaFunction{slide₁}\AgdaSpace{}%
\AgdaSymbol{(}\AgdaInductiveConstructor{ι}\AgdaSpace{}%
\AgdaBound{i}\AgdaSymbol{)}\AgdaSpace{}%
\AgdaBound{v}\AgdaSpace{}%
\AgdaSymbol{(}\AgdaInductiveConstructor{ι}\AgdaSpace{}%
\AgdaBound{j}\AgdaSymbol{)}\AgdaSpace{}%
\AgdaSymbol{=}\AgdaSpace{}%
\AgdaBound{v}\AgdaSpace{}%
\AgdaSymbol{(}\AgdaInductiveConstructor{ι}\AgdaSpace{}%
\AgdaSymbol{(}\AgdaBound{i}\AgdaSpace{}%
\AgdaOperator{\AgdaFunction{⊕}}\AgdaSpace{}%
\AgdaBound{j}\AgdaSymbol{))}\<%
\\
\>[0]\<%
\\
\>[2]\AgdaFunction{conv₁}\AgdaSpace{}%
\AgdaSymbol{:}\AgdaSpace{}%
\AgdaFunction{Vec}\AgdaSpace{}%
\AgdaSymbol{(}\AgdaGeneralizable{m}\AgdaSpace{}%
\AgdaOperator{\AgdaPrimitive{+}}\AgdaSpace{}%
\AgdaGeneralizable{n}\AgdaSymbol{)}\AgdaSpace{}%
\AgdaDatatype{ℕ}\AgdaSpace{}%
\AgdaSymbol{→}\AgdaSpace{}%
\AgdaFunction{Vec}\AgdaSpace{}%
\AgdaGeneralizable{m}\AgdaSpace{}%
\AgdaDatatype{ℕ}\AgdaSpace{}%
\AgdaSymbol{→}\AgdaSpace{}%
\AgdaFunction{Vec}\AgdaSpace{}%
\AgdaSymbol{(}\AgdaNumber{1}\AgdaSpace{}%
\AgdaOperator{\AgdaPrimitive{+}}\AgdaSpace{}%
\AgdaGeneralizable{n}\AgdaSymbol{)}\AgdaSpace{}%
\AgdaDatatype{ℕ}\<%
\\
\>[2]\AgdaFunction{conv₁}\AgdaSpace{}%
\AgdaBound{a}\AgdaSpace{}%
\AgdaBound{w}\AgdaSpace{}%
\AgdaSymbol{=}\AgdaSpace{}%
\AgdaFunction{sum}\AgdaSpace{}%
\AgdaSymbol{(}\AgdaFunction{zipWith}\AgdaSpace{}%
\AgdaOperator{\AgdaPrimitive{\AgdaUnderscore{}+\AgdaUnderscore{}}}\AgdaSymbol{)}\AgdaSpace{}%
\AgdaSymbol{(}\AgdaFunction{K}\AgdaSpace{}%
\AgdaNumber{0}\AgdaSymbol{)}\AgdaSpace{}%
\AgdaSymbol{(λ}\AgdaSpace{}%
\AgdaBound{i}\AgdaSpace{}%
\AgdaSymbol{→}\AgdaSpace{}%
\AgdaFunction{map}\AgdaSpace{}%
\AgdaSymbol{(}\AgdaBound{w}\AgdaSpace{}%
\AgdaBound{i}\AgdaSpace{}%
\AgdaOperator{\AgdaPrimitive{*\AgdaUnderscore{}}}\AgdaSymbol{)}\AgdaSpace{}%
\AgdaSymbol{(}\AgdaFunction{slide₁}\AgdaSpace{}%
\AgdaBound{i}\AgdaSpace{}%
\AgdaBound{a}\AgdaSymbol{))}\<%
\end{code}}
\end{mathpar}

\subsection{Generalisation\label{sec:general-ix-ops}}
We want to define convolution for arrays of higher ranks.  The first thing
to do is to express $m + n$ and $1 + n$ where $m$ and $n$ become arbitrary
shape trees.  In case of addition, we need a witness that both arguments
have the same tree structure.  If they do, we can simply add their nodes point-wise.
We define the three-way relation \AF{\_+\_≈\_} that combines the witness and
the action.  That is, the type \AB{p} \AF{+} \AB{q} \AF{≈} \AB{r} says that
$p$ and $q$ have the same tree structure and that $q$ is a point-wise addition
of $p$ and $q$.  We introduce a similar relation \AF{suc\_≈\_} for $1 + n$
case, and we introduce the relation \AF{\_*\_≈\_} that witnesses point-wise
multiplication that will be needed for blocking.
\begin{code}[hide]%
\>[2]\AgdaKeyword{infix}\AgdaSpace{}%
\AgdaNumber{5}\AgdaSpace{}%
\AgdaOperator{\AgdaDatatype{\AgdaUnderscore{}+\AgdaUnderscore{}≈\AgdaUnderscore{}}}\<%
\\
\>[2]\AgdaKeyword{infix}\AgdaSpace{}%
\AgdaNumber{5}\AgdaSpace{}%
\AgdaOperator{\AgdaDatatype{suc\AgdaUnderscore{}≈\AgdaUnderscore{}}}\<%
\\
\>[2]\AgdaKeyword{infix}\AgdaSpace{}%
\AgdaNumber{5}\AgdaSpace{}%
\AgdaOperator{\AgdaDatatype{\AgdaUnderscore{}*\AgdaUnderscore{}≈\AgdaUnderscore{}}}\<%
\\
\>[2]\AgdaKeyword{infixl}\AgdaSpace{}%
\AgdaNumber{8}\AgdaSpace{}%
\AgdaOperator{\AgdaFunction{\AgdaUnderscore{}⊝ₚ\AgdaUnderscore{}}}\<%
\\
\>[0]\<%
\\
\>[2]\AgdaFunction{ι-injective}\AgdaSpace{}%
\AgdaSymbol{:}\AgdaSpace{}%
\AgdaSymbol{\{}\AgdaBound{i}\AgdaSpace{}%
\AgdaBound{j}\AgdaSpace{}%
\AgdaSymbol{:}\AgdaSpace{}%
\AgdaDatatype{Fin}\AgdaSpace{}%
\AgdaGeneralizable{n}\AgdaSymbol{\}}\AgdaSpace{}%
\AgdaSymbol{→}\AgdaSpace{}%
\AgdaOperator{\AgdaDatatype{\AgdaUnderscore{}≡\AgdaUnderscore{}}}\AgdaSpace{}%
\AgdaSymbol{\{}\AgdaArgument{A}\AgdaSpace{}%
\AgdaSymbol{=}\AgdaSpace{}%
\AgdaDatatype{P}\AgdaSpace{}%
\AgdaSymbol{(}\AgdaInductiveConstructor{ι}\AgdaSpace{}%
\AgdaGeneralizable{n}\AgdaSymbol{)\}}\AgdaSpace{}%
\AgdaSymbol{(}\AgdaInductiveConstructor{ι}\AgdaSpace{}%
\AgdaBound{i}\AgdaSymbol{)}\AgdaSpace{}%
\AgdaSymbol{(}\AgdaInductiveConstructor{ι}\AgdaSpace{}%
\AgdaBound{j}\AgdaSymbol{)}\AgdaSpace{}%
\AgdaSymbol{→}\AgdaSpace{}%
\AgdaBound{i}\AgdaSpace{}%
\AgdaOperator{\AgdaDatatype{≡}}\AgdaSpace{}%
\AgdaBound{j}\<%
\\
\>[2]\AgdaFunction{ι-injective}\AgdaSpace{}%
\AgdaInductiveConstructor{refl}\AgdaSpace{}%
\AgdaSymbol{=}\AgdaSpace{}%
\AgdaInductiveConstructor{refl}\<%
\\
\>[0]\<%
\\
\>[2]\AgdaFunction{⊗-fst-≡}\AgdaSpace{}%
\AgdaSymbol{:}\AgdaSpace{}%
\AgdaSymbol{\{}\AgdaBound{i}\AgdaSpace{}%
\AgdaBound{i′}\AgdaSpace{}%
\AgdaSymbol{:}\AgdaSpace{}%
\AgdaDatatype{P}\AgdaSpace{}%
\AgdaGeneralizable{s}\AgdaSymbol{\}\{}\AgdaBound{j}\AgdaSpace{}%
\AgdaBound{j′}\AgdaSpace{}%
\AgdaSymbol{:}\AgdaSpace{}%
\AgdaDatatype{P}\AgdaSpace{}%
\AgdaGeneralizable{p}\AgdaSymbol{\}}\AgdaSpace{}%
\AgdaSymbol{→}\AgdaSpace{}%
\AgdaOperator{\AgdaDatatype{\AgdaUnderscore{}≡\AgdaUnderscore{}}}\AgdaSpace{}%
\AgdaSymbol{\{}\AgdaArgument{A}\AgdaSpace{}%
\AgdaSymbol{=}\AgdaSpace{}%
\AgdaDatatype{P}\AgdaSpace{}%
\AgdaSymbol{(}\AgdaGeneralizable{s}\AgdaSpace{}%
\AgdaOperator{\AgdaInductiveConstructor{⊗}}\AgdaSpace{}%
\AgdaGeneralizable{p}\AgdaSymbol{)\}}\AgdaSpace{}%
\AgdaSymbol{(}\AgdaBound{i}\AgdaSpace{}%
\AgdaOperator{\AgdaInductiveConstructor{⊗}}\AgdaSpace{}%
\AgdaBound{j}\AgdaSymbol{)}\AgdaSpace{}%
\AgdaSymbol{(}\AgdaBound{i′}\AgdaSpace{}%
\AgdaOperator{\AgdaInductiveConstructor{⊗}}\AgdaSpace{}%
\AgdaBound{j′}\AgdaSymbol{)}\AgdaSpace{}%
\AgdaSymbol{→}\AgdaSpace{}%
\AgdaBound{i}\AgdaSpace{}%
\AgdaOperator{\AgdaDatatype{≡}}\AgdaSpace{}%
\AgdaBound{i′}\<%
\\
\>[2]\AgdaFunction{⊗-fst-≡}\AgdaSpace{}%
\AgdaInductiveConstructor{refl}\AgdaSpace{}%
\AgdaSymbol{=}\AgdaSpace{}%
\AgdaInductiveConstructor{refl}\<%
\\
\>[0]\<%
\\
\>[2]\AgdaFunction{⊗-snd-≡}\AgdaSpace{}%
\AgdaSymbol{:}\AgdaSpace{}%
\AgdaSymbol{\{}\AgdaBound{i}\AgdaSpace{}%
\AgdaBound{i′}\AgdaSpace{}%
\AgdaSymbol{:}\AgdaSpace{}%
\AgdaDatatype{P}\AgdaSpace{}%
\AgdaGeneralizable{s}\AgdaSymbol{\}\{}\AgdaBound{j}\AgdaSpace{}%
\AgdaBound{j′}\AgdaSpace{}%
\AgdaSymbol{:}\AgdaSpace{}%
\AgdaDatatype{P}\AgdaSpace{}%
\AgdaGeneralizable{p}\AgdaSymbol{\}}\AgdaSpace{}%
\AgdaSymbol{→}\AgdaSpace{}%
\AgdaOperator{\AgdaDatatype{\AgdaUnderscore{}≡\AgdaUnderscore{}}}\AgdaSpace{}%
\AgdaSymbol{\{}\AgdaArgument{A}\AgdaSpace{}%
\AgdaSymbol{=}\AgdaSpace{}%
\AgdaDatatype{P}\AgdaSpace{}%
\AgdaSymbol{(}\AgdaGeneralizable{s}\AgdaSpace{}%
\AgdaOperator{\AgdaInductiveConstructor{⊗}}\AgdaSpace{}%
\AgdaGeneralizable{p}\AgdaSymbol{)\}}\AgdaSpace{}%
\AgdaSymbol{(}\AgdaBound{i}\AgdaSpace{}%
\AgdaOperator{\AgdaInductiveConstructor{⊗}}\AgdaSpace{}%
\AgdaBound{j}\AgdaSymbol{)}\AgdaSpace{}%
\AgdaSymbol{(}\AgdaBound{i′}\AgdaSpace{}%
\AgdaOperator{\AgdaInductiveConstructor{⊗}}\AgdaSpace{}%
\AgdaBound{j′}\AgdaSymbol{)}\AgdaSpace{}%
\AgdaSymbol{→}\AgdaSpace{}%
\AgdaBound{j}\AgdaSpace{}%
\AgdaOperator{\AgdaDatatype{≡}}\AgdaSpace{}%
\AgdaBound{j′}\<%
\\
\>[2]\AgdaFunction{⊗-snd-≡}\AgdaSpace{}%
\AgdaInductiveConstructor{refl}\AgdaSpace{}%
\AgdaSymbol{=}\AgdaSpace{}%
\AgdaInductiveConstructor{refl}\<%
\\
\>[0]\<%
\end{code}
\begin{mathpar}
\codeblock{\begin{code}%
\>[0][@{}l@{\AgdaIndent{1}}]%
\>[2]\AgdaKeyword{data}\AgdaSpace{}%
\AgdaOperator{\AgdaDatatype{\AgdaUnderscore{}+\AgdaUnderscore{}≈\AgdaUnderscore{}}}\AgdaSpace{}%
\AgdaSymbol{:}\AgdaSpace{}%
\AgdaDatatype{S}\AgdaSpace{}%
\AgdaSymbol{→}\AgdaSpace{}%
\AgdaDatatype{S}\AgdaSpace{}%
\AgdaSymbol{→}\AgdaSpace{}%
\AgdaDatatype{S}\AgdaSpace{}%
\AgdaSymbol{→}\AgdaSpace{}%
\AgdaPrimitive{Set}\AgdaSpace{}%
\AgdaKeyword{where}\<%
\\
\>[2][@{}l@{\AgdaIndent{0}}]%
\>[4]\AgdaInductiveConstructor{ι}%
\>[9]\AgdaSymbol{:}\AgdaSpace{}%
\AgdaInductiveConstructor{ι}\AgdaSpace{}%
\AgdaGeneralizable{m}\AgdaSpace{}%
\AgdaOperator{\AgdaDatatype{+}}\AgdaSpace{}%
\AgdaInductiveConstructor{ι}\AgdaSpace{}%
\AgdaGeneralizable{n}\AgdaSpace{}%
\AgdaOperator{\AgdaDatatype{≈}}\AgdaSpace{}%
\AgdaInductiveConstructor{ι}\AgdaSpace{}%
\AgdaSymbol{(}\AgdaGeneralizable{m}\AgdaSpace{}%
\AgdaOperator{\AgdaPrimitive{+}}\AgdaSpace{}%
\AgdaGeneralizable{n}\AgdaSymbol{)}\<%
\\
\>[4]\AgdaOperator{\AgdaInductiveConstructor{\AgdaUnderscore{}⊗\AgdaUnderscore{}}}%
\>[9]\AgdaSymbol{:}\AgdaSpace{}%
\AgdaGeneralizable{s}\AgdaSpace{}%
\AgdaOperator{\AgdaDatatype{+}}\AgdaSpace{}%
\AgdaGeneralizable{q}\AgdaSpace{}%
\AgdaOperator{\AgdaDatatype{≈}}\AgdaSpace{}%
\AgdaGeneralizable{u}\AgdaSpace{}%
\AgdaSymbol{→}\AgdaSpace{}%
\AgdaGeneralizable{p}\AgdaSpace{}%
\AgdaOperator{\AgdaDatatype{+}}\AgdaSpace{}%
\AgdaGeneralizable{r}\AgdaSpace{}%
\AgdaOperator{\AgdaDatatype{≈}}\AgdaSpace{}%
\AgdaGeneralizable{w}\<%
\\
\>[9]\AgdaSymbol{→}\AgdaSpace{}%
\AgdaGeneralizable{s}\AgdaSpace{}%
\AgdaOperator{\AgdaInductiveConstructor{⊗}}\AgdaSpace{}%
\AgdaGeneralizable{p}\AgdaSpace{}%
\AgdaOperator{\AgdaDatatype{+}}\AgdaSpace{}%
\AgdaGeneralizable{q}\AgdaSpace{}%
\AgdaOperator{\AgdaInductiveConstructor{⊗}}\AgdaSpace{}%
\AgdaGeneralizable{r}\AgdaSpace{}%
\AgdaOperator{\AgdaDatatype{≈}}\AgdaSpace{}%
\AgdaGeneralizable{u}\AgdaSpace{}%
\AgdaOperator{\AgdaInductiveConstructor{⊗}}\AgdaSpace{}%
\AgdaGeneralizable{w}\<%
\end{code}}
\and
\codeblock{\begin{code}%
\>[2]\AgdaKeyword{data}\AgdaSpace{}%
\AgdaOperator{\AgdaDatatype{suc\AgdaUnderscore{}≈\AgdaUnderscore{}}}\AgdaSpace{}%
\AgdaSymbol{:}\AgdaSpace{}%
\AgdaDatatype{S}\AgdaSpace{}%
\AgdaSymbol{→}\AgdaSpace{}%
\AgdaDatatype{S}\AgdaSpace{}%
\AgdaSymbol{→}\AgdaSpace{}%
\AgdaPrimitive{Set}\AgdaSpace{}%
\AgdaKeyword{where}\<%
\\
\>[2][@{}l@{\AgdaIndent{0}}]%
\>[4]\AgdaInductiveConstructor{ι}%
\>[9]\AgdaSymbol{:}\AgdaSpace{}%
\AgdaOperator{\AgdaDatatype{suc}}\AgdaSpace{}%
\AgdaSymbol{(}\AgdaInductiveConstructor{ι}\AgdaSpace{}%
\AgdaGeneralizable{n}\AgdaSymbol{)}\AgdaSpace{}%
\AgdaOperator{\AgdaDatatype{≈}}\AgdaSpace{}%
\AgdaInductiveConstructor{ι}\AgdaSpace{}%
\AgdaSymbol{(}\AgdaInductiveConstructor{suc}\AgdaSpace{}%
\AgdaGeneralizable{n}\AgdaSymbol{)}\<%
\\
\>[4]\AgdaOperator{\AgdaInductiveConstructor{\AgdaUnderscore{}⊗\AgdaUnderscore{}}}%
\>[9]\AgdaSymbol{:}\AgdaSpace{}%
\AgdaOperator{\AgdaDatatype{suc}}\AgdaSpace{}%
\AgdaGeneralizable{s}\AgdaSpace{}%
\AgdaOperator{\AgdaDatatype{≈}}\AgdaSpace{}%
\AgdaGeneralizable{u}\AgdaSpace{}%
\AgdaSymbol{→}\AgdaSpace{}%
\AgdaOperator{\AgdaDatatype{suc}}\AgdaSpace{}%
\AgdaGeneralizable{p}\AgdaSpace{}%
\AgdaOperator{\AgdaDatatype{≈}}\AgdaSpace{}%
\AgdaGeneralizable{w}\<%
\\
\>[9]\AgdaSymbol{→}\AgdaSpace{}%
\AgdaOperator{\AgdaDatatype{suc}}\AgdaSpace{}%
\AgdaSymbol{(}\AgdaGeneralizable{s}\AgdaSpace{}%
\AgdaOperator{\AgdaInductiveConstructor{⊗}}\AgdaSpace{}%
\AgdaGeneralizable{p}\AgdaSymbol{)}\AgdaSpace{}%
\AgdaOperator{\AgdaDatatype{≈}}\AgdaSpace{}%
\AgdaGeneralizable{u}\AgdaSpace{}%
\AgdaOperator{\AgdaInductiveConstructor{⊗}}\AgdaSpace{}%
\AgdaGeneralizable{w}\<%
\end{code}}
\and
\codeblock{\begin{code}%
\>[2]\AgdaKeyword{data}\AgdaSpace{}%
\AgdaOperator{\AgdaDatatype{\AgdaUnderscore{}*\AgdaUnderscore{}≈\AgdaUnderscore{}}}\AgdaSpace{}%
\AgdaSymbol{:}\AgdaSpace{}%
\AgdaDatatype{S}\AgdaSpace{}%
\AgdaSymbol{→}\AgdaSpace{}%
\AgdaDatatype{S}\AgdaSpace{}%
\AgdaSymbol{→}\AgdaSpace{}%
\AgdaDatatype{S}\AgdaSpace{}%
\AgdaSymbol{→}\AgdaSpace{}%
\AgdaPrimitive{Set}\AgdaSpace{}%
\AgdaKeyword{where}\<%
\\
\>[2][@{}l@{\AgdaIndent{0}}]%
\>[4]\AgdaInductiveConstructor{ι}%
\>[9]\AgdaSymbol{:}\AgdaSpace{}%
\AgdaInductiveConstructor{ι}\AgdaSpace{}%
\AgdaGeneralizable{m}\AgdaSpace{}%
\AgdaOperator{\AgdaDatatype{*}}\AgdaSpace{}%
\AgdaInductiveConstructor{ι}\AgdaSpace{}%
\AgdaGeneralizable{n}\AgdaSpace{}%
\AgdaOperator{\AgdaDatatype{≈}}\AgdaSpace{}%
\AgdaInductiveConstructor{ι}\AgdaSpace{}%
\AgdaSymbol{(}\AgdaGeneralizable{m}\AgdaSpace{}%
\AgdaOperator{\AgdaPrimitive{*}}\AgdaSpace{}%
\AgdaGeneralizable{n}\AgdaSymbol{)}\<%
\\
\>[4]\AgdaOperator{\AgdaInductiveConstructor{\AgdaUnderscore{}⊗\AgdaUnderscore{}}}%
\>[9]\AgdaSymbol{:}\AgdaSpace{}%
\AgdaGeneralizable{s}\AgdaSpace{}%
\AgdaOperator{\AgdaDatatype{*}}\AgdaSpace{}%
\AgdaGeneralizable{q}\AgdaSpace{}%
\AgdaOperator{\AgdaDatatype{≈}}\AgdaSpace{}%
\AgdaGeneralizable{u}\AgdaSpace{}%
\AgdaSymbol{→}\AgdaSpace{}%
\AgdaGeneralizable{p}\AgdaSpace{}%
\AgdaOperator{\AgdaDatatype{*}}\AgdaSpace{}%
\AgdaGeneralizable{r}\AgdaSpace{}%
\AgdaOperator{\AgdaDatatype{≈}}\AgdaSpace{}%
\AgdaGeneralizable{w}\AgdaSpace{}%
\AgdaSymbol{→}\AgdaSpace{}%
\AgdaSymbol{(}\AgdaGeneralizable{s}\AgdaSpace{}%
\AgdaOperator{\AgdaInductiveConstructor{⊗}}\AgdaSpace{}%
\AgdaGeneralizable{p}\AgdaSymbol{)}\AgdaSpace{}%
\AgdaOperator{\AgdaDatatype{*}}\AgdaSpace{}%
\AgdaSymbol{(}\AgdaGeneralizable{q}\AgdaSpace{}%
\AgdaOperator{\AgdaInductiveConstructor{⊗}}\AgdaSpace{}%
\AgdaGeneralizable{r}\AgdaSymbol{)}\AgdaSpace{}%
\AgdaOperator{\AgdaDatatype{≈}}\AgdaSpace{}%
\AgdaGeneralizable{u}\AgdaSpace{}%
\AgdaOperator{\AgdaInductiveConstructor{⊗}}\AgdaSpace{}%
\AgdaGeneralizable{w}\<%
\end{code}}
\end{mathpar}

With these relations in place, how do we define generalised convolution?  One
possible way is to use the \AF{sum} approach where we recurse over the shape
tree and perform one operation at a time.  However, there is a good point made
in~\cite{cnn-array} that we can shift the shape recursion into index computation.
Therefore we define \AF{\_⊕ₚ\_} and \AF{\_⊝ₚ\_} which generalise \AF{\_⊕\_} and
\AF{\_⊝\_} for higher ranks.  Once again, \AD{Dec} type forces \AF{⊝ₚ} to justify
the cases when the inverse does not exist.
\begin{code}[hide]%
\>[0]\AgdaComment{--\ \ \ XXX\ this\ implementation\ is\ needed\ if\ we\ switch\ to\ irrelevant\ Fin\ from\ the\ code.}\<%
\\
\>[0]\AgdaComment{--\ \ \ data\ \AgdaUnderscore{}≈ₚ\AgdaUnderscore{}\ :\ P\ s\ →\ P\ s\ →\ Set\ where}\<%
\\
\>[0]\AgdaComment{--\ \ \ \ \ ι\ :\ ∀\ \{i\ j\ :\ Fin\ n\}\ →\ Fin.index\ i\ ≡\ Fin.index\ j\ →\ ι\ i\ ≈ₚ\ ι\ j}\<%
\\
\>[0]\AgdaComment{--\ \ \ \ \ \AgdaUnderscore{}⊗\AgdaUnderscore{}\ :\ ∀\ \{i\ k\ :\ P\ s\}\{j\ l\ :\ P\ p\}\ →\ i\ ≈ₚ\ k\ →\ j\ ≈ₚ\ l\ →\ (i\ ⊗\ j)\ ≈ₚ\ (k\ ⊗\ l)\ }\<%
\\
\\[\AgdaEmptyExtraSkip]%
\>[0]\AgdaComment{--\ \ \ \AgdaUnderscore{}⊝ₚ\AgdaUnderscore{}\ :\ (i\ :\ P\ r)\ (j\ :\ P\ s)\ (su\ :\ suc\ p\ ≈\ u)\ (sp\ :\ s\ +\ p\ ≈\ r)\ }\<%
\\
\>[0]\AgdaComment{--\ \ \ \ \ \ \ \ →\ Dec\ (∃\ λ\ k\ →\ (j\ ⊕ₚ\ k)\ su\ sp\ ≈ₚ\ i)}\<%
\\
\>[0]\AgdaComment{--\ \ \ (ι\ i\ ⊝ₚ\ ι\ j)\ ι\ ι\ with\ i\ ⊝\ j}\<%
\\
\>[0]\AgdaComment{--\ \ \ ...\ |\ yes\ \ (k\ ,\ p)\ \ \ =\ yes\ (ι\ k\ ,\ ι\ p)}\<%
\\
\>[0]\AgdaComment{--\ \ \ ...\ |\ no\ \ \ ¬p\ \ \ \ \ \ \ \ =\ no\ λ\ \{\ (ι\ k\ ,\ ι\ p)\ →\ ¬p\ (k\ ,\ p)\ \}}\<%
\\
\>[0]\AgdaComment{--\ \ \ ((i\ ⊗\ i′)\ ⊝ₚ\ (j\ ⊗\ j′))\ (su\ ⊗\ su′)\ (sp\ ⊗\ sp′)\ with\ (i\ ⊝ₚ\ j)\ su\ sp}\<%
\\
\>[0]\AgdaComment{--\ \ \ ...\ |\ no\ \ \ ¬p\ \ \ \ \ \ \ \ =\ no\ λ\ \{\ (k\ ⊗\ k′\ ,\ p\ ⊗\ p′)\ →\ ¬p\ (k\ ,\ p)\}}\<%
\\
\>[0]\AgdaComment{--\ \ \ ...\ |\ yes\ \ (k\ ,\ \ p)\ \ with\ (i′\ ⊝ₚ\ j′)\ su′\ sp′}\<%
\\
\>[0]\AgdaComment{--\ \ \ ...\ |\ no\ \ \ ¬p\ \ \ \ \ \ \ \ =\ no\ λ\ \{\ (k\ ⊗\ k′\ ,\ p\ ⊗\ p′)\ →\ ¬p\ (k′\ ,\ p′)\}}\<%
\\
\>[0]\AgdaComment{--\ \ \ ...\ |\ yes\ \ (k′\ ,\ p′)\ =\ yes\ (k\ ⊗\ k′\ ,\ p\ ⊗\ p′)}\<%
\\
\\[\AgdaEmptyExtraSkip]%
\\[\AgdaEmptyExtraSkip]%
\>[0][@{}l@{\AgdaIndent{0}}]%
\>[2]\AgdaComment{--\ data\ \AgdaUnderscore{}≈ₚ\AgdaUnderscore{}\ :\ P\ s\ →\ P\ s\ →\ Set\ where}\<%
\\
\>[2]\AgdaComment{--\ \ \ ι\ :\ \{i\ j\ :\ Fin\ n\}\ →\ i\ ≡\ j\ →\ ι\ i\ ≈ₚ\ ι\ j}\<%
\\
\>[2]\AgdaComment{--\ \ \ \AgdaUnderscore{}⊗\AgdaUnderscore{}\ :\ \{i\ k\ :\ P\ s\}\{j\ l\ :\ P\ p\}\ →\ i\ ≈ₚ\ k\ →\ j\ ≈ₚ\ l\ →\ (i\ ⊗\ j)\ ≈ₚ\ (k\ ⊗\ l)\ }\<%
\\
\\[\AgdaEmptyExtraSkip]%
\>[2]\AgdaFunction{p-eq-proj₁}\AgdaSpace{}%
\AgdaSymbol{:}\AgdaSpace{}%
\AgdaSymbol{∀}\AgdaSpace{}%
\AgdaSymbol{\{}\AgdaBound{i}\AgdaSpace{}%
\AgdaBound{j}\AgdaSpace{}%
\AgdaBound{k}\AgdaSpace{}%
\AgdaBound{l}\AgdaSymbol{\}}\AgdaSpace{}%
\AgdaSymbol{→}\AgdaSpace{}%
\AgdaOperator{\AgdaDatatype{\AgdaUnderscore{}≡\AgdaUnderscore{}}}\AgdaSpace{}%
\AgdaSymbol{\{}\AgdaArgument{A}\AgdaSpace{}%
\AgdaSymbol{=}\AgdaSpace{}%
\AgdaDatatype{P}\AgdaSpace{}%
\AgdaSymbol{(}\AgdaGeneralizable{s}\AgdaSpace{}%
\AgdaOperator{\AgdaInductiveConstructor{⊗}}\AgdaSpace{}%
\AgdaGeneralizable{p}\AgdaSymbol{)\}}\AgdaSpace{}%
\AgdaSymbol{(}\AgdaBound{i}\AgdaSpace{}%
\AgdaOperator{\AgdaInductiveConstructor{⊗}}\AgdaSpace{}%
\AgdaBound{j}\AgdaSymbol{)}\AgdaSpace{}%
\AgdaSymbol{(}\AgdaBound{k}\AgdaSpace{}%
\AgdaOperator{\AgdaInductiveConstructor{⊗}}\AgdaSpace{}%
\AgdaBound{l}\AgdaSymbol{)}\AgdaSpace{}%
\AgdaSymbol{→}\AgdaSpace{}%
\AgdaBound{i}\AgdaSpace{}%
\AgdaOperator{\AgdaDatatype{≡}}\AgdaSpace{}%
\AgdaBound{k}\<%
\\
\>[2]\AgdaFunction{p-eq-proj₁}\AgdaSpace{}%
\AgdaInductiveConstructor{refl}\AgdaSpace{}%
\AgdaSymbol{=}\AgdaSpace{}%
\AgdaInductiveConstructor{refl}\<%
\\
\>[0]\<%
\\
\>[2]\AgdaFunction{p-eq-proj₂}\AgdaSpace{}%
\AgdaSymbol{:}\AgdaSpace{}%
\AgdaSymbol{∀}\AgdaSpace{}%
\AgdaSymbol{\{}\AgdaBound{i}\AgdaSpace{}%
\AgdaBound{j}\AgdaSpace{}%
\AgdaBound{k}\AgdaSpace{}%
\AgdaBound{l}\AgdaSymbol{\}}\AgdaSpace{}%
\AgdaSymbol{→}\AgdaSpace{}%
\AgdaOperator{\AgdaDatatype{\AgdaUnderscore{}≡\AgdaUnderscore{}}}\AgdaSpace{}%
\AgdaSymbol{\{}\AgdaArgument{A}\AgdaSpace{}%
\AgdaSymbol{=}\AgdaSpace{}%
\AgdaDatatype{P}\AgdaSpace{}%
\AgdaSymbol{(}\AgdaGeneralizable{s}\AgdaSpace{}%
\AgdaOperator{\AgdaInductiveConstructor{⊗}}\AgdaSpace{}%
\AgdaGeneralizable{p}\AgdaSymbol{)\}}\AgdaSpace{}%
\AgdaSymbol{(}\AgdaBound{i}\AgdaSpace{}%
\AgdaOperator{\AgdaInductiveConstructor{⊗}}\AgdaSpace{}%
\AgdaBound{j}\AgdaSymbol{)}\AgdaSpace{}%
\AgdaSymbol{(}\AgdaBound{k}\AgdaSpace{}%
\AgdaOperator{\AgdaInductiveConstructor{⊗}}\AgdaSpace{}%
\AgdaBound{l}\AgdaSymbol{)}\AgdaSpace{}%
\AgdaSymbol{→}\AgdaSpace{}%
\AgdaBound{j}\AgdaSpace{}%
\AgdaOperator{\AgdaDatatype{≡}}\AgdaSpace{}%
\AgdaBound{l}\<%
\\
\>[2]\AgdaFunction{p-eq-proj₂}\AgdaSpace{}%
\AgdaInductiveConstructor{refl}\AgdaSpace{}%
\AgdaSymbol{=}\AgdaSpace{}%
\AgdaInductiveConstructor{refl}\<%
\\
\\[\AgdaEmptyExtraSkip]%
\>[2]\AgdaOperator{\AgdaFunction{\AgdaUnderscore{}≟ₚ\AgdaUnderscore{}}}\AgdaSpace{}%
\AgdaSymbol{:}\AgdaSpace{}%
\AgdaSymbol{(}\AgdaBound{i}\AgdaSpace{}%
\AgdaBound{j}\AgdaSpace{}%
\AgdaSymbol{:}\AgdaSpace{}%
\AgdaDatatype{P}\AgdaSpace{}%
\AgdaGeneralizable{s}\AgdaSymbol{)}\AgdaSpace{}%
\AgdaSymbol{→}\AgdaSpace{}%
\AgdaRecord{Dec}\AgdaSpace{}%
\AgdaSymbol{(}\AgdaBound{i}\AgdaSpace{}%
\AgdaOperator{\AgdaDatatype{≡}}\AgdaSpace{}%
\AgdaBound{j}\AgdaSymbol{)}\<%
\\
\>[2]\AgdaInductiveConstructor{ι}\AgdaSpace{}%
\AgdaBound{i}\AgdaSpace{}%
\AgdaOperator{\AgdaFunction{≟ₚ}}\AgdaSpace{}%
\AgdaInductiveConstructor{ι}\AgdaSpace{}%
\AgdaBound{j}\AgdaSpace{}%
\AgdaKeyword{with}\AgdaSpace{}%
\AgdaBound{i}\AgdaSpace{}%
\AgdaOperator{\AgdaFunction{Data.Fin.≟}}\AgdaSpace{}%
\AgdaBound{j}\<%
\\
\>[2]\AgdaSymbol{...}\AgdaSpace{}%
\AgdaSymbol{|}\AgdaSpace{}%
\AgdaInductiveConstructor{yes}\AgdaSpace{}%
\AgdaBound{p}\AgdaSpace{}%
\AgdaSymbol{=}\AgdaSpace{}%
\AgdaInductiveConstructor{yes}\AgdaSpace{}%
\AgdaSymbol{(}\AgdaFunction{cong}\AgdaSpace{}%
\AgdaInductiveConstructor{ι}\AgdaSpace{}%
\AgdaBound{p}\AgdaSymbol{)}\<%
\\
\>[2]\AgdaSymbol{...}\AgdaSpace{}%
\AgdaSymbol{|}\AgdaSpace{}%
\AgdaInductiveConstructor{no}\AgdaSpace{}%
\AgdaBound{¬p}\AgdaSpace{}%
\AgdaSymbol{=}\AgdaSpace{}%
\AgdaInductiveConstructor{no}\AgdaSpace{}%
\AgdaSymbol{λ}\AgdaSpace{}%
\AgdaBound{p}\AgdaSpace{}%
\AgdaSymbol{→}\AgdaSpace{}%
\AgdaBound{¬p}\AgdaSpace{}%
\AgdaSymbol{(}\AgdaFunction{ι-injective}\AgdaSpace{}%
\AgdaBound{p}\AgdaSymbol{)}\<%
\\
\>[2]\AgdaSymbol{(}\AgdaBound{i}\AgdaSpace{}%
\AgdaOperator{\AgdaInductiveConstructor{⊗}}\AgdaSpace{}%
\AgdaBound{j}\AgdaSymbol{)}\AgdaSpace{}%
\AgdaOperator{\AgdaFunction{≟ₚ}}\AgdaSpace{}%
\AgdaSymbol{(}\AgdaBound{i′}\AgdaSpace{}%
\AgdaOperator{\AgdaInductiveConstructor{⊗}}\AgdaSpace{}%
\AgdaBound{j′}\AgdaSymbol{)}\AgdaSpace{}%
\AgdaKeyword{with}\AgdaSpace{}%
\AgdaBound{i}\AgdaSpace{}%
\AgdaOperator{\AgdaFunction{≟ₚ}}\AgdaSpace{}%
\AgdaBound{i′}\AgdaSpace{}%
\AgdaSymbol{|}\AgdaSpace{}%
\AgdaBound{j}\AgdaSpace{}%
\AgdaOperator{\AgdaFunction{≟ₚ}}\AgdaSpace{}%
\AgdaBound{j′}\<%
\\
\>[2]\AgdaSymbol{...}\AgdaSpace{}%
\AgdaSymbol{|}\AgdaSpace{}%
\AgdaInductiveConstructor{yes}\AgdaSpace{}%
\AgdaBound{p}\AgdaSpace{}%
\AgdaSymbol{|}\AgdaSpace{}%
\AgdaInductiveConstructor{yes}\AgdaSpace{}%
\AgdaBound{q}\AgdaSpace{}%
\AgdaSymbol{=}\AgdaSpace{}%
\AgdaInductiveConstructor{yes}\AgdaSpace{}%
\AgdaSymbol{(}\AgdaFunction{cong₂}\AgdaSpace{}%
\AgdaOperator{\AgdaInductiveConstructor{\AgdaUnderscore{}⊗\AgdaUnderscore{}}}\AgdaSpace{}%
\AgdaBound{p}\AgdaSpace{}%
\AgdaBound{q}\AgdaSymbol{)}\<%
\\
\>[2]\AgdaSymbol{...}\AgdaSpace{}%
\AgdaSymbol{|}\AgdaSpace{}%
\AgdaInductiveConstructor{yes}\AgdaSpace{}%
\AgdaBound{p}\AgdaSpace{}%
\AgdaSymbol{|}\AgdaSpace{}%
\AgdaInductiveConstructor{no}\AgdaSpace{}%
\AgdaBound{¬q}\AgdaSpace{}%
\AgdaSymbol{=}\AgdaSpace{}%
\AgdaInductiveConstructor{no}\AgdaSpace{}%
\AgdaSymbol{λ}\AgdaSpace{}%
\AgdaBound{r}\AgdaSpace{}%
\AgdaSymbol{→}\AgdaSpace{}%
\AgdaBound{¬q}\AgdaSpace{}%
\AgdaSymbol{(}\AgdaFunction{p-eq-proj₂}\AgdaSpace{}%
\AgdaBound{r}\AgdaSymbol{)}\<%
\\
\>[2]\AgdaSymbol{...}\AgdaSpace{}%
\AgdaSymbol{|}\AgdaSpace{}%
\AgdaInductiveConstructor{no}\AgdaSpace{}%
\AgdaBound{¬p}\AgdaSpace{}%
\AgdaSymbol{|}\AgdaSpace{}%
\AgdaSymbol{\AgdaUnderscore{}}%
\>[22]\AgdaSymbol{=}\AgdaSpace{}%
\AgdaInductiveConstructor{no}\AgdaSpace{}%
\AgdaSymbol{λ}\AgdaSpace{}%
\AgdaBound{r}\AgdaSpace{}%
\AgdaSymbol{→}\AgdaSpace{}%
\AgdaBound{¬p}\AgdaSpace{}%
\AgdaSymbol{(}\AgdaFunction{p-eq-proj₁}\AgdaSpace{}%
\AgdaBound{r}\AgdaSymbol{)}\<%
\\
\>[0]\<%
\end{code}
\begin{mathpar}
\codeblock{\begin{code}%
\>[0][@{}l@{\AgdaIndent{1}}]%
\>[2]\AgdaOperator{\AgdaFunction{\AgdaUnderscore{}⊕ₚ\AgdaUnderscore{}}}\AgdaSpace{}%
\AgdaSymbol{:}\AgdaSpace{}%
\AgdaDatatype{P}\AgdaSpace{}%
\AgdaGeneralizable{s}\AgdaSpace{}%
\AgdaSymbol{→}\AgdaSpace{}%
\AgdaDatatype{P}\AgdaSpace{}%
\AgdaGeneralizable{u}\AgdaSpace{}%
\AgdaSymbol{→}\AgdaSpace{}%
\AgdaOperator{\AgdaDatatype{suc}}\AgdaSpace{}%
\AgdaGeneralizable{p}\AgdaSpace{}%
\AgdaOperator{\AgdaDatatype{≈}}\AgdaSpace{}%
\AgdaGeneralizable{u}\AgdaSpace{}%
\AgdaSymbol{→}\AgdaSpace{}%
\AgdaGeneralizable{s}\AgdaSpace{}%
\AgdaOperator{\AgdaDatatype{+}}\AgdaSpace{}%
\AgdaGeneralizable{p}\AgdaSpace{}%
\AgdaOperator{\AgdaDatatype{≈}}\AgdaSpace{}%
\AgdaGeneralizable{r}\AgdaSpace{}%
\AgdaSymbol{→}\AgdaSpace{}%
\AgdaDatatype{P}\AgdaSpace{}%
\AgdaGeneralizable{r}\<%
\\
\>[2]\AgdaSymbol{(}\AgdaInductiveConstructor{ι}\AgdaSpace{}%
\AgdaBound{i}%
\>[13]\AgdaOperator{\AgdaFunction{⊕ₚ}}\AgdaSpace{}%
\AgdaInductiveConstructor{ι}\AgdaSpace{}%
\AgdaBound{j}\AgdaSymbol{)}%
\>[27]\AgdaInductiveConstructor{ι}%
\>[37]\AgdaInductiveConstructor{ι}%
\>[47]\AgdaSymbol{=}\AgdaSpace{}%
\AgdaInductiveConstructor{ι}\AgdaSpace{}%
\AgdaSymbol{(}\AgdaBound{i}\AgdaSpace{}%
\AgdaOperator{\AgdaFunction{⊕}}\AgdaSpace{}%
\AgdaBound{j}\AgdaSymbol{)}\<%
\\
\>[2]\AgdaSymbol{((}\AgdaBound{a}\AgdaSpace{}%
\AgdaOperator{\AgdaInductiveConstructor{⊗}}\AgdaSpace{}%
\AgdaBound{a₁}\AgdaSymbol{)}%
\>[13]\AgdaOperator{\AgdaFunction{⊕ₚ}}\AgdaSpace{}%
\AgdaSymbol{(}\AgdaBound{b}\AgdaSpace{}%
\AgdaOperator{\AgdaInductiveConstructor{⊗}}\AgdaSpace{}%
\AgdaBound{b₁}\AgdaSymbol{))}%
\>[27]\AgdaSymbol{(}\AgdaBound{s}\AgdaSpace{}%
\AgdaOperator{\AgdaInductiveConstructor{⊗}}\AgdaSpace{}%
\AgdaBound{s₁}\AgdaSymbol{)}%
\>[37]\AgdaSymbol{(}\AgdaBound{p}\AgdaSpace{}%
\AgdaOperator{\AgdaInductiveConstructor{⊗}}\AgdaSpace{}%
\AgdaBound{p₁}\AgdaSymbol{)}%
\>[47]\AgdaSymbol{=}\AgdaSpace{}%
\AgdaSymbol{(}\AgdaBound{a}\AgdaSpace{}%
\AgdaOperator{\AgdaFunction{⊕ₚ}}\AgdaSpace{}%
\AgdaBound{b}\AgdaSymbol{)}\AgdaSpace{}%
\AgdaBound{s}\AgdaSpace{}%
\AgdaBound{p}\AgdaSpace{}%
\AgdaOperator{\AgdaInductiveConstructor{⊗}}\AgdaSpace{}%
\AgdaSymbol{(}\AgdaBound{a₁}\AgdaSpace{}%
\AgdaOperator{\AgdaFunction{⊕ₚ}}\AgdaSpace{}%
\AgdaBound{b₁}\AgdaSymbol{)}\AgdaSpace{}%
\AgdaBound{s₁}\AgdaSpace{}%
\AgdaBound{p₁}\<%
\\
\>[0]\<%
\\
\>[2]\AgdaOperator{\AgdaFunction{\AgdaUnderscore{}⊝ₚ\AgdaUnderscore{}}}\AgdaSpace{}%
\AgdaSymbol{:}\AgdaSpace{}%
\AgdaSymbol{(}\AgdaBound{i}\AgdaSpace{}%
\AgdaSymbol{:}\AgdaSpace{}%
\AgdaDatatype{P}\AgdaSpace{}%
\AgdaGeneralizable{r}\AgdaSymbol{)}\AgdaSpace{}%
\AgdaSymbol{(}\AgdaBound{j}\AgdaSpace{}%
\AgdaSymbol{:}\AgdaSpace{}%
\AgdaDatatype{P}\AgdaSpace{}%
\AgdaGeneralizable{s}\AgdaSymbol{)}\AgdaSpace{}%
\AgdaSymbol{(}\AgdaBound{su}\AgdaSpace{}%
\AgdaSymbol{:}\AgdaSpace{}%
\AgdaOperator{\AgdaDatatype{suc}}\AgdaSpace{}%
\AgdaGeneralizable{p}\AgdaSpace{}%
\AgdaOperator{\AgdaDatatype{≈}}\AgdaSpace{}%
\AgdaGeneralizable{u}\AgdaSymbol{)}\AgdaSpace{}%
\AgdaSymbol{(}\AgdaBound{sp}\AgdaSpace{}%
\AgdaSymbol{:}\AgdaSpace{}%
\AgdaGeneralizable{s}\AgdaSpace{}%
\AgdaOperator{\AgdaDatatype{+}}\AgdaSpace{}%
\AgdaGeneralizable{p}\AgdaSpace{}%
\AgdaOperator{\AgdaDatatype{≈}}\AgdaSpace{}%
\AgdaGeneralizable{r}\AgdaSymbol{)}\AgdaSpace{}%
\AgdaSymbol{→}\AgdaSpace{}%
\AgdaRecord{Dec}\AgdaSpace{}%
\AgdaSymbol{(}\AgdaFunction{∃}\AgdaSpace{}%
\AgdaSymbol{λ}\AgdaSpace{}%
\AgdaBound{k}\AgdaSpace{}%
\AgdaSymbol{→}\AgdaSpace{}%
\AgdaSymbol{(}\AgdaBound{j}\AgdaSpace{}%
\AgdaOperator{\AgdaFunction{⊕ₚ}}\AgdaSpace{}%
\AgdaBound{k}\AgdaSymbol{)}\AgdaSpace{}%
\AgdaBound{su}\AgdaSpace{}%
\AgdaBound{sp}\AgdaSpace{}%
\AgdaOperator{\AgdaDatatype{≡}}\AgdaSpace{}%
\AgdaBound{i}\AgdaSymbol{)}\<%
\end{code}}
\end{mathpar}
We do not show the implementation of the \AF{⊝ₚ}, but it very much follows the
structure of \AF{⊕ₚ}: we apply \AF{⊝} on the leaves and we recurse on the product
shape with a little bit plumbing to construct the proof of (non-)existence of the
inverse.
\begin{code}[hide]%
\>[2]\AgdaSymbol{(}\AgdaInductiveConstructor{ι}\AgdaSpace{}%
\AgdaBound{i}\AgdaSpace{}%
\AgdaOperator{\AgdaFunction{⊝ₚ}}\AgdaSpace{}%
\AgdaInductiveConstructor{ι}\AgdaSpace{}%
\AgdaBound{j}\AgdaSymbol{)}\AgdaSpace{}%
\AgdaInductiveConstructor{ι}\AgdaSpace{}%
\AgdaInductiveConstructor{ι}\AgdaSpace{}%
\AgdaKeyword{with}\AgdaSpace{}%
\AgdaBound{i}\AgdaSpace{}%
\AgdaOperator{\AgdaFunction{⊝}}\AgdaSpace{}%
\AgdaBound{j}\<%
\\
\>[2]\AgdaSymbol{...}\AgdaSpace{}%
\AgdaSymbol{|}\AgdaSpace{}%
\AgdaInductiveConstructor{yes}%
\>[13]\AgdaSymbol{(}\AgdaBound{k}\AgdaSpace{}%
\AgdaOperator{\AgdaInductiveConstructor{,}}\AgdaSpace{}%
\AgdaBound{p}\AgdaSymbol{)}%
\>[23]\AgdaSymbol{=}\AgdaSpace{}%
\AgdaInductiveConstructor{yes}\AgdaSpace{}%
\AgdaSymbol{(}\AgdaInductiveConstructor{ι}\AgdaSpace{}%
\AgdaBound{k}\AgdaSpace{}%
\AgdaOperator{\AgdaInductiveConstructor{,}}\AgdaSpace{}%
\AgdaFunction{cong}\AgdaSpace{}%
\AgdaInductiveConstructor{ι}\AgdaSpace{}%
\AgdaBound{p}\AgdaSymbol{)}\<%
\\
\>[2]\AgdaSymbol{...}\AgdaSpace{}%
\AgdaSymbol{|}\AgdaSpace{}%
\AgdaInductiveConstructor{no}%
\>[13]\AgdaBound{¬p}%
\>[23]\AgdaSymbol{=}\AgdaSpace{}%
\AgdaInductiveConstructor{no}\AgdaSpace{}%
\AgdaSymbol{λ}\AgdaSpace{}%
\AgdaSymbol{\{}\AgdaSpace{}%
\AgdaSymbol{(}\AgdaInductiveConstructor{ι}\AgdaSpace{}%
\AgdaBound{k}\AgdaSpace{}%
\AgdaOperator{\AgdaInductiveConstructor{,}}\AgdaSpace{}%
\AgdaBound{p}\AgdaSymbol{)}\AgdaSpace{}%
\AgdaSymbol{→}\AgdaSpace{}%
\AgdaBound{¬p}\AgdaSpace{}%
\AgdaSymbol{(}\AgdaBound{k}\AgdaSpace{}%
\AgdaOperator{\AgdaInductiveConstructor{,}}\AgdaSpace{}%
\AgdaFunction{ι-injective}\AgdaSpace{}%
\AgdaBound{p}\AgdaSymbol{)}\AgdaSpace{}%
\AgdaSymbol{\}}\<%
\\
\>[2]\AgdaSymbol{((}\AgdaBound{i}\AgdaSpace{}%
\AgdaOperator{\AgdaInductiveConstructor{⊗}}\AgdaSpace{}%
\AgdaBound{i′}\AgdaSymbol{)}\AgdaSpace{}%
\AgdaOperator{\AgdaFunction{⊝ₚ}}\AgdaSpace{}%
\AgdaSymbol{(}\AgdaBound{j}\AgdaSpace{}%
\AgdaOperator{\AgdaInductiveConstructor{⊗}}\AgdaSpace{}%
\AgdaBound{j′}\AgdaSymbol{))}\AgdaSpace{}%
\AgdaSymbol{(}\AgdaBound{su}\AgdaSpace{}%
\AgdaOperator{\AgdaInductiveConstructor{⊗}}\AgdaSpace{}%
\AgdaBound{su′}\AgdaSymbol{)}\AgdaSpace{}%
\AgdaSymbol{(}\AgdaBound{sp}\AgdaSpace{}%
\AgdaOperator{\AgdaInductiveConstructor{⊗}}\AgdaSpace{}%
\AgdaBound{sp′}\AgdaSymbol{)}\AgdaSpace{}%
\AgdaKeyword{with}\AgdaSpace{}%
\AgdaSymbol{(}\AgdaBound{i}\AgdaSpace{}%
\AgdaOperator{\AgdaFunction{⊝ₚ}}\AgdaSpace{}%
\AgdaBound{j}\AgdaSymbol{)}\AgdaSpace{}%
\AgdaBound{su}\AgdaSpace{}%
\AgdaBound{sp}\<%
\\
\>[2]\AgdaSymbol{...}\AgdaSpace{}%
\AgdaSymbol{|}\AgdaSpace{}%
\AgdaInductiveConstructor{no}%
\>[13]\AgdaBound{¬p}%
\>[23]\AgdaSymbol{=}\AgdaSpace{}%
\AgdaInductiveConstructor{no}\AgdaSpace{}%
\AgdaSymbol{λ}\AgdaSpace{}%
\AgdaSymbol{\{}\AgdaSpace{}%
\AgdaSymbol{(}\AgdaBound{k}\AgdaSpace{}%
\AgdaOperator{\AgdaInductiveConstructor{⊗}}\AgdaSpace{}%
\AgdaBound{k′}\AgdaSpace{}%
\AgdaOperator{\AgdaInductiveConstructor{,}}\AgdaSpace{}%
\AgdaBound{p}\AgdaSymbol{)}\AgdaSpace{}%
\AgdaSymbol{→}\AgdaSpace{}%
\AgdaBound{¬p}\AgdaSpace{}%
\AgdaSymbol{(}\AgdaBound{k}\AgdaSpace{}%
\AgdaOperator{\AgdaInductiveConstructor{,}}\AgdaSpace{}%
\AgdaFunction{⊗-fst-≡}\AgdaSpace{}%
\AgdaBound{p}\AgdaSymbol{)\}}\<%
\\
\>[2]\AgdaSymbol{...}\AgdaSpace{}%
\AgdaSymbol{|}\AgdaSpace{}%
\AgdaInductiveConstructor{yes}%
\>[13]\AgdaSymbol{(}\AgdaBound{k}\AgdaSpace{}%
\AgdaOperator{\AgdaInductiveConstructor{,}}%
\>[19]\AgdaBound{p}\AgdaSymbol{)}%
\>[23]\AgdaKeyword{with}\AgdaSpace{}%
\AgdaSymbol{(}\AgdaBound{i′}\AgdaSpace{}%
\AgdaOperator{\AgdaFunction{⊝ₚ}}\AgdaSpace{}%
\AgdaBound{j′}\AgdaSymbol{)}\AgdaSpace{}%
\AgdaBound{su′}\AgdaSpace{}%
\AgdaBound{sp′}\<%
\\
\>[2]\AgdaSymbol{...}\AgdaSpace{}%
\AgdaSymbol{|}\AgdaSpace{}%
\AgdaInductiveConstructor{no}%
\>[13]\AgdaBound{¬p}%
\>[23]\AgdaSymbol{=}\AgdaSpace{}%
\AgdaInductiveConstructor{no}\AgdaSpace{}%
\AgdaSymbol{λ}\AgdaSpace{}%
\AgdaSymbol{\{}\AgdaSpace{}%
\AgdaSymbol{(}\AgdaBound{k}\AgdaSpace{}%
\AgdaOperator{\AgdaInductiveConstructor{⊗}}\AgdaSpace{}%
\AgdaBound{k′}\AgdaSpace{}%
\AgdaOperator{\AgdaInductiveConstructor{,}}\AgdaSpace{}%
\AgdaBound{p}\AgdaSymbol{)}\AgdaSpace{}%
\AgdaSymbol{→}\AgdaSpace{}%
\AgdaBound{¬p}\AgdaSpace{}%
\AgdaSymbol{(}\AgdaBound{k′}\AgdaSpace{}%
\AgdaOperator{\AgdaInductiveConstructor{,}}\AgdaSpace{}%
\AgdaFunction{⊗-snd-≡}\AgdaSpace{}%
\AgdaBound{p}\AgdaSymbol{)\}}\<%
\\
\>[2]\AgdaSymbol{...}\AgdaSpace{}%
\AgdaSymbol{|}\AgdaSpace{}%
\AgdaInductiveConstructor{yes}%
\>[13]\AgdaSymbol{(}\AgdaBound{k′}\AgdaSpace{}%
\AgdaOperator{\AgdaInductiveConstructor{,}}\AgdaSpace{}%
\AgdaBound{p′}\AgdaSymbol{)}\AgdaSpace{}%
\AgdaSymbol{=}\AgdaSpace{}%
\AgdaInductiveConstructor{yes}\AgdaSpace{}%
\AgdaSymbol{(}\AgdaBound{k}\AgdaSpace{}%
\AgdaOperator{\AgdaInductiveConstructor{⊗}}\AgdaSpace{}%
\AgdaBound{k′}\AgdaSpace{}%
\AgdaOperator{\AgdaInductiveConstructor{,}}\AgdaSpace{}%
\AgdaFunction{cong₂}\AgdaSpace{}%
\AgdaOperator{\AgdaInductiveConstructor{\AgdaUnderscore{}⊗\AgdaUnderscore{}}}\AgdaSpace{}%
\AgdaBound{p}\AgdaSpace{}%
\AgdaBound{p′}\AgdaSymbol{)}\<%
\end{code}

Our generalised \AF{slide} looks very much the same as its 1-dimensional
counterpart.  All the difference lies in the index computation.  We also
introduce a section of the slide that we call \AF{backslide} which embed
a $(1+p)$-dimensional array into a $(s+p)$-dimensional one at offset $i$
using some the provided default element \AB{def}.
\begin{mathpar}
\codeblock{\begin{code}%
\>[2]\AgdaFunction{slide}\AgdaSpace{}%
\AgdaSymbol{:}\AgdaSpace{}%
\AgdaDatatype{P}\AgdaSpace{}%
\AgdaGeneralizable{s}\AgdaSpace{}%
\AgdaSymbol{→}\AgdaSpace{}%
\AgdaGeneralizable{s}\AgdaSpace{}%
\AgdaOperator{\AgdaDatatype{+}}\AgdaSpace{}%
\AgdaGeneralizable{p}\AgdaSpace{}%
\AgdaOperator{\AgdaDatatype{≈}}\AgdaSpace{}%
\AgdaGeneralizable{r}\AgdaSpace{}%
\AgdaSymbol{→}\AgdaSpace{}%
\AgdaFunction{Ar}\AgdaSpace{}%
\AgdaGeneralizable{r}\AgdaSpace{}%
\AgdaGeneralizable{X}\AgdaSpace{}%
\AgdaSymbol{→}\AgdaSpace{}%
\AgdaOperator{\AgdaDatatype{suc}}\AgdaSpace{}%
\AgdaGeneralizable{p}\AgdaSpace{}%
\AgdaOperator{\AgdaDatatype{≈}}\AgdaSpace{}%
\AgdaGeneralizable{u}\AgdaSpace{}%
\AgdaSymbol{→}\AgdaSpace{}%
\AgdaFunction{Ar}\AgdaSpace{}%
\AgdaGeneralizable{u}\AgdaSpace{}%
\AgdaGeneralizable{X}\<%
\\
\>[2]\AgdaFunction{slide}\AgdaSpace{}%
\AgdaBound{i}\AgdaSpace{}%
\AgdaBound{pl}\AgdaSpace{}%
\AgdaBound{a}\AgdaSpace{}%
\AgdaBound{su}\AgdaSpace{}%
\AgdaBound{j}\AgdaSpace{}%
\AgdaSymbol{=}\AgdaSpace{}%
\AgdaBound{a}\AgdaSpace{}%
\AgdaSymbol{((}\AgdaBound{i}\AgdaSpace{}%
\AgdaOperator{\AgdaFunction{⊕ₚ}}\AgdaSpace{}%
\AgdaBound{j}\AgdaSymbol{)}\AgdaSpace{}%
\AgdaBound{su}\AgdaSpace{}%
\AgdaBound{pl}\AgdaSymbol{)}\<%
\\
\>[0]\<%
\\
\>[2]\AgdaFunction{backslide}\AgdaSpace{}%
\AgdaSymbol{:}\AgdaSpace{}%
\AgdaDatatype{P}\AgdaSpace{}%
\AgdaGeneralizable{s}\AgdaSpace{}%
\AgdaSymbol{→}\AgdaSpace{}%
\AgdaFunction{Ar}\AgdaSpace{}%
\AgdaGeneralizable{u}\AgdaSpace{}%
\AgdaGeneralizable{X}\AgdaSpace{}%
\AgdaSymbol{→}\AgdaSpace{}%
\AgdaOperator{\AgdaDatatype{suc}}\AgdaSpace{}%
\AgdaGeneralizable{p}\AgdaSpace{}%
\AgdaOperator{\AgdaDatatype{≈}}\AgdaSpace{}%
\AgdaGeneralizable{u}\AgdaSpace{}%
\AgdaSymbol{→}\AgdaSpace{}%
\AgdaSymbol{(}\AgdaBound{def}\AgdaSpace{}%
\AgdaSymbol{:}\AgdaSpace{}%
\AgdaGeneralizable{X}\AgdaSymbol{)}\AgdaSpace{}%
\AgdaSymbol{→}\AgdaSpace{}%
\AgdaGeneralizable{s}\AgdaSpace{}%
\AgdaOperator{\AgdaDatatype{+}}\AgdaSpace{}%
\AgdaGeneralizable{p}\AgdaSpace{}%
\AgdaOperator{\AgdaDatatype{≈}}\AgdaSpace{}%
\AgdaGeneralizable{r}\AgdaSpace{}%
\AgdaSymbol{→}\AgdaSpace{}%
\AgdaFunction{Ar}\AgdaSpace{}%
\AgdaGeneralizable{r}\AgdaSpace{}%
\AgdaGeneralizable{X}\<%
\\
\>[2]\AgdaFunction{backslide}\AgdaSpace{}%
\AgdaBound{i}\AgdaSpace{}%
\AgdaBound{a}\AgdaSpace{}%
\AgdaBound{su}\AgdaSpace{}%
\AgdaBound{def}\AgdaSpace{}%
\AgdaBound{pl}\AgdaSpace{}%
\AgdaBound{j}\AgdaSpace{}%
\AgdaKeyword{with}\AgdaSpace{}%
\AgdaSymbol{((}\AgdaBound{j}\AgdaSpace{}%
\AgdaOperator{\AgdaFunction{⊝ₚ}}\AgdaSpace{}%
\AgdaBound{i}\AgdaSymbol{)}\AgdaSpace{}%
\AgdaBound{su}\AgdaSpace{}%
\AgdaBound{pl}\AgdaSymbol{)}\<%
\\
\>[2]\AgdaSymbol{...}\AgdaSpace{}%
\AgdaSymbol{|}\AgdaSpace{}%
\AgdaInductiveConstructor{yes}\AgdaSpace{}%
\AgdaSymbol{(}\AgdaBound{k}\AgdaSpace{}%
\AgdaOperator{\AgdaInductiveConstructor{,}}\AgdaSpace{}%
\AgdaSymbol{\AgdaUnderscore{})}%
\>[21]\AgdaSymbol{=}\AgdaSpace{}%
\AgdaBound{a}\AgdaSpace{}%
\AgdaBound{k}\<%
\\
\>[2]\AgdaCatchallClause{\AgdaSymbol{...}}\AgdaSpace{}%
\AgdaCatchallClause{\AgdaSymbol{|}}\AgdaSpace{}%
\AgdaCatchallClause{\AgdaSymbol{\AgdaUnderscore{}}}%
\>[21]\AgdaSymbol{=}\AgdaSpace{}%
\AgdaBound{def}\<%
\end{code}}
\end{mathpar}

\subsection{Remaining primitives}
In the rest of this section we implement the remaining CNN-specific primitives.
We are going to use the builtin Float type that we call \AD{ℝ} so that we can
run our specification with concrete values.  However, all we require from \AD{R}
is a set of standard arithmetic operations.  Therefore, \AD{ℝ} can be abstracted
out as a parameter.

Generalised convolution is given by \AF{conv}, and it is almost identical to its
1-dimensional counterpart (except it used \AF{slide} instead of \AF{slide₁}).
The \AF{mconv} runs $u$ \AF{conv}s adds biases to each of them from the array $b$.
\begin{code}[hide]%
\>[0]\AgdaKeyword{module}\AgdaSpace{}%
\AgdaModule{CNN}\AgdaSpace{}%
\AgdaKeyword{where}\<%
\\
\>[0][@{}l@{\AgdaIndent{0}}]%
\>[2]\AgdaKeyword{open}\AgdaSpace{}%
\AgdaKeyword{import}\AgdaSpace{}%
\AgdaModule{Data.Nat}\AgdaSpace{}%
\AgdaSymbol{as}\AgdaSpace{}%
\AgdaModule{ℕ}\AgdaSpace{}%
\AgdaKeyword{using}\AgdaSpace{}%
\AgdaSymbol{(}\AgdaDatatype{ℕ}\AgdaSymbol{)}\<%
\\
\>[2]\AgdaKeyword{open}\AgdaSpace{}%
\AgdaKeyword{import}\AgdaSpace{}%
\AgdaModule{Data.Float}\AgdaSpace{}%
\AgdaSymbol{as}\AgdaSpace{}%
\AgdaModule{F}\AgdaSpace{}%
\AgdaKeyword{using}\AgdaSpace{}%
\AgdaSymbol{(}\AgdaPrimitive{\AgdaUnderscore{}+\AgdaUnderscore{}}\AgdaSymbol{;}\AgdaSpace{}%
\AgdaPrimitive{\AgdaUnderscore{}*\AgdaUnderscore{}}\AgdaSymbol{;}\AgdaSpace{}%
\AgdaPrimitive{\AgdaUnderscore{}÷\AgdaUnderscore{}}\AgdaSymbol{;}\AgdaSpace{}%
\AgdaPrimitive{e\textasciicircum{}\AgdaUnderscore{}}\AgdaSymbol{;}\AgdaSpace{}%
\AgdaPrimitive{-\AgdaUnderscore{}}\AgdaSymbol{;}\AgdaSpace{}%
\AgdaPrimitive{fromℕ}\AgdaSymbol{)}\AgdaSpace{}%
\AgdaKeyword{renaming}\AgdaSpace{}%
\AgdaSymbol{(}\AgdaPostulate{Float}\AgdaSpace{}%
\AgdaSymbol{to}\AgdaSpace{}%
\AgdaPostulate{ℝ}\AgdaSymbol{)}\<%
\\
\>[2]\AgdaKeyword{open}\AgdaSpace{}%
\AgdaModule{Array}\<%
\end{code}
\begin{mathpar}
\codeblock{\begin{code}%
\>[2]\AgdaFunction{conv}\AgdaSpace{}%
\AgdaSymbol{:}\AgdaSpace{}%
\AgdaGeneralizable{s}\AgdaSpace{}%
\AgdaOperator{\AgdaDatatype{+}}\AgdaSpace{}%
\AgdaGeneralizable{p}\AgdaSpace{}%
\AgdaOperator{\AgdaDatatype{≈}}\AgdaSpace{}%
\AgdaGeneralizable{r}\AgdaSpace{}%
\AgdaSymbol{→}\AgdaSpace{}%
\AgdaFunction{Ar}\AgdaSpace{}%
\AgdaGeneralizable{r}\AgdaSpace{}%
\AgdaPostulate{ℝ}\AgdaSpace{}%
\AgdaSymbol{→}\AgdaSpace{}%
\AgdaFunction{Ar}\AgdaSpace{}%
\AgdaGeneralizable{s}\AgdaSpace{}%
\AgdaPostulate{ℝ}\AgdaSpace{}%
\AgdaSymbol{→}\AgdaSpace{}%
\AgdaOperator{\AgdaDatatype{suc}}\AgdaSpace{}%
\AgdaGeneralizable{p}\AgdaSpace{}%
\AgdaOperator{\AgdaDatatype{≈}}\AgdaSpace{}%
\AgdaGeneralizable{u}\AgdaSpace{}%
\AgdaSymbol{→}\AgdaSpace{}%
\AgdaFunction{Ar}\AgdaSpace{}%
\AgdaGeneralizable{u}\AgdaSpace{}%
\AgdaPostulate{ℝ}\<%
\\
\>[2]\AgdaFunction{conv}\AgdaSpace{}%
\AgdaBound{sp}\AgdaSpace{}%
\AgdaBound{a}\AgdaSpace{}%
\AgdaBound{w}\AgdaSpace{}%
\AgdaBound{su}\AgdaSpace{}%
\AgdaSymbol{=}\AgdaSpace{}%
\AgdaFunction{sum}\AgdaSpace{}%
\AgdaSymbol{(}\AgdaFunction{zipWith}\AgdaSpace{}%
\AgdaOperator{\AgdaPrimitive{\AgdaUnderscore{}+\AgdaUnderscore{}}}\AgdaSymbol{)}\AgdaSpace{}%
\AgdaSymbol{(}\AgdaFunction{K}\AgdaSpace{}%
\AgdaNumber{0.0}\AgdaSymbol{)}\AgdaSpace{}%
\AgdaSymbol{λ}\AgdaSpace{}%
\AgdaBound{i}\AgdaSpace{}%
\AgdaSymbol{→}\AgdaSpace{}%
\AgdaFunction{map}\AgdaSpace{}%
\AgdaSymbol{(}\AgdaBound{w}\AgdaSpace{}%
\AgdaBound{i}\AgdaSpace{}%
\AgdaOperator{\AgdaPrimitive{*\AgdaUnderscore{}}}\AgdaSymbol{)}\AgdaSpace{}%
\AgdaSymbol{(}\AgdaFunction{slide}\AgdaSpace{}%
\AgdaBound{i}\AgdaSpace{}%
\AgdaBound{sp}\AgdaSpace{}%
\AgdaBound{a}\AgdaSpace{}%
\AgdaBound{su}\AgdaSymbol{)}\<%
\end{code}}
\and
\codeblock{\begin{code}%
\>[2]\AgdaFunction{mconv}%
\>[1771I]\AgdaSymbol{:}\AgdaSpace{}%
\AgdaGeneralizable{s}\AgdaSpace{}%
\AgdaOperator{\AgdaDatatype{+}}\AgdaSpace{}%
\AgdaGeneralizable{p}\AgdaSpace{}%
\AgdaOperator{\AgdaDatatype{≈}}\AgdaSpace{}%
\AgdaGeneralizable{r}\AgdaSpace{}%
\AgdaSymbol{→}\AgdaSpace{}%
\AgdaSymbol{(}\AgdaBound{inp}\AgdaSpace{}%
\AgdaSymbol{:}\AgdaSpace{}%
\AgdaFunction{Ar}\AgdaSpace{}%
\AgdaGeneralizable{r}\AgdaSpace{}%
\AgdaPostulate{ℝ}\AgdaSymbol{)}\AgdaSpace{}%
\AgdaSymbol{(}\AgdaBound{w}\AgdaSpace{}%
\AgdaSymbol{:}\AgdaSpace{}%
\AgdaFunction{Ar}\AgdaSpace{}%
\AgdaSymbol{(}\AgdaGeneralizable{u}\AgdaSpace{}%
\AgdaOperator{\AgdaInductiveConstructor{⊗}}\AgdaSpace{}%
\AgdaGeneralizable{s}\AgdaSymbol{)}\AgdaSpace{}%
\AgdaPostulate{ℝ}\AgdaSymbol{)}\AgdaSpace{}%
\AgdaSymbol{(}\AgdaBound{b}\AgdaSpace{}%
\AgdaSymbol{:}\AgdaSpace{}%
\AgdaFunction{Ar}\AgdaSpace{}%
\AgdaGeneralizable{u}\AgdaSpace{}%
\AgdaPostulate{ℝ}\AgdaSymbol{)}\<%
\\
\>[.][@{}l@{}]\<[1771I]%
\>[8]\AgdaSymbol{→}\AgdaSpace{}%
\AgdaOperator{\AgdaDatatype{suc}}\AgdaSpace{}%
\AgdaGeneralizable{p}\AgdaSpace{}%
\AgdaOperator{\AgdaDatatype{≈}}\AgdaSpace{}%
\AgdaGeneralizable{q}\AgdaSpace{}%
\AgdaSymbol{→}\AgdaSpace{}%
\AgdaFunction{Ar}\AgdaSpace{}%
\AgdaSymbol{(}\AgdaGeneralizable{u}\AgdaSpace{}%
\AgdaOperator{\AgdaInductiveConstructor{⊗}}\AgdaSpace{}%
\AgdaGeneralizable{q}\AgdaSymbol{)}\AgdaSpace{}%
\AgdaPostulate{ℝ}\<%
\\
\>[2]\AgdaFunction{mconv}\AgdaSpace{}%
\AgdaBound{sp}\AgdaSpace{}%
\AgdaBound{inp}\AgdaSpace{}%
\AgdaBound{w}\AgdaSpace{}%
\AgdaBound{b}\AgdaSpace{}%
\AgdaBound{su}\AgdaSpace{}%
\AgdaSymbol{=}\AgdaSpace{}%
\AgdaFunction{unnest}\AgdaSpace{}%
\AgdaSymbol{λ}\AgdaSpace{}%
\AgdaBound{i}\AgdaSpace{}%
\AgdaSymbol{→}\AgdaSpace{}%
\AgdaFunction{map}\AgdaSpace{}%
\AgdaSymbol{(}\AgdaBound{b}\AgdaSpace{}%
\AgdaBound{i}\AgdaSpace{}%
\AgdaOperator{\AgdaPrimitive{+\AgdaUnderscore{}}}\AgdaSymbol{)}\AgdaSpace{}%
\AgdaSymbol{(}\AgdaFunction{conv}\AgdaSpace{}%
\AgdaBound{sp}\AgdaSpace{}%
\AgdaBound{inp}\AgdaSpace{}%
\AgdaSymbol{(}\AgdaFunction{nest}\AgdaSpace{}%
\AgdaBound{w}\AgdaSpace{}%
\AgdaBound{i}\AgdaSymbol{)}\AgdaSpace{}%
\AgdaBound{su}\AgdaSymbol{)}\<%
\end{code}}
\end{mathpar}
The logistic function computes ${1}/(1 + e^{-x})$ for every element in the array.
\begin{mathpar}
\codeblock{\begin{code}%
\>[2]\AgdaFunction{logistic}\AgdaSpace{}%
\AgdaSymbol{:}\AgdaSpace{}%
\AgdaFunction{Ar}\AgdaSpace{}%
\AgdaGeneralizable{s}\AgdaSpace{}%
\AgdaPostulate{ℝ}\AgdaSpace{}%
\AgdaSymbol{→}\AgdaSpace{}%
\AgdaFunction{Ar}\AgdaSpace{}%
\AgdaGeneralizable{s}\AgdaSpace{}%
\AgdaPostulate{ℝ}\<%
\\
\>[2]\AgdaFunction{logistic}\AgdaSpace{}%
\AgdaSymbol{=}\AgdaSpace{}%
\AgdaFunction{map}\AgdaSpace{}%
\AgdaSymbol{λ}\AgdaSpace{}%
\AgdaBound{x}\AgdaSpace{}%
\AgdaSymbol{→}\AgdaSpace{}%
\AgdaNumber{1.0}\AgdaSpace{}%
\AgdaOperator{\AgdaPrimitive{÷}}\AgdaSpace{}%
\AgdaSymbol{(}\AgdaNumber{1.0}\AgdaSpace{}%
\AgdaOperator{\AgdaPrimitive{+}}\AgdaSpace{}%
\AgdaOperator{\AgdaPrimitive{e\textasciicircum{}}}\AgdaSpace{}%
\AgdaSymbol{(}\AgdaOperator{\AgdaPrimitive{-}}\AgdaSpace{}%
\AgdaBound{x}\AgdaSymbol{))}\<%
\end{code}}
\end{mathpar}

\paragraph{Average Pooling}
One of the steps of the machine learning algorithm is average pooling which
splits an array into sub-blocks and computes the average for every such
block.  While this sounds almost trivial, implementing this generally is
quite tricky.  In the proposed framework it is
not straight-forward to block an array of shape (\AC{ι} 12 \AC{⊗} \AC{ι} 12)
into an array of shape  ((\AC{ι} 6 \AC{⊗} \AC{ι} 6) \AC{⊗} (\AC{ι} 2 \AC{⊗} \AC{ι} 2)).
The difficulty is in preserving local neighbourhood within the blocks (for example
it would be wrong to flatten the array and then reshape it into the desired shape
as the local neighbourhood will be lost).  At the same time, it would be inconvenient
to block arrays beforehand as this does not go well with \AF{slides}.  Therefore
we introduce the mechanism to \AF{block} and \AF{unblock} arrays of shape $(s * p)$
and ($s$ \AF{⊗} $p$).

The key to array blocking is a reshaping operation that turns arrays
of shape (($s$ \AC{⊗} $p$) \AC{⊗} ($q$ \AC{⊗} $r$)) into arrays of shape 
(($s$ \AC{⊗} $q$) \AC{⊗} ($p$ \AC{⊗} $r$)), by swapping $p$ and $q$.
We express a \AF{Reshape} relation, and as it follows from the type
this reshape is a self inverse, as can be also seen from the theorem below:
\begin{mathpar}
\codeblock{\begin{code}%
\>[2]\AgdaFunction{rblock}\AgdaSpace{}%
\AgdaSymbol{:}\AgdaSpace{}%
\AgdaDatatype{Reshape}\AgdaSpace{}%
\AgdaSymbol{((}\AgdaGeneralizable{s}\AgdaSpace{}%
\AgdaOperator{\AgdaInductiveConstructor{⊗}}\AgdaSpace{}%
\AgdaGeneralizable{p}\AgdaSymbol{)}\AgdaSpace{}%
\AgdaOperator{\AgdaInductiveConstructor{⊗}}\AgdaSpace{}%
\AgdaSymbol{(}\AgdaGeneralizable{q}\AgdaSpace{}%
\AgdaOperator{\AgdaInductiveConstructor{⊗}}\AgdaSpace{}%
\AgdaGeneralizable{r}\AgdaSymbol{))}\AgdaSpace{}%
\AgdaSymbol{((}\AgdaGeneralizable{s}\AgdaSpace{}%
\AgdaOperator{\AgdaInductiveConstructor{⊗}}\AgdaSpace{}%
\AgdaGeneralizable{q}\AgdaSymbol{)}\AgdaSpace{}%
\AgdaOperator{\AgdaInductiveConstructor{⊗}}\AgdaSpace{}%
\AgdaSymbol{(}\AgdaGeneralizable{p}\AgdaSpace{}%
\AgdaOperator{\AgdaInductiveConstructor{⊗}}\AgdaSpace{}%
\AgdaGeneralizable{r}\AgdaSymbol{))}\<%
\\
\>[2]\AgdaFunction{rblock}\AgdaSpace{}%
\AgdaSymbol{=}\AgdaSpace{}%
\AgdaInductiveConstructor{assocl}\AgdaSpace{}%
\AgdaOperator{\AgdaInductiveConstructor{∙}}\AgdaSpace{}%
\AgdaInductiveConstructor{eq}\AgdaSpace{}%
\AgdaOperator{\AgdaInductiveConstructor{,}}\AgdaSpace{}%
\AgdaSymbol{(}\AgdaInductiveConstructor{assocr}\AgdaSpace{}%
\AgdaOperator{\AgdaInductiveConstructor{∙}}\AgdaSpace{}%
\AgdaInductiveConstructor{swap}\AgdaSpace{}%
\AgdaOperator{\AgdaInductiveConstructor{,}}\AgdaSpace{}%
\AgdaInductiveConstructor{eq}\AgdaSpace{}%
\AgdaOperator{\AgdaInductiveConstructor{∙}}\AgdaSpace{}%
\AgdaInductiveConstructor{assocl}\AgdaSymbol{)}\AgdaSpace{}%
\AgdaOperator{\AgdaInductiveConstructor{∙}}\AgdaSpace{}%
\AgdaInductiveConstructor{assocr}\<%
\\
\\[\AgdaEmptyExtraSkip]%
\>[2]\AgdaFunction{rblock-selfinv}\AgdaSpace{}%
\AgdaSymbol{:}\AgdaSpace{}%
\AgdaSymbol{∀}\AgdaSpace{}%
\AgdaBound{i}\AgdaSpace{}%
\AgdaSymbol{→}\AgdaSpace{}%
\AgdaBound{i}\AgdaSpace{}%
\AgdaOperator{\AgdaFunction{⟨}}\AgdaSpace{}%
\AgdaFunction{rev}\AgdaSpace{}%
\AgdaSymbol{(}\AgdaFunction{rblock}\AgdaSpace{}%
\AgdaSymbol{\{}\AgdaGeneralizable{s}\AgdaSymbol{\}}\AgdaSpace{}%
\AgdaSymbol{\{}\AgdaGeneralizable{p}\AgdaSymbol{\}}\AgdaSpace{}%
\AgdaSymbol{\{}\AgdaGeneralizable{q}\AgdaSymbol{\}}\AgdaSpace{}%
\AgdaSymbol{\{}\AgdaGeneralizable{r}\AgdaSymbol{\})}\AgdaSpace{}%
\AgdaOperator{\AgdaFunction{⟩}}\AgdaSpace{}%
\AgdaOperator{\AgdaDatatype{≡}}\AgdaSpace{}%
\AgdaBound{i}\AgdaSpace{}%
\AgdaOperator{\AgdaFunction{⟨}}\AgdaSpace{}%
\AgdaFunction{rblock}\AgdaSpace{}%
\AgdaOperator{\AgdaFunction{⟩}}\<%
\\
\>[2]\AgdaFunction{rblock-selfinv}\AgdaSpace{}%
\AgdaSymbol{((}\AgdaBound{i}\AgdaSpace{}%
\AgdaOperator{\AgdaInductiveConstructor{⊗}}\AgdaSpace{}%
\AgdaBound{j}\AgdaSymbol{)}\AgdaSpace{}%
\AgdaOperator{\AgdaInductiveConstructor{⊗}}\AgdaSpace{}%
\AgdaSymbol{(}\AgdaBound{k}\AgdaSpace{}%
\AgdaOperator{\AgdaInductiveConstructor{⊗}}\AgdaSpace{}%
\AgdaBound{l}\AgdaSymbol{))}\AgdaSpace{}%
\AgdaSymbol{=}\AgdaSpace{}%
\AgdaInductiveConstructor{refl}\<%
\end{code}}
\end{mathpar}

With this primitive we define \AF{block} and \AF{unblock} as follow:
\begin{mathpar}
\codeblock{\begin{code}%
\>[2]\AgdaFunction{block}\AgdaSpace{}%
\AgdaSymbol{:}\AgdaSpace{}%
\AgdaGeneralizable{s}\AgdaSpace{}%
\AgdaOperator{\AgdaDatatype{*}}\AgdaSpace{}%
\AgdaGeneralizable{p}\AgdaSpace{}%
\AgdaOperator{\AgdaDatatype{≈}}\AgdaSpace{}%
\AgdaGeneralizable{q}\AgdaSpace{}%
\AgdaSymbol{→}\AgdaSpace{}%
\AgdaFunction{Ar}\AgdaSpace{}%
\AgdaGeneralizable{q}\AgdaSpace{}%
\AgdaGeneralizable{X}\AgdaSpace{}%
\AgdaSymbol{→}\AgdaSpace{}%
\AgdaFunction{Ar}\AgdaSpace{}%
\AgdaSymbol{(}\AgdaGeneralizable{s}\AgdaSpace{}%
\AgdaOperator{\AgdaInductiveConstructor{⊗}}\AgdaSpace{}%
\AgdaGeneralizable{p}\AgdaSymbol{)}\AgdaSpace{}%
\AgdaGeneralizable{X}\<%
\\
\>[2]\AgdaFunction{block}\AgdaSpace{}%
\AgdaInductiveConstructor{ι}%
\>[19]\AgdaSymbol{=}\AgdaSpace{}%
\AgdaFunction{reshape}\AgdaSpace{}%
\AgdaInductiveConstructor{split}\<%
\\
\>[2]\AgdaFunction{block}\AgdaSpace{}%
\AgdaSymbol{(}\AgdaBound{l}\AgdaSpace{}%
\AgdaOperator{\AgdaInductiveConstructor{⊗}}\AgdaSpace{}%
\AgdaBound{r}\AgdaSymbol{)}%
\>[19]\AgdaSymbol{=}\AgdaSpace{}%
\AgdaFunction{reshape}\AgdaSpace{}%
\AgdaFunction{rblock}\AgdaSpace{}%
\AgdaOperator{\AgdaFunction{∘}}\AgdaSpace{}%
\AgdaFunction{unnest}\AgdaSpace{}%
\AgdaOperator{\AgdaFunction{∘}}\AgdaSpace{}%
\AgdaFunction{block}\AgdaSpace{}%
\AgdaBound{l}\AgdaSpace{}%
\AgdaOperator{\AgdaFunction{∘}}\AgdaSpace{}%
\AgdaFunction{map}\AgdaSpace{}%
\AgdaSymbol{(}\AgdaFunction{block}\AgdaSpace{}%
\AgdaBound{r}\AgdaSymbol{)}\AgdaSpace{}%
\AgdaOperator{\AgdaFunction{∘}}\AgdaSpace{}%
\AgdaFunction{nest}\<%
\\
\>[0]\<%
\\
\>[2]\AgdaFunction{unblock}\AgdaSpace{}%
\AgdaSymbol{:}\AgdaSpace{}%
\AgdaGeneralizable{s}\AgdaSpace{}%
\AgdaOperator{\AgdaDatatype{*}}\AgdaSpace{}%
\AgdaGeneralizable{p}\AgdaSpace{}%
\AgdaOperator{\AgdaDatatype{≈}}\AgdaSpace{}%
\AgdaGeneralizable{q}\AgdaSpace{}%
\AgdaSymbol{→}\AgdaSpace{}%
\AgdaFunction{Ar}\AgdaSpace{}%
\AgdaSymbol{(}\AgdaGeneralizable{s}\AgdaSpace{}%
\AgdaOperator{\AgdaInductiveConstructor{⊗}}\AgdaSpace{}%
\AgdaGeneralizable{p}\AgdaSymbol{)}\AgdaSpace{}%
\AgdaGeneralizable{X}\AgdaSpace{}%
\AgdaSymbol{→}\AgdaSpace{}%
\AgdaFunction{Ar}\AgdaSpace{}%
\AgdaGeneralizable{q}\AgdaSpace{}%
\AgdaGeneralizable{X}\<%
\\
\>[2]\AgdaFunction{unblock}\AgdaSpace{}%
\AgdaInductiveConstructor{ι}%
\>[19]\AgdaSymbol{=}\AgdaSpace{}%
\AgdaFunction{reshape}\AgdaSpace{}%
\AgdaInductiveConstructor{flat}\<%
\\
\>[2]\AgdaFunction{unblock}\AgdaSpace{}%
\AgdaSymbol{(}\AgdaBound{l}\AgdaSpace{}%
\AgdaOperator{\AgdaInductiveConstructor{⊗}}\AgdaSpace{}%
\AgdaBound{r}\AgdaSymbol{)}%
\>[19]\AgdaSymbol{=}\AgdaSpace{}%
\AgdaFunction{unnest}\AgdaSpace{}%
\AgdaOperator{\AgdaFunction{∘}}\AgdaSpace{}%
\AgdaFunction{unblock}\AgdaSpace{}%
\AgdaBound{l}\AgdaSpace{}%
\AgdaOperator{\AgdaFunction{∘}}\AgdaSpace{}%
\AgdaFunction{map}\AgdaSpace{}%
\AgdaSymbol{(}\AgdaFunction{unblock}\AgdaSpace{}%
\AgdaBound{r}\AgdaSymbol{)}\AgdaSpace{}%
\AgdaOperator{\AgdaFunction{∘}}\AgdaSpace{}%
\AgdaFunction{nest}\AgdaSpace{}%
\AgdaOperator{\AgdaFunction{∘}}\AgdaSpace{}%
\AgdaFunction{reshape}\AgdaSpace{}%
\AgdaFunction{rblock}\<%
\end{code}}
\end{mathpar}
\begin{code}[hide]  %
\>[2]\AgdaFunction{block₂}%
\>[1973I]\AgdaSymbol{:}\AgdaSpace{}%
\AgdaSymbol{∀}\AgdaSpace{}%
\AgdaSymbol{\{}\AgdaBound{X}\AgdaSymbol{\}}\AgdaSpace{}%
\AgdaSymbol{→}\AgdaSpace{}%
\AgdaSymbol{(}\AgdaBound{m}\AgdaSpace{}%
\AgdaBound{n}\AgdaSpace{}%
\AgdaSymbol{:}\AgdaSpace{}%
\AgdaDatatype{ℕ}\AgdaSymbol{)}\AgdaSpace{}%
\AgdaSymbol{→}\AgdaSpace{}%
\AgdaFunction{Ar}\AgdaSpace{}%
\AgdaSymbol{(}\AgdaInductiveConstructor{ι}\AgdaSpace{}%
\AgdaSymbol{(}\AgdaBound{m}\AgdaSpace{}%
\AgdaOperator{\AgdaPrimitive{ℕ.*}}\AgdaSpace{}%
\AgdaNumber{2}\AgdaSymbol{)}\AgdaSpace{}%
\AgdaOperator{\AgdaInductiveConstructor{⊗}}\AgdaSpace{}%
\AgdaSymbol{(}\AgdaInductiveConstructor{ι}\AgdaSpace{}%
\AgdaSymbol{(}\AgdaBound{n}\AgdaSpace{}%
\AgdaOperator{\AgdaPrimitive{ℕ.*}}\AgdaSpace{}%
\AgdaNumber{2}\AgdaSymbol{)))}\AgdaSpace{}%
\AgdaBound{X}\<%
\\
\>[.][@{}l@{}]\<[1973I]%
\>[9]\AgdaSymbol{→}\AgdaSpace{}%
\AgdaFunction{Ar}\AgdaSpace{}%
\AgdaSymbol{((}\AgdaInductiveConstructor{ι}\AgdaSpace{}%
\AgdaBound{m}\AgdaSpace{}%
\AgdaOperator{\AgdaInductiveConstructor{⊗}}\AgdaSpace{}%
\AgdaInductiveConstructor{ι}\AgdaSpace{}%
\AgdaBound{n}\AgdaSymbol{)}\AgdaSpace{}%
\AgdaOperator{\AgdaInductiveConstructor{⊗}}\AgdaSpace{}%
\AgdaSymbol{(}\AgdaInductiveConstructor{ι}\AgdaSpace{}%
\AgdaNumber{2}\AgdaSpace{}%
\AgdaOperator{\AgdaInductiveConstructor{⊗}}\AgdaSpace{}%
\AgdaInductiveConstructor{ι}\AgdaSpace{}%
\AgdaNumber{2}\AgdaSymbol{))}\AgdaSpace{}%
\AgdaBound{X}\<%
\\
\>[2]\AgdaFunction{block₂}\AgdaSpace{}%
\AgdaBound{m}\AgdaSpace{}%
\AgdaBound{n}\AgdaSpace{}%
\AgdaSymbol{=}\AgdaSpace{}%
\AgdaFunction{reshape}\AgdaSpace{}%
\AgdaSymbol{(}\AgdaFunction{rblock}\AgdaSpace{}%
\AgdaOperator{\AgdaInductiveConstructor{∙}}\AgdaSpace{}%
\AgdaSymbol{(}\AgdaInductiveConstructor{split}\AgdaSpace{}%
\AgdaSymbol{\{}\AgdaBound{m}\AgdaSymbol{\}}\AgdaSpace{}%
\AgdaOperator{\AgdaInductiveConstructor{,}}\AgdaSpace{}%
\AgdaInductiveConstructor{split}\AgdaSpace{}%
\AgdaSymbol{\{}\AgdaBound{n}\AgdaSymbol{\}))}\<%
\end{code}
We specialise average pooling to the 2-dimensional case, that is needed in
our running example.
\begin{mathpar}
\codeblock{\begin{code}%
\>[2]\AgdaFunction{avgp₂}\AgdaSpace{}%
\AgdaSymbol{:}\AgdaSpace{}%
\AgdaSymbol{(}\AgdaBound{m}\AgdaSpace{}%
\AgdaBound{n}\AgdaSpace{}%
\AgdaSymbol{:}\AgdaSpace{}%
\AgdaDatatype{ℕ}\AgdaSymbol{)}\AgdaSpace{}%
\AgdaSymbol{→}\AgdaSpace{}%
\AgdaFunction{Ar}\AgdaSpace{}%
\AgdaSymbol{(}\AgdaInductiveConstructor{ι}\AgdaSpace{}%
\AgdaSymbol{(}\AgdaBound{m}\AgdaSpace{}%
\AgdaOperator{\AgdaPrimitive{ℕ.*}}\AgdaSpace{}%
\AgdaNumber{2}\AgdaSymbol{)}\AgdaSpace{}%
\AgdaOperator{\AgdaInductiveConstructor{⊗}}\AgdaSpace{}%
\AgdaSymbol{(}\AgdaInductiveConstructor{ι}\AgdaSpace{}%
\AgdaSymbol{(}\AgdaBound{n}\AgdaSpace{}%
\AgdaOperator{\AgdaPrimitive{ℕ.*}}\AgdaSpace{}%
\AgdaNumber{2}\AgdaSymbol{)))}\AgdaSpace{}%
\AgdaPostulate{ℝ}\AgdaSpace{}%
\AgdaSymbol{→}\AgdaSpace{}%
\AgdaFunction{Ar}\AgdaSpace{}%
\AgdaSymbol{(}\AgdaInductiveConstructor{ι}\AgdaSpace{}%
\AgdaBound{m}\AgdaSpace{}%
\AgdaOperator{\AgdaInductiveConstructor{⊗}}\AgdaSpace{}%
\AgdaInductiveConstructor{ι}\AgdaSpace{}%
\AgdaBound{n}\AgdaSymbol{)}\AgdaSpace{}%
\AgdaPostulate{ℝ}\<%
\\
\>[2]\AgdaFunction{avgp₂}\AgdaSpace{}%
\AgdaBound{m}\AgdaSpace{}%
\AgdaBound{n}\AgdaSpace{}%
\AgdaBound{a}\AgdaSpace{}%
\AgdaSymbol{=}\AgdaSpace{}%
\AgdaFunction{map}\AgdaSpace{}%
\AgdaSymbol{((}\AgdaOperator{\AgdaPrimitive{\AgdaUnderscore{}÷}}\AgdaSpace{}%
\AgdaPrimitive{fromℕ}\AgdaSpace{}%
\AgdaNumber{4}\AgdaSymbol{)}\AgdaSpace{}%
\AgdaOperator{\AgdaFunction{∘}}\AgdaSpace{}%
\AgdaFunction{sum}\AgdaSpace{}%
\AgdaOperator{\AgdaPrimitive{\AgdaUnderscore{}+\AgdaUnderscore{}}}\AgdaSpace{}%
\AgdaNumber{0.0}\AgdaSymbol{)}\AgdaSpace{}%
\AgdaSymbol{(}\AgdaFunction{nest}\AgdaSpace{}%
\AgdaOperator{\AgdaFunction{\$}}\AgdaSpace{}%
\AgdaFunction{block}\AgdaSpace{}%
\AgdaSymbol{(}\AgdaInductiveConstructor{ι}\AgdaSpace{}%
\AgdaOperator{\AgdaInductiveConstructor{⊗}}\AgdaSpace{}%
\AgdaInductiveConstructor{ι}\AgdaSymbol{)}\AgdaSpace{}%
\AgdaBound{a}\AgdaSymbol{)}\<%
\end{code}}
\end{mathpar}

We are now ready to provide the implementation of the forward part of the CNN
as follows.  The \AB{inp} argument is the image of a hand-written digit, all
the other arguments are weights, and the function returns the 10-element vector
with probabilities which digit that is.
\begin{mathpar}
\codeblock{\begin{code}%
\>[2]\AgdaFunction{forward}%
\>[2061I]\AgdaSymbol{:}\AgdaSpace{}%
\AgdaSymbol{(}\AgdaBound{inp}%
\>[18]\AgdaSymbol{:}%
\>[21]\AgdaFunction{Ar}\AgdaSpace{}%
\AgdaSymbol{(}\AgdaInductiveConstructor{ι}\AgdaSpace{}%
\AgdaNumber{28}\AgdaSpace{}%
\AgdaOperator{\AgdaInductiveConstructor{⊗}}\AgdaSpace{}%
\AgdaInductiveConstructor{ι}\AgdaSpace{}%
\AgdaNumber{28}\AgdaSymbol{)}\AgdaSpace{}%
\AgdaPostulate{ℝ}\AgdaSymbol{)}%
\>[49]\AgdaSymbol{→}\AgdaSpace{}%
\AgdaSymbol{(}\AgdaBound{k₁}%
\>[57]\AgdaSymbol{:}%
\>[60]\AgdaFunction{Ar}\AgdaSpace{}%
\AgdaSymbol{(}\AgdaInductiveConstructor{ι}\AgdaSpace{}%
\AgdaNumber{6}\AgdaSpace{}%
\AgdaOperator{\AgdaInductiveConstructor{⊗}}\AgdaSpace{}%
\AgdaSymbol{(}\AgdaInductiveConstructor{ι}\AgdaSpace{}%
\AgdaNumber{5}\AgdaSpace{}%
\AgdaOperator{\AgdaInductiveConstructor{⊗}}\AgdaSpace{}%
\AgdaInductiveConstructor{ι}\AgdaSpace{}%
\AgdaNumber{5}\AgdaSymbol{))}\AgdaSpace{}%
\AgdaPostulate{ℝ}\AgdaSymbol{)}\<%
\\
\>[.][@{}l@{}]\<[2061I]%
\>[10]\AgdaSymbol{→}\AgdaSpace{}%
\AgdaSymbol{(}\AgdaBound{b₁}%
\>[18]\AgdaSymbol{:}%
\>[21]\AgdaFunction{Ar}\AgdaSpace{}%
\AgdaSymbol{(}\AgdaInductiveConstructor{ι}\AgdaSpace{}%
\AgdaNumber{6}\AgdaSymbol{)}\AgdaSpace{}%
\AgdaPostulate{ℝ}\AgdaSymbol{)}%
\>[49]\AgdaSymbol{→}\AgdaSpace{}%
\AgdaSymbol{(}\AgdaBound{k₂}%
\>[57]\AgdaSymbol{:}%
\>[60]\AgdaFunction{Ar}\AgdaSpace{}%
\AgdaSymbol{(}\AgdaInductiveConstructor{ι}\AgdaSpace{}%
\AgdaNumber{12}\AgdaSpace{}%
\AgdaOperator{\AgdaInductiveConstructor{⊗}}\AgdaSpace{}%
\AgdaSymbol{(}\AgdaInductiveConstructor{ι}\AgdaSpace{}%
\AgdaNumber{6}\AgdaSpace{}%
\AgdaOperator{\AgdaInductiveConstructor{⊗}}\AgdaSpace{}%
\AgdaSymbol{(}\AgdaInductiveConstructor{ι}\AgdaSpace{}%
\AgdaNumber{5}\AgdaSpace{}%
\AgdaOperator{\AgdaInductiveConstructor{⊗}}\AgdaSpace{}%
\AgdaInductiveConstructor{ι}\AgdaSpace{}%
\AgdaNumber{5}\AgdaSymbol{)))}\AgdaSpace{}%
\AgdaPostulate{ℝ}\AgdaSymbol{)}\<%
\\
\>[10]\AgdaSymbol{→}\AgdaSpace{}%
\AgdaSymbol{(}\AgdaBound{b₂}%
\>[18]\AgdaSymbol{:}%
\>[21]\AgdaFunction{Ar}\AgdaSpace{}%
\AgdaSymbol{(}\AgdaInductiveConstructor{ι}\AgdaSpace{}%
\AgdaNumber{12}\AgdaSymbol{)}\AgdaSpace{}%
\AgdaPostulate{ℝ}\AgdaSymbol{)}%
\>[49]\AgdaSymbol{→}\AgdaSpace{}%
\AgdaSymbol{(}\AgdaBound{fc}%
\>[57]\AgdaSymbol{:}%
\>[60]\AgdaFunction{Ar}\AgdaSpace{}%
\AgdaSymbol{(}\AgdaInductiveConstructor{ι}\AgdaSpace{}%
\AgdaNumber{10}\AgdaSpace{}%
\AgdaOperator{\AgdaInductiveConstructor{⊗}}\AgdaSpace{}%
\AgdaSymbol{(}\AgdaInductiveConstructor{ι}\AgdaSpace{}%
\AgdaNumber{12}\AgdaSpace{}%
\AgdaOperator{\AgdaInductiveConstructor{⊗}}\AgdaSpace{}%
\AgdaSymbol{(}\AgdaInductiveConstructor{ι}\AgdaSpace{}%
\AgdaNumber{1}\AgdaSpace{}%
\AgdaOperator{\AgdaInductiveConstructor{⊗}}\AgdaSpace{}%
\AgdaSymbol{(}\AgdaInductiveConstructor{ι}\AgdaSpace{}%
\AgdaNumber{4}\AgdaSpace{}%
\AgdaOperator{\AgdaInductiveConstructor{⊗}}\AgdaSpace{}%
\AgdaInductiveConstructor{ι}\AgdaSpace{}%
\AgdaNumber{4}\AgdaSymbol{))))}\AgdaSpace{}%
\AgdaPostulate{ℝ}\AgdaSymbol{)}\<%
\\
\>[10]\AgdaSymbol{→}\AgdaSpace{}%
\AgdaSymbol{(}\AgdaBound{b}%
\>[18]\AgdaSymbol{:}%
\>[21]\AgdaFunction{Ar}\AgdaSpace{}%
\AgdaSymbol{(}\AgdaInductiveConstructor{ι}\AgdaSpace{}%
\AgdaNumber{10}\AgdaSymbol{)}\AgdaSpace{}%
\AgdaPostulate{ℝ}\AgdaSymbol{)}%
\>[49]\AgdaSymbol{→}\AgdaSpace{}%
\AgdaFunction{Ar}\AgdaSpace{}%
\AgdaSymbol{(}\AgdaInductiveConstructor{ι}\AgdaSpace{}%
\AgdaNumber{10}\AgdaSpace{}%
\AgdaOperator{\AgdaInductiveConstructor{⊗}}\AgdaSpace{}%
\AgdaSymbol{(}\AgdaInductiveConstructor{ι}\AgdaSpace{}%
\AgdaNumber{1}\AgdaSpace{}%
\AgdaOperator{\AgdaInductiveConstructor{⊗}}\AgdaSpace{}%
\AgdaSymbol{(}\AgdaInductiveConstructor{ι}\AgdaSpace{}%
\AgdaNumber{1}\AgdaSpace{}%
\AgdaOperator{\AgdaInductiveConstructor{⊗}}\AgdaSpace{}%
\AgdaSymbol{(}\AgdaInductiveConstructor{ι}\AgdaSpace{}%
\AgdaNumber{1}\AgdaSpace{}%
\AgdaOperator{\AgdaInductiveConstructor{⊗}}\AgdaSpace{}%
\AgdaInductiveConstructor{ι}\AgdaSpace{}%
\AgdaNumber{1}\AgdaSymbol{))))}\AgdaSpace{}%
\AgdaPostulate{ℝ}\<%
\\
\>[2]\AgdaFunction{forward}\AgdaSpace{}%
\AgdaBound{inp}\AgdaSpace{}%
\AgdaBound{k₁}\AgdaSpace{}%
\AgdaBound{b₁}\AgdaSpace{}%
\AgdaBound{k₂}\AgdaSpace{}%
\AgdaBound{b₂}\AgdaSpace{}%
\AgdaBound{fc}\AgdaSpace{}%
\AgdaBound{b}\AgdaSpace{}%
\AgdaSymbol{=}\AgdaSpace{}%
\AgdaKeyword{let}\<%
\\
\>[2][@{}l@{\AgdaIndent{0}}]%
\>[6]\AgdaBound{c₁}\AgdaSpace{}%
\AgdaSymbol{:}\AgdaSpace{}%
\AgdaFunction{Ar}\AgdaSpace{}%
\AgdaSymbol{(}\AgdaInductiveConstructor{ι}\AgdaSpace{}%
\AgdaNumber{6}\AgdaSpace{}%
\AgdaOperator{\AgdaInductiveConstructor{⊗}}\AgdaSpace{}%
\AgdaSymbol{(}\AgdaInductiveConstructor{ι}\AgdaSpace{}%
\AgdaNumber{24}\AgdaSpace{}%
\AgdaOperator{\AgdaInductiveConstructor{⊗}}\AgdaSpace{}%
\AgdaInductiveConstructor{ι}\AgdaSpace{}%
\AgdaNumber{24}\AgdaSymbol{))}\AgdaSpace{}%
\AgdaPostulate{ℝ}\<%
\\
\>[6]\AgdaBound{c₁}\AgdaSpace{}%
\AgdaSymbol{=}\AgdaSpace{}%
\AgdaFunction{logistic}\AgdaSpace{}%
\AgdaOperator{\AgdaFunction{\$}}\AgdaSpace{}%
\AgdaFunction{mconv}\AgdaSpace{}%
\AgdaSymbol{(}\AgdaInductiveConstructor{ι}\AgdaSpace{}%
\AgdaOperator{\AgdaInductiveConstructor{⊗}}\AgdaSpace{}%
\AgdaInductiveConstructor{ι}\AgdaSymbol{)}\AgdaSpace{}%
\AgdaBound{inp}\AgdaSpace{}%
\AgdaBound{k₁}\AgdaSpace{}%
\AgdaBound{b₁}\AgdaSpace{}%
\AgdaSymbol{(}\AgdaInductiveConstructor{ι}\AgdaSpace{}%
\AgdaOperator{\AgdaInductiveConstructor{⊗}}\AgdaSpace{}%
\AgdaInductiveConstructor{ι}\AgdaSymbol{)}\<%
\\
\\[\AgdaEmptyExtraSkip]%
\>[6]\AgdaBound{s₁}\AgdaSpace{}%
\AgdaSymbol{:}\AgdaSpace{}%
\AgdaFunction{Ar}\AgdaSpace{}%
\AgdaSymbol{(}\AgdaInductiveConstructor{ι}\AgdaSpace{}%
\AgdaNumber{6}\AgdaSpace{}%
\AgdaOperator{\AgdaInductiveConstructor{⊗}}\AgdaSpace{}%
\AgdaSymbol{(}\AgdaInductiveConstructor{ι}\AgdaSpace{}%
\AgdaNumber{12}\AgdaSpace{}%
\AgdaOperator{\AgdaInductiveConstructor{⊗}}\AgdaSpace{}%
\AgdaInductiveConstructor{ι}\AgdaSpace{}%
\AgdaNumber{12}\AgdaSymbol{))}\AgdaSpace{}%
\AgdaPostulate{ℝ}\<%
\\
\>[6]\AgdaBound{s₁}\AgdaSpace{}%
\AgdaSymbol{=}\AgdaSpace{}%
\AgdaFunction{unnest}\AgdaSpace{}%
\AgdaOperator{\AgdaFunction{\$}}\AgdaSpace{}%
\AgdaFunction{map}\AgdaSpace{}%
\AgdaSymbol{(}\AgdaFunction{avgp₂}\AgdaSpace{}%
\AgdaNumber{12}\AgdaSpace{}%
\AgdaNumber{12}\AgdaSymbol{)}\AgdaSpace{}%
\AgdaSymbol{(}\AgdaFunction{nest}\AgdaSpace{}%
\AgdaBound{c₁}\AgdaSymbol{)}\<%
\\
\\[\AgdaEmptyExtraSkip]%
\>[6]\AgdaBound{c₂}\AgdaSpace{}%
\AgdaSymbol{:}\AgdaSpace{}%
\AgdaFunction{Ar}\AgdaSpace{}%
\AgdaSymbol{(}\AgdaInductiveConstructor{ι}\AgdaSpace{}%
\AgdaNumber{12}\AgdaSpace{}%
\AgdaOperator{\AgdaInductiveConstructor{⊗}}\AgdaSpace{}%
\AgdaSymbol{(}\AgdaInductiveConstructor{ι}\AgdaSpace{}%
\AgdaNumber{1}\AgdaSpace{}%
\AgdaOperator{\AgdaInductiveConstructor{⊗}}\AgdaSpace{}%
\AgdaSymbol{(}\AgdaInductiveConstructor{ι}\AgdaSpace{}%
\AgdaNumber{8}\AgdaSpace{}%
\AgdaOperator{\AgdaInductiveConstructor{⊗}}\AgdaSpace{}%
\AgdaInductiveConstructor{ι}\AgdaSpace{}%
\AgdaNumber{8}\AgdaSymbol{)))}\AgdaSpace{}%
\AgdaPostulate{ℝ}\<%
\\
\>[6]\AgdaBound{c₂}\AgdaSpace{}%
\AgdaSymbol{=}\AgdaSpace{}%
\AgdaFunction{logistic}\AgdaSpace{}%
\AgdaOperator{\AgdaFunction{\$}}\AgdaSpace{}%
\AgdaFunction{mconv}\AgdaSpace{}%
\AgdaSymbol{(}\AgdaInductiveConstructor{ι}\AgdaSpace{}%
\AgdaOperator{\AgdaInductiveConstructor{⊗}}\AgdaSpace{}%
\AgdaSymbol{(}\AgdaInductiveConstructor{ι}\AgdaSpace{}%
\AgdaOperator{\AgdaInductiveConstructor{⊗}}\AgdaSpace{}%
\AgdaInductiveConstructor{ι}\AgdaSymbol{))}\AgdaSpace{}%
\AgdaBound{s₁}\AgdaSpace{}%
\AgdaBound{k₂}\AgdaSpace{}%
\AgdaBound{b₂}\AgdaSpace{}%
\AgdaSymbol{(}\AgdaInductiveConstructor{ι}\AgdaSpace{}%
\AgdaOperator{\AgdaInductiveConstructor{⊗}}\AgdaSpace{}%
\AgdaSymbol{(}\AgdaInductiveConstructor{ι}\AgdaSpace{}%
\AgdaOperator{\AgdaInductiveConstructor{⊗}}\AgdaSpace{}%
\AgdaInductiveConstructor{ι}\AgdaSymbol{))}\<%
\\
\\[\AgdaEmptyExtraSkip]%
\>[6]\AgdaBound{s₂}\AgdaSpace{}%
\AgdaSymbol{:}\AgdaSpace{}%
\AgdaFunction{Ar}\AgdaSpace{}%
\AgdaSymbol{(}\AgdaInductiveConstructor{ι}\AgdaSpace{}%
\AgdaNumber{12}\AgdaSpace{}%
\AgdaOperator{\AgdaInductiveConstructor{⊗}}\AgdaSpace{}%
\AgdaSymbol{(}\AgdaInductiveConstructor{ι}\AgdaSpace{}%
\AgdaNumber{1}\AgdaSpace{}%
\AgdaOperator{\AgdaInductiveConstructor{⊗}}\AgdaSpace{}%
\AgdaSymbol{(}\AgdaInductiveConstructor{ι}\AgdaSpace{}%
\AgdaNumber{4}\AgdaSpace{}%
\AgdaOperator{\AgdaInductiveConstructor{⊗}}\AgdaSpace{}%
\AgdaInductiveConstructor{ι}\AgdaSpace{}%
\AgdaNumber{4}\AgdaSymbol{)))}\AgdaSpace{}%
\AgdaPostulate{ℝ}\<%
\\
\>[6]\AgdaBound{s₂}\AgdaSpace{}%
\AgdaSymbol{=}\AgdaSpace{}%
\AgdaFunction{unnest}\AgdaSpace{}%
\AgdaOperator{\AgdaFunction{\$}}\AgdaSpace{}%
\AgdaFunction{map}\AgdaSpace{}%
\AgdaSymbol{(}\AgdaFunction{unnest}\AgdaSpace{}%
\AgdaOperator{\AgdaFunction{∘}}\AgdaSpace{}%
\AgdaFunction{map}\AgdaSpace{}%
\AgdaSymbol{(}\AgdaFunction{avgp₂}\AgdaSpace{}%
\AgdaNumber{4}\AgdaSpace{}%
\AgdaNumber{4}\AgdaSymbol{)}\AgdaSpace{}%
\AgdaOperator{\AgdaFunction{∘}}\AgdaSpace{}%
\AgdaFunction{nest}\AgdaSymbol{)}\AgdaSpace{}%
\AgdaSymbol{(}\AgdaFunction{nest}\AgdaSpace{}%
\AgdaBound{c₂}\AgdaSymbol{)}\<%
\\
\\[\AgdaEmptyExtraSkip]%
\>[6]\AgdaBound{r}\AgdaSpace{}%
\AgdaSymbol{=}\AgdaSpace{}%
\AgdaFunction{logistic}\AgdaSpace{}%
\AgdaOperator{\AgdaFunction{\$}}\AgdaSpace{}%
\AgdaFunction{mconv}\AgdaSpace{}%
\AgdaSymbol{(}\AgdaInductiveConstructor{ι}\AgdaSpace{}%
\AgdaOperator{\AgdaInductiveConstructor{⊗}}\AgdaSpace{}%
\AgdaSymbol{(}\AgdaInductiveConstructor{ι}\AgdaSpace{}%
\AgdaOperator{\AgdaInductiveConstructor{⊗}}\AgdaSpace{}%
\AgdaSymbol{(}\AgdaInductiveConstructor{ι}\AgdaSpace{}%
\AgdaOperator{\AgdaInductiveConstructor{⊗}}\AgdaSpace{}%
\AgdaInductiveConstructor{ι}\AgdaSymbol{)))}\AgdaSpace{}%
\AgdaBound{s₂}\AgdaSpace{}%
\AgdaBound{fc}\AgdaSpace{}%
\AgdaBound{b}\AgdaSpace{}%
\AgdaSymbol{(}\AgdaInductiveConstructor{ι}\AgdaSpace{}%
\AgdaOperator{\AgdaInductiveConstructor{⊗}}\AgdaSpace{}%
\AgdaSymbol{(}\AgdaInductiveConstructor{ι}\AgdaSpace{}%
\AgdaOperator{\AgdaInductiveConstructor{⊗}}\AgdaSpace{}%
\AgdaSymbol{(}\AgdaInductiveConstructor{ι}\AgdaSpace{}%
\AgdaOperator{\AgdaInductiveConstructor{⊗}}\AgdaSpace{}%
\AgdaInductiveConstructor{ι}\AgdaSymbol{)))}\<%
\\
\>[2][@{}l@{\AgdaIndent{0}}]%
\>[4]\AgdaKeyword{in}\AgdaSpace{}%
\AgdaBound{r}\<%
\end{code}}
\end{mathpar}

\begin{code}[hide]%
\>[0]\AgdaKeyword{module}\AgdaSpace{}%
\AgdaModule{Tests}\AgdaSpace{}%
\AgdaKeyword{where}\<%
\\
\>[0][@{}l@{\AgdaIndent{0}}]%
\>[2]\AgdaKeyword{open}\AgdaSpace{}%
\AgdaKeyword{import}\AgdaSpace{}%
\AgdaModule{Data.String}\<%
\\
\>[2]\AgdaKeyword{open}\AgdaSpace{}%
\AgdaKeyword{import}\AgdaSpace{}%
\AgdaModule{Data.Nat.Show}\AgdaSpace{}%
\AgdaKeyword{renaming}\AgdaSpace{}%
\AgdaSymbol{(}\AgdaFunction{show}\AgdaSpace{}%
\AgdaSymbol{to}\AgdaSpace{}%
\AgdaFunction{shownat}\AgdaSymbol{)}\<%
\\
\>[2]\AgdaKeyword{open}\AgdaSpace{}%
\AgdaKeyword{import}\AgdaSpace{}%
\AgdaModule{Data.Nat}\AgdaSpace{}%
\AgdaKeyword{using}\AgdaSpace{}%
\AgdaSymbol{(}\AgdaDatatype{ℕ}\AgdaSymbol{;}\AgdaSpace{}%
\AgdaInductiveConstructor{zero}\AgdaSymbol{;}\AgdaSpace{}%
\AgdaInductiveConstructor{suc}\AgdaSymbol{;}\AgdaSpace{}%
\AgdaOperator{\AgdaPrimitive{\AgdaUnderscore{}+\AgdaUnderscore{}}}\AgdaSymbol{;}\AgdaSpace{}%
\AgdaOperator{\AgdaPrimitive{\AgdaUnderscore{}*\AgdaUnderscore{}}}\AgdaSymbol{)}\<%
\\
\>[2]\AgdaKeyword{open}\AgdaSpace{}%
\AgdaKeyword{import}\AgdaSpace{}%
\AgdaModule{Data.Fin}\AgdaSpace{}%
\AgdaKeyword{using}\AgdaSpace{}%
\AgdaSymbol{(}\AgdaDatatype{Fin}\AgdaSymbol{;}\AgdaSpace{}%
\AgdaInductiveConstructor{zero}\AgdaSymbol{;}\AgdaSpace{}%
\AgdaInductiveConstructor{suc}\AgdaSymbol{;}\AgdaSpace{}%
\AgdaFunction{toℕ}\AgdaSymbol{;}\AgdaSpace{}%
\AgdaOperator{\AgdaFunction{\#\AgdaUnderscore{}}}\AgdaSymbol{)}\<%
\\
\>[2]\AgdaComment{--open\ import\ Fin2\ using\ (Fin;\ \#\AgdaUnderscore{};\ combine;\ remQuot;\ zerof;\ sucf;\ \AgdaUnderscore{}⊕\AgdaUnderscore{};\ \AgdaUnderscore{}⊝\AgdaUnderscore{})}\<%
\\
\>[2]\AgdaKeyword{open}\AgdaSpace{}%
\AgdaKeyword{import}\AgdaSpace{}%
\AgdaModule{Data.Product}\AgdaSpace{}%
\AgdaKeyword{using}\AgdaSpace{}%
\AgdaSymbol{(}\AgdaRecord{Σ}\AgdaSymbol{;}\AgdaSpace{}%
\AgdaOperator{\AgdaInductiveConstructor{\AgdaUnderscore{},\AgdaUnderscore{}}}\AgdaSymbol{;}\AgdaSpace{}%
\AgdaField{proj₁}\AgdaSymbol{)}\<%
\\
\>[2]\AgdaKeyword{open}\AgdaSpace{}%
\AgdaModule{Array}\<%
\\
\\[\AgdaEmptyExtraSkip]%
\>[2]\AgdaFunction{tab}\AgdaSpace{}%
\AgdaSymbol{:}\AgdaSpace{}%
\AgdaSymbol{∀}\AgdaSpace{}%
\AgdaSymbol{\{}\AgdaBound{X}\AgdaSymbol{\}}\AgdaSpace{}%
\AgdaSymbol{→}\AgdaSpace{}%
\AgdaFunction{Vec}\AgdaSpace{}%
\AgdaGeneralizable{m}\AgdaSpace{}%
\AgdaBound{X}\AgdaSpace{}%
\AgdaSymbol{→}\AgdaSpace{}%
\AgdaDatatype{List}\AgdaSpace{}%
\AgdaBound{X}\<%
\\
\>[2]\AgdaFunction{tab}\AgdaSpace{}%
\AgdaSymbol{\{}\AgdaArgument{m}\AgdaSpace{}%
\AgdaSymbol{=}\AgdaSpace{}%
\AgdaInductiveConstructor{zero}\AgdaSymbol{\}}%
\>[18]\AgdaBound{f}\AgdaSpace{}%
\AgdaSymbol{=}\AgdaSpace{}%
\AgdaInductiveConstructor{[]}\<%
\\
\>[2]\AgdaFunction{tab}\AgdaSpace{}%
\AgdaSymbol{\{}\AgdaArgument{m}\AgdaSpace{}%
\AgdaSymbol{=}\AgdaSpace{}%
\AgdaInductiveConstructor{suc}\AgdaSpace{}%
\AgdaBound{m}\AgdaSymbol{\}}\AgdaSpace{}%
\AgdaBound{f}\AgdaSpace{}%
\AgdaSymbol{=}\AgdaSpace{}%
\AgdaBound{f}\AgdaSpace{}%
\AgdaSymbol{(}\AgdaInductiveConstructor{ι}\AgdaSpace{}%
\AgdaSymbol{(}\AgdaOperator{\AgdaFunction{\#}}\AgdaSpace{}%
\AgdaNumber{0}\AgdaSymbol{))}\AgdaSpace{}%
\AgdaOperator{\AgdaInductiveConstructor{∷}}\AgdaSpace{}%
\AgdaFunction{tab}\AgdaSpace{}%
\AgdaSymbol{(}\AgdaBound{f}\AgdaSpace{}%
\AgdaOperator{\AgdaFunction{∘}}\AgdaSpace{}%
\AgdaFunction{ιsuc}\AgdaSymbol{)}\<%
\\
\\[\AgdaEmptyExtraSkip]%
\>[2]\AgdaFunction{ListNest}\AgdaSpace{}%
\AgdaSymbol{:}\AgdaSpace{}%
\AgdaDatatype{S}\AgdaSpace{}%
\AgdaSymbol{→}\AgdaSpace{}%
\AgdaPrimitive{Set}\AgdaSpace{}%
\AgdaSymbol{→}\AgdaSpace{}%
\AgdaPrimitive{Set}\<%
\\
\>[2]\AgdaFunction{ListNest}\AgdaSpace{}%
\AgdaSymbol{(}\AgdaInductiveConstructor{ι}\AgdaSpace{}%
\AgdaSymbol{\AgdaUnderscore{})}\AgdaSpace{}%
\AgdaSymbol{=}\AgdaSpace{}%
\AgdaDatatype{List}\<%
\\
\>[2]\AgdaFunction{ListNest}\AgdaSpace{}%
\AgdaSymbol{(}\AgdaBound{s}\AgdaSpace{}%
\AgdaOperator{\AgdaInductiveConstructor{⊗}}\AgdaSpace{}%
\AgdaBound{p}\AgdaSymbol{)}\AgdaSpace{}%
\AgdaSymbol{=}\AgdaSpace{}%
\AgdaFunction{ListNest}\AgdaSpace{}%
\AgdaBound{s}\AgdaSpace{}%
\AgdaOperator{\AgdaFunction{∘}}\AgdaSpace{}%
\AgdaFunction{ListNest}\AgdaSpace{}%
\AgdaBound{p}\<%
\\
\\[\AgdaEmptyExtraSkip]%
\>[2]\AgdaFunction{atab}\AgdaSpace{}%
\AgdaSymbol{:}\AgdaSpace{}%
\AgdaFunction{Ar}\AgdaSpace{}%
\AgdaGeneralizable{s}\AgdaSpace{}%
\AgdaGeneralizable{X}\AgdaSpace{}%
\AgdaSymbol{→}\AgdaSpace{}%
\AgdaFunction{ListNest}\AgdaSpace{}%
\AgdaGeneralizable{s}\AgdaSpace{}%
\AgdaGeneralizable{X}\<%
\\
\>[2]\AgdaFunction{atab}\AgdaSpace{}%
\AgdaSymbol{\{}\AgdaInductiveConstructor{ι}\AgdaSpace{}%
\AgdaSymbol{\AgdaUnderscore{}\}}\AgdaSpace{}%
\AgdaBound{a}\AgdaSpace{}%
\AgdaSymbol{=}\AgdaSpace{}%
\AgdaFunction{tab}\AgdaSpace{}%
\AgdaBound{a}\<%
\\
\>[2]\AgdaFunction{atab}\AgdaSpace{}%
\AgdaSymbol{\{}\AgdaBound{s}\AgdaSpace{}%
\AgdaOperator{\AgdaInductiveConstructor{⊗}}\AgdaSpace{}%
\AgdaBound{p}\AgdaSymbol{\}}\AgdaSpace{}%
\AgdaBound{a}\AgdaSpace{}%
\AgdaSymbol{=}\AgdaSpace{}%
\AgdaFunction{atab}\AgdaSpace{}%
\AgdaSymbol{(}\AgdaFunction{Array.map}\AgdaSpace{}%
\AgdaFunction{atab}\AgdaSpace{}%
\AgdaOperator{\AgdaFunction{\$}}\AgdaSpace{}%
\AgdaFunction{nest}\AgdaSpace{}%
\AgdaBound{a}\AgdaSymbol{)}\<%
\\
\\[\AgdaEmptyExtraSkip]%
\>[2]\AgdaFunction{viota}\AgdaSpace{}%
\AgdaSymbol{:}\AgdaSpace{}%
\AgdaSymbol{(}\AgdaBound{n}\AgdaSpace{}%
\AgdaSymbol{:}\AgdaSpace{}%
\AgdaDatatype{ℕ}\AgdaSymbol{)}\AgdaSpace{}%
\AgdaSymbol{→}\AgdaSpace{}%
\AgdaFunction{Vec}\AgdaSpace{}%
\AgdaBound{n}\AgdaSpace{}%
\AgdaDatatype{ℕ}\<%
\\
\>[2]\AgdaFunction{viota}\AgdaSpace{}%
\AgdaBound{n}\AgdaSpace{}%
\AgdaSymbol{(}\AgdaInductiveConstructor{ι}\AgdaSpace{}%
\AgdaBound{i}\AgdaSymbol{)}\AgdaSpace{}%
\AgdaSymbol{=}\AgdaSpace{}%
\AgdaFunction{toℕ}\AgdaSpace{}%
\AgdaBound{i}\<%
\\
\\[\AgdaEmptyExtraSkip]%
\>[2]\AgdaFunction{size}\AgdaSpace{}%
\AgdaSymbol{:}\AgdaSpace{}%
\AgdaDatatype{S}\AgdaSpace{}%
\AgdaSymbol{→}\AgdaSpace{}%
\AgdaDatatype{ℕ}\<%
\\
\>[2]\AgdaFunction{size}\AgdaSpace{}%
\AgdaSymbol{(}\AgdaInductiveConstructor{ι}\AgdaSpace{}%
\AgdaBound{n}\AgdaSymbol{)}\AgdaSpace{}%
\AgdaSymbol{=}\AgdaSpace{}%
\AgdaBound{n}\<%
\\
\>[2]\AgdaFunction{size}\AgdaSpace{}%
\AgdaSymbol{(}\AgdaBound{s}\AgdaSpace{}%
\AgdaOperator{\AgdaInductiveConstructor{⊗}}\AgdaSpace{}%
\AgdaBound{p}\AgdaSymbol{)}\AgdaSpace{}%
\AgdaSymbol{=}\AgdaSpace{}%
\AgdaFunction{size}\AgdaSpace{}%
\AgdaBound{s}\AgdaSpace{}%
\AgdaOperator{\AgdaPrimitive{*}}\AgdaSpace{}%
\AgdaFunction{size}\AgdaSpace{}%
\AgdaBound{p}\<%
\\
\\[\AgdaEmptyExtraSkip]%
\>[2]\AgdaFunction{iota}\AgdaSpace{}%
\AgdaSymbol{:}\AgdaSpace{}%
\AgdaFunction{Ar}\AgdaSpace{}%
\AgdaGeneralizable{s}\AgdaSpace{}%
\AgdaDatatype{ℕ}\<%
\\
\>[2]\AgdaFunction{iota}\AgdaSpace{}%
\AgdaSymbol{\{}\AgdaArgument{s}\AgdaSpace{}%
\AgdaSymbol{=}\AgdaSpace{}%
\AgdaInductiveConstructor{ι}\AgdaSpace{}%
\AgdaBound{n}\AgdaSymbol{\}}\AgdaSpace{}%
\AgdaSymbol{=}\AgdaSpace{}%
\AgdaFunction{viota}\AgdaSpace{}%
\AgdaBound{n}\<%
\\
\>[2]\AgdaFunction{iota}\AgdaSpace{}%
\AgdaSymbol{\{}\AgdaArgument{s}\AgdaSpace{}%
\AgdaSymbol{=}\AgdaSpace{}%
\AgdaBound{s}\AgdaSpace{}%
\AgdaOperator{\AgdaInductiveConstructor{⊗}}\AgdaSpace{}%
\AgdaBound{p}\AgdaSymbol{\}}\AgdaSpace{}%
\AgdaSymbol{=}\AgdaSpace{}%
\AgdaFunction{unnest}\AgdaSpace{}%
\AgdaSymbol{(}\AgdaFunction{Array.map}\AgdaSpace{}%
\AgdaSymbol{(λ}\AgdaSpace{}%
\AgdaBound{i}\AgdaSpace{}%
\AgdaSymbol{→}\AgdaSpace{}%
\AgdaFunction{Array.map}\AgdaSpace{}%
\AgdaSymbol{(λ}\AgdaSpace{}%
\AgdaBound{j}\AgdaSpace{}%
\AgdaSymbol{→}\AgdaSpace{}%
\AgdaBound{i}\AgdaSpace{}%
\AgdaOperator{\AgdaPrimitive{*}}\AgdaSpace{}%
\AgdaFunction{size}\AgdaSpace{}%
\AgdaBound{p}\AgdaSpace{}%
\AgdaOperator{\AgdaPrimitive{+}}\AgdaSpace{}%
\AgdaBound{j}\AgdaSymbol{)}\AgdaSpace{}%
\AgdaFunction{iota}\AgdaSymbol{)}\AgdaSpace{}%
\AgdaFunction{iota}\AgdaSymbol{)}\<%
\\
\\[\AgdaEmptyExtraSkip]%
\>[2]\AgdaFunction{test-block}\AgdaSpace{}%
\AgdaSymbol{:}\AgdaSpace{}%
\AgdaSymbol{\AgdaUnderscore{}}\<%
\\
\>[2]\AgdaFunction{test-block}\AgdaSpace{}%
\AgdaSymbol{=}\AgdaSpace{}%
\AgdaFunction{atab}\AgdaSpace{}%
\AgdaOperator{\AgdaFunction{\$}}\AgdaSpace{}%
\AgdaFunction{CNN.block₂}\AgdaSpace{}%
\AgdaNumber{2}\AgdaSpace{}%
\AgdaNumber{2}\AgdaSpace{}%
\AgdaFunction{iota}\<%
\\
\\[\AgdaEmptyExtraSkip]%
\>[2]\AgdaFunction{\AgdaUnderscore{}}\AgdaSpace{}%
\AgdaSymbol{:}%
\>[2452I]\AgdaFunction{test-block}\<%
\\
\>[.][@{}l@{}]\<[2452I]%
\>[6]\AgdaOperator{\AgdaDatatype{≡}}%
\>[2453I]\AgdaSymbol{(((}\AgdaNumber{0}\AgdaSpace{}%
\AgdaOperator{\AgdaInductiveConstructor{∷}}\AgdaSpace{}%
\AgdaNumber{1}\AgdaSpace{}%
\AgdaOperator{\AgdaInductiveConstructor{∷}}\AgdaSpace{}%
\AgdaInductiveConstructor{[]}\AgdaSymbol{)}\AgdaSpace{}%
\AgdaOperator{\AgdaInductiveConstructor{∷}}\AgdaSpace{}%
\AgdaSymbol{(}\AgdaNumber{4}\AgdaSpace{}%
\AgdaOperator{\AgdaInductiveConstructor{∷}}\AgdaSpace{}%
\AgdaNumber{5}\AgdaSpace{}%
\AgdaOperator{\AgdaInductiveConstructor{∷}}\AgdaSpace{}%
\AgdaInductiveConstructor{[]}\AgdaSymbol{)}\AgdaSpace{}%
\AgdaOperator{\AgdaInductiveConstructor{∷}}\AgdaSpace{}%
\AgdaInductiveConstructor{[]}\AgdaSymbol{)}\AgdaSpace{}%
\AgdaOperator{\AgdaInductiveConstructor{∷}}\<%
\\
\>[2453I][@{}l@{\AgdaIndent{0}}]%
\>[9]\AgdaSymbol{((}\AgdaNumber{2}\AgdaSpace{}%
\AgdaOperator{\AgdaInductiveConstructor{∷}}\AgdaSpace{}%
\AgdaNumber{3}\AgdaSpace{}%
\AgdaOperator{\AgdaInductiveConstructor{∷}}\AgdaSpace{}%
\AgdaInductiveConstructor{[]}\AgdaSymbol{)}\AgdaSpace{}%
\AgdaOperator{\AgdaInductiveConstructor{∷}}\AgdaSpace{}%
\AgdaSymbol{(}\AgdaNumber{6}\AgdaSpace{}%
\AgdaOperator{\AgdaInductiveConstructor{∷}}\AgdaSpace{}%
\AgdaNumber{7}\AgdaSpace{}%
\AgdaOperator{\AgdaInductiveConstructor{∷}}\AgdaSpace{}%
\AgdaInductiveConstructor{[]}\AgdaSymbol{)}\AgdaSpace{}%
\AgdaOperator{\AgdaInductiveConstructor{∷}}\AgdaSpace{}%
\AgdaInductiveConstructor{[]}\AgdaSymbol{)}\AgdaSpace{}%
\AgdaOperator{\AgdaInductiveConstructor{∷}}\AgdaSpace{}%
\AgdaInductiveConstructor{[]}\AgdaSymbol{)}\<%
\\
\>[.][@{}l@{}]\<[2453I]%
\>[8]\AgdaOperator{\AgdaInductiveConstructor{∷}}\<%
\\
\>[8]\AgdaSymbol{(((}\AgdaNumber{8}\AgdaSpace{}%
\AgdaOperator{\AgdaInductiveConstructor{∷}}\AgdaSpace{}%
\AgdaNumber{9}\AgdaSpace{}%
\AgdaOperator{\AgdaInductiveConstructor{∷}}\AgdaSpace{}%
\AgdaInductiveConstructor{[]}\AgdaSymbol{)}\AgdaSpace{}%
\AgdaOperator{\AgdaInductiveConstructor{∷}}\AgdaSpace{}%
\AgdaSymbol{(}\AgdaNumber{12}\AgdaSpace{}%
\AgdaOperator{\AgdaInductiveConstructor{∷}}\AgdaSpace{}%
\AgdaNumber{13}\AgdaSpace{}%
\AgdaOperator{\AgdaInductiveConstructor{∷}}\AgdaSpace{}%
\AgdaInductiveConstructor{[]}\AgdaSymbol{)}\AgdaSpace{}%
\AgdaOperator{\AgdaInductiveConstructor{∷}}\AgdaSpace{}%
\AgdaInductiveConstructor{[]}\AgdaSymbol{)}\AgdaSpace{}%
\AgdaOperator{\AgdaInductiveConstructor{∷}}\<%
\\
\>[8][@{}l@{\AgdaIndent{0}}]%
\>[9]\AgdaSymbol{((}\AgdaNumber{10}\AgdaSpace{}%
\AgdaOperator{\AgdaInductiveConstructor{∷}}\AgdaSpace{}%
\AgdaNumber{11}\AgdaSpace{}%
\AgdaOperator{\AgdaInductiveConstructor{∷}}\AgdaSpace{}%
\AgdaInductiveConstructor{[]}\AgdaSymbol{)}\AgdaSpace{}%
\AgdaOperator{\AgdaInductiveConstructor{∷}}\AgdaSpace{}%
\AgdaSymbol{(}\AgdaNumber{14}\AgdaSpace{}%
\AgdaOperator{\AgdaInductiveConstructor{∷}}\AgdaSpace{}%
\AgdaNumber{15}\AgdaSpace{}%
\AgdaOperator{\AgdaInductiveConstructor{∷}}\AgdaSpace{}%
\AgdaInductiveConstructor{[]}\AgdaSymbol{)}\AgdaSpace{}%
\AgdaOperator{\AgdaInductiveConstructor{∷}}\AgdaSpace{}%
\AgdaInductiveConstructor{[]}\AgdaSymbol{)}\AgdaSpace{}%
\AgdaOperator{\AgdaInductiveConstructor{∷}}\AgdaSpace{}%
\AgdaInductiveConstructor{[]}\AgdaSymbol{)}\<%
\\
\>[8]\AgdaOperator{\AgdaInductiveConstructor{∷}}\AgdaSpace{}%
\AgdaInductiveConstructor{[]}\<%
\\
\>[2]\AgdaSymbol{\AgdaUnderscore{}}\AgdaSpace{}%
\AgdaSymbol{=}\AgdaSpace{}%
\AgdaInductiveConstructor{refl}\<%
\\
\\[\AgdaEmptyExtraSkip]%
\>[2]\AgdaFunction{\AgdaUnderscore{}}\AgdaSpace{}%
\AgdaSymbol{:}\AgdaSpace{}%
\AgdaFunction{tab}\AgdaSpace{}%
\AgdaSymbol{(}\AgdaFunction{slide₁}\AgdaSpace{}%
\AgdaSymbol{\{}\AgdaArgument{m}\AgdaSpace{}%
\AgdaSymbol{=}\AgdaSpace{}%
\AgdaNumber{1}\AgdaSymbol{\}\{}\AgdaArgument{n}\AgdaSpace{}%
\AgdaSymbol{=}\AgdaSpace{}%
\AgdaNumber{0}\AgdaSymbol{\}}\AgdaSpace{}%
\AgdaSymbol{(}\AgdaInductiveConstructor{ι}\AgdaSpace{}%
\AgdaSymbol{(}\AgdaOperator{\AgdaFunction{\#}}\AgdaSpace{}%
\AgdaNumber{0}\AgdaSymbol{))}\AgdaSpace{}%
\AgdaFunction{iota}\AgdaSymbol{)}\AgdaSpace{}%
\AgdaOperator{\AgdaDatatype{≡}}\AgdaSpace{}%
\AgdaNumber{0}\AgdaSpace{}%
\AgdaOperator{\AgdaInductiveConstructor{∷}}\AgdaSpace{}%
\AgdaInductiveConstructor{[]}\<%
\\
\>[2]\AgdaSymbol{\AgdaUnderscore{}}\AgdaSpace{}%
\AgdaSymbol{=}\AgdaSpace{}%
\AgdaInductiveConstructor{refl}\<%
\\
\>[0]\<%
\\
\>[2]\AgdaFunction{\AgdaUnderscore{}}\AgdaSpace{}%
\AgdaSymbol{:}\AgdaSpace{}%
\AgdaFunction{tab}\AgdaSpace{}%
\AgdaSymbol{(}\AgdaFunction{slide₁}\AgdaSpace{}%
\AgdaSymbol{\{}\AgdaArgument{m}\AgdaSpace{}%
\AgdaSymbol{=}\AgdaSpace{}%
\AgdaNumber{2}\AgdaSymbol{\}\{}\AgdaArgument{n}\AgdaSpace{}%
\AgdaSymbol{=}\AgdaSpace{}%
\AgdaNumber{0}\AgdaSymbol{\}}\AgdaSpace{}%
\AgdaSymbol{(}\AgdaInductiveConstructor{ι}\AgdaSpace{}%
\AgdaSymbol{(}\AgdaOperator{\AgdaFunction{\#}}\AgdaSpace{}%
\AgdaNumber{1}\AgdaSymbol{))}\AgdaSpace{}%
\AgdaFunction{iota}\AgdaSymbol{)}\AgdaSpace{}%
\AgdaOperator{\AgdaDatatype{≡}}\AgdaSpace{}%
\AgdaNumber{1}\AgdaSpace{}%
\AgdaOperator{\AgdaInductiveConstructor{∷}}\AgdaSpace{}%
\AgdaInductiveConstructor{[]}\<%
\\
\>[2]\AgdaSymbol{\AgdaUnderscore{}}\AgdaSpace{}%
\AgdaSymbol{=}\AgdaSpace{}%
\AgdaInductiveConstructor{refl}\<%
\\
\\[\AgdaEmptyExtraSkip]%
\>[0]\<%
\\
\>[2]\AgdaFunction{backslide₁}\AgdaSpace{}%
\AgdaSymbol{:}\AgdaSpace{}%
\AgdaFunction{Ix}\AgdaSpace{}%
\AgdaGeneralizable{m}\AgdaSpace{}%
\AgdaSymbol{→}\AgdaSpace{}%
\AgdaFunction{Vec}\AgdaSpace{}%
\AgdaSymbol{(}\AgdaInductiveConstructor{suc}\AgdaSpace{}%
\AgdaGeneralizable{n}\AgdaSymbol{)}\AgdaSpace{}%
\AgdaGeneralizable{X}\AgdaSpace{}%
\AgdaSymbol{→}\AgdaSpace{}%
\AgdaGeneralizable{X}\AgdaSpace{}%
\AgdaSymbol{→}\AgdaSpace{}%
\AgdaFunction{Vec}\AgdaSpace{}%
\AgdaSymbol{(}\AgdaGeneralizable{m}\AgdaSpace{}%
\AgdaOperator{\AgdaPrimitive{+}}\AgdaSpace{}%
\AgdaGeneralizable{n}\AgdaSymbol{)}\AgdaSpace{}%
\AgdaGeneralizable{X}\<%
\\
\>[2]\AgdaFunction{backslide₁}\AgdaSpace{}%
\AgdaSymbol{(}\AgdaInductiveConstructor{ι}\AgdaSpace{}%
\AgdaBound{i}\AgdaSymbol{)}\AgdaSpace{}%
\AgdaBound{v}\AgdaSpace{}%
\AgdaBound{e}\AgdaSpace{}%
\AgdaSymbol{(}\AgdaInductiveConstructor{ι}\AgdaSpace{}%
\AgdaBound{j}\AgdaSymbol{)}\AgdaSpace{}%
\AgdaKeyword{with}\AgdaSpace{}%
\AgdaBound{j}\AgdaSpace{}%
\AgdaOperator{\AgdaFunction{⊝}}\AgdaSpace{}%
\AgdaBound{i}\<%
\\
\>[2]\AgdaSymbol{...}\AgdaSpace{}%
\AgdaSymbol{|}\AgdaSpace{}%
\AgdaInductiveConstructor{yes}\AgdaSpace{}%
\AgdaSymbol{(}\AgdaBound{k}\AgdaSpace{}%
\AgdaOperator{\AgdaInductiveConstructor{,}}\AgdaSpace{}%
\AgdaSymbol{\AgdaUnderscore{})}\AgdaSpace{}%
\AgdaSymbol{=}\AgdaSpace{}%
\AgdaBound{v}\AgdaSpace{}%
\AgdaSymbol{(}\AgdaInductiveConstructor{ι}\AgdaSpace{}%
\AgdaBound{k}\AgdaSymbol{)}\<%
\\
\>[2]\AgdaSymbol{...}\AgdaSpace{}%
\AgdaSymbol{|}\AgdaSpace{}%
\AgdaInductiveConstructor{no}\AgdaSpace{}%
\AgdaSymbol{\AgdaUnderscore{}}\AgdaSpace{}%
\AgdaSymbol{=}\AgdaSpace{}%
\AgdaBound{e}\<%
\\
\\[\AgdaEmptyExtraSkip]%
\>[2]\AgdaFunction{test-⊕}\AgdaSpace{}%
\AgdaSymbol{=}\AgdaSpace{}%
\AgdaOperator{\AgdaFunction{\AgdaUnderscore{}⊕\AgdaUnderscore{}}}\AgdaSpace{}%
\AgdaSymbol{\{}\AgdaArgument{m}\AgdaSpace{}%
\AgdaSymbol{=}\AgdaSpace{}%
\AgdaNumber{3}\AgdaSymbol{\}}\AgdaSpace{}%
\AgdaSymbol{\{}\AgdaArgument{n}\AgdaSpace{}%
\AgdaSymbol{=}\AgdaSpace{}%
\AgdaNumber{4}\AgdaSymbol{\}}\AgdaSpace{}%
\AgdaSymbol{(}\AgdaOperator{\AgdaFunction{\#}}\AgdaSpace{}%
\AgdaNumber{1}\AgdaSymbol{)}\AgdaSpace{}%
\AgdaSymbol{(}\AgdaOperator{\AgdaFunction{\#}}\AgdaSpace{}%
\AgdaNumber{2}\AgdaSymbol{)}\<%
\\
\>[2]\AgdaFunction{test-⊝}\AgdaSpace{}%
\AgdaSymbol{=}\AgdaSpace{}%
\AgdaOperator{\AgdaFunction{\AgdaUnderscore{}⊝\AgdaUnderscore{}}}\AgdaSpace{}%
\AgdaSymbol{\{}\AgdaArgument{m}\AgdaSpace{}%
\AgdaSymbol{=}\AgdaSpace{}%
\AgdaNumber{3}\AgdaSymbol{\}}\AgdaSpace{}%
\AgdaSymbol{\{}\AgdaArgument{n}\AgdaSpace{}%
\AgdaSymbol{=}\AgdaSpace{}%
\AgdaNumber{4}\AgdaSymbol{\}}\AgdaSpace{}%
\AgdaSymbol{(}\AgdaOperator{\AgdaFunction{\#}}\AgdaSpace{}%
\AgdaNumber{3}\AgdaSymbol{)}\AgdaSpace{}%
\AgdaSymbol{(}\AgdaOperator{\AgdaFunction{\#}}\AgdaSpace{}%
\AgdaNumber{1}\AgdaSymbol{)}\<%
\\
\\[\AgdaEmptyExtraSkip]%
\>[2]\AgdaFunction{\AgdaUnderscore{}}\AgdaSpace{}%
\AgdaSymbol{:}\AgdaSpace{}%
\AgdaFunction{test-⊕}\AgdaSpace{}%
\AgdaOperator{\AgdaDatatype{≡}}\AgdaSpace{}%
\AgdaOperator{\AgdaFunction{\#}}\AgdaSpace{}%
\AgdaNumber{3}\<%
\\
\>[2]\AgdaSymbol{\AgdaUnderscore{}}\AgdaSpace{}%
\AgdaSymbol{=}\AgdaSpace{}%
\AgdaInductiveConstructor{refl}\<%
\\
\\[\AgdaEmptyExtraSkip]%
\>[2]\AgdaFunction{\AgdaUnderscore{}}\AgdaSpace{}%
\AgdaSymbol{:}\AgdaSpace{}%
\AgdaFunction{test-⊝}\AgdaSpace{}%
\AgdaOperator{\AgdaDatatype{≡}}\AgdaSpace{}%
\AgdaInductiveConstructor{yes}\AgdaSpace{}%
\AgdaSymbol{(}\AgdaOperator{\AgdaFunction{\#}}\AgdaSpace{}%
\AgdaNumber{2}\AgdaSpace{}%
\AgdaOperator{\AgdaInductiveConstructor{,}}\AgdaSpace{}%
\AgdaInductiveConstructor{refl}\AgdaSymbol{)}\<%
\\
\>[2]\AgdaSymbol{\AgdaUnderscore{}}\AgdaSpace{}%
\AgdaSymbol{=}\AgdaSpace{}%
\AgdaInductiveConstructor{refl}\<%
\\
\\[\AgdaEmptyExtraSkip]%
\>[2]\AgdaFunction{test-backslide}\AgdaSpace{}%
\AgdaSymbol{=}\AgdaSpace{}%
\AgdaFunction{tab}\AgdaSpace{}%
\AgdaOperator{\AgdaFunction{\$}}\AgdaSpace{}%
\AgdaFunction{backslide₁}\AgdaSpace{}%
\AgdaSymbol{\{}\AgdaArgument{m}\AgdaSpace{}%
\AgdaSymbol{=}\AgdaSpace{}%
\AgdaNumber{3}\AgdaSymbol{\}}\AgdaSpace{}%
\AgdaSymbol{\{}\AgdaArgument{n}\AgdaSpace{}%
\AgdaSymbol{=}\AgdaSpace{}%
\AgdaNumber{4}\AgdaSymbol{\}}\AgdaSpace{}%
\AgdaSymbol{(}\AgdaInductiveConstructor{ι}\AgdaSpace{}%
\AgdaSymbol{(}\AgdaOperator{\AgdaFunction{\#}}\AgdaSpace{}%
\AgdaNumber{2}\AgdaSymbol{))}\AgdaSpace{}%
\AgdaSymbol{(}\AgdaFunction{viota}\AgdaSpace{}%
\AgdaNumber{5}\AgdaSymbol{)}\AgdaSpace{}%
\AgdaNumber{121}\<%
\\
\\[\AgdaEmptyExtraSkip]%
\>[2]\AgdaFunction{\AgdaUnderscore{}}\AgdaSpace{}%
\AgdaSymbol{:}\AgdaSpace{}%
\AgdaFunction{test-backslide}\AgdaSpace{}%
\AgdaOperator{\AgdaDatatype{≡}}\AgdaSpace{}%
\AgdaNumber{121}\AgdaSpace{}%
\AgdaOperator{\AgdaInductiveConstructor{∷}}\AgdaSpace{}%
\AgdaNumber{121}\AgdaSpace{}%
\AgdaOperator{\AgdaInductiveConstructor{∷}}\AgdaSpace{}%
\AgdaNumber{0}\AgdaSpace{}%
\AgdaOperator{\AgdaInductiveConstructor{∷}}\AgdaSpace{}%
\AgdaNumber{1}\AgdaSpace{}%
\AgdaOperator{\AgdaInductiveConstructor{∷}}\AgdaSpace{}%
\AgdaNumber{2}\AgdaSpace{}%
\AgdaOperator{\AgdaInductiveConstructor{∷}}\AgdaSpace{}%
\AgdaNumber{3}\AgdaSpace{}%
\AgdaOperator{\AgdaInductiveConstructor{∷}}\AgdaSpace{}%
\AgdaNumber{4}\AgdaSpace{}%
\AgdaOperator{\AgdaInductiveConstructor{∷}}\AgdaSpace{}%
\AgdaInductiveConstructor{[]}\<%
\\
\>[2]\AgdaSymbol{\AgdaUnderscore{}}\AgdaSpace{}%
\AgdaSymbol{=}\AgdaSpace{}%
\AgdaInductiveConstructor{refl}\<%
\\
\>[0]\<%
\end{code}

\begin{code}[hide]%
\>[0]\AgdaKeyword{open}\AgdaSpace{}%
\AgdaKeyword{import}\AgdaSpace{}%
\AgdaModule{Relation.Binary.PropositionalEquality}\<%
\\
\>[0]\AgdaKeyword{open}\AgdaSpace{}%
\AgdaKeyword{import}\AgdaSpace{}%
\AgdaModule{Relation.Nullary}\<%
\\
\>[0]\AgdaKeyword{open}\AgdaSpace{}%
\AgdaKeyword{import}\AgdaSpace{}%
\AgdaModule{Data.List}\AgdaSpace{}%
\AgdaKeyword{using}\AgdaSpace{}%
\AgdaSymbol{(}\AgdaDatatype{List}\AgdaSymbol{;}\AgdaSpace{}%
\AgdaInductiveConstructor{[]}\AgdaSymbol{;}\AgdaSpace{}%
\AgdaOperator{\AgdaInductiveConstructor{\AgdaUnderscore{}∷\AgdaUnderscore{}}}\AgdaSymbol{)}\<%
\\
\>[0]\AgdaKeyword{open}\AgdaSpace{}%
\AgdaKeyword{import}\AgdaSpace{}%
\AgdaModule{Data.Empty}\<%
\\
\>[0]\AgdaKeyword{open}\AgdaSpace{}%
\AgdaKeyword{import}\AgdaSpace{}%
\AgdaModule{Function}\<%
\\
\\[\AgdaEmptyExtraSkip]%
\>[0]\AgdaComment{--\ Our\ local\ files.}\<%
\\
\>[0]\AgdaKeyword{open}\AgdaSpace{}%
\AgdaKeyword{import}\AgdaSpace{}%
\AgdaModule{arrays}\<%
\\
\>[0]\AgdaKeyword{module}\AgdaSpace{}%
\AgdaModule{\AgdaUnderscore{}}\AgdaSpace{}%
\AgdaKeyword{where}\<%
\end{code}
\section{Embedded DSL}

Any implementation of automatic differentiation has to decide which operations
are supported.  Surely, it does not make sense to compute derivatives
of a function that opens a file.  This choice, no matter how it is implemented,
can be seen as a definition of an embedded language.
Once we accept to identify an embedded language, the idea of embedding it in a way
that facilitates extraction actually appears rather naturally and thus advances
the approach that we propose in this paper.

Coming back to our example, we have to choose the primitives that the embedded language
should support. They need to be sufficient to express AD as well as to define CNNs.
The main trade-off here is the choice of the level of abstraction of these primitives:
low-level primitives are easier to differentiate, but they make the overall expressions
more complex which also adds to the challenge of optimising code.
Making this choice is difficult and, most likely, requires quite some adjustment 
when striving for performance.
Here we see a key benefit of the approach we propose in this paper:
the use of a single framework for the embedding, the optimisation, and the extraction
makes the implementation comparatively small, allowing for quick adjustments in the
level of abstraction, code optimisation and its extraction.
We start with a pragmatic approach; we include the primitives that are either shared
by the model and the back-end
or that can be easily implemented in the back-end language.

It turns out that it is possible to choose the primitives in a way that the derivatives 
can be expressed in the very same embedded language. 
While this at first glance may just seem to be just a nice coincidence, it turns out
that this has several tangible benefits: high-order derivatives can be computed 
by the same transformation and we can share all optimisations between the code itself
and its derivatives.
\begin{code}[hide]%
\>[0]\AgdaKeyword{module}\AgdaSpace{}%
\AgdaModule{Lang}\AgdaSpace{}%
\AgdaKeyword{where}\<%
\\
\>[0][@{}l@{\AgdaIndent{0}}]%
\>[2]\AgdaKeyword{open}\AgdaSpace{}%
\AgdaModule{Array}\AgdaSpace{}%
\AgdaKeyword{hiding}\AgdaSpace{}%
\AgdaSymbol{(}\AgdaFunction{sum}\AgdaSymbol{;}\AgdaSpace{}%
\AgdaFunction{slide}\AgdaSymbol{;}\AgdaSpace{}%
\AgdaFunction{backslide}\AgdaSymbol{)}\<%
\\
\>[2]\AgdaKeyword{open}\AgdaSpace{}%
\AgdaKeyword{import}\AgdaSpace{}%
\AgdaModule{Data.Nat}\AgdaSpace{}%
\AgdaKeyword{using}\AgdaSpace{}%
\AgdaSymbol{(}\AgdaDatatype{ℕ}\AgdaSymbol{;}\AgdaSpace{}%
\AgdaInductiveConstructor{zero}\AgdaSymbol{;}\AgdaSpace{}%
\AgdaInductiveConstructor{suc}\AgdaSymbol{)}\<%
\\
\>[2]\AgdaKeyword{infixl}\AgdaSpace{}%
\AgdaNumber{15}\AgdaSpace{}%
\AgdaOperator{\AgdaInductiveConstructor{\AgdaUnderscore{}▹\AgdaUnderscore{}}}\<%
\end{code}

As we operate within a dependently-typed proof-assistant, we can easily make our
embedded language well-scoped and intrinsically typed (shaped in our case).  Our
context \AF{Ctx} is a snoc-list of shapes where each shape has a tag indicating whether
it is an index or an array.  We use de Bruijn variables given by the relation
\AF{\_∈\_} in the usual way.  We also define variables \AB{v₁}, \AB{v₂}, \etc{}
by iteratively applying \AC{vₛ} to \AC{v₀} (definition not shown).
\begin{mathpar}
\codeblock{\begin{code}%
\>[2]\AgdaKeyword{data}\AgdaSpace{}%
\AgdaDatatype{IS}\AgdaSpace{}%
\AgdaSymbol{:}\AgdaSpace{}%
\AgdaPrimitive{Set}\AgdaSpace{}%
\AgdaKeyword{where}\<%
\\
\>[2][@{}l@{\AgdaIndent{0}}]%
\>[4]\AgdaInductiveConstructor{ix}%
\>[8]\AgdaSymbol{:}\AgdaSpace{}%
\AgdaDatatype{S}\AgdaSpace{}%
\AgdaSymbol{→}\AgdaSpace{}%
\AgdaDatatype{IS}\<%
\\
\>[4]\AgdaInductiveConstructor{ar}%
\>[8]\AgdaSymbol{:}\AgdaSpace{}%
\AgdaDatatype{S}\AgdaSpace{}%
\AgdaSymbol{→}\AgdaSpace{}%
\AgdaDatatype{IS}\<%
\end{code}}
\and
\codeblock{\begin{code}%
\>[2]\AgdaKeyword{data}\AgdaSpace{}%
\AgdaDatatype{Ctx}\AgdaSpace{}%
\AgdaSymbol{:}\AgdaSpace{}%
\AgdaPrimitive{Set}\AgdaSpace{}%
\AgdaKeyword{where}\<%
\\
\>[2][@{}l@{\AgdaIndent{0}}]%
\>[4]\AgdaInductiveConstructor{ε}%
\>[9]\AgdaSymbol{:}\AgdaSpace{}%
\AgdaDatatype{Ctx}\<%
\\
\>[4]\AgdaOperator{\AgdaInductiveConstructor{\AgdaUnderscore{}▹\AgdaUnderscore{}}}%
\>[9]\AgdaSymbol{:}\AgdaSpace{}%
\AgdaDatatype{Ctx}\AgdaSpace{}%
\AgdaSymbol{→}\AgdaSpace{}%
\AgdaDatatype{IS}\AgdaSpace{}%
\AgdaSymbol{→}\AgdaSpace{}%
\AgdaDatatype{Ctx}\<%
\end{code}}
\and
\codeblock{\begin{code}[hide]%
\>[2]\AgdaKeyword{variable}\<%
\\
\>[2][@{}l@{\AgdaIndent{0}}]%
\>[4]\AgdaGeneralizable{Γ}\AgdaSpace{}%
\AgdaGeneralizable{Δ}\AgdaSpace{}%
\AgdaGeneralizable{Ξ}\AgdaSpace{}%
\AgdaGeneralizable{Ψ}\AgdaSpace{}%
\AgdaSymbol{:}\AgdaSpace{}%
\AgdaDatatype{Ctx}\<%
\\
\>[4]\AgdaGeneralizable{is}\AgdaSpace{}%
\AgdaGeneralizable{ip}\AgdaSpace{}%
\AgdaGeneralizable{iq}\AgdaSpace{}%
\AgdaGeneralizable{ir}\AgdaSpace{}%
\AgdaSymbol{:}\AgdaSpace{}%
\AgdaDatatype{IS}\<%
\end{code}
\begin{code}%
\>[2]\AgdaKeyword{data}\AgdaSpace{}%
\AgdaOperator{\AgdaDatatype{\AgdaUnderscore{}∈\AgdaUnderscore{}}}\AgdaSpace{}%
\AgdaSymbol{:}\AgdaSpace{}%
\AgdaDatatype{IS}\AgdaSpace{}%
\AgdaSymbol{→}\AgdaSpace{}%
\AgdaDatatype{Ctx}\AgdaSpace{}%
\AgdaSymbol{→}\AgdaSpace{}%
\AgdaPrimitive{Set}\AgdaSpace{}%
\AgdaKeyword{where}\<%
\\
\>[2][@{}l@{\AgdaIndent{0}}]%
\>[4]\AgdaInductiveConstructor{v₀}%
\>[8]\AgdaSymbol{:}\AgdaSpace{}%
\AgdaGeneralizable{is}\AgdaSpace{}%
\AgdaOperator{\AgdaDatatype{∈}}\AgdaSpace{}%
\AgdaSymbol{(}\AgdaGeneralizable{Γ}\AgdaSpace{}%
\AgdaOperator{\AgdaInductiveConstructor{▹}}\AgdaSpace{}%
\AgdaGeneralizable{is}\AgdaSymbol{)}\<%
\\
\>[4]\AgdaInductiveConstructor{vₛ}%
\>[8]\AgdaSymbol{:}\AgdaSpace{}%
\AgdaGeneralizable{is}\AgdaSpace{}%
\AgdaOperator{\AgdaDatatype{∈}}\AgdaSpace{}%
\AgdaGeneralizable{Γ}\AgdaSpace{}%
\AgdaSymbol{→}\AgdaSpace{}%
\AgdaGeneralizable{is}\AgdaSpace{}%
\AgdaOperator{\AgdaDatatype{∈}}\AgdaSpace{}%
\AgdaSymbol{(}\AgdaGeneralizable{Γ}\AgdaSpace{}%
\AgdaOperator{\AgdaInductiveConstructor{▹}}\AgdaSpace{}%
\AgdaGeneralizable{ip}\AgdaSymbol{)}\<%
\end{code}}
\end{mathpar}
Note that while our contexts are non-dependent (\ie{} the shapes do not depend on the
terms), we use non-trivial shape dependencies within the constructors.  The embedded
language does not have a notion of shape as a value, therefore all the shape dependencies
are handled by Agda, keeping our language simply typed (shaped).  This separation is
very helpful when it comes to writing embedded programs.
\begin{code}[hide]%
\>[2]\AgdaComment{--pattern\ v₀\ =\ v₀}\<%
\\
\>[2]\AgdaKeyword{pattern}\AgdaSpace{}%
\AgdaInductiveConstructor{v₁}\AgdaSpace{}%
\AgdaSymbol{=}\AgdaSpace{}%
\AgdaInductiveConstructor{vₛ}\AgdaSpace{}%
\AgdaInductiveConstructor{v₀}\<%
\\
\>[2]\AgdaKeyword{pattern}\AgdaSpace{}%
\AgdaInductiveConstructor{v₂}\AgdaSpace{}%
\AgdaSymbol{=}\AgdaSpace{}%
\AgdaInductiveConstructor{vₛ}\AgdaSpace{}%
\AgdaInductiveConstructor{v₁}\<%
\\
\>[2]\AgdaKeyword{pattern}\AgdaSpace{}%
\AgdaInductiveConstructor{v₃}\AgdaSpace{}%
\AgdaSymbol{=}\AgdaSpace{}%
\AgdaInductiveConstructor{vₛ}\AgdaSpace{}%
\AgdaInductiveConstructor{v₂}\<%
\\
\>[2]\AgdaKeyword{pattern}\AgdaSpace{}%
\AgdaInductiveConstructor{v₄}\AgdaSpace{}%
\AgdaSymbol{=}\AgdaSpace{}%
\AgdaInductiveConstructor{vₛ}\AgdaSpace{}%
\AgdaInductiveConstructor{v₃}\<%
\\
\>[2]\AgdaKeyword{pattern}\AgdaSpace{}%
\AgdaInductiveConstructor{v₅}\AgdaSpace{}%
\AgdaSymbol{=}\AgdaSpace{}%
\AgdaInductiveConstructor{vₛ}\AgdaSpace{}%
\AgdaInductiveConstructor{v₄}\<%
\\
\>[2]\AgdaKeyword{pattern}\AgdaSpace{}%
\AgdaInductiveConstructor{v₆}\AgdaSpace{}%
\AgdaSymbol{=}\AgdaSpace{}%
\AgdaInductiveConstructor{vₛ}\AgdaSpace{}%
\AgdaInductiveConstructor{v₅}\<%
\\
\>[2]\AgdaKeyword{pattern}\AgdaSpace{}%
\AgdaInductiveConstructor{v₇}\AgdaSpace{}%
\AgdaSymbol{=}\AgdaSpace{}%
\AgdaInductiveConstructor{vₛ}\AgdaSpace{}%
\AgdaInductiveConstructor{v₆}\<%
\\
\>[2]\AgdaKeyword{pattern}\AgdaSpace{}%
\AgdaInductiveConstructor{v₈}\AgdaSpace{}%
\AgdaSymbol{=}\AgdaSpace{}%
\AgdaInductiveConstructor{vₛ}\AgdaSpace{}%
\AgdaInductiveConstructor{v₇}\<%
\\
\>[2]\AgdaKeyword{pattern}\AgdaSpace{}%
\AgdaInductiveConstructor{v₉}\AgdaSpace{}%
\AgdaSymbol{=}\AgdaSpace{}%
\AgdaInductiveConstructor{vₛ}\AgdaSpace{}%
\AgdaInductiveConstructor{v₈}\<%
\\
\\[\AgdaEmptyExtraSkip]%
\>[2]\AgdaKeyword{infixl}\AgdaSpace{}%
\AgdaNumber{10}\AgdaSpace{}%
\AgdaOperator{\AgdaInductiveConstructor{\AgdaUnderscore{}⊞\AgdaUnderscore{}}}\<%
\\
\>[2]\AgdaKeyword{infixl}\AgdaSpace{}%
\AgdaNumber{15}\AgdaSpace{}%
\AgdaOperator{\AgdaInductiveConstructor{\AgdaUnderscore{}⊠\AgdaUnderscore{}}}\<%
\end{code}
We start with two helper definitions: a singleton shape that we call \AF{unit}
and the type for binary operations that we support (for now only addition and
multiplication).
\begin{mathpar}
\codeblock{\begin{code}%
\>[2]\AgdaFunction{unit}\AgdaSpace{}%
\AgdaSymbol{:}\AgdaSpace{}%
\AgdaDatatype{S}\<%
\\
\>[2]\AgdaFunction{unit}\AgdaSpace{}%
\AgdaSymbol{=}\AgdaSpace{}%
\AgdaInductiveConstructor{ι}\AgdaSpace{}%
\AgdaNumber{1}\<%
\end{code}}
\and
\codeblock{\begin{code}%
\>[2]\AgdaKeyword{data}\AgdaSpace{}%
\AgdaDatatype{Bop}\AgdaSpace{}%
\AgdaSymbol{:}\AgdaSpace{}%
\AgdaPrimitive{Set}\AgdaSpace{}%
\AgdaKeyword{where}\<%
\\
\>[2][@{}l@{\AgdaIndent{0}}]%
\>[4]\AgdaInductiveConstructor{plus}\AgdaSpace{}%
\AgdaInductiveConstructor{mul}\AgdaSpace{}%
\AgdaSymbol{:}\AgdaSpace{}%
\AgdaDatatype{Bop}\<%
\end{code}}
\end{mathpar}

The embedded language \AF{E} includes: variables \AC{var}; constants 0 and 1 given
by \AC{zero} and \AC{one} correspondingly; three flavours of array constructor/eliminator
pairs given by \AC{imapₛ}/\AC{selₛ}, \AC{imap}/\AC{sel} and \AC{imapb}/\AC{selb};
summation \AC{sum}; conditional \AC{zero-but} where the predicate is fixed to equality
of two indices and the else branch is zero; \AC{slide} and \AC{backslide} exactly
as described before; and numerical operations.  The latter includes \AC{logistic},
plus and multiplication, division by a constant \AC{scaledown}, and unary \AC{minus}.
The definition of the embedded language \AF{E} follows.  We also introduce the
syntax for infix plus and multiplication denoted \AC{⊞} and \AC{⊠} correspondingly.
\begin{code}%
\>[2]\AgdaKeyword{data}\AgdaSpace{}%
\AgdaDatatype{E}\AgdaSpace{}%
\AgdaSymbol{:}\AgdaSpace{}%
\AgdaDatatype{Ctx}\AgdaSpace{}%
\AgdaSymbol{→}\AgdaSpace{}%
\AgdaDatatype{IS}\AgdaSpace{}%
\AgdaSymbol{→}\AgdaSpace{}%
\AgdaPrimitive{Set}\AgdaSpace{}%
\AgdaKeyword{where}\<%
\\
\>[2][@{}l@{\AgdaIndent{0}}]%
\>[4]\AgdaInductiveConstructor{var}%
\>[15]\AgdaSymbol{:}\AgdaSpace{}%
\AgdaGeneralizable{is}\AgdaSpace{}%
\AgdaOperator{\AgdaDatatype{∈}}\AgdaSpace{}%
\AgdaGeneralizable{Γ}\AgdaSpace{}%
\AgdaSymbol{→}\AgdaSpace{}%
\AgdaDatatype{E}\AgdaSpace{}%
\AgdaGeneralizable{Γ}\AgdaSpace{}%
\AgdaGeneralizable{is}\<%
\\
\>[4]\AgdaInductiveConstructor{zero}%
\>[15]\AgdaSymbol{:}\AgdaSpace{}%
\AgdaDatatype{E}\AgdaSpace{}%
\AgdaGeneralizable{Γ}\AgdaSpace{}%
\AgdaSymbol{(}\AgdaInductiveConstructor{ar}\AgdaSpace{}%
\AgdaGeneralizable{s}\AgdaSymbol{)}\<%
\\
\>[4]\AgdaInductiveConstructor{one}%
\>[15]\AgdaSymbol{:}\AgdaSpace{}%
\AgdaDatatype{E}\AgdaSpace{}%
\AgdaGeneralizable{Γ}\AgdaSpace{}%
\AgdaSymbol{(}\AgdaInductiveConstructor{ar}\AgdaSpace{}%
\AgdaGeneralizable{s}\AgdaSymbol{)}\<%
\\
\\[\AgdaEmptyExtraSkip]%
\>[4]\AgdaInductiveConstructor{imapₛ}%
\>[15]\AgdaSymbol{:}\AgdaSpace{}%
\AgdaDatatype{E}\AgdaSpace{}%
\AgdaSymbol{(}\AgdaGeneralizable{Γ}\AgdaSpace{}%
\AgdaOperator{\AgdaInductiveConstructor{▹}}\AgdaSpace{}%
\AgdaInductiveConstructor{ix}\AgdaSpace{}%
\AgdaGeneralizable{s}\AgdaSymbol{)}\AgdaSpace{}%
\AgdaSymbol{(}\AgdaInductiveConstructor{ar}\AgdaSpace{}%
\AgdaFunction{unit}\AgdaSymbol{)}\AgdaSpace{}%
\AgdaSymbol{→}\AgdaSpace{}%
\AgdaDatatype{E}\AgdaSpace{}%
\AgdaGeneralizable{Γ}\AgdaSpace{}%
\AgdaSymbol{(}\AgdaInductiveConstructor{ar}\AgdaSpace{}%
\AgdaGeneralizable{s}\AgdaSymbol{)}\<%
\\
\>[4]\AgdaInductiveConstructor{selₛ}%
\>[15]\AgdaSymbol{:}\AgdaSpace{}%
\AgdaDatatype{E}\AgdaSpace{}%
\AgdaGeneralizable{Γ}\AgdaSpace{}%
\AgdaSymbol{(}\AgdaInductiveConstructor{ar}\AgdaSpace{}%
\AgdaGeneralizable{s}\AgdaSymbol{)}\AgdaSpace{}%
\AgdaSymbol{→}\AgdaSpace{}%
\AgdaDatatype{E}\AgdaSpace{}%
\AgdaGeneralizable{Γ}\AgdaSpace{}%
\AgdaSymbol{(}\AgdaInductiveConstructor{ix}\AgdaSpace{}%
\AgdaGeneralizable{s}\AgdaSymbol{)}\AgdaSpace{}%
\AgdaSymbol{→}\AgdaSpace{}%
\AgdaDatatype{E}\AgdaSpace{}%
\AgdaGeneralizable{Γ}\AgdaSpace{}%
\AgdaSymbol{(}\AgdaInductiveConstructor{ar}\AgdaSpace{}%
\AgdaFunction{unit}\AgdaSymbol{)}\<%
\\
\\[\AgdaEmptyExtraSkip]%
\>[4]\AgdaInductiveConstructor{imap}%
\>[15]\AgdaSymbol{:}\AgdaSpace{}%
\AgdaDatatype{E}\AgdaSpace{}%
\AgdaSymbol{(}\AgdaGeneralizable{Γ}\AgdaSpace{}%
\AgdaOperator{\AgdaInductiveConstructor{▹}}\AgdaSpace{}%
\AgdaInductiveConstructor{ix}\AgdaSpace{}%
\AgdaGeneralizable{s}\AgdaSymbol{)}\AgdaSpace{}%
\AgdaSymbol{(}\AgdaInductiveConstructor{ar}\AgdaSpace{}%
\AgdaGeneralizable{p}\AgdaSymbol{)}\AgdaSpace{}%
\AgdaSymbol{→}\AgdaSpace{}%
\AgdaDatatype{E}\AgdaSpace{}%
\AgdaGeneralizable{Γ}\AgdaSpace{}%
\AgdaSymbol{(}\AgdaInductiveConstructor{ar}\AgdaSpace{}%
\AgdaSymbol{(}\AgdaGeneralizable{s}\AgdaSpace{}%
\AgdaOperator{\AgdaInductiveConstructor{⊗}}\AgdaSpace{}%
\AgdaGeneralizable{p}\AgdaSymbol{))}\<%
\\
\>[4]\AgdaInductiveConstructor{sel}%
\>[15]\AgdaSymbol{:}\AgdaSpace{}%
\AgdaDatatype{E}\AgdaSpace{}%
\AgdaGeneralizable{Γ}\AgdaSpace{}%
\AgdaSymbol{(}\AgdaInductiveConstructor{ar}\AgdaSpace{}%
\AgdaSymbol{(}\AgdaGeneralizable{s}\AgdaSpace{}%
\AgdaOperator{\AgdaInductiveConstructor{⊗}}\AgdaSpace{}%
\AgdaGeneralizable{p}\AgdaSymbol{))}\AgdaSpace{}%
\AgdaSymbol{→}\AgdaSpace{}%
\AgdaDatatype{E}\AgdaSpace{}%
\AgdaGeneralizable{Γ}\AgdaSpace{}%
\AgdaSymbol{(}\AgdaInductiveConstructor{ix}\AgdaSpace{}%
\AgdaGeneralizable{s}\AgdaSymbol{)}\AgdaSpace{}%
\AgdaSymbol{→}\AgdaSpace{}%
\AgdaDatatype{E}\AgdaSpace{}%
\AgdaGeneralizable{Γ}\AgdaSpace{}%
\AgdaSymbol{(}\AgdaInductiveConstructor{ar}\AgdaSpace{}%
\AgdaGeneralizable{p}\AgdaSymbol{)}\<%
\\
\\[\AgdaEmptyExtraSkip]%
\>[4]\AgdaInductiveConstructor{imapb}%
\>[15]\AgdaSymbol{:}\AgdaSpace{}%
\AgdaGeneralizable{s}\AgdaSpace{}%
\AgdaOperator{\AgdaDatatype{*}}\AgdaSpace{}%
\AgdaGeneralizable{p}\AgdaSpace{}%
\AgdaOperator{\AgdaDatatype{≈}}\AgdaSpace{}%
\AgdaGeneralizable{q}\AgdaSpace{}%
\AgdaSymbol{→}\AgdaSpace{}%
\AgdaDatatype{E}\AgdaSpace{}%
\AgdaSymbol{(}\AgdaGeneralizable{Γ}\AgdaSpace{}%
\AgdaOperator{\AgdaInductiveConstructor{▹}}\AgdaSpace{}%
\AgdaInductiveConstructor{ix}\AgdaSpace{}%
\AgdaGeneralizable{s}\AgdaSymbol{)}\AgdaSpace{}%
\AgdaSymbol{(}\AgdaInductiveConstructor{ar}\AgdaSpace{}%
\AgdaGeneralizable{p}\AgdaSymbol{)}\AgdaSpace{}%
\AgdaSymbol{→}\AgdaSpace{}%
\AgdaDatatype{E}\AgdaSpace{}%
\AgdaGeneralizable{Γ}\AgdaSpace{}%
\AgdaSymbol{(}\AgdaInductiveConstructor{ar}\AgdaSpace{}%
\AgdaGeneralizable{q}\AgdaSymbol{)}\<%
\\
\>[4]\AgdaInductiveConstructor{selb}%
\>[15]\AgdaSymbol{:}\AgdaSpace{}%
\AgdaGeneralizable{s}\AgdaSpace{}%
\AgdaOperator{\AgdaDatatype{*}}\AgdaSpace{}%
\AgdaGeneralizable{p}\AgdaSpace{}%
\AgdaOperator{\AgdaDatatype{≈}}\AgdaSpace{}%
\AgdaGeneralizable{q}\AgdaSpace{}%
\AgdaSymbol{→}\AgdaSpace{}%
\AgdaDatatype{E}\AgdaSpace{}%
\AgdaGeneralizable{Γ}\AgdaSpace{}%
\AgdaSymbol{(}\AgdaInductiveConstructor{ar}\AgdaSpace{}%
\AgdaGeneralizable{q}\AgdaSymbol{)}\AgdaSpace{}%
\AgdaSymbol{→}\AgdaSpace{}%
\AgdaDatatype{E}\AgdaSpace{}%
\AgdaGeneralizable{Γ}\AgdaSpace{}%
\AgdaSymbol{(}\AgdaInductiveConstructor{ix}\AgdaSpace{}%
\AgdaGeneralizable{s}\AgdaSymbol{)}\AgdaSpace{}%
\AgdaSymbol{→}\AgdaSpace{}%
\AgdaDatatype{E}\AgdaSpace{}%
\AgdaGeneralizable{Γ}\AgdaSpace{}%
\AgdaSymbol{(}\AgdaInductiveConstructor{ar}\AgdaSpace{}%
\AgdaGeneralizable{p}\AgdaSymbol{)}\<%
\\
\\[\AgdaEmptyExtraSkip]%
\>[4]\AgdaInductiveConstructor{sum}%
\>[15]\AgdaSymbol{:}\AgdaSpace{}%
\AgdaDatatype{E}\AgdaSpace{}%
\AgdaSymbol{(}\AgdaGeneralizable{Γ}\AgdaSpace{}%
\AgdaOperator{\AgdaInductiveConstructor{▹}}\AgdaSpace{}%
\AgdaInductiveConstructor{ix}\AgdaSpace{}%
\AgdaGeneralizable{s}\AgdaSymbol{)}\AgdaSpace{}%
\AgdaSymbol{(}\AgdaInductiveConstructor{ar}\AgdaSpace{}%
\AgdaGeneralizable{p}\AgdaSymbol{)}\AgdaSpace{}%
\AgdaSymbol{→}\AgdaSpace{}%
\AgdaDatatype{E}\AgdaSpace{}%
\AgdaGeneralizable{Γ}\AgdaSpace{}%
\AgdaSymbol{(}\AgdaInductiveConstructor{ar}\AgdaSpace{}%
\AgdaGeneralizable{p}\AgdaSymbol{)}\<%
\\
\>[4]\AgdaInductiveConstructor{zero-but}%
\>[15]\AgdaSymbol{:}\AgdaSpace{}%
\AgdaDatatype{E}\AgdaSpace{}%
\AgdaGeneralizable{Γ}\AgdaSpace{}%
\AgdaSymbol{(}\AgdaInductiveConstructor{ix}\AgdaSpace{}%
\AgdaGeneralizable{s}\AgdaSymbol{)}\AgdaSpace{}%
\AgdaSymbol{→}\AgdaSpace{}%
\AgdaDatatype{E}\AgdaSpace{}%
\AgdaGeneralizable{Γ}\AgdaSpace{}%
\AgdaSymbol{(}\AgdaInductiveConstructor{ix}\AgdaSpace{}%
\AgdaGeneralizable{s}\AgdaSymbol{)}\AgdaSpace{}%
\AgdaSymbol{→}\AgdaSpace{}%
\AgdaDatatype{E}\AgdaSpace{}%
\AgdaGeneralizable{Γ}\AgdaSpace{}%
\AgdaSymbol{(}\AgdaInductiveConstructor{ar}\AgdaSpace{}%
\AgdaGeneralizable{p}\AgdaSymbol{)}\AgdaSpace{}%
\AgdaSymbol{→}\AgdaSpace{}%
\AgdaDatatype{E}\AgdaSpace{}%
\AgdaGeneralizable{Γ}\AgdaSpace{}%
\AgdaSymbol{(}\AgdaInductiveConstructor{ar}\AgdaSpace{}%
\AgdaGeneralizable{p}\AgdaSymbol{)}\<%
\\
\\[\AgdaEmptyExtraSkip]%
\>[4]\AgdaInductiveConstructor{slide}%
\>[15]\AgdaSymbol{:}\AgdaSpace{}%
\AgdaDatatype{E}\AgdaSpace{}%
\AgdaGeneralizable{Γ}\AgdaSpace{}%
\AgdaSymbol{(}\AgdaInductiveConstructor{ix}\AgdaSpace{}%
\AgdaGeneralizable{s}\AgdaSymbol{)}\AgdaSpace{}%
\AgdaSymbol{→}\AgdaSpace{}%
\AgdaGeneralizable{s}\AgdaSpace{}%
\AgdaOperator{\AgdaDatatype{+}}\AgdaSpace{}%
\AgdaGeneralizable{p}\AgdaSpace{}%
\AgdaOperator{\AgdaDatatype{≈}}\AgdaSpace{}%
\AgdaGeneralizable{r}\AgdaSpace{}%
\AgdaSymbol{→}\AgdaSpace{}%
\AgdaDatatype{E}\AgdaSpace{}%
\AgdaGeneralizable{Γ}\AgdaSpace{}%
\AgdaSymbol{(}\AgdaInductiveConstructor{ar}\AgdaSpace{}%
\AgdaGeneralizable{r}\AgdaSymbol{)}\AgdaSpace{}%
\AgdaSymbol{→}\AgdaSpace{}%
\AgdaOperator{\AgdaDatatype{suc}}\AgdaSpace{}%
\AgdaGeneralizable{p}\AgdaSpace{}%
\AgdaOperator{\AgdaDatatype{≈}}\AgdaSpace{}%
\AgdaGeneralizable{u}\AgdaSpace{}%
\AgdaSymbol{→}\AgdaSpace{}%
\AgdaDatatype{E}\AgdaSpace{}%
\AgdaGeneralizable{Γ}\AgdaSpace{}%
\AgdaSymbol{(}\AgdaInductiveConstructor{ar}\AgdaSpace{}%
\AgdaGeneralizable{u}\AgdaSymbol{)}\<%
\\
\>[4]\AgdaInductiveConstructor{backslide}%
\>[15]\AgdaSymbol{:}\AgdaSpace{}%
\AgdaDatatype{E}\AgdaSpace{}%
\AgdaGeneralizable{Γ}\AgdaSpace{}%
\AgdaSymbol{(}\AgdaInductiveConstructor{ix}\AgdaSpace{}%
\AgdaGeneralizable{s}\AgdaSymbol{)}\AgdaSpace{}%
\AgdaSymbol{→}\AgdaSpace{}%
\AgdaDatatype{E}\AgdaSpace{}%
\AgdaGeneralizable{Γ}\AgdaSpace{}%
\AgdaSymbol{(}\AgdaInductiveConstructor{ar}\AgdaSpace{}%
\AgdaGeneralizable{u}\AgdaSymbol{)}\AgdaSpace{}%
\AgdaSymbol{→}\AgdaSpace{}%
\AgdaOperator{\AgdaDatatype{suc}}\AgdaSpace{}%
\AgdaGeneralizable{p}\AgdaSpace{}%
\AgdaOperator{\AgdaDatatype{≈}}\AgdaSpace{}%
\AgdaGeneralizable{u}\AgdaSpace{}%
\AgdaSymbol{→}\AgdaSpace{}%
\AgdaGeneralizable{s}\AgdaSpace{}%
\AgdaOperator{\AgdaDatatype{+}}\AgdaSpace{}%
\AgdaGeneralizable{p}\AgdaSpace{}%
\AgdaOperator{\AgdaDatatype{≈}}\AgdaSpace{}%
\AgdaGeneralizable{r}\AgdaSpace{}%
\AgdaSymbol{→}\AgdaSpace{}%
\AgdaDatatype{E}\AgdaSpace{}%
\AgdaGeneralizable{Γ}\AgdaSpace{}%
\AgdaSymbol{(}\AgdaInductiveConstructor{ar}\AgdaSpace{}%
\AgdaGeneralizable{r}\AgdaSymbol{)}\<%
\\
\\[\AgdaEmptyExtraSkip]%
\>[4]\AgdaInductiveConstructor{logistic}%
\>[15]\AgdaSymbol{:}\AgdaSpace{}%
\AgdaDatatype{E}\AgdaSpace{}%
\AgdaGeneralizable{Γ}\AgdaSpace{}%
\AgdaSymbol{(}\AgdaInductiveConstructor{ar}\AgdaSpace{}%
\AgdaGeneralizable{s}\AgdaSymbol{)}\AgdaSpace{}%
\AgdaSymbol{→}\AgdaSpace{}%
\AgdaDatatype{E}\AgdaSpace{}%
\AgdaGeneralizable{Γ}\AgdaSpace{}%
\AgdaSymbol{(}\AgdaInductiveConstructor{ar}\AgdaSpace{}%
\AgdaGeneralizable{s}\AgdaSymbol{)}\<%
\\
\>[4]\AgdaInductiveConstructor{bin}%
\>[15]\AgdaSymbol{:}\AgdaSpace{}%
\AgdaDatatype{Bop}\AgdaSpace{}%
\AgdaSymbol{→}\AgdaSpace{}%
\AgdaDatatype{E}\AgdaSpace{}%
\AgdaGeneralizable{Γ}\AgdaSpace{}%
\AgdaSymbol{(}\AgdaInductiveConstructor{ar}\AgdaSpace{}%
\AgdaGeneralizable{s}\AgdaSymbol{)}\AgdaSpace{}%
\AgdaSymbol{→}\AgdaSpace{}%
\AgdaDatatype{E}\AgdaSpace{}%
\AgdaGeneralizable{Γ}\AgdaSpace{}%
\AgdaSymbol{(}\AgdaInductiveConstructor{ar}\AgdaSpace{}%
\AgdaGeneralizable{s}\AgdaSymbol{)}\AgdaSpace{}%
\AgdaSymbol{→}\AgdaSpace{}%
\AgdaDatatype{E}\AgdaSpace{}%
\AgdaGeneralizable{Γ}\AgdaSpace{}%
\AgdaSymbol{(}\AgdaInductiveConstructor{ar}\AgdaSpace{}%
\AgdaGeneralizable{s}\AgdaSymbol{)}\<%
\\
\>[4]\AgdaInductiveConstructor{scaledown}%
\>[15]\AgdaSymbol{:}\AgdaSpace{}%
\AgdaDatatype{ℕ}\AgdaSpace{}%
\AgdaSymbol{→}\AgdaSpace{}%
\AgdaDatatype{E}\AgdaSpace{}%
\AgdaGeneralizable{Γ}\AgdaSpace{}%
\AgdaSymbol{(}\AgdaInductiveConstructor{ar}\AgdaSpace{}%
\AgdaGeneralizable{s}\AgdaSymbol{)}\AgdaSpace{}%
\AgdaSymbol{→}\AgdaSpace{}%
\AgdaDatatype{E}\AgdaSpace{}%
\AgdaGeneralizable{Γ}\AgdaSpace{}%
\AgdaSymbol{(}\AgdaInductiveConstructor{ar}\AgdaSpace{}%
\AgdaGeneralizable{s}\AgdaSymbol{)}\<%
\\
\>[4]\AgdaInductiveConstructor{minus}%
\>[15]\AgdaSymbol{:}\AgdaSpace{}%
\AgdaDatatype{E}\AgdaSpace{}%
\AgdaGeneralizable{Γ}\AgdaSpace{}%
\AgdaSymbol{(}\AgdaInductiveConstructor{ar}\AgdaSpace{}%
\AgdaGeneralizable{s}\AgdaSymbol{)}\AgdaSpace{}%
\AgdaSymbol{→}\AgdaSpace{}%
\AgdaDatatype{E}\AgdaSpace{}%
\AgdaGeneralizable{Γ}\AgdaSpace{}%
\AgdaSymbol{(}\AgdaInductiveConstructor{ar}\AgdaSpace{}%
\AgdaGeneralizable{s}\AgdaSymbol{)}\<%
\\
\\[\AgdaEmptyExtraSkip]%
\\[\AgdaEmptyExtraSkip]%
\>[2]\AgdaKeyword{pattern}\AgdaSpace{}%
\AgdaOperator{\AgdaInductiveConstructor{\AgdaUnderscore{}⊠\AgdaUnderscore{}}}\AgdaSpace{}%
\AgdaBound{a}\AgdaSpace{}%
\AgdaBound{b}\AgdaSpace{}%
\AgdaSymbol{=}\AgdaSpace{}%
\AgdaInductiveConstructor{bin}\AgdaSpace{}%
\AgdaInductiveConstructor{mul}\AgdaSpace{}%
\AgdaBound{a}\AgdaSpace{}%
\AgdaBound{b}\<%
\\
\>[2]\AgdaKeyword{pattern}\AgdaSpace{}%
\AgdaOperator{\AgdaInductiveConstructor{\AgdaUnderscore{}⊞\AgdaUnderscore{}}}\AgdaSpace{}%
\AgdaBound{a}\AgdaSpace{}%
\AgdaBound{b}\AgdaSpace{}%
\AgdaSymbol{=}\AgdaSpace{}%
\AgdaInductiveConstructor{bin}\AgdaSpace{}%
\AgdaInductiveConstructor{plus}\AgdaSpace{}%
\AgdaBound{a}\AgdaSpace{}%
\AgdaBound{b}\<%
\end{code}

\subsection{Evaluation}
\begin{code}[hide]%
\>[0]\AgdaKeyword{module}\AgdaSpace{}%
\AgdaModule{Eval}\AgdaSpace{}%
\AgdaKeyword{where}\<%
\\
\>[0][@{}l@{\AgdaIndent{0}}]%
\>[2]\AgdaKeyword{open}\AgdaSpace{}%
\AgdaModule{Lang}\<%
\\
\>[2]\AgdaKeyword{open}\AgdaSpace{}%
\AgdaModule{Array}\<%
\\
\>[2]\AgdaKeyword{open}\AgdaSpace{}%
\AgdaKeyword{import}\AgdaSpace{}%
\AgdaModule{Data.Float}\AgdaSpace{}%
\AgdaSymbol{as}\AgdaSpace{}%
\AgdaModule{F}\AgdaSpace{}%
\AgdaKeyword{renaming}\AgdaSpace{}%
\AgdaSymbol{(}\AgdaPostulate{Float}\AgdaSpace{}%
\AgdaSymbol{to}\AgdaSpace{}%
\AgdaPostulate{ℝ}\AgdaSymbol{)}\AgdaSpace{}%
\AgdaKeyword{hiding}\AgdaSpace{}%
\AgdaSymbol{(}\AgdaPrimitive{⌊\AgdaUnderscore{}⌋}\AgdaSymbol{)}\<%
\\
\>[2]\AgdaKeyword{open}\AgdaSpace{}%
\AgdaKeyword{import}\AgdaSpace{}%
\AgdaModule{Data.Unit}\<%
\\
\>[2]\AgdaKeyword{open}\AgdaSpace{}%
\AgdaKeyword{import}\AgdaSpace{}%
\AgdaModule{Data.Product}\<%
\\
\>[2]\AgdaKeyword{open}\AgdaSpace{}%
\AgdaKeyword{import}\AgdaSpace{}%
\AgdaModule{Data.Fin}\AgdaSpace{}%
\AgdaKeyword{using}\AgdaSpace{}%
\AgdaSymbol{(}\AgdaDatatype{Fin}\AgdaSymbol{;}\AgdaSpace{}%
\AgdaInductiveConstructor{zero}\AgdaSymbol{;}\AgdaSpace{}%
\AgdaInductiveConstructor{suc}\AgdaSymbol{;}\AgdaSpace{}%
\AgdaOperator{\AgdaFunction{\#\AgdaUnderscore{}}}\AgdaSymbol{)}\<%
\\
\>[2]\AgdaKeyword{open}\AgdaSpace{}%
\AgdaKeyword{import}\AgdaSpace{}%
\AgdaModule{Relation.Nullary.Decidable}\<%
\\
\>[2]\AgdaKeyword{open}\AgdaSpace{}%
\AgdaKeyword{import}\AgdaSpace{}%
\AgdaModule{Data.Bool}\<%
\end{code}

We define the interpretation \AF{⟦\_⟧} for (\AF{E} \AB{Γ} \AB{is}) into the value
(\AF{Val} \AB{is}) in the environment (\AF{Env} \AB{Γ}).  The values are either
arrays or positions of the corresponding shape.  Environments for the given context
\AB{Γ} are tuples of values of the corresponding shapes.  The \AF{lookup} function
translates variables within the context into variables within the environment.
\begin{mathpar}
\codeblock{\begin{code}%
\>[2]\AgdaFunction{Val}\AgdaSpace{}%
\AgdaSymbol{:}\AgdaSpace{}%
\AgdaDatatype{IS}\AgdaSpace{}%
\AgdaSymbol{→}\AgdaSpace{}%
\AgdaPrimitive{Set}\<%
\\
\>[2]\AgdaFunction{Val}\AgdaSpace{}%
\AgdaSymbol{(}\AgdaInductiveConstructor{ar}\AgdaSpace{}%
\AgdaBound{s}\AgdaSymbol{)}%
\>[14]\AgdaSymbol{=}\AgdaSpace{}%
\AgdaFunction{Ar}\AgdaSpace{}%
\AgdaBound{s}\AgdaSpace{}%
\AgdaPostulate{ℝ}\<%
\\
\>[2]\AgdaFunction{Val}\AgdaSpace{}%
\AgdaSymbol{(}\AgdaInductiveConstructor{ix}\AgdaSpace{}%
\AgdaBound{s}\AgdaSymbol{)}%
\>[14]\AgdaSymbol{=}\AgdaSpace{}%
\AgdaDatatype{P}\AgdaSpace{}%
\AgdaBound{s}\<%
\end{code}}
\and
\codeblock{\begin{code}%
\>[2]\AgdaFunction{Env}\AgdaSpace{}%
\AgdaSymbol{:}\AgdaSpace{}%
\AgdaDatatype{Ctx}\AgdaSpace{}%
\AgdaSymbol{→}\AgdaSpace{}%
\AgdaPrimitive{Set}\<%
\\
\>[2]\AgdaFunction{Env}\AgdaSpace{}%
\AgdaInductiveConstructor{ε}%
\>[16]\AgdaSymbol{=}\AgdaSpace{}%
\AgdaRecord{⊤}\<%
\\
\>[2]\AgdaFunction{Env}\AgdaSpace{}%
\AgdaSymbol{(}\AgdaBound{Γ}\AgdaSpace{}%
\AgdaOperator{\AgdaInductiveConstructor{▹}}\AgdaSpace{}%
\AgdaBound{is}\AgdaSymbol{)}%
\>[16]\AgdaSymbol{=}\AgdaSpace{}%
\AgdaFunction{Env}\AgdaSpace{}%
\AgdaBound{Γ}\AgdaSpace{}%
\AgdaOperator{\AgdaFunction{×}}\AgdaSpace{}%
\AgdaFunction{Val}\AgdaSpace{}%
\AgdaBound{is}\<%
\end{code}}
\and
\codeblock{\begin{code}%
\>[2]\AgdaFunction{lookup}\AgdaSpace{}%
\AgdaSymbol{:}\AgdaSpace{}%
\AgdaGeneralizable{is}\AgdaSpace{}%
\AgdaOperator{\AgdaDatatype{∈}}\AgdaSpace{}%
\AgdaGeneralizable{Γ}\AgdaSpace{}%
\AgdaSymbol{→}\AgdaSpace{}%
\AgdaFunction{Env}\AgdaSpace{}%
\AgdaGeneralizable{Γ}\AgdaSpace{}%
\AgdaSymbol{→}\AgdaSpace{}%
\AgdaFunction{Val}\AgdaSpace{}%
\AgdaGeneralizable{is}\<%
\\
\>[2]\AgdaFunction{lookup}\AgdaSpace{}%
\AgdaInductiveConstructor{v₀}%
\>[17]\AgdaSymbol{(}\AgdaBound{ρ}\AgdaSpace{}%
\AgdaOperator{\AgdaInductiveConstructor{,}}\AgdaSpace{}%
\AgdaBound{x}\AgdaSymbol{)}%
\>[26]\AgdaSymbol{=}\AgdaSpace{}%
\AgdaBound{x}\<%
\\
\>[2]\AgdaFunction{lookup}\AgdaSpace{}%
\AgdaSymbol{(}\AgdaInductiveConstructor{vₛ}\AgdaSpace{}%
\AgdaBound{v}\AgdaSymbol{)}%
\>[17]\AgdaSymbol{(}\AgdaBound{ρ}\AgdaSpace{}%
\AgdaOperator{\AgdaInductiveConstructor{,}}\AgdaSpace{}%
\AgdaSymbol{\AgdaUnderscore{})}%
\>[26]\AgdaSymbol{=}\AgdaSpace{}%
\AgdaFunction{lookup}\AgdaSpace{}%
\AgdaBound{v}\AgdaSpace{}%
\AgdaBound{ρ}\<%
\end{code}}
\end{mathpar}

In the definition of \AF{⟦\_⟧} we wrap the environment argument into double braces.
This is an Agda-specific syntax for instance arguments\footnote{%
  For more details on instance arguments see:
  \url{https://agda.readthedocs.io/en/v2.6.3/language/instance-arguments.html}}
which behave similarly to hidden arguments, but they have a more powerful resolution
algorithm.  As a result we can omit mentioning the environment in recursive calls
when it is passed unchanged.
\begin{code}%
\>[2]\AgdaOperator{\AgdaFunction{⟦\AgdaUnderscore{}⟧}}\AgdaSpace{}%
\AgdaSymbol{:}\AgdaSpace{}%
\AgdaDatatype{E}\AgdaSpace{}%
\AgdaGeneralizable{Γ}\AgdaSpace{}%
\AgdaGeneralizable{is}\AgdaSpace{}%
\AgdaSymbol{→}\AgdaSpace{}%
\AgdaSymbol{⦃}\AgdaSpace{}%
\AgdaFunction{Env}\AgdaSpace{}%
\AgdaGeneralizable{Γ}\AgdaSpace{}%
\AgdaSymbol{⦄}\AgdaSpace{}%
\AgdaSymbol{→}\AgdaSpace{}%
\AgdaFunction{Val}\AgdaSpace{}%
\AgdaGeneralizable{is}\<%
\\
\>[2]\AgdaOperator{\AgdaFunction{⟦}}\AgdaSpace{}%
\AgdaInductiveConstructor{var}\AgdaSpace{}%
\AgdaBound{x}%
\>[25]\AgdaOperator{\AgdaFunction{⟧}}\AgdaSpace{}%
\AgdaSymbol{⦃}\AgdaSpace{}%
\AgdaBound{ρ}\AgdaSpace{}%
\AgdaSymbol{⦄}%
\>[34]\AgdaSymbol{=}\AgdaSpace{}%
\AgdaFunction{lookup}\AgdaSpace{}%
\AgdaBound{x}\AgdaSpace{}%
\AgdaBound{ρ}\<%
\\
\>[2]\AgdaOperator{\AgdaFunction{⟦}}\AgdaSpace{}%
\AgdaInductiveConstructor{zero}%
\>[25]\AgdaOperator{\AgdaFunction{⟧}}\AgdaSpace{}%
\AgdaSymbol{⦃}\AgdaSpace{}%
\AgdaBound{ρ}\AgdaSpace{}%
\AgdaSymbol{⦄}%
\>[34]\AgdaSymbol{=}\AgdaSpace{}%
\AgdaFunction{K}\AgdaSpace{}%
\AgdaNumber{0.0}\<%
\\
\>[2]\AgdaOperator{\AgdaFunction{⟦}}\AgdaSpace{}%
\AgdaInductiveConstructor{one}%
\>[25]\AgdaOperator{\AgdaFunction{⟧}}\AgdaSpace{}%
\AgdaSymbol{⦃}\AgdaSpace{}%
\AgdaBound{ρ}\AgdaSpace{}%
\AgdaSymbol{⦄}%
\>[34]\AgdaSymbol{=}\AgdaSpace{}%
\AgdaFunction{K}\AgdaSpace{}%
\AgdaNumber{1.0}\<%
\\
\>[2]\AgdaOperator{\AgdaFunction{⟦}}\AgdaSpace{}%
\AgdaInductiveConstructor{imapₛ}\AgdaSpace{}%
\AgdaBound{e}%
\>[25]\AgdaOperator{\AgdaFunction{⟧}}\AgdaSpace{}%
\AgdaSymbol{⦃}\AgdaSpace{}%
\AgdaBound{ρ}\AgdaSpace{}%
\AgdaSymbol{⦄}%
\>[34]\AgdaSymbol{=}\AgdaSpace{}%
\AgdaSymbol{λ}\AgdaSpace{}%
\AgdaBound{i}\AgdaSpace{}%
\AgdaSymbol{→}\AgdaSpace{}%
\AgdaOperator{\AgdaFunction{⟦}}\AgdaSpace{}%
\AgdaBound{e}\AgdaSpace{}%
\AgdaOperator{\AgdaFunction{⟧}}\AgdaSpace{}%
\AgdaSymbol{⦃}\AgdaSpace{}%
\AgdaBound{ρ}\AgdaSpace{}%
\AgdaOperator{\AgdaInductiveConstructor{,}}\AgdaSpace{}%
\AgdaBound{i}\AgdaSpace{}%
\AgdaSymbol{⦄}\AgdaSpace{}%
\AgdaSymbol{(}\AgdaInductiveConstructor{ι}\AgdaSpace{}%
\AgdaSymbol{(}\AgdaOperator{\AgdaFunction{\#}}\AgdaSpace{}%
\AgdaNumber{0}\AgdaSymbol{))}\<%
\\
\>[2]\AgdaOperator{\AgdaFunction{⟦}}\AgdaSpace{}%
\AgdaInductiveConstructor{selₛ}\AgdaSpace{}%
\AgdaBound{e}\AgdaSpace{}%
\AgdaBound{e₁}%
\>[25]\AgdaOperator{\AgdaFunction{⟧}}\AgdaSpace{}%
\AgdaSymbol{⦃}\AgdaSpace{}%
\AgdaBound{ρ}\AgdaSpace{}%
\AgdaSymbol{⦄}%
\>[34]\AgdaSymbol{=}\AgdaSpace{}%
\AgdaFunction{K}\AgdaSpace{}%
\AgdaOperator{\AgdaFunction{\$}}\AgdaSpace{}%
\AgdaOperator{\AgdaFunction{⟦}}\AgdaSpace{}%
\AgdaBound{e}\AgdaSpace{}%
\AgdaOperator{\AgdaFunction{⟧}}\AgdaSpace{}%
\AgdaOperator{\AgdaFunction{⟦}}\AgdaSpace{}%
\AgdaBound{e₁}\AgdaSpace{}%
\AgdaOperator{\AgdaFunction{⟧}}\<%
\\
\>[2]\AgdaOperator{\AgdaFunction{⟦}}\AgdaSpace{}%
\AgdaInductiveConstructor{imap}\AgdaSpace{}%
\AgdaBound{e}%
\>[25]\AgdaOperator{\AgdaFunction{⟧}}\AgdaSpace{}%
\AgdaSymbol{⦃}\AgdaSpace{}%
\AgdaBound{ρ}\AgdaSpace{}%
\AgdaSymbol{⦄}%
\>[34]\AgdaSymbol{=}\AgdaSpace{}%
\AgdaFunction{unnest}\AgdaSpace{}%
\AgdaSymbol{λ}\AgdaSpace{}%
\AgdaBound{i}\AgdaSpace{}%
\AgdaSymbol{→}\AgdaSpace{}%
\AgdaOperator{\AgdaFunction{⟦}}\AgdaSpace{}%
\AgdaBound{e}\AgdaSpace{}%
\AgdaOperator{\AgdaFunction{⟧}}\AgdaSpace{}%
\AgdaSymbol{⦃}\AgdaSpace{}%
\AgdaBound{ρ}\AgdaSpace{}%
\AgdaOperator{\AgdaInductiveConstructor{,}}\AgdaSpace{}%
\AgdaBound{i}\AgdaSpace{}%
\AgdaSymbol{⦄}\<%
\\
\>[2]\AgdaOperator{\AgdaFunction{⟦}}\AgdaSpace{}%
\AgdaInductiveConstructor{sel}\AgdaSpace{}%
\AgdaBound{e}\AgdaSpace{}%
\AgdaBound{e₁}%
\>[25]\AgdaOperator{\AgdaFunction{⟧}}\AgdaSpace{}%
\AgdaSymbol{⦃}\AgdaSpace{}%
\AgdaBound{ρ}\AgdaSpace{}%
\AgdaSymbol{⦄}%
\>[34]\AgdaSymbol{=}\AgdaSpace{}%
\AgdaFunction{nest}\AgdaSpace{}%
\AgdaOperator{\AgdaFunction{⟦}}\AgdaSpace{}%
\AgdaBound{e}\AgdaSpace{}%
\AgdaOperator{\AgdaFunction{⟧}}\AgdaSpace{}%
\AgdaOperator{\AgdaFunction{⟦}}\AgdaSpace{}%
\AgdaBound{e₁}\AgdaSpace{}%
\AgdaOperator{\AgdaFunction{⟧}}\<%
\\
\>[2]\AgdaOperator{\AgdaFunction{⟦}}\AgdaSpace{}%
\AgdaInductiveConstructor{imapb}\AgdaSpace{}%
\AgdaBound{m}\AgdaSpace{}%
\AgdaBound{e}%
\>[25]\AgdaOperator{\AgdaFunction{⟧}}\AgdaSpace{}%
\AgdaSymbol{⦃}\AgdaSpace{}%
\AgdaBound{ρ}\AgdaSpace{}%
\AgdaSymbol{⦄}%
\>[34]\AgdaSymbol{=}\AgdaSpace{}%
\AgdaFunction{CNN.unblock}\AgdaSpace{}%
\AgdaBound{m}\AgdaSpace{}%
\AgdaOperator{\AgdaFunction{\$}}\AgdaSpace{}%
\AgdaFunction{unnest}\AgdaSpace{}%
\AgdaSymbol{λ}\AgdaSpace{}%
\AgdaBound{i}\AgdaSpace{}%
\AgdaSymbol{→}\AgdaSpace{}%
\AgdaOperator{\AgdaFunction{⟦}}\AgdaSpace{}%
\AgdaBound{e}\AgdaSpace{}%
\AgdaOperator{\AgdaFunction{⟧}}\AgdaSpace{}%
\AgdaSymbol{⦃}\AgdaSpace{}%
\AgdaBound{ρ}\AgdaSpace{}%
\AgdaOperator{\AgdaInductiveConstructor{,}}\AgdaSpace{}%
\AgdaBound{i}\AgdaSpace{}%
\AgdaSymbol{⦄}\<%
\\
\>[2]\AgdaOperator{\AgdaFunction{⟦}}\AgdaSpace{}%
\AgdaInductiveConstructor{selb}\AgdaSpace{}%
\AgdaBound{m}\AgdaSpace{}%
\AgdaBound{e}\AgdaSpace{}%
\AgdaBound{e₁}%
\>[25]\AgdaOperator{\AgdaFunction{⟧}}\AgdaSpace{}%
\AgdaSymbol{⦃}\AgdaSpace{}%
\AgdaBound{ρ}\AgdaSpace{}%
\AgdaSymbol{⦄}%
\>[34]\AgdaSymbol{=}\AgdaSpace{}%
\AgdaFunction{nest}\AgdaSpace{}%
\AgdaSymbol{(}\AgdaFunction{CNN.block}\AgdaSpace{}%
\AgdaBound{m}\AgdaSpace{}%
\AgdaOperator{\AgdaFunction{⟦}}\AgdaSpace{}%
\AgdaBound{e}\AgdaSpace{}%
\AgdaOperator{\AgdaFunction{⟧}}\AgdaSymbol{)}\AgdaSpace{}%
\AgdaOperator{\AgdaFunction{⟦}}\AgdaSpace{}%
\AgdaBound{e₁}\AgdaSpace{}%
\AgdaOperator{\AgdaFunction{⟧}}\<%
\\
\>[2]\AgdaOperator{\AgdaFunction{⟦}}\AgdaSpace{}%
\AgdaInductiveConstructor{zero-but}\AgdaSpace{}%
\AgdaBound{i}\AgdaSpace{}%
\AgdaBound{j}\AgdaSpace{}%
\AgdaBound{e}%
\>[25]\AgdaOperator{\AgdaFunction{⟧}}\AgdaSpace{}%
\AgdaSymbol{⦃}\AgdaSpace{}%
\AgdaBound{ρ}\AgdaSpace{}%
\AgdaSymbol{⦄}%
\>[34]\AgdaSymbol{=}\AgdaSpace{}%
\AgdaOperator{\AgdaFunction{if}}\AgdaSpace{}%
\AgdaOperator{\AgdaFunction{⌊}}\AgdaSpace{}%
\AgdaOperator{\AgdaFunction{⟦}}\AgdaSpace{}%
\AgdaBound{i}\AgdaSpace{}%
\AgdaOperator{\AgdaFunction{⟧}}\AgdaSpace{}%
\AgdaOperator{\AgdaFunction{≟ₚ}}\AgdaSpace{}%
\AgdaOperator{\AgdaFunction{⟦}}\AgdaSpace{}%
\AgdaBound{j}\AgdaSpace{}%
\AgdaOperator{\AgdaFunction{⟧}}\AgdaSpace{}%
\AgdaOperator{\AgdaFunction{⌋}}\AgdaSpace{}%
\AgdaOperator{\AgdaFunction{then}}\AgdaSpace{}%
\AgdaOperator{\AgdaFunction{⟦}}\AgdaSpace{}%
\AgdaBound{e}\AgdaSpace{}%
\AgdaOperator{\AgdaFunction{⟧}}\AgdaSpace{}%
\AgdaOperator{\AgdaFunction{else}}\AgdaSpace{}%
\AgdaFunction{K}\AgdaSpace{}%
\AgdaNumber{0.0}\<%
\\
\>[2]\AgdaOperator{\AgdaFunction{⟦}}\AgdaSpace{}%
\AgdaInductiveConstructor{sum}\AgdaSpace{}%
\AgdaBound{e}%
\>[25]\AgdaOperator{\AgdaFunction{⟧}}\AgdaSpace{}%
\AgdaSymbol{⦃}\AgdaSpace{}%
\AgdaBound{ρ}\AgdaSpace{}%
\AgdaSymbol{⦄}%
\>[34]\AgdaSymbol{=}\AgdaSpace{}%
\AgdaFunction{Array.sum}\AgdaSpace{}%
\AgdaSymbol{(}\AgdaFunction{zipWith}\AgdaSpace{}%
\AgdaOperator{\AgdaPrimitive{\AgdaUnderscore{}+\AgdaUnderscore{}}}\AgdaSymbol{)}\AgdaSpace{}%
\AgdaSymbol{(}\AgdaFunction{K}\AgdaSpace{}%
\AgdaNumber{0.0}\AgdaSymbol{)}\AgdaSpace{}%
\AgdaSymbol{λ}\AgdaSpace{}%
\AgdaBound{i}\AgdaSpace{}%
\AgdaSymbol{→}\AgdaSpace{}%
\AgdaOperator{\AgdaFunction{⟦}}\AgdaSpace{}%
\AgdaBound{e}\AgdaSpace{}%
\AgdaOperator{\AgdaFunction{⟧}}\AgdaSpace{}%
\AgdaSymbol{⦃}\AgdaSpace{}%
\AgdaBound{ρ}\AgdaSpace{}%
\AgdaOperator{\AgdaInductiveConstructor{,}}\AgdaSpace{}%
\AgdaBound{i}\AgdaSpace{}%
\AgdaSymbol{⦄}\<%
\\
\>[2]\AgdaOperator{\AgdaFunction{⟦}}\AgdaSpace{}%
\AgdaBound{e}\AgdaSpace{}%
\AgdaOperator{\AgdaInductiveConstructor{⊞}}\AgdaSpace{}%
\AgdaBound{e₁}%
\>[25]\AgdaOperator{\AgdaFunction{⟧}}\AgdaSpace{}%
\AgdaSymbol{⦃}\AgdaSpace{}%
\AgdaBound{ρ}\AgdaSpace{}%
\AgdaSymbol{⦄}%
\>[34]\AgdaSymbol{=}\AgdaSpace{}%
\AgdaFunction{Array.zipWith}\AgdaSpace{}%
\AgdaOperator{\AgdaPrimitive{\AgdaUnderscore{}+\AgdaUnderscore{}}}\AgdaSpace{}%
\AgdaOperator{\AgdaFunction{⟦}}\AgdaSpace{}%
\AgdaBound{e}\AgdaSpace{}%
\AgdaOperator{\AgdaFunction{⟧}}\AgdaSpace{}%
\AgdaOperator{\AgdaFunction{⟦}}\AgdaSpace{}%
\AgdaBound{e₁}\AgdaSpace{}%
\AgdaOperator{\AgdaFunction{⟧}}\<%
\\
\>[2]\AgdaOperator{\AgdaFunction{⟦}}\AgdaSpace{}%
\AgdaBound{e}\AgdaSpace{}%
\AgdaOperator{\AgdaInductiveConstructor{⊠}}\AgdaSpace{}%
\AgdaBound{e₁}%
\>[25]\AgdaOperator{\AgdaFunction{⟧}}\AgdaSpace{}%
\AgdaSymbol{⦃}\AgdaSpace{}%
\AgdaBound{ρ}\AgdaSpace{}%
\AgdaSymbol{⦄}%
\>[34]\AgdaSymbol{=}\AgdaSpace{}%
\AgdaFunction{Array.zipWith}\AgdaSpace{}%
\AgdaOperator{\AgdaPrimitive{\AgdaUnderscore{}*\AgdaUnderscore{}}}\AgdaSpace{}%
\AgdaOperator{\AgdaFunction{⟦}}\AgdaSpace{}%
\AgdaBound{e}\AgdaSpace{}%
\AgdaOperator{\AgdaFunction{⟧}}\AgdaSpace{}%
\AgdaOperator{\AgdaFunction{⟦}}\AgdaSpace{}%
\AgdaBound{e₁}\AgdaSpace{}%
\AgdaOperator{\AgdaFunction{⟧}}\<%
\\
\>[2]\AgdaOperator{\AgdaFunction{⟦}}\AgdaSpace{}%
\AgdaInductiveConstructor{slide}\AgdaSpace{}%
\AgdaBound{i}\AgdaSpace{}%
\AgdaBound{pl}\AgdaSpace{}%
\AgdaBound{e}\AgdaSpace{}%
\AgdaBound{su}%
\>[25]\AgdaOperator{\AgdaFunction{⟧}}\AgdaSpace{}%
\AgdaSymbol{⦃}\AgdaSpace{}%
\AgdaBound{ρ}\AgdaSpace{}%
\AgdaSymbol{⦄}%
\>[34]\AgdaSymbol{=}\AgdaSpace{}%
\AgdaFunction{Array.slide}\AgdaSpace{}%
\AgdaOperator{\AgdaFunction{⟦}}\AgdaSpace{}%
\AgdaBound{i}\AgdaSpace{}%
\AgdaOperator{\AgdaFunction{⟧}}\AgdaSpace{}%
\AgdaBound{pl}\AgdaSpace{}%
\AgdaOperator{\AgdaFunction{⟦}}\AgdaSpace{}%
\AgdaBound{e}\AgdaSpace{}%
\AgdaOperator{\AgdaFunction{⟧}}\AgdaSpace{}%
\AgdaBound{su}\<%
\\
\>[2]\AgdaOperator{\AgdaFunction{⟦}}\AgdaSpace{}%
\AgdaInductiveConstructor{backslide}\AgdaSpace{}%
\AgdaBound{i}\AgdaSpace{}%
\AgdaBound{e}\AgdaSpace{}%
\AgdaBound{su}\AgdaSpace{}%
\AgdaBound{pl}%
\>[25]\AgdaOperator{\AgdaFunction{⟧}}\AgdaSpace{}%
\AgdaSymbol{⦃}\AgdaSpace{}%
\AgdaBound{ρ}\AgdaSpace{}%
\AgdaSymbol{⦄}%
\>[34]\AgdaSymbol{=}\AgdaSpace{}%
\AgdaFunction{Array.backslide}\AgdaSpace{}%
\AgdaOperator{\AgdaFunction{⟦}}\AgdaSpace{}%
\AgdaBound{i}\AgdaSpace{}%
\AgdaOperator{\AgdaFunction{⟧}}\AgdaSpace{}%
\AgdaOperator{\AgdaFunction{⟦}}\AgdaSpace{}%
\AgdaBound{e}\AgdaSpace{}%
\AgdaOperator{\AgdaFunction{⟧}}\AgdaSpace{}%
\AgdaBound{su}\AgdaSpace{}%
\AgdaNumber{0.0}\AgdaSpace{}%
\AgdaBound{pl}\<%
\\
\>[2]\AgdaOperator{\AgdaFunction{⟦}}\AgdaSpace{}%
\AgdaInductiveConstructor{scaledown}\AgdaSpace{}%
\AgdaBound{n}\AgdaSpace{}%
\AgdaBound{e}%
\>[25]\AgdaOperator{\AgdaFunction{⟧}}\AgdaSpace{}%
\AgdaSymbol{⦃}\AgdaSpace{}%
\AgdaBound{ρ}\AgdaSpace{}%
\AgdaSymbol{⦄}%
\>[34]\AgdaSymbol{=}\AgdaSpace{}%
\AgdaFunction{Array.map}\AgdaSpace{}%
\AgdaSymbol{(}\AgdaOperator{\AgdaPrimitive{\AgdaUnderscore{}÷}}\AgdaSpace{}%
\AgdaPrimitive{fromℕ}\AgdaSpace{}%
\AgdaBound{n}\AgdaSymbol{)}\AgdaSpace{}%
\AgdaOperator{\AgdaFunction{⟦}}\AgdaSpace{}%
\AgdaBound{e}\AgdaSpace{}%
\AgdaOperator{\AgdaFunction{⟧}}\<%
\\
\>[2]\AgdaOperator{\AgdaFunction{⟦}}\AgdaSpace{}%
\AgdaInductiveConstructor{minus}\AgdaSpace{}%
\AgdaBound{e}%
\>[25]\AgdaOperator{\AgdaFunction{⟧}}\AgdaSpace{}%
\AgdaSymbol{⦃}\AgdaSpace{}%
\AgdaBound{ρ}\AgdaSpace{}%
\AgdaSymbol{⦄}%
\>[34]\AgdaSymbol{=}\AgdaSpace{}%
\AgdaFunction{Array.map}\AgdaSpace{}%
\AgdaSymbol{(}\AgdaOperator{\AgdaPrimitive{-\AgdaUnderscore{}}}\AgdaSymbol{)}\AgdaSpace{}%
\AgdaOperator{\AgdaFunction{⟦}}\AgdaSpace{}%
\AgdaBound{e}\AgdaSpace{}%
\AgdaOperator{\AgdaFunction{⟧}}\<%
\\
\>[2]\AgdaOperator{\AgdaFunction{⟦}}\AgdaSpace{}%
\AgdaInductiveConstructor{logistic}\AgdaSpace{}%
\AgdaBound{e}%
\>[25]\AgdaOperator{\AgdaFunction{⟧}}\AgdaSpace{}%
\AgdaSymbol{⦃}\AgdaSpace{}%
\AgdaBound{ρ}\AgdaSpace{}%
\AgdaSymbol{⦄}%
\>[34]\AgdaSymbol{=}\AgdaSpace{}%
\AgdaFunction{CNN.logistic}\AgdaSpace{}%
\AgdaOperator{\AgdaFunction{⟦}}\AgdaSpace{}%
\AgdaBound{e}\AgdaSpace{}%
\AgdaOperator{\AgdaFunction{⟧}}\<%
\end{code}
With the above definition we can better explain the choices of language constructors.
The most important question to clarify is why do we have three array
constructors/eliminators.  As the only conceptual datatype of our language is
an array (of some shape), we do not have any direct way to talk about array elements.
Therefore, we model the type of array elements (scalars) as arrays of a singleton shape.
As can be seen, scalar selection \AC{selₛ} returns a singleton array
(application of \AF{K}) where all the element(s) are equal to the element we are
selecting.  The corresponding array constructor \AC{imapₛ} makes sure that if we
compute \AB{s} elements of the shape \AF{unit}, we produce an array of shape \AB{s}
(and not \AB{s} \AC{⊗} \AF{unit}).  This soves the problem of constructing arays
from scalars, but how do we construct an array of a product shape?  Given that we
have an expression in the context (\AB{Γ} \AC{▹} \AC{ix} \AB{s} \AC{▹} \AC{ix} \AB{p}),
we need to produce an array of \AB{s} \AC{⊗} \AB{p}.  There are several ways how to
solve this (\eg{} introducing nest/unnest or projections and pairing on indices),
but it is clear that we need something more than just an \AC{imapₛ}.
This is the reason to introduce \AC{imap}/\AC{sel} pair which operates on arrays
of product shapes.  As average pooling operates on blocked arrays, we need
a construction to express this in \AF{E}.  One could introduce explicit 
\AF{block}/\AF{unblock}, but we merge blocking/unlocking action with
imap/sel obtaining \AC{imapb}/\AC{selb}.  Our \AC{sum} constructor 
gets an argument in the extended context which is summation index, so 
conceptually we generate the values at every summation index before
summing these values together.  As a result, we only need
one instance of \AC{sum} which makes our expressions a little tidier.

\subsection{Weakening and Substitution}

As our language has explicit de Bruin variables (as opposed to HOAS~\cite{hoas} approaches),
we need the means to do weakening and substitution when we optimise expressions in \AF{E}.
Our language is intrinsically typed(shaped) which
makes the definition of both operations challenging.  However, this problem has
been well-understood, and we adopt the solution from~\cite{subst}.  We only show the
basic mechanisms of the definition, for full details refer to~\cite{subst}.

The key structure that gives rise to weakening and substitution is a function that
computes the context \AB{Γ} \emph{without} the variable \AB{v} 
(denoted \AB{Γ} \AF{/} \AB{v}).  Then we define the weakening for
variables (\AF{wkv}) and expressions (\AF{wk}) that take a variable or expression
in the context without the variable \AF{v} and return this variable or expression
in the context where \AB{v} is present.
\begin{code}[hide]%
\>[0]\AgdaKeyword{module}\AgdaSpace{}%
\AgdaModule{SubWk}\AgdaSpace{}%
\AgdaKeyword{where}\<%
\\
\>[0][@{}l@{\AgdaIndent{0}}]%
\>[2]\AgdaKeyword{open}\AgdaSpace{}%
\AgdaModule{Lang}\<%
\end{code}
\begin{mathpar}
\codeblock{\begin{code}%
\>[2]\AgdaOperator{\AgdaFunction{\AgdaUnderscore{}/\AgdaUnderscore{}}}\AgdaSpace{}%
\AgdaSymbol{:}\AgdaSpace{}%
\AgdaSymbol{(}\AgdaBound{Γ}\AgdaSpace{}%
\AgdaSymbol{:}\AgdaSpace{}%
\AgdaDatatype{Ctx}\AgdaSymbol{)}\AgdaSpace{}%
\AgdaSymbol{→}\AgdaSpace{}%
\AgdaGeneralizable{is}\AgdaSpace{}%
\AgdaOperator{\AgdaDatatype{∈}}\AgdaSpace{}%
\AgdaBound{Γ}\AgdaSpace{}%
\AgdaSymbol{→}\AgdaSpace{}%
\AgdaDatatype{Ctx}\<%
\\
\>[2]\AgdaSymbol{(}\AgdaBound{Γ}\AgdaSpace{}%
\AgdaOperator{\AgdaInductiveConstructor{▹}}\AgdaSpace{}%
\AgdaBound{x}\AgdaSymbol{)}\AgdaSpace{}%
\AgdaOperator{\AgdaFunction{/}}\AgdaSpace{}%
\AgdaInductiveConstructor{v₀}%
\>[18]\AgdaSymbol{=}\AgdaSpace{}%
\AgdaBound{Γ}\<%
\\
\>[2]\AgdaSymbol{(}\AgdaBound{Γ}\AgdaSpace{}%
\AgdaOperator{\AgdaInductiveConstructor{▹}}\AgdaSpace{}%
\AgdaBound{x}\AgdaSymbol{)}\AgdaSpace{}%
\AgdaOperator{\AgdaFunction{/}}\AgdaSpace{}%
\AgdaInductiveConstructor{vₛ}\AgdaSpace{}%
\AgdaBound{v}%
\>[18]\AgdaSymbol{=}\AgdaSpace{}%
\AgdaSymbol{(}\AgdaBound{Γ}\AgdaSpace{}%
\AgdaOperator{\AgdaFunction{/}}\AgdaSpace{}%
\AgdaBound{v}\AgdaSymbol{)}\AgdaSpace{}%
\AgdaOperator{\AgdaInductiveConstructor{▹}}\AgdaSpace{}%
\AgdaBound{x}\<%
\end{code}}
\and
\codeblock{\begin{code}  %
\>[2]\AgdaFunction{wkv}%
\>[7]\AgdaSymbol{:}\AgdaSpace{}%
\AgdaSymbol{(}\AgdaBound{v}\AgdaSpace{}%
\AgdaSymbol{:}\AgdaSpace{}%
\AgdaGeneralizable{is}\AgdaSpace{}%
\AgdaOperator{\AgdaDatatype{∈}}\AgdaSpace{}%
\AgdaGeneralizable{Γ}\AgdaSymbol{)}\AgdaSpace{}%
\AgdaSymbol{→}\AgdaSpace{}%
\AgdaGeneralizable{ip}\AgdaSpace{}%
\AgdaOperator{\AgdaDatatype{∈}}\AgdaSpace{}%
\AgdaSymbol{(}\AgdaGeneralizable{Γ}\AgdaSpace{}%
\AgdaOperator{\AgdaFunction{/}}\AgdaSpace{}%
\AgdaBound{v}\AgdaSymbol{)}\AgdaSpace{}%
\AgdaSymbol{→}\AgdaSpace{}%
\AgdaGeneralizable{ip}\AgdaSpace{}%
\AgdaOperator{\AgdaDatatype{∈}}\AgdaSpace{}%
\AgdaGeneralizable{Γ}\<%
\\
\>[2]\AgdaFunction{wk}%
\>[7]\AgdaSymbol{:}\AgdaSpace{}%
\AgdaSymbol{(}\AgdaBound{v}\AgdaSpace{}%
\AgdaSymbol{:}\AgdaSpace{}%
\AgdaGeneralizable{is}\AgdaSpace{}%
\AgdaOperator{\AgdaDatatype{∈}}\AgdaSpace{}%
\AgdaGeneralizable{Γ}\AgdaSymbol{)}\AgdaSpace{}%
\AgdaSymbol{→}\AgdaSpace{}%
\AgdaDatatype{E}\AgdaSpace{}%
\AgdaSymbol{(}\AgdaGeneralizable{Γ}\AgdaSpace{}%
\AgdaOperator{\AgdaFunction{/}}\AgdaSpace{}%
\AgdaBound{v}\AgdaSymbol{)}\AgdaSpace{}%
\AgdaGeneralizable{ip}\AgdaSpace{}%
\AgdaSymbol{→}\AgdaSpace{}%
\AgdaDatatype{E}\AgdaSpace{}%
\AgdaGeneralizable{Γ}\AgdaSpace{}%
\AgdaGeneralizable{ip}\<%
\end{code}}
\end{mathpar}
We give ourselves a nicer syntax for common cases when expressions
are lifted into the context with extra one or two variables:
\begin{code}[hide]%
\>[2]\AgdaKeyword{infixr}\AgdaSpace{}%
\AgdaNumber{18}\AgdaSpace{}%
\AgdaOperator{\AgdaFunction{↑\AgdaUnderscore{}}}\<%
\\
\>[2]\AgdaKeyword{infixr}\AgdaSpace{}%
\AgdaNumber{18}\AgdaSpace{}%
\AgdaOperator{\AgdaFunction{↑↑\AgdaUnderscore{}}}\<%
\end{code}
\begin{mathpar}
\codeblock{\begin{code}%
\>[2]\AgdaOperator{\AgdaFunction{↑\AgdaUnderscore{}}}\AgdaSpace{}%
\AgdaSymbol{:}\AgdaSpace{}%
\AgdaDatatype{E}\AgdaSpace{}%
\AgdaGeneralizable{Γ}\AgdaSpace{}%
\AgdaGeneralizable{is}\AgdaSpace{}%
\AgdaSymbol{→}\AgdaSpace{}%
\AgdaDatatype{E}\AgdaSpace{}%
\AgdaSymbol{(}\AgdaGeneralizable{Γ}\AgdaSpace{}%
\AgdaOperator{\AgdaInductiveConstructor{▹}}\AgdaSpace{}%
\AgdaGeneralizable{ip}\AgdaSymbol{)}\AgdaSpace{}%
\AgdaGeneralizable{is}\<%
\\
\>[2]\AgdaOperator{\AgdaFunction{↑\AgdaUnderscore{}}}\AgdaSpace{}%
\AgdaSymbol{=}\AgdaSpace{}%
\AgdaFunction{wk}\AgdaSpace{}%
\AgdaInductiveConstructor{v₀}\<%
\end{code}}
\and
\codeblock{\begin{code}%
\>[2]\AgdaOperator{\AgdaFunction{↑↑\AgdaUnderscore{}}}\AgdaSpace{}%
\AgdaSymbol{:}\AgdaSpace{}%
\AgdaDatatype{E}\AgdaSpace{}%
\AgdaGeneralizable{Γ}\AgdaSpace{}%
\AgdaGeneralizable{is}\AgdaSpace{}%
\AgdaSymbol{→}\AgdaSpace{}%
\AgdaDatatype{E}\AgdaSpace{}%
\AgdaSymbol{(}\AgdaGeneralizable{Γ}\AgdaSpace{}%
\AgdaOperator{\AgdaInductiveConstructor{▹}}\AgdaSpace{}%
\AgdaGeneralizable{ip}\AgdaSpace{}%
\AgdaOperator{\AgdaInductiveConstructor{▹}}\AgdaSpace{}%
\AgdaGeneralizable{iq}\AgdaSymbol{)}\AgdaSpace{}%
\AgdaGeneralizable{is}\<%
\\
\>[2]\AgdaOperator{\AgdaFunction{↑↑\AgdaUnderscore{}}}\AgdaSpace{}%
\AgdaSymbol{=}\AgdaSpace{}%
\AgdaOperator{\AgdaFunction{↑\AgdaUnderscore{}}}\AgdaSpace{}%
\AgdaOperator{\AgdaFunction{∘}}\AgdaSpace{}%
\AgdaOperator{\AgdaFunction{↑\AgdaUnderscore{}}}\<%
\end{code}}
\end{mathpar}
\begin{code}[hide]%
\>[2]\AgdaFunction{wkv}\AgdaSpace{}%
\AgdaInductiveConstructor{v₀}\AgdaSpace{}%
\AgdaBound{w}\AgdaSpace{}%
\AgdaSymbol{=}\AgdaSpace{}%
\AgdaInductiveConstructor{vₛ}\AgdaSpace{}%
\AgdaBound{w}\<%
\\
\>[2]\AgdaFunction{wkv}\AgdaSpace{}%
\AgdaSymbol{(}\AgdaInductiveConstructor{vₛ}\AgdaSpace{}%
\AgdaBound{v}\AgdaSymbol{)}\AgdaSpace{}%
\AgdaInductiveConstructor{v₀}\AgdaSpace{}%
\AgdaSymbol{=}\AgdaSpace{}%
\AgdaInductiveConstructor{v₀}\<%
\\
\>[2]\AgdaFunction{wkv}\AgdaSpace{}%
\AgdaSymbol{(}\AgdaInductiveConstructor{vₛ}\AgdaSpace{}%
\AgdaBound{v}\AgdaSymbol{)}\AgdaSpace{}%
\AgdaSymbol{(}\AgdaInductiveConstructor{vₛ}\AgdaSpace{}%
\AgdaBound{w}\AgdaSymbol{)}\AgdaSpace{}%
\AgdaSymbol{=}\AgdaSpace{}%
\AgdaInductiveConstructor{vₛ}\AgdaSpace{}%
\AgdaSymbol{(}\AgdaFunction{wkv}\AgdaSpace{}%
\AgdaBound{v}\AgdaSpace{}%
\AgdaBound{w}\AgdaSymbol{)}\<%
\\
\>[0]\<%
\\
\>[2]\AgdaFunction{wk}\AgdaSpace{}%
\AgdaBound{v}\AgdaSpace{}%
\AgdaSymbol{(}\AgdaInductiveConstructor{var}\AgdaSpace{}%
\AgdaBound{x}\AgdaSymbol{)}\AgdaSpace{}%
\AgdaSymbol{=}\AgdaSpace{}%
\AgdaSymbol{(}\AgdaInductiveConstructor{var}\AgdaSpace{}%
\AgdaSymbol{(}\AgdaFunction{wkv}\AgdaSpace{}%
\AgdaBound{v}\AgdaSpace{}%
\AgdaBound{x}\AgdaSymbol{))}\<%
\\
\>[2]\AgdaFunction{wk}\AgdaSpace{}%
\AgdaBound{v}\AgdaSpace{}%
\AgdaInductiveConstructor{zero}\AgdaSpace{}%
\AgdaSymbol{=}\AgdaSpace{}%
\AgdaInductiveConstructor{zero}\<%
\\
\>[2]\AgdaFunction{wk}\AgdaSpace{}%
\AgdaBound{v}\AgdaSpace{}%
\AgdaInductiveConstructor{one}\AgdaSpace{}%
\AgdaSymbol{=}\AgdaSpace{}%
\AgdaInductiveConstructor{one}\<%
\\
\>[0]\<%
\\
\>[2]\AgdaFunction{wk}\AgdaSpace{}%
\AgdaBound{v}\AgdaSpace{}%
\AgdaSymbol{(}\AgdaInductiveConstructor{imapₛ}\AgdaSpace{}%
\AgdaBound{e}\AgdaSymbol{)}\AgdaSpace{}%
\AgdaSymbol{=}\AgdaSpace{}%
\AgdaInductiveConstructor{imapₛ}\AgdaSpace{}%
\AgdaSymbol{(}\AgdaFunction{wk}\AgdaSpace{}%
\AgdaSymbol{(}\AgdaInductiveConstructor{vₛ}\AgdaSpace{}%
\AgdaBound{v}\AgdaSymbol{)}\AgdaSpace{}%
\AgdaBound{e}\AgdaSymbol{)}\<%
\\
\>[2]\AgdaFunction{wk}\AgdaSpace{}%
\AgdaBound{v}\AgdaSpace{}%
\AgdaSymbol{(}\AgdaInductiveConstructor{selₛ}\AgdaSpace{}%
\AgdaBound{e}\AgdaSpace{}%
\AgdaBound{e₁}\AgdaSymbol{)}\AgdaSpace{}%
\AgdaSymbol{=}\AgdaSpace{}%
\AgdaInductiveConstructor{selₛ}\AgdaSpace{}%
\AgdaSymbol{(}\AgdaFunction{wk}\AgdaSpace{}%
\AgdaBound{v}\AgdaSpace{}%
\AgdaBound{e}\AgdaSymbol{)}\AgdaSpace{}%
\AgdaSymbol{(}\AgdaFunction{wk}\AgdaSpace{}%
\AgdaBound{v}\AgdaSpace{}%
\AgdaBound{e₁}\AgdaSymbol{)}\<%
\\
\>[2]\AgdaComment{--wk\ v\ (zero-butₛ\ idx\ e)\ =\ zero-butₛ\ (wk\ v\ idx)\ (wk\ v\ e)}\<%
\\
\>[0]\<%
\\
\>[2]\AgdaComment{--\ Copy-paste\ from\ scalar\ versions}\<%
\\
\>[2]\AgdaFunction{wk}\AgdaSpace{}%
\AgdaBound{v}\AgdaSpace{}%
\AgdaSymbol{(}\AgdaInductiveConstructor{imap}\AgdaSpace{}%
\AgdaBound{e}\AgdaSymbol{)}\AgdaSpace{}%
\AgdaSymbol{=}\AgdaSpace{}%
\AgdaInductiveConstructor{imap}\AgdaSpace{}%
\AgdaSymbol{(}\AgdaFunction{wk}\AgdaSpace{}%
\AgdaSymbol{(}\AgdaInductiveConstructor{vₛ}\AgdaSpace{}%
\AgdaBound{v}\AgdaSymbol{)}\AgdaSpace{}%
\AgdaBound{e}\AgdaSymbol{)}\<%
\\
\>[2]\AgdaFunction{wk}\AgdaSpace{}%
\AgdaBound{v}\AgdaSpace{}%
\AgdaSymbol{(}\AgdaInductiveConstructor{sel}\AgdaSpace{}%
\AgdaBound{e}\AgdaSpace{}%
\AgdaBound{e₁}\AgdaSymbol{)}\AgdaSpace{}%
\AgdaSymbol{=}\AgdaSpace{}%
\AgdaInductiveConstructor{sel}\AgdaSpace{}%
\AgdaSymbol{(}\AgdaFunction{wk}\AgdaSpace{}%
\AgdaBound{v}\AgdaSpace{}%
\AgdaBound{e}\AgdaSymbol{)}\AgdaSpace{}%
\AgdaSymbol{(}\AgdaFunction{wk}\AgdaSpace{}%
\AgdaBound{v}\AgdaSpace{}%
\AgdaBound{e₁}\AgdaSymbol{)}\<%
\\
\>[2]\AgdaComment{--wk\ v\ (zero-but\ idx\ e)\ =\ zero-but\ (wk\ v\ idx)\ (wk\ v\ e)}\<%
\\
\>[0]\<%
\\
\>[2]\AgdaFunction{wk}\AgdaSpace{}%
\AgdaBound{v}\AgdaSpace{}%
\AgdaSymbol{(}\AgdaInductiveConstructor{zero-but}\AgdaSpace{}%
\AgdaBound{i}\AgdaSpace{}%
\AgdaBound{j}\AgdaSpace{}%
\AgdaBound{e}\AgdaSymbol{)}\AgdaSpace{}%
\AgdaSymbol{=}\AgdaSpace{}%
\AgdaInductiveConstructor{zero-but}\AgdaSpace{}%
\AgdaSymbol{(}\AgdaFunction{wk}\AgdaSpace{}%
\AgdaBound{v}\AgdaSpace{}%
\AgdaBound{i}\AgdaSymbol{)}\AgdaSpace{}%
\AgdaSymbol{(}\AgdaFunction{wk}\AgdaSpace{}%
\AgdaBound{v}\AgdaSpace{}%
\AgdaBound{j}\AgdaSymbol{)}\AgdaSpace{}%
\AgdaSymbol{(}\AgdaFunction{wk}\AgdaSpace{}%
\AgdaBound{v}\AgdaSpace{}%
\AgdaBound{e}\AgdaSymbol{)}\<%
\\
\>[2]\AgdaFunction{wk}\AgdaSpace{}%
\AgdaBound{v}\AgdaSpace{}%
\AgdaSymbol{(}\AgdaInductiveConstructor{sum}\AgdaSpace{}%
\AgdaBound{e}\AgdaSymbol{)}\AgdaSpace{}%
\AgdaSymbol{=}\AgdaSpace{}%
\AgdaInductiveConstructor{E.sum}\AgdaSpace{}%
\AgdaSymbol{(}\AgdaFunction{wk}\AgdaSpace{}%
\AgdaSymbol{(}\AgdaInductiveConstructor{vₛ}\AgdaSpace{}%
\AgdaBound{v}\AgdaSymbol{)}\AgdaSpace{}%
\AgdaBound{e}\AgdaSymbol{)}\<%
\\
\>[2]\AgdaFunction{wk}\AgdaSpace{}%
\AgdaBound{v}\AgdaSpace{}%
\AgdaSymbol{(}\AgdaInductiveConstructor{bin}\AgdaSpace{}%
\AgdaBound{x}\AgdaSpace{}%
\AgdaBound{e}\AgdaSpace{}%
\AgdaBound{e₁}\AgdaSymbol{)}\AgdaSpace{}%
\AgdaSymbol{=}\AgdaSpace{}%
\AgdaInductiveConstructor{bin}\AgdaSpace{}%
\AgdaBound{x}\AgdaSpace{}%
\AgdaSymbol{(}\AgdaFunction{wk}\AgdaSpace{}%
\AgdaBound{v}\AgdaSpace{}%
\AgdaBound{e}\AgdaSymbol{)}\AgdaSpace{}%
\AgdaSymbol{(}\AgdaFunction{wk}\AgdaSpace{}%
\AgdaBound{v}\AgdaSpace{}%
\AgdaBound{e₁}\AgdaSymbol{)}\<%
\\
\>[2]\AgdaFunction{wk}\AgdaSpace{}%
\AgdaBound{v}\AgdaSpace{}%
\AgdaSymbol{(}\AgdaInductiveConstructor{slide}\AgdaSpace{}%
\AgdaBound{i}\AgdaSpace{}%
\AgdaBound{pl}\AgdaSpace{}%
\AgdaBound{e}\AgdaSpace{}%
\AgdaBound{su}\AgdaSymbol{)}\AgdaSpace{}%
\AgdaSymbol{=}\AgdaSpace{}%
\AgdaInductiveConstructor{E.slide}\AgdaSpace{}%
\AgdaSymbol{(}\AgdaFunction{wk}\AgdaSpace{}%
\AgdaBound{v}\AgdaSpace{}%
\AgdaBound{i}\AgdaSymbol{)}\AgdaSpace{}%
\AgdaBound{pl}\AgdaSpace{}%
\AgdaSymbol{(}\AgdaFunction{wk}\AgdaSpace{}%
\AgdaBound{v}\AgdaSpace{}%
\AgdaBound{e}\AgdaSymbol{)}\AgdaSpace{}%
\AgdaBound{su}\<%
\\
\>[2]\AgdaFunction{wk}\AgdaSpace{}%
\AgdaBound{v}\AgdaSpace{}%
\AgdaSymbol{(}\AgdaInductiveConstructor{backslide}\AgdaSpace{}%
\AgdaBound{i}\AgdaSpace{}%
\AgdaBound{e}\AgdaSpace{}%
\AgdaBound{su}\AgdaSpace{}%
\AgdaBound{pl}\AgdaSymbol{)}\AgdaSpace{}%
\AgdaSymbol{=}\AgdaSpace{}%
\AgdaInductiveConstructor{E.backslide}\AgdaSpace{}%
\AgdaSymbol{(}\AgdaFunction{wk}\AgdaSpace{}%
\AgdaBound{v}\AgdaSpace{}%
\AgdaBound{i}\AgdaSymbol{)}\AgdaSpace{}%
\AgdaSymbol{(}\AgdaFunction{wk}\AgdaSpace{}%
\AgdaBound{v}\AgdaSpace{}%
\AgdaBound{e}\AgdaSymbol{)}\AgdaSpace{}%
\AgdaBound{su}\AgdaSpace{}%
\AgdaBound{pl}\<%
\\
\>[0]\<%
\\
\>[2]\AgdaFunction{wk}\AgdaSpace{}%
\AgdaBound{v}\AgdaSpace{}%
\AgdaSymbol{(}\AgdaInductiveConstructor{scaledown}\AgdaSpace{}%
\AgdaBound{x}\AgdaSpace{}%
\AgdaBound{e}\AgdaSymbol{)}\AgdaSpace{}%
\AgdaSymbol{=}\AgdaSpace{}%
\AgdaInductiveConstructor{scaledown}\AgdaSpace{}%
\AgdaBound{x}\AgdaSpace{}%
\AgdaSymbol{(}\AgdaFunction{wk}\AgdaSpace{}%
\AgdaBound{v}\AgdaSpace{}%
\AgdaBound{e}\AgdaSymbol{)}\<%
\\
\>[2]\AgdaFunction{wk}\AgdaSpace{}%
\AgdaBound{v}\AgdaSpace{}%
\AgdaSymbol{(}\AgdaInductiveConstructor{minus}\AgdaSpace{}%
\AgdaBound{e}\AgdaSymbol{)}\AgdaSpace{}%
\AgdaSymbol{=}\AgdaSpace{}%
\AgdaInductiveConstructor{minus}\AgdaSpace{}%
\AgdaSymbol{(}\AgdaFunction{wk}\AgdaSpace{}%
\AgdaBound{v}\AgdaSpace{}%
\AgdaBound{e}\AgdaSymbol{)}\<%
\\
\>[2]\AgdaFunction{wk}\AgdaSpace{}%
\AgdaBound{v}\AgdaSpace{}%
\AgdaSymbol{(}\AgdaInductiveConstructor{logistic}\AgdaSpace{}%
\AgdaBound{e}\AgdaSymbol{)}\AgdaSpace{}%
\AgdaSymbol{=}\AgdaSpace{}%
\AgdaInductiveConstructor{logistic}\AgdaSpace{}%
\AgdaSymbol{(}\AgdaFunction{wk}\AgdaSpace{}%
\AgdaBound{v}\AgdaSpace{}%
\AgdaBound{e}\AgdaSymbol{)}\<%
\\
\>[2]\AgdaComment{--wk\ v\ (block\ x\ e)\ =\ block\ x\ (wk\ v\ e)}\<%
\\
\>[2]\AgdaComment{--wk\ v\ (unblock\ x\ e)\ =\ unblock\ x\ (wk\ v\ e)}\<%
\\
\>[2]\AgdaFunction{wk}\AgdaSpace{}%
\AgdaBound{v}\AgdaSpace{}%
\AgdaSymbol{(}\AgdaInductiveConstructor{imapb}\AgdaSpace{}%
\AgdaBound{m}\AgdaSpace{}%
\AgdaBound{e}\AgdaSymbol{)}\AgdaSpace{}%
\AgdaSymbol{=}\AgdaSpace{}%
\AgdaInductiveConstructor{imapb}\AgdaSpace{}%
\AgdaBound{m}\AgdaSpace{}%
\AgdaSymbol{(}\AgdaFunction{wk}\AgdaSpace{}%
\AgdaSymbol{(}\AgdaInductiveConstructor{vₛ}\AgdaSpace{}%
\AgdaBound{v}\AgdaSymbol{)}\AgdaSpace{}%
\AgdaBound{e}\AgdaSymbol{)}\<%
\\
\>[2]\AgdaFunction{wk}\AgdaSpace{}%
\AgdaBound{v}\AgdaSpace{}%
\AgdaSymbol{(}\AgdaInductiveConstructor{selb}\AgdaSpace{}%
\AgdaBound{m}\AgdaSpace{}%
\AgdaBound{e}\AgdaSpace{}%
\AgdaBound{e₁}\AgdaSymbol{)}\AgdaSpace{}%
\AgdaSymbol{=}\AgdaSpace{}%
\AgdaInductiveConstructor{selb}\AgdaSpace{}%
\AgdaBound{m}\AgdaSpace{}%
\AgdaSymbol{(}\AgdaFunction{wk}\AgdaSpace{}%
\AgdaBound{v}\AgdaSpace{}%
\AgdaBound{e}\AgdaSymbol{)}\AgdaSpace{}%
\AgdaSymbol{(}\AgdaFunction{wk}\AgdaSpace{}%
\AgdaBound{v}\AgdaSpace{}%
\AgdaBound{e₁}\AgdaSymbol{)}\<%
\\
\>[2]\AgdaComment{--wk\ v\ (zero-butb\ m\ e\ e₁)\ =\ zero-butb\ m\ (wk\ v\ e)\ (wk\ v\ e₁)}\<%
\end{code} 

A prerequisite for substitution is decidable equality
of variables which will be also useful during optimisations.  The code below
is a copy-paste from~\cite{subst}, but we reiterate its wonderfully mind-twisting
mechanics here.  The relation for variable equality is given by the type \AD{Eq}
which has two constructors.  In case variables are equal (\AC{eq} constructor)
they literally have to match.  In case variables $x$ and $y$ are different
(\AC{neq} constructor), we would like to know where to find $y$ in the context
without $x$.  After that, \AF{eq?} shows that variable equality is decidable.
The substitution \AF{sub} explains how to substitute the variable $v$ in the
expression $e$ with the expression $e₁$. 
\begin{mathpar}
\codeblock{\begin{code}%
\>[2]\AgdaKeyword{data}\AgdaSpace{}%
\AgdaDatatype{Eq}\AgdaSpace{}%
\AgdaSymbol{:}\AgdaSpace{}%
\AgdaGeneralizable{is}\AgdaSpace{}%
\AgdaOperator{\AgdaDatatype{∈}}\AgdaSpace{}%
\AgdaGeneralizable{Γ}\AgdaSpace{}%
\AgdaSymbol{→}\AgdaSpace{}%
\AgdaGeneralizable{ip}\AgdaSpace{}%
\AgdaOperator{\AgdaDatatype{∈}}\AgdaSpace{}%
\AgdaGeneralizable{Γ}\AgdaSpace{}%
\AgdaSymbol{→}\AgdaSpace{}%
\AgdaPrimitive{Set}\AgdaSpace{}%
\AgdaKeyword{where}\<%
\\
\>[2][@{}l@{\AgdaIndent{0}}]%
\>[4]\AgdaInductiveConstructor{eq}%
\>[9]\AgdaSymbol{:}\AgdaSpace{}%
\AgdaSymbol{\{}\AgdaBound{x}\AgdaSpace{}%
\AgdaSymbol{:}\AgdaSpace{}%
\AgdaGeneralizable{is}\AgdaSpace{}%
\AgdaOperator{\AgdaDatatype{∈}}\AgdaSpace{}%
\AgdaGeneralizable{Γ}\AgdaSymbol{\}}\AgdaSpace{}%
\AgdaSymbol{→}\AgdaSpace{}%
\AgdaDatatype{Eq}\AgdaSpace{}%
\AgdaBound{x}\AgdaSpace{}%
\AgdaBound{x}\<%
\\
\>[4]\AgdaInductiveConstructor{neq}%
\>[9]\AgdaSymbol{:}\AgdaSpace{}%
\AgdaSymbol{(}\AgdaBound{x}\AgdaSpace{}%
\AgdaSymbol{:}\AgdaSpace{}%
\AgdaGeneralizable{is}\AgdaSpace{}%
\AgdaOperator{\AgdaDatatype{∈}}\AgdaSpace{}%
\AgdaGeneralizable{Γ}\AgdaSymbol{)}\AgdaSpace{}%
\AgdaSymbol{→}\AgdaSpace{}%
\AgdaSymbol{(}\AgdaBound{y}\AgdaSpace{}%
\AgdaSymbol{:}\AgdaSpace{}%
\AgdaGeneralizable{ip}\AgdaSpace{}%
\AgdaOperator{\AgdaDatatype{∈}}\AgdaSpace{}%
\AgdaSymbol{(}\AgdaGeneralizable{Γ}\AgdaSpace{}%
\AgdaOperator{\AgdaFunction{/}}\AgdaSpace{}%
\AgdaBound{x}\AgdaSymbol{))}\<%
\\
\>[9]\AgdaSymbol{→}\AgdaSpace{}%
\AgdaDatatype{Eq}\AgdaSpace{}%
\AgdaBound{x}\AgdaSpace{}%
\AgdaSymbol{(}\AgdaFunction{wkv}\AgdaSpace{}%
\AgdaBound{x}\AgdaSpace{}%
\AgdaBound{y}\AgdaSymbol{)}\<%
\\
\\[\AgdaEmptyExtraSkip]%
\>[2]\AgdaFunction{sub}%
\>[1074I]\AgdaSymbol{:}\AgdaSpace{}%
\AgdaSymbol{(}\AgdaBound{v}\AgdaSpace{}%
\AgdaSymbol{:}\AgdaSpace{}%
\AgdaGeneralizable{is}\AgdaSpace{}%
\AgdaOperator{\AgdaDatatype{∈}}\AgdaSpace{}%
\AgdaGeneralizable{Γ}\AgdaSymbol{)}\AgdaSpace{}%
\AgdaSymbol{(}\AgdaBound{e}\AgdaSpace{}%
\AgdaSymbol{:}\AgdaSpace{}%
\AgdaDatatype{E}\AgdaSpace{}%
\AgdaGeneralizable{Γ}\AgdaSpace{}%
\AgdaGeneralizable{ip}\AgdaSymbol{)}\AgdaSpace{}%
\AgdaSymbol{(}\AgdaBound{e₁}\AgdaSpace{}%
\AgdaSymbol{:}\AgdaSpace{}%
\AgdaDatatype{E}\AgdaSpace{}%
\AgdaSymbol{(}\AgdaGeneralizable{Γ}\AgdaSpace{}%
\AgdaOperator{\AgdaFunction{/}}\AgdaSpace{}%
\AgdaBound{v}\AgdaSymbol{)}\AgdaSpace{}%
\AgdaGeneralizable{is}\AgdaSymbol{)}\<%
\\
\>[.][@{}l@{}]\<[1074I]%
\>[6]\AgdaSymbol{→}\AgdaSpace{}%
\AgdaDatatype{E}\AgdaSpace{}%
\AgdaSymbol{(}\AgdaGeneralizable{Γ}\AgdaSpace{}%
\AgdaOperator{\AgdaFunction{/}}\AgdaSpace{}%
\AgdaBound{v}\AgdaSymbol{)}\AgdaSpace{}%
\AgdaGeneralizable{ip}\<%
\end{code}}
\and
\codeblock{\begin{code}%
\>[2]\AgdaFunction{eq?}\AgdaSpace{}%
\AgdaSymbol{:}\AgdaSpace{}%
\AgdaSymbol{(}\AgdaBound{x}\AgdaSpace{}%
\AgdaSymbol{:}\AgdaSpace{}%
\AgdaGeneralizable{is}\AgdaSpace{}%
\AgdaOperator{\AgdaDatatype{∈}}\AgdaSpace{}%
\AgdaGeneralizable{Γ}\AgdaSymbol{)}\AgdaSpace{}%
\AgdaSymbol{→}\AgdaSpace{}%
\AgdaSymbol{(}\AgdaBound{y}\AgdaSpace{}%
\AgdaSymbol{:}\AgdaSpace{}%
\AgdaGeneralizable{ip}\AgdaSpace{}%
\AgdaOperator{\AgdaDatatype{∈}}\AgdaSpace{}%
\AgdaGeneralizable{Γ}\AgdaSymbol{)}\AgdaSpace{}%
\AgdaSymbol{→}\AgdaSpace{}%
\AgdaDatatype{Eq}\AgdaSpace{}%
\AgdaBound{x}\AgdaSpace{}%
\AgdaBound{y}\<%
\\
\>[2]\AgdaFunction{eq?}\AgdaSpace{}%
\AgdaInductiveConstructor{v₀}%
\>[14]\AgdaInductiveConstructor{v₀}%
\>[22]\AgdaSymbol{=}\AgdaSpace{}%
\AgdaInductiveConstructor{eq}\<%
\\
\>[2]\AgdaFunction{eq?}\AgdaSpace{}%
\AgdaInductiveConstructor{v₀}%
\>[14]\AgdaSymbol{(}\AgdaInductiveConstructor{vₛ}\AgdaSpace{}%
\AgdaBound{y}\AgdaSymbol{)}%
\>[22]\AgdaSymbol{=}\AgdaSpace{}%
\AgdaInductiveConstructor{neq}\AgdaSpace{}%
\AgdaInductiveConstructor{v₀}\AgdaSpace{}%
\AgdaBound{y}\<%
\\
\>[2]\AgdaFunction{eq?}\AgdaSpace{}%
\AgdaSymbol{(}\AgdaInductiveConstructor{vₛ}\AgdaSpace{}%
\AgdaBound{x}\AgdaSymbol{)}%
\>[14]\AgdaInductiveConstructor{v₀}%
\>[22]\AgdaSymbol{=}\AgdaSpace{}%
\AgdaInductiveConstructor{neq}\AgdaSpace{}%
\AgdaSymbol{(}\AgdaInductiveConstructor{vₛ}\AgdaSpace{}%
\AgdaBound{x}\AgdaSymbol{)}\AgdaSpace{}%
\AgdaInductiveConstructor{v₀}\<%
\\
\>[2]\AgdaFunction{eq?}\AgdaSpace{}%
\AgdaSymbol{(}\AgdaInductiveConstructor{vₛ}\AgdaSpace{}%
\AgdaBound{x}\AgdaSymbol{)}%
\>[14]\AgdaSymbol{(}\AgdaInductiveConstructor{vₛ}\AgdaSpace{}%
\AgdaBound{y}\AgdaSymbol{)}\AgdaSpace{}%
\AgdaKeyword{with}\AgdaSpace{}%
\AgdaFunction{eq?}\AgdaSpace{}%
\AgdaBound{x}\AgdaSpace{}%
\AgdaBound{y}\<%
\\
\>[2]\AgdaSymbol{...}\AgdaSpace{}%
\AgdaSymbol{|}\AgdaSpace{}%
\AgdaInductiveConstructor{eq}%
\>[18]\AgdaSymbol{=}\AgdaSpace{}%
\AgdaInductiveConstructor{eq}\<%
\\
\>[2]\AgdaSymbol{...}\AgdaSpace{}%
\AgdaSymbol{|}\AgdaSpace{}%
\AgdaInductiveConstructor{neq}\AgdaSpace{}%
\AgdaDottedPattern{\AgdaSymbol{.}}\AgdaDottedPattern{\AgdaBound{x}}\AgdaSpace{}%
\AgdaBound{y}%
\>[18]\AgdaSymbol{=}\AgdaSpace{}%
\AgdaInductiveConstructor{neq}\AgdaSpace{}%
\AgdaSymbol{(}\AgdaInductiveConstructor{vₛ}\AgdaSpace{}%
\AgdaBound{x}\AgdaSymbol{)}\AgdaSpace{}%
\AgdaSymbol{(}\AgdaInductiveConstructor{vₛ}\AgdaSpace{}%
\AgdaBound{y}\AgdaSymbol{)}\<%
\end{code}}
\end{mathpar}
\begin{code}[hide]%
\>[2]\AgdaFunction{sub-var}\AgdaSpace{}%
\AgdaSymbol{:}\AgdaSpace{}%
\AgdaSymbol{(}\AgdaBound{v}\AgdaSpace{}%
\AgdaSymbol{:}\AgdaSpace{}%
\AgdaGeneralizable{is}\AgdaSpace{}%
\AgdaOperator{\AgdaDatatype{∈}}\AgdaSpace{}%
\AgdaGeneralizable{Γ}\AgdaSymbol{)}\AgdaSpace{}%
\AgdaSymbol{→}\AgdaSpace{}%
\AgdaGeneralizable{ip}\AgdaSpace{}%
\AgdaOperator{\AgdaDatatype{∈}}\AgdaSpace{}%
\AgdaGeneralizable{Γ}\AgdaSpace{}%
\AgdaSymbol{→}\AgdaSpace{}%
\AgdaDatatype{E}\AgdaSpace{}%
\AgdaSymbol{(}\AgdaGeneralizable{Γ}\AgdaSpace{}%
\AgdaOperator{\AgdaFunction{/}}\AgdaSpace{}%
\AgdaBound{v}\AgdaSymbol{)}\AgdaSpace{}%
\AgdaGeneralizable{is}\AgdaSpace{}%
\AgdaSymbol{→}\AgdaSpace{}%
\AgdaDatatype{E}\AgdaSpace{}%
\AgdaSymbol{(}\AgdaGeneralizable{Γ}\AgdaSpace{}%
\AgdaOperator{\AgdaFunction{/}}\AgdaSpace{}%
\AgdaBound{v}\AgdaSymbol{)}\AgdaSpace{}%
\AgdaGeneralizable{ip}\<%
\\
\>[2]\AgdaFunction{sub-var}\AgdaSpace{}%
\AgdaBound{x}\AgdaSpace{}%
\AgdaBound{y}\AgdaSpace{}%
\AgdaBound{e}\AgdaSpace{}%
\AgdaKeyword{with}\AgdaSpace{}%
\AgdaFunction{eq?}\AgdaSpace{}%
\AgdaBound{x}\AgdaSpace{}%
\AgdaBound{y}\<%
\\
\>[2]\AgdaSymbol{...}\AgdaSpace{}%
\AgdaSymbol{|}\AgdaSpace{}%
\AgdaInductiveConstructor{eq}\AgdaSpace{}%
\AgdaSymbol{=}\AgdaSpace{}%
\AgdaBound{e}\<%
\\
\>[2]\AgdaSymbol{...}\AgdaSpace{}%
\AgdaSymbol{|}\AgdaSpace{}%
\AgdaInductiveConstructor{neq}\AgdaSpace{}%
\AgdaDottedPattern{\AgdaSymbol{.}}\AgdaDottedPattern{\AgdaBound{x}}\AgdaSpace{}%
\AgdaBound{y}\AgdaSpace{}%
\AgdaSymbol{=}\AgdaSpace{}%
\AgdaInductiveConstructor{var}\AgdaSpace{}%
\AgdaBound{y}\<%
\\
\>[0]\<%
\\
\>[2]\AgdaFunction{sub}\AgdaSpace{}%
\AgdaBound{v}\AgdaSpace{}%
\AgdaInductiveConstructor{zero}\AgdaSpace{}%
\AgdaBound{e₂}\AgdaSpace{}%
\AgdaSymbol{=}\AgdaSpace{}%
\AgdaInductiveConstructor{zero}\<%
\\
\>[2]\AgdaFunction{sub}\AgdaSpace{}%
\AgdaBound{v}\AgdaSpace{}%
\AgdaInductiveConstructor{one}\AgdaSpace{}%
\AgdaBound{e₂}\AgdaSpace{}%
\AgdaSymbol{=}\AgdaSpace{}%
\AgdaInductiveConstructor{one}\<%
\\
\>[0]\<%
\\
\>[2]\AgdaFunction{sub}\AgdaSpace{}%
\AgdaBound{v}\AgdaSpace{}%
\AgdaSymbol{(}\AgdaInductiveConstructor{var}\AgdaSpace{}%
\AgdaBound{x}\AgdaSymbol{)}\AgdaSpace{}%
\AgdaBound{e₂}\AgdaSpace{}%
\AgdaSymbol{=}\AgdaSpace{}%
\AgdaFunction{sub-var}\AgdaSpace{}%
\AgdaBound{v}\AgdaSpace{}%
\AgdaBound{x}\AgdaSpace{}%
\AgdaBound{e₂}\<%
\\
\>[2]\AgdaFunction{sub}\AgdaSpace{}%
\AgdaBound{v}\AgdaSpace{}%
\AgdaSymbol{(}\AgdaInductiveConstructor{imapₛ}\AgdaSpace{}%
\AgdaBound{e₁}\AgdaSymbol{)}\AgdaSpace{}%
\AgdaBound{e₂}\AgdaSpace{}%
\AgdaSymbol{=}\AgdaSpace{}%
\AgdaInductiveConstructor{imapₛ}\AgdaSpace{}%
\AgdaSymbol{(}\AgdaFunction{sub}\AgdaSpace{}%
\AgdaSymbol{(}\AgdaInductiveConstructor{vₛ}\AgdaSpace{}%
\AgdaBound{v}\AgdaSymbol{)}\AgdaSpace{}%
\AgdaBound{e₁}\AgdaSpace{}%
\AgdaSymbol{(}\AgdaFunction{wk}\AgdaSpace{}%
\AgdaInductiveConstructor{v₀}\AgdaSpace{}%
\AgdaBound{e₂}\AgdaSymbol{))}\<%
\\
\>[2]\AgdaFunction{sub}\AgdaSpace{}%
\AgdaBound{v}\AgdaSpace{}%
\AgdaSymbol{(}\AgdaInductiveConstructor{selₛ}\AgdaSpace{}%
\AgdaBound{e₁}\AgdaSpace{}%
\AgdaBound{e₃}\AgdaSymbol{)}\AgdaSpace{}%
\AgdaBound{e₂}\AgdaSpace{}%
\AgdaSymbol{=}\AgdaSpace{}%
\AgdaInductiveConstructor{selₛ}\AgdaSpace{}%
\AgdaSymbol{(}\AgdaFunction{sub}\AgdaSpace{}%
\AgdaBound{v}\AgdaSpace{}%
\AgdaBound{e₁}\AgdaSpace{}%
\AgdaBound{e₂}\AgdaSymbol{)}\AgdaSpace{}%
\AgdaSymbol{(}\AgdaFunction{sub}\AgdaSpace{}%
\AgdaBound{v}\AgdaSpace{}%
\AgdaBound{e₃}\AgdaSpace{}%
\AgdaBound{e₂}\AgdaSymbol{)}\<%
\\
\>[0]\<%
\\
\>[2]\AgdaFunction{sub}\AgdaSpace{}%
\AgdaBound{v}\AgdaSpace{}%
\AgdaSymbol{(}\AgdaInductiveConstructor{imap}\AgdaSpace{}%
\AgdaBound{e₁}\AgdaSymbol{)}\AgdaSpace{}%
\AgdaBound{e₂}\AgdaSpace{}%
\AgdaSymbol{=}\AgdaSpace{}%
\AgdaInductiveConstructor{imap}\AgdaSpace{}%
\AgdaSymbol{(}\AgdaFunction{sub}\AgdaSpace{}%
\AgdaSymbol{(}\AgdaInductiveConstructor{vₛ}\AgdaSpace{}%
\AgdaBound{v}\AgdaSymbol{)}\AgdaSpace{}%
\AgdaBound{e₁}\AgdaSpace{}%
\AgdaSymbol{(}\AgdaFunction{wk}\AgdaSpace{}%
\AgdaInductiveConstructor{v₀}\AgdaSpace{}%
\AgdaBound{e₂}\AgdaSymbol{))}\<%
\\
\>[2]\AgdaFunction{sub}\AgdaSpace{}%
\AgdaBound{v}\AgdaSpace{}%
\AgdaSymbol{(}\AgdaInductiveConstructor{sel}\AgdaSpace{}%
\AgdaBound{e₁}\AgdaSpace{}%
\AgdaBound{e₃}\AgdaSymbol{)}\AgdaSpace{}%
\AgdaBound{e₂}\AgdaSpace{}%
\AgdaSymbol{=}\AgdaSpace{}%
\AgdaInductiveConstructor{sel}\AgdaSpace{}%
\AgdaSymbol{(}\AgdaFunction{sub}\AgdaSpace{}%
\AgdaBound{v}\AgdaSpace{}%
\AgdaBound{e₁}\AgdaSpace{}%
\AgdaBound{e₂}\AgdaSymbol{)}\AgdaSpace{}%
\AgdaSymbol{(}\AgdaFunction{sub}\AgdaSpace{}%
\AgdaBound{v}\AgdaSpace{}%
\AgdaBound{e₃}\AgdaSpace{}%
\AgdaBound{e₂}\AgdaSymbol{)}\<%
\\
\>[0]\<%
\\
\>[2]\AgdaFunction{sub}\AgdaSpace{}%
\AgdaBound{v}\AgdaSpace{}%
\AgdaSymbol{(}\AgdaInductiveConstructor{zero-but}\AgdaSpace{}%
\AgdaBound{i}\AgdaSpace{}%
\AgdaBound{j}\AgdaSpace{}%
\AgdaBound{e}\AgdaSymbol{)}\AgdaSpace{}%
\AgdaBound{e₂}\AgdaSpace{}%
\AgdaSymbol{=}\AgdaSpace{}%
\AgdaInductiveConstructor{zero-but}\AgdaSpace{}%
\AgdaSymbol{(}\AgdaFunction{sub}\AgdaSpace{}%
\AgdaBound{v}\AgdaSpace{}%
\AgdaBound{i}\AgdaSpace{}%
\AgdaBound{e₂}\AgdaSymbol{)}\AgdaSpace{}%
\AgdaSymbol{(}\AgdaFunction{sub}\AgdaSpace{}%
\AgdaBound{v}\AgdaSpace{}%
\AgdaBound{j}\AgdaSpace{}%
\AgdaBound{e₂}\AgdaSymbol{)}\AgdaSpace{}%
\AgdaSymbol{(}\AgdaFunction{sub}\AgdaSpace{}%
\AgdaBound{v}\AgdaSpace{}%
\AgdaBound{e}\AgdaSpace{}%
\AgdaBound{e₂}\AgdaSymbol{)}\<%
\\
\>[2]\AgdaFunction{sub}\AgdaSpace{}%
\AgdaBound{v}\AgdaSpace{}%
\AgdaSymbol{(}\AgdaInductiveConstructor{sum}\AgdaSpace{}%
\AgdaBound{e₁}\AgdaSymbol{)}\AgdaSpace{}%
\AgdaBound{e₂}\AgdaSpace{}%
\AgdaSymbol{=}\AgdaSpace{}%
\AgdaInductiveConstructor{E.sum}\AgdaSpace{}%
\AgdaSymbol{(}\AgdaFunction{sub}\AgdaSpace{}%
\AgdaSymbol{(}\AgdaInductiveConstructor{vₛ}\AgdaSpace{}%
\AgdaBound{v}\AgdaSymbol{)}\AgdaSpace{}%
\AgdaBound{e₁}\AgdaSpace{}%
\AgdaSymbol{(}\AgdaFunction{wk}\AgdaSpace{}%
\AgdaInductiveConstructor{v₀}\AgdaSpace{}%
\AgdaBound{e₂}\AgdaSymbol{))}\<%
\\
\>[2]\AgdaFunction{sub}\AgdaSpace{}%
\AgdaBound{v}\AgdaSpace{}%
\AgdaSymbol{(}\AgdaInductiveConstructor{bin}\AgdaSpace{}%
\AgdaBound{x}\AgdaSpace{}%
\AgdaBound{e₁}\AgdaSpace{}%
\AgdaBound{e₃}\AgdaSymbol{)}\AgdaSpace{}%
\AgdaBound{e₂}\AgdaSpace{}%
\AgdaSymbol{=}\AgdaSpace{}%
\AgdaInductiveConstructor{bin}\AgdaSpace{}%
\AgdaBound{x}\AgdaSpace{}%
\AgdaSymbol{(}\AgdaFunction{sub}\AgdaSpace{}%
\AgdaBound{v}\AgdaSpace{}%
\AgdaBound{e₁}\AgdaSpace{}%
\AgdaBound{e₂}\AgdaSymbol{)}\AgdaSpace{}%
\AgdaSymbol{(}\AgdaFunction{sub}\AgdaSpace{}%
\AgdaBound{v}\AgdaSpace{}%
\AgdaBound{e₃}\AgdaSpace{}%
\AgdaBound{e₂}\AgdaSymbol{)}\<%
\\
\>[2]\AgdaFunction{sub}\AgdaSpace{}%
\AgdaBound{v}\AgdaSpace{}%
\AgdaSymbol{(}\AgdaInductiveConstructor{slide}\AgdaSpace{}%
\AgdaBound{i}\AgdaSpace{}%
\AgdaBound{pl}\AgdaSpace{}%
\AgdaBound{e}\AgdaSpace{}%
\AgdaBound{su}\AgdaSymbol{)}\AgdaSpace{}%
\AgdaBound{e₂}\AgdaSpace{}%
\AgdaSymbol{=}\AgdaSpace{}%
\AgdaInductiveConstructor{E.slide}\AgdaSpace{}%
\AgdaSymbol{(}\AgdaFunction{sub}\AgdaSpace{}%
\AgdaBound{v}\AgdaSpace{}%
\AgdaBound{i}\AgdaSpace{}%
\AgdaBound{e₂}\AgdaSymbol{)}\AgdaSpace{}%
\AgdaBound{pl}\AgdaSpace{}%
\AgdaSymbol{(}\AgdaFunction{sub}\AgdaSpace{}%
\AgdaBound{v}\AgdaSpace{}%
\AgdaBound{e}\AgdaSpace{}%
\AgdaBound{e₂}\AgdaSymbol{)}\AgdaSpace{}%
\AgdaBound{su}\<%
\\
\>[2]\AgdaFunction{sub}\AgdaSpace{}%
\AgdaBound{v}\AgdaSpace{}%
\AgdaSymbol{(}\AgdaInductiveConstructor{backslide}\AgdaSpace{}%
\AgdaBound{i}\AgdaSpace{}%
\AgdaBound{e}\AgdaSpace{}%
\AgdaBound{su}\AgdaSpace{}%
\AgdaBound{pl}\AgdaSymbol{)}\AgdaSpace{}%
\AgdaBound{e₂}\AgdaSpace{}%
\AgdaSymbol{=}\AgdaSpace{}%
\AgdaInductiveConstructor{E.backslide}\AgdaSpace{}%
\AgdaSymbol{(}\AgdaFunction{sub}\AgdaSpace{}%
\AgdaBound{v}\AgdaSpace{}%
\AgdaBound{i}\AgdaSpace{}%
\AgdaBound{e₂}\AgdaSymbol{)}\AgdaSpace{}%
\AgdaSymbol{(}\AgdaFunction{sub}\AgdaSpace{}%
\AgdaBound{v}\AgdaSpace{}%
\AgdaBound{e}\AgdaSpace{}%
\AgdaBound{e₂}\AgdaSymbol{)}\AgdaSpace{}%
\AgdaBound{su}\AgdaSpace{}%
\AgdaBound{pl}\<%
\\
\>[0]\<%
\\
\>[2]\AgdaFunction{sub}\AgdaSpace{}%
\AgdaBound{v}\AgdaSpace{}%
\AgdaSymbol{(}\AgdaInductiveConstructor{scaledown}\AgdaSpace{}%
\AgdaBound{x}\AgdaSpace{}%
\AgdaBound{e}\AgdaSymbol{)}\AgdaSpace{}%
\AgdaBound{e₂}\AgdaSpace{}%
\AgdaSymbol{=}\AgdaSpace{}%
\AgdaInductiveConstructor{scaledown}\AgdaSpace{}%
\AgdaBound{x}\AgdaSpace{}%
\AgdaSymbol{(}\AgdaFunction{sub}\AgdaSpace{}%
\AgdaBound{v}\AgdaSpace{}%
\AgdaBound{e}\AgdaSpace{}%
\AgdaBound{e₂}\AgdaSymbol{)}\<%
\\
\>[2]\AgdaFunction{sub}\AgdaSpace{}%
\AgdaBound{v}\AgdaSpace{}%
\AgdaSymbol{(}\AgdaInductiveConstructor{minus}\AgdaSpace{}%
\AgdaBound{e}\AgdaSymbol{)}\AgdaSpace{}%
\AgdaBound{e₂}\AgdaSpace{}%
\AgdaSymbol{=}\AgdaSpace{}%
\AgdaInductiveConstructor{minus}\AgdaSpace{}%
\AgdaSymbol{(}\AgdaFunction{sub}\AgdaSpace{}%
\AgdaBound{v}\AgdaSpace{}%
\AgdaBound{e}\AgdaSpace{}%
\AgdaBound{e₂}\AgdaSymbol{)}\<%
\\
\>[2]\AgdaFunction{sub}\AgdaSpace{}%
\AgdaBound{v}\AgdaSpace{}%
\AgdaSymbol{(}\AgdaInductiveConstructor{logistic}\AgdaSpace{}%
\AgdaBound{e}\AgdaSymbol{)}\AgdaSpace{}%
\AgdaBound{e₂}\AgdaSpace{}%
\AgdaSymbol{=}\AgdaSpace{}%
\AgdaInductiveConstructor{logistic}\AgdaSpace{}%
\AgdaSymbol{(}\AgdaFunction{sub}\AgdaSpace{}%
\AgdaBound{v}\AgdaSpace{}%
\AgdaBound{e}\AgdaSpace{}%
\AgdaBound{e₂}\AgdaSymbol{)}\<%
\\
\>[0]\<%
\\
\>[2]\AgdaFunction{sub}\AgdaSpace{}%
\AgdaBound{v}\AgdaSpace{}%
\AgdaSymbol{(}\AgdaInductiveConstructor{imapb}\AgdaSpace{}%
\AgdaBound{m}\AgdaSpace{}%
\AgdaBound{e}\AgdaSymbol{)}\AgdaSpace{}%
\AgdaBound{e₂}\AgdaSpace{}%
\AgdaSymbol{=}\AgdaSpace{}%
\AgdaInductiveConstructor{imapb}\AgdaSpace{}%
\AgdaBound{m}\AgdaSpace{}%
\AgdaSymbol{(}\AgdaFunction{sub}\AgdaSpace{}%
\AgdaSymbol{(}\AgdaInductiveConstructor{vₛ}\AgdaSpace{}%
\AgdaBound{v}\AgdaSymbol{)}\AgdaSpace{}%
\AgdaBound{e}\AgdaSpace{}%
\AgdaSymbol{(}\AgdaFunction{wk}\AgdaSpace{}%
\AgdaInductiveConstructor{v₀}\AgdaSpace{}%
\AgdaBound{e₂}\AgdaSymbol{))}\<%
\\
\>[2]\AgdaFunction{sub}\AgdaSpace{}%
\AgdaBound{v}\AgdaSpace{}%
\AgdaSymbol{(}\AgdaInductiveConstructor{selb}\AgdaSpace{}%
\AgdaBound{m}\AgdaSpace{}%
\AgdaBound{e}\AgdaSpace{}%
\AgdaBound{e₁}\AgdaSymbol{)}\AgdaSpace{}%
\AgdaBound{e₂}\AgdaSpace{}%
\AgdaSymbol{=}\AgdaSpace{}%
\AgdaInductiveConstructor{selb}\AgdaSpace{}%
\AgdaBound{m}\AgdaSpace{}%
\AgdaSymbol{(}\AgdaFunction{sub}\AgdaSpace{}%
\AgdaBound{v}\AgdaSpace{}%
\AgdaBound{e}\AgdaSpace{}%
\AgdaBound{e₂}\AgdaSymbol{)}\AgdaSpace{}%
\AgdaSymbol{(}\AgdaFunction{sub}\AgdaSpace{}%
\AgdaBound{v}\AgdaSpace{}%
\AgdaBound{e₁}\AgdaSpace{}%
\AgdaBound{e₂}\AgdaSymbol{)}\<%
\end{code}

As our context do not encode explicit dependencies between the variables,
we can define the operation that swaps two consequent variables at any given
position in the context.  Similarly to (\AB{Γ} \AF{/} \AB{v}), we define the
function \AF{SwapAt} that computes the context where $x$ and its successor are
swapped.  Then we define the operation \AF{ctx-swap} that translates the expression
$e$ into the context where $x$ is swapped with its successor.
\begin{code}%
\>[2]\AgdaFunction{SwapAt}\AgdaSpace{}%
\AgdaSymbol{:}\AgdaSpace{}%
\AgdaSymbol{(}\AgdaBound{Γ}\AgdaSpace{}%
\AgdaSymbol{:}\AgdaSpace{}%
\AgdaDatatype{Ctx}\AgdaSymbol{)}\AgdaSpace{}%
\AgdaSymbol{→}\AgdaSpace{}%
\AgdaGeneralizable{is}\AgdaSpace{}%
\AgdaOperator{\AgdaDatatype{∈}}\AgdaSpace{}%
\AgdaBound{Γ}\AgdaSpace{}%
\AgdaSymbol{→}\AgdaSpace{}%
\AgdaDatatype{Ctx}\<%
\\
\>[2]\AgdaFunction{SwapAt}\AgdaSpace{}%
\AgdaSymbol{(}\AgdaBound{Γ}\AgdaSpace{}%
\AgdaOperator{\AgdaInductiveConstructor{▹}}\AgdaSpace{}%
\AgdaBound{is}\AgdaSymbol{)}%
\>[24]\AgdaInductiveConstructor{v₀}%
\>[37]\AgdaSymbol{=}\AgdaSpace{}%
\AgdaBound{Γ}\AgdaSpace{}%
\AgdaOperator{\AgdaInductiveConstructor{▹}}\AgdaSpace{}%
\AgdaBound{is}\<%
\\
\>[2]\AgdaFunction{SwapAt}\AgdaSpace{}%
\AgdaSymbol{(}\AgdaBound{Γ}\AgdaSpace{}%
\AgdaOperator{\AgdaInductiveConstructor{▹}}\AgdaSpace{}%
\AgdaBound{ip}\AgdaSpace{}%
\AgdaOperator{\AgdaInductiveConstructor{▹}}\AgdaSpace{}%
\AgdaBound{is}\AgdaSymbol{)}%
\>[24]\AgdaInductiveConstructor{v₁}%
\>[37]\AgdaSymbol{=}\AgdaSpace{}%
\AgdaBound{Γ}\AgdaSpace{}%
\AgdaOperator{\AgdaInductiveConstructor{▹}}\AgdaSpace{}%
\AgdaBound{is}\AgdaSpace{}%
\AgdaOperator{\AgdaInductiveConstructor{▹}}\AgdaSpace{}%
\AgdaBound{ip}\<%
\\
\>[2]\AgdaFunction{SwapAt}\AgdaSpace{}%
\AgdaSymbol{(}\AgdaBound{Γ}\AgdaSpace{}%
\AgdaOperator{\AgdaInductiveConstructor{▹}}\AgdaSpace{}%
\AgdaBound{ip}\AgdaSpace{}%
\AgdaOperator{\AgdaInductiveConstructor{▹}}\AgdaSpace{}%
\AgdaBound{is}\AgdaSymbol{)}%
\>[24]\AgdaSymbol{(}\AgdaInductiveConstructor{vₛ}\AgdaSpace{}%
\AgdaSymbol{(}\AgdaInductiveConstructor{vₛ}\AgdaSpace{}%
\AgdaBound{x}\AgdaSymbol{))}%
\>[37]\AgdaSymbol{=}\AgdaSpace{}%
\AgdaFunction{SwapAt}\AgdaSpace{}%
\AgdaSymbol{(}\AgdaBound{Γ}\AgdaSpace{}%
\AgdaOperator{\AgdaInductiveConstructor{▹}}\AgdaSpace{}%
\AgdaBound{ip}\AgdaSymbol{)}\AgdaSpace{}%
\AgdaSymbol{(}\AgdaInductiveConstructor{vₛ}\AgdaSpace{}%
\AgdaBound{x}\AgdaSymbol{)}\AgdaSpace{}%
\AgdaOperator{\AgdaInductiveConstructor{▹}}\AgdaSpace{}%
\AgdaBound{is}\<%
\\
\>[0]\<%
\\
\>[2]\AgdaFunction{ctx-swap}\AgdaSpace{}%
\AgdaSymbol{:}\AgdaSpace{}%
\AgdaSymbol{(}\AgdaBound{x}\AgdaSpace{}%
\AgdaSymbol{:}\AgdaSpace{}%
\AgdaGeneralizable{is}\AgdaSpace{}%
\AgdaOperator{\AgdaDatatype{∈}}\AgdaSpace{}%
\AgdaGeneralizable{Γ}\AgdaSymbol{)}\AgdaSpace{}%
\AgdaSymbol{(}\AgdaBound{e}\AgdaSpace{}%
\AgdaSymbol{:}\AgdaSpace{}%
\AgdaDatatype{E}\AgdaSpace{}%
\AgdaGeneralizable{Γ}\AgdaSpace{}%
\AgdaGeneralizable{ip}\AgdaSymbol{)}\AgdaSpace{}%
\AgdaSymbol{→}\AgdaSpace{}%
\AgdaDatatype{E}\AgdaSpace{}%
\AgdaSymbol{(}\AgdaFunction{SwapAt}\AgdaSpace{}%
\AgdaGeneralizable{Γ}\AgdaSpace{}%
\AgdaBound{x}\AgdaSymbol{)}\AgdaSpace{}%
\AgdaGeneralizable{ip}\<%
\end{code}
\begin{code}[hide]%
\>[2]\AgdaFunction{var-swap}\AgdaSpace{}%
\AgdaSymbol{:}\AgdaSpace{}%
\AgdaSymbol{(}\AgdaBound{x}\AgdaSpace{}%
\AgdaSymbol{:}\AgdaSpace{}%
\AgdaGeneralizable{is}\AgdaSpace{}%
\AgdaOperator{\AgdaDatatype{∈}}\AgdaSpace{}%
\AgdaGeneralizable{Γ}\AgdaSymbol{)}\AgdaSpace{}%
\AgdaSymbol{→}\AgdaSpace{}%
\AgdaGeneralizable{ip}\AgdaSpace{}%
\AgdaOperator{\AgdaDatatype{∈}}\AgdaSpace{}%
\AgdaGeneralizable{Γ}\AgdaSpace{}%
\AgdaSymbol{→}\AgdaSpace{}%
\AgdaGeneralizable{ip}\AgdaSpace{}%
\AgdaOperator{\AgdaDatatype{∈}}\AgdaSpace{}%
\AgdaFunction{SwapAt}\AgdaSpace{}%
\AgdaGeneralizable{Γ}\AgdaSpace{}%
\AgdaBound{x}\<%
\\
\>[2]\AgdaFunction{var-swap}\AgdaSpace{}%
\AgdaInductiveConstructor{v₀}\AgdaSpace{}%
\AgdaBound{y}\AgdaSpace{}%
\AgdaSymbol{=}\AgdaSpace{}%
\AgdaBound{y}\<%
\\
\>[2]\AgdaFunction{var-swap}\AgdaSpace{}%
\AgdaInductiveConstructor{v₁}\AgdaSpace{}%
\AgdaInductiveConstructor{v₀}\AgdaSpace{}%
\AgdaSymbol{=}\AgdaSpace{}%
\AgdaInductiveConstructor{v₁}\<%
\\
\>[2]\AgdaFunction{var-swap}\AgdaSpace{}%
\AgdaInductiveConstructor{v₁}\AgdaSpace{}%
\AgdaInductiveConstructor{v₁}\AgdaSpace{}%
\AgdaSymbol{=}\AgdaSpace{}%
\AgdaInductiveConstructor{v₀}\<%
\\
\>[2]\AgdaFunction{var-swap}\AgdaSpace{}%
\AgdaInductiveConstructor{v₁}\AgdaSpace{}%
\AgdaSymbol{(}\AgdaInductiveConstructor{vₛ}\AgdaSpace{}%
\AgdaSymbol{(}\AgdaInductiveConstructor{vₛ}\AgdaSpace{}%
\AgdaBound{y}\AgdaSymbol{))}\AgdaSpace{}%
\AgdaSymbol{=}\AgdaSpace{}%
\AgdaInductiveConstructor{vₛ}\AgdaSpace{}%
\AgdaSymbol{(}\AgdaInductiveConstructor{vₛ}\AgdaSpace{}%
\AgdaBound{y}\AgdaSymbol{)}\<%
\\
\>[2]\AgdaFunction{var-swap}\AgdaSpace{}%
\AgdaSymbol{(}\AgdaInductiveConstructor{vₛ}\AgdaSpace{}%
\AgdaSymbol{(}\AgdaInductiveConstructor{vₛ}\AgdaSpace{}%
\AgdaBound{x}\AgdaSymbol{))}\AgdaSpace{}%
\AgdaInductiveConstructor{v₀}\AgdaSpace{}%
\AgdaSymbol{=}\AgdaSpace{}%
\AgdaInductiveConstructor{v₀}\<%
\\
\>[2]\AgdaFunction{var-swap}\AgdaSpace{}%
\AgdaSymbol{(}\AgdaInductiveConstructor{vₛ}\AgdaSpace{}%
\AgdaSymbol{(}\AgdaInductiveConstructor{vₛ}\AgdaSpace{}%
\AgdaBound{x}\AgdaSymbol{))}\AgdaSpace{}%
\AgdaSymbol{(}\AgdaInductiveConstructor{vₛ}\AgdaSpace{}%
\AgdaBound{y}\AgdaSymbol{)}\AgdaSpace{}%
\AgdaSymbol{=}\AgdaSpace{}%
\AgdaInductiveConstructor{vₛ}\AgdaSpace{}%
\AgdaSymbol{(}\AgdaFunction{var-swap}\AgdaSpace{}%
\AgdaSymbol{(}\AgdaInductiveConstructor{vₛ}\AgdaSpace{}%
\AgdaBound{x}\AgdaSymbol{)}\AgdaSpace{}%
\AgdaBound{y}\AgdaSymbol{)}\<%
\\
\>[0]\<%
\\
\>[2]\AgdaFunction{ctx-swap}\AgdaSpace{}%
\AgdaBound{x}\AgdaSpace{}%
\AgdaInductiveConstructor{zero}\AgdaSpace{}%
\AgdaSymbol{=}\AgdaSpace{}%
\AgdaInductiveConstructor{zero}\<%
\\
\>[2]\AgdaFunction{ctx-swap}\AgdaSpace{}%
\AgdaBound{x}\AgdaSpace{}%
\AgdaInductiveConstructor{one}\AgdaSpace{}%
\AgdaSymbol{=}\AgdaSpace{}%
\AgdaInductiveConstructor{one}\<%
\\
\>[0]\<%
\\
\>[2]\AgdaFunction{ctx-swap}\AgdaSpace{}%
\AgdaBound{x}\AgdaSpace{}%
\AgdaSymbol{(}\AgdaInductiveConstructor{var}\AgdaSpace{}%
\AgdaBound{y}\AgdaSymbol{)}\AgdaSpace{}%
\AgdaSymbol{=}\AgdaSpace{}%
\AgdaInductiveConstructor{var}\AgdaSpace{}%
\AgdaSymbol{(}\AgdaFunction{var-swap}\AgdaSpace{}%
\AgdaBound{x}\AgdaSpace{}%
\AgdaBound{y}\AgdaSymbol{)}\<%
\\
\>[2]\AgdaFunction{ctx-swap}\AgdaSpace{}%
\AgdaInductiveConstructor{v₀}\AgdaSpace{}%
\AgdaSymbol{(}\AgdaInductiveConstructor{imapₛ}\AgdaSpace{}%
\AgdaBound{e}\AgdaSymbol{)}\AgdaSpace{}%
\AgdaSymbol{=}\AgdaSpace{}%
\AgdaInductiveConstructor{imapₛ}\AgdaSpace{}%
\AgdaBound{e}\<%
\\
\>[2]\AgdaFunction{ctx-swap}\AgdaSpace{}%
\AgdaSymbol{(}\AgdaInductiveConstructor{vₛ}\AgdaSpace{}%
\AgdaBound{x}\AgdaSymbol{)}\AgdaSpace{}%
\AgdaSymbol{(}\AgdaInductiveConstructor{imapₛ}\AgdaSpace{}%
\AgdaBound{e}\AgdaSymbol{)}\AgdaSpace{}%
\AgdaSymbol{=}\AgdaSpace{}%
\AgdaInductiveConstructor{imapₛ}\AgdaSpace{}%
\AgdaSymbol{(}\AgdaFunction{ctx-swap}\AgdaSpace{}%
\AgdaSymbol{(}\AgdaInductiveConstructor{vₛ}\AgdaSpace{}%
\AgdaSymbol{(}\AgdaInductiveConstructor{vₛ}\AgdaSpace{}%
\AgdaBound{x}\AgdaSymbol{))}\AgdaSpace{}%
\AgdaBound{e}\AgdaSymbol{)}\<%
\\
\>[2]\AgdaFunction{ctx-swap}\AgdaSpace{}%
\AgdaBound{x}\AgdaSpace{}%
\AgdaSymbol{(}\AgdaInductiveConstructor{selₛ}\AgdaSpace{}%
\AgdaBound{e}\AgdaSpace{}%
\AgdaBound{e₁}\AgdaSymbol{)}\AgdaSpace{}%
\AgdaSymbol{=}\AgdaSpace{}%
\AgdaInductiveConstructor{selₛ}\AgdaSpace{}%
\AgdaSymbol{(}\AgdaFunction{ctx-swap}\AgdaSpace{}%
\AgdaBound{x}\AgdaSpace{}%
\AgdaBound{e}\AgdaSymbol{)}\AgdaSpace{}%
\AgdaSymbol{(}\AgdaFunction{ctx-swap}\AgdaSpace{}%
\AgdaBound{x}\AgdaSpace{}%
\AgdaBound{e₁}\AgdaSymbol{)}\<%
\\
\>[2]\AgdaFunction{ctx-swap}\AgdaSpace{}%
\AgdaInductiveConstructor{v₀}\AgdaSpace{}%
\AgdaSymbol{(}\AgdaInductiveConstructor{imap}\AgdaSpace{}%
\AgdaBound{e}\AgdaSymbol{)}\AgdaSpace{}%
\AgdaSymbol{=}\AgdaSpace{}%
\AgdaInductiveConstructor{imap}\AgdaSpace{}%
\AgdaBound{e}\<%
\\
\>[2]\AgdaFunction{ctx-swap}\AgdaSpace{}%
\AgdaSymbol{(}\AgdaInductiveConstructor{vₛ}\AgdaSpace{}%
\AgdaBound{x}\AgdaSymbol{)}\AgdaSpace{}%
\AgdaSymbol{(}\AgdaInductiveConstructor{imap}\AgdaSpace{}%
\AgdaBound{e}\AgdaSymbol{)}\AgdaSpace{}%
\AgdaSymbol{=}\AgdaSpace{}%
\AgdaInductiveConstructor{imap}\AgdaSpace{}%
\AgdaSymbol{(}\AgdaFunction{ctx-swap}\AgdaSpace{}%
\AgdaSymbol{(}\AgdaInductiveConstructor{vₛ}\AgdaSpace{}%
\AgdaSymbol{(}\AgdaInductiveConstructor{vₛ}\AgdaSpace{}%
\AgdaBound{x}\AgdaSymbol{))}\AgdaSpace{}%
\AgdaBound{e}\AgdaSymbol{)}\<%
\\
\>[2]\AgdaFunction{ctx-swap}\AgdaSpace{}%
\AgdaBound{x}\AgdaSpace{}%
\AgdaSymbol{(}\AgdaInductiveConstructor{sel}\AgdaSpace{}%
\AgdaBound{e}\AgdaSpace{}%
\AgdaBound{e₁}\AgdaSymbol{)}\AgdaSpace{}%
\AgdaSymbol{=}\AgdaSpace{}%
\AgdaInductiveConstructor{sel}\AgdaSpace{}%
\AgdaSymbol{(}\AgdaFunction{ctx-swap}\AgdaSpace{}%
\AgdaBound{x}\AgdaSpace{}%
\AgdaBound{e}\AgdaSymbol{)}\AgdaSpace{}%
\AgdaSymbol{(}\AgdaFunction{ctx-swap}\AgdaSpace{}%
\AgdaBound{x}\AgdaSpace{}%
\AgdaBound{e₁}\AgdaSymbol{)}\<%
\\
\>[2]\AgdaFunction{ctx-swap}\AgdaSpace{}%
\AgdaInductiveConstructor{v₀}\AgdaSpace{}%
\AgdaSymbol{(}\AgdaInductiveConstructor{imapb}\AgdaSpace{}%
\AgdaBound{m}\AgdaSpace{}%
\AgdaBound{e}\AgdaSymbol{)}\AgdaSpace{}%
\AgdaSymbol{=}\AgdaSpace{}%
\AgdaInductiveConstructor{imapb}\AgdaSpace{}%
\AgdaBound{m}\AgdaSpace{}%
\AgdaBound{e}\<%
\\
\>[2]\AgdaFunction{ctx-swap}\AgdaSpace{}%
\AgdaSymbol{(}\AgdaInductiveConstructor{vₛ}\AgdaSpace{}%
\AgdaBound{x}\AgdaSymbol{)}\AgdaSpace{}%
\AgdaSymbol{(}\AgdaInductiveConstructor{imapb}\AgdaSpace{}%
\AgdaBound{m}\AgdaSpace{}%
\AgdaBound{e}\AgdaSymbol{)}\AgdaSpace{}%
\AgdaSymbol{=}\AgdaSpace{}%
\AgdaInductiveConstructor{imapb}\AgdaSpace{}%
\AgdaBound{m}\AgdaSpace{}%
\AgdaSymbol{(}\AgdaFunction{ctx-swap}\AgdaSpace{}%
\AgdaSymbol{(}\AgdaInductiveConstructor{vₛ}\AgdaSpace{}%
\AgdaSymbol{(}\AgdaInductiveConstructor{vₛ}\AgdaSpace{}%
\AgdaBound{x}\AgdaSymbol{))}\AgdaSpace{}%
\AgdaBound{e}\AgdaSymbol{)}\<%
\\
\>[2]\AgdaFunction{ctx-swap}\AgdaSpace{}%
\AgdaBound{x}\AgdaSpace{}%
\AgdaSymbol{(}\AgdaInductiveConstructor{selb}\AgdaSpace{}%
\AgdaBound{m}\AgdaSpace{}%
\AgdaBound{e}\AgdaSpace{}%
\AgdaBound{e₁}\AgdaSymbol{)}\AgdaSpace{}%
\AgdaSymbol{=}\AgdaSpace{}%
\AgdaInductiveConstructor{selb}\AgdaSpace{}%
\AgdaBound{m}\AgdaSpace{}%
\AgdaSymbol{(}\AgdaFunction{ctx-swap}\AgdaSpace{}%
\AgdaBound{x}\AgdaSpace{}%
\AgdaBound{e}\AgdaSymbol{)}\AgdaSpace{}%
\AgdaSymbol{(}\AgdaFunction{ctx-swap}\AgdaSpace{}%
\AgdaBound{x}\AgdaSpace{}%
\AgdaBound{e₁}\AgdaSymbol{)}\<%
\\
\>[2]\AgdaFunction{ctx-swap}\AgdaSpace{}%
\AgdaBound{x}\AgdaSpace{}%
\AgdaSymbol{(}\AgdaInductiveConstructor{zero-but}\AgdaSpace{}%
\AgdaBound{e}\AgdaSpace{}%
\AgdaBound{e₁}\AgdaSpace{}%
\AgdaBound{e₂}\AgdaSymbol{)}\AgdaSpace{}%
\AgdaSymbol{=}\AgdaSpace{}%
\AgdaInductiveConstructor{zero-but}\AgdaSpace{}%
\AgdaSymbol{(}\AgdaFunction{ctx-swap}\AgdaSpace{}%
\AgdaBound{x}\AgdaSpace{}%
\AgdaBound{e}\AgdaSymbol{)}\AgdaSpace{}%
\AgdaSymbol{(}\AgdaFunction{ctx-swap}\AgdaSpace{}%
\AgdaBound{x}\AgdaSpace{}%
\AgdaBound{e₁}\AgdaSymbol{)}\AgdaSpace{}%
\AgdaSymbol{(}\AgdaFunction{ctx-swap}\AgdaSpace{}%
\AgdaBound{x}\AgdaSpace{}%
\AgdaBound{e₂}\AgdaSymbol{)}\<%
\\
\>[2]\AgdaFunction{ctx-swap}\AgdaSpace{}%
\AgdaInductiveConstructor{v₀}\AgdaSpace{}%
\AgdaSymbol{(}\AgdaInductiveConstructor{sum}\AgdaSpace{}%
\AgdaBound{e}\AgdaSymbol{)}\AgdaSpace{}%
\AgdaSymbol{=}\AgdaSpace{}%
\AgdaInductiveConstructor{E.sum}\AgdaSpace{}%
\AgdaBound{e}\<%
\\
\>[2]\AgdaFunction{ctx-swap}\AgdaSpace{}%
\AgdaSymbol{(}\AgdaInductiveConstructor{vₛ}\AgdaSpace{}%
\AgdaBound{x}\AgdaSymbol{)}\AgdaSpace{}%
\AgdaSymbol{(}\AgdaInductiveConstructor{sum}\AgdaSpace{}%
\AgdaBound{e}\AgdaSymbol{)}\AgdaSpace{}%
\AgdaSymbol{=}\AgdaSpace{}%
\AgdaInductiveConstructor{E.sum}\AgdaSpace{}%
\AgdaSymbol{(}\AgdaFunction{ctx-swap}\AgdaSpace{}%
\AgdaSymbol{(}\AgdaInductiveConstructor{vₛ}\AgdaSpace{}%
\AgdaSymbol{(}\AgdaInductiveConstructor{vₛ}\AgdaSpace{}%
\AgdaBound{x}\AgdaSymbol{))}\AgdaSpace{}%
\AgdaBound{e}\AgdaSymbol{)}\<%
\\
\>[2]\AgdaFunction{ctx-swap}\AgdaSpace{}%
\AgdaBound{x}\AgdaSpace{}%
\AgdaSymbol{(}\AgdaInductiveConstructor{bin}\AgdaSpace{}%
\AgdaBound{op}\AgdaSpace{}%
\AgdaBound{e}\AgdaSpace{}%
\AgdaBound{e₁}\AgdaSymbol{)}\AgdaSpace{}%
\AgdaSymbol{=}\AgdaSpace{}%
\AgdaInductiveConstructor{bin}\AgdaSpace{}%
\AgdaBound{op}\AgdaSpace{}%
\AgdaSymbol{(}\AgdaFunction{ctx-swap}\AgdaSpace{}%
\AgdaBound{x}\AgdaSpace{}%
\AgdaBound{e}\AgdaSymbol{)}\AgdaSpace{}%
\AgdaSymbol{(}\AgdaFunction{ctx-swap}\AgdaSpace{}%
\AgdaBound{x}\AgdaSpace{}%
\AgdaBound{e₁}\AgdaSymbol{)}\<%
\\
\>[2]\AgdaFunction{ctx-swap}\AgdaSpace{}%
\AgdaBound{x}\AgdaSpace{}%
\AgdaSymbol{(}\AgdaInductiveConstructor{slide}\AgdaSpace{}%
\AgdaBound{e}\AgdaSpace{}%
\AgdaBound{x₁}\AgdaSpace{}%
\AgdaBound{e₁}\AgdaSpace{}%
\AgdaBound{x₂}\AgdaSymbol{)}\AgdaSpace{}%
\AgdaSymbol{=}\AgdaSpace{}%
\AgdaInductiveConstructor{E.slide}\AgdaSpace{}%
\AgdaSymbol{(}\AgdaFunction{ctx-swap}\AgdaSpace{}%
\AgdaBound{x}\AgdaSpace{}%
\AgdaBound{e}\AgdaSymbol{)}\AgdaSpace{}%
\AgdaBound{x₁}\AgdaSpace{}%
\AgdaSymbol{(}\AgdaFunction{ctx-swap}\AgdaSpace{}%
\AgdaBound{x}\AgdaSpace{}%
\AgdaBound{e₁}\AgdaSymbol{)}\AgdaSpace{}%
\AgdaBound{x₂}\<%
\\
\>[2]\AgdaFunction{ctx-swap}\AgdaSpace{}%
\AgdaBound{x}\AgdaSpace{}%
\AgdaSymbol{(}\AgdaInductiveConstructor{backslide}\AgdaSpace{}%
\AgdaBound{e}\AgdaSpace{}%
\AgdaBound{e₁}\AgdaSpace{}%
\AgdaBound{x₁}\AgdaSpace{}%
\AgdaBound{x₂}\AgdaSymbol{)}\AgdaSpace{}%
\AgdaSymbol{=}\AgdaSpace{}%
\AgdaInductiveConstructor{E.backslide}\AgdaSpace{}%
\AgdaSymbol{(}\AgdaFunction{ctx-swap}\AgdaSpace{}%
\AgdaBound{x}\AgdaSpace{}%
\AgdaBound{e}\AgdaSymbol{)}\AgdaSpace{}%
\AgdaSymbol{(}\AgdaFunction{ctx-swap}\AgdaSpace{}%
\AgdaBound{x}\AgdaSpace{}%
\AgdaBound{e₁}\AgdaSymbol{)}\AgdaSpace{}%
\AgdaBound{x₁}\AgdaSpace{}%
\AgdaBound{x₂}\<%
\\
\>[2]\AgdaFunction{ctx-swap}\AgdaSpace{}%
\AgdaBound{x}\AgdaSpace{}%
\AgdaSymbol{(}\AgdaInductiveConstructor{scaledown}\AgdaSpace{}%
\AgdaBound{x₁}\AgdaSpace{}%
\AgdaBound{e}\AgdaSymbol{)}\AgdaSpace{}%
\AgdaSymbol{=}\AgdaSpace{}%
\AgdaInductiveConstructor{scaledown}\AgdaSpace{}%
\AgdaBound{x₁}\AgdaSpace{}%
\AgdaSymbol{(}\AgdaFunction{ctx-swap}\AgdaSpace{}%
\AgdaBound{x}\AgdaSpace{}%
\AgdaBound{e}\AgdaSymbol{)}\<%
\\
\>[2]\AgdaFunction{ctx-swap}\AgdaSpace{}%
\AgdaBound{x}\AgdaSpace{}%
\AgdaSymbol{(}\AgdaInductiveConstructor{minus}\AgdaSpace{}%
\AgdaBound{e}\AgdaSymbol{)}\AgdaSpace{}%
\AgdaSymbol{=}\AgdaSpace{}%
\AgdaInductiveConstructor{minus}\AgdaSpace{}%
\AgdaSymbol{(}\AgdaFunction{ctx-swap}\AgdaSpace{}%
\AgdaBound{x}\AgdaSpace{}%
\AgdaBound{e}\AgdaSymbol{)}\<%
\\
\>[2]\AgdaFunction{ctx-swap}\AgdaSpace{}%
\AgdaBound{x}\AgdaSpace{}%
\AgdaSymbol{(}\AgdaInductiveConstructor{logistic}\AgdaSpace{}%
\AgdaBound{e}\AgdaSymbol{)}\AgdaSpace{}%
\AgdaSymbol{=}\AgdaSpace{}%
\AgdaInductiveConstructor{logistic}\AgdaSpace{}%
\AgdaSymbol{(}\AgdaFunction{ctx-swap}\AgdaSpace{}%
\AgdaBound{x}\AgdaSpace{}%
\AgdaBound{e}\AgdaSymbol{)}\<%
\end{code}

\paragraph{Building Blocks}
Now we implement the remaining building blocks in \AD{E} that are needed
to define our CNN.
\begin{code}[hide]%
\>[0]\AgdaKeyword{module}\AgdaSpace{}%
\AgdaModule{BB}\AgdaSpace{}%
\AgdaKeyword{where}\<%
\\
\>[0][@{}l@{\AgdaIndent{0}}]%
\>[2]\AgdaKeyword{open}\AgdaSpace{}%
\AgdaKeyword{import}\AgdaSpace{}%
\AgdaModule{Data.Nat}\AgdaSpace{}%
\AgdaSymbol{as}\AgdaSpace{}%
\AgdaModule{ℕ}\AgdaSpace{}%
\AgdaKeyword{using}\AgdaSpace{}%
\AgdaSymbol{(}\AgdaDatatype{ℕ}\AgdaSymbol{;}\AgdaSpace{}%
\AgdaInductiveConstructor{zero}\AgdaSymbol{;}\AgdaSpace{}%
\AgdaInductiveConstructor{suc}\AgdaSymbol{)}\<%
\\
\>[2]\AgdaKeyword{open}\AgdaSpace{}%
\AgdaModule{Array}\AgdaSpace{}%
\AgdaKeyword{hiding}\AgdaSpace{}%
\AgdaSymbol{(}\AgdaFunction{sum}\AgdaSymbol{;}\AgdaSpace{}%
\AgdaFunction{slide}\AgdaSymbol{;}\AgdaSpace{}%
\AgdaFunction{backslide}\AgdaSymbol{)}\<%
\\
\>[2]\AgdaKeyword{open}\AgdaSpace{}%
\AgdaModule{Lang}\<%
\\
\>[2]\AgdaKeyword{open}\AgdaSpace{}%
\AgdaModule{SubWk}\AgdaSpace{}%
\AgdaKeyword{using}\AgdaSpace{}%
\AgdaSymbol{(}\AgdaFunction{wk}\AgdaSymbol{;}\AgdaSpace{}%
\AgdaOperator{\AgdaFunction{↑\AgdaUnderscore{}}}\AgdaSymbol{;}\AgdaSpace{}%
\AgdaOperator{\AgdaFunction{↑↑\AgdaUnderscore{}}}\AgdaSymbol{)}\<%
\\
\\[\AgdaEmptyExtraSkip]%
\>[2]\AgdaComment{--\AgdaUnderscore{}⊞\AgdaUnderscore{}\ \AgdaUnderscore{}⊠\AgdaUnderscore{}\ :\ (a\ b\ :\ E\ Γ\ (ar\ s))\ →\ E\ Γ\ (ar\ s)}\<%
\\
\>[2]\AgdaFunction{Imapₛ}\AgdaSpace{}%
\AgdaSymbol{:}\AgdaSpace{}%
\AgdaSymbol{(}\AgdaDatatype{E}\AgdaSpace{}%
\AgdaSymbol{(}\AgdaGeneralizable{Γ}\AgdaSpace{}%
\AgdaOperator{\AgdaInductiveConstructor{▹}}\AgdaSpace{}%
\AgdaInductiveConstructor{ix}\AgdaSpace{}%
\AgdaGeneralizable{s}\AgdaSymbol{)}\AgdaSpace{}%
\AgdaSymbol{(}\AgdaInductiveConstructor{ix}\AgdaSpace{}%
\AgdaGeneralizable{s}\AgdaSymbol{)}\AgdaSpace{}%
\AgdaSymbol{→}\AgdaSpace{}%
\AgdaDatatype{E}\AgdaSpace{}%
\AgdaSymbol{(}\AgdaGeneralizable{Γ}\AgdaSpace{}%
\AgdaOperator{\AgdaInductiveConstructor{▹}}\AgdaSpace{}%
\AgdaInductiveConstructor{ix}\AgdaSpace{}%
\AgdaGeneralizable{s}\AgdaSymbol{)}\AgdaSpace{}%
\AgdaSymbol{(}\AgdaInductiveConstructor{ar}\AgdaSpace{}%
\AgdaFunction{unit}\AgdaSymbol{))}\AgdaSpace{}%
\AgdaSymbol{→}\AgdaSpace{}%
\AgdaDatatype{E}\AgdaSpace{}%
\AgdaGeneralizable{Γ}\AgdaSpace{}%
\AgdaSymbol{(}\AgdaInductiveConstructor{ar}\AgdaSpace{}%
\AgdaGeneralizable{s}\AgdaSymbol{)}\<%
\\
\>[2]\AgdaFunction{Imap}\AgdaSpace{}%
\AgdaSymbol{:}\AgdaSpace{}%
\AgdaSymbol{(}\AgdaDatatype{E}\AgdaSpace{}%
\AgdaSymbol{(}\AgdaGeneralizable{Γ}\AgdaSpace{}%
\AgdaOperator{\AgdaInductiveConstructor{▹}}\AgdaSpace{}%
\AgdaInductiveConstructor{ix}\AgdaSpace{}%
\AgdaGeneralizable{s}\AgdaSymbol{)}\AgdaSpace{}%
\AgdaSymbol{(}\AgdaInductiveConstructor{ix}\AgdaSpace{}%
\AgdaGeneralizable{s}\AgdaSymbol{)}\AgdaSpace{}%
\AgdaSymbol{→}\AgdaSpace{}%
\AgdaDatatype{E}\AgdaSpace{}%
\AgdaSymbol{(}\AgdaGeneralizable{Γ}\AgdaSpace{}%
\AgdaOperator{\AgdaInductiveConstructor{▹}}\AgdaSpace{}%
\AgdaInductiveConstructor{ix}\AgdaSpace{}%
\AgdaGeneralizable{s}\AgdaSymbol{)}\AgdaSpace{}%
\AgdaSymbol{(}\AgdaInductiveConstructor{ar}\AgdaSpace{}%
\AgdaGeneralizable{p}\AgdaSymbol{))}\AgdaSpace{}%
\AgdaSymbol{→}\AgdaSpace{}%
\AgdaDatatype{E}\AgdaSpace{}%
\AgdaGeneralizable{Γ}\AgdaSpace{}%
\AgdaSymbol{(}\AgdaInductiveConstructor{ar}\AgdaSpace{}%
\AgdaSymbol{(}\AgdaGeneralizable{s}\AgdaSpace{}%
\AgdaOperator{\AgdaInductiveConstructor{⊗}}\AgdaSpace{}%
\AgdaGeneralizable{p}\AgdaSymbol{))}\<%
\\
\>[2]\AgdaFunction{Sum}\AgdaSpace{}%
\AgdaSymbol{:}\AgdaSpace{}%
\AgdaSymbol{(}\AgdaDatatype{E}\AgdaSpace{}%
\AgdaSymbol{(}\AgdaGeneralizable{Γ}\AgdaSpace{}%
\AgdaOperator{\AgdaInductiveConstructor{▹}}\AgdaSpace{}%
\AgdaInductiveConstructor{ix}\AgdaSpace{}%
\AgdaGeneralizable{s}\AgdaSymbol{)}\AgdaSpace{}%
\AgdaSymbol{(}\AgdaInductiveConstructor{ix}\AgdaSpace{}%
\AgdaGeneralizable{s}\AgdaSymbol{)}\AgdaSpace{}%
\AgdaSymbol{→}\AgdaSpace{}%
\AgdaDatatype{E}\AgdaSpace{}%
\AgdaSymbol{(}\AgdaGeneralizable{Γ}\AgdaSpace{}%
\AgdaOperator{\AgdaInductiveConstructor{▹}}\AgdaSpace{}%
\AgdaInductiveConstructor{ix}\AgdaSpace{}%
\AgdaGeneralizable{s}\AgdaSymbol{)}\AgdaSpace{}%
\AgdaSymbol{(}\AgdaInductiveConstructor{ar}\AgdaSpace{}%
\AgdaGeneralizable{p}\AgdaSymbol{))}\AgdaSpace{}%
\AgdaSymbol{→}\AgdaSpace{}%
\AgdaDatatype{E}\AgdaSpace{}%
\AgdaGeneralizable{Γ}\AgdaSpace{}%
\AgdaSymbol{(}\AgdaInductiveConstructor{ar}\AgdaSpace{}%
\AgdaGeneralizable{p}\AgdaSymbol{)}\<%
\end{code}
We start with a several convenience functions that wrap \AC{imap}s and \AC{sum}
such that when we write (\AF{Imap} \AB{λ} \AB{i} \AB{→} \AB{⋯}), Agda's variable
$i$ is mapped to the \AF{E}'s variable \AC{v₀}.
\begin{mathpar}
\codeblock{\begin{code}%
\>[2]\AgdaFunction{Imapₛ}\AgdaSpace{}%
\AgdaBound{f}\AgdaSpace{}%
\AgdaSymbol{=}\AgdaSpace{}%
\AgdaInductiveConstructor{imapₛ}\AgdaSpace{}%
\AgdaSymbol{(}\AgdaBound{f}\AgdaSpace{}%
\AgdaSymbol{(}\AgdaInductiveConstructor{var}\AgdaSpace{}%
\AgdaInductiveConstructor{v₀}\AgdaSymbol{))}\<%
\end{code}}
\and
\codeblock{\begin{code}%
\>[2]\AgdaFunction{Imap}\AgdaSpace{}%
\AgdaBound{f}\AgdaSpace{}%
\AgdaSymbol{=}\AgdaSpace{}%
\AgdaInductiveConstructor{imap}\AgdaSpace{}%
\AgdaSymbol{(}\AgdaBound{f}\AgdaSpace{}%
\AgdaSymbol{(}\AgdaInductiveConstructor{var}\AgdaSpace{}%
\AgdaInductiveConstructor{v₀}\AgdaSymbol{))}\<%
\end{code}}
\and
\codeblock{\begin{code}%
\>[2]\AgdaFunction{Sum}\AgdaSpace{}%
\AgdaBound{f}\AgdaSpace{}%
\AgdaSymbol{=}\AgdaSpace{}%
\AgdaInductiveConstructor{sum}\AgdaSpace{}%
\AgdaSymbol{(}\AgdaBound{f}\AgdaSpace{}%
\AgdaSymbol{(}\AgdaInductiveConstructor{var}\AgdaSpace{}%
\AgdaInductiveConstructor{v₀}\AgdaSymbol{))}\<%
\end{code}}
\end{mathpar}

The remaining operations are \AF{conv}, \AF{mconv} and \AF{avgp₂} which
can be defined as functions on \AF{E} as follows.
\begin{code}%
\>[2]\AgdaFunction{conv}\AgdaSpace{}%
\AgdaSymbol{:}\AgdaSpace{}%
\AgdaDatatype{E}\AgdaSpace{}%
\AgdaGeneralizable{Γ}\AgdaSpace{}%
\AgdaSymbol{(}\AgdaInductiveConstructor{ar}\AgdaSpace{}%
\AgdaGeneralizable{r}\AgdaSymbol{)}\AgdaSpace{}%
\AgdaSymbol{→}\AgdaSpace{}%
\AgdaGeneralizable{s}\AgdaSpace{}%
\AgdaOperator{\AgdaDatatype{+}}\AgdaSpace{}%
\AgdaGeneralizable{p}\AgdaSpace{}%
\AgdaOperator{\AgdaDatatype{≈}}\AgdaSpace{}%
\AgdaGeneralizable{r}\AgdaSpace{}%
\AgdaSymbol{→}\AgdaSpace{}%
\AgdaDatatype{E}\AgdaSpace{}%
\AgdaGeneralizable{Γ}\AgdaSpace{}%
\AgdaSymbol{(}\AgdaInductiveConstructor{ar}\AgdaSpace{}%
\AgdaGeneralizable{s}\AgdaSymbol{)}\AgdaSpace{}%
\AgdaSymbol{→}\AgdaSpace{}%
\AgdaOperator{\AgdaDatatype{suc}}\AgdaSpace{}%
\AgdaGeneralizable{p}\AgdaSpace{}%
\AgdaOperator{\AgdaDatatype{≈}}\AgdaSpace{}%
\AgdaGeneralizable{u}\AgdaSpace{}%
\AgdaSymbol{→}\AgdaSpace{}%
\AgdaDatatype{E}\AgdaSpace{}%
\AgdaGeneralizable{Γ}\AgdaSpace{}%
\AgdaSymbol{(}\AgdaInductiveConstructor{ar}\AgdaSpace{}%
\AgdaGeneralizable{u}\AgdaSymbol{)}\<%
\\
\>[2]\AgdaFunction{conv}\AgdaSpace{}%
\AgdaBound{f}\AgdaSpace{}%
\AgdaBound{sp}\AgdaSpace{}%
\AgdaBound{g}\AgdaSpace{}%
\AgdaBound{su}\AgdaSpace{}%
\AgdaSymbol{=}\AgdaSpace{}%
\AgdaFunction{Sum}\AgdaSpace{}%
\AgdaSymbol{λ}\AgdaSpace{}%
\AgdaBound{i}\AgdaSpace{}%
\AgdaSymbol{→}\AgdaSpace{}%
\AgdaInductiveConstructor{slide}\AgdaSpace{}%
\AgdaBound{i}\AgdaSpace{}%
\AgdaBound{sp}\AgdaSpace{}%
\AgdaSymbol{(}\AgdaOperator{\AgdaFunction{↑}}\AgdaSpace{}%
\AgdaBound{f}\AgdaSymbol{)}\AgdaSpace{}%
\AgdaBound{su}\AgdaSpace{}%
\AgdaOperator{\AgdaInductiveConstructor{⊠}}\AgdaSpace{}%
\AgdaFunction{Imapₛ}\AgdaSpace{}%
\AgdaSymbol{λ}\AgdaSpace{}%
\AgdaBound{\AgdaUnderscore{}}\AgdaSpace{}%
\AgdaSymbol{→}\AgdaSpace{}%
\AgdaInductiveConstructor{selₛ}\AgdaSpace{}%
\AgdaSymbol{(}\AgdaOperator{\AgdaFunction{↑↑}}\AgdaSpace{}%
\AgdaBound{g}\AgdaSymbol{)}\AgdaSpace{}%
\AgdaSymbol{(}\AgdaOperator{\AgdaFunction{↑}}\AgdaSpace{}%
\AgdaBound{i}\AgdaSymbol{)}\<%
\\
\\[\AgdaEmptyExtraSkip]%
\>[2]\AgdaFunction{mconv}%
\>[1892I]\AgdaSymbol{:}\AgdaSpace{}%
\AgdaGeneralizable{s}\AgdaSpace{}%
\AgdaOperator{\AgdaDatatype{+}}\AgdaSpace{}%
\AgdaGeneralizable{p}\AgdaSpace{}%
\AgdaOperator{\AgdaDatatype{≈}}\AgdaSpace{}%
\AgdaGeneralizable{r}\AgdaSpace{}%
\AgdaSymbol{→}\AgdaSpace{}%
\AgdaSymbol{(}\AgdaBound{inp}\AgdaSpace{}%
\AgdaSymbol{:}\AgdaSpace{}%
\AgdaDatatype{E}\AgdaSpace{}%
\AgdaGeneralizable{Γ}\AgdaSpace{}%
\AgdaSymbol{(}\AgdaInductiveConstructor{ar}\AgdaSpace{}%
\AgdaGeneralizable{r}\AgdaSymbol{))}\AgdaSpace{}%
\AgdaSymbol{(}\AgdaBound{we}\AgdaSpace{}%
\AgdaSymbol{:}\AgdaSpace{}%
\AgdaDatatype{E}\AgdaSpace{}%
\AgdaGeneralizable{Γ}\AgdaSpace{}%
\AgdaSymbol{(}\AgdaInductiveConstructor{ar}\AgdaSpace{}%
\AgdaSymbol{(}\AgdaGeneralizable{u}\AgdaSpace{}%
\AgdaOperator{\AgdaInductiveConstructor{⊗}}\AgdaSpace{}%
\AgdaGeneralizable{s}\AgdaSymbol{)))}\AgdaSpace{}%
\AgdaSymbol{(}\AgdaBound{b}\AgdaSpace{}%
\AgdaSymbol{:}\AgdaSpace{}%
\AgdaDatatype{E}\AgdaSpace{}%
\AgdaGeneralizable{Γ}\AgdaSpace{}%
\AgdaSymbol{(}\AgdaInductiveConstructor{ar}\AgdaSpace{}%
\AgdaGeneralizable{u}\AgdaSymbol{))}\<%
\\
\>[.][@{}l@{}]\<[1892I]%
\>[8]\AgdaSymbol{→}\AgdaSpace{}%
\AgdaOperator{\AgdaDatatype{suc}}\AgdaSpace{}%
\AgdaGeneralizable{p}\AgdaSpace{}%
\AgdaOperator{\AgdaDatatype{≈}}\AgdaSpace{}%
\AgdaGeneralizable{w}\AgdaSpace{}%
\AgdaSymbol{→}\AgdaSpace{}%
\AgdaDatatype{E}\AgdaSpace{}%
\AgdaGeneralizable{Γ}\AgdaSpace{}%
\AgdaSymbol{(}\AgdaInductiveConstructor{ar}\AgdaSpace{}%
\AgdaSymbol{(}\AgdaGeneralizable{u}\AgdaSpace{}%
\AgdaOperator{\AgdaInductiveConstructor{⊗}}\AgdaSpace{}%
\AgdaGeneralizable{w}\AgdaSymbol{))}\<%
\\
\>[2]\AgdaFunction{mconv}\AgdaSpace{}%
\AgdaBound{sp}\AgdaSpace{}%
\AgdaBound{inp}\AgdaSpace{}%
\AgdaBound{we}\AgdaSpace{}%
\AgdaBound{b}\AgdaSpace{}%
\AgdaBound{su}\AgdaSpace{}%
\AgdaSymbol{=}\AgdaSpace{}%
\AgdaFunction{Imap}\AgdaSpace{}%
\AgdaSymbol{λ}\AgdaSpace{}%
\AgdaBound{i}\AgdaSpace{}%
\AgdaSymbol{→}\AgdaSpace{}%
\AgdaFunction{conv}\AgdaSpace{}%
\AgdaSymbol{(}\AgdaOperator{\AgdaFunction{↑}}\AgdaSpace{}%
\AgdaBound{inp}\AgdaSymbol{)}\AgdaSpace{}%
\AgdaBound{sp}\AgdaSpace{}%
\AgdaSymbol{(}\AgdaInductiveConstructor{sel}\AgdaSpace{}%
\AgdaSymbol{(}\AgdaOperator{\AgdaFunction{↑}}\AgdaSpace{}%
\AgdaBound{we}\AgdaSymbol{)}\AgdaSpace{}%
\AgdaBound{i}\AgdaSymbol{)}\AgdaSpace{}%
\AgdaBound{su}\AgdaSpace{}%
\AgdaOperator{\AgdaInductiveConstructor{⊞}}\AgdaSpace{}%
\AgdaFunction{Imapₛ}\AgdaSpace{}%
\AgdaSymbol{λ}\AgdaSpace{}%
\AgdaBound{\AgdaUnderscore{}}\AgdaSpace{}%
\AgdaSymbol{→}\AgdaSpace{}%
\AgdaInductiveConstructor{selₛ}\AgdaSpace{}%
\AgdaSymbol{(}\AgdaOperator{\AgdaFunction{↑↑}}\AgdaSpace{}%
\AgdaBound{b}\AgdaSymbol{)}\AgdaSpace{}%
\AgdaSymbol{(}\AgdaOperator{\AgdaFunction{↑}}\AgdaSpace{}%
\AgdaBound{i}\AgdaSymbol{)}\<%
\\
\\[\AgdaEmptyExtraSkip]%
\>[2]\AgdaFunction{avgp₂}\AgdaSpace{}%
\AgdaSymbol{:}\AgdaSpace{}%
\AgdaSymbol{∀}\AgdaSpace{}%
\AgdaBound{m}\AgdaSpace{}%
\AgdaBound{n}\AgdaSpace{}%
\AgdaSymbol{→}\AgdaSpace{}%
\AgdaSymbol{(}\AgdaBound{a}\AgdaSpace{}%
\AgdaSymbol{:}\AgdaSpace{}%
\AgdaDatatype{E}\AgdaSpace{}%
\AgdaGeneralizable{Γ}\AgdaSpace{}%
\AgdaSymbol{(}\AgdaInductiveConstructor{ar}\AgdaSpace{}%
\AgdaSymbol{(}\AgdaInductiveConstructor{ι}\AgdaSpace{}%
\AgdaSymbol{(}\AgdaBound{m}\AgdaSpace{}%
\AgdaOperator{\AgdaPrimitive{ℕ.*}}\AgdaSpace{}%
\AgdaNumber{2}\AgdaSymbol{)}\AgdaSpace{}%
\AgdaOperator{\AgdaInductiveConstructor{⊗}}\AgdaSpace{}%
\AgdaInductiveConstructor{ι}\AgdaSpace{}%
\AgdaSymbol{(}\AgdaBound{n}\AgdaSpace{}%
\AgdaOperator{\AgdaPrimitive{ℕ.*}}\AgdaSpace{}%
\AgdaNumber{2}\AgdaSymbol{))))}\AgdaSpace{}%
\AgdaSymbol{→}\AgdaSpace{}%
\AgdaDatatype{E}\AgdaSpace{}%
\AgdaGeneralizable{Γ}\AgdaSpace{}%
\AgdaSymbol{(}\AgdaInductiveConstructor{ar}\AgdaSpace{}%
\AgdaSymbol{(}\AgdaInductiveConstructor{ι}\AgdaSpace{}%
\AgdaBound{m}\AgdaSpace{}%
\AgdaOperator{\AgdaInductiveConstructor{⊗}}\AgdaSpace{}%
\AgdaInductiveConstructor{ι}\AgdaSpace{}%
\AgdaBound{n}\AgdaSymbol{))}\<%
\\
\>[2]\AgdaFunction{avgp₂}\AgdaSpace{}%
\AgdaBound{m}\AgdaSpace{}%
\AgdaBound{n}\AgdaSpace{}%
\AgdaBound{a}\AgdaSpace{}%
\AgdaSymbol{=}\AgdaSpace{}%
\AgdaFunction{Imapₛ}\AgdaSpace{}%
\AgdaSymbol{λ}\AgdaSpace{}%
\AgdaBound{i}\AgdaSpace{}%
\AgdaSymbol{→}\AgdaSpace{}%
\AgdaInductiveConstructor{scaledown}\AgdaSpace{}%
\AgdaNumber{4}\AgdaSpace{}%
\AgdaOperator{\AgdaFunction{\$}}\AgdaSpace{}%
\AgdaFunction{Sum}\AgdaSpace{}%
\AgdaSymbol{λ}\AgdaSpace{}%
\AgdaBound{j}\AgdaSpace{}%
\AgdaSymbol{→}\AgdaSpace{}%
\AgdaInductiveConstructor{selₛ}\AgdaSpace{}%
\AgdaSymbol{(}\AgdaInductiveConstructor{selb}\AgdaSpace{}%
\AgdaSymbol{(}\AgdaInductiveConstructor{ι}\AgdaSpace{}%
\AgdaOperator{\AgdaInductiveConstructor{⊗}}\AgdaSpace{}%
\AgdaInductiveConstructor{ι}\AgdaSymbol{)}\AgdaSpace{}%
\AgdaSymbol{(}\AgdaOperator{\AgdaFunction{↑↑}}\AgdaSpace{}%
\AgdaBound{a}\AgdaSymbol{)}\AgdaSpace{}%
\AgdaSymbol{(}\AgdaOperator{\AgdaFunction{↑}}\AgdaSpace{}%
\AgdaBound{i}\AgdaSymbol{))}\AgdaSpace{}%
\AgdaBound{j}\<%
\\
\>[0]\<%
\end{code}
Note that these definitions are not very different from those found in
Section~\ref{sec:array-theory}.  Some operations such as \AF{nest} and \AF{unnest}
got inlined into \AF{E}'s operators, and all we really have to take care of is 
weakening of the expressions whenever we go under binders.

\begin{code}[hide]%
\>[0]\AgdaKeyword{open}\AgdaSpace{}%
\AgdaKeyword{import}\AgdaSpace{}%
\AgdaModule{Relation.Binary.PropositionalEquality}\<%
\\
\>[0]\AgdaKeyword{open}\AgdaSpace{}%
\AgdaKeyword{import}\AgdaSpace{}%
\AgdaModule{Relation.Nullary}\<%
\\
\>[0]\AgdaKeyword{open}\AgdaSpace{}%
\AgdaKeyword{import}\AgdaSpace{}%
\AgdaModule{Data.List}\AgdaSpace{}%
\AgdaKeyword{using}\AgdaSpace{}%
\AgdaSymbol{(}\AgdaDatatype{List}\AgdaSymbol{;}\AgdaSpace{}%
\AgdaInductiveConstructor{[]}\AgdaSymbol{;}\AgdaSpace{}%
\AgdaOperator{\AgdaInductiveConstructor{\AgdaUnderscore{}∷\AgdaUnderscore{}}}\AgdaSymbol{)}\<%
\\
\>[0]\AgdaKeyword{open}\AgdaSpace{}%
\AgdaKeyword{import}\AgdaSpace{}%
\AgdaModule{Data.Empty}\<%
\\
\>[0]\AgdaKeyword{open}\AgdaSpace{}%
\AgdaKeyword{import}\AgdaSpace{}%
\AgdaModule{Function}\<%
\\
\\[\AgdaEmptyExtraSkip]%
\>[0]\AgdaComment{--\ Our\ local\ files.}\<%
\\
\>[0]\AgdaKeyword{open}\AgdaSpace{}%
\AgdaKeyword{import}\AgdaSpace{}%
\AgdaModule{arrays}\<%
\\
\>[0]\AgdaKeyword{open}\AgdaSpace{}%
\AgdaKeyword{import}\AgdaSpace{}%
\AgdaModule{lang}\<%
\\
\>[0]\AgdaKeyword{module}\AgdaSpace{}%
\AgdaModule{\AgdaUnderscore{}}\AgdaSpace{}%
\AgdaKeyword{where}\<%
\end{code}

\section{Automatic Differentiation\label{sec:ad}}

We implement automatic differentiation in reverse mode
for expressions in \AF{E}.  We focus on reverse mode because it is
of most interest in machine learning, and it is more challenging to implement.
We start with a brief introduction of the AD, for much more in-depth
explanations refer to~\cite{autodiff-survey, backprop-stlc}.   Consider differentiating
a function composition consisting of three functions:
\[ 
   y = (f \circ g \circ h)\ x
\]
rewrite it using temporary variables:
\begin{eqnarray*}
  w_0 &=& x \\
  w_1 &=& h\ w_0 \\
  w_2 &=& g\ w_1 \\
  w_3 &=& f\ w_2 = y
\end{eqnarray*}
The chain rule gives us 
$\frac{\partial y}{\partial x} 
  = \frac{\partial y}{\partial w_2}
    \frac{\partial w_2}{\partial w_1}
    \frac{\partial w_1}{\partial x}$.  The difference between the forward and reverse
    mode lies in the direction that we traverse the chain rule.  In forward mode we
    traverse the chain inside-out, and the revers mode traverses the chain outside-in
    thus computing recursive relation:
$\frac{\partial y}{\partial w_i}
  = \frac{\partial y}{\partial w_{i+1}}
    \frac{\partial w_{i+1}}{\partial w_i}$.  For our example, we compute
$\frac{\partial y}{\partial w_2}$, then $\frac{\partial w_2}{\partial w_1}$ and
finally $\frac{\partial w_1}{\partial x}$.  While there seem to be no difference for
functions of one variable, there is a big difference for functions of $n$ variables
as we can compute derivatives of all the non-dependent variables simultaneously.
Consider an example of the $z = f\ x\ y = sin(xy + x)$:
\begin{eqnarray*}
  w_0 &=& x \\
  w_1 &=& y \\
  w_2 &=& w_1w_2\\
  w_3 &=& w_2 + w_0 \\
  w_4 &=& \sin w_3 = z
\end{eqnarray*}
We compute the adjoints $\bar{w}_i = \frac{\partial y}{\partial w_i}$ using the following
rule.  If $w_i$ has successors in the computational graph, we can apply the chain rule
as follows:
\[ 
    \bar{w}_i = \sum_{j \in succ\ i} \bar{w}_j\frac{\partial w_j}{\partial w_i}
\]
For our example:
\begin{eqnarray*}
  \bar{w}_4 &=& 1 = \frac{\partial z}{\partial z} \\
  \bar{w}_3 &=& \bar{w}_4 \cos w_3\\
  \bar{w}_2 &=& \bar{w}_3 \cdot 1 \\
  \bar{w}_1 &=& \bar{w}_2 w_0 \\
  \bar{w}_0 &=& \bar{w}_3 + \bar{w}_2 w_1
\end{eqnarray*}
If we inline all the $\bar{w}_i$ definitions and inspect the values of partial derivatives
with respect to $x$ and $y$ we obtain expected results:
$\frac{\partial z}{\partial x} = \cos (xy + x)(y + 1)$ and
$\frac{\partial z}{\partial y} = \cos (xy + x)x$.

In the implementation of the AD for \AF{E} in some context \AB{Γ}, we would like to obtain
all the partial derivatives with respect to the variables in context \AB{Γ}.  Each partial
derivative is itself an expression \AF{E} in context \AF{Γ}.  Therefore, we need to define
a data type for an environment of \AB{Γ}-many expressions in context \AB{Γ}.  We call this
environment \AF{Env} defined as follows:
\begin{code}[hide]%
\>[0]\AgdaKeyword{module}\AgdaSpace{}%
\AgdaModule{AD}\AgdaSpace{}%
\AgdaKeyword{where}\<%
\\
\>[0][@{}l@{\AgdaIndent{0}}]%
\>[2]\AgdaKeyword{open}\AgdaSpace{}%
\AgdaKeyword{import}\AgdaSpace{}%
\AgdaModule{Data.Unit}\<%
\\
\>[2]\AgdaKeyword{open}\AgdaSpace{}%
\AgdaKeyword{import}\AgdaSpace{}%
\AgdaModule{Data.Product}\AgdaSpace{}%
\AgdaSymbol{as}\AgdaSpace{}%
\AgdaModule{Prod}\<%
\\
\>[2]\AgdaKeyword{open}\AgdaSpace{}%
\AgdaModule{Array}\AgdaSpace{}%
\AgdaKeyword{hiding}\AgdaSpace{}%
\AgdaSymbol{(}\AgdaFunction{sum}\AgdaSymbol{;}\AgdaSpace{}%
\AgdaFunction{backslide}\AgdaSymbol{;}\AgdaSpace{}%
\AgdaFunction{slide}\AgdaSymbol{)}\<%
\\
\>[2]\AgdaKeyword{open}\AgdaSpace{}%
\AgdaModule{SubWk}\AgdaSpace{}%
\AgdaKeyword{using}\AgdaSpace{}%
\AgdaSymbol{(}\AgdaFunction{wk}\AgdaSymbol{;}\AgdaSpace{}%
\AgdaOperator{\AgdaFunction{↑\AgdaUnderscore{}}}\AgdaSymbol{;}\AgdaSpace{}%
\AgdaOperator{\AgdaFunction{↑↑\AgdaUnderscore{}}}\AgdaSymbol{)}\<%
\\
\>[2]\AgdaKeyword{open}\AgdaSpace{}%
\AgdaModule{Lang}\<%
\\
\\[\AgdaEmptyExtraSkip]%
\>[2]\AgdaComment{--\ Left-associated\ pairing}\<%
\\
\>[2]\AgdaKeyword{infixl}\AgdaSpace{}%
\AgdaNumber{4}\AgdaSpace{}%
\AgdaOperator{\AgdaFunction{\AgdaUnderscore{},,\AgdaUnderscore{}}}\<%
\\
\>[2]\AgdaOperator{\AgdaFunction{\AgdaUnderscore{},,\AgdaUnderscore{}}}\AgdaSpace{}%
\AgdaSymbol{:}\AgdaSpace{}%
\AgdaGeneralizable{X}\AgdaSpace{}%
\AgdaSymbol{→}\AgdaSpace{}%
\AgdaGeneralizable{Y}\AgdaSpace{}%
\AgdaSymbol{→}\AgdaSpace{}%
\AgdaGeneralizable{X}\AgdaSpace{}%
\AgdaOperator{\AgdaFunction{×}}\AgdaSpace{}%
\AgdaGeneralizable{Y}\<%
\\
\>[2]\AgdaOperator{\AgdaFunction{\AgdaUnderscore{},,\AgdaUnderscore{}}}\AgdaSpace{}%
\AgdaSymbol{=}\AgdaSpace{}%
\AgdaOperator{\AgdaFunction{Prod.\AgdaUnderscore{},′\AgdaUnderscore{}}}\<%
\end{code}
\begin{code}%
\>[2]\AgdaFunction{Env}\AgdaSpace{}%
\AgdaSymbol{:}\AgdaSpace{}%
\AgdaDatatype{Ctx}\AgdaSpace{}%
\AgdaSymbol{→}\AgdaSpace{}%
\AgdaDatatype{Ctx}\AgdaSpace{}%
\AgdaSymbol{→}\AgdaSpace{}%
\AgdaPrimitive{Set}\<%
\\
\>[2]\AgdaFunction{Env}\AgdaSpace{}%
\AgdaInductiveConstructor{ε}%
\>[18]\AgdaBound{Δ}%
\>[21]\AgdaSymbol{=}\AgdaSpace{}%
\AgdaRecord{⊤}\<%
\\
\>[2]\AgdaFunction{Env}\AgdaSpace{}%
\AgdaSymbol{(}\AgdaBound{Γ}\AgdaSpace{}%
\AgdaOperator{\AgdaInductiveConstructor{▹}}\AgdaSpace{}%
\AgdaInductiveConstructor{ar}\AgdaSpace{}%
\AgdaBound{s}\AgdaSymbol{)}%
\>[18]\AgdaBound{Δ}%
\>[21]\AgdaSymbol{=}\AgdaSpace{}%
\AgdaFunction{Env}\AgdaSpace{}%
\AgdaBound{Γ}\AgdaSpace{}%
\AgdaBound{Δ}\AgdaSpace{}%
\AgdaOperator{\AgdaFunction{×}}\AgdaSpace{}%
\AgdaDatatype{E}\AgdaSpace{}%
\AgdaBound{Δ}\AgdaSpace{}%
\AgdaSymbol{(}\AgdaInductiveConstructor{ar}\AgdaSpace{}%
\AgdaBound{s}\AgdaSymbol{)}\<%
\\
\>[2]\AgdaFunction{Env}\AgdaSpace{}%
\AgdaSymbol{(}\AgdaBound{Γ}\AgdaSpace{}%
\AgdaOperator{\AgdaInductiveConstructor{▹}}\AgdaSpace{}%
\AgdaInductiveConstructor{ix}\AgdaSpace{}%
\AgdaBound{s}\AgdaSymbol{)}%
\>[18]\AgdaBound{Δ}%
\>[21]\AgdaSymbol{=}\AgdaSpace{}%
\AgdaFunction{Env}\AgdaSpace{}%
\AgdaBound{Γ}\AgdaSpace{}%
\AgdaBound{Δ}\<%
\end{code}
Note that \AF{Env} only keeps array expressions, as (i) derivatives for indices do
not exist; and (ii) we can always make an initial environment by populating all the
elements with \AC{zero}s.  

We define several helper operations to manipulate environments: \AF{env-zero} is 
an environment where all the values are \AC{zero}s; \AF{update} modifies the 
expression at the $v$-th position by applying $f$ to it; \AF{env-map} applies the function
$f$ from \AF{E} to \AF{E} to all the elements of the environment; and \AF{env-zipWith}
applies the binary function $f$ on two environments point-wise.  The types of these
helper functions follow.  As environments are very similar to lists, the implementation
of the above functions are straight-forward.
\begin{code}%
\>[2]\AgdaFunction{env-zero}\AgdaSpace{}%
\AgdaSymbol{:}\AgdaSpace{}%
\AgdaFunction{Env}\AgdaSpace{}%
\AgdaGeneralizable{Γ}\AgdaSpace{}%
\AgdaGeneralizable{Δ}\<%
\\
\>[2]\AgdaFunction{update}\AgdaSpace{}%
\AgdaSymbol{:}\AgdaSpace{}%
\AgdaFunction{Env}\AgdaSpace{}%
\AgdaGeneralizable{Γ}\AgdaSpace{}%
\AgdaGeneralizable{Δ}\AgdaSpace{}%
\AgdaSymbol{→}\AgdaSpace{}%
\AgdaSymbol{(}\AgdaBound{v}\AgdaSpace{}%
\AgdaSymbol{:}\AgdaSpace{}%
\AgdaInductiveConstructor{ar}\AgdaSpace{}%
\AgdaGeneralizable{s}\AgdaSpace{}%
\AgdaOperator{\AgdaDatatype{∈}}\AgdaSpace{}%
\AgdaGeneralizable{Γ}\AgdaSymbol{)}\AgdaSpace{}%
\AgdaSymbol{→}\AgdaSpace{}%
\AgdaSymbol{(}\AgdaBound{f}\AgdaSpace{}%
\AgdaSymbol{:}\AgdaSpace{}%
\AgdaDatatype{E}\AgdaSpace{}%
\AgdaGeneralizable{Δ}\AgdaSpace{}%
\AgdaSymbol{(}\AgdaInductiveConstructor{ar}\AgdaSpace{}%
\AgdaGeneralizable{s}\AgdaSymbol{)}\AgdaSpace{}%
\AgdaSymbol{→}\AgdaSpace{}%
\AgdaDatatype{E}\AgdaSpace{}%
\AgdaGeneralizable{Δ}\AgdaSpace{}%
\AgdaSymbol{(}\AgdaInductiveConstructor{ar}\AgdaSpace{}%
\AgdaGeneralizable{s}\AgdaSymbol{))}\AgdaSpace{}%
\AgdaSymbol{→}\AgdaSpace{}%
\AgdaFunction{Env}\AgdaSpace{}%
\AgdaGeneralizable{Γ}\AgdaSpace{}%
\AgdaGeneralizable{Δ}\<%
\\
\>[2]\AgdaFunction{env-map}\AgdaSpace{}%
\AgdaSymbol{:}\AgdaSpace{}%
\AgdaSymbol{∀}\AgdaSpace{}%
\AgdaSymbol{\{}\AgdaBound{Γ}\AgdaSpace{}%
\AgdaBound{Δ}\AgdaSpace{}%
\AgdaBound{Ψ}\AgdaSymbol{\}}\AgdaSpace{}%
\AgdaSymbol{→}\AgdaSpace{}%
\AgdaSymbol{(}\AgdaBound{f}\AgdaSpace{}%
\AgdaSymbol{:}\AgdaSpace{}%
\AgdaSymbol{∀}\AgdaSpace{}%
\AgdaSymbol{\{}\AgdaBound{s}\AgdaSymbol{\}}\AgdaSpace{}%
\AgdaSymbol{→}\AgdaSpace{}%
\AgdaDatatype{E}\AgdaSpace{}%
\AgdaBound{Δ}\AgdaSpace{}%
\AgdaSymbol{(}\AgdaInductiveConstructor{ar}\AgdaSpace{}%
\AgdaBound{s}\AgdaSymbol{)}\AgdaSpace{}%
\AgdaSymbol{→}\AgdaSpace{}%
\AgdaDatatype{E}\AgdaSpace{}%
\AgdaBound{Ψ}\AgdaSpace{}%
\AgdaSymbol{(}\AgdaInductiveConstructor{ar}\AgdaSpace{}%
\AgdaBound{s}\AgdaSymbol{))}\AgdaSpace{}%
\AgdaSymbol{→}\AgdaSpace{}%
\AgdaFunction{Env}\AgdaSpace{}%
\AgdaBound{Γ}\AgdaSpace{}%
\AgdaBound{Δ}\AgdaSpace{}%
\AgdaSymbol{→}\AgdaSpace{}%
\AgdaFunction{Env}\AgdaSpace{}%
\AgdaBound{Γ}\AgdaSpace{}%
\AgdaBound{Ψ}\<%
\\
\>[2]\AgdaFunction{env-zipWith}%
\>[15]\AgdaSymbol{:}\AgdaSpace{}%
\AgdaSymbol{∀}\AgdaSpace{}%
\AgdaSymbol{\{}\AgdaBound{Γ}\AgdaSpace{}%
\AgdaBound{Δ}\AgdaSpace{}%
\AgdaBound{Ψ}\AgdaSpace{}%
\AgdaBound{Ξ}\AgdaSymbol{\}}\AgdaSpace{}%
\AgdaSymbol{→}\AgdaSpace{}%
\AgdaSymbol{(}\AgdaBound{f}\AgdaSpace{}%
\AgdaSymbol{:}\AgdaSpace{}%
\AgdaSymbol{∀}\AgdaSpace{}%
\AgdaSymbol{\{}\AgdaBound{s}\AgdaSymbol{\}}\AgdaSpace{}%
\AgdaSymbol{→}\AgdaSpace{}%
\AgdaDatatype{E}\AgdaSpace{}%
\AgdaBound{Δ}\AgdaSpace{}%
\AgdaSymbol{(}\AgdaInductiveConstructor{ar}\AgdaSpace{}%
\AgdaBound{s}\AgdaSymbol{)}\AgdaSpace{}%
\AgdaSymbol{→}\AgdaSpace{}%
\AgdaDatatype{E}\AgdaSpace{}%
\AgdaBound{Ψ}\AgdaSpace{}%
\AgdaSymbol{(}\AgdaInductiveConstructor{ar}\AgdaSpace{}%
\AgdaBound{s}\AgdaSymbol{)}\AgdaSpace{}%
\AgdaSymbol{→}\AgdaSpace{}%
\AgdaDatatype{E}\AgdaSpace{}%
\AgdaBound{Ξ}\AgdaSpace{}%
\AgdaSymbol{(}\AgdaInductiveConstructor{ar}\AgdaSpace{}%
\AgdaBound{s}\AgdaSymbol{))}\<%
\\
\>[15]\AgdaSymbol{→}\AgdaSpace{}%
\AgdaFunction{Env}\AgdaSpace{}%
\AgdaBound{Γ}\AgdaSpace{}%
\AgdaBound{Δ}\AgdaSpace{}%
\AgdaSymbol{→}\AgdaSpace{}%
\AgdaFunction{Env}\AgdaSpace{}%
\AgdaBound{Γ}\AgdaSpace{}%
\AgdaBound{Ψ}\AgdaSpace{}%
\AgdaSymbol{→}\AgdaSpace{}%
\AgdaFunction{Env}\AgdaSpace{}%
\AgdaBound{Γ}\AgdaSpace{}%
\AgdaBound{Ξ}\<%
\end{code}
\begin{code}[hide]%
\>[2]\AgdaFunction{update}\AgdaSpace{}%
\AgdaSymbol{\{}\AgdaBound{Γ}\AgdaSpace{}%
\AgdaOperator{\AgdaInductiveConstructor{▹}}\AgdaSpace{}%
\AgdaInductiveConstructor{ar}\AgdaSpace{}%
\AgdaBound{s}\AgdaSymbol{\}}\AgdaSpace{}%
\AgdaSymbol{(}\AgdaBound{ρ}\AgdaSpace{}%
\AgdaOperator{\AgdaInductiveConstructor{,}}\AgdaSpace{}%
\AgdaBound{e}\AgdaSymbol{)}\AgdaSpace{}%
\AgdaInductiveConstructor{v₀}\AgdaSpace{}%
\AgdaBound{f}\AgdaSpace{}%
\AgdaSymbol{=}\AgdaSpace{}%
\AgdaBound{ρ}\AgdaSpace{}%
\AgdaOperator{\AgdaInductiveConstructor{,}}\AgdaSpace{}%
\AgdaBound{f}\AgdaSpace{}%
\AgdaBound{e}\<%
\\
\>[2]\AgdaFunction{update}\AgdaSpace{}%
\AgdaSymbol{\{}\AgdaBound{Γ}\AgdaSpace{}%
\AgdaOperator{\AgdaInductiveConstructor{▹}}\AgdaSpace{}%
\AgdaInductiveConstructor{ix}\AgdaSpace{}%
\AgdaBound{s}\AgdaSymbol{\}}\AgdaSpace{}%
\AgdaBound{ρ}\AgdaSpace{}%
\AgdaSymbol{(}\AgdaInductiveConstructor{vₛ}\AgdaSpace{}%
\AgdaBound{x}\AgdaSymbol{)}\AgdaSpace{}%
\AgdaBound{f}\AgdaSpace{}%
\AgdaSymbol{=}\AgdaSpace{}%
\AgdaFunction{update}\AgdaSpace{}%
\AgdaBound{ρ}\AgdaSpace{}%
\AgdaBound{x}\AgdaSpace{}%
\AgdaBound{f}\<%
\\
\>[2]\AgdaFunction{update}\AgdaSpace{}%
\AgdaSymbol{\{}\AgdaBound{Γ}\AgdaSpace{}%
\AgdaOperator{\AgdaInductiveConstructor{▹}}\AgdaSpace{}%
\AgdaInductiveConstructor{ar}\AgdaSpace{}%
\AgdaBound{s}\AgdaSymbol{\}}\AgdaSpace{}%
\AgdaSymbol{(}\AgdaBound{ρ}\AgdaSpace{}%
\AgdaOperator{\AgdaInductiveConstructor{,}}\AgdaSpace{}%
\AgdaBound{e}\AgdaSymbol{)}\AgdaSpace{}%
\AgdaSymbol{(}\AgdaInductiveConstructor{vₛ}\AgdaSpace{}%
\AgdaBound{x}\AgdaSymbol{)}\AgdaSpace{}%
\AgdaBound{f}\AgdaSpace{}%
\AgdaSymbol{=}\AgdaSpace{}%
\AgdaFunction{update}\AgdaSpace{}%
\AgdaBound{ρ}\AgdaSpace{}%
\AgdaBound{x}\AgdaSpace{}%
\AgdaBound{f}\AgdaSpace{}%
\AgdaOperator{\AgdaInductiveConstructor{,}}\AgdaSpace{}%
\AgdaBound{e}\<%
\\
\\[\AgdaEmptyExtraSkip]%
\>[2]\AgdaFunction{env-ix}\AgdaSpace{}%
\AgdaSymbol{:}\AgdaSpace{}%
\AgdaFunction{Env}\AgdaSpace{}%
\AgdaGeneralizable{Γ}\AgdaSpace{}%
\AgdaGeneralizable{Δ}\AgdaSpace{}%
\AgdaSymbol{→}\AgdaSpace{}%
\AgdaSymbol{(}\AgdaBound{ix}\AgdaSpace{}%
\AgdaSymbol{:}\AgdaSpace{}%
\AgdaSymbol{(}\AgdaInductiveConstructor{ar}\AgdaSpace{}%
\AgdaGeneralizable{s}\AgdaSymbol{)}\AgdaSpace{}%
\AgdaOperator{\AgdaDatatype{∈}}\AgdaSpace{}%
\AgdaGeneralizable{Γ}\AgdaSymbol{)}\AgdaSpace{}%
\AgdaSymbol{→}\AgdaSpace{}%
\AgdaDatatype{E}\AgdaSpace{}%
\AgdaGeneralizable{Δ}\AgdaSpace{}%
\AgdaSymbol{(}\AgdaInductiveConstructor{ar}\AgdaSpace{}%
\AgdaGeneralizable{s}\AgdaSymbol{)}\<%
\\
\>[2]\AgdaFunction{env-ix}\AgdaSpace{}%
\AgdaSymbol{\{}\AgdaBound{Γ}\AgdaSpace{}%
\AgdaOperator{\AgdaInductiveConstructor{▹}}\AgdaSpace{}%
\AgdaInductiveConstructor{ix}\AgdaSpace{}%
\AgdaBound{s}\AgdaSymbol{\}}\AgdaSpace{}%
\AgdaBound{ρ}\AgdaSpace{}%
\AgdaSymbol{(}\AgdaInductiveConstructor{vₛ}\AgdaSpace{}%
\AgdaBound{x}\AgdaSymbol{)}\AgdaSpace{}%
\AgdaSymbol{=}\AgdaSpace{}%
\AgdaFunction{env-ix}\AgdaSpace{}%
\AgdaBound{ρ}\AgdaSpace{}%
\AgdaBound{x}\<%
\\
\>[2]\AgdaFunction{env-ix}\AgdaSpace{}%
\AgdaSymbol{\{}\AgdaBound{Γ}\AgdaSpace{}%
\AgdaOperator{\AgdaInductiveConstructor{▹}}\AgdaSpace{}%
\AgdaInductiveConstructor{ar}\AgdaSpace{}%
\AgdaBound{s}\AgdaSymbol{\}}\AgdaSpace{}%
\AgdaSymbol{(}\AgdaBound{ρ}\AgdaSpace{}%
\AgdaOperator{\AgdaInductiveConstructor{,}}\AgdaSpace{}%
\AgdaBound{e}\AgdaSymbol{)}\AgdaSpace{}%
\AgdaInductiveConstructor{v₀}\AgdaSpace{}%
\AgdaSymbol{=}\AgdaSpace{}%
\AgdaBound{e}\<%
\\
\>[2]\AgdaFunction{env-ix}\AgdaSpace{}%
\AgdaSymbol{\{}\AgdaBound{Γ}\AgdaSpace{}%
\AgdaOperator{\AgdaInductiveConstructor{▹}}\AgdaSpace{}%
\AgdaInductiveConstructor{ar}\AgdaSpace{}%
\AgdaBound{s}\AgdaSymbol{\}}\AgdaSpace{}%
\AgdaSymbol{(}\AgdaBound{ρ}\AgdaSpace{}%
\AgdaOperator{\AgdaInductiveConstructor{,}}\AgdaSpace{}%
\AgdaBound{e}\AgdaSymbol{)}\AgdaSpace{}%
\AgdaSymbol{(}\AgdaInductiveConstructor{vₛ}\AgdaSpace{}%
\AgdaBound{x}\AgdaSymbol{)}\AgdaSpace{}%
\AgdaSymbol{=}\AgdaSpace{}%
\AgdaFunction{env-ix}\AgdaSpace{}%
\AgdaBound{ρ}\AgdaSpace{}%
\AgdaBound{x}\<%
\\
\\[\AgdaEmptyExtraSkip]%
\>[2]\AgdaComment{--\ Update\ array\ values\ in\ the\ environment}\<%
\\
\>[2]\AgdaFunction{env-imap}\AgdaSpace{}%
\AgdaSymbol{:}\AgdaSpace{}%
\AgdaSymbol{(∀}\AgdaSpace{}%
\AgdaSymbol{\{}\AgdaBound{s}\AgdaSymbol{\}}\AgdaSpace{}%
\AgdaSymbol{→}\AgdaSpace{}%
\AgdaSymbol{(}\AgdaInductiveConstructor{ar}\AgdaSpace{}%
\AgdaBound{s}\AgdaSymbol{)}\AgdaSpace{}%
\AgdaOperator{\AgdaDatatype{∈}}\AgdaSpace{}%
\AgdaGeneralizable{Γ}\AgdaSpace{}%
\AgdaSymbol{→}\AgdaSpace{}%
\AgdaDatatype{E}\AgdaSpace{}%
\AgdaGeneralizable{Δ}\AgdaSpace{}%
\AgdaSymbol{(}\AgdaInductiveConstructor{ar}\AgdaSpace{}%
\AgdaBound{s}\AgdaSymbol{))}\AgdaSpace{}%
\AgdaSymbol{→}\AgdaSpace{}%
\AgdaFunction{Env}\AgdaSpace{}%
\AgdaGeneralizable{Γ}\AgdaSpace{}%
\AgdaGeneralizable{Δ}\AgdaSpace{}%
\AgdaComment{--→\ Env\ Γ\ Δ}\<%
\\
\>[2]\AgdaFunction{env-imap}\AgdaSpace{}%
\AgdaSymbol{\{}\AgdaArgument{Γ}\AgdaSpace{}%
\AgdaSymbol{=}\AgdaSpace{}%
\AgdaInductiveConstructor{ε}\AgdaSymbol{\}}%
\>[23]\AgdaBound{f}\AgdaSpace{}%
\AgdaSymbol{=}\AgdaSpace{}%
\AgdaInductiveConstructor{tt}\<%
\\
\>[2]\AgdaFunction{env-imap}\AgdaSpace{}%
\AgdaSymbol{\{}\AgdaArgument{Γ}\AgdaSpace{}%
\AgdaSymbol{=}\AgdaSpace{}%
\AgdaBound{Γ}\AgdaSpace{}%
\AgdaOperator{\AgdaInductiveConstructor{▹}}\AgdaSpace{}%
\AgdaInductiveConstructor{ar}\AgdaSpace{}%
\AgdaBound{s}\AgdaSymbol{\}}\AgdaSpace{}%
\AgdaBound{f}\AgdaSpace{}%
\AgdaSymbol{=}\AgdaSpace{}%
\AgdaFunction{env-imap}\AgdaSpace{}%
\AgdaSymbol{(}\AgdaBound{f}\AgdaSpace{}%
\AgdaOperator{\AgdaFunction{∘}}\AgdaSpace{}%
\AgdaInductiveConstructor{vₛ}\AgdaSymbol{)}\AgdaSpace{}%
\AgdaOperator{\AgdaInductiveConstructor{,}}\AgdaSpace{}%
\AgdaBound{f}\AgdaSpace{}%
\AgdaInductiveConstructor{v₀}\<%
\\
\>[2]\AgdaFunction{env-imap}\AgdaSpace{}%
\AgdaSymbol{\{}\AgdaArgument{Γ}\AgdaSpace{}%
\AgdaSymbol{=}\AgdaSpace{}%
\AgdaBound{Γ}\AgdaSpace{}%
\AgdaOperator{\AgdaInductiveConstructor{▹}}\AgdaSpace{}%
\AgdaInductiveConstructor{ix}\AgdaSpace{}%
\AgdaBound{s}\AgdaSymbol{\}}\AgdaSpace{}%
\AgdaBound{f}\AgdaSpace{}%
\AgdaSymbol{=}\AgdaSpace{}%
\AgdaFunction{env-imap}\AgdaSpace{}%
\AgdaSymbol{(}\AgdaBound{f}\AgdaSpace{}%
\AgdaOperator{\AgdaFunction{∘}}\AgdaSpace{}%
\AgdaInductiveConstructor{vₛ}\AgdaSymbol{)}\<%
\\
\\[\AgdaEmptyExtraSkip]%
\>[2]\AgdaFunction{env-map}\AgdaSpace{}%
\AgdaSymbol{\{}\AgdaArgument{Γ}\AgdaSpace{}%
\AgdaSymbol{=}\AgdaSpace{}%
\AgdaInductiveConstructor{ε}\AgdaSymbol{\}}\AgdaSpace{}%
\AgdaBound{f}\AgdaSpace{}%
\AgdaBound{ρ}\AgdaSpace{}%
\AgdaSymbol{=}\AgdaSpace{}%
\AgdaInductiveConstructor{tt}\<%
\\
\>[2]\AgdaFunction{env-map}\AgdaSpace{}%
\AgdaSymbol{\{}\AgdaArgument{Γ}\AgdaSpace{}%
\AgdaSymbol{=}\AgdaSpace{}%
\AgdaBound{Γ}\AgdaSpace{}%
\AgdaOperator{\AgdaInductiveConstructor{▹}}\AgdaSpace{}%
\AgdaInductiveConstructor{ix}\AgdaSpace{}%
\AgdaBound{s}\AgdaSymbol{\}}\AgdaSpace{}%
\AgdaBound{f}\AgdaSpace{}%
\AgdaBound{ρ}\AgdaSpace{}%
\AgdaSymbol{=}\AgdaSpace{}%
\AgdaFunction{env-map}\AgdaSpace{}%
\AgdaSymbol{\{}\AgdaArgument{Γ}\AgdaSpace{}%
\AgdaSymbol{=}\AgdaSpace{}%
\AgdaBound{Γ}\AgdaSymbol{\}}\AgdaSpace{}%
\AgdaBound{f}\AgdaSpace{}%
\AgdaBound{ρ}\<%
\\
\>[2]\AgdaFunction{env-map}\AgdaSpace{}%
\AgdaSymbol{\{}\AgdaArgument{Γ}\AgdaSpace{}%
\AgdaSymbol{=}\AgdaSpace{}%
\AgdaBound{Γ}\AgdaSpace{}%
\AgdaOperator{\AgdaInductiveConstructor{▹}}\AgdaSpace{}%
\AgdaInductiveConstructor{ar}\AgdaSpace{}%
\AgdaBound{s}\AgdaSymbol{\}}\AgdaSpace{}%
\AgdaBound{f}\AgdaSpace{}%
\AgdaSymbol{(}\AgdaBound{ρ}\AgdaSpace{}%
\AgdaOperator{\AgdaInductiveConstructor{,}}\AgdaSpace{}%
\AgdaBound{e}\AgdaSymbol{)}\AgdaSpace{}%
\AgdaSymbol{=}\AgdaSpace{}%
\AgdaFunction{env-map}\AgdaSpace{}%
\AgdaSymbol{\{}\AgdaArgument{Γ}\AgdaSpace{}%
\AgdaSymbol{=}\AgdaSpace{}%
\AgdaBound{Γ}\AgdaSymbol{\}}\AgdaSpace{}%
\AgdaBound{f}\AgdaSpace{}%
\AgdaBound{ρ}\AgdaSpace{}%
\AgdaOperator{\AgdaInductiveConstructor{,}}\AgdaSpace{}%
\AgdaBound{f}\AgdaSpace{}%
\AgdaBound{e}\<%
\\
\\[\AgdaEmptyExtraSkip]%
\>[2]\AgdaFunction{env-zero}\AgdaSpace{}%
\AgdaSymbol{\{}\AgdaInductiveConstructor{ε}\AgdaSymbol{\}}\AgdaSpace{}%
\AgdaSymbol{=}\AgdaSpace{}%
\AgdaSymbol{\AgdaUnderscore{}}\<%
\\
\>[2]\AgdaFunction{env-zero}\AgdaSpace{}%
\AgdaSymbol{\{}\AgdaBound{Γ}\AgdaSpace{}%
\AgdaOperator{\AgdaInductiveConstructor{▹}}\AgdaSpace{}%
\AgdaInductiveConstructor{ix}\AgdaSpace{}%
\AgdaBound{x}\AgdaSymbol{\}}\AgdaSpace{}%
\AgdaSymbol{=}\AgdaSpace{}%
\AgdaFunction{env-zero}\AgdaSpace{}%
\AgdaSymbol{\{}\AgdaBound{Γ}\AgdaSymbol{\}}\<%
\\
\>[2]\AgdaFunction{env-zero}\AgdaSpace{}%
\AgdaSymbol{\{}\AgdaBound{Γ}\AgdaSpace{}%
\AgdaOperator{\AgdaInductiveConstructor{▹}}\AgdaSpace{}%
\AgdaInductiveConstructor{ar}\AgdaSpace{}%
\AgdaBound{x}\AgdaSymbol{\}}\AgdaSpace{}%
\AgdaSymbol{=}\AgdaSpace{}%
\AgdaFunction{env-zero}\AgdaSpace{}%
\AgdaSymbol{\{}\AgdaBound{Γ}\AgdaSymbol{\}}\AgdaSpace{}%
\AgdaOperator{\AgdaInductiveConstructor{,}}\AgdaSpace{}%
\AgdaInductiveConstructor{zero}\<%
\\
\\[\AgdaEmptyExtraSkip]%
\>[2]\AgdaFunction{env-zipWith}\AgdaSpace{}%
\AgdaSymbol{\{}\AgdaInductiveConstructor{ε}\AgdaSymbol{\}}\AgdaSpace{}%
\AgdaBound{f}\AgdaSpace{}%
\AgdaBound{l}\AgdaSpace{}%
\AgdaBound{r}\AgdaSpace{}%
\AgdaSymbol{=}\AgdaSpace{}%
\AgdaSymbol{\AgdaUnderscore{}}\<%
\\
\>[2]\AgdaFunction{env-zipWith}\AgdaSpace{}%
\AgdaSymbol{\{}\AgdaBound{Γ}\AgdaSpace{}%
\AgdaOperator{\AgdaInductiveConstructor{▹}}\AgdaSpace{}%
\AgdaInductiveConstructor{ix}\AgdaSpace{}%
\AgdaBound{x}\AgdaSymbol{\}}\AgdaSpace{}%
\AgdaBound{f}\AgdaSpace{}%
\AgdaBound{l}\AgdaSpace{}%
\AgdaBound{r}\AgdaSpace{}%
\AgdaSymbol{=}\AgdaSpace{}%
\AgdaFunction{env-zipWith}\AgdaSpace{}%
\AgdaSymbol{\{}\AgdaBound{Γ}\AgdaSymbol{\}}\AgdaSpace{}%
\AgdaBound{f}\AgdaSpace{}%
\AgdaBound{l}\AgdaSpace{}%
\AgdaBound{r}\<%
\\
\>[2]\AgdaFunction{env-zipWith}\AgdaSpace{}%
\AgdaSymbol{\{}\AgdaBound{Γ}\AgdaSpace{}%
\AgdaOperator{\AgdaInductiveConstructor{▹}}\AgdaSpace{}%
\AgdaInductiveConstructor{ar}\AgdaSpace{}%
\AgdaBound{x}\AgdaSymbol{\}}\AgdaSpace{}%
\AgdaBound{f}\AgdaSpace{}%
\AgdaSymbol{(}\AgdaBound{l}\AgdaSpace{}%
\AgdaOperator{\AgdaInductiveConstructor{,}}\AgdaSpace{}%
\AgdaBound{e₁}\AgdaSymbol{)}\AgdaSpace{}%
\AgdaSymbol{(}\AgdaBound{r}\AgdaSpace{}%
\AgdaOperator{\AgdaInductiveConstructor{,}}\AgdaSpace{}%
\AgdaBound{e₂}\AgdaSymbol{)}\AgdaSpace{}%
\AgdaSymbol{=}\AgdaSpace{}%
\AgdaFunction{env-zipWith}\AgdaSpace{}%
\AgdaSymbol{\{}\AgdaBound{Γ}\AgdaSymbol{\}}\AgdaSpace{}%
\AgdaBound{f}\AgdaSpace{}%
\AgdaBound{l}\AgdaSpace{}%
\AgdaBound{r}\AgdaSpace{}%
\AgdaOperator{\AgdaInductiveConstructor{,}}\AgdaSpace{}%
\AgdaBound{f}\AgdaSpace{}%
\AgdaBound{e₁}\AgdaSpace{}%
\AgdaBound{e₂}\<%
\end{code}

We define the function \AF{∇} that takes an expression \AF{E} and the seed
which is the multiplier on the left of the chain, and we compute a function
from that updates the environment.
\begin{code}%
\>[2]\AgdaFunction{∇}\AgdaSpace{}%
\AgdaSymbol{:}\AgdaSpace{}%
\AgdaDatatype{E}\AgdaSpace{}%
\AgdaGeneralizable{Δ}\AgdaSpace{}%
\AgdaGeneralizable{is}\AgdaSpace{}%
\AgdaSymbol{→}\AgdaSpace{}%
\AgdaSymbol{(}\AgdaBound{seed}\AgdaSpace{}%
\AgdaSymbol{:}\AgdaSpace{}%
\AgdaDatatype{E}\AgdaSpace{}%
\AgdaGeneralizable{Δ}\AgdaSpace{}%
\AgdaGeneralizable{is}\AgdaSymbol{)}\AgdaSpace{}%
\AgdaSymbol{→}\AgdaSpace{}%
\AgdaFunction{Env}\AgdaSpace{}%
\AgdaGeneralizable{Δ}\AgdaSpace{}%
\AgdaGeneralizable{Δ}\AgdaSpace{}%
\AgdaSymbol{→}\AgdaSpace{}%
\AgdaFunction{Env}\AgdaSpace{}%
\AgdaGeneralizable{Δ}\AgdaSpace{}%
\AgdaGeneralizable{Δ}\<%
\\
\\[\AgdaEmptyExtraSkip]%
\>[2]\AgdaFunction{map-sum}\AgdaSpace{}%
\AgdaSymbol{:}\AgdaSpace{}%
\AgdaSymbol{(}\AgdaBound{e}\AgdaSpace{}%
\AgdaBound{s}\AgdaSpace{}%
\AgdaSymbol{:}\AgdaSpace{}%
\AgdaDatatype{E}\AgdaSpace{}%
\AgdaSymbol{(}\AgdaGeneralizable{Δ}\AgdaSpace{}%
\AgdaOperator{\AgdaInductiveConstructor{▹}}\AgdaSpace{}%
\AgdaInductiveConstructor{ix}\AgdaSpace{}%
\AgdaGeneralizable{s}\AgdaSymbol{)}\AgdaSpace{}%
\AgdaGeneralizable{ip}\AgdaSymbol{)}\AgdaSpace{}%
\AgdaSymbol{→}\AgdaSpace{}%
\AgdaFunction{Env}\AgdaSpace{}%
\AgdaGeneralizable{Δ}\AgdaSpace{}%
\AgdaGeneralizable{Δ}\AgdaSpace{}%
\AgdaSymbol{→}\AgdaSpace{}%
\AgdaFunction{Env}\AgdaSpace{}%
\AgdaGeneralizable{Δ}\AgdaSpace{}%
\AgdaGeneralizable{Δ}\<%
\\
\>[2]\AgdaFunction{map-sum}\AgdaSpace{}%
\AgdaSymbol{\{}\AgdaBound{Δ}\AgdaSymbol{\}}\AgdaSpace{}%
\AgdaBound{e}\AgdaSpace{}%
\AgdaBound{s}\AgdaSpace{}%
\AgdaBound{δ}\AgdaSpace{}%
\AgdaSymbol{=}\AgdaSpace{}%
\AgdaFunction{env-zipWith}\AgdaSpace{}%
\AgdaSymbol{\{}\AgdaBound{Δ}\AgdaSymbol{\}}\AgdaSpace{}%
\AgdaOperator{\AgdaInductiveConstructor{\AgdaUnderscore{}⊞\AgdaUnderscore{}}}\AgdaSpace{}%
\AgdaSymbol{(}\AgdaFunction{env-map}\AgdaSpace{}%
\AgdaSymbol{\{}\AgdaBound{Δ}\AgdaSymbol{\}}\AgdaSpace{}%
\AgdaInductiveConstructor{sum}\AgdaSpace{}%
\AgdaSymbol{(}\AgdaFunction{∇}\AgdaSpace{}%
\AgdaBound{e}\AgdaSpace{}%
\AgdaBound{s}\AgdaSpace{}%
\AgdaSymbol{(}\AgdaFunction{env-zero}\AgdaSpace{}%
\AgdaSymbol{\{}\AgdaBound{Δ}\AgdaSymbol{\})))}\AgdaSpace{}%
\AgdaBound{δ}\<%
\\
\\[\AgdaEmptyExtraSkip]%
\>[2]\AgdaFunction{∇}\AgdaSpace{}%
\AgdaSymbol{(}\AgdaInductiveConstructor{zero}\AgdaSymbol{)}%
\>[27]\AgdaBound{s}\AgdaSpace{}%
\AgdaBound{δ}\AgdaSpace{}%
\AgdaSymbol{=}\AgdaSpace{}%
\AgdaBound{δ}\<%
\\
\>[2]\AgdaFunction{∇}\AgdaSpace{}%
\AgdaSymbol{(}\AgdaInductiveConstructor{one}\AgdaSymbol{)}%
\>[27]\AgdaBound{s}\AgdaSpace{}%
\AgdaBound{δ}\AgdaSpace{}%
\AgdaSymbol{=}\AgdaSpace{}%
\AgdaBound{δ}\<%
\\
\>[2]\AgdaFunction{∇}\AgdaSpace{}%
\AgdaSymbol{(}\AgdaInductiveConstructor{var}\AgdaSpace{}%
\AgdaSymbol{\{}\AgdaInductiveConstructor{ix}\AgdaSpace{}%
\AgdaSymbol{\AgdaUnderscore{}\}}\AgdaSpace{}%
\AgdaBound{x}\AgdaSymbol{)}%
\>[27]\AgdaBound{s}\AgdaSpace{}%
\AgdaBound{δ}\AgdaSpace{}%
\AgdaSymbol{=}\AgdaSpace{}%
\AgdaBound{δ}\<%
\\
\>[2]\AgdaFunction{∇}\AgdaSpace{}%
\AgdaSymbol{(}\AgdaInductiveConstructor{var}\AgdaSpace{}%
\AgdaSymbol{\{}\AgdaInductiveConstructor{ar}\AgdaSpace{}%
\AgdaSymbol{\AgdaUnderscore{}\}}\AgdaSpace{}%
\AgdaBound{x}\AgdaSymbol{)}%
\>[27]\AgdaBound{s}\AgdaSpace{}%
\AgdaBound{δ}\AgdaSpace{}%
\AgdaSymbol{=}\AgdaSpace{}%
\AgdaFunction{update}\AgdaSpace{}%
\AgdaBound{δ}\AgdaSpace{}%
\AgdaBound{x}\AgdaSpace{}%
\AgdaSymbol{(}\AgdaOperator{\AgdaInductiveConstructor{\AgdaUnderscore{}⊞}}\AgdaSpace{}%
\AgdaBound{s}\AgdaSymbol{)}\<%
\\
\\[\AgdaEmptyExtraSkip]%
\>[2]\AgdaFunction{∇}\AgdaSpace{}%
\AgdaSymbol{(}\AgdaInductiveConstructor{imapₛ}\AgdaSpace{}%
\AgdaBound{e}\AgdaSymbol{)}%
\>[27]\AgdaBound{s}%
\>[31]\AgdaSymbol{=}\AgdaSpace{}%
\AgdaFunction{map-sum}\AgdaSpace{}%
\AgdaBound{e}\AgdaSpace{}%
\AgdaSymbol{(}\AgdaInductiveConstructor{selₛ}%
\>[52]\AgdaSymbol{(}\AgdaOperator{\AgdaFunction{↑}}\AgdaSpace{}%
\AgdaBound{s}\AgdaSymbol{)}\AgdaSpace{}%
\AgdaSymbol{(}\AgdaInductiveConstructor{var}\AgdaSpace{}%
\AgdaInductiveConstructor{v₀}\AgdaSymbol{))}\<%
\\
\>[2]\AgdaFunction{∇}\AgdaSpace{}%
\AgdaSymbol{(}\AgdaInductiveConstructor{imap}\AgdaSpace{}%
\AgdaBound{e}\AgdaSymbol{)}%
\>[27]\AgdaBound{s}%
\>[31]\AgdaSymbol{=}\AgdaSpace{}%
\AgdaFunction{map-sum}\AgdaSpace{}%
\AgdaBound{e}\AgdaSpace{}%
\AgdaSymbol{(}\AgdaInductiveConstructor{sel}%
\>[52]\AgdaSymbol{(}\AgdaOperator{\AgdaFunction{↑}}\AgdaSpace{}%
\AgdaBound{s}\AgdaSymbol{)}\AgdaSpace{}%
\AgdaSymbol{(}\AgdaInductiveConstructor{var}\AgdaSpace{}%
\AgdaInductiveConstructor{v₀}\AgdaSymbol{))}\<%
\\
\>[2]\AgdaFunction{∇}\AgdaSpace{}%
\AgdaSymbol{(}\AgdaInductiveConstructor{imapb}\AgdaSpace{}%
\AgdaBound{m}\AgdaSpace{}%
\AgdaBound{e}\AgdaSymbol{)}%
\>[27]\AgdaBound{s}%
\>[31]\AgdaSymbol{=}\AgdaSpace{}%
\AgdaFunction{map-sum}\AgdaSpace{}%
\AgdaBound{e}\AgdaSpace{}%
\AgdaSymbol{(}\AgdaInductiveConstructor{selb}\AgdaSpace{}%
\AgdaBound{m}%
\>[52]\AgdaSymbol{(}\AgdaOperator{\AgdaFunction{↑}}\AgdaSpace{}%
\AgdaBound{s}\AgdaSymbol{)}\AgdaSpace{}%
\AgdaSymbol{(}\AgdaInductiveConstructor{var}\AgdaSpace{}%
\AgdaInductiveConstructor{v₀}\AgdaSymbol{))}\<%
\\
\\[\AgdaEmptyExtraSkip]%
\>[2]\AgdaFunction{∇}\AgdaSpace{}%
\AgdaSymbol{(}\AgdaInductiveConstructor{selₛ}\AgdaSpace{}%
\AgdaBound{e}\AgdaSpace{}%
\AgdaBound{i}\AgdaSymbol{)}%
\>[27]\AgdaBound{s}%
\>[31]\AgdaSymbol{=}\AgdaSpace{}%
\AgdaFunction{∇}\AgdaSpace{}%
\AgdaBound{e}\AgdaSpace{}%
\AgdaSymbol{(}\AgdaInductiveConstructor{imapₛ}%
\>[47]\AgdaSymbol{(}\AgdaInductiveConstructor{zero-but}\AgdaSpace{}%
\AgdaSymbol{(}\AgdaInductiveConstructor{var}\AgdaSpace{}%
\AgdaInductiveConstructor{v₀}\AgdaSymbol{)}\AgdaSpace{}%
\AgdaSymbol{(}\AgdaOperator{\AgdaFunction{↑}}\AgdaSpace{}%
\AgdaBound{i}\AgdaSymbol{)}\AgdaSpace{}%
\AgdaSymbol{(}\AgdaOperator{\AgdaFunction{↑}}\AgdaSpace{}%
\AgdaBound{s}\AgdaSymbol{)))}\<%
\\
\>[2]\AgdaFunction{∇}\AgdaSpace{}%
\AgdaSymbol{(}\AgdaInductiveConstructor{sel}\AgdaSpace{}%
\AgdaBound{e}\AgdaSpace{}%
\AgdaBound{i}\AgdaSymbol{)}%
\>[27]\AgdaBound{s}%
\>[31]\AgdaSymbol{=}\AgdaSpace{}%
\AgdaFunction{∇}\AgdaSpace{}%
\AgdaBound{e}\AgdaSpace{}%
\AgdaSymbol{(}\AgdaInductiveConstructor{imap}%
\>[47]\AgdaSymbol{(}\AgdaInductiveConstructor{zero-but}\AgdaSpace{}%
\AgdaSymbol{(}\AgdaInductiveConstructor{var}\AgdaSpace{}%
\AgdaInductiveConstructor{v₀}\AgdaSymbol{)}\AgdaSpace{}%
\AgdaSymbol{(}\AgdaOperator{\AgdaFunction{↑}}\AgdaSpace{}%
\AgdaBound{i}\AgdaSymbol{)}\AgdaSpace{}%
\AgdaSymbol{(}\AgdaOperator{\AgdaFunction{↑}}\AgdaSpace{}%
\AgdaBound{s}\AgdaSymbol{)))}\<%
\\
\>[2]\AgdaFunction{∇}\AgdaSpace{}%
\AgdaSymbol{(}\AgdaInductiveConstructor{selb}\AgdaSpace{}%
\AgdaBound{m}\AgdaSpace{}%
\AgdaBound{e}\AgdaSpace{}%
\AgdaBound{i}\AgdaSymbol{)}%
\>[27]\AgdaBound{s}%
\>[31]\AgdaSymbol{=}\AgdaSpace{}%
\AgdaFunction{∇}\AgdaSpace{}%
\AgdaBound{e}\AgdaSpace{}%
\AgdaSymbol{(}\AgdaInductiveConstructor{imapb}\AgdaSpace{}%
\AgdaBound{m}%
\>[47]\AgdaSymbol{(}\AgdaInductiveConstructor{zero-but}\AgdaSpace{}%
\AgdaSymbol{(}\AgdaInductiveConstructor{var}\AgdaSpace{}%
\AgdaInductiveConstructor{v₀}\AgdaSymbol{)}\AgdaSpace{}%
\AgdaSymbol{(}\AgdaOperator{\AgdaFunction{↑}}\AgdaSpace{}%
\AgdaBound{i}\AgdaSymbol{)}\AgdaSpace{}%
\AgdaSymbol{(}\AgdaOperator{\AgdaFunction{↑}}\AgdaSpace{}%
\AgdaBound{s}\AgdaSymbol{)))}\<%
\\
\\[\AgdaEmptyExtraSkip]%
\>[2]\AgdaFunction{∇}\AgdaSpace{}%
\AgdaSymbol{(}\AgdaInductiveConstructor{zero-but}\AgdaSpace{}%
\AgdaBound{i}\AgdaSpace{}%
\AgdaBound{j}\AgdaSpace{}%
\AgdaBound{e}\AgdaSymbol{)}%
\>[27]\AgdaBound{s}%
\>[31]\AgdaSymbol{=}\AgdaSpace{}%
\AgdaFunction{∇}\AgdaSpace{}%
\AgdaBound{e}\AgdaSpace{}%
\AgdaSymbol{(}\AgdaInductiveConstructor{zero-but}\AgdaSpace{}%
\AgdaBound{i}\AgdaSpace{}%
\AgdaBound{j}\AgdaSpace{}%
\AgdaBound{s}\AgdaSymbol{)}\<%
\\
\>[2]\AgdaFunction{∇}\AgdaSpace{}%
\AgdaSymbol{(}\AgdaInductiveConstructor{sum}\AgdaSpace{}%
\AgdaBound{e}\AgdaSymbol{)}%
\>[27]\AgdaBound{s}%
\>[31]\AgdaSymbol{=}\AgdaSpace{}%
\AgdaFunction{map-sum}\AgdaSpace{}%
\AgdaBound{e}\AgdaSpace{}%
\AgdaSymbol{(}\AgdaOperator{\AgdaFunction{↑}}\AgdaSpace{}%
\AgdaBound{s}\AgdaSymbol{)}\<%
\\
\\[\AgdaEmptyExtraSkip]%
\>[2]\AgdaFunction{∇}\AgdaSpace{}%
\AgdaSymbol{(}\AgdaBound{e}\AgdaSpace{}%
\AgdaOperator{\AgdaInductiveConstructor{⊞}}\AgdaSpace{}%
\AgdaBound{e₁}\AgdaSymbol{)}%
\>[27]\AgdaBound{s}%
\>[31]\AgdaSymbol{=}\AgdaSpace{}%
\AgdaFunction{∇}\AgdaSpace{}%
\AgdaBound{e}\AgdaSpace{}%
\AgdaBound{s}\AgdaSpace{}%
\AgdaOperator{\AgdaFunction{∘}}\AgdaSpace{}%
\AgdaFunction{∇}\AgdaSpace{}%
\AgdaBound{e₁}\AgdaSpace{}%
\AgdaBound{s}\<%
\\
\>[2]\AgdaFunction{∇}\AgdaSpace{}%
\AgdaSymbol{(}\AgdaBound{e}\AgdaSpace{}%
\AgdaOperator{\AgdaInductiveConstructor{⊠}}\AgdaSpace{}%
\AgdaBound{e₁}\AgdaSymbol{)}%
\>[27]\AgdaBound{s}%
\>[31]\AgdaSymbol{=}\AgdaSpace{}%
\AgdaFunction{∇}\AgdaSpace{}%
\AgdaBound{e}\AgdaSpace{}%
\AgdaSymbol{(}\AgdaBound{s}\AgdaSpace{}%
\AgdaOperator{\AgdaInductiveConstructor{⊠}}\AgdaSpace{}%
\AgdaBound{e₁}\AgdaSymbol{)}\AgdaSpace{}%
\AgdaOperator{\AgdaFunction{∘}}\AgdaSpace{}%
\AgdaFunction{∇}\AgdaSpace{}%
\AgdaBound{e₁}\AgdaSpace{}%
\AgdaSymbol{(}\AgdaBound{s}\AgdaSpace{}%
\AgdaOperator{\AgdaInductiveConstructor{⊠}}\AgdaSpace{}%
\AgdaBound{e}\AgdaSymbol{)}\<%
\\
\>[2]\AgdaFunction{∇}\AgdaSpace{}%
\AgdaSymbol{(}\AgdaInductiveConstructor{slide}\AgdaSpace{}%
\AgdaBound{i}\AgdaSpace{}%
\AgdaBound{pl}\AgdaSpace{}%
\AgdaBound{e}\AgdaSpace{}%
\AgdaBound{su}\AgdaSymbol{)}%
\>[27]\AgdaBound{s}%
\>[31]\AgdaSymbol{=}\AgdaSpace{}%
\AgdaFunction{∇}\AgdaSpace{}%
\AgdaBound{e}\AgdaSpace{}%
\AgdaSymbol{(}\AgdaInductiveConstructor{backslide}\AgdaSpace{}%
\AgdaBound{i}\AgdaSpace{}%
\AgdaBound{s}\AgdaSpace{}%
\AgdaBound{su}\AgdaSpace{}%
\AgdaBound{pl}\AgdaSymbol{)}\<%
\\
\>[2]\AgdaFunction{∇}\AgdaSpace{}%
\AgdaSymbol{(}\AgdaInductiveConstructor{backslide}\AgdaSpace{}%
\AgdaBound{i}\AgdaSpace{}%
\AgdaBound{e}\AgdaSpace{}%
\AgdaBound{su}\AgdaSpace{}%
\AgdaBound{pl}\AgdaSymbol{)}%
\>[27]\AgdaBound{s}%
\>[31]\AgdaSymbol{=}\AgdaSpace{}%
\AgdaFunction{∇}\AgdaSpace{}%
\AgdaBound{e}\AgdaSpace{}%
\AgdaSymbol{(}\AgdaInductiveConstructor{slide}\AgdaSpace{}%
\AgdaBound{i}\AgdaSpace{}%
\AgdaBound{pl}\AgdaSpace{}%
\AgdaBound{s}\AgdaSpace{}%
\AgdaBound{su}\AgdaSymbol{)}\<%
\\
\\[\AgdaEmptyExtraSkip]%
\>[2]\AgdaFunction{∇}\AgdaSpace{}%
\AgdaSymbol{(}\AgdaInductiveConstructor{scaledown}\AgdaSpace{}%
\AgdaBound{x}\AgdaSpace{}%
\AgdaBound{e}\AgdaSymbol{)}%
\>[27]\AgdaBound{s}%
\>[31]\AgdaSymbol{=}\AgdaSpace{}%
\AgdaFunction{∇}\AgdaSpace{}%
\AgdaBound{e}\AgdaSpace{}%
\AgdaSymbol{(}\AgdaInductiveConstructor{scaledown}\AgdaSpace{}%
\AgdaBound{x}\AgdaSpace{}%
\AgdaBound{s}\AgdaSymbol{)}\<%
\\
\>[2]\AgdaFunction{∇}\AgdaSpace{}%
\AgdaSymbol{(}\AgdaInductiveConstructor{minus}\AgdaSpace{}%
\AgdaBound{e}\AgdaSymbol{)}%
\>[27]\AgdaBound{s}%
\>[31]\AgdaSymbol{=}\AgdaSpace{}%
\AgdaFunction{∇}\AgdaSpace{}%
\AgdaBound{e}\AgdaSpace{}%
\AgdaSymbol{(}\AgdaInductiveConstructor{minus}\AgdaSpace{}%
\AgdaBound{s}\AgdaSymbol{)}\<%
\\
\>[2]\AgdaFunction{∇}\AgdaSpace{}%
\AgdaSymbol{(}\AgdaInductiveConstructor{logistic}\AgdaSpace{}%
\AgdaBound{e}\AgdaSymbol{)}%
\>[27]\AgdaBound{s}%
\>[31]\AgdaSymbol{=}\AgdaSpace{}%
\AgdaFunction{∇}\AgdaSpace{}%
\AgdaBound{e}\AgdaSpace{}%
\AgdaSymbol{(}\AgdaBound{s}\AgdaSpace{}%
\AgdaOperator{\AgdaInductiveConstructor{⊠}}\AgdaSpace{}%
\AgdaInductiveConstructor{logistic}\AgdaSpace{}%
\AgdaBound{e}\AgdaSpace{}%
\AgdaOperator{\AgdaInductiveConstructor{⊠}}\AgdaSpace{}%
\AgdaSymbol{(}\AgdaInductiveConstructor{one}\AgdaSpace{}%
\AgdaOperator{\AgdaInductiveConstructor{⊞}}\AgdaSpace{}%
\AgdaInductiveConstructor{minus}\AgdaSpace{}%
\AgdaSymbol{(}\AgdaInductiveConstructor{logistic}\AgdaSpace{}%
\AgdaBound{e}\AgdaSymbol{)))}\<%
\end{code}
Let us now walk through the cases.  Derivative of constants (\AC{zero} and \AC{one})
is zero, so nothing needs to be updated in the environment.  Index variables are
not stored in the environment, so no updates are needed either.  If we differentiate
the variable $x$ with some seed \AB{s}, we update the $x$-th position in the environment
by adding \AB{s} to it.  Differentiation of \AC{imap}s proceeds as follows: we
recursively apply \AF{∇} to $e$ (in the context \AB{Γ} \AC{▹} (\AC{ix} \AB{p}))
with the element of the original seed \AB{s} selected at the top variable.  This
gives us the environment in the extended context, then we map \AC{sum} to every
element of the environment to accumulate the derivatives at every index.
When differentiating selections we recurse on the array we are
selecting from with the seed that is zero everywhere except the index we were
selecting at.  Differentiating
conditional is straight-forward, as $i$ and $j$ must be in the context, we can
simply differentiate $e$ with the condition on seed.  If indices were equal, we will
compute the update, otherwise we will differentiate with seed \AC{zero} which
has no effect.  As we are operating in a total language, there is no need to worry
about pulling the expression out of conditional.  The argument of \AC{sum}
lives is in the extended context, so we apply the same rules as for the \AC{imap} family,
except we propagate the original seed to all the summands.  Addition and multiplication
rules are straight-forward application of differentiation rules.  The \AC{slide}/\AC{backslide}
pair forms a satisfying \AF{∇}-symmetry.  Finally, \AC{scaledown}, \AC{minus} and
\AC{logistic} follow the rules of differentiation.

\subsection{Optimisation\label{sec:opt}}

Our algorithm often delivers expressions that are not computationally efficient.
While we can hope for the backend to take care of this, it is relatively
easy to implement a number of rewriting rules prior to extraction.  
We constructed \AF{E} such that no computation is happening in the shape
or context positions.  As a result, dependent pattern-matching is always
applicable on \AF{E} expressions, and our optimisations can be formulated
very concisely.  We omit constant-folding like rewrites such as addition
with zero and multiplication by one and focus on less trivial cases that have
to do with selections a+nd sum.  Consider the snippet of the optimiser for
\AF{selₛ} and \AF{sum}.

\begin{code}[hide]%
\>[0]\AgdaKeyword{module}\AgdaSpace{}%
\AgdaModule{Opt}\AgdaSpace{}%
\AgdaKeyword{where}\<%
\\
\>[0][@{}l@{\AgdaIndent{0}}]%
\>[2]\AgdaKeyword{open}\AgdaSpace{}%
\AgdaKeyword{import}\AgdaSpace{}%
\AgdaModule{Data.Nat}\AgdaSpace{}%
\AgdaSymbol{as}\AgdaSpace{}%
\AgdaModule{ℕ}\AgdaSpace{}%
\AgdaKeyword{using}\AgdaSpace{}%
\AgdaSymbol{(}\AgdaDatatype{ℕ}\AgdaSymbol{;}\AgdaSpace{}%
\AgdaInductiveConstructor{zero}\AgdaSymbol{;}\AgdaSpace{}%
\AgdaInductiveConstructor{suc}\AgdaSymbol{)}\<%
\\
\>[2]\AgdaKeyword{open}\AgdaSpace{}%
\AgdaModule{Lang}\<%
\\
\>[2]\AgdaKeyword{open}\AgdaSpace{}%
\AgdaModule{SubWk}\<%
\\
\>[2]\AgdaComment{--open\ Eval\ using\ (sub;\ ctx-swap;\ ↑\AgdaUnderscore{};\ ↑↑\AgdaUnderscore{};\ eq?)}\<%
\\
\>[2]\AgdaKeyword{open}\AgdaSpace{}%
\AgdaModule{Array}\AgdaSpace{}%
\AgdaKeyword{hiding}\AgdaSpace{}%
\AgdaSymbol{(}\AgdaFunction{sum}\AgdaSymbol{;}\AgdaSpace{}%
\AgdaFunction{slide}\AgdaSymbol{;}\AgdaFunction{backslide}\AgdaSymbol{)}\<%
\\
\>[2]\AgdaKeyword{open}\AgdaSpace{}%
\AgdaModule{BB}\<%
\\
\>[2]\AgdaKeyword{open}\AgdaSpace{}%
\AgdaModule{AD}\<%
\end{code}
\begin{code}%
\>[2]\AgdaFunction{opt}\AgdaSpace{}%
\AgdaSymbol{:}\AgdaSpace{}%
\AgdaDatatype{E}\AgdaSpace{}%
\AgdaGeneralizable{Γ}\AgdaSpace{}%
\AgdaGeneralizable{is}\AgdaSpace{}%
\AgdaSymbol{→}\AgdaSpace{}%
\AgdaDatatype{E}\AgdaSpace{}%
\AgdaGeneralizable{Γ}\AgdaSpace{}%
\AgdaGeneralizable{is}\<%
\\
\>[2]\AgdaFunction{opt}\AgdaSpace{}%
\AgdaSymbol{(}\AgdaInductiveConstructor{selₛ}\AgdaSpace{}%
\AgdaBound{e}\AgdaSpace{}%
\AgdaBound{e₁}\AgdaSymbol{)}\AgdaSpace{}%
\AgdaKeyword{with}\AgdaSpace{}%
\AgdaFunction{opt}\AgdaSpace{}%
\AgdaBound{e}\AgdaSpace{}%
\AgdaSymbol{|}\AgdaSpace{}%
\AgdaFunction{opt}\AgdaSpace{}%
\AgdaBound{e₁}\<%
\\
\>[2]\AgdaSymbol{...}\AgdaSpace{}%
\AgdaSymbol{|}\AgdaSpace{}%
\AgdaInductiveConstructor{zero}%
\>[24]\AgdaSymbol{|}\AgdaSpace{}%
\AgdaBound{i}\AgdaSpace{}%
\AgdaSymbol{=}\AgdaSpace{}%
\AgdaInductiveConstructor{zero}\<%
\\
\>[2]\AgdaSymbol{...}\AgdaSpace{}%
\AgdaSymbol{|}\AgdaSpace{}%
\AgdaInductiveConstructor{one}%
\>[24]\AgdaSymbol{|}\AgdaSpace{}%
\AgdaBound{i}\AgdaSpace{}%
\AgdaSymbol{=}\AgdaSpace{}%
\AgdaInductiveConstructor{one}\<%
\\
\>[2]\AgdaSymbol{...}\AgdaSpace{}%
\AgdaSymbol{|}\AgdaSpace{}%
\AgdaInductiveConstructor{imapₛ}\AgdaSpace{}%
\AgdaBound{e}%
\>[24]\AgdaSymbol{|}\AgdaSpace{}%
\AgdaBound{i}\AgdaSpace{}%
\AgdaSymbol{=}\AgdaSpace{}%
\AgdaFunction{sub}\AgdaSpace{}%
\AgdaInductiveConstructor{v₀}\AgdaSpace{}%
\AgdaBound{e}\AgdaSpace{}%
\AgdaBound{i}\<%
\\
\>[2]\AgdaSymbol{...}\AgdaSpace{}%
\AgdaSymbol{|}\AgdaSpace{}%
\AgdaInductiveConstructor{bin}\AgdaSpace{}%
\AgdaBound{op}\AgdaSpace{}%
\AgdaBound{a}\AgdaSpace{}%
\AgdaBound{b}%
\>[24]\AgdaSymbol{|}\AgdaSpace{}%
\AgdaBound{i}\AgdaSpace{}%
\AgdaSymbol{=}\AgdaSpace{}%
\AgdaInductiveConstructor{bin}\AgdaSpace{}%
\AgdaBound{op}\AgdaSpace{}%
\AgdaSymbol{(}\AgdaInductiveConstructor{selₛ}\AgdaSpace{}%
\AgdaBound{a}\AgdaSpace{}%
\AgdaBound{i}\AgdaSymbol{)}\AgdaSpace{}%
\AgdaSymbol{(}\AgdaInductiveConstructor{selₛ}\AgdaSpace{}%
\AgdaBound{b}\AgdaSpace{}%
\AgdaBound{i}\AgdaSymbol{)}\<%
\\
\>[2]\AgdaSymbol{...}\AgdaSpace{}%
\AgdaSymbol{|}\AgdaSpace{}%
\AgdaInductiveConstructor{sum}\AgdaSpace{}%
\AgdaBound{e}%
\>[24]\AgdaSymbol{|}\AgdaSpace{}%
\AgdaBound{i}\AgdaSpace{}%
\AgdaSymbol{=}\AgdaSpace{}%
\AgdaInductiveConstructor{sum}\AgdaSpace{}%
\AgdaSymbol{(}\AgdaInductiveConstructor{selₛ}\AgdaSpace{}%
\AgdaBound{e}\AgdaSpace{}%
\AgdaSymbol{(}\AgdaOperator{\AgdaFunction{↑}}\AgdaSpace{}%
\AgdaBound{i}\AgdaSymbol{))}\<%
\\
\>[2]\AgdaSymbol{...}\AgdaSpace{}%
\AgdaSymbol{|}\AgdaSpace{}%
\AgdaInductiveConstructor{zero-but}\AgdaSpace{}%
\AgdaBound{i}\AgdaSpace{}%
\AgdaBound{j}\AgdaSpace{}%
\AgdaBound{a}%
\>[24]\AgdaSymbol{|}\AgdaSpace{}%
\AgdaBound{k}\AgdaSpace{}%
\AgdaSymbol{=}\AgdaSpace{}%
\AgdaInductiveConstructor{zero-but}\AgdaSpace{}%
\AgdaBound{i}\AgdaSpace{}%
\AgdaBound{j}\AgdaSpace{}%
\AgdaSymbol{(}\AgdaInductiveConstructor{selₛ}\AgdaSpace{}%
\AgdaBound{a}\AgdaSpace{}%
\AgdaBound{k}\AgdaSymbol{)}\<%
\\
\>[2]\AgdaCatchallClause{\AgdaSymbol{...}}\AgdaSpace{}%
\AgdaCatchallClause{\AgdaSymbol{|}}\AgdaSpace{}%
\AgdaCatchallClause{\AgdaBound{a}}%
\>[24]\AgdaCatchallClause{\AgdaSymbol{|}}\AgdaSpace{}%
\AgdaCatchallClause{\AgdaBound{i}}\AgdaSpace{}%
\AgdaSymbol{=}\AgdaSpace{}%
\AgdaInductiveConstructor{selₛ}\AgdaSpace{}%
\AgdaBound{a}\AgdaSpace{}%
\AgdaBound{i}\<%
\\
\\[\AgdaEmptyExtraSkip]%
\>[2]\AgdaFunction{opt}\AgdaSpace{}%
\AgdaSymbol{(}\AgdaInductiveConstructor{sum}\AgdaSpace{}%
\AgdaBound{e}\AgdaSymbol{)}\AgdaSpace{}%
\AgdaKeyword{with}\AgdaSpace{}%
\AgdaFunction{opt}\AgdaSpace{}%
\AgdaBound{e}\<%
\\
\>[2]\AgdaSymbol{...}\AgdaSpace{}%
\AgdaSymbol{|}\AgdaSpace{}%
\AgdaInductiveConstructor{zero}%
\>[24]\AgdaSymbol{=}\AgdaSpace{}%
\AgdaInductiveConstructor{zero}\<%
\\
\>[2]\AgdaSymbol{...}\AgdaSpace{}%
\AgdaSymbol{|}\AgdaSpace{}%
\AgdaInductiveConstructor{imap}\AgdaSpace{}%
\AgdaBound{a}%
\>[24]\AgdaSymbol{=}\AgdaSpace{}%
\AgdaInductiveConstructor{imap}%
\>[35]\AgdaSymbol{(}\AgdaInductiveConstructor{sum}\AgdaSpace{}%
\AgdaSymbol{(}\AgdaFunction{ctx-swap}\AgdaSpace{}%
\AgdaInductiveConstructor{v₁}\AgdaSpace{}%
\AgdaBound{a}\AgdaSymbol{))}\<%
\\
\>[2]\AgdaSymbol{...}\AgdaSpace{}%
\AgdaSymbol{|}\AgdaSpace{}%
\AgdaInductiveConstructor{imapₛ}\AgdaSpace{}%
\AgdaBound{a}%
\>[24]\AgdaSymbol{=}\AgdaSpace{}%
\AgdaInductiveConstructor{imapₛ}%
\>[35]\AgdaSymbol{(}\AgdaInductiveConstructor{sum}\AgdaSpace{}%
\AgdaSymbol{(}\AgdaFunction{ctx-swap}\AgdaSpace{}%
\AgdaInductiveConstructor{v₁}\AgdaSpace{}%
\AgdaBound{a}\AgdaSymbol{))}\<%
\\
\>[2]\AgdaSymbol{...}\AgdaSpace{}%
\AgdaSymbol{|}\AgdaSpace{}%
\AgdaInductiveConstructor{imapb}\AgdaSpace{}%
\AgdaBound{m}\AgdaSpace{}%
\AgdaBound{a}%
\>[24]\AgdaSymbol{=}\AgdaSpace{}%
\AgdaInductiveConstructor{imapb}\AgdaSpace{}%
\AgdaBound{m}%
\>[35]\AgdaSymbol{(}\AgdaInductiveConstructor{sum}\AgdaSpace{}%
\AgdaSymbol{(}\AgdaFunction{ctx-swap}\AgdaSpace{}%
\AgdaInductiveConstructor{v₁}\AgdaSpace{}%
\AgdaBound{a}\AgdaSymbol{))}\<%
\\
\>[2]\AgdaSymbol{...}\AgdaSpace{}%
\AgdaSymbol{|}\AgdaSpace{}%
\AgdaInductiveConstructor{zero-but}\AgdaSpace{}%
\AgdaSymbol{(}\AgdaInductiveConstructor{var}\AgdaSpace{}%
\AgdaBound{i}\AgdaSymbol{)}\AgdaSpace{}%
\AgdaSymbol{(}\AgdaInductiveConstructor{var}\AgdaSpace{}%
\AgdaBound{j}\AgdaSymbol{)}\AgdaSpace{}%
\AgdaBound{a}\AgdaSpace{}%
\AgdaKeyword{with}\AgdaSpace{}%
\AgdaFunction{eq?}\AgdaSpace{}%
\AgdaInductiveConstructor{v₀}\AgdaSpace{}%
\AgdaBound{i}\AgdaSpace{}%
\AgdaSymbol{|}\AgdaSpace{}%
\AgdaFunction{eq?}\AgdaSpace{}%
\AgdaInductiveConstructor{v₀}\AgdaSpace{}%
\AgdaBound{j}\<%
\\
\>[2]\AgdaSymbol{...}\AgdaSpace{}%
\AgdaSymbol{|}\AgdaSpace{}%
\AgdaInductiveConstructor{eq}%
\>[18]\AgdaSymbol{|}\AgdaSpace{}%
\AgdaInductiveConstructor{eq}%
\>[30]\AgdaSymbol{=}\AgdaSpace{}%
\AgdaInductiveConstructor{sum}\AgdaSpace{}%
\AgdaBound{a}\<%
\\
\>[2]\AgdaSymbol{...}\AgdaSpace{}%
\AgdaSymbol{|}\AgdaSpace{}%
\AgdaInductiveConstructor{neq}\AgdaSpace{}%
\AgdaSymbol{\AgdaUnderscore{}}\AgdaSpace{}%
\AgdaBound{i′}%
\>[18]\AgdaSymbol{|}\AgdaSpace{}%
\AgdaInductiveConstructor{eq}%
\>[30]\AgdaSymbol{=}\AgdaSpace{}%
\AgdaFunction{sub}\AgdaSpace{}%
\AgdaInductiveConstructor{v₀}\AgdaSpace{}%
\AgdaBound{a}\AgdaSpace{}%
\AgdaSymbol{(}\AgdaInductiveConstructor{var}\AgdaSpace{}%
\AgdaBound{i′}\AgdaSymbol{)}\<%
\\
\>[2]\AgdaSymbol{...}\AgdaSpace{}%
\AgdaSymbol{|}\AgdaSpace{}%
\AgdaInductiveConstructor{eq}%
\>[18]\AgdaSymbol{|}\AgdaSpace{}%
\AgdaInductiveConstructor{neq}\AgdaSpace{}%
\AgdaSymbol{\AgdaUnderscore{}}\AgdaSpace{}%
\AgdaBound{j′}%
\>[30]\AgdaSymbol{=}\AgdaSpace{}%
\AgdaFunction{sub}\AgdaSpace{}%
\AgdaInductiveConstructor{v₀}\AgdaSpace{}%
\AgdaBound{a}\AgdaSpace{}%
\AgdaSymbol{(}\AgdaInductiveConstructor{var}\AgdaSpace{}%
\AgdaBound{j′}\AgdaSymbol{)}\<%
\\
\>[2]\AgdaSymbol{...}\AgdaSpace{}%
\AgdaSymbol{|}\AgdaSpace{}%
\AgdaInductiveConstructor{neq}\AgdaSpace{}%
\AgdaSymbol{\AgdaUnderscore{}}\AgdaSpace{}%
\AgdaBound{i′}%
\>[18]\AgdaSymbol{|}\AgdaSpace{}%
\AgdaInductiveConstructor{neq}\AgdaSpace{}%
\AgdaSymbol{\AgdaUnderscore{}}\AgdaSpace{}%
\AgdaBound{j′}%
\>[30]\AgdaSymbol{=}\AgdaSpace{}%
\AgdaInductiveConstructor{zero-but}\AgdaSpace{}%
\AgdaSymbol{(}\AgdaInductiveConstructor{var}\AgdaSpace{}%
\AgdaBound{i′}\AgdaSymbol{)}\AgdaSpace{}%
\AgdaSymbol{(}\AgdaInductiveConstructor{var}\AgdaSpace{}%
\AgdaBound{j′}\AgdaSymbol{)}\AgdaSpace{}%
\AgdaSymbol{(}\AgdaInductiveConstructor{sum}\AgdaSpace{}%
\AgdaBound{a}\AgdaSymbol{)}\<%
\\
\>[2]\AgdaCatchallClause{\AgdaFunction{opt}}\AgdaSpace{}%
\AgdaCatchallClause{\AgdaSymbol{(}}\AgdaCatchallClause{\AgdaInductiveConstructor{sum}}\AgdaSpace{}%
\AgdaCatchallClause{\AgdaBound{e}}\AgdaCatchallClause{\AgdaSymbol{)}}\AgdaSpace{}%
\AgdaCatchallClause{\AgdaSymbol{|}}\AgdaSpace{}%
\AgdaCatchallClause{\AgdaBound{a}}\AgdaSpace{}%
\AgdaSymbol{=}\AgdaSpace{}%
\AgdaInductiveConstructor{sum}\AgdaSpace{}%
\AgdaBound{a}\<%
\\
\>[2]\AgdaComment{--\ ⋯}\<%
\end{code}
Selection into \AC{zero} and \AC{one} is \AF{zero} and \AC{one}, as our constants
are shape-polymorphic.  Selection into an \AF{imapₛ} is evaluation of the \AC{imapₛ}
body at the given index (this is an array version of the $\beta$-rule).  Selection
from the binary operation is a binary operation of selections.  Selection into \AC{sum}
is the \AC{sum} of selections.  Selection into conditional is the same as conditional
over selection.  Summing \AC{zero} is \AC{zero}.  Summing $s$-many $p$-shaped arrays
is the same as computing the sum of $i$-th index of every array for all $p$ indices.
If we have a sum of the conditional with the predicate is the equality of indices
$i$ and $j$ and we know that $i$ and $j$ are variables, we can compare the index
variable of the \AC{sum} with $i$ and $j$.  If they match, then conditional will
be triggered at every iteration so it can be removed.  If only one of them match,
and we are comparing variables of the same shape, there will be exactly one case
(for non-empty shapes) where this conditional will be triggered.  Therefore, all
the iterations except the one at the non-matching variable will turn to zero, and
we can simply return the expressions substituted at this variable.  If the shape
of the index variables is empty, we are in the absurd case, as we cannot possibly
create an element of an empty type.  Finally, if none of the variables match,
the iteration within the \AC{sum} do not affect the result of the predicate ---
it will be either true or false for all the iterations.  Therefore, we can lift
the conditional outside of the sum.
\begin{code}[hide]%
\>[2]\AgdaFunction{opt}\AgdaSpace{}%
\AgdaInductiveConstructor{zero}\AgdaSpace{}%
\AgdaSymbol{=}\AgdaSpace{}%
\AgdaInductiveConstructor{zero}\<%
\\
\>[2]\AgdaFunction{opt}\AgdaSpace{}%
\AgdaInductiveConstructor{one}\AgdaSpace{}%
\AgdaSymbol{=}\AgdaSpace{}%
\AgdaInductiveConstructor{one}\<%
\\
\>[0]\<%
\\
\>[2]\AgdaFunction{opt}\AgdaSpace{}%
\AgdaSymbol{(}\AgdaInductiveConstructor{var}\AgdaSpace{}%
\AgdaBound{x}\AgdaSymbol{)}\AgdaSpace{}%
\AgdaSymbol{=}\AgdaSpace{}%
\AgdaInductiveConstructor{var}\AgdaSpace{}%
\AgdaBound{x}\<%
\\
\>[0]\<%
\\
\>[2]\AgdaFunction{opt}\AgdaSpace{}%
\AgdaSymbol{(}\AgdaInductiveConstructor{imapₛ}\AgdaSpace{}%
\AgdaBound{e}\AgdaSymbol{)}\AgdaSpace{}%
\AgdaSymbol{=}\AgdaSpace{}%
\AgdaInductiveConstructor{imapₛ}\AgdaSpace{}%
\AgdaSymbol{(}\AgdaFunction{opt}\AgdaSpace{}%
\AgdaBound{e}\AgdaSymbol{)}\<%
\\
\>[0]\<%
\\
\>[2]\AgdaComment{--\ Literal\ copy\ of\ the\ above,\ replaing\ scalar\ versions}\<%
\\
\>[2]\AgdaComment{--\ with\ normal\ one}\<%
\\
\>[2]\AgdaFunction{opt}\AgdaSpace{}%
\AgdaSymbol{(}\AgdaInductiveConstructor{imap}\AgdaSpace{}%
\AgdaBound{e}\AgdaSymbol{)}\AgdaSpace{}%
\AgdaSymbol{=}\AgdaSpace{}%
\AgdaInductiveConstructor{imap}\AgdaSpace{}%
\AgdaSymbol{(}\AgdaFunction{opt}\AgdaSpace{}%
\AgdaBound{e}\AgdaSymbol{)}\<%
\\
\>[2]\AgdaFunction{opt}\AgdaSpace{}%
\AgdaSymbol{(}\AgdaInductiveConstructor{sel}\AgdaSpace{}%
\AgdaBound{e}\AgdaSpace{}%
\AgdaBound{e₁}\AgdaSymbol{)}\AgdaSpace{}%
\AgdaKeyword{with}\AgdaSpace{}%
\AgdaFunction{opt}\AgdaSpace{}%
\AgdaBound{e}\AgdaSpace{}%
\AgdaSymbol{|}\AgdaSpace{}%
\AgdaFunction{opt}\AgdaSpace{}%
\AgdaBound{e₁}\<%
\\
\>[2]\AgdaSymbol{...}\AgdaSpace{}%
\AgdaSymbol{|}\AgdaSpace{}%
\AgdaInductiveConstructor{zero}\AgdaSpace{}%
\AgdaSymbol{|}\AgdaSpace{}%
\AgdaBound{i}\AgdaSpace{}%
\AgdaSymbol{=}\AgdaSpace{}%
\AgdaInductiveConstructor{zero}\<%
\\
\>[2]\AgdaSymbol{...}\AgdaSpace{}%
\AgdaSymbol{|}\AgdaSpace{}%
\AgdaInductiveConstructor{one}\AgdaSpace{}%
\AgdaSymbol{|}\AgdaSpace{}%
\AgdaBound{i}\AgdaSpace{}%
\AgdaSymbol{=}\AgdaSpace{}%
\AgdaInductiveConstructor{one}\<%
\\
\>[2]\AgdaSymbol{...}\AgdaSpace{}%
\AgdaSymbol{|}\AgdaSpace{}%
\AgdaInductiveConstructor{imap}\AgdaSpace{}%
\AgdaBound{e}\AgdaSpace{}%
\AgdaSymbol{|}\AgdaSpace{}%
\AgdaBound{i}\AgdaSpace{}%
\AgdaSymbol{=}\AgdaSpace{}%
\AgdaFunction{sub}\AgdaSpace{}%
\AgdaInductiveConstructor{v₀}\AgdaSpace{}%
\AgdaBound{e}\AgdaSpace{}%
\AgdaBound{i}\<%
\\
\>[2]\AgdaComment{--...\ |\ imapb\ m\ e\ |\ i\ =\ ?}\<%
\\
\>[2]\AgdaSymbol{...}\AgdaSpace{}%
\AgdaSymbol{|}\AgdaSpace{}%
\AgdaInductiveConstructor{bin}\AgdaSpace{}%
\AgdaBound{op}\AgdaSpace{}%
\AgdaBound{a}\AgdaSpace{}%
\AgdaBound{b}\AgdaSpace{}%
\AgdaSymbol{|}\AgdaSpace{}%
\AgdaBound{i}\AgdaSpace{}%
\AgdaSymbol{=}\AgdaSpace{}%
\AgdaInductiveConstructor{bin}\AgdaSpace{}%
\AgdaBound{op}\AgdaSpace{}%
\AgdaSymbol{(}\AgdaInductiveConstructor{sel}\AgdaSpace{}%
\AgdaBound{a}\AgdaSpace{}%
\AgdaBound{i}\AgdaSymbol{)}\AgdaSpace{}%
\AgdaSymbol{(}\AgdaInductiveConstructor{sel}\AgdaSpace{}%
\AgdaBound{b}\AgdaSpace{}%
\AgdaBound{i}\AgdaSymbol{)}\<%
\\
\>[2]\AgdaSymbol{...}\AgdaSpace{}%
\AgdaSymbol{|}\AgdaSpace{}%
\AgdaInductiveConstructor{sum}\AgdaSpace{}%
\AgdaBound{e}\AgdaSpace{}%
\AgdaSymbol{|}\AgdaSpace{}%
\AgdaBound{i}\AgdaSpace{}%
\AgdaSymbol{=}\AgdaSpace{}%
\AgdaInductiveConstructor{sum}\AgdaSpace{}%
\AgdaSymbol{(}\AgdaInductiveConstructor{sel}\AgdaSpace{}%
\AgdaBound{e}\AgdaSpace{}%
\AgdaSymbol{(}\AgdaFunction{wk}\AgdaSpace{}%
\AgdaInductiveConstructor{v₀}\AgdaSpace{}%
\AgdaBound{i}\AgdaSymbol{))}\<%
\\
\>[2]\AgdaSymbol{...}\AgdaSpace{}%
\AgdaSymbol{|}\AgdaSpace{}%
\AgdaInductiveConstructor{zero-but}\AgdaSpace{}%
\AgdaBound{i}\AgdaSpace{}%
\AgdaBound{j}\AgdaSpace{}%
\AgdaBound{a}\AgdaSpace{}%
\AgdaSymbol{|}\AgdaSpace{}%
\AgdaBound{k}\AgdaSpace{}%
\AgdaSymbol{=}\AgdaSpace{}%
\AgdaInductiveConstructor{zero-but}\AgdaSpace{}%
\AgdaBound{i}\AgdaSpace{}%
\AgdaBound{j}\AgdaSpace{}%
\AgdaSymbol{(}\AgdaInductiveConstructor{sel}\AgdaSpace{}%
\AgdaBound{a}\AgdaSpace{}%
\AgdaBound{k}\AgdaSymbol{)}\<%
\\
\>[2]\AgdaCatchallClause{\AgdaSymbol{...}}\AgdaSpace{}%
\AgdaCatchallClause{\AgdaSymbol{|}}\AgdaSpace{}%
\AgdaCatchallClause{\AgdaBound{a}}\AgdaSpace{}%
\AgdaCatchallClause{\AgdaSymbol{|}}\AgdaSpace{}%
\AgdaCatchallClause{\AgdaBound{i}}\AgdaSpace{}%
\AgdaSymbol{=}\AgdaSpace{}%
\AgdaInductiveConstructor{sel}\AgdaSpace{}%
\AgdaBound{a}\AgdaSpace{}%
\AgdaBound{i}\<%
\\
\>[0]\<%
\\
\>[2]\AgdaComment{--\ Literal\ copy\ of\ the\ above\ for\ the\ blocked\ version}\<%
\\
\>[2]\AgdaFunction{opt}\AgdaSpace{}%
\AgdaSymbol{(}\AgdaInductiveConstructor{imapb}\AgdaSpace{}%
\AgdaBound{m}\AgdaSpace{}%
\AgdaBound{e}\AgdaSymbol{)}\AgdaSpace{}%
\AgdaSymbol{=}\AgdaSpace{}%
\AgdaInductiveConstructor{imapb}\AgdaSpace{}%
\AgdaBound{m}\AgdaSpace{}%
\AgdaSymbol{(}\AgdaFunction{opt}\AgdaSpace{}%
\AgdaBound{e}\AgdaSymbol{)}\<%
\\
\>[2]\AgdaFunction{opt}\AgdaSpace{}%
\AgdaSymbol{(}\AgdaInductiveConstructor{selb}\AgdaSpace{}%
\AgdaBound{m}\AgdaSpace{}%
\AgdaBound{e}\AgdaSpace{}%
\AgdaBound{k}\AgdaSymbol{)}\AgdaSpace{}%
\AgdaKeyword{with}\AgdaSpace{}%
\AgdaFunction{opt}\AgdaSpace{}%
\AgdaBound{e}\<%
\\
\>[2]\AgdaSymbol{...}\AgdaSpace{}%
\AgdaSymbol{|}\AgdaSpace{}%
\AgdaInductiveConstructor{zero}\AgdaSpace{}%
\AgdaSymbol{=}\AgdaSpace{}%
\AgdaInductiveConstructor{zero}\<%
\\
\>[2]\AgdaSymbol{...}\AgdaSpace{}%
\AgdaSymbol{|}\AgdaSpace{}%
\AgdaInductiveConstructor{one}\AgdaSpace{}%
\AgdaSymbol{=}\AgdaSpace{}%
\AgdaInductiveConstructor{one}\<%
\\
\>[2]\AgdaSymbol{...}\AgdaSpace{}%
\AgdaSymbol{|}\AgdaSpace{}%
\AgdaInductiveConstructor{sum}\AgdaSpace{}%
\AgdaBound{e}\AgdaSpace{}%
\AgdaSymbol{=}\AgdaSpace{}%
\AgdaInductiveConstructor{sum}\AgdaSpace{}%
\AgdaSymbol{(}\AgdaInductiveConstructor{selb}\AgdaSpace{}%
\AgdaBound{m}\AgdaSpace{}%
\AgdaBound{e}\AgdaSpace{}%
\AgdaSymbol{(}\AgdaOperator{\AgdaFunction{↑}}\AgdaSpace{}%
\AgdaBound{k}\AgdaSpace{}%
\AgdaComment{\{-\ var\ \$\ vₛ\ k-\}}\AgdaSymbol{))}\<%
\\
\>[2]\AgdaSymbol{...}\AgdaSpace{}%
\AgdaSymbol{|}\AgdaSpace{}%
\AgdaInductiveConstructor{zero-but}\AgdaSpace{}%
\AgdaBound{i}\AgdaSpace{}%
\AgdaBound{j}\AgdaSpace{}%
\AgdaBound{a}\AgdaSpace{}%
\AgdaSymbol{=}\AgdaSpace{}%
\AgdaInductiveConstructor{zero-but}\AgdaSpace{}%
\AgdaBound{i}\AgdaSpace{}%
\AgdaBound{j}\AgdaSpace{}%
\AgdaSymbol{(}\AgdaInductiveConstructor{selb}\AgdaSpace{}%
\AgdaBound{m}\AgdaSpace{}%
\AgdaBound{a}\AgdaSpace{}%
\AgdaBound{k}\AgdaSymbol{)}\<%
\\
\>[2]\AgdaSymbol{...}\AgdaSpace{}%
\AgdaSymbol{|}\AgdaSpace{}%
\AgdaInductiveConstructor{bin}\AgdaSpace{}%
\AgdaBound{op}\AgdaSpace{}%
\AgdaBound{a}\AgdaSpace{}%
\AgdaBound{b}\AgdaSpace{}%
\AgdaSymbol{=}\AgdaSpace{}%
\AgdaInductiveConstructor{bin}\AgdaSpace{}%
\AgdaBound{op}\AgdaSpace{}%
\AgdaSymbol{(}\AgdaInductiveConstructor{selb}\AgdaSpace{}%
\AgdaBound{m}\AgdaSpace{}%
\AgdaBound{a}\AgdaSpace{}%
\AgdaBound{k}\AgdaSymbol{)}\AgdaSpace{}%
\AgdaSymbol{(}\AgdaInductiveConstructor{selb}\AgdaSpace{}%
\AgdaBound{m}\AgdaSpace{}%
\AgdaBound{b}\AgdaSpace{}%
\AgdaBound{k}\AgdaSymbol{)}\<%
\\
\>[2]\AgdaCatchallClause{\AgdaFunction{opt}}\AgdaSpace{}%
\AgdaCatchallClause{\AgdaSymbol{(}}\AgdaCatchallClause{\AgdaInductiveConstructor{selb}}\AgdaSpace{}%
\AgdaCatchallClause{\AgdaBound{m}}\AgdaSpace{}%
\AgdaCatchallClause{\AgdaBound{e}}\AgdaSpace{}%
\AgdaCatchallClause{\AgdaBound{j}}\AgdaCatchallClause{\AgdaSymbol{)}}\AgdaSpace{}%
\AgdaCatchallClause{\AgdaSymbol{|}}\AgdaSpace{}%
\AgdaCatchallClause{\AgdaBound{a}}\AgdaSpace{}%
\AgdaSymbol{=}\AgdaSpace{}%
\AgdaInductiveConstructor{selb}\AgdaSpace{}%
\AgdaBound{m}\AgdaSpace{}%
\AgdaBound{a}\AgdaSpace{}%
\AgdaBound{j}\<%
\\
\>[0]\<%
\\
\>[0]\<%
\\
\>[2]\AgdaFunction{opt}\AgdaSpace{}%
\AgdaSymbol{(}\AgdaInductiveConstructor{zero-but}\AgdaSpace{}%
\AgdaSymbol{(}\AgdaInductiveConstructor{var}\AgdaSpace{}%
\AgdaBound{i}\AgdaSymbol{)}\AgdaSpace{}%
\AgdaSymbol{(}\AgdaInductiveConstructor{var}\AgdaSpace{}%
\AgdaBound{j}\AgdaSymbol{)}\AgdaSpace{}%
\AgdaBound{e}\AgdaSymbol{)}\AgdaSpace{}%
\AgdaKeyword{with}\AgdaSpace{}%
\AgdaFunction{opt}\AgdaSpace{}%
\AgdaBound{e}\<%
\\
\>[2]\AgdaSymbol{...}\AgdaSpace{}%
\AgdaSymbol{|}\AgdaSpace{}%
\AgdaBound{a}\AgdaSpace{}%
\AgdaKeyword{with}\AgdaSpace{}%
\AgdaFunction{eq?}\AgdaSpace{}%
\AgdaBound{i}\AgdaSpace{}%
\AgdaBound{j}\<%
\\
\>[2]\AgdaSymbol{...}\AgdaSpace{}%
\AgdaSymbol{|}\AgdaSpace{}%
\AgdaInductiveConstructor{eq}\AgdaSpace{}%
\AgdaSymbol{=}\AgdaSpace{}%
\AgdaBound{a}\<%
\\
\>[2]\AgdaSymbol{...}\AgdaSpace{}%
\AgdaSymbol{|}\AgdaSpace{}%
\AgdaInductiveConstructor{neq}\AgdaSpace{}%
\AgdaSymbol{\AgdaUnderscore{}}\AgdaSpace{}%
\AgdaSymbol{\AgdaUnderscore{}}\AgdaSpace{}%
\AgdaSymbol{=}\AgdaSpace{}%
\AgdaInductiveConstructor{zero-but}\AgdaSpace{}%
\AgdaSymbol{(}\AgdaInductiveConstructor{var}\AgdaSpace{}%
\AgdaBound{i}\AgdaSymbol{)}\AgdaSpace{}%
\AgdaSymbol{(}\AgdaInductiveConstructor{var}\AgdaSpace{}%
\AgdaBound{j}\AgdaSymbol{)}\AgdaSpace{}%
\AgdaBound{a}\<%
\\
\>[2]\AgdaComment{--opt\ (zero-but\ i\ j\ e)\ =\ zero-but\ i\ j\ (opt\ e)}\<%
\\
\>[0]\<%
\\
\>[2]\AgdaFunction{opt}\AgdaSpace{}%
\AgdaSymbol{(}\AgdaInductiveConstructor{bin}\AgdaSpace{}%
\AgdaInductiveConstructor{plus}\AgdaSpace{}%
\AgdaBound{e}\AgdaSpace{}%
\AgdaBound{e₁}\AgdaSymbol{)}\AgdaSpace{}%
\AgdaKeyword{with}\AgdaSpace{}%
\AgdaFunction{opt}\AgdaSpace{}%
\AgdaBound{e}\AgdaSpace{}%
\AgdaSymbol{|}\AgdaSpace{}%
\AgdaFunction{opt}\AgdaSpace{}%
\AgdaBound{e₁}\<%
\\
\>[2]\AgdaSymbol{...}\AgdaSpace{}%
\AgdaSymbol{|}\AgdaSpace{}%
\AgdaInductiveConstructor{zero}\AgdaSpace{}%
\AgdaSymbol{|}\AgdaSpace{}%
\AgdaBound{b}\AgdaSpace{}%
\AgdaSymbol{=}\AgdaSpace{}%
\AgdaBound{b}\<%
\\
\>[2]\AgdaCatchallClause{\AgdaSymbol{...}}\AgdaSpace{}%
\AgdaCatchallClause{\AgdaSymbol{|}}\AgdaSpace{}%
\AgdaCatchallClause{\AgdaBound{a}}\AgdaSpace{}%
\AgdaCatchallClause{\AgdaSymbol{|}}\AgdaSpace{}%
\AgdaCatchallClause{\AgdaInductiveConstructor{zero}}\AgdaSpace{}%
\AgdaSymbol{=}\AgdaSpace{}%
\AgdaBound{a}\<%
\\
\>[2]\AgdaCatchallClause{\AgdaSymbol{...}}\AgdaSpace{}%
\AgdaCatchallClause{\AgdaSymbol{|}}\AgdaSpace{}%
\AgdaCatchallClause{\AgdaSymbol{(}}\AgdaCatchallClause{\AgdaInductiveConstructor{zero-but}}\AgdaSpace{}%
\AgdaCatchallClause{\AgdaBound{i}}\AgdaSpace{}%
\AgdaCatchallClause{\AgdaBound{j}}\AgdaSpace{}%
\AgdaCatchallClause{\AgdaBound{e}}\AgdaCatchallClause{\AgdaSymbol{)}}\AgdaSpace{}%
\AgdaCatchallClause{\AgdaSymbol{|}}\AgdaSpace{}%
\AgdaCatchallClause{\AgdaBound{b}}\AgdaSpace{}%
\AgdaSymbol{=}\AgdaSpace{}%
\AgdaInductiveConstructor{zero-but}\AgdaSpace{}%
\AgdaBound{i}\AgdaSpace{}%
\AgdaBound{j}\AgdaSpace{}%
\AgdaSymbol{(}\AgdaInductiveConstructor{bin}\AgdaSpace{}%
\AgdaInductiveConstructor{plus}\AgdaSpace{}%
\AgdaBound{e}\AgdaSpace{}%
\AgdaBound{b}\AgdaSymbol{)}\<%
\\
\>[2]\AgdaCatchallClause{\AgdaSymbol{...}}\AgdaSpace{}%
\AgdaCatchallClause{\AgdaSymbol{|}}\AgdaSpace{}%
\AgdaCatchallClause{\AgdaBound{a}}\AgdaSpace{}%
\AgdaCatchallClause{\AgdaSymbol{|}}\AgdaSpace{}%
\AgdaCatchallClause{\AgdaSymbol{(}}\AgdaCatchallClause{\AgdaInductiveConstructor{zero-but}}\AgdaSpace{}%
\AgdaCatchallClause{\AgdaBound{i}}\AgdaSpace{}%
\AgdaCatchallClause{\AgdaBound{j}}\AgdaSpace{}%
\AgdaCatchallClause{\AgdaBound{e}}\AgdaCatchallClause{\AgdaSymbol{)}}\AgdaSpace{}%
\AgdaSymbol{=}\AgdaSpace{}%
\AgdaInductiveConstructor{zero-but}\AgdaSpace{}%
\AgdaBound{i}\AgdaSpace{}%
\AgdaBound{j}\AgdaSpace{}%
\AgdaSymbol{(}\AgdaInductiveConstructor{bin}\AgdaSpace{}%
\AgdaInductiveConstructor{plus}\AgdaSpace{}%
\AgdaBound{a}\AgdaSpace{}%
\AgdaBound{e}\AgdaSymbol{)}\<%
\\
\\[\AgdaEmptyExtraSkip]%
\>[2]\AgdaCatchallClause{\AgdaSymbol{...}}\AgdaSpace{}%
\AgdaCatchallClause{\AgdaSymbol{|}}\AgdaSpace{}%
\AgdaCatchallClause{\AgdaInductiveConstructor{imapₛ}}\AgdaSpace{}%
\AgdaCatchallClause{\AgdaBound{a}}\AgdaSpace{}%
\AgdaCatchallClause{\AgdaSymbol{|}}\AgdaSpace{}%
\AgdaCatchallClause{\AgdaBound{b}}\AgdaSpace{}%
\AgdaSymbol{=}\AgdaSpace{}%
\AgdaInductiveConstructor{imapₛ}\AgdaSpace{}%
\AgdaSymbol{(}\AgdaInductiveConstructor{bin}\AgdaSpace{}%
\AgdaInductiveConstructor{plus}\AgdaSpace{}%
\AgdaBound{a}\AgdaSpace{}%
\AgdaSymbol{(}\AgdaInductiveConstructor{selₛ}\AgdaSpace{}%
\AgdaSymbol{(}\AgdaOperator{\AgdaFunction{↑}}\AgdaSpace{}%
\AgdaBound{b}\AgdaSymbol{)}\AgdaSpace{}%
\AgdaSymbol{(}\AgdaInductiveConstructor{var}\AgdaSpace{}%
\AgdaInductiveConstructor{v₀}\AgdaSymbol{)))}\<%
\\
\>[2]\AgdaCatchallClause{\AgdaSymbol{...}}\AgdaSpace{}%
\AgdaCatchallClause{\AgdaSymbol{|}}\AgdaSpace{}%
\AgdaCatchallClause{\AgdaBound{a}}\AgdaSpace{}%
\AgdaCatchallClause{\AgdaSymbol{|}}\AgdaSpace{}%
\AgdaCatchallClause{\AgdaInductiveConstructor{imapₛ}}\AgdaSpace{}%
\AgdaCatchallClause{\AgdaBound{b}}\AgdaSpace{}%
\AgdaSymbol{=}\AgdaSpace{}%
\AgdaInductiveConstructor{imapₛ}\AgdaSpace{}%
\AgdaSymbol{(}\AgdaInductiveConstructor{bin}\AgdaSpace{}%
\AgdaInductiveConstructor{plus}\AgdaSpace{}%
\AgdaSymbol{(}\AgdaInductiveConstructor{selₛ}\AgdaSpace{}%
\AgdaSymbol{(}\AgdaOperator{\AgdaFunction{↑}}\AgdaSpace{}%
\AgdaBound{a}\AgdaSymbol{)}\AgdaSpace{}%
\AgdaSymbol{(}\AgdaInductiveConstructor{var}\AgdaSpace{}%
\AgdaInductiveConstructor{v₀}\AgdaSymbol{))}\AgdaSpace{}%
\AgdaBound{b}\AgdaSymbol{)}\<%
\\
\>[2]\AgdaCatchallClause{\AgdaSymbol{...}}\AgdaSpace{}%
\AgdaCatchallClause{\AgdaSymbol{|}}\AgdaSpace{}%
\AgdaCatchallClause{\AgdaInductiveConstructor{imap}}\AgdaSpace{}%
\AgdaCatchallClause{\AgdaBound{a}}\AgdaSpace{}%
\AgdaCatchallClause{\AgdaSymbol{|}}\AgdaSpace{}%
\AgdaCatchallClause{\AgdaBound{b}}\AgdaSpace{}%
\AgdaSymbol{=}\AgdaSpace{}%
\AgdaInductiveConstructor{imap}\AgdaSpace{}%
\AgdaSymbol{(}\AgdaInductiveConstructor{bin}\AgdaSpace{}%
\AgdaInductiveConstructor{plus}\AgdaSpace{}%
\AgdaBound{a}\AgdaSpace{}%
\AgdaSymbol{(}\AgdaInductiveConstructor{sel}\AgdaSpace{}%
\AgdaSymbol{(}\AgdaOperator{\AgdaFunction{↑}}\AgdaSpace{}%
\AgdaBound{b}\AgdaSymbol{)}\AgdaSpace{}%
\AgdaSymbol{(}\AgdaInductiveConstructor{var}\AgdaSpace{}%
\AgdaInductiveConstructor{v₀}\AgdaSymbol{)))}\<%
\\
\>[2]\AgdaCatchallClause{\AgdaSymbol{...}}\AgdaSpace{}%
\AgdaCatchallClause{\AgdaSymbol{|}}\AgdaSpace{}%
\AgdaCatchallClause{\AgdaBound{a}}\AgdaSpace{}%
\AgdaCatchallClause{\AgdaSymbol{|}}\AgdaSpace{}%
\AgdaCatchallClause{\AgdaInductiveConstructor{imap}}\AgdaSpace{}%
\AgdaCatchallClause{\AgdaBound{b}}\AgdaSpace{}%
\AgdaSymbol{=}\AgdaSpace{}%
\AgdaInductiveConstructor{imap}\AgdaSpace{}%
\AgdaSymbol{(}\AgdaInductiveConstructor{bin}\AgdaSpace{}%
\AgdaInductiveConstructor{plus}\AgdaSpace{}%
\AgdaSymbol{(}\AgdaInductiveConstructor{sel}\AgdaSpace{}%
\AgdaSymbol{(}\AgdaOperator{\AgdaFunction{↑}}\AgdaSpace{}%
\AgdaBound{a}\AgdaSymbol{)}\AgdaSpace{}%
\AgdaSymbol{(}\AgdaInductiveConstructor{var}\AgdaSpace{}%
\AgdaInductiveConstructor{v₀}\AgdaSymbol{))}\AgdaSpace{}%
\AgdaBound{b}\AgdaSymbol{)}\<%
\\
\>[2]\AgdaCatchallClause{\AgdaSymbol{...}}\AgdaSpace{}%
\AgdaCatchallClause{\AgdaSymbol{|}}\AgdaSpace{}%
\AgdaCatchallClause{\AgdaInductiveConstructor{imapb}}\AgdaSpace{}%
\AgdaCatchallClause{\AgdaBound{m}}\AgdaSpace{}%
\AgdaCatchallClause{\AgdaBound{a}}\AgdaSpace{}%
\AgdaCatchallClause{\AgdaSymbol{|}}\AgdaSpace{}%
\AgdaCatchallClause{\AgdaBound{b}}\AgdaSpace{}%
\AgdaSymbol{=}\AgdaSpace{}%
\AgdaInductiveConstructor{imapb}\AgdaSpace{}%
\AgdaBound{m}\AgdaSpace{}%
\AgdaSymbol{(}\AgdaInductiveConstructor{bin}\AgdaSpace{}%
\AgdaInductiveConstructor{plus}\AgdaSpace{}%
\AgdaBound{a}\AgdaSpace{}%
\AgdaSymbol{(}\AgdaInductiveConstructor{selb}\AgdaSpace{}%
\AgdaBound{m}\AgdaSpace{}%
\AgdaSymbol{(}\AgdaOperator{\AgdaFunction{↑}}\AgdaSpace{}%
\AgdaBound{b}\AgdaSymbol{)}\AgdaSpace{}%
\AgdaSymbol{(}\AgdaInductiveConstructor{var}\AgdaSpace{}%
\AgdaInductiveConstructor{v₀}\AgdaSymbol{)))}\<%
\\
\>[2]\AgdaCatchallClause{\AgdaSymbol{...}}\AgdaSpace{}%
\AgdaCatchallClause{\AgdaSymbol{|}}\AgdaSpace{}%
\AgdaCatchallClause{\AgdaBound{a}}\AgdaSpace{}%
\AgdaCatchallClause{\AgdaSymbol{|}}\AgdaSpace{}%
\AgdaCatchallClause{\AgdaInductiveConstructor{imapb}}\AgdaSpace{}%
\AgdaCatchallClause{\AgdaBound{m}}\AgdaSpace{}%
\AgdaCatchallClause{\AgdaBound{b}}\AgdaSpace{}%
\AgdaSymbol{=}\AgdaSpace{}%
\AgdaInductiveConstructor{imapb}\AgdaSpace{}%
\AgdaBound{m}\AgdaSpace{}%
\AgdaSymbol{(}\AgdaInductiveConstructor{bin}\AgdaSpace{}%
\AgdaInductiveConstructor{plus}\AgdaSpace{}%
\AgdaSymbol{(}\AgdaInductiveConstructor{selb}\AgdaSpace{}%
\AgdaBound{m}\AgdaSpace{}%
\AgdaSymbol{(}\AgdaOperator{\AgdaFunction{↑}}\AgdaSpace{}%
\AgdaBound{a}\AgdaSymbol{)}\AgdaSpace{}%
\AgdaSymbol{(}\AgdaInductiveConstructor{var}\AgdaSpace{}%
\AgdaInductiveConstructor{v₀}\AgdaSymbol{))}\AgdaSpace{}%
\AgdaBound{b}\AgdaSymbol{)}\<%
\\
\\[\AgdaEmptyExtraSkip]%
\>[2]\AgdaCatchallClause{\AgdaSymbol{...}}\AgdaSpace{}%
\AgdaCatchallClause{\AgdaSymbol{|}}\AgdaSpace{}%
\AgdaCatchallClause{\AgdaBound{a}}\AgdaSpace{}%
\AgdaCatchallClause{\AgdaSymbol{|}}\AgdaSpace{}%
\AgdaCatchallClause{\AgdaBound{b}}\AgdaSpace{}%
\AgdaSymbol{=}\AgdaSpace{}%
\AgdaInductiveConstructor{bin}\AgdaSpace{}%
\AgdaInductiveConstructor{plus}\AgdaSpace{}%
\AgdaBound{a}\AgdaSpace{}%
\AgdaBound{b}\<%
\\
\>[2]\AgdaFunction{opt}\AgdaSpace{}%
\AgdaSymbol{(}\AgdaInductiveConstructor{bin}\AgdaSpace{}%
\AgdaInductiveConstructor{mul}\AgdaSpace{}%
\AgdaBound{e}\AgdaSpace{}%
\AgdaBound{e₁}\AgdaSymbol{)}\AgdaSpace{}%
\AgdaKeyword{with}\AgdaSpace{}%
\AgdaFunction{opt}\AgdaSpace{}%
\AgdaBound{e}\AgdaSpace{}%
\AgdaSymbol{|}\AgdaSpace{}%
\AgdaFunction{opt}\AgdaSpace{}%
\AgdaBound{e₁}\<%
\\
\>[2]\AgdaSymbol{...}\AgdaSpace{}%
\AgdaSymbol{|}\AgdaSpace{}%
\AgdaInductiveConstructor{zero}\AgdaSpace{}%
\AgdaSymbol{|}\AgdaSpace{}%
\AgdaBound{b}\AgdaSpace{}%
\AgdaSymbol{=}\AgdaSpace{}%
\AgdaInductiveConstructor{zero}\<%
\\
\>[2]\AgdaCatchallClause{\AgdaSymbol{...}}\AgdaSpace{}%
\AgdaCatchallClause{\AgdaSymbol{|}}\AgdaSpace{}%
\AgdaCatchallClause{\AgdaBound{a}}\AgdaSpace{}%
\AgdaCatchallClause{\AgdaSymbol{|}}\AgdaSpace{}%
\AgdaCatchallClause{\AgdaInductiveConstructor{zero}}\AgdaSpace{}%
\AgdaSymbol{=}\AgdaSpace{}%
\AgdaInductiveConstructor{zero}\<%
\\
\>[2]\AgdaCatchallClause{\AgdaSymbol{...}}\AgdaSpace{}%
\AgdaCatchallClause{\AgdaSymbol{|}}\AgdaSpace{}%
\AgdaCatchallClause{\AgdaInductiveConstructor{one}}\AgdaSpace{}%
\AgdaCatchallClause{\AgdaSymbol{|}}\AgdaSpace{}%
\AgdaCatchallClause{\AgdaBound{b}}\AgdaSpace{}%
\AgdaSymbol{=}\AgdaSpace{}%
\AgdaBound{b}\<%
\\
\>[2]\AgdaCatchallClause{\AgdaSymbol{...}}\AgdaSpace{}%
\AgdaCatchallClause{\AgdaSymbol{|}}\AgdaSpace{}%
\AgdaCatchallClause{\AgdaBound{a}}\AgdaSpace{}%
\AgdaCatchallClause{\AgdaSymbol{|}}\AgdaSpace{}%
\AgdaCatchallClause{\AgdaInductiveConstructor{one}}\AgdaSpace{}%
\AgdaSymbol{=}\AgdaSpace{}%
\AgdaBound{a}\<%
\\
\>[2]\AgdaCatchallClause{\AgdaSymbol{...}}\AgdaSpace{}%
\AgdaCatchallClause{\AgdaSymbol{|}}\AgdaSpace{}%
\AgdaCatchallClause{\AgdaSymbol{(}}\AgdaCatchallClause{\AgdaInductiveConstructor{zero-but}}\AgdaSpace{}%
\AgdaCatchallClause{\AgdaBound{i}}\AgdaSpace{}%
\AgdaCatchallClause{\AgdaBound{j}}\AgdaSpace{}%
\AgdaCatchallClause{\AgdaBound{e}}\AgdaCatchallClause{\AgdaSymbol{)}}\AgdaSpace{}%
\AgdaCatchallClause{\AgdaSymbol{|}}\AgdaSpace{}%
\AgdaCatchallClause{\AgdaBound{b}}\AgdaSpace{}%
\AgdaSymbol{=}\AgdaSpace{}%
\AgdaInductiveConstructor{zero-but}\AgdaSpace{}%
\AgdaBound{i}\AgdaSpace{}%
\AgdaBound{j}\AgdaSpace{}%
\AgdaSymbol{(}\AgdaInductiveConstructor{bin}\AgdaSpace{}%
\AgdaInductiveConstructor{mul}\AgdaSpace{}%
\AgdaBound{e}\AgdaSpace{}%
\AgdaBound{b}\AgdaSymbol{)}\<%
\\
\>[2]\AgdaCatchallClause{\AgdaSymbol{...}}\AgdaSpace{}%
\AgdaCatchallClause{\AgdaSymbol{|}}\AgdaSpace{}%
\AgdaCatchallClause{\AgdaBound{a}}\AgdaSpace{}%
\AgdaCatchallClause{\AgdaSymbol{|}}\AgdaSpace{}%
\AgdaCatchallClause{\AgdaSymbol{(}}\AgdaCatchallClause{\AgdaInductiveConstructor{zero-but}}\AgdaSpace{}%
\AgdaCatchallClause{\AgdaBound{i}}\AgdaSpace{}%
\AgdaCatchallClause{\AgdaBound{j}}\AgdaSpace{}%
\AgdaCatchallClause{\AgdaBound{e}}\AgdaCatchallClause{\AgdaSymbol{)}}\AgdaSpace{}%
\AgdaSymbol{=}\AgdaSpace{}%
\AgdaInductiveConstructor{zero-but}\AgdaSpace{}%
\AgdaBound{i}\AgdaSpace{}%
\AgdaBound{j}\AgdaSpace{}%
\AgdaSymbol{(}\AgdaInductiveConstructor{bin}\AgdaSpace{}%
\AgdaInductiveConstructor{mul}\AgdaSpace{}%
\AgdaBound{a}\AgdaSpace{}%
\AgdaBound{e}\AgdaSymbol{)}\<%
\\
\>[0]\<%
\\
\>[2]\AgdaCatchallClause{\AgdaSymbol{...}}\AgdaSpace{}%
\AgdaCatchallClause{\AgdaSymbol{|}}\AgdaSpace{}%
\AgdaCatchallClause{\AgdaInductiveConstructor{imapₛ}}\AgdaSpace{}%
\AgdaCatchallClause{\AgdaBound{a}}\AgdaSpace{}%
\AgdaCatchallClause{\AgdaSymbol{|}}\AgdaSpace{}%
\AgdaCatchallClause{\AgdaBound{b}}\AgdaSpace{}%
\AgdaSymbol{=}\AgdaSpace{}%
\AgdaInductiveConstructor{imapₛ}\AgdaSpace{}%
\AgdaSymbol{(}\AgdaInductiveConstructor{bin}\AgdaSpace{}%
\AgdaInductiveConstructor{mul}\AgdaSpace{}%
\AgdaBound{a}\AgdaSpace{}%
\AgdaSymbol{(}\AgdaInductiveConstructor{selₛ}\AgdaSpace{}%
\AgdaSymbol{(}\AgdaOperator{\AgdaFunction{↑}}\AgdaSpace{}%
\AgdaBound{b}\AgdaSymbol{)}\AgdaSpace{}%
\AgdaSymbol{(}\AgdaInductiveConstructor{var}\AgdaSpace{}%
\AgdaInductiveConstructor{v₀}\AgdaSymbol{)))}\<%
\\
\>[2]\AgdaCatchallClause{\AgdaSymbol{...}}\AgdaSpace{}%
\AgdaCatchallClause{\AgdaSymbol{|}}\AgdaSpace{}%
\AgdaCatchallClause{\AgdaBound{a}}\AgdaSpace{}%
\AgdaCatchallClause{\AgdaSymbol{|}}\AgdaSpace{}%
\AgdaCatchallClause{\AgdaInductiveConstructor{imapₛ}}\AgdaSpace{}%
\AgdaCatchallClause{\AgdaBound{b}}\AgdaSpace{}%
\AgdaSymbol{=}\AgdaSpace{}%
\AgdaInductiveConstructor{imapₛ}\AgdaSpace{}%
\AgdaSymbol{(}\AgdaInductiveConstructor{bin}\AgdaSpace{}%
\AgdaInductiveConstructor{mul}\AgdaSpace{}%
\AgdaSymbol{(}\AgdaInductiveConstructor{selₛ}\AgdaSpace{}%
\AgdaSymbol{(}\AgdaOperator{\AgdaFunction{↑}}\AgdaSpace{}%
\AgdaBound{a}\AgdaSymbol{)}\AgdaSpace{}%
\AgdaSymbol{(}\AgdaInductiveConstructor{var}\AgdaSpace{}%
\AgdaInductiveConstructor{v₀}\AgdaSymbol{))}\AgdaSpace{}%
\AgdaBound{b}\AgdaSymbol{)}\<%
\\
\>[2]\AgdaCatchallClause{\AgdaSymbol{...}}\AgdaSpace{}%
\AgdaCatchallClause{\AgdaSymbol{|}}\AgdaSpace{}%
\AgdaCatchallClause{\AgdaInductiveConstructor{imap}}\AgdaSpace{}%
\AgdaCatchallClause{\AgdaBound{a}}\AgdaSpace{}%
\AgdaCatchallClause{\AgdaSymbol{|}}\AgdaSpace{}%
\AgdaCatchallClause{\AgdaBound{b}}\AgdaSpace{}%
\AgdaSymbol{=}\AgdaSpace{}%
\AgdaInductiveConstructor{imap}\AgdaSpace{}%
\AgdaSymbol{(}\AgdaInductiveConstructor{bin}\AgdaSpace{}%
\AgdaInductiveConstructor{mul}\AgdaSpace{}%
\AgdaBound{a}\AgdaSpace{}%
\AgdaSymbol{(}\AgdaInductiveConstructor{sel}\AgdaSpace{}%
\AgdaSymbol{(}\AgdaOperator{\AgdaFunction{↑}}\AgdaSpace{}%
\AgdaBound{b}\AgdaSymbol{)}\AgdaSpace{}%
\AgdaSymbol{(}\AgdaInductiveConstructor{var}\AgdaSpace{}%
\AgdaInductiveConstructor{v₀}\AgdaSymbol{)))}\<%
\\
\>[2]\AgdaCatchallClause{\AgdaSymbol{...}}\AgdaSpace{}%
\AgdaCatchallClause{\AgdaSymbol{|}}\AgdaSpace{}%
\AgdaCatchallClause{\AgdaBound{a}}\AgdaSpace{}%
\AgdaCatchallClause{\AgdaSymbol{|}}\AgdaSpace{}%
\AgdaCatchallClause{\AgdaInductiveConstructor{imap}}\AgdaSpace{}%
\AgdaCatchallClause{\AgdaBound{b}}\AgdaSpace{}%
\AgdaSymbol{=}\AgdaSpace{}%
\AgdaInductiveConstructor{imap}\AgdaSpace{}%
\AgdaSymbol{(}\AgdaInductiveConstructor{bin}\AgdaSpace{}%
\AgdaInductiveConstructor{mul}\AgdaSpace{}%
\AgdaSymbol{(}\AgdaInductiveConstructor{sel}\AgdaSpace{}%
\AgdaSymbol{(}\AgdaOperator{\AgdaFunction{↑}}\AgdaSpace{}%
\AgdaBound{a}\AgdaSymbol{)}\AgdaSpace{}%
\AgdaSymbol{(}\AgdaInductiveConstructor{var}\AgdaSpace{}%
\AgdaInductiveConstructor{v₀}\AgdaSymbol{))}\AgdaSpace{}%
\AgdaBound{b}\AgdaSymbol{)}\<%
\\
\>[2]\AgdaCatchallClause{\AgdaSymbol{...}}\AgdaSpace{}%
\AgdaCatchallClause{\AgdaSymbol{|}}\AgdaSpace{}%
\AgdaCatchallClause{\AgdaInductiveConstructor{imapb}}\AgdaSpace{}%
\AgdaCatchallClause{\AgdaBound{m}}\AgdaSpace{}%
\AgdaCatchallClause{\AgdaBound{a}}\AgdaSpace{}%
\AgdaCatchallClause{\AgdaSymbol{|}}\AgdaSpace{}%
\AgdaCatchallClause{\AgdaBound{b}}\AgdaSpace{}%
\AgdaSymbol{=}\AgdaSpace{}%
\AgdaInductiveConstructor{imapb}\AgdaSpace{}%
\AgdaBound{m}\AgdaSpace{}%
\AgdaSymbol{(}\AgdaInductiveConstructor{bin}\AgdaSpace{}%
\AgdaInductiveConstructor{mul}\AgdaSpace{}%
\AgdaBound{a}\AgdaSpace{}%
\AgdaSymbol{(}\AgdaInductiveConstructor{selb}\AgdaSpace{}%
\AgdaBound{m}\AgdaSpace{}%
\AgdaSymbol{(}\AgdaOperator{\AgdaFunction{↑}}\AgdaSpace{}%
\AgdaBound{b}\AgdaSymbol{)}\AgdaSpace{}%
\AgdaSymbol{(}\AgdaInductiveConstructor{var}\AgdaSpace{}%
\AgdaInductiveConstructor{v₀}\AgdaSymbol{)))}\<%
\\
\>[2]\AgdaCatchallClause{\AgdaSymbol{...}}\AgdaSpace{}%
\AgdaCatchallClause{\AgdaSymbol{|}}\AgdaSpace{}%
\AgdaCatchallClause{\AgdaBound{a}}\AgdaSpace{}%
\AgdaCatchallClause{\AgdaSymbol{|}}\AgdaSpace{}%
\AgdaCatchallClause{\AgdaInductiveConstructor{imapb}}\AgdaSpace{}%
\AgdaCatchallClause{\AgdaBound{m}}\AgdaSpace{}%
\AgdaCatchallClause{\AgdaBound{b}}\AgdaSpace{}%
\AgdaSymbol{=}\AgdaSpace{}%
\AgdaInductiveConstructor{imapb}\AgdaSpace{}%
\AgdaBound{m}\AgdaSpace{}%
\AgdaSymbol{(}\AgdaInductiveConstructor{bin}\AgdaSpace{}%
\AgdaInductiveConstructor{mul}\AgdaSpace{}%
\AgdaSymbol{(}\AgdaInductiveConstructor{selb}\AgdaSpace{}%
\AgdaBound{m}\AgdaSpace{}%
\AgdaSymbol{(}\AgdaOperator{\AgdaFunction{↑}}\AgdaSpace{}%
\AgdaBound{a}\AgdaSymbol{)}\AgdaSpace{}%
\AgdaSymbol{(}\AgdaInductiveConstructor{var}\AgdaSpace{}%
\AgdaInductiveConstructor{v₀}\AgdaSymbol{))}\AgdaSpace{}%
\AgdaBound{b}\AgdaSymbol{)}\<%
\\
\>[0]\<%
\\
\>[2]\AgdaCatchallClause{\AgdaSymbol{...}}\AgdaSpace{}%
\AgdaCatchallClause{\AgdaSymbol{|}}\AgdaSpace{}%
\AgdaCatchallClause{\AgdaBound{a}}\AgdaSpace{}%
\AgdaCatchallClause{\AgdaSymbol{|}}\AgdaSpace{}%
\AgdaCatchallClause{\AgdaBound{b}}\AgdaSpace{}%
\AgdaSymbol{=}\AgdaSpace{}%
\AgdaInductiveConstructor{bin}\AgdaSpace{}%
\AgdaInductiveConstructor{mul}\AgdaSpace{}%
\AgdaBound{a}\AgdaSpace{}%
\AgdaBound{b}\<%
\\
\>[0]\<%
\\
\>[2]\AgdaComment{--\ XXX:\ not\ calling\ opt\ on\ e,\ as\ this\ is\ index}\<%
\\
\>[2]\AgdaFunction{opt}\AgdaSpace{}%
\AgdaSymbol{(}\AgdaInductiveConstructor{slide}\AgdaSpace{}%
\AgdaBound{i}\AgdaSpace{}%
\AgdaBound{pl}\AgdaSpace{}%
\AgdaBound{e}\AgdaSpace{}%
\AgdaBound{su}\AgdaSymbol{)}\AgdaSpace{}%
\AgdaKeyword{with}\AgdaSpace{}%
\AgdaFunction{opt}\AgdaSpace{}%
\AgdaBound{e}\<%
\\
\>[2]\AgdaSymbol{...}\AgdaSpace{}%
\AgdaSymbol{|}\AgdaSpace{}%
\AgdaInductiveConstructor{zero}\AgdaSpace{}%
\AgdaSymbol{=}\AgdaSpace{}%
\AgdaInductiveConstructor{zero}\<%
\\
\>[2]\AgdaCatchallClause{\AgdaSymbol{...}}\AgdaSpace{}%
\AgdaCatchallClause{\AgdaSymbol{|}}\AgdaSpace{}%
\AgdaCatchallClause{\AgdaBound{a}}\AgdaSpace{}%
\AgdaSymbol{=}\AgdaSpace{}%
\AgdaInductiveConstructor{slide}\AgdaSpace{}%
\AgdaBound{i}\AgdaSpace{}%
\AgdaBound{pl}\AgdaSpace{}%
\AgdaBound{a}\AgdaSpace{}%
\AgdaBound{su}\<%
\\
\>[2]\AgdaFunction{opt}\AgdaSpace{}%
\AgdaSymbol{(}\AgdaInductiveConstructor{backslide}\AgdaSpace{}%
\AgdaBound{i}\AgdaSpace{}%
\AgdaBound{e}\AgdaSpace{}%
\AgdaBound{su}\AgdaSpace{}%
\AgdaBound{pl}\AgdaSymbol{)}\AgdaSpace{}%
\AgdaKeyword{with}\AgdaSpace{}%
\AgdaFunction{opt}\AgdaSpace{}%
\AgdaBound{e}\<%
\\
\>[2]\AgdaSymbol{...}\AgdaSpace{}%
\AgdaSymbol{|}\AgdaSpace{}%
\AgdaInductiveConstructor{zero}\AgdaSpace{}%
\AgdaSymbol{=}\AgdaSpace{}%
\AgdaInductiveConstructor{zero}\<%
\\
\>[2]\AgdaCatchallClause{\AgdaSymbol{...}}\AgdaSpace{}%
\AgdaCatchallClause{\AgdaSymbol{|}}\AgdaSpace{}%
\AgdaCatchallClause{\AgdaBound{a}}\AgdaSpace{}%
\AgdaSymbol{=}\AgdaSpace{}%
\AgdaInductiveConstructor{backslide}\AgdaSpace{}%
\AgdaBound{i}\AgdaSpace{}%
\AgdaBound{a}\AgdaSpace{}%
\AgdaBound{su}\AgdaSpace{}%
\AgdaBound{pl}\<%
\\
\>[2]\AgdaFunction{opt}\AgdaSpace{}%
\AgdaSymbol{(}\AgdaInductiveConstructor{scaledown}\AgdaSpace{}%
\AgdaBound{x}\AgdaSpace{}%
\AgdaBound{e}\AgdaSymbol{)}\AgdaSpace{}%
\AgdaKeyword{with}\AgdaSpace{}%
\AgdaFunction{opt}\AgdaSpace{}%
\AgdaBound{e}\<%
\\
\>[2]\AgdaSymbol{...}\AgdaSpace{}%
\AgdaSymbol{|}\AgdaSpace{}%
\AgdaInductiveConstructor{scaledown}\AgdaSpace{}%
\AgdaBound{y}\AgdaSpace{}%
\AgdaBound{a}\AgdaSpace{}%
\AgdaSymbol{=}\AgdaSpace{}%
\AgdaInductiveConstructor{scaledown}\AgdaSpace{}%
\AgdaSymbol{(}\AgdaBound{x}\AgdaSpace{}%
\AgdaOperator{\AgdaPrimitive{ℕ.*}}\AgdaSpace{}%
\AgdaBound{y}\AgdaSymbol{)}\AgdaSpace{}%
\AgdaBound{a}\<%
\\
\>[2]\AgdaCatchallClause{\AgdaSymbol{...}}\AgdaSpace{}%
\AgdaCatchallClause{\AgdaSymbol{|}}\AgdaSpace{}%
\AgdaCatchallClause{\AgdaBound{a}}\AgdaSpace{}%
\AgdaSymbol{=}\AgdaSpace{}%
\AgdaInductiveConstructor{scaledown}\AgdaSpace{}%
\AgdaBound{x}\AgdaSpace{}%
\AgdaBound{a}\<%
\\
\>[2]\AgdaComment{--\ TODO:\ propogate\ minues\ inside\ of\ +,\ *,\ imap,\ etc.}\<%
\\
\>[2]\AgdaFunction{opt}\AgdaSpace{}%
\AgdaSymbol{(}\AgdaInductiveConstructor{minus}\AgdaSpace{}%
\AgdaBound{e}\AgdaSymbol{)}\AgdaSpace{}%
\AgdaKeyword{with}\AgdaSpace{}%
\AgdaFunction{opt}\AgdaSpace{}%
\AgdaBound{e}\<%
\\
\>[2]\AgdaSymbol{...}\AgdaSpace{}%
\AgdaSymbol{|}\AgdaSpace{}%
\AgdaInductiveConstructor{minus}\AgdaSpace{}%
\AgdaBound{a}\AgdaSpace{}%
\AgdaSymbol{=}\AgdaSpace{}%
\AgdaBound{a}\<%
\\
\>[2]\AgdaSymbol{...}\AgdaSpace{}%
\AgdaSymbol{|}\AgdaSpace{}%
\AgdaInductiveConstructor{imapₛ}\AgdaSpace{}%
\AgdaBound{a}\AgdaSpace{}%
\AgdaSymbol{=}\AgdaSpace{}%
\AgdaInductiveConstructor{imapₛ}\AgdaSpace{}%
\AgdaSymbol{(}\AgdaInductiveConstructor{minus}\AgdaSpace{}%
\AgdaBound{a}\AgdaSymbol{)}\<%
\\
\>[2]\AgdaSymbol{...}\AgdaSpace{}%
\AgdaSymbol{|}\AgdaSpace{}%
\AgdaInductiveConstructor{imap}\AgdaSpace{}%
\AgdaBound{a}\AgdaSpace{}%
\AgdaSymbol{=}\AgdaSpace{}%
\AgdaInductiveConstructor{imap}\AgdaSpace{}%
\AgdaSymbol{(}\AgdaInductiveConstructor{minus}\AgdaSpace{}%
\AgdaBound{a}\AgdaSymbol{)}\<%
\\
\>[2]\AgdaSymbol{...}\AgdaSpace{}%
\AgdaSymbol{|}\AgdaSpace{}%
\AgdaInductiveConstructor{imapb}\AgdaSpace{}%
\AgdaBound{m}\AgdaSpace{}%
\AgdaBound{a}\AgdaSpace{}%
\AgdaSymbol{=}\AgdaSpace{}%
\AgdaInductiveConstructor{imapb}\AgdaSpace{}%
\AgdaBound{m}\AgdaSpace{}%
\AgdaSymbol{(}\AgdaInductiveConstructor{minus}\AgdaSpace{}%
\AgdaBound{a}\AgdaSymbol{)}\<%
\\
\>[2]\AgdaSymbol{...}\AgdaSpace{}%
\AgdaSymbol{|}\AgdaSpace{}%
\AgdaInductiveConstructor{sum}\AgdaSpace{}%
\AgdaBound{e}\AgdaSpace{}%
\AgdaSymbol{=}\AgdaSpace{}%
\AgdaInductiveConstructor{sum}\AgdaSpace{}%
\AgdaSymbol{(}\AgdaInductiveConstructor{minus}\AgdaSpace{}%
\AgdaBound{e}\AgdaSymbol{)}\<%
\\
\>[2]\AgdaSymbol{...}\AgdaSpace{}%
\AgdaSymbol{|}\AgdaSpace{}%
\AgdaInductiveConstructor{bin}\AgdaSpace{}%
\AgdaInductiveConstructor{plus}\AgdaSpace{}%
\AgdaBound{a}\AgdaSpace{}%
\AgdaBound{b}\AgdaSpace{}%
\AgdaSymbol{=}\AgdaSpace{}%
\AgdaInductiveConstructor{bin}\AgdaSpace{}%
\AgdaInductiveConstructor{plus}\AgdaSpace{}%
\AgdaSymbol{(}\AgdaInductiveConstructor{minus}\AgdaSpace{}%
\AgdaBound{a}\AgdaSymbol{)}\AgdaSpace{}%
\AgdaSymbol{(}\AgdaInductiveConstructor{minus}\AgdaSpace{}%
\AgdaBound{b}\AgdaSymbol{)}\<%
\\
\>[2]\AgdaSymbol{...}\AgdaSpace{}%
\AgdaSymbol{|}\AgdaSpace{}%
\AgdaInductiveConstructor{bin}\AgdaSpace{}%
\AgdaInductiveConstructor{mul}\AgdaSpace{}%
\AgdaBound{a}\AgdaSpace{}%
\AgdaBound{b}\AgdaSpace{}%
\AgdaSymbol{=}\AgdaSpace{}%
\AgdaInductiveConstructor{bin}\AgdaSpace{}%
\AgdaInductiveConstructor{plus}\AgdaSpace{}%
\AgdaSymbol{(}\AgdaInductiveConstructor{minus}\AgdaSpace{}%
\AgdaBound{a}\AgdaSymbol{)}\AgdaSpace{}%
\AgdaBound{b}\<%
\\
\>[2]\AgdaCatchallClause{\AgdaSymbol{...}}\AgdaSpace{}%
\AgdaCatchallClause{\AgdaSymbol{|}}\AgdaSpace{}%
\AgdaCatchallClause{\AgdaBound{a}}\AgdaSpace{}%
\AgdaSymbol{=}\AgdaSpace{}%
\AgdaInductiveConstructor{minus}\AgdaSpace{}%
\AgdaBound{a}\<%
\\
\>[2]\AgdaFunction{opt}\AgdaSpace{}%
\AgdaSymbol{(}\AgdaInductiveConstructor{logistic}\AgdaSpace{}%
\AgdaBound{e}\AgdaSymbol{)}\AgdaSpace{}%
\AgdaKeyword{with}\AgdaSpace{}%
\AgdaFunction{opt}\AgdaSpace{}%
\AgdaBound{e}\<%
\\
\>[2]\AgdaSymbol{...}\AgdaSpace{}%
\AgdaSymbol{|}\AgdaSpace{}%
\AgdaInductiveConstructor{imapₛ}\AgdaSpace{}%
\AgdaBound{a}\AgdaSpace{}%
\AgdaSymbol{=}\AgdaSpace{}%
\AgdaInductiveConstructor{imapₛ}\AgdaSpace{}%
\AgdaSymbol{(}\AgdaInductiveConstructor{logistic}\AgdaSpace{}%
\AgdaBound{a}\AgdaSymbol{)}\<%
\\
\>[2]\AgdaSymbol{...}\AgdaSpace{}%
\AgdaSymbol{|}\AgdaSpace{}%
\AgdaInductiveConstructor{imap}\AgdaSpace{}%
\AgdaBound{a}\AgdaSpace{}%
\AgdaSymbol{=}\AgdaSpace{}%
\AgdaInductiveConstructor{imap}\AgdaSpace{}%
\AgdaSymbol{(}\AgdaInductiveConstructor{logistic}\AgdaSpace{}%
\AgdaBound{a}\AgdaSymbol{)}\<%
\\
\>[2]\AgdaCatchallClause{\AgdaSymbol{...}}\AgdaSpace{}%
\AgdaCatchallClause{\AgdaSymbol{|}}\AgdaSpace{}%
\AgdaCatchallClause{\AgdaBound{a}}\AgdaSpace{}%
\AgdaSymbol{=}\AgdaSpace{}%
\AgdaInductiveConstructor{logistic}\AgdaSpace{}%
\AgdaBound{a}\<%
\\
\\[\AgdaEmptyExtraSkip]%
\\[\AgdaEmptyExtraSkip]%
\>[2]\AgdaFunction{multiopt}\AgdaSpace{}%
\AgdaSymbol{:}\AgdaSpace{}%
\AgdaDatatype{ℕ}\AgdaSpace{}%
\AgdaSymbol{→}\AgdaSpace{}%
\AgdaDatatype{E}\AgdaSpace{}%
\AgdaGeneralizable{Γ}\AgdaSpace{}%
\AgdaGeneralizable{is}\AgdaSpace{}%
\AgdaSymbol{→}\AgdaSpace{}%
\AgdaDatatype{E}\AgdaSpace{}%
\AgdaGeneralizable{Γ}\AgdaSpace{}%
\AgdaGeneralizable{is}\<%
\\
\>[2]\AgdaFunction{multiopt}\AgdaSpace{}%
\AgdaInductiveConstructor{zero}\AgdaSpace{}%
\AgdaBound{e}\AgdaSpace{}%
\AgdaSymbol{=}\AgdaSpace{}%
\AgdaBound{e}\<%
\\
\>[2]\AgdaFunction{multiopt}\AgdaSpace{}%
\AgdaSymbol{(}\AgdaInductiveConstructor{suc}\AgdaSpace{}%
\AgdaBound{n}\AgdaSymbol{)}\AgdaSpace{}%
\AgdaBound{e}\AgdaSpace{}%
\AgdaSymbol{=}\AgdaSpace{}%
\AgdaFunction{opt}\AgdaSpace{}%
\AgdaSymbol{(}\AgdaFunction{multiopt}\AgdaSpace{}%
\AgdaBound{n}\AgdaSpace{}%
\AgdaBound{e}\AgdaSymbol{)}\<%
\\
\\[\AgdaEmptyExtraSkip]%
\>[2]\AgdaKeyword{module}\AgdaSpace{}%
\AgdaModule{TryOpt}\AgdaSpace{}%
\AgdaKeyword{where}\<%
\end{code}

Let us observe optimisation effects when computing derivatives of
the scalar dot-product defined as follows.
\begin{code}%
\>[2][@{}l@{\AgdaIndent{1}}]%
\>[4]\AgdaFunction{dotp}\AgdaSpace{}%
\AgdaSymbol{:}\AgdaSpace{}%
\AgdaDatatype{E}\AgdaSpace{}%
\AgdaGeneralizable{Γ}\AgdaSpace{}%
\AgdaSymbol{(}\AgdaInductiveConstructor{ar}\AgdaSpace{}%
\AgdaGeneralizable{s}\AgdaSymbol{)}\AgdaSpace{}%
\AgdaSymbol{→}\AgdaSpace{}%
\AgdaDatatype{E}\AgdaSpace{}%
\AgdaGeneralizable{Γ}\AgdaSpace{}%
\AgdaSymbol{(}\AgdaInductiveConstructor{ar}\AgdaSpace{}%
\AgdaGeneralizable{s}\AgdaSymbol{)}\AgdaSpace{}%
\AgdaSymbol{→}\AgdaSpace{}%
\AgdaDatatype{E}\AgdaSpace{}%
\AgdaGeneralizable{Γ}\AgdaSpace{}%
\AgdaSymbol{(}\AgdaInductiveConstructor{ar}\AgdaSpace{}%
\AgdaFunction{unit}\AgdaSymbol{)}\<%
\\
\>[4]\AgdaFunction{dotp}\AgdaSpace{}%
\AgdaBound{a}\AgdaSpace{}%
\AgdaBound{b}\AgdaSpace{}%
\AgdaSymbol{=}\AgdaSpace{}%
\AgdaFunction{Sum}\AgdaSpace{}%
\AgdaSymbol{λ}\AgdaSpace{}%
\AgdaBound{i}\AgdaSpace{}%
\AgdaSymbol{→}\AgdaSpace{}%
\AgdaInductiveConstructor{selₛ}\AgdaSpace{}%
\AgdaSymbol{(}\AgdaOperator{\AgdaFunction{↑}}\AgdaSpace{}%
\AgdaBound{a}\AgdaSymbol{)}\AgdaSpace{}%
\AgdaBound{i}\AgdaSpace{}%
\AgdaOperator{\AgdaInductiveConstructor{⊠}}\AgdaSpace{}%
\AgdaInductiveConstructor{selₛ}\AgdaSpace{}%
\AgdaSymbol{(}\AgdaOperator{\AgdaFunction{↑}}\AgdaSpace{}%
\AgdaBound{b}\AgdaSymbol{)}\AgdaSpace{}%
\AgdaBound{i}\<%
\end{code}
\begin{code}[hide]%
\>[4]\AgdaFunction{C}\AgdaSpace{}%
\AgdaSymbol{:}\AgdaSpace{}%
\AgdaDatatype{Ctx}\<%
\\
\>[4]\AgdaFunction{a}\AgdaSpace{}%
\AgdaSymbol{:}\AgdaSpace{}%
\AgdaDatatype{E}\AgdaSpace{}%
\AgdaFunction{C}\AgdaSpace{}%
\AgdaSymbol{\AgdaUnderscore{}}\<%
\\
\>[4]\AgdaFunction{b}\AgdaSpace{}%
\AgdaSymbol{:}\AgdaSpace{}%
\AgdaDatatype{E}\AgdaSpace{}%
\AgdaFunction{C}\AgdaSpace{}%
\AgdaSymbol{\AgdaUnderscore{}}\<%
\\
\>[4]\AgdaFunction{seed}\AgdaSpace{}%
\AgdaSymbol{:}\AgdaSpace{}%
\AgdaDatatype{E}\AgdaSpace{}%
\AgdaFunction{C}\AgdaSpace{}%
\AgdaSymbol{\AgdaUnderscore{}}\<%
\end{code}
We define the context \AF{C} where two top variables are of 5-element vector shape
and the last variable (\AC{v₂}) is of scalar shape.  We bind these variables to Agda
variables for convenience.
\begin{code}%
\>[4]\AgdaFunction{C}\AgdaSpace{}%
\AgdaSymbol{=}\AgdaSpace{}%
\AgdaInductiveConstructor{ε}\AgdaSpace{}%
\AgdaOperator{\AgdaInductiveConstructor{▹}}%
\>[13]\AgdaInductiveConstructor{ar}\AgdaSpace{}%
\AgdaSymbol{(}\AgdaInductiveConstructor{ι}\AgdaSpace{}%
\AgdaNumber{1}\AgdaSymbol{)}%
\>[28]\AgdaOperator{\AgdaInductiveConstructor{▹}}%
\>[31]\AgdaInductiveConstructor{ar}\AgdaSpace{}%
\AgdaSymbol{(}\AgdaInductiveConstructor{ι}\AgdaSpace{}%
\AgdaNumber{5}\AgdaSymbol{)}%
\>[43]\AgdaOperator{\AgdaInductiveConstructor{▹}}%
\>[46]\AgdaInductiveConstructor{ar}\AgdaSpace{}%
\AgdaSymbol{(}\AgdaInductiveConstructor{ι}\AgdaSpace{}%
\AgdaNumber{5}\AgdaSymbol{);}\<%
\\
\>[13]\AgdaFunction{seed}\AgdaSpace{}%
\AgdaSymbol{=}\AgdaSpace{}%
\AgdaInductiveConstructor{var}\AgdaSpace{}%
\AgdaInductiveConstructor{v₂}%
\>[28]\AgdaSymbol{;}%
\>[31]\AgdaFunction{a}\AgdaSpace{}%
\AgdaSymbol{=}\AgdaSpace{}%
\AgdaInductiveConstructor{var}\AgdaSpace{}%
\AgdaInductiveConstructor{v₁}%
\>[43]\AgdaSymbol{;}%
\>[46]\AgdaFunction{b}%
\>[49]\AgdaSymbol{=}\AgdaSpace{}%
\AgdaInductiveConstructor{var}\AgdaSpace{}%
\AgdaInductiveConstructor{v₀}\<%
\end{code}
\begin{code}[hide]%
\>[4]\AgdaFunction{∂a}%
\>[11]\AgdaSymbol{=}\AgdaSpace{}%
\AgdaFunction{env-ix}\AgdaSpace{}%
\AgdaSymbol{\{}\AgdaFunction{C}\AgdaSymbol{\}}\AgdaSpace{}%
\AgdaSymbol{(}\AgdaFunction{∇}\AgdaSpace{}%
\AgdaSymbol{\{}\AgdaFunction{C}\AgdaSymbol{\}}\AgdaSpace{}%
\AgdaSymbol{(}\AgdaFunction{dotp}\AgdaSpace{}%
\AgdaFunction{a}\AgdaSpace{}%
\AgdaFunction{b}\AgdaSymbol{)}\AgdaSpace{}%
\AgdaFunction{seed}\AgdaSpace{}%
\AgdaSymbol{(}\AgdaFunction{env-zero}\AgdaSpace{}%
\AgdaSymbol{\{}\AgdaFunction{C}\AgdaSymbol{\}))}\AgdaSpace{}%
\AgdaInductiveConstructor{v₁}\<%
\\
\>[4]\AgdaFunction{∂a′}%
\>[11]\AgdaSymbol{=}\AgdaSpace{}%
\AgdaFunction{multiopt}\AgdaSpace{}%
\AgdaNumber{3}\AgdaSpace{}%
\AgdaFunction{∂a}\<%
\end{code}
We compute the derivatives of \AF{dotp a b} with seed \AF{seed} and we inspect
the $a$-th position in the returned environment that we call \AF{∂a}.  Then we repeatedly
apply \AF{opt} (three times) to \AF{∂a} and save it in \AF{∂a′}.  We force Agda to
verify that the content of the variables is as follows:
\begin{code}%
\>[4]\AgdaFunction{non-opt}%
\>[14]\AgdaSymbol{:}\AgdaSpace{}%
\AgdaFunction{∂a}%
\>[21]\AgdaOperator{\AgdaDatatype{≡}}\AgdaSpace{}%
\AgdaSymbol{(}\AgdaFunction{Sum}\AgdaSpace{}%
\AgdaSymbol{λ}\AgdaSpace{}%
\AgdaBound{i}\AgdaSpace{}%
\AgdaSymbol{→}\AgdaSpace{}%
\AgdaInductiveConstructor{zero}\AgdaSpace{}%
\AgdaOperator{\AgdaInductiveConstructor{⊞}}\AgdaSpace{}%
\AgdaFunction{Imapₛ}\AgdaSpace{}%
\AgdaSymbol{λ}\AgdaSpace{}%
\AgdaBound{j}\AgdaSpace{}%
\AgdaSymbol{→}\AgdaSpace{}%
\AgdaInductiveConstructor{zero-but}\AgdaSpace{}%
\AgdaBound{j}\AgdaSpace{}%
\AgdaSymbol{(}\AgdaOperator{\AgdaFunction{↑}}\AgdaSpace{}%
\AgdaBound{i}\AgdaSymbol{)}\AgdaSpace{}%
\AgdaSymbol{(}\AgdaOperator{\AgdaFunction{↑↑}}\AgdaSpace{}%
\AgdaFunction{seed}\AgdaSpace{}%
\AgdaOperator{\AgdaInductiveConstructor{⊠}}\AgdaSpace{}%
\AgdaInductiveConstructor{selₛ}\AgdaSpace{}%
\AgdaSymbol{(}\AgdaOperator{\AgdaFunction{↑↑}}\AgdaSpace{}%
\AgdaFunction{b}\AgdaSymbol{)}\AgdaSpace{}%
\AgdaSymbol{(}\AgdaOperator{\AgdaFunction{↑}}\AgdaSpace{}%
\AgdaBound{i}\AgdaSymbol{)))}\AgdaSpace{}%
\AgdaOperator{\AgdaInductiveConstructor{⊞}}\AgdaSpace{}%
\AgdaInductiveConstructor{zero}\<%
\\
\>[4]\AgdaFunction{with-opt}%
\>[14]\AgdaSymbol{:}\AgdaSpace{}%
\AgdaFunction{∂a′}%
\>[21]\AgdaOperator{\AgdaDatatype{≡}}\AgdaSpace{}%
\AgdaFunction{Imapₛ}\AgdaSpace{}%
\AgdaSymbol{λ}\AgdaSpace{}%
\AgdaBound{i}\AgdaSpace{}%
\AgdaSymbol{→}\AgdaSpace{}%
\AgdaSymbol{(}\AgdaOperator{\AgdaFunction{↑}}\AgdaSpace{}%
\AgdaFunction{seed}\AgdaSpace{}%
\AgdaOperator{\AgdaInductiveConstructor{⊠}}\AgdaSpace{}%
\AgdaInductiveConstructor{selₛ}\AgdaSpace{}%
\AgdaSymbol{(}\AgdaOperator{\AgdaFunction{↑}}\AgdaSpace{}%
\AgdaFunction{b}\AgdaSymbol{)}\AgdaSpace{}%
\AgdaBound{i}\AgdaSymbol{)}\<%
\end{code}
\begin{code}[hide]%
\>[4]\AgdaFunction{non-opt}\AgdaSpace{}%
\AgdaSymbol{=}\AgdaSpace{}%
\AgdaInductiveConstructor{refl}\<%
\\
\>[4]\AgdaFunction{with-opt}\AgdaSpace{}%
\AgdaSymbol{=}\AgdaSpace{}%
\AgdaInductiveConstructor{refl}\<%
\\
\>[0]\AgdaComment{--\ open\ Lang}\<%
\\
\>[0]\AgdaComment{--\ open\ SubWk}\<%
\end{code}
As it can be seen, \AF{∂a} sums-up the arrays, where only one element is non-zero at
every iteration.  Such a computation is highly inefficient when executed directly,
as it needs to compute all the inner arrays before summing them up.  However, the
optimised version correctly rewrites \AF{∂a} into \AC{imap} that multiplies
the \AB{seed} by $b$, which is the expected answer.  This reduces complexity
of the expression form squared to linear.

\subsection{Extraction}

Extraction from \AF{E} into SaC translates constructors of \AF{E} into
corresponding SaC expressions or function calls.  The translation starts with
a definition of an environment (\AF{SEnv} \AB{Γ}) that assigns SaC variable names
to all positions in \AB{Γ}.  The assumption here is that when we compile
expressions in context \AF{Γ}, variable names of the corresponding shapes are
available in SaC.

Next, we have to take care of shapes.  Array shapes in \AF{E} are binary trees,
but array shapes in SaC are 1-dimensional arrays (flattened binary trees).
When some expression in \AF{E} is of product shape, we usually have to
supply left or right subshapes of the product to SaC. These are always available
through implicit arguments of \AF{E} constructors. Assuming that by the
time we come to extraction, all the \AF{E} shapes are constants, we can
always generate shape expressions in SaC.  This is implemented in \AF{show-shape}.
Relaxing the assumption about constant shapes is possible but requires
extension of \AF{E} so that we can always bind the shapes used in \AF{E}
to some expressions in SaC.

We also need a source of fresh variables so that we can generate indices
for \AC{imap} expressions.  We define a stateful function \AF{iv} that
generates a fresh index variable.  

Extraction is given by \AF{to-sac} that translates the expression $e$ in
the environment $\rho$.  The function is stateful so that we can generate
fresh variables when needed.

The definitions of \AF{SEnv}, \AF{iv}, {\AF{show-shape}, and \AF{to-sac} follow.
\begin{code}[hide]%
\>[0]\AgdaKeyword{module}\AgdaSpace{}%
\AgdaModule{Sac}\AgdaSpace{}%
\AgdaKeyword{where}\<%
\\
\>[0][@{}l@{\AgdaIndent{0}}]%
\>[2]\AgdaKeyword{open}\AgdaSpace{}%
\AgdaKeyword{import}\AgdaSpace{}%
\AgdaModule{Data.Unit}\<%
\\
\>[2]\AgdaKeyword{open}\AgdaSpace{}%
\AgdaKeyword{import}\AgdaSpace{}%
\AgdaModule{Data.Product}\<%
\\
\>[2]\AgdaKeyword{open}\AgdaSpace{}%
\AgdaKeyword{import}\AgdaSpace{}%
\AgdaModule{Data.List}\AgdaSpace{}%
\AgdaSymbol{as}\AgdaSpace{}%
\AgdaModule{L}\AgdaSpace{}%
\AgdaKeyword{using}\AgdaSpace{}%
\AgdaSymbol{(}\AgdaDatatype{List}\AgdaSymbol{;}\AgdaSpace{}%
\AgdaInductiveConstructor{[]}\AgdaSymbol{;}\AgdaSpace{}%
\AgdaOperator{\AgdaInductiveConstructor{\AgdaUnderscore{}∷\AgdaUnderscore{}}}\AgdaSymbol{;}\AgdaSpace{}%
\AgdaOperator{\AgdaFunction{\AgdaUnderscore{}++\AgdaUnderscore{}}}\AgdaSymbol{)}\<%
\\
\>[2]\AgdaKeyword{open}\AgdaSpace{}%
\AgdaKeyword{import}\AgdaSpace{}%
\AgdaModule{Data.Nat}\AgdaSpace{}%
\AgdaSymbol{as}\AgdaSpace{}%
\AgdaModule{ℕ}\AgdaSpace{}%
\AgdaKeyword{using}\AgdaSpace{}%
\AgdaSymbol{(}\AgdaDatatype{ℕ}\AgdaSymbol{;}\AgdaSpace{}%
\AgdaInductiveConstructor{zero}\AgdaSymbol{;}\AgdaSpace{}%
\AgdaInductiveConstructor{suc}\AgdaSymbol{)}\<%
\\
\>[2]\AgdaKeyword{open}\AgdaSpace{}%
\AgdaKeyword{import}\AgdaSpace{}%
\AgdaModule{Data.Nat.Show}\AgdaSpace{}%
\AgdaKeyword{using}\AgdaSpace{}%
\AgdaSymbol{()}\AgdaSpace{}%
\AgdaKeyword{renaming}\AgdaSpace{}%
\AgdaSymbol{(}\AgdaFunction{show}\AgdaSpace{}%
\AgdaSymbol{to}\AgdaSpace{}%
\AgdaFunction{show-nat}\AgdaSymbol{)}\<%
\\
\>[2]\AgdaKeyword{open}\AgdaSpace{}%
\AgdaKeyword{import}\AgdaSpace{}%
\AgdaModule{Data.String}\AgdaSpace{}%
\AgdaKeyword{hiding}\AgdaSpace{}%
\AgdaSymbol{(}\AgdaOperator{\AgdaFunction{\AgdaUnderscore{}++\AgdaUnderscore{}}}\AgdaSymbol{)}\<%
\\
\>[2]\AgdaKeyword{open}\AgdaSpace{}%
\AgdaKeyword{import}\AgdaSpace{}%
\AgdaModule{Text.Printf}\<%
\\
\>[2]\AgdaKeyword{open}\AgdaSpace{}%
\AgdaKeyword{import}\AgdaSpace{}%
\AgdaModule{Category.Monad.State}\AgdaSpace{}%
\AgdaComment{--using\ (State;\ StateMonad;\ RawMonadState)}\<%
\\
\>[2]\AgdaKeyword{open}\AgdaSpace{}%
\AgdaKeyword{import}\AgdaSpace{}%
\AgdaModule{Category.Monad}\AgdaSpace{}%
\AgdaKeyword{using}\AgdaSpace{}%
\AgdaSymbol{(}\AgdaFunction{RawMonad}\AgdaSymbol{)}\<%
\\
\>[2]\AgdaComment{--open\ RawMonad\ \{\{...\}\}\ public}\<%
\\
\>[2]\AgdaKeyword{open}\AgdaSpace{}%
\AgdaModule{RawMonadState}\AgdaSpace{}%
\AgdaSymbol{\{\{...\}\}}\AgdaSpace{}%
\AgdaKeyword{public}\<%
\\
\>[2]\AgdaKeyword{open}\AgdaSpace{}%
\AgdaModule{Lang}\<%
\\
\>[2]\AgdaKeyword{open}\AgdaSpace{}%
\AgdaModule{Array}\AgdaSpace{}%
\AgdaKeyword{hiding}\AgdaSpace{}%
\AgdaSymbol{(}\AgdaFunction{sum}\AgdaSymbol{;}\AgdaSpace{}%
\AgdaFunction{slide}\AgdaSymbol{;}\AgdaSpace{}%
\AgdaFunction{backslide}\AgdaSymbol{)}\<%
\\
\>[2]\AgdaKeyword{open}\AgdaSpace{}%
\AgdaModule{SubWk}\<%
\\
\\[\AgdaEmptyExtraSkip]%
\>[2]\AgdaKeyword{instance}\<%
\\
\>[2][@{}l@{\AgdaIndent{0}}]%
\>[4]\AgdaComment{--\ stateMon\ :\ ∀\ \{S\ :\ Set\}\ →\ RawMonad\ (State\ S)}\<%
\\
\>[4]\AgdaComment{--\ stateMon\ \{S\}\ =\ StateMonad\ S}\<%
\\
\\[\AgdaEmptyExtraSkip]%
\>[4]\AgdaFunction{stateMonState}\AgdaSpace{}%
\AgdaSymbol{:}\AgdaSpace{}%
\AgdaSymbol{∀}\AgdaSpace{}%
\AgdaSymbol{\{}\AgdaBound{S}\AgdaSpace{}%
\AgdaSymbol{:}\AgdaSpace{}%
\AgdaPrimitive{Set}\AgdaSymbol{\}}\AgdaSpace{}%
\AgdaSymbol{→}\AgdaSpace{}%
\AgdaFunction{RawMonadState}\AgdaSpace{}%
\AgdaBound{S}\AgdaSpace{}%
\AgdaSymbol{(}\AgdaFunction{State}\AgdaSpace{}%
\AgdaBound{S}\AgdaSymbol{)}\<%
\\
\>[4]\AgdaFunction{stateMonState}\AgdaSpace{}%
\AgdaSymbol{\{}\AgdaBound{S}\AgdaSymbol{\}}\AgdaSpace{}%
\AgdaSymbol{=}\AgdaSpace{}%
\AgdaFunction{StateMonadState}\AgdaSpace{}%
\AgdaBound{S}\<%
\end{code}
\begin{mathpar}
\codeblock{\begin{code}%
\>[2]\AgdaFunction{SEnv}\AgdaSpace{}%
\AgdaSymbol{:}\AgdaSpace{}%
\AgdaDatatype{Ctx}\AgdaSpace{}%
\AgdaSymbol{→}\AgdaSpace{}%
\AgdaPrimitive{Set}\<%
\\
\>[2]\AgdaFunction{SEnv}\AgdaSpace{}%
\AgdaInductiveConstructor{ε}%
\>[17]\AgdaSymbol{=}\AgdaSpace{}%
\AgdaRecord{⊤}\<%
\\
\>[2]\AgdaFunction{SEnv}\AgdaSpace{}%
\AgdaSymbol{(}\AgdaBound{Γ}\AgdaSpace{}%
\AgdaOperator{\AgdaInductiveConstructor{▹}}\AgdaSpace{}%
\AgdaBound{is}\AgdaSymbol{)}%
\>[17]\AgdaSymbol{=}\AgdaSpace{}%
\AgdaFunction{SEnv}\AgdaSpace{}%
\AgdaBound{Γ}\AgdaSpace{}%
\AgdaOperator{\AgdaFunction{×}}\AgdaSpace{}%
\AgdaPostulate{String}\<%
\end{code}}
\and
\codeblock{\begin{code}%
\>[2]\AgdaFunction{iv}\AgdaSpace{}%
\AgdaSymbol{:}\AgdaSpace{}%
\AgdaDatatype{S}\AgdaSpace{}%
\AgdaSymbol{→}\AgdaSpace{}%
\AgdaFunction{State}\AgdaSpace{}%
\AgdaDatatype{ℕ}\AgdaSpace{}%
\AgdaPostulate{String}\<%
\\
\>[2]\AgdaFunction{iv}\AgdaSpace{}%
\AgdaBound{s}\AgdaSpace{}%
\AgdaSymbol{=}\AgdaSpace{}%
\AgdaKeyword{do}%
\>[13]\AgdaBound{v}\AgdaSpace{}%
\AgdaOperator{\AgdaFunction{←}}\AgdaSpace{}%
\AgdaField{get}\<%
\\
\>[13]\AgdaFunction{modify}\AgdaSpace{}%
\AgdaInductiveConstructor{suc}\<%
\\
\>[13]\AgdaFunction{return}\AgdaSpace{}%
\AgdaOperator{\AgdaFunction{\$}}\AgdaSpace{}%
\AgdaFunction{printf}\AgdaSpace{}%
\AgdaString{"x\%u"}\AgdaSpace{}%
\AgdaBound{v}\<%
\end{code}
\begin{code}[hide]%
\>[0]\<%
\\
\>[2]\AgdaFunction{lookup}\AgdaSpace{}%
\AgdaSymbol{:}\AgdaSpace{}%
\AgdaGeneralizable{is}\AgdaSpace{}%
\AgdaOperator{\AgdaDatatype{∈}}\AgdaSpace{}%
\AgdaGeneralizable{Γ}\AgdaSpace{}%
\AgdaSymbol{→}\AgdaSpace{}%
\AgdaFunction{SEnv}\AgdaSpace{}%
\AgdaGeneralizable{Γ}\AgdaSpace{}%
\AgdaSymbol{→}\AgdaSpace{}%
\AgdaPostulate{String}\<%
\\
\>[2]\AgdaFunction{lookup}\AgdaSpace{}%
\AgdaInductiveConstructor{v₀}%
\>[17]\AgdaSymbol{(}\AgdaBound{ρ}\AgdaSpace{}%
\AgdaOperator{\AgdaInductiveConstructor{,}}\AgdaSpace{}%
\AgdaBound{e}\AgdaSymbol{)}\AgdaSpace{}%
\AgdaSymbol{=}\AgdaSpace{}%
\AgdaBound{e}\<%
\\
\>[2]\AgdaFunction{lookup}\AgdaSpace{}%
\AgdaSymbol{(}\AgdaInductiveConstructor{vₛ}\AgdaSpace{}%
\AgdaBound{x}\AgdaSymbol{)}%
\>[17]\AgdaSymbol{(}\AgdaBound{ρ}\AgdaSpace{}%
\AgdaOperator{\AgdaInductiveConstructor{,}}\AgdaSpace{}%
\AgdaBound{e}\AgdaSymbol{)}\AgdaSpace{}%
\AgdaSymbol{=}\AgdaSpace{}%
\AgdaFunction{lookup}\AgdaSpace{}%
\AgdaBound{x}\AgdaSpace{}%
\AgdaBound{ρ}\<%
\\
\\[\AgdaEmptyExtraSkip]%
\\[\AgdaEmptyExtraSkip]%
\>[2]\AgdaComment{--\ show-shape\ :\ S\ →\ String}\<%
\\
\>[2]\AgdaComment{--\ show-shape\ (ι\ x)\ =\ show-nat\ x}\<%
\\
\>[2]\AgdaComment{--\ show-shape\ (s\ S.⊗\ p)\ =\ printf\ "⟨\%s,\ \%s⟩"\ (show-shape\ s)\ (show-shape\ p)}\<%
\\
\\[\AgdaEmptyExtraSkip]%
\>[2]\AgdaFunction{fresh-var}\AgdaSpace{}%
\AgdaSymbol{:}\AgdaSpace{}%
\AgdaDatatype{ℕ}\AgdaSpace{}%
\AgdaSymbol{→}\AgdaSpace{}%
\AgdaPostulate{String}\<%
\\
\>[2]\AgdaFunction{fresh-var}\AgdaSpace{}%
\AgdaBound{n}\AgdaSpace{}%
\AgdaSymbol{=}\AgdaSpace{}%
\AgdaFunction{printf}\AgdaSpace{}%
\AgdaString{"x\%u"}\AgdaSpace{}%
\AgdaBound{n}\<%
\\
\\[\AgdaEmptyExtraSkip]%
\>[2]\AgdaFunction{bop}\AgdaSpace{}%
\AgdaSymbol{:}\AgdaSpace{}%
\AgdaDatatype{Bop}\AgdaSpace{}%
\AgdaSymbol{->}\AgdaSpace{}%
\AgdaPostulate{String}\<%
\\
\>[2]\AgdaFunction{bop}\AgdaSpace{}%
\AgdaInductiveConstructor{plus}\AgdaSpace{}%
\AgdaSymbol{=}\AgdaSpace{}%
\AgdaString{"+"}\<%
\\
\>[2]\AgdaFunction{bop}\AgdaSpace{}%
\AgdaInductiveConstructor{mul}\AgdaSpace{}%
\AgdaSymbol{=}\AgdaSpace{}%
\AgdaString{"*"}\<%
\\
\\[\AgdaEmptyExtraSkip]%
\>[2]\AgdaFunction{dim}\AgdaSpace{}%
\AgdaSymbol{:}\AgdaSpace{}%
\AgdaDatatype{S}\AgdaSpace{}%
\AgdaSymbol{→}\AgdaSpace{}%
\AgdaDatatype{ℕ}\<%
\\
\>[2]\AgdaFunction{dim}\AgdaSpace{}%
\AgdaSymbol{(}\AgdaInductiveConstructor{ι}\AgdaSpace{}%
\AgdaSymbol{\AgdaUnderscore{})}\AgdaSpace{}%
\AgdaSymbol{=}\AgdaSpace{}%
\AgdaNumber{1}\<%
\\
\>[2]\AgdaFunction{dim}\AgdaSpace{}%
\AgdaSymbol{(}\AgdaBound{s}\AgdaSpace{}%
\AgdaOperator{\AgdaInductiveConstructor{Array.⊗}}\AgdaSpace{}%
\AgdaBound{p}\AgdaSymbol{)}\AgdaSpace{}%
\AgdaSymbol{=}\AgdaSpace{}%
\AgdaFunction{dim}\AgdaSpace{}%
\AgdaBound{s}\AgdaSpace{}%
\AgdaOperator{\AgdaPrimitive{ℕ.+}}\AgdaSpace{}%
\AgdaFunction{dim}\AgdaSpace{}%
\AgdaBound{p}\<%
\\
\\[\AgdaEmptyExtraSkip]%
\>[2]\AgdaFunction{ivl}\AgdaSpace{}%
\AgdaSymbol{:}\AgdaSpace{}%
\AgdaDatatype{S}\AgdaSpace{}%
\AgdaSymbol{→}\AgdaSpace{}%
\AgdaFunction{State}\AgdaSpace{}%
\AgdaDatatype{ℕ}\AgdaSpace{}%
\AgdaSymbol{(}\AgdaDatatype{List}\AgdaSpace{}%
\AgdaPostulate{String}\AgdaSymbol{)}\<%
\\
\>[2]\AgdaFunction{ivl}\AgdaSpace{}%
\AgdaSymbol{(}\AgdaInductiveConstructor{ι}\AgdaSpace{}%
\AgdaSymbol{\AgdaUnderscore{})}\AgdaSpace{}%
\AgdaSymbol{=}\AgdaSpace{}%
\AgdaKeyword{do}\<%
\\
\>[2][@{}l@{\AgdaIndent{0}}]%
\>[4]\AgdaBound{v}\AgdaSpace{}%
\AgdaOperator{\AgdaFunction{←}}\AgdaSpace{}%
\AgdaField{get}\<%
\\
\>[4]\AgdaFunction{modify}\AgdaSpace{}%
\AgdaInductiveConstructor{suc}\<%
\\
\>[4]\AgdaFunction{return}\AgdaSpace{}%
\AgdaOperator{\AgdaFunction{\$}}\AgdaSpace{}%
\AgdaSymbol{(}\AgdaFunction{fresh-var}\AgdaSpace{}%
\AgdaBound{v}\AgdaSpace{}%
\AgdaOperator{\AgdaInductiveConstructor{∷}}\AgdaSpace{}%
\AgdaInductiveConstructor{[]}\AgdaSymbol{)}\<%
\\
\>[2]\AgdaFunction{ivl}\AgdaSpace{}%
\AgdaSymbol{(}\AgdaBound{s}\AgdaSpace{}%
\AgdaOperator{\AgdaInductiveConstructor{S.⊗}}\AgdaSpace{}%
\AgdaBound{p}\AgdaSymbol{)}\AgdaSpace{}%
\AgdaSymbol{=}\AgdaSpace{}%
\AgdaKeyword{do}\<%
\\
\>[2][@{}l@{\AgdaIndent{0}}]%
\>[4]\AgdaBound{l}\AgdaSpace{}%
\AgdaOperator{\AgdaFunction{←}}\AgdaSpace{}%
\AgdaFunction{ivl}\AgdaSpace{}%
\AgdaBound{s}\<%
\\
\>[4]\AgdaBound{r}\AgdaSpace{}%
\AgdaOperator{\AgdaFunction{←}}\AgdaSpace{}%
\AgdaFunction{ivl}\AgdaSpace{}%
\AgdaBound{p}\<%
\\
\>[4]\AgdaFunction{return}\AgdaSpace{}%
\AgdaOperator{\AgdaFunction{\$}}\AgdaSpace{}%
\AgdaBound{l}\AgdaSpace{}%
\AgdaOperator{\AgdaFunction{L.++}}\AgdaSpace{}%
\AgdaBound{r}\<%
\\
\>[0]\<%
\\
\>[2]\AgdaComment{--iv\ s\ =\ printf\ "[\%s]"\ ∘\ intersperse\ ",\ "\ <\$>\ ivl\ s}\<%
\end{code}}
\and
\codeblock{\begin{code}%
\>[2]\AgdaFunction{show-shape}\AgdaSpace{}%
\AgdaSymbol{:}\AgdaSpace{}%
\AgdaDatatype{S}\AgdaSpace{}%
\AgdaSymbol{→}\AgdaSpace{}%
\AgdaPostulate{String}\<%
\\
\>[2]\AgdaFunction{show-shape}\AgdaSpace{}%
\AgdaBound{s}%
\>[1867I]\AgdaSymbol{=}\AgdaSpace{}%
\AgdaFunction{printf}\AgdaSpace{}%
\AgdaString{"[\%s]"}\<%
\\
\>[.][@{}l@{}]\<[1867I]%
\>[15]\AgdaOperator{\AgdaFunction{\$}}\AgdaSpace{}%
\AgdaFunction{intersperse}\AgdaSpace{}%
\AgdaString{",\ "}\<%
\\
\>[15]\AgdaOperator{\AgdaFunction{\$}}\AgdaSpace{}%
\AgdaFunction{go}\AgdaSpace{}%
\AgdaBound{s}\<%
\\
\>[2][@{}l@{\AgdaIndent{0}}]%
\>[4]\AgdaKeyword{where}\<%
\\
\>[4][@{}l@{\AgdaIndent{0}}]%
\>[6]\AgdaFunction{go}\AgdaSpace{}%
\AgdaSymbol{:}\AgdaSpace{}%
\AgdaDatatype{S}\AgdaSpace{}%
\AgdaSymbol{→}\AgdaSpace{}%
\AgdaDatatype{List}\AgdaSpace{}%
\AgdaPostulate{String}\<%
\\
\>[6]\AgdaFunction{go}\AgdaSpace{}%
\AgdaSymbol{(}\AgdaInductiveConstructor{ι}\AgdaSpace{}%
\AgdaBound{x}\AgdaSymbol{)}%
\>[18]\AgdaSymbol{=}\AgdaSpace{}%
\AgdaFunction{show-nat}\AgdaSpace{}%
\AgdaBound{x}\AgdaSpace{}%
\AgdaOperator{\AgdaInductiveConstructor{∷}}\AgdaSpace{}%
\AgdaInductiveConstructor{[]}\<%
\\
\>[6]\AgdaFunction{go}\AgdaSpace{}%
\AgdaSymbol{(}\AgdaBound{s}\AgdaSpace{}%
\AgdaOperator{\AgdaInductiveConstructor{⊗}}\AgdaSpace{}%
\AgdaBound{p}\AgdaSymbol{)}%
\>[18]\AgdaSymbol{=}\AgdaSpace{}%
\AgdaFunction{go}\AgdaSpace{}%
\AgdaBound{s}\AgdaSpace{}%
\AgdaOperator{\AgdaFunction{++}}\AgdaSpace{}%
\AgdaFunction{go}\AgdaSpace{}%
\AgdaBound{p}\<%
\end{code}}
\and
\codeblock{\begin{code}%
\>[2]\AgdaFunction{to-sac}\AgdaSpace{}%
\AgdaSymbol{:}\AgdaSpace{}%
\AgdaSymbol{(}\AgdaBound{e}\AgdaSpace{}%
\AgdaSymbol{:}\AgdaSpace{}%
\AgdaDatatype{E}\AgdaSpace{}%
\AgdaGeneralizable{Γ}\AgdaSpace{}%
\AgdaGeneralizable{is}\AgdaSymbol{)}\AgdaSpace{}%
\AgdaSymbol{→}\AgdaSpace{}%
\AgdaSymbol{(}\AgdaBound{ρ}\AgdaSpace{}%
\AgdaSymbol{:}\AgdaSpace{}%
\AgdaFunction{SEnv}\AgdaSpace{}%
\AgdaGeneralizable{Γ}\AgdaSymbol{)}\AgdaSpace{}%
\AgdaSymbol{→}\AgdaSpace{}%
\AgdaFunction{State}\AgdaSpace{}%
\AgdaDatatype{ℕ}\AgdaSpace{}%
\AgdaPostulate{String}\<%
\\
\>[2]\AgdaFunction{to-sac}\AgdaSpace{}%
\AgdaSymbol{(}\AgdaInductiveConstructor{imap}\AgdaSpace{}%
\AgdaSymbol{\{}\AgdaArgument{s}\AgdaSpace{}%
\AgdaSymbol{=}\AgdaSpace{}%
\AgdaBound{s}\AgdaSymbol{\}}\AgdaSpace{}%
\AgdaBound{e}\AgdaSymbol{)}\AgdaSpace{}%
\AgdaBound{ρ}\AgdaSpace{}%
\AgdaSymbol{=}\AgdaSpace{}%
\AgdaKeyword{do}\<%
\\
\>[2][@{}l@{\AgdaIndent{0}}]%
\>[5]\AgdaBound{i}\AgdaSpace{}%
\AgdaOperator{\AgdaFunction{←}}\AgdaSpace{}%
\AgdaFunction{iv}\AgdaSpace{}%
\AgdaBound{s}\<%
\\
\>[5]\AgdaBound{b}\AgdaSpace{}%
\AgdaOperator{\AgdaFunction{←}}\AgdaSpace{}%
\AgdaFunction{to-sac}\AgdaSpace{}%
\AgdaBound{e}\AgdaSpace{}%
\AgdaSymbol{(}\AgdaBound{ρ}\AgdaSpace{}%
\AgdaOperator{\AgdaInductiveConstructor{,}}\AgdaSpace{}%
\AgdaBound{i}\AgdaSymbol{)}\<%
\\
\>[5]\AgdaFunction{return}\AgdaSpace{}%
\AgdaOperator{\AgdaFunction{\$}}\AgdaSpace{}%
\AgdaFunction{printf}%
\>[1927I]\AgdaString{"\{\ \%s\ ->\ \%s\ |\ \%s\ <\ \%s\ \}"}\<%
\\
\>[.][@{}l@{}]\<[1927I]%
\>[21]\AgdaBound{i}\AgdaSpace{}%
\AgdaBound{b}\AgdaSpace{}%
\AgdaBound{i}\AgdaSpace{}%
\AgdaSymbol{(}\AgdaFunction{show-shape}\AgdaSpace{}%
\AgdaBound{s}\AgdaSymbol{)}\<%
\\
\>[2]\AgdaFunction{to-sac}\AgdaSpace{}%
\AgdaSymbol{(}\AgdaInductiveConstructor{sel}\AgdaSpace{}%
\AgdaBound{e}\AgdaSpace{}%
\AgdaBound{e₁}\AgdaSymbol{)}\AgdaSpace{}%
\AgdaBound{ρ}\AgdaSpace{}%
\AgdaSymbol{=}\<%
\\
\>[2][@{}l@{\AgdaIndent{0}}]%
\>[5]\AgdaFunction{printf}\AgdaSpace{}%
\AgdaString{"(\%s)[\%s]"}\AgdaSpace{}%
\AgdaOperator{\AgdaFunction{<\$>}}\AgdaSpace{}%
\AgdaFunction{to-sac}\AgdaSpace{}%
\AgdaBound{e}\AgdaSpace{}%
\AgdaBound{ρ}\AgdaSpace{}%
\AgdaOperator{\AgdaFunction{⊛}}\AgdaSpace{}%
\AgdaFunction{to-sac}\AgdaSpace{}%
\AgdaBound{e₁}\AgdaSpace{}%
\AgdaBound{ρ}\<%
\\
\>[2]\AgdaComment{--\ ⋯}\<%
\end{code}}
\end{mathpar}
\begin{code}[hide]%
\>[2]\AgdaFunction{to-sac}\AgdaSpace{}%
\AgdaInductiveConstructor{zero}\AgdaSpace{}%
\AgdaBound{ρ}\AgdaSpace{}%
\AgdaSymbol{=}\AgdaSpace{}%
\AgdaFunction{return}\AgdaSpace{}%
\AgdaString{"zero"}\<%
\\
\>[2]\AgdaFunction{to-sac}\AgdaSpace{}%
\AgdaInductiveConstructor{one}\AgdaSpace{}%
\AgdaBound{ρ}\AgdaSpace{}%
\AgdaSymbol{=}\AgdaSpace{}%
\AgdaFunction{return}\AgdaSpace{}%
\AgdaString{"one"}\<%
\\
\>[2]\AgdaFunction{to-sac}\AgdaSpace{}%
\AgdaSymbol{(}\AgdaInductiveConstructor{var}\AgdaSpace{}%
\AgdaBound{x}\AgdaSymbol{)}\AgdaSpace{}%
\AgdaBound{ρ}\AgdaSpace{}%
\AgdaSymbol{=}\AgdaSpace{}%
\AgdaFunction{return}\AgdaSpace{}%
\AgdaOperator{\AgdaFunction{\$}}\AgdaSpace{}%
\AgdaFunction{lookup}\AgdaSpace{}%
\AgdaBound{x}\AgdaSpace{}%
\AgdaBound{ρ}\<%
\\
\>[2]\AgdaFunction{to-sac}\AgdaSpace{}%
\AgdaSymbol{(}\AgdaInductiveConstructor{imapₛ}\AgdaSpace{}%
\AgdaSymbol{\{}\AgdaArgument{s}\AgdaSpace{}%
\AgdaSymbol{=}\AgdaSpace{}%
\AgdaBound{s}\AgdaSymbol{\}}\AgdaSpace{}%
\AgdaBound{e}\AgdaSymbol{)}\AgdaSpace{}%
\AgdaBound{ρ}\AgdaSpace{}%
\AgdaSymbol{=}\AgdaSpace{}%
\AgdaKeyword{do}\<%
\\
\>[2][@{}l@{\AgdaIndent{0}}]%
\>[5]\AgdaBound{i}\AgdaSpace{}%
\AgdaOperator{\AgdaFunction{←}}\AgdaSpace{}%
\AgdaFunction{iv}\AgdaSpace{}%
\AgdaBound{s}\<%
\\
\>[5]\AgdaBound{b}\AgdaSpace{}%
\AgdaOperator{\AgdaFunction{←}}\AgdaSpace{}%
\AgdaFunction{to-sac}\AgdaSpace{}%
\AgdaBound{e}\AgdaSpace{}%
\AgdaSymbol{(}\AgdaBound{ρ}\AgdaSpace{}%
\AgdaOperator{\AgdaInductiveConstructor{,}}\AgdaSpace{}%
\AgdaBound{i}\AgdaSymbol{)}\<%
\\
\>[5]\AgdaKeyword{let}\AgdaSpace{}%
\AgdaBound{sh}\AgdaSpace{}%
\AgdaSymbol{=}\AgdaSpace{}%
\AgdaFunction{show-shape}\AgdaSpace{}%
\AgdaBound{s}\<%
\\
\>[5]\AgdaComment{--return\ \$\ printf\ "\{\ \%s\ ->\ \%s\ |\ \%s\ <\ \%s\ \}"\ i\ b\ i\ sh}\<%
\\
\>[5]\AgdaFunction{return}\AgdaSpace{}%
\AgdaOperator{\AgdaFunction{\$}}\AgdaSpace{}%
\AgdaFunction{printf}\AgdaSpace{}%
\AgdaString{"IMAPS(\%s,\ (\%s),\ (\%s))"}\AgdaSpace{}%
\AgdaBound{i}\AgdaSpace{}%
\AgdaBound{b}\AgdaSpace{}%
\AgdaBound{sh}\<%
\\
\>[2]\AgdaFunction{to-sac}\AgdaSpace{}%
\AgdaSymbol{(}\AgdaInductiveConstructor{selₛ}\AgdaSpace{}%
\AgdaBound{e}\AgdaSpace{}%
\AgdaBound{e₁}\AgdaSymbol{)}\AgdaSpace{}%
\AgdaBound{ρ}\AgdaSpace{}%
\AgdaSymbol{=}\AgdaSpace{}%
\AgdaKeyword{do}\<%
\\
\>[2][@{}l@{\AgdaIndent{0}}]%
\>[5]\AgdaBound{a}\AgdaSpace{}%
\AgdaOperator{\AgdaFunction{←}}\AgdaSpace{}%
\AgdaFunction{to-sac}\AgdaSpace{}%
\AgdaBound{e}\AgdaSpace{}%
\AgdaBound{ρ}\<%
\\
\>[5]\AgdaBound{i}\AgdaSpace{}%
\AgdaOperator{\AgdaFunction{←}}\AgdaSpace{}%
\AgdaFunction{to-sac}\AgdaSpace{}%
\AgdaBound{e₁}\AgdaSpace{}%
\AgdaBound{ρ}\<%
\\
\>[5]\AgdaComment{--return\ \$\ printf\ "(\%s)[\%s]"\ a\ i}\<%
\\
\>[5]\AgdaFunction{return}\AgdaSpace{}%
\AgdaOperator{\AgdaFunction{\$}}\AgdaSpace{}%
\AgdaFunction{printf}\AgdaSpace{}%
\AgdaString{"sels(\%s,\ \%s)"}\AgdaSpace{}%
\AgdaBound{a}\AgdaSpace{}%
\AgdaBound{i}\<%
\\
\\[\AgdaEmptyExtraSkip]%
\>[2]\AgdaComment{--\ Copy-paste\ from\ scalar\ versions}\<%
\\
\\[\AgdaEmptyExtraSkip]%
\>[2]\AgdaComment{--\ Copy-paste\ from\ scalar\ versions}\<%
\\
\>[2]\AgdaFunction{to-sac}\AgdaSpace{}%
\AgdaSymbol{(}\AgdaInductiveConstructor{imapb}\AgdaSpace{}%
\AgdaSymbol{\{}\AgdaArgument{s}\AgdaSpace{}%
\AgdaSymbol{=}\AgdaSpace{}%
\AgdaBound{s}\AgdaSymbol{\}\{}\AgdaBound{p}\AgdaSymbol{\}}\AgdaSpace{}%
\AgdaBound{m}\AgdaSpace{}%
\AgdaBound{e}\AgdaSymbol{)}\AgdaSpace{}%
\AgdaBound{ρ}\AgdaSpace{}%
\AgdaSymbol{=}\AgdaSpace{}%
\AgdaKeyword{do}\<%
\\
\>[2][@{}l@{\AgdaIndent{0}}]%
\>[5]\AgdaBound{i}\AgdaSpace{}%
\AgdaOperator{\AgdaFunction{←}}\AgdaSpace{}%
\AgdaFunction{iv}\AgdaSpace{}%
\AgdaBound{s}\<%
\\
\>[5]\AgdaBound{b}\AgdaSpace{}%
\AgdaOperator{\AgdaFunction{←}}\AgdaSpace{}%
\AgdaFunction{to-sac}\AgdaSpace{}%
\AgdaBound{e}\AgdaSpace{}%
\AgdaSymbol{(}\AgdaBound{ρ}\AgdaSpace{}%
\AgdaOperator{\AgdaInductiveConstructor{,}}\AgdaSpace{}%
\AgdaBound{i}\AgdaSymbol{)}\<%
\\
\>[5]\AgdaKeyword{let}\AgdaSpace{}%
\AgdaBound{sh-s}\AgdaSpace{}%
\AgdaSymbol{=}\AgdaSpace{}%
\AgdaFunction{show-shape}\AgdaSpace{}%
\AgdaBound{s}\<%
\\
\>[5]\AgdaKeyword{let}\AgdaSpace{}%
\AgdaBound{sh-p}\AgdaSpace{}%
\AgdaSymbol{=}\AgdaSpace{}%
\AgdaFunction{show-shape}\AgdaSpace{}%
\AgdaBound{p}\<%
\\
\>[5]\AgdaFunction{return}\AgdaSpace{}%
\AgdaOperator{\AgdaFunction{\$}}\AgdaSpace{}%
\AgdaFunction{printf}\AgdaSpace{}%
\AgdaString{"unblock(\{\ \%s\ ->\ \%s\ |\ \%s\ <\ \%s\ \},\ \%s)"}\AgdaSpace{}%
\AgdaBound{i}\AgdaSpace{}%
\AgdaBound{b}\AgdaSpace{}%
\AgdaBound{i}\AgdaSpace{}%
\AgdaBound{sh-s}\AgdaSpace{}%
\AgdaBound{sh-p}\<%
\\
\>[2]\AgdaFunction{to-sac}\AgdaSpace{}%
\AgdaSymbol{(}\AgdaInductiveConstructor{selb}\AgdaSpace{}%
\AgdaSymbol{\{}\AgdaArgument{p}\AgdaSpace{}%
\AgdaSymbol{=}\AgdaSpace{}%
\AgdaBound{p}\AgdaSymbol{\}}\AgdaSpace{}%
\AgdaBound{m}\AgdaSpace{}%
\AgdaBound{e}\AgdaSpace{}%
\AgdaBound{e₁}\AgdaSymbol{)}\AgdaSpace{}%
\AgdaBound{ρ}\AgdaSpace{}%
\AgdaSymbol{=}\AgdaSpace{}%
\AgdaKeyword{do}\<%
\\
\>[2][@{}l@{\AgdaIndent{0}}]%
\>[5]\AgdaBound{a}\AgdaSpace{}%
\AgdaOperator{\AgdaFunction{←}}\AgdaSpace{}%
\AgdaFunction{to-sac}\AgdaSpace{}%
\AgdaBound{e}\AgdaSpace{}%
\AgdaBound{ρ}\<%
\\
\>[5]\AgdaBound{i}\AgdaSpace{}%
\AgdaOperator{\AgdaFunction{←}}\AgdaSpace{}%
\AgdaFunction{to-sac}\AgdaSpace{}%
\AgdaBound{e₁}\AgdaSpace{}%
\AgdaBound{ρ}\<%
\\
\>[5]\AgdaKeyword{let}\AgdaSpace{}%
\AgdaBound{sh-p}\AgdaSpace{}%
\AgdaSymbol{=}\AgdaSpace{}%
\AgdaFunction{show-shape}\AgdaSpace{}%
\AgdaBound{p}\<%
\\
\>[5]\AgdaFunction{return}\AgdaSpace{}%
\AgdaOperator{\AgdaFunction{\$}}\AgdaSpace{}%
\AgdaFunction{printf}\AgdaSpace{}%
\AgdaString{"selb(\%s,\ \%s,\ \%s)"}\AgdaSpace{}%
\AgdaBound{a}\AgdaSpace{}%
\AgdaBound{i}\AgdaSpace{}%
\AgdaBound{sh-p}\<%
\\
\\[\AgdaEmptyExtraSkip]%
\>[2]\AgdaFunction{to-sac}\AgdaSpace{}%
\AgdaSymbol{(}\AgdaInductiveConstructor{zero-but}\AgdaSpace{}%
\AgdaBound{i}\AgdaSpace{}%
\AgdaBound{j}\AgdaSpace{}%
\AgdaBound{e}\AgdaSymbol{)}\AgdaSpace{}%
\AgdaBound{ρ}\<%
\\
\>[2][@{}l@{\AgdaIndent{0}}]%
\>[5]\AgdaSymbol{=}\AgdaSpace{}%
\AgdaFunction{printf}\AgdaSpace{}%
\AgdaString{"\%s\ ==\ \%s\ ?\ \%s\ :\ zero"}\AgdaSpace{}%
\AgdaOperator{\AgdaFunction{<\$>}}\AgdaSpace{}%
\AgdaSymbol{(}\AgdaFunction{to-sac}\AgdaSpace{}%
\AgdaBound{i}\AgdaSpace{}%
\AgdaBound{ρ}\AgdaSymbol{)}\AgdaSpace{}%
\AgdaOperator{\AgdaFunction{⊛}}\AgdaSpace{}%
\AgdaSymbol{(}\AgdaFunction{to-sac}\AgdaSpace{}%
\AgdaBound{j}\AgdaSpace{}%
\AgdaBound{ρ}\AgdaSymbol{)}\AgdaSpace{}%
\AgdaOperator{\AgdaFunction{⊛}}\AgdaSpace{}%
\AgdaSymbol{(}\AgdaFunction{to-sac}\AgdaSpace{}%
\AgdaBound{e}\AgdaSpace{}%
\AgdaBound{ρ}\AgdaSymbol{)}\<%
\\
\>[2]\AgdaFunction{to-sac}\AgdaSpace{}%
\AgdaSymbol{(}\AgdaInductiveConstructor{sum}\AgdaSpace{}%
\AgdaSymbol{\{}\AgdaArgument{s}\AgdaSpace{}%
\AgdaSymbol{=}\AgdaSpace{}%
\AgdaBound{s}\AgdaSymbol{\}}\AgdaSpace{}%
\AgdaSymbol{\{}\AgdaArgument{p}\AgdaSpace{}%
\AgdaSymbol{=}\AgdaSpace{}%
\AgdaBound{p}\AgdaSymbol{\}}\AgdaSpace{}%
\AgdaBound{e}\AgdaSymbol{)}\AgdaSpace{}%
\AgdaBound{ρ}\AgdaSpace{}%
\AgdaSymbol{=}\AgdaSpace{}%
\AgdaKeyword{do}\<%
\\
\>[2][@{}l@{\AgdaIndent{0}}]%
\>[5]\AgdaComment{--\ outer\ index\ }\<%
\\
\>[5]\AgdaBound{i}\AgdaSpace{}%
\AgdaOperator{\AgdaFunction{←}}\AgdaSpace{}%
\AgdaFunction{iv}\AgdaSpace{}%
\AgdaBound{s}\<%
\\
\>[5]\AgdaComment{--\ inner\ index\ which\ is\ juts\ a\ fresh\ name}\<%
\\
\>[5]\AgdaBound{j}\AgdaSpace{}%
\AgdaOperator{\AgdaFunction{←}}\AgdaSpace{}%
\AgdaFunction{iv}\AgdaSpace{}%
\AgdaBound{p}\<%
\\
\>[5]\AgdaBound{b}\AgdaSpace{}%
\AgdaOperator{\AgdaFunction{←}}\AgdaSpace{}%
\AgdaFunction{to-sac}\AgdaSpace{}%
\AgdaBound{e}\AgdaSpace{}%
\AgdaSymbol{(}\AgdaBound{ρ}\AgdaSpace{}%
\AgdaOperator{\AgdaInductiveConstructor{,}}\AgdaSpace{}%
\AgdaBound{i}\AgdaSymbol{)}\<%
\\
\>[5]\AgdaComment{--\ `s`\ is\ outer\ shape,\ and\ `p`\ is\ the\ inner\ one}\<%
\\
\>[5]\AgdaKeyword{let}\AgdaSpace{}%
\AgdaBound{sh-s}\AgdaSpace{}%
\AgdaSymbol{=}\AgdaSpace{}%
\AgdaFunction{show-shape}\AgdaSpace{}%
\AgdaBound{s}\<%
\\
\>[5]\AgdaKeyword{let}\AgdaSpace{}%
\AgdaBound{sh-p}\AgdaSpace{}%
\AgdaSymbol{=}\AgdaSpace{}%
\AgdaFunction{show-shape}\AgdaSpace{}%
\AgdaBound{p}\<%
\\
\>[5]\AgdaComment{--return\ \$\ printf\ "sumOuter(\%u,\ \{\ \%s\ ->\ \%s\ |\ \%s\ <\ \%s\})"\ (dim\ s)\ i\ b\ i\ sh-s}\<%
\\
\>[5]\AgdaComment{--\ sumOuter(ivOuter,\ ivInner,\ e,\ shOuter,\ shInner)}\<%
\\
\>[5]\AgdaFunction{return}\AgdaSpace{}%
\AgdaOperator{\AgdaFunction{\$}}\AgdaSpace{}%
\AgdaFunction{printf}\AgdaSpace{}%
\AgdaString{"sumOuter(\%s,\ \%s,\ \%s,\ (\%s),\ (\%s))"}\AgdaSpace{}%
\AgdaBound{i}\AgdaSpace{}%
\AgdaBound{j}\AgdaSpace{}%
\AgdaBound{b}\AgdaSpace{}%
\AgdaBound{sh-s}\AgdaSpace{}%
\AgdaBound{sh-p}\<%
\\
\>[2]\AgdaFunction{to-sac}\AgdaSpace{}%
\AgdaSymbol{(}\AgdaInductiveConstructor{bin}\AgdaSpace{}%
\AgdaBound{x}\AgdaSpace{}%
\AgdaBound{e}\AgdaSpace{}%
\AgdaBound{e₁}\AgdaSymbol{)}\AgdaSpace{}%
\AgdaBound{ρ}\AgdaSpace{}%
\AgdaSymbol{=}\AgdaSpace{}%
\AgdaKeyword{do}\<%
\\
\>[2][@{}l@{\AgdaIndent{0}}]%
\>[5]\AgdaBound{a}\AgdaSpace{}%
\AgdaOperator{\AgdaFunction{←}}\AgdaSpace{}%
\AgdaFunction{to-sac}\AgdaSpace{}%
\AgdaBound{e}\AgdaSpace{}%
\AgdaBound{ρ}\<%
\\
\>[5]\AgdaBound{b}\AgdaSpace{}%
\AgdaOperator{\AgdaFunction{←}}\AgdaSpace{}%
\AgdaFunction{to-sac}\AgdaSpace{}%
\AgdaBound{e₁}\AgdaSpace{}%
\AgdaBound{ρ}\<%
\\
\>[5]\AgdaFunction{return}\AgdaSpace{}%
\AgdaOperator{\AgdaFunction{\$}}\AgdaSpace{}%
\AgdaFunction{printf}\AgdaSpace{}%
\AgdaString{"(\%s)\ \%s\ (\%s)"}\AgdaSpace{}%
\AgdaBound{a}\AgdaSpace{}%
\AgdaSymbol{(}\AgdaFunction{bop}\AgdaSpace{}%
\AgdaBound{x}\AgdaSymbol{)}\AgdaSpace{}%
\AgdaBound{b}\<%
\\
\>[2]\AgdaFunction{to-sac}\AgdaSpace{}%
\AgdaSymbol{(}\AgdaInductiveConstructor{slide}\AgdaSpace{}%
\AgdaSymbol{\{}\AgdaArgument{p}\AgdaSpace{}%
\AgdaSymbol{=}\AgdaSpace{}%
\AgdaBound{p}\AgdaSymbol{\}}\AgdaSpace{}%
\AgdaBound{e}\AgdaSpace{}%
\AgdaBound{pl}\AgdaSpace{}%
\AgdaBound{e₁}\AgdaSpace{}%
\AgdaBound{su}\AgdaSymbol{)}\AgdaSpace{}%
\AgdaBound{ρ}\AgdaSpace{}%
\AgdaSymbol{=}\AgdaSpace{}%
\AgdaKeyword{do}\<%
\\
\>[2][@{}l@{\AgdaIndent{0}}]%
\>[5]\AgdaBound{i}\AgdaSpace{}%
\AgdaOperator{\AgdaFunction{←}}\AgdaSpace{}%
\AgdaFunction{to-sac}\AgdaSpace{}%
\AgdaBound{e}\AgdaSpace{}%
\AgdaBound{ρ}\<%
\\
\>[5]\AgdaBound{a}\AgdaSpace{}%
\AgdaOperator{\AgdaFunction{←}}\AgdaSpace{}%
\AgdaFunction{to-sac}\AgdaSpace{}%
\AgdaBound{e₁}\AgdaSpace{}%
\AgdaBound{ρ}\<%
\\
\>[5]\AgdaKeyword{let}\AgdaSpace{}%
\AgdaBound{sh-p}\AgdaSpace{}%
\AgdaSymbol{=}\AgdaSpace{}%
\AgdaFunction{show-shape}\AgdaSpace{}%
\AgdaBound{p}\<%
\\
\>[5]\AgdaFunction{return}\AgdaSpace{}%
\AgdaOperator{\AgdaFunction{\$}}\AgdaSpace{}%
\AgdaFunction{printf}\AgdaSpace{}%
\AgdaString{"slide(\%s,\ \%s,\ \%s)"}\AgdaSpace{}%
\AgdaBound{i}\AgdaSpace{}%
\AgdaBound{a}\AgdaSpace{}%
\AgdaBound{sh-p}\<%
\\
\>[2]\AgdaFunction{to-sac}\AgdaSpace{}%
\AgdaSymbol{(}\AgdaInductiveConstructor{backslide}\AgdaSpace{}%
\AgdaSymbol{\{}\AgdaArgument{r}\AgdaSpace{}%
\AgdaSymbol{=}\AgdaSpace{}%
\AgdaBound{r}\AgdaSymbol{\}}\AgdaSpace{}%
\AgdaBound{e}\AgdaSpace{}%
\AgdaBound{e₁}\AgdaSpace{}%
\AgdaBound{su}\AgdaSpace{}%
\AgdaBound{pl}\AgdaSymbol{)}\AgdaSpace{}%
\AgdaBound{ρ}\AgdaSpace{}%
\AgdaSymbol{=}\AgdaSpace{}%
\AgdaKeyword{do}\<%
\\
\>[2][@{}l@{\AgdaIndent{0}}]%
\>[5]\AgdaBound{i}\AgdaSpace{}%
\AgdaOperator{\AgdaFunction{←}}\AgdaSpace{}%
\AgdaFunction{to-sac}\AgdaSpace{}%
\AgdaBound{e}\AgdaSpace{}%
\AgdaBound{ρ}\<%
\\
\>[5]\AgdaBound{a}\AgdaSpace{}%
\AgdaOperator{\AgdaFunction{←}}\AgdaSpace{}%
\AgdaFunction{to-sac}\AgdaSpace{}%
\AgdaBound{e₁}\AgdaSpace{}%
\AgdaBound{ρ}\<%
\\
\>[5]\AgdaKeyword{let}\AgdaSpace{}%
\AgdaBound{sh-sp}\AgdaSpace{}%
\AgdaSymbol{=}\AgdaSpace{}%
\AgdaFunction{show-shape}\AgdaSpace{}%
\AgdaBound{r}\<%
\\
\>[5]\AgdaFunction{return}\AgdaSpace{}%
\AgdaOperator{\AgdaFunction{\$}}\AgdaSpace{}%
\AgdaFunction{printf}\AgdaSpace{}%
\AgdaString{"backlide(\%s,\ \%s,\ \%s)"}\AgdaSpace{}%
\AgdaBound{i}\AgdaSpace{}%
\AgdaBound{a}\AgdaSpace{}%
\AgdaBound{sh-sp}\<%
\\
\\[\AgdaEmptyExtraSkip]%
\>[2]\AgdaFunction{to-sac}\AgdaSpace{}%
\AgdaSymbol{(}\AgdaInductiveConstructor{scaledown}\AgdaSpace{}%
\AgdaBound{x}\AgdaSpace{}%
\AgdaBound{e}\AgdaSymbol{)}\AgdaSpace{}%
\AgdaBound{ρ}\AgdaSpace{}%
\AgdaSymbol{=}\AgdaSpace{}%
\AgdaKeyword{do}\<%
\\
\>[2][@{}l@{\AgdaIndent{0}}]%
\>[5]\AgdaBound{a}\AgdaSpace{}%
\AgdaOperator{\AgdaFunction{←}}\AgdaSpace{}%
\AgdaFunction{to-sac}\AgdaSpace{}%
\AgdaBound{e}\AgdaSpace{}%
\AgdaBound{ρ}\<%
\\
\>[5]\AgdaFunction{return}\AgdaSpace{}%
\AgdaOperator{\AgdaFunction{\$}}\AgdaSpace{}%
\AgdaFunction{printf}\AgdaSpace{}%
\AgdaString{"(\%s)\ /\ \%s"}\AgdaSpace{}%
\AgdaBound{a}\AgdaSpace{}%
\AgdaSymbol{(}\AgdaFunction{show-nat}\AgdaSpace{}%
\AgdaBound{x}\AgdaSymbol{)}\<%
\\
\\[\AgdaEmptyExtraSkip]%
\>[2]\AgdaFunction{to-sac}\AgdaSpace{}%
\AgdaSymbol{(}\AgdaInductiveConstructor{minus}\AgdaSpace{}%
\AgdaBound{e}\AgdaSymbol{)}\AgdaSpace{}%
\AgdaBound{ρ}\AgdaSpace{}%
\AgdaSymbol{=}\AgdaSpace{}%
\AgdaFunction{printf}\AgdaSpace{}%
\AgdaString{"-(\%s)"}\AgdaSpace{}%
\AgdaOperator{\AgdaFunction{<\$>}}\AgdaSpace{}%
\AgdaFunction{to-sac}\AgdaSpace{}%
\AgdaBound{e}\AgdaSpace{}%
\AgdaBound{ρ}\<%
\\
\>[2]\AgdaFunction{to-sac}\AgdaSpace{}%
\AgdaSymbol{(}\AgdaInductiveConstructor{logistic}\AgdaSpace{}%
\AgdaBound{e}\AgdaSymbol{)}\AgdaSpace{}%
\AgdaBound{ρ}\AgdaSpace{}%
\AgdaSymbol{=}\AgdaSpace{}%
\AgdaFunction{printf}\AgdaSpace{}%
\AgdaString{"logistics(\%s)"}\AgdaSpace{}%
\AgdaOperator{\AgdaFunction{<\$>}}\AgdaSpace{}%
\AgdaFunction{to-sac}\AgdaSpace{}%
\AgdaBound{e}\AgdaSpace{}%
\AgdaBound{ρ}\<%
\\
\\[\AgdaEmptyExtraSkip]%
\\[\AgdaEmptyExtraSkip]%
\>[2]\AgdaComment{--\ This\ can\ be\ made\ stateful,\ but\ we\ are\ assuming\ that}\<%
\\
\>[2]\AgdaComment{--\ vₛ\ is\ no\ need\ to\ make\ imap/sum\ index\ variables\ unique.}\<%
\\
\>[2]\AgdaFunction{env-sac}\AgdaSpace{}%
\AgdaSymbol{:}\AgdaSpace{}%
\AgdaFunction{AD.Env}\AgdaSpace{}%
\AgdaGeneralizable{Γ}\AgdaSpace{}%
\AgdaGeneralizable{Δ}\AgdaSpace{}%
\AgdaSymbol{→}\AgdaSpace{}%
\AgdaSymbol{(}\AgdaBound{vars}\AgdaSpace{}%
\AgdaSymbol{:}\AgdaSpace{}%
\AgdaFunction{SEnv}\AgdaSpace{}%
\AgdaGeneralizable{Δ}\AgdaSymbol{)}\AgdaSpace{}%
\AgdaSymbol{→}\AgdaSpace{}%
\AgdaFunction{SEnv}\AgdaSpace{}%
\AgdaGeneralizable{Γ}\<%
\\
\>[2]\AgdaFunction{env-sac}\AgdaSpace{}%
\AgdaSymbol{\{}\AgdaInductiveConstructor{ε}\AgdaSymbol{\}}\AgdaSpace{}%
\AgdaBound{ρ}\AgdaSpace{}%
\AgdaBound{σ}\AgdaSpace{}%
\AgdaSymbol{=}\AgdaSpace{}%
\AgdaSymbol{\AgdaUnderscore{}}\<%
\\
\>[2]\AgdaFunction{env-sac}\AgdaSpace{}%
\AgdaSymbol{\{}\AgdaBound{Γ}\AgdaSpace{}%
\AgdaOperator{\AgdaInductiveConstructor{▹}}\AgdaSpace{}%
\AgdaInductiveConstructor{ix}\AgdaSpace{}%
\AgdaBound{s}\AgdaSymbol{\}}\AgdaSpace{}%
\AgdaBound{ρ}\AgdaSpace{}%
\AgdaBound{σ}\AgdaSpace{}%
\AgdaSymbol{=}\AgdaSpace{}%
\AgdaFunction{env-sac}\AgdaSpace{}%
\AgdaBound{ρ}\AgdaSpace{}%
\AgdaBound{σ}\AgdaSpace{}%
\AgdaOperator{\AgdaInductiveConstructor{,}}\AgdaSpace{}%
\AgdaString{"--"}\<%
\\
\>[2]\AgdaFunction{env-sac}\AgdaSpace{}%
\AgdaSymbol{\{}\AgdaBound{Γ}\AgdaSpace{}%
\AgdaOperator{\AgdaInductiveConstructor{▹}}\AgdaSpace{}%
\AgdaInductiveConstructor{ar}\AgdaSpace{}%
\AgdaBound{s}\AgdaSymbol{\}}\AgdaSpace{}%
\AgdaSymbol{(}\AgdaBound{ρ}\AgdaSpace{}%
\AgdaOperator{\AgdaInductiveConstructor{,}}\AgdaSpace{}%
\AgdaBound{e}\AgdaSymbol{)}\AgdaSpace{}%
\AgdaBound{σ}\AgdaSpace{}%
\AgdaSymbol{=}\AgdaSpace{}%
\AgdaFunction{env-sac}\AgdaSpace{}%
\AgdaBound{ρ}\AgdaSpace{}%
\AgdaBound{σ}\AgdaSpace{}%
\AgdaOperator{\AgdaInductiveConstructor{,}}\AgdaSpace{}%
\AgdaField{proj₁}\AgdaSpace{}%
\AgdaSymbol{(}\AgdaFunction{to-sac}\AgdaSpace{}%
\AgdaBound{e}\AgdaSpace{}%
\AgdaBound{σ}\AgdaSpace{}%
\AgdaNumber{1}\AgdaSymbol{)}\<%
\\
\\[\AgdaEmptyExtraSkip]%
\>[2]\AgdaComment{--\ Reversed\ environment\ to\ list}\<%
\\
\>[2]\AgdaFunction{env-rev-list}\AgdaSpace{}%
\AgdaSymbol{:}\AgdaSpace{}%
\AgdaFunction{SEnv}\AgdaSpace{}%
\AgdaGeneralizable{Γ}\AgdaSpace{}%
\AgdaSymbol{→}\AgdaSpace{}%
\AgdaDatatype{List}\AgdaSpace{}%
\AgdaPostulate{String}\<%
\\
\>[2]\AgdaFunction{env-rev-list}\AgdaSpace{}%
\AgdaSymbol{\{}\AgdaInductiveConstructor{ε}\AgdaSymbol{\}}%
\>[23]\AgdaBound{ρ}\AgdaSpace{}%
\AgdaSymbol{=}\AgdaSpace{}%
\AgdaInductiveConstructor{[]}\<%
\\
\>[2]\AgdaFunction{env-rev-list}\AgdaSpace{}%
\AgdaSymbol{\{}\AgdaBound{Γ}\AgdaSpace{}%
\AgdaOperator{\AgdaInductiveConstructor{▹}}\AgdaSpace{}%
\AgdaSymbol{\AgdaUnderscore{}\}}\AgdaSpace{}%
\AgdaSymbol{(}\AgdaBound{ρ}\AgdaSpace{}%
\AgdaOperator{\AgdaInductiveConstructor{,}}\AgdaSpace{}%
\AgdaBound{x}\AgdaSymbol{)}\AgdaSpace{}%
\AgdaSymbol{=}\AgdaSpace{}%
\AgdaBound{x}\AgdaSpace{}%
\AgdaOperator{\AgdaInductiveConstructor{∷}}\AgdaSpace{}%
\AgdaFunction{env-rev-list}\AgdaSpace{}%
\AgdaBound{ρ}\<%
\\
\>[0]\<%
\\
\>[2]\AgdaComment{--\ zipWith\ for\ Environments}\<%
\\
\>[2]\AgdaFunction{zip-env}\AgdaSpace{}%
\AgdaSymbol{:}\AgdaSpace{}%
\AgdaSymbol{(}\AgdaPostulate{String}\AgdaSpace{}%
\AgdaSymbol{→}\AgdaSpace{}%
\AgdaPostulate{String}\AgdaSpace{}%
\AgdaSymbol{→}\AgdaSpace{}%
\AgdaPostulate{String}\AgdaSymbol{)}\AgdaSpace{}%
\AgdaSymbol{→}\AgdaSpace{}%
\AgdaFunction{SEnv}\AgdaSpace{}%
\AgdaGeneralizable{Γ}\AgdaSpace{}%
\AgdaSymbol{→}\AgdaSpace{}%
\AgdaFunction{SEnv}\AgdaSpace{}%
\AgdaGeneralizable{Γ}\AgdaSpace{}%
\AgdaSymbol{→}\AgdaSpace{}%
\AgdaFunction{SEnv}\AgdaSpace{}%
\AgdaGeneralizable{Γ}\<%
\\
\>[2]\AgdaFunction{zip-env}\AgdaSpace{}%
\AgdaSymbol{\{}\AgdaInductiveConstructor{ε}\AgdaSymbol{\}}%
\>[18]\AgdaBound{f}\AgdaSpace{}%
\AgdaInductiveConstructor{tt}%
\>[28]\AgdaInductiveConstructor{tt}%
\>[36]\AgdaSymbol{=}\AgdaSpace{}%
\AgdaInductiveConstructor{tt}\<%
\\
\>[2]\AgdaFunction{zip-env}\AgdaSpace{}%
\AgdaSymbol{\{}\AgdaBound{Γ}\AgdaSpace{}%
\AgdaOperator{\AgdaInductiveConstructor{▹}}\AgdaSpace{}%
\AgdaBound{x}\AgdaSymbol{\}}\AgdaSpace{}%
\AgdaBound{f}\AgdaSpace{}%
\AgdaSymbol{(}\AgdaBound{ν}\AgdaSpace{}%
\AgdaOperator{\AgdaInductiveConstructor{,}}\AgdaSpace{}%
\AgdaBound{n}\AgdaSymbol{)}\AgdaSpace{}%
\AgdaSymbol{(}\AgdaBound{ρ}\AgdaSpace{}%
\AgdaOperator{\AgdaInductiveConstructor{,}}\AgdaSpace{}%
\AgdaBound{e}\AgdaSymbol{)}\AgdaSpace{}%
\AgdaSymbol{=}\AgdaSpace{}%
\AgdaFunction{zip-env}\AgdaSpace{}%
\AgdaBound{f}\AgdaSpace{}%
\AgdaBound{ν}\AgdaSpace{}%
\AgdaBound{ρ}\AgdaSpace{}%
\AgdaOperator{\AgdaInductiveConstructor{,}}\AgdaSpace{}%
\AgdaBound{f}\AgdaSpace{}%
\AgdaBound{n}\AgdaSpace{}%
\AgdaBound{e}\<%
\end{code}

\subsubsection{SaC Primitives\label{sec:sac-primitives}}
As can be seen from the two cases of \AF{to-sac}, the extraction process is
not complicated. In essence, we define a small snippet of SaC code for 
each \AF{E} constructor.  Consider the \AC{imap}/\AC{sel}
family from the code snippet.  The \AC{imap} constructor maps directly to SaC's
tensor comprehensions~\cite{tensor-comp} expressed as: \texttt{\{ iv -> e | iv < s \}}.
This expression constructs arrays by evaluating \texttt{e} for every array non-negative index
vector
\texttt{iv} whose components are element-wise smaller than the shape \texttt{s}.  The shape of the resulting
array is concatenation of \texttt{s} and whatever the shape of \texttt{e} is.
Selections \AC{sel} correspond to the built-in array selection using
C-like syntax \texttt{e[iv]} where \texttt{e} is the array we are selecting
from and \texttt{iv} is the index vector.   Shape constraints are exactly as in
\AF{E}: if \texttt{e} is of shape \texttt{s ++ p}, and \texttt{iv} is bounded
by \texttt{s} then \texttt{e[iv]} is of shape \texttt{p}.

Scalar versions of imap/sel require a little wrapping.  For \AC{imapₛ} we
generate a tensor comprehension that selects inner expressions (they are
1-element vectors) at zero-th position.  For \AC{selₛ} we make selection into
an array and we wrap the result in a 1-d vector:
\begin{mathpar}
{\begin{varwidth}{0.9\textwidth}
\begin{lstlisting}[linewidth=.4\textwidth]
#define IMAPS(iv, e, shp) \
  {iv -> (e)[[0]] | iv < shp}
\end{lstlisting}
\end{varwidth}}
\and
{\begin{varwidth}{0.9\textwidth}
\begin{lstlisting}[linewidth=.55\textwidth]
inline float[1]
sels(float[d:shp] x, int[d] iv)
{
  return [x[iv]];
}
\end{lstlisting}
\end{varwidth}}
\end{mathpar}
When translating (\AC{imapₛ} \{ \AB{s} \} \AB{e}) we pick a fresh index variable
\texttt{iv}, then we translate \AB{e} (in the environment extended with \texttt{iv})
into \texttt{e'} and we generate \texttt{IMAPS(iv, e', shp)}, where \texttt{shp} is
a translation of \texttt{s}.  On the side of SaC we expand this macro as shown
above.  We could have expanded this macro on the Agda side, but this abstraction
makes it possible to make adjustments in the generated code without running Agda.
We map \AC{selₛ} into the \texttt{sels} function.  Consider the type of \texttt{sels}
which uses the recently added feature of SaC that makes it possible to encode
shape constraints in types~\cite{type-pattern}.  While these constraints are potentially checked at runtime,
they are very useful for readability and they provide some confidence about the
generated code.  The meaning of the type \texttt{float[d:shp]} is that it is
an array of base type \texttt{float} of rank \texttt{d} and shape \texttt{shp}.
When a variable of the same name is used within different arguments, it automatically
triggers the equality constraint between the corresponding ranks/shapes.

\paragraph{Blocking} Implementation of \AC{selb}/\AC{imapb} pair relies on
the notion of blocking, so we introduce the analogue to \AF{block}/\AF{unblock}
functionality in SaC as follows:
\begin{mathpar}
{\begin{varwidth}{0.9\textwidth}
\begin{lstlisting}[linewidth=.44\textwidth]
inline float[n:s,n:p]
block(float[n:sp] x, int[n] p)
     | all(s*p == sp)
     , all(p   >= 0)
{
  return { iv -> tile(p, iv * p, x) 
         | iv < sp / p};
}
\end{lstlisting}
\end{varwidth}}
\and
{\begin{varwidth}{0.9\textwidth}
\begin{lstlisting}[linewidth=.55\textwidth]
inline float[n:sp] 
unblock(float[n:s,n:p] a, int[n] p)
       | all(s*p == sp)
       , all(p   >= 0)
{
  return { iv -> a[(iv / p) ++ mod (iv, p)]
         | iv < s*p};
}
\end{lstlisting}
\end{varwidth}}
\end{mathpar}
The type \texttt{float[n:s,n:p]} denotes an array of the shape \texttt{s ++ p}
where \texttt{s} and \texttt{p} are of length \texttt{n}.  This is a product
shape in terms of our array theory.  As \texttt{sp} is just a variable that
is not related to \texttt{s} or \texttt{p}, we add two constraints (expressions
behind the bar after the function definition) saying that: (i) \texttt{sp} is
a point-wise product of \texttt{s} and \texttt{p}; (ii) all the elements of
the \texttt{p}-shape are greater than zero.  Keep in mind that these are potential
runtime constraints, they may be proved or flagged as disproved during compilation
but they do not provide a static guarantee. The implementation of block uses the \texttt{tile}
operation from the standard library of SaC. It selects a sub-array of the given shape at the given position.
In \texttt{unblock} we use a division and a modulo operation to remap the indices.
When translating \AC{selb}, we simply select into \texttt{block}-ed array.
When translating \AC{imapb}, we use the tensor comprehension as in case of
\AC{imap} to compute blocked array and then we call \texttt{unblock} on it.

\paragraph{Sliding} Slides and backslides are translated into calls to
the following SaC functions:
\begin{mathpar}
{\begin{varwidth}{0.9\textwidth}
\begin{lstlisting}
inline float[d:n1] 
slide(int[d] i, float[d:mn] x, int[d] n)       | all(n1        == n + 1)
                                               , all(n + 1 + i <= mn)
{
  return { iv -> x[iv + i] | iv < n + 1 };
}

inline float[d:mn]
backslide(int[d] i, float[d:n1] y, int[d] mn)  | all(i < 1 + mn - n1)
{
  return { iv -> y[iv - i] | i <= iv < n1 + i;
           iv -> 0f        |      iv < mn };
}
\end{lstlisting}
\end{varwidth}}
\end{mathpar}
Shape constraints become a little bit involved here because we implicitly
reconstruct the proof objects such as \AB{m} \AF{+} \AB{n} \AF{≈} \AB{mn}
and \AF{suc} \AB{n} \AF{≈} \AB{n1}.  Otherwise, \texttt{slide} selects a
sub-array of the shape (\texttt{n+1}) starting at the index \texttt{i}.
The \texttt{backslide} populates the sub-array with the elements of
\texttt{y} and the second partition of the tensor comprehension specifies
that all the other indices evaluate to zero.  Translation of \AC{slide}
and \AC{backslide} maps the arguments one-to-one, additionally providing
the $n$-shape in case of slide and the $(m+n)$ shape in case of backslide.

\paragraph{Summation} When translating (\AC{sum} \{\AB{s}\} \AB{e}), where
\AB{e} is of shape \AB{p} (and the index variable within the \AC{sum} is
bounded by \AB{s}), we map these arguments into the following SaC function:
\begin{lstlisting}
inline float[n:p] sumOuter(float[m:s,n:p] a, int[m] s, int[n] p) {
  return { jv -> sum({iv -> a[iv++jv] | iv < s}) | jv < p };
}
\end{lstlisting}
We use SaC's builtin \texttt{sum} function that sums-up all the elements
of the given array.

The rest of the constructions are mapped into regular arithmetic operations
that are provided by SaC.

\subsection{Local Variables}

The framework that we built so far computes derivatives of the variables in
the context.  This means that for complex expressions in \AF{E} (such as \AF{forward}),
all the let bindings will be inlined.  This is often not desirable both for performance
and readability.  Here we present a mechanism that introduce local variables
and preserves them during AD.
\begin{code}[hide]%
\>[0]\AgdaKeyword{module}\AgdaSpace{}%
\AgdaModule{DoubleChain}\AgdaSpace{}%
\AgdaKeyword{where}\<%
\\
\>[0][@{}l@{\AgdaIndent{0}}]%
\>[2]\AgdaComment{--\ In\ this\ module\ I\ want\ to\ preserve\ derivatives}\<%
\\
\>[2]\AgdaComment{--\ of\ the\ local\ variables\ in\ the\ chain\ (instead\ of\ inlining\ them)}\<%
\\
\>[2]\AgdaKeyword{open}\AgdaSpace{}%
\AgdaKeyword{import}\AgdaSpace{}%
\AgdaModule{Data.String}\<%
\\
\>[2]\AgdaKeyword{open}\AgdaSpace{}%
\AgdaKeyword{import}\AgdaSpace{}%
\AgdaModule{Text.Printf}\<%
\\
\>[2]\AgdaKeyword{open}\AgdaSpace{}%
\AgdaKeyword{import}\AgdaSpace{}%
\AgdaModule{Data.Product}\AgdaSpace{}%
\AgdaComment{--using\ (Σ;\ \AgdaUnderscore{}×\AgdaUnderscore{};\ \AgdaUnderscore{},\AgdaUnderscore{})}\<%
\\
\>[2]\AgdaKeyword{open}\AgdaSpace{}%
\AgdaKeyword{import}\AgdaSpace{}%
\AgdaModule{Data.Unit}\<%
\\
\>[2]\AgdaKeyword{open}\AgdaSpace{}%
\AgdaKeyword{import}\AgdaSpace{}%
\AgdaModule{Data.Nat}\AgdaSpace{}%
\AgdaSymbol{as}\AgdaSpace{}%
\AgdaModule{ℕ}\AgdaSpace{}%
\AgdaKeyword{using}\AgdaSpace{}%
\AgdaSymbol{(}\AgdaDatatype{ℕ}\AgdaSymbol{;}\AgdaSpace{}%
\AgdaInductiveConstructor{zero}\AgdaSymbol{;}\AgdaSpace{}%
\AgdaInductiveConstructor{suc}\AgdaSymbol{)}\<%
\\
\>[2]\AgdaKeyword{open}\AgdaSpace{}%
\AgdaKeyword{import}\AgdaSpace{}%
\AgdaModule{Data.List}\AgdaSpace{}%
\AgdaSymbol{as}\AgdaSpace{}%
\AgdaModule{L}\AgdaSpace{}%
\AgdaKeyword{using}\AgdaSpace{}%
\AgdaSymbol{(}\AgdaDatatype{List}\AgdaSymbol{;}\AgdaSpace{}%
\AgdaInductiveConstructor{[]}\AgdaSymbol{;}\AgdaSpace{}%
\AgdaOperator{\AgdaInductiveConstructor{\AgdaUnderscore{}∷\AgdaUnderscore{}}}\AgdaSymbol{)}\<%
\\
\>[2]\AgdaKeyword{open}\AgdaSpace{}%
\AgdaModule{Array}\AgdaSpace{}%
\AgdaKeyword{hiding}\AgdaSpace{}%
\AgdaSymbol{(}\AgdaFunction{sum}\AgdaSymbol{;}\AgdaSpace{}%
\AgdaFunction{slide}\AgdaSymbol{;}\AgdaSpace{}%
\AgdaFunction{backslide}\AgdaSymbol{)}\<%
\\
\>[2]\AgdaKeyword{open}\AgdaSpace{}%
\AgdaModule{Lang}\<%
\\
\>[2]\AgdaKeyword{open}\AgdaSpace{}%
\AgdaModule{SubWk}\<%
\\
\>[2]\AgdaKeyword{open}\AgdaSpace{}%
\AgdaModule{AD}\<%
\\
\>[2]\AgdaKeyword{open}\AgdaSpace{}%
\AgdaModule{Opt}\<%
\\
\>[2]\AgdaKeyword{open}\AgdaSpace{}%
\AgdaModule{BB}\<%
\\
\\[\AgdaEmptyExtraSkip]%
\>[2]\AgdaFunction{Env′}\AgdaSpace{}%
\AgdaSymbol{:}\AgdaSpace{}%
\AgdaDatatype{Ctx}\AgdaSpace{}%
\AgdaSymbol{→}\AgdaSpace{}%
\AgdaPrimitive{Set}\<%
\\
\>[2]\AgdaFunction{Env′}\AgdaSpace{}%
\AgdaBound{Γ}\AgdaSpace{}%
\AgdaSymbol{=}\AgdaSpace{}%
\AgdaFunction{Env}\AgdaSpace{}%
\AgdaBound{Γ}\AgdaSpace{}%
\AgdaBound{Γ}\<%
\end{code}

The key data structure that makes it possible to introduce local variables
is called \AF{Chain} which has two constructors.  The empty chain consists
of the names for all the variables in the context \AB{Γ}.  This represents the
case where no local variables have been introduced.  The \AC{\_▹\_} constructor
takes a chain in context \AB{Δ} and the array expression of shape \AB{p} in
the same context together with the variable name.  This produces the chain
in the context extended by two variables.  One variable is a place-holder
for the expression and the other variable is a placeholder for the derivative
of that expression.
\begin{code}%
\>[2]\AgdaKeyword{data}\AgdaSpace{}%
\AgdaDatatype{Chain}\AgdaSpace{}%
\AgdaSymbol{:}\AgdaSpace{}%
\AgdaDatatype{Ctx}\AgdaSpace{}%
\AgdaSymbol{→}\AgdaSpace{}%
\AgdaPrimitive{Set}\AgdaSpace{}%
\AgdaKeyword{where}\<%
\\
\>[2][@{}l@{\AgdaIndent{0}}]%
\>[4]\AgdaInductiveConstructor{ε}%
\>[9]\AgdaSymbol{:}\AgdaSpace{}%
\AgdaFunction{Sac.SEnv}\AgdaSpace{}%
\AgdaGeneralizable{Γ}\AgdaSpace{}%
\AgdaSymbol{→}\AgdaSpace{}%
\AgdaDatatype{Chain}\AgdaSpace{}%
\AgdaGeneralizable{Γ}\<%
\\
\>[4]\AgdaOperator{\AgdaInductiveConstructor{\AgdaUnderscore{}▹\AgdaUnderscore{}}}%
\>[9]\AgdaSymbol{:}\AgdaSpace{}%
\AgdaDatatype{Chain}\AgdaSpace{}%
\AgdaGeneralizable{Δ}\AgdaSpace{}%
\AgdaSymbol{→}\AgdaSpace{}%
\AgdaSymbol{(}\AgdaPostulate{String}\AgdaSpace{}%
\AgdaOperator{\AgdaFunction{×}}\AgdaSpace{}%
\AgdaDatatype{E}\AgdaSpace{}%
\AgdaGeneralizable{Δ}\AgdaSpace{}%
\AgdaSymbol{(}\AgdaInductiveConstructor{ar}\AgdaSpace{}%
\AgdaGeneralizable{p}\AgdaSymbol{))}\AgdaSpace{}%
\AgdaSymbol{→}\AgdaSpace{}%
\AgdaDatatype{Chain}\AgdaSpace{}%
\AgdaSymbol{(}\AgdaGeneralizable{Δ}\AgdaSpace{}%
\AgdaOperator{\AgdaInductiveConstructor{▹}}\AgdaSpace{}%
\AgdaInductiveConstructor{ar}\AgdaSpace{}%
\AgdaGeneralizable{p}\AgdaSpace{}%
\AgdaOperator{\AgdaInductiveConstructor{▹}}\AgdaSpace{}%
\AgdaInductiveConstructor{ar}\AgdaSpace{}%
\AgdaGeneralizable{p}\AgdaSymbol{)}\<%
\end{code}

The computation of the derivative in \AF{Chain}s follows the following
simple idea.  Consider the chain with two variables $a$ and
$b$ in the initial context \AB{Γ}, and two local variables $x$ and $y$.
Here is what happens when we compute the derivative of some expression
$e$ (that may depend on $a$, $b$, $x$, $y$) with some seed $s$ in the
empty $\delta_0$ environment. 

\begin{center}
\begin{tabular}{cc|cccc|l}
   $a$         &$b$         &$\partial{x}$& $x$         &$\partial{y}$&$y$       & \text{compute $\nabla\ e\ s\ \delta_0$}\\
   \hline
   $\delta_a$  &$\delta_b$  &-            & $\delta_x$  &-            &$\delta_y$& \text{assign $\delta_y$ to $\partial{y}$}\\
   $\delta_a$  &$\delta_b$  &-            & $\delta_x$  &$\delta_y$   &$\delta_y$& \text{compute $\nabla\ y_e\ \partial{y}$}\\
   $\delta'_a$ &$\delta'_b$ &-            & $\delta'_x$ &$\delta_y$   &$\delta_y$& \text{assign $\delta'_x$ to $\partial{x}$}\\
   $\delta'_a$ &$\delta'_b$ &-            & $\delta'_x$ &$\delta_y$   &$\delta_y$& \text{compute $\nabla\ x_e\ \partial{x}$}\\
   $\delta''_a$ &$\delta''_b$ &$\delta'_x$  & $\delta'_x$ &$\delta_y$   &$\delta_y$& \text{done}
\end{tabular}
\end{center}

First of all, the computation of $e$ returns the environment $\delta$ that can
be found in the first line of the table.  Then we repeat the following steps while
traversing the chain backwards: we copy the $y$-th position of the $\delta$-environment
to the $\partial{y}$-th position, and we compute the expression $y_e$ that is assigned to $y$
($xx$ in this case) with the seed $\partial{y}$-th variable.  Just to clarify, the seed
is the variable $\partial{y}$ and not its value.  Then we repeat the same process
for $x$ and potentially all the other remaining local variables (not in this case) until
we hit the beginning of the chain.

At the end of the process we obtain an environment where derivatives for $a$ and
$b$ are expressed in terms of $\partial{x}$ and $\partial{y}$.  The remaining step
is to collect the values of $\partial{x}$ and $\partial{y}$ which can be found
at the corresponding positions in the $\delta$-environment.
\begin{code}[hide]%
\>[2]\AgdaKeyword{data}\AgdaSpace{}%
\AgdaDatatype{LCtx}\AgdaSpace{}%
\AgdaSymbol{:}\AgdaSpace{}%
\AgdaPrimitive{Set}\AgdaSpace{}%
\AgdaKeyword{where}\<%
\\
\>[2][@{}l@{\AgdaIndent{0}}]%
\>[4]\AgdaInductiveConstructor{[]}%
\>[8]\AgdaSymbol{:}\AgdaSpace{}%
\AgdaDatatype{LCtx}\<%
\\
\>[4]\AgdaOperator{\AgdaInductiveConstructor{\AgdaUnderscore{}◃\AgdaUnderscore{}}}\AgdaSpace{}%
\AgdaSymbol{:}\AgdaSpace{}%
\AgdaDatatype{IS}\AgdaSpace{}%
\AgdaSymbol{→}\AgdaSpace{}%
\AgdaDatatype{LCtx}\AgdaSpace{}%
\AgdaSymbol{→}\AgdaSpace{}%
\AgdaDatatype{LCtx}\<%
\\
\\[\AgdaEmptyExtraSkip]%
\>[2]\AgdaOperator{\AgdaFunction{\AgdaUnderscore{}<><\AgdaUnderscore{}}}\AgdaSpace{}%
\AgdaSymbol{:}\AgdaSpace{}%
\AgdaDatatype{Ctx}\AgdaSpace{}%
\AgdaSymbol{→}\AgdaSpace{}%
\AgdaDatatype{LCtx}\AgdaSpace{}%
\AgdaSymbol{→}\AgdaSpace{}%
\AgdaDatatype{Ctx}\<%
\\
\>[2]\AgdaBound{Γ}\AgdaSpace{}%
\AgdaOperator{\AgdaFunction{<><}}\AgdaSpace{}%
\AgdaInductiveConstructor{[]}\AgdaSpace{}%
\AgdaSymbol{=}\AgdaSpace{}%
\AgdaBound{Γ}\<%
\\
\>[2]\AgdaBound{Γ}\AgdaSpace{}%
\AgdaOperator{\AgdaFunction{<><}}\AgdaSpace{}%
\AgdaSymbol{(}\AgdaBound{x}\AgdaSpace{}%
\AgdaOperator{\AgdaInductiveConstructor{◃}}\AgdaSpace{}%
\AgdaBound{Δ}\AgdaSymbol{)}\AgdaSpace{}%
\AgdaSymbol{=}\AgdaSpace{}%
\AgdaSymbol{(}\AgdaBound{Γ}\AgdaSpace{}%
\AgdaOperator{\AgdaInductiveConstructor{▹}}\AgdaSpace{}%
\AgdaBound{x}\AgdaSymbol{)}\AgdaSpace{}%
\AgdaOperator{\AgdaFunction{<><}}\AgdaSpace{}%
\AgdaBound{Δ}\<%
\\
\\[\AgdaEmptyExtraSkip]%
\>[2]\AgdaKeyword{data}\AgdaSpace{}%
\AgdaDatatype{LEnv}\AgdaSpace{}%
\AgdaSymbol{:}\AgdaSpace{}%
\AgdaDatatype{LCtx}\AgdaSpace{}%
\AgdaSymbol{→}\AgdaSpace{}%
\AgdaDatatype{Ctx}\AgdaSpace{}%
\AgdaSymbol{→}\AgdaSpace{}%
\AgdaPrimitive{Set}\AgdaSpace{}%
\AgdaKeyword{where}\<%
\\
\>[2][@{}l@{\AgdaIndent{0}}]%
\>[4]\AgdaInductiveConstructor{[]}%
\>[8]\AgdaSymbol{:}\AgdaSpace{}%
\AgdaDatatype{LEnv}\AgdaSpace{}%
\AgdaInductiveConstructor{[]}\AgdaSpace{}%
\AgdaGeneralizable{Γ}\<%
\\
\>[4]\AgdaOperator{\AgdaInductiveConstructor{\AgdaUnderscore{}◃\AgdaUnderscore{}}}\AgdaSpace{}%
\AgdaSymbol{:}\AgdaSpace{}%
\AgdaSymbol{∀}\AgdaSpace{}%
\AgdaSymbol{\{}\AgdaBound{Δ′}\AgdaSymbol{\}}\AgdaSpace{}%
\AgdaSymbol{→}\AgdaSpace{}%
\AgdaDatatype{E}\AgdaSpace{}%
\AgdaGeneralizable{Γ}\AgdaSpace{}%
\AgdaSymbol{(}\AgdaInductiveConstructor{ar}\AgdaSpace{}%
\AgdaGeneralizable{s}\AgdaSymbol{)}\AgdaSpace{}%
\AgdaSymbol{→}\AgdaSpace{}%
\AgdaDatatype{LEnv}\AgdaSpace{}%
\AgdaBound{Δ′}\AgdaSpace{}%
\AgdaGeneralizable{Γ}\AgdaSpace{}%
\AgdaSymbol{→}\AgdaSpace{}%
\AgdaDatatype{LEnv}\AgdaSpace{}%
\AgdaSymbol{(}\AgdaInductiveConstructor{ar}\AgdaSpace{}%
\AgdaGeneralizable{s}\AgdaSpace{}%
\AgdaOperator{\AgdaInductiveConstructor{◃}}\AgdaSpace{}%
\AgdaBound{Δ′}\AgdaSymbol{)}\AgdaSpace{}%
\AgdaGeneralizable{Γ}\<%
\\
\\[\AgdaEmptyExtraSkip]%
\>[2]\AgdaKeyword{data}\AgdaSpace{}%
\AgdaDatatype{Postfix}\AgdaSpace{}%
\AgdaSymbol{:}\AgdaSpace{}%
\AgdaDatatype{Ctx}\AgdaSpace{}%
\AgdaSymbol{→}\AgdaSpace{}%
\AgdaDatatype{Ctx}\AgdaSpace{}%
\AgdaSymbol{→}\AgdaSpace{}%
\AgdaPrimitive{Set}\AgdaSpace{}%
\AgdaKeyword{where}\<%
\\
\>[2][@{}l@{\AgdaIndent{0}}]%
\>[4]\AgdaInductiveConstructor{done}\AgdaSpace{}%
\AgdaSymbol{:}\AgdaSpace{}%
\AgdaDatatype{Postfix}\AgdaSpace{}%
\AgdaInductiveConstructor{ε}\AgdaSpace{}%
\AgdaGeneralizable{Γ}\<%
\\
\>[4]\AgdaInductiveConstructor{next}\AgdaSpace{}%
\AgdaSymbol{:}\AgdaSpace{}%
\AgdaDatatype{Postfix}\AgdaSpace{}%
\AgdaGeneralizable{Γ}\AgdaSpace{}%
\AgdaGeneralizable{Δ}\AgdaSpace{}%
\AgdaSymbol{→}\AgdaSpace{}%
\AgdaDatatype{Postfix}\AgdaSpace{}%
\AgdaSymbol{(}\AgdaGeneralizable{Γ}\AgdaSpace{}%
\AgdaOperator{\AgdaInductiveConstructor{▹}}\AgdaSpace{}%
\AgdaInductiveConstructor{ar}\AgdaSpace{}%
\AgdaGeneralizable{s}\AgdaSymbol{)}\AgdaSpace{}%
\AgdaSymbol{(}\AgdaGeneralizable{Δ}\AgdaSpace{}%
\AgdaOperator{\AgdaInductiveConstructor{▹}}\AgdaSpace{}%
\AgdaInductiveConstructor{ar}\AgdaSpace{}%
\AgdaGeneralizable{s}\AgdaSymbol{)}\<%
\\
\\[\AgdaEmptyExtraSkip]%
\>[2]\AgdaFunction{double-ctx}\AgdaSpace{}%
\AgdaSymbol{:}\AgdaSpace{}%
\AgdaDatatype{Ctx}\AgdaSpace{}%
\AgdaSymbol{→}\AgdaSpace{}%
\AgdaDatatype{Ctx}\<%
\\
\>[2]\AgdaFunction{double-ctx}\AgdaSpace{}%
\AgdaInductiveConstructor{ε}\AgdaSpace{}%
\AgdaSymbol{=}\AgdaSpace{}%
\AgdaInductiveConstructor{ε}\<%
\\
\>[2]\AgdaFunction{double-ctx}\AgdaSpace{}%
\AgdaSymbol{(}\AgdaBound{Γ}\AgdaSpace{}%
\AgdaOperator{\AgdaInductiveConstructor{▹}}\AgdaSpace{}%
\AgdaBound{x}\AgdaSymbol{)}\AgdaSpace{}%
\AgdaSymbol{=}\AgdaSpace{}%
\AgdaFunction{double-ctx}\AgdaSpace{}%
\AgdaBound{Γ}\AgdaSpace{}%
\AgdaOperator{\AgdaInductiveConstructor{▹}}\AgdaSpace{}%
\AgdaBound{x}\AgdaSpace{}%
\AgdaOperator{\AgdaInductiveConstructor{▹}}\AgdaSpace{}%
\AgdaBound{x}\<%
\\
\\[\AgdaEmptyExtraSkip]%
\>[2]\AgdaFunction{chain-to-env}\AgdaSpace{}%
\AgdaSymbol{:}\AgdaSpace{}%
\AgdaDatatype{Chain}\AgdaSpace{}%
\AgdaGeneralizable{Γ}\AgdaSpace{}%
\AgdaSymbol{→}\AgdaSpace{}%
\AgdaRecord{Σ}\AgdaSpace{}%
\AgdaDatatype{Ctx}\AgdaSpace{}%
\AgdaSymbol{λ}\AgdaSpace{}%
\AgdaBound{Δ}\AgdaSpace{}%
\AgdaSymbol{→}\AgdaSpace{}%
\AgdaFunction{Env}\AgdaSpace{}%
\AgdaSymbol{(}\AgdaFunction{double-ctx}\AgdaSpace{}%
\AgdaBound{Δ}\AgdaSymbol{)}\AgdaSpace{}%
\AgdaGeneralizable{Γ}\AgdaSpace{}%
\AgdaOperator{\AgdaFunction{×}}\AgdaSpace{}%
\AgdaDatatype{Postfix}\AgdaSpace{}%
\AgdaSymbol{(}\AgdaFunction{double-ctx}\AgdaSpace{}%
\AgdaBound{Δ}\AgdaSymbol{)}\AgdaSpace{}%
\AgdaGeneralizable{Γ}\<%
\\
\>[2]\AgdaFunction{chain-to-env}\AgdaSpace{}%
\AgdaSymbol{(}\AgdaInductiveConstructor{ε}\AgdaSpace{}%
\AgdaBound{x}\AgdaSymbol{)}%
\>[23]\AgdaSymbol{=}\AgdaSpace{}%
\AgdaInductiveConstructor{ε}\AgdaSpace{}%
\AgdaOperator{\AgdaInductiveConstructor{,}}\AgdaSpace{}%
\AgdaInductiveConstructor{tt}\AgdaSpace{}%
\AgdaOperator{\AgdaInductiveConstructor{,}}\AgdaSpace{}%
\AgdaInductiveConstructor{done}\<%
\\
\>[2]\AgdaFunction{chain-to-env}\AgdaSpace{}%
\AgdaSymbol{(}\AgdaOperator{\AgdaInductiveConstructor{\AgdaUnderscore{}▹\AgdaUnderscore{}}}\AgdaSpace{}%
\AgdaSymbol{\{}\AgdaArgument{p}\AgdaSpace{}%
\AgdaSymbol{=}\AgdaSpace{}%
\AgdaBound{p}\AgdaSymbol{\}}\AgdaSpace{}%
\AgdaBound{c}\AgdaSpace{}%
\AgdaSymbol{(\AgdaUnderscore{}}\AgdaSpace{}%
\AgdaOperator{\AgdaInductiveConstructor{,}}\AgdaSpace{}%
\AgdaBound{x}\AgdaSymbol{))}\AgdaSpace{}%
\AgdaSymbol{=}\AgdaSpace{}%
\AgdaKeyword{let}\<%
\\
\>[2][@{}l@{\AgdaIndent{0}}]%
\>[4]\AgdaBound{Δ}\AgdaSpace{}%
\AgdaOperator{\AgdaInductiveConstructor{,}}\AgdaSpace{}%
\AgdaBound{ρ}\AgdaSpace{}%
\AgdaOperator{\AgdaInductiveConstructor{,}}\AgdaSpace{}%
\AgdaBound{po}\AgdaSpace{}%
\AgdaSymbol{=}\AgdaSpace{}%
\AgdaFunction{chain-to-env}\AgdaSpace{}%
\AgdaBound{c}\<%
\\
\>[4]\AgdaKeyword{in}\AgdaSpace{}%
\AgdaSymbol{(}\AgdaBound{Δ}\AgdaSpace{}%
\AgdaOperator{\AgdaInductiveConstructor{▹}}\AgdaSpace{}%
\AgdaInductiveConstructor{ar}\AgdaSpace{}%
\AgdaBound{p}\AgdaSymbol{)}\AgdaSpace{}%
\AgdaOperator{\AgdaInductiveConstructor{,}}\AgdaSpace{}%
\AgdaSymbol{((}\AgdaFunction{env-map}\AgdaSpace{}%
\AgdaSymbol{\{}\AgdaArgument{Γ}\AgdaSpace{}%
\AgdaSymbol{=}\AgdaSpace{}%
\AgdaFunction{double-ctx}\AgdaSpace{}%
\AgdaBound{Δ}\AgdaSymbol{\}}\AgdaSpace{}%
\AgdaSymbol{(}\AgdaOperator{\AgdaFunction{↑↑\AgdaUnderscore{}}}\AgdaSymbol{)}\AgdaSpace{}%
\AgdaBound{ρ}\AgdaSpace{}%
\AgdaOperator{\AgdaInductiveConstructor{,}}\AgdaSpace{}%
\AgdaInductiveConstructor{zero}\AgdaSymbol{)}\AgdaSpace{}%
\AgdaOperator{\AgdaInductiveConstructor{,}}\AgdaSpace{}%
\AgdaSymbol{(}\AgdaOperator{\AgdaFunction{↑}}\AgdaSpace{}%
\AgdaOperator{\AgdaFunction{↑}}\AgdaSpace{}%
\AgdaBound{x}\AgdaSymbol{))}\AgdaSpace{}%
\AgdaOperator{\AgdaInductiveConstructor{,}}\AgdaSpace{}%
\AgdaSymbol{(}\AgdaInductiveConstructor{next}\AgdaSpace{}%
\AgdaSymbol{(}\AgdaInductiveConstructor{next}\AgdaSpace{}%
\AgdaBound{po}\AgdaSymbol{))}\<%
\\
\\[\AgdaEmptyExtraSkip]%
\>[2]\AgdaFunction{pstep}\AgdaSpace{}%
\AgdaSymbol{:}\AgdaSpace{}%
\AgdaSymbol{∀}\AgdaSpace{}%
\AgdaSymbol{\{}\AgdaBound{Δ′}\AgdaSymbol{\}}\AgdaSpace{}%
\AgdaSymbol{→}\AgdaSpace{}%
\AgdaDatatype{Postfix}\AgdaSpace{}%
\AgdaSymbol{((}\AgdaGeneralizable{Δ}\AgdaSpace{}%
\AgdaOperator{\AgdaInductiveConstructor{▹}}\AgdaSpace{}%
\AgdaInductiveConstructor{ar}\AgdaSpace{}%
\AgdaGeneralizable{s}\AgdaSymbol{)}\AgdaSpace{}%
\AgdaOperator{\AgdaFunction{<><}}\AgdaSpace{}%
\AgdaBound{Δ′}\AgdaSymbol{)}\AgdaSpace{}%
\AgdaGeneralizable{Γ}\AgdaSpace{}%
\AgdaSymbol{→}\AgdaSpace{}%
\AgdaDatatype{Postfix}\AgdaSpace{}%
\AgdaSymbol{(}\AgdaGeneralizable{Δ}\AgdaSpace{}%
\AgdaOperator{\AgdaFunction{<><}}\AgdaSpace{}%
\AgdaSymbol{(}\AgdaInductiveConstructor{ar}\AgdaSpace{}%
\AgdaGeneralizable{s}\AgdaSpace{}%
\AgdaOperator{\AgdaInductiveConstructor{◃}}\AgdaSpace{}%
\AgdaBound{Δ′}\AgdaSymbol{))}\AgdaSpace{}%
\AgdaGeneralizable{Γ}\<%
\\
\>[2]\AgdaFunction{pstep}\AgdaSpace{}%
\AgdaSymbol{\{}\AgdaArgument{Δ′}\AgdaSpace{}%
\AgdaSymbol{=}\AgdaSpace{}%
\AgdaInductiveConstructor{[]}\AgdaSymbol{\}}\AgdaSpace{}%
\AgdaSymbol{(}\AgdaInductiveConstructor{next}\AgdaSpace{}%
\AgdaBound{p}\AgdaSymbol{)}\AgdaSpace{}%
\AgdaSymbol{=}\AgdaSpace{}%
\AgdaInductiveConstructor{next}\AgdaSpace{}%
\AgdaBound{p}\<%
\\
\>[2]\AgdaFunction{pstep}\AgdaSpace{}%
\AgdaSymbol{\{}\AgdaArgument{Δ′}\AgdaSpace{}%
\AgdaSymbol{=}\AgdaSpace{}%
\AgdaBound{x}\AgdaSpace{}%
\AgdaOperator{\AgdaInductiveConstructor{◃}}\AgdaSpace{}%
\AgdaBound{Δ′}\AgdaSymbol{\}}\AgdaSpace{}%
\AgdaBound{p}\AgdaSpace{}%
\AgdaSymbol{=}\AgdaSpace{}%
\AgdaBound{p}\<%
\\
\\[\AgdaEmptyExtraSkip]%
\>[2]\AgdaFunction{post-var}\AgdaSpace{}%
\AgdaSymbol{:}\AgdaSpace{}%
\AgdaSymbol{∀}\AgdaSpace{}%
\AgdaSymbol{\{}\AgdaBound{Δ′}\AgdaSymbol{\}}\AgdaSpace{}%
\AgdaSymbol{→}\AgdaSpace{}%
\AgdaDatatype{Postfix}\AgdaSpace{}%
\AgdaSymbol{(}\AgdaGeneralizable{Δ}\AgdaSpace{}%
\AgdaOperator{\AgdaFunction{<><}}\AgdaSpace{}%
\AgdaBound{Δ′}\AgdaSymbol{)}\AgdaSpace{}%
\AgdaGeneralizable{Γ}\AgdaSpace{}%
\AgdaSymbol{→}\AgdaSpace{}%
\AgdaGeneralizable{is}\AgdaSpace{}%
\AgdaOperator{\AgdaDatatype{∈}}\AgdaSpace{}%
\AgdaGeneralizable{Δ}\AgdaSpace{}%
\AgdaSymbol{→}\AgdaSpace{}%
\AgdaGeneralizable{is}\AgdaSpace{}%
\AgdaOperator{\AgdaDatatype{∈}}\AgdaSpace{}%
\AgdaGeneralizable{Γ}\<%
\\
\>[2]\AgdaFunction{post-var}\AgdaSpace{}%
\AgdaSymbol{\{}\AgdaArgument{Δ′}\AgdaSpace{}%
\AgdaSymbol{=}\AgdaSpace{}%
\AgdaInductiveConstructor{[]}\AgdaSymbol{\}}\AgdaSpace{}%
\AgdaSymbol{(}\AgdaInductiveConstructor{next}\AgdaSpace{}%
\AgdaBound{p}\AgdaSymbol{)}\AgdaSpace{}%
\AgdaInductiveConstructor{v₀}\AgdaSpace{}%
\AgdaSymbol{=}\AgdaSpace{}%
\AgdaInductiveConstructor{v₀}\<%
\\
\>[2]\AgdaFunction{post-var}\AgdaSpace{}%
\AgdaSymbol{\{}\AgdaArgument{Δ′}\AgdaSpace{}%
\AgdaSymbol{=}\AgdaSpace{}%
\AgdaInductiveConstructor{[]}\AgdaSymbol{\}}\AgdaSpace{}%
\AgdaSymbol{(}\AgdaInductiveConstructor{next}\AgdaSpace{}%
\AgdaBound{p}\AgdaSymbol{)}\AgdaSpace{}%
\AgdaSymbol{(}\AgdaInductiveConstructor{vₛ}\AgdaSpace{}%
\AgdaBound{x}\AgdaSymbol{)}\AgdaSpace{}%
\AgdaSymbol{=}\AgdaSpace{}%
\AgdaInductiveConstructor{vₛ}\AgdaSpace{}%
\AgdaSymbol{(}\AgdaFunction{post-var}\AgdaSpace{}%
\AgdaSymbol{\{}\AgdaArgument{Δ′}\AgdaSpace{}%
\AgdaSymbol{=}\AgdaSpace{}%
\AgdaInductiveConstructor{[]}\AgdaSymbol{\}}\AgdaSpace{}%
\AgdaBound{p}\AgdaSpace{}%
\AgdaBound{x}\AgdaSymbol{)}\<%
\\
\>[2]\AgdaFunction{post-var}\AgdaSpace{}%
\AgdaSymbol{\{}\AgdaArgument{Δ′}\AgdaSpace{}%
\AgdaSymbol{=}\AgdaSpace{}%
\AgdaBound{is}\AgdaSpace{}%
\AgdaOperator{\AgdaInductiveConstructor{◃}}\AgdaSpace{}%
\AgdaBound{Δ′}\AgdaSymbol{\}}\AgdaSpace{}%
\AgdaBound{p}\AgdaSpace{}%
\AgdaBound{x}\AgdaSpace{}%
\AgdaSymbol{=}\AgdaSpace{}%
\AgdaFunction{post-var}\AgdaSpace{}%
\AgdaSymbol{\{}\AgdaArgument{Δ′}\AgdaSpace{}%
\AgdaSymbol{=}\AgdaSpace{}%
\AgdaBound{Δ′}\AgdaSymbol{\}}\AgdaSpace{}%
\AgdaBound{p}\AgdaSpace{}%
\AgdaSymbol{(}\AgdaInductiveConstructor{vₛ}\AgdaSpace{}%
\AgdaBound{x}\AgdaSymbol{)}\<%
\\
\\[\AgdaEmptyExtraSkip]%
\>[2]\AgdaFunction{no-ix}\AgdaSpace{}%
\AgdaSymbol{:}\AgdaSpace{}%
\AgdaInductiveConstructor{ix}\AgdaSpace{}%
\AgdaGeneralizable{s}\AgdaSpace{}%
\AgdaOperator{\AgdaDatatype{∈}}\AgdaSpace{}%
\AgdaGeneralizable{Δ}\AgdaSpace{}%
\AgdaSymbol{→}\AgdaSpace{}%
\AgdaOperator{\AgdaFunction{¬}}\AgdaSpace{}%
\AgdaDatatype{Postfix}\AgdaSpace{}%
\AgdaGeneralizable{Δ}\AgdaSpace{}%
\AgdaGeneralizable{Γ}\<%
\\
\>[2]\AgdaFunction{no-ix}\AgdaSpace{}%
\AgdaInductiveConstructor{v₀}\AgdaSpace{}%
\AgdaSymbol{=}\AgdaSpace{}%
\AgdaSymbol{λ}\AgdaSpace{}%
\AgdaSymbol{()}\<%
\\
\>[2]\AgdaFunction{no-ix}\AgdaSpace{}%
\AgdaSymbol{(}\AgdaInductiveConstructor{vₛ}\AgdaSpace{}%
\AgdaBound{v}\AgdaSymbol{)}\AgdaSpace{}%
\AgdaSymbol{(}\AgdaInductiveConstructor{next}\AgdaSpace{}%
\AgdaBound{p}\AgdaSymbol{)}\AgdaSpace{}%
\AgdaSymbol{=}\AgdaSpace{}%
\AgdaFunction{no-ix}\AgdaSpace{}%
\AgdaBound{v}\AgdaSpace{}%
\AgdaBound{p}\<%
\\
\\[\AgdaEmptyExtraSkip]%
\>[2]\AgdaFunction{post-fish}\AgdaSpace{}%
\AgdaSymbol{:}\AgdaSpace{}%
\AgdaSymbol{∀}\AgdaSpace{}%
\AgdaBound{Δ′}\AgdaSpace{}%
\AgdaSymbol{→}\AgdaSpace{}%
\AgdaGeneralizable{is}\AgdaSpace{}%
\AgdaOperator{\AgdaDatatype{∈}}\AgdaSpace{}%
\AgdaGeneralizable{Δ}\AgdaSpace{}%
\AgdaSymbol{→}\AgdaSpace{}%
\AgdaGeneralizable{is}\AgdaSpace{}%
\AgdaOperator{\AgdaDatatype{∈}}\AgdaSpace{}%
\AgdaSymbol{(}\AgdaGeneralizable{Δ}\AgdaSpace{}%
\AgdaOperator{\AgdaFunction{<><}}\AgdaSpace{}%
\AgdaBound{Δ′}\AgdaSymbol{)}\<%
\\
\>[2]\AgdaFunction{post-fish}\AgdaSpace{}%
\AgdaInductiveConstructor{[]}\AgdaSpace{}%
\AgdaBound{v}\AgdaSpace{}%
\AgdaSymbol{=}\AgdaSpace{}%
\AgdaBound{v}\<%
\\
\>[2]\AgdaFunction{post-fish}\AgdaSpace{}%
\AgdaSymbol{(}\AgdaBound{x}\AgdaSpace{}%
\AgdaOperator{\AgdaInductiveConstructor{◃}}\AgdaSpace{}%
\AgdaBound{Δ′}\AgdaSymbol{)}\AgdaSpace{}%
\AgdaBound{v}\AgdaSpace{}%
\AgdaSymbol{=}\AgdaSpace{}%
\AgdaFunction{post-fish}\AgdaSpace{}%
\AgdaBound{Δ′}\AgdaSpace{}%
\AgdaSymbol{(}\AgdaInductiveConstructor{vₛ}\AgdaSpace{}%
\AgdaBound{v}\AgdaSymbol{)}\<%
\\
\\[\AgdaEmptyExtraSkip]%
\>[2]\AgdaFunction{gradc}\AgdaSpace{}%
\AgdaSymbol{:}\AgdaSpace{}%
\AgdaSymbol{∀}%
\>[2736I]\AgdaSymbol{\{}\AgdaBound{Δ′}\AgdaSymbol{\}}\AgdaSpace{}%
\AgdaSymbol{→}\AgdaSpace{}%
\AgdaFunction{Env}\AgdaSpace{}%
\AgdaSymbol{(}\AgdaFunction{double-ctx}\AgdaSpace{}%
\AgdaGeneralizable{Δ}\AgdaSymbol{)}\AgdaSpace{}%
\AgdaGeneralizable{Γ}\AgdaSpace{}%
\AgdaSymbol{→}\AgdaSpace{}%
\AgdaDatatype{LEnv}\AgdaSpace{}%
\AgdaBound{Δ′}\AgdaSpace{}%
\AgdaGeneralizable{Γ}\<%
\\
\>[.][@{}l@{}]\<[2736I]%
\>[12]\AgdaSymbol{→}\AgdaSpace{}%
\AgdaDatatype{Postfix}\AgdaSpace{}%
\AgdaSymbol{((}\AgdaFunction{double-ctx}\AgdaSpace{}%
\AgdaGeneralizable{Δ}\AgdaSymbol{)}\AgdaSpace{}%
\AgdaOperator{\AgdaFunction{<><}}\AgdaSpace{}%
\AgdaBound{Δ′}\AgdaSymbol{)}\AgdaSpace{}%
\AgdaGeneralizable{Γ}\AgdaSpace{}%
\AgdaSymbol{→}%
\>[51]\AgdaFunction{Env′}\AgdaSpace{}%
\AgdaGeneralizable{Γ}\AgdaSpace{}%
\AgdaSymbol{→}\AgdaSpace{}%
\AgdaFunction{Env′}\AgdaSpace{}%
\AgdaGeneralizable{Γ}\<%
\\
\>[2]\AgdaFunction{gradc}\AgdaSpace{}%
\AgdaSymbol{\{}\AgdaInductiveConstructor{ε}\AgdaSymbol{\}}%
\>[19]\AgdaSymbol{\{}\AgdaBound{Γ}\AgdaSymbol{\}}\AgdaSpace{}%
\AgdaSymbol{\{}\AgdaBound{Δ′}\AgdaSymbol{\}}\AgdaSpace{}%
\AgdaBound{ρ}\AgdaSpace{}%
\AgdaBound{ρ′}\AgdaSpace{}%
\AgdaBound{p}\AgdaSpace{}%
\AgdaBound{δ}\AgdaSpace{}%
\AgdaSymbol{=}\AgdaSpace{}%
\AgdaBound{δ}\<%
\\
\>[2]\AgdaFunction{gradc}\AgdaSpace{}%
\AgdaSymbol{\{}\AgdaBound{Δ}\AgdaSpace{}%
\AgdaOperator{\AgdaInductiveConstructor{▹}}\AgdaSpace{}%
\AgdaInductiveConstructor{ix}\AgdaSpace{}%
\AgdaBound{x}\AgdaSymbol{\}}\AgdaSpace{}%
\AgdaSymbol{\{}\AgdaBound{Γ}\AgdaSymbol{\}}\AgdaSpace{}%
\AgdaSymbol{\{}\AgdaBound{Δ′}\AgdaSymbol{\}}\AgdaSpace{}%
\AgdaBound{ρ}\AgdaSpace{}%
\AgdaBound{ρ′}\AgdaSpace{}%
\AgdaBound{p}\AgdaSpace{}%
\AgdaBound{δ}\AgdaSpace{}%
\AgdaSymbol{=}\AgdaSpace{}%
\AgdaFunction{⊥-elim}\AgdaSpace{}%
\AgdaSymbol{(}\AgdaFunction{no-ix}\AgdaSpace{}%
\AgdaSymbol{(}\AgdaFunction{post-fish}\AgdaSpace{}%
\AgdaBound{Δ′}\AgdaSpace{}%
\AgdaInductiveConstructor{v₀}\AgdaSymbol{)}\AgdaSpace{}%
\AgdaBound{p}\AgdaSymbol{)}\<%
\\
\>[2]\AgdaFunction{gradc}\AgdaSpace{}%
\AgdaSymbol{\{}\AgdaBound{Δ}\AgdaSpace{}%
\AgdaOperator{\AgdaInductiveConstructor{▹}}\AgdaSpace{}%
\AgdaInductiveConstructor{ar}\AgdaSpace{}%
\AgdaBound{x}\AgdaSymbol{\}}\AgdaSpace{}%
\AgdaSymbol{\{}\AgdaBound{Γ}\AgdaSymbol{\}}\AgdaSpace{}%
\AgdaSymbol{\{}\AgdaBound{Δ′}\AgdaSymbol{\}}\AgdaSpace{}%
\AgdaSymbol{((}\AgdaBound{ρ}\AgdaSpace{}%
\AgdaOperator{\AgdaInductiveConstructor{,}}\AgdaSpace{}%
\AgdaBound{z}\AgdaSymbol{)}\AgdaSpace{}%
\AgdaOperator{\AgdaInductiveConstructor{,}}\AgdaSpace{}%
\AgdaBound{e}\AgdaSymbol{)}\AgdaSpace{}%
\AgdaBound{ρ′}\AgdaSpace{}%
\AgdaBound{p}\AgdaSpace{}%
\AgdaBound{δ}\AgdaSpace{}%
\AgdaSymbol{=}\<%
\\
\>[2][@{}l@{\AgdaIndent{0}}]%
\>[4]\AgdaKeyword{let}\<%
\\
\>[4]\AgdaBound{ve}\AgdaSpace{}%
\AgdaSymbol{=}\AgdaSpace{}%
\AgdaFunction{post-var}\AgdaSpace{}%
\AgdaSymbol{\{}\AgdaArgument{Δ′}\AgdaSpace{}%
\AgdaSymbol{=}\AgdaSpace{}%
\AgdaBound{Δ′}\AgdaSymbol{\}}\AgdaSpace{}%
\AgdaBound{p}\AgdaSpace{}%
\AgdaInductiveConstructor{v₀}%
\>[34]\AgdaComment{--\ variable\ for\ e\ in\ Γ}\<%
\\
\>[4]\AgdaBound{vz}\AgdaSpace{}%
\AgdaSymbol{=}\AgdaSpace{}%
\AgdaFunction{post-var}\AgdaSpace{}%
\AgdaSymbol{\{}\AgdaArgument{Δ′}\AgdaSpace{}%
\AgdaSymbol{=}\AgdaSpace{}%
\AgdaBound{Δ′}\AgdaSymbol{\}}\AgdaSpace{}%
\AgdaBound{p}\AgdaSpace{}%
\AgdaInductiveConstructor{v₁}%
\>[34]\AgdaComment{--\ variable\ for\ z\ in\ Γ}\<%
\\
\>[4]\AgdaBound{s}%
\>[7]\AgdaSymbol{=}\AgdaSpace{}%
\AgdaFunction{env-ix}\AgdaSpace{}%
\AgdaBound{δ}\AgdaSpace{}%
\AgdaBound{ve}\<%
\\
\>[4]\AgdaBound{δ₁}\AgdaSpace{}%
\AgdaSymbol{=}\AgdaSpace{}%
\AgdaFunction{update}\AgdaSpace{}%
\AgdaBound{δ}\AgdaSpace{}%
\AgdaBound{vz}\AgdaSpace{}%
\AgdaSymbol{(}\AgdaFunction{const}\AgdaSpace{}%
\AgdaBound{s}\AgdaSymbol{)}%
\>[34]\AgdaComment{--\ save\ s\ in\ the\ z's\ position}\<%
\\
\>[4]\AgdaBound{δ₂}\AgdaSpace{}%
\AgdaSymbol{=}\AgdaSpace{}%
\AgdaFunction{∇}\AgdaSpace{}%
\AgdaBound{e}\AgdaSpace{}%
\AgdaSymbol{(}\AgdaInductiveConstructor{var}\AgdaSpace{}%
\AgdaBound{vz}\AgdaSymbol{)}\AgdaSpace{}%
\AgdaBound{δ₁}%
\>[34]\AgdaComment{--\ use\ vz\ position\ as\ seed}\<%
\\
\>[4]\AgdaKeyword{in}\AgdaSpace{}%
\AgdaFunction{gradc}\AgdaSpace{}%
\AgdaSymbol{\{}\AgdaBound{Δ}\AgdaSymbol{\}}\AgdaSpace{}%
\AgdaBound{ρ}\AgdaSpace{}%
\AgdaSymbol{(}\AgdaBound{z}\AgdaSpace{}%
\AgdaOperator{\AgdaInductiveConstructor{◃}}\AgdaSpace{}%
\AgdaSymbol{(}\AgdaBound{e}\AgdaSpace{}%
\AgdaOperator{\AgdaInductiveConstructor{◃}}\AgdaSpace{}%
\AgdaBound{ρ′}\AgdaSymbol{))}\AgdaSpace{}%
\AgdaSymbol{(}\AgdaFunction{pstep}\AgdaSpace{}%
\AgdaSymbol{\{}\AgdaArgument{Δ′}\AgdaSpace{}%
\AgdaSymbol{=}\AgdaSpace{}%
\AgdaInductiveConstructor{ar}\AgdaSpace{}%
\AgdaBound{x}\AgdaSpace{}%
\AgdaOperator{\AgdaInductiveConstructor{◃}}\AgdaSpace{}%
\AgdaBound{Δ′}\AgdaSymbol{\}}\AgdaSpace{}%
\AgdaSymbol{(}\AgdaFunction{pstep}\AgdaSpace{}%
\AgdaSymbol{\{}\AgdaArgument{Δ′}\AgdaSpace{}%
\AgdaSymbol{=}\AgdaSpace{}%
\AgdaBound{Δ′}\AgdaSymbol{\}}\AgdaSpace{}%
\AgdaBound{p}\AgdaSymbol{))}\AgdaSpace{}%
\AgdaBound{δ₂}\<%
\\
\\[\AgdaEmptyExtraSkip]%
\>[2]\AgdaFunction{chain-grad}\AgdaSpace{}%
\AgdaSymbol{:}\AgdaSpace{}%
\AgdaDatatype{Chain}\AgdaSpace{}%
\AgdaSymbol{(}\AgdaGeneralizable{Γ}\AgdaSpace{}%
\AgdaOperator{\AgdaInductiveConstructor{▹}}\AgdaSpace{}%
\AgdaInductiveConstructor{ar}\AgdaSpace{}%
\AgdaGeneralizable{s}\AgdaSymbol{)}\AgdaSpace{}%
\AgdaSymbol{→}\AgdaSpace{}%
\AgdaDatatype{E}\AgdaSpace{}%
\AgdaSymbol{(}\AgdaGeneralizable{Γ}\AgdaSpace{}%
\AgdaOperator{\AgdaInductiveConstructor{▹}}\AgdaSpace{}%
\AgdaInductiveConstructor{ar}\AgdaSpace{}%
\AgdaGeneralizable{s}\AgdaSymbol{)}\AgdaSpace{}%
\AgdaSymbol{(}\AgdaInductiveConstructor{ar}\AgdaSpace{}%
\AgdaGeneralizable{s}\AgdaSymbol{)}\AgdaSpace{}%
\AgdaSymbol{→}\AgdaSpace{}%
\AgdaFunction{Env′}\AgdaSpace{}%
\AgdaSymbol{(}\AgdaGeneralizable{Γ}\AgdaSpace{}%
\AgdaOperator{\AgdaInductiveConstructor{▹}}\AgdaSpace{}%
\AgdaInductiveConstructor{ar}\AgdaSpace{}%
\AgdaGeneralizable{s}\AgdaSymbol{)}\<%
\\
\>[2]\AgdaFunction{chain-grad}\AgdaSpace{}%
\AgdaSymbol{\{}\AgdaBound{Γ}\AgdaSymbol{\}}\AgdaSpace{}%
\AgdaSymbol{\{}\AgdaBound{s}\AgdaSymbol{\}}\AgdaSpace{}%
\AgdaBound{c}\AgdaSpace{}%
\AgdaBound{seed}\AgdaSpace{}%
\AgdaSymbol{=}\AgdaSpace{}%
\AgdaKeyword{let}\<%
\\
\>[2][@{}l@{\AgdaIndent{0}}]%
\>[4]\AgdaComment{--\ Well,\ this\ is\ a\ choice\ I\ suppose}\<%
\\
\>[4]\AgdaComment{--δ\ =\ ∇\ seed\ one\ (env-imap\ \{Γ\ =\ Γ\ ▹\ ar\ s\}\ (const\ zero))}\<%
\\
\>[4]\AgdaBound{δ}\AgdaSpace{}%
\AgdaSymbol{=}\AgdaSpace{}%
\AgdaFunction{env-imap}\AgdaSpace{}%
\AgdaSymbol{\{}\AgdaArgument{Γ}\AgdaSpace{}%
\AgdaSymbol{=}\AgdaSpace{}%
\AgdaBound{Γ}\AgdaSymbol{\}}\AgdaSpace{}%
\AgdaSymbol{(}\AgdaFunction{const}\AgdaSpace{}%
\AgdaInductiveConstructor{zero}\AgdaSymbol{)}\AgdaSpace{}%
\AgdaOperator{\AgdaInductiveConstructor{,}}\AgdaSpace{}%
\AgdaBound{seed}\<%
\\
\>[4]\AgdaBound{Δ}\AgdaSpace{}%
\AgdaOperator{\AgdaInductiveConstructor{,}}\AgdaSpace{}%
\AgdaBound{ρ}\AgdaSpace{}%
\AgdaOperator{\AgdaInductiveConstructor{,}}\AgdaSpace{}%
\AgdaBound{po}\AgdaSpace{}%
\AgdaSymbol{=}\AgdaSpace{}%
\AgdaFunction{chain-to-env}\AgdaSpace{}%
\AgdaBound{c}\<%
\\
\>[4]\AgdaKeyword{in}\AgdaSpace{}%
\AgdaFunction{env-map}\AgdaSpace{}%
\AgdaSymbol{\{}\AgdaArgument{Γ}\AgdaSpace{}%
\AgdaSymbol{=}\AgdaSpace{}%
\AgdaBound{Γ}\AgdaSpace{}%
\AgdaOperator{\AgdaInductiveConstructor{▹}}\AgdaSpace{}%
\AgdaInductiveConstructor{ar}\AgdaSpace{}%
\AgdaBound{s}\AgdaSymbol{\}}\AgdaSpace{}%
\AgdaSymbol{(}\AgdaFunction{multiopt}\AgdaSpace{}%
\AgdaNumber{10}\AgdaSymbol{)}\AgdaSpace{}%
\AgdaOperator{\AgdaFunction{\$}}\AgdaSpace{}%
\AgdaFunction{gradc}\AgdaSpace{}%
\AgdaBound{ρ}\AgdaSpace{}%
\AgdaInductiveConstructor{[]}\AgdaSpace{}%
\AgdaBound{po}\AgdaSpace{}%
\AgdaBound{δ}\<%
\\
\\[\AgdaEmptyExtraSkip]%
\>[2]\AgdaFunction{chain-sac-ctx}\AgdaSpace{}%
\AgdaSymbol{:}\AgdaSpace{}%
\AgdaDatatype{Chain}\AgdaSpace{}%
\AgdaGeneralizable{Γ}\AgdaSpace{}%
\AgdaSymbol{→}\AgdaSpace{}%
\AgdaFunction{Sac.SEnv}\AgdaSpace{}%
\AgdaGeneralizable{Γ}\<%
\\
\>[2]\AgdaFunction{chain-sac-ctx}\AgdaSpace{}%
\AgdaSymbol{(}\AgdaInductiveConstructor{ε}\AgdaSpace{}%
\AgdaBound{x}\AgdaSymbol{)}\AgdaSpace{}%
\AgdaSymbol{=}\AgdaSpace{}%
\AgdaBound{x}\<%
\\
\>[2]\AgdaFunction{chain-sac-ctx}\AgdaSpace{}%
\AgdaSymbol{(}\AgdaBound{c}\AgdaSpace{}%
\AgdaOperator{\AgdaInductiveConstructor{▹}}\AgdaSpace{}%
\AgdaSymbol{(}\AgdaBound{v}\AgdaSpace{}%
\AgdaOperator{\AgdaInductiveConstructor{,}}\AgdaSpace{}%
\AgdaSymbol{\AgdaUnderscore{}))}\AgdaSpace{}%
\AgdaSymbol{=}\AgdaSpace{}%
\AgdaFunction{chain-sac-ctx}\AgdaSpace{}%
\AgdaBound{c}\AgdaSpace{}%
\AgdaOperator{\AgdaFunction{,,}}\AgdaSpace{}%
\AgdaSymbol{(}\AgdaString{"∂/∂"}\AgdaSpace{}%
\AgdaOperator{\AgdaFunction{++}}\AgdaSpace{}%
\AgdaBound{v}\AgdaSymbol{)}\AgdaSpace{}%
\AgdaOperator{\AgdaFunction{,,}}\AgdaSpace{}%
\AgdaBound{v}\<%
\\
\>[0]\<%
\\
\>[2]\AgdaFunction{filter-grad}\AgdaSpace{}%
\AgdaSymbol{:}\AgdaSpace{}%
\AgdaDatatype{Chain}\AgdaSpace{}%
\AgdaGeneralizable{Γ}\AgdaSpace{}%
\AgdaSymbol{→}\AgdaSpace{}%
\AgdaFunction{Sac.SEnv}\AgdaSpace{}%
\AgdaGeneralizable{Γ}\AgdaSpace{}%
\AgdaSymbol{→}\AgdaSpace{}%
\AgdaDatatype{List}\AgdaSpace{}%
\AgdaPostulate{String}\<%
\\
\>[2]\AgdaFunction{filter-grad}\AgdaSpace{}%
\AgdaSymbol{(}\AgdaInductiveConstructor{ε}\AgdaSpace{}%
\AgdaBound{x}\AgdaSymbol{)}%
\>[22]\AgdaBound{δ}\AgdaSpace{}%
\AgdaSymbol{=}\AgdaSpace{}%
\AgdaFunction{Sac.env-rev-list}\AgdaSpace{}%
\AgdaBound{δ}\<%
\\
\>[2]\AgdaFunction{filter-grad}\AgdaSpace{}%
\AgdaSymbol{(}\AgdaBound{c}\AgdaSpace{}%
\AgdaOperator{\AgdaInductiveConstructor{▹}}\AgdaSpace{}%
\AgdaSymbol{\AgdaUnderscore{})}\AgdaSpace{}%
\AgdaSymbol{((}\AgdaBound{δ}\AgdaSpace{}%
\AgdaOperator{\AgdaInductiveConstructor{,}}\AgdaSpace{}%
\AgdaSymbol{\AgdaUnderscore{})}\AgdaOperator{\AgdaInductiveConstructor{,}}\AgdaSpace{}%
\AgdaBound{x}\AgdaSymbol{)}\AgdaSpace{}%
\AgdaSymbol{=}\AgdaSpace{}%
\AgdaBound{x}\AgdaSpace{}%
\AgdaOperator{\AgdaInductiveConstructor{∷}}\AgdaSpace{}%
\AgdaFunction{filter-grad}\AgdaSpace{}%
\AgdaBound{c}\AgdaSpace{}%
\AgdaBound{δ}\<%
\\
\\[\AgdaEmptyExtraSkip]%
\>[2]\AgdaFunction{chain-grad-sac}\AgdaSpace{}%
\AgdaSymbol{:}\AgdaSpace{}%
\AgdaDatatype{Chain}\AgdaSpace{}%
\AgdaGeneralizable{Γ}\AgdaSpace{}%
\AgdaSymbol{→}\AgdaSpace{}%
\AgdaFunction{Env′}\AgdaSpace{}%
\AgdaGeneralizable{Γ}\AgdaSpace{}%
\AgdaSymbol{→}\AgdaSpace{}%
\AgdaPostulate{String}\<%
\\
\>[2]\AgdaFunction{chain-grad-sac}\AgdaSpace{}%
\AgdaSymbol{\{}\AgdaBound{Γ}\AgdaSymbol{\}}\AgdaSpace{}%
\AgdaBound{c}\AgdaSpace{}%
\AgdaBound{δ}\AgdaSpace{}%
\AgdaSymbol{=}\AgdaSpace{}%
\AgdaKeyword{let}\<%
\\
\>[2][@{}l@{\AgdaIndent{0}}]%
\>[4]\AgdaBound{vars}\AgdaSpace{}%
\AgdaSymbol{=}\AgdaSpace{}%
\AgdaFunction{chain-sac-ctx}\AgdaSpace{}%
\AgdaBound{c}\<%
\\
\>[4]\AgdaBound{vals}\AgdaSpace{}%
\AgdaSymbol{=}\AgdaSpace{}%
\AgdaFunction{Sac.env-sac}\AgdaSpace{}%
\AgdaSymbol{\{}\AgdaBound{Γ}\AgdaSymbol{\}}\AgdaSpace{}%
\AgdaBound{δ}\AgdaSpace{}%
\AgdaBound{vars}\<%
\\
\>[4]\AgdaBound{assignments}\AgdaSpace{}%
\AgdaSymbol{=}\AgdaSpace{}%
\AgdaFunction{filter-grad}\AgdaSpace{}%
\AgdaBound{c}\AgdaSpace{}%
\AgdaOperator{\AgdaFunction{\$}}\AgdaSpace{}%
\AgdaFunction{Sac.zip-env}\AgdaSpace{}%
\AgdaSymbol{(}\AgdaFunction{printf}\AgdaSpace{}%
\AgdaString{"∂/∂\%s\ =\ \%s;"}\AgdaSymbol{)}\AgdaSpace{}%
\AgdaBound{vars}\AgdaSpace{}%
\AgdaBound{vals}\<%
\\
\>[4]\AgdaKeyword{in}\AgdaSpace{}%
\AgdaFunction{intersperse}\AgdaSpace{}%
\AgdaString{"\textbackslash{}n"}\AgdaSpace{}%
\AgdaBound{assignments}\<%
\\
\\[\AgdaEmptyExtraSkip]%
\>[2]\AgdaFunction{chain-sac-l}\AgdaSpace{}%
\AgdaSymbol{:}\AgdaSpace{}%
\AgdaDatatype{Chain}\AgdaSpace{}%
\AgdaGeneralizable{Γ}\AgdaSpace{}%
\AgdaSymbol{→}\AgdaSpace{}%
\AgdaDatatype{ℕ}\AgdaSpace{}%
\AgdaSymbol{→}\AgdaSpace{}%
\AgdaDatatype{List}\AgdaSpace{}%
\AgdaPostulate{String}\<%
\\
\>[2]\AgdaFunction{chain-sac-l}\AgdaSpace{}%
\AgdaSymbol{(}\AgdaInductiveConstructor{ε}\AgdaSpace{}%
\AgdaBound{x}\AgdaSymbol{)}\AgdaSpace{}%
\AgdaSymbol{\AgdaUnderscore{}}\AgdaSpace{}%
\AgdaSymbol{=}\AgdaSpace{}%
\AgdaInductiveConstructor{[]}\<%
\\
\>[2]\AgdaFunction{chain-sac-l}\AgdaSpace{}%
\AgdaSymbol{(}\AgdaBound{c}\AgdaSpace{}%
\AgdaOperator{\AgdaInductiveConstructor{▹}}\AgdaSpace{}%
\AgdaSymbol{(}\AgdaBound{v}\AgdaSpace{}%
\AgdaOperator{\AgdaInductiveConstructor{,}}\AgdaSpace{}%
\AgdaBound{e}\AgdaSymbol{))}\AgdaSpace{}%
\AgdaBound{n}\AgdaSpace{}%
\AgdaSymbol{=}%
\>[3008I]\AgdaKeyword{let}\AgdaSpace{}%
\AgdaBound{r}\AgdaSpace{}%
\AgdaOperator{\AgdaInductiveConstructor{,}}\AgdaSpace{}%
\AgdaBound{n′}\AgdaSpace{}%
\AgdaSymbol{=}\AgdaSpace{}%
\AgdaFunction{Sac.to-sac}\AgdaSpace{}%
\AgdaSymbol{(}\AgdaFunction{multiopt}\AgdaSpace{}%
\AgdaNumber{10}\AgdaSpace{}%
\AgdaBound{e}\AgdaSymbol{)}\AgdaSpace{}%
\AgdaSymbol{(}\AgdaFunction{chain-sac-ctx}\AgdaSpace{}%
\AgdaBound{c}\AgdaSymbol{)}\AgdaSpace{}%
\AgdaBound{n}\<%
\\
\>[.][@{}l@{}]\<[3008I]%
\>[32]\AgdaKeyword{in}\AgdaSpace{}%
\AgdaFunction{printf}\AgdaSpace{}%
\AgdaString{"\%s\ =\ \%s;"}\AgdaSpace{}%
\AgdaBound{v}\AgdaSpace{}%
\AgdaBound{r}\AgdaSpace{}%
\AgdaOperator{\AgdaInductiveConstructor{∷}}\AgdaSpace{}%
\AgdaFunction{chain-sac-l}\AgdaSpace{}%
\AgdaBound{c}\AgdaSpace{}%
\AgdaBound{n′}\<%
\\
\\[\AgdaEmptyExtraSkip]%
\>[2]\AgdaFunction{chain-sac}\AgdaSpace{}%
\AgdaSymbol{:}\AgdaSpace{}%
\AgdaDatatype{Chain}\AgdaSpace{}%
\AgdaGeneralizable{Γ}\AgdaSpace{}%
\AgdaSymbol{→}\AgdaSpace{}%
\AgdaPostulate{String}\<%
\\
\>[2]\AgdaFunction{chain-sac}\AgdaSpace{}%
\AgdaBound{c}\AgdaSpace{}%
\AgdaSymbol{=}\AgdaSpace{}%
\AgdaFunction{intersperse}\AgdaSpace{}%
\AgdaString{"\textbackslash{}n"}\AgdaSpace{}%
\AgdaOperator{\AgdaFunction{\$}}\AgdaSpace{}%
\AgdaFunction{L.reverse}\AgdaSpace{}%
\AgdaOperator{\AgdaFunction{\$}}\AgdaSpace{}%
\AgdaFunction{chain-sac-l}\AgdaSpace{}%
\AgdaBound{c}\AgdaSpace{}%
\AgdaNumber{1}\<%
\\
\\[\AgdaEmptyExtraSkip]%
\\[\AgdaEmptyExtraSkip]%
\>[2]\AgdaComment{--\ test-chain\ :\ Chain\ \AgdaUnderscore{}\ --(ε\ ▹\ ar\ (ι\ 3))}\<%
\\
\>[2]\AgdaComment{--\ test-chain\ =\ ε\ \{Γ\ =\ ε\ ▹\ ar\ (ι\ 3)\}\ (\AgdaUnderscore{}\ ,,\ "a")\ }\<%
\\
\>[2]\AgdaComment{--\ \ \ \ \ \ \ \ \ \ \ \ ▹\ ("r"\ ,\ mul-test)}\<%
\\
\>[2]\AgdaComment{--\ \ \ \ \ \ \ \ \ \ \ \ ▹\ ("r₁"\ ,\ (var\ v₀)\ ⊠\ (var\ v₂))}\<%
\\
\\[\AgdaEmptyExtraSkip]%
\>[2]\AgdaComment{--\ test-grad\ :\ String}\<%
\\
\>[2]\AgdaComment{--\ test-grad\ =\ chain-sac\ test-chain\ }\<%
\\
\>[2]\AgdaComment{--\ \ \ \ \ \ \ \ \ \ \ \ \ ++\ "\textbackslash{}n"\ ++\ chain-grad-sac\ test-chain\ (chain-grad\ test-chain\ one)}\<%
\end{code}

Let us consider a small example to see this in action.  We start with a little
convenience data structure \AF{ChainCtx} that keeps the shapes and the variable names
together.  We also define the function \AF{ce-split} that splits 
\AF{ChainCtx} into the context and the environment with variable names in that context:
\begin{code}%
\>[2]\AgdaKeyword{data}\AgdaSpace{}%
\AgdaDatatype{ChainCtx}\AgdaSpace{}%
\AgdaSymbol{:}\AgdaSpace{}%
\AgdaPrimitive{Set}\AgdaSpace{}%
\AgdaKeyword{where}\<%
\\
\>[2][@{}l@{\AgdaIndent{0}}]%
\>[4]\AgdaInductiveConstructor{ε}\AgdaSpace{}%
\AgdaSymbol{:}\AgdaSpace{}%
\AgdaDatatype{ChainCtx}\<%
\\
\>[4]\AgdaOperator{\AgdaInductiveConstructor{\AgdaUnderscore{}▹\AgdaUnderscore{}}}\AgdaSpace{}%
\AgdaSymbol{:}\AgdaSpace{}%
\AgdaDatatype{ChainCtx}\AgdaSpace{}%
\AgdaSymbol{→}\AgdaSpace{}%
\AgdaPostulate{String}\AgdaSpace{}%
\AgdaOperator{\AgdaFunction{×}}\AgdaSpace{}%
\AgdaDatatype{S}\AgdaSpace{}%
\AgdaSymbol{→}\AgdaSpace{}%
\AgdaDatatype{ChainCtx}\<%
\\
\\[\AgdaEmptyExtraSkip]%
\>[2]\AgdaFunction{ce-split}\AgdaSpace{}%
\AgdaSymbol{:}\AgdaSpace{}%
\AgdaDatatype{ChainCtx}\AgdaSpace{}%
\AgdaSymbol{→}\AgdaSpace{}%
\AgdaRecord{Σ}\AgdaSpace{}%
\AgdaDatatype{Ctx}\AgdaSpace{}%
\AgdaFunction{Sac.SEnv}\<%
\end{code}
\begin{code}[hide]%
\>[2]\AgdaFunction{ce-split}\AgdaSpace{}%
\AgdaInductiveConstructor{ε}\AgdaSpace{}%
\AgdaSymbol{=}\AgdaSpace{}%
\AgdaInductiveConstructor{ε}\AgdaSpace{}%
\AgdaOperator{\AgdaInductiveConstructor{,}}\AgdaSpace{}%
\AgdaInductiveConstructor{tt}\<%
\\
\>[2]\AgdaFunction{ce-split}\AgdaSpace{}%
\AgdaSymbol{(}\AgdaBound{cx}\AgdaSpace{}%
\AgdaOperator{\AgdaInductiveConstructor{▹}}\AgdaSpace{}%
\AgdaSymbol{(}\AgdaBound{v}\AgdaSpace{}%
\AgdaOperator{\AgdaInductiveConstructor{,}}\AgdaSpace{}%
\AgdaBound{s}\AgdaSymbol{))}\AgdaSpace{}%
\AgdaSymbol{=}\AgdaSpace{}%
\AgdaKeyword{let}\AgdaSpace{}%
\AgdaBound{Δ}\AgdaSpace{}%
\AgdaOperator{\AgdaInductiveConstructor{,}}\AgdaSpace{}%
\AgdaBound{ρ}\AgdaSpace{}%
\AgdaSymbol{=}\AgdaSpace{}%
\AgdaFunction{ce-split}\AgdaSpace{}%
\AgdaBound{cx}\AgdaSpace{}%
\AgdaKeyword{in}\AgdaSpace{}%
\AgdaSymbol{(}\AgdaBound{Δ}\AgdaSpace{}%
\AgdaOperator{\AgdaInductiveConstructor{▹}}\AgdaSpace{}%
\AgdaInductiveConstructor{ar}\AgdaSpace{}%
\AgdaBound{s}\AgdaSymbol{)}\AgdaSpace{}%
\AgdaOperator{\AgdaInductiveConstructor{,}}\AgdaSpace{}%
\AgdaSymbol{(}\AgdaBound{ρ}\AgdaSpace{}%
\AgdaOperator{\AgdaInductiveConstructor{,}}\AgdaSpace{}%
\AgdaBound{v}\AgdaSymbol{)}\<%
\\
\\[\AgdaEmptyExtraSkip]%
\>[2]\AgdaFunction{Product}\AgdaSpace{}%
\AgdaSymbol{:}\AgdaSpace{}%
\AgdaDatatype{ℕ}\AgdaSpace{}%
\AgdaSymbol{→}\AgdaSpace{}%
\AgdaPrimitive{Set}\AgdaSpace{}%
\AgdaSymbol{→}\AgdaSpace{}%
\AgdaPrimitive{Set}\<%
\\
\>[2]\AgdaFunction{Product}\AgdaSpace{}%
\AgdaNumber{0}%
\>[18]\AgdaBound{A}\AgdaSpace{}%
\AgdaSymbol{=}\AgdaSpace{}%
\AgdaRecord{⊤}\<%
\\
\>[2]\AgdaFunction{Product}\AgdaSpace{}%
\AgdaNumber{1}%
\>[18]\AgdaBound{A}\AgdaSpace{}%
\AgdaSymbol{=}\AgdaSpace{}%
\AgdaBound{A}\<%
\\
\>[2]\AgdaCatchallClause{\AgdaFunction{Product}}\AgdaSpace{}%
\AgdaCatchallClause{\AgdaSymbol{(}}\AgdaCatchallClause{\AgdaInductiveConstructor{suc}}\AgdaSpace{}%
\AgdaCatchallClause{\AgdaBound{n}}\AgdaCatchallClause{\AgdaSymbol{)}}\AgdaSpace{}%
\AgdaCatchallClause{\AgdaBound{A}}\AgdaSpace{}%
\AgdaSymbol{=}\AgdaSpace{}%
\AgdaBound{A}\AgdaSpace{}%
\AgdaOperator{\AgdaFunction{×}}\AgdaSpace{}%
\AgdaFunction{Product}\AgdaSpace{}%
\AgdaBound{n}\AgdaSpace{}%
\AgdaBound{A}\<%
\\
\\[\AgdaEmptyExtraSkip]%
\>[2]\AgdaFunction{Es}\AgdaSpace{}%
\AgdaSymbol{:}\AgdaSpace{}%
\AgdaSymbol{∀}\AgdaSpace{}%
\AgdaSymbol{\{}\AgdaBound{Γ}\AgdaSpace{}%
\AgdaSymbol{:}\AgdaSpace{}%
\AgdaDatatype{Ctx}\AgdaSymbol{\}}\AgdaSpace{}%
\AgdaSymbol{→}\AgdaSpace{}%
\AgdaSymbol{(}\AgdaBound{n}\AgdaSpace{}%
\AgdaSymbol{:}\AgdaSpace{}%
\AgdaDatatype{ℕ}\AgdaSymbol{)}\AgdaSpace{}%
\AgdaSymbol{→}\AgdaSpace{}%
\AgdaSymbol{\{}\AgdaFunction{Product}\AgdaSpace{}%
\AgdaBound{n}\AgdaSpace{}%
\AgdaDatatype{IS}\AgdaSymbol{\}}\AgdaSpace{}%
\AgdaSymbol{→}\AgdaSpace{}%
\AgdaPrimitive{Set}\<%
\\
\>[2]\AgdaFunction{Es}\AgdaSpace{}%
\AgdaSymbol{\{}\AgdaBound{Γ}\AgdaSymbol{\}}\AgdaSpace{}%
\AgdaNumber{0}%
\>[23]\AgdaSymbol{\{}\AgdaBound{is}\AgdaSymbol{\}}\AgdaSpace{}%
\AgdaSymbol{=}\AgdaSpace{}%
\AgdaRecord{⊤}\<%
\\
\>[2]\AgdaFunction{Es}\AgdaSpace{}%
\AgdaSymbol{\{}\AgdaBound{Γ}\AgdaSymbol{\}}\AgdaSpace{}%
\AgdaNumber{1}%
\>[23]\AgdaSymbol{\{}\AgdaBound{is}\AgdaSymbol{\}}\AgdaSpace{}%
\AgdaSymbol{=}\AgdaSpace{}%
\AgdaDatatype{E}\AgdaSpace{}%
\AgdaBound{Γ}\AgdaSpace{}%
\AgdaBound{is}\<%
\\
\>[2]\AgdaFunction{Es}\AgdaSpace{}%
\AgdaSymbol{\{}\AgdaBound{Γ}\AgdaSymbol{\}}\AgdaSpace{}%
\AgdaSymbol{(}\AgdaInductiveConstructor{suc}\AgdaSpace{}%
\AgdaSymbol{(}\AgdaInductiveConstructor{suc}\AgdaSpace{}%
\AgdaBound{n}\AgdaSymbol{))}\AgdaSpace{}%
\AgdaSymbol{\{}\AgdaBound{is}\AgdaSpace{}%
\AgdaOperator{\AgdaInductiveConstructor{,}}\AgdaSpace{}%
\AgdaBound{p}\AgdaSymbol{\}}%
\>[33]\AgdaSymbol{=}\AgdaSpace{}%
\AgdaDatatype{E}\AgdaSpace{}%
\AgdaBound{Γ}\AgdaSpace{}%
\AgdaBound{is}\AgdaSpace{}%
\AgdaOperator{\AgdaFunction{×}}\AgdaSpace{}%
\AgdaFunction{Es}\AgdaSpace{}%
\AgdaSymbol{\{}\AgdaBound{Γ}\AgdaSymbol{\}}\AgdaSpace{}%
\AgdaSymbol{(}\AgdaInductiveConstructor{suc}\AgdaSpace{}%
\AgdaBound{n}\AgdaSymbol{)}\AgdaSpace{}%
\AgdaSymbol{\{}\AgdaBound{p}\AgdaSymbol{\}}\<%
\\
\\[\AgdaEmptyExtraSkip]%
\>[2]\AgdaFunction{↑↑ₙ}\AgdaSpace{}%
\AgdaSymbol{:}\AgdaSpace{}%
\AgdaSymbol{∀}\AgdaSpace{}%
\AgdaSymbol{\{}\AgdaBound{Γ}\AgdaSpace{}%
\AgdaSymbol{:}\AgdaSpace{}%
\AgdaDatatype{Ctx}\AgdaSymbol{\}}\AgdaSpace{}%
\AgdaSymbol{\{}\AgdaBound{is}\AgdaSymbol{\}}\AgdaSpace{}%
\AgdaBound{n}\AgdaSpace{}%
\AgdaSymbol{\{}\AgdaBound{p}\AgdaSpace{}%
\AgdaSymbol{:}\AgdaSpace{}%
\AgdaFunction{Product}\AgdaSpace{}%
\AgdaBound{n}\AgdaSpace{}%
\AgdaDatatype{IS}\AgdaSymbol{\}}\AgdaSpace{}%
\AgdaSymbol{→}\AgdaSpace{}%
\AgdaFunction{Es}\AgdaSpace{}%
\AgdaSymbol{\{}\AgdaBound{Γ}\AgdaSymbol{\}}\AgdaSpace{}%
\AgdaBound{n}\AgdaSpace{}%
\AgdaSymbol{\{}\AgdaBound{p}\AgdaSymbol{\}}\AgdaSpace{}%
\AgdaSymbol{→}\AgdaSpace{}%
\AgdaFunction{Es}\AgdaSpace{}%
\AgdaSymbol{\{}\AgdaBound{Γ}\AgdaSpace{}%
\AgdaOperator{\AgdaInductiveConstructor{▹}}\AgdaSpace{}%
\AgdaBound{is}\AgdaSpace{}%
\AgdaOperator{\AgdaInductiveConstructor{▹}}\AgdaSpace{}%
\AgdaBound{is}\AgdaSymbol{\}}\AgdaSpace{}%
\AgdaBound{n}\AgdaSpace{}%
\AgdaSymbol{\{}\AgdaBound{p}\AgdaSymbol{\}}\<%
\\
\>[2]\AgdaFunction{↑↑ₙ}\AgdaSpace{}%
\AgdaNumber{0}\AgdaSpace{}%
\AgdaBound{es}\AgdaSpace{}%
\AgdaSymbol{=}\AgdaSpace{}%
\AgdaSymbol{\AgdaUnderscore{}}\<%
\\
\>[2]\AgdaFunction{↑↑ₙ}\AgdaSpace{}%
\AgdaNumber{1}\AgdaSpace{}%
\AgdaBound{e}%
\>[11]\AgdaSymbol{=}\AgdaSpace{}%
\AgdaOperator{\AgdaFunction{↑↑}}\AgdaSpace{}%
\AgdaBound{e}\<%
\\
\>[2]\AgdaFunction{↑↑ₙ}\AgdaSpace{}%
\AgdaSymbol{(}\AgdaInductiveConstructor{suc}\AgdaSpace{}%
\AgdaSymbol{(}\AgdaInductiveConstructor{suc}\AgdaSpace{}%
\AgdaBound{n}\AgdaSymbol{))}\AgdaSpace{}%
\AgdaSymbol{(}\AgdaBound{e}\AgdaSpace{}%
\AgdaOperator{\AgdaInductiveConstructor{,}}\AgdaSpace{}%
\AgdaBound{es}\AgdaSymbol{)}\AgdaSpace{}%
\AgdaSymbol{=}\AgdaSpace{}%
\AgdaOperator{\AgdaFunction{↑↑}}\AgdaSpace{}%
\AgdaBound{e}\AgdaSpace{}%
\AgdaOperator{\AgdaInductiveConstructor{,}}\AgdaSpace{}%
\AgdaFunction{↑↑ₙ}\AgdaSpace{}%
\AgdaSymbol{(}\AgdaInductiveConstructor{suc}\AgdaSpace{}%
\AgdaBound{n}\AgdaSymbol{)}\AgdaSpace{}%
\AgdaBound{es}\<%
\end{code}
Consider an initial environment of two 5-element vectors $a$ and $b$; local
computations $x = ab$ and $y = xx$; and the generated code when computing derivative
of $y$ (\AC{var v₀}) on the right.
\begin{mathpar}
\codeblock{\begin{code}%
\>[2]\AgdaFunction{test-chain}\AgdaSpace{}%
\AgdaSymbol{:}\AgdaSpace{}%
\AgdaDatatype{Chain}\AgdaSpace{}%
\AgdaSymbol{\AgdaUnderscore{}}\<%
\\
\>[2]\AgdaFunction{test-chain}\AgdaSpace{}%
\AgdaSymbol{=}\AgdaSpace{}%
\AgdaKeyword{let}\<%
\\
\>[2][@{}l@{\AgdaIndent{0}}]%
\>[4]\AgdaBound{Γ}\AgdaSpace{}%
\AgdaOperator{\AgdaInductiveConstructor{,}}\AgdaSpace{}%
\AgdaBound{ρ}\AgdaSpace{}%
\AgdaSymbol{=}\AgdaSpace{}%
\AgdaFunction{ce-split}\AgdaSpace{}%
\AgdaSymbol{(}\AgdaInductiveConstructor{ε}\AgdaSpace{}%
\AgdaOperator{\AgdaInductiveConstructor{▹}}\AgdaSpace{}%
\AgdaSymbol{(}\AgdaString{"a"}\AgdaSpace{}%
\AgdaOperator{\AgdaInductiveConstructor{,}}\AgdaSpace{}%
\AgdaInductiveConstructor{ι}\AgdaSpace{}%
\AgdaNumber{5}\AgdaSymbol{)}\AgdaSpace{}%
\AgdaOperator{\AgdaInductiveConstructor{▹}}\AgdaSpace{}%
\AgdaSymbol{(}\AgdaString{"b"}\AgdaSpace{}%
\AgdaOperator{\AgdaInductiveConstructor{,}}\AgdaSpace{}%
\AgdaInductiveConstructor{ι}\AgdaSpace{}%
\AgdaNumber{5}\AgdaSymbol{))}\<%
\\
\>[4]\AgdaBound{a}\AgdaSpace{}%
\AgdaSymbol{=}\AgdaSpace{}%
\AgdaInductiveConstructor{var}\AgdaSpace{}%
\AgdaInductiveConstructor{v₁}\AgdaSymbol{;}\AgdaSpace{}%
\AgdaBound{b}\AgdaSpace{}%
\AgdaSymbol{=}\AgdaSpace{}%
\AgdaInductiveConstructor{var}\AgdaSpace{}%
\AgdaInductiveConstructor{v₀}\<%
\\
\>[4]\AgdaBound{C₁}\AgdaSpace{}%
\AgdaSymbol{=}\AgdaSpace{}%
\AgdaInductiveConstructor{ε}\AgdaSpace{}%
\AgdaSymbol{\{}\AgdaBound{Γ}\AgdaSymbol{\}}\AgdaSpace{}%
\AgdaBound{ρ}%
\>[18]\AgdaOperator{\AgdaInductiveConstructor{▹}}\AgdaSpace{}%
\AgdaSymbol{(}\AgdaString{"x"}\AgdaSpace{}%
\AgdaOperator{\AgdaInductiveConstructor{,}}\AgdaSpace{}%
\AgdaBound{a}\AgdaSpace{}%
\AgdaOperator{\AgdaInductiveConstructor{⊠}}\AgdaSpace{}%
\AgdaBound{b}\AgdaSymbol{)}\<%
\\
\>[4]\AgdaBound{x}\AgdaSpace{}%
\AgdaSymbol{=}\AgdaSpace{}%
\AgdaInductiveConstructor{var}\AgdaSpace{}%
\AgdaInductiveConstructor{v₀}\<%
\\
\>[4]\AgdaBound{C₂}\AgdaSpace{}%
\AgdaSymbol{=}\AgdaSpace{}%
\AgdaBound{C₁}%
\>[18]\AgdaOperator{\AgdaInductiveConstructor{▹}}\AgdaSpace{}%
\AgdaSymbol{(}\AgdaString{"y"}\AgdaSpace{}%
\AgdaOperator{\AgdaInductiveConstructor{,}}\AgdaSpace{}%
\AgdaBound{x}\AgdaSpace{}%
\AgdaOperator{\AgdaInductiveConstructor{⊠}}\AgdaSpace{}%
\AgdaBound{x}\AgdaSymbol{)}\<%
\\
\>[4]\AgdaKeyword{in}\AgdaSpace{}%
\AgdaBound{C₂}\<%
\end{code}}
\and
{\begin{varwidth}{0.9\textwidth}
\begin{lstlisting}[linewidth=.44\textwidth]
x = (a) * (b);
y = (x) * (x);
ddy = one;
ddx = ((ddy) * (x)) + ((ddy) * (x));
ddb = (ddx) * (a);
dda = (ddx) * (b);
\end{lstlisting}
\end{varwidth}}
\end{mathpar}
Let us convince ourselves that the result is correct.  Our expression is $abab = a^2b^2$,
and its partial derivatives $\frac{\partial}{\partial a} = 2ab^2$,
$\frac{\partial}{\partial b} = 2ba^2$.  If we fold the assignments, we get:
\begin{eqnarray*}
   \text{dda} &= (x + x)b = (ab + ab)b = 2ab^2\\
   \text{ddb} &= (x + x)a = (ab + ab)a = 2ba^2
\end{eqnarray*}
Note that computations in $x$ and \texttt{ddx} are shared in further computations
which was the main goal of introducing this mechanism.

There are two inconveniences in the above implementation that we would like to
mention:
\begin{enumerate}
\item There is no restriction on using the placeholders for derivatives in the 
chain expressions, so in principle, one could write expression in terms of
variables and their derivatives.  However, this is not being handled and likely
to generate bogus terms.  If this is a useful feature, it requires more thinking
on how exactly it should work.  Otherwise it is easy to introduce restrictions
that rule out such cases.
\item If we define variables in the chain that do not contribute to the final
expression, we may introduce extra computations.  We do not compromise correctness,
as all inaccessible terms will get zero value.  However, direct execution of the
resulting expressions may introduce redundant computations.
\end{enumerate}
Both of these are future work.  For now, we make an assumption that placeholders
are not used in the expressions and that we do not insert bindings that do not
contribute to the final result.

\begin{code}[hide]%
\>[2]\AgdaFunction{test-chain-sac}\AgdaSpace{}%
\AgdaSymbol{:}\AgdaSpace{}%
\AgdaPostulate{String}\<%
\\
\>[2]\AgdaFunction{test-chain-sac}\<%
\\
\>[2][@{}l@{\AgdaIndent{0}}]%
\>[4]\AgdaSymbol{=}%
\>[3249I]\AgdaFunction{chain-sac}\AgdaSpace{}%
\AgdaFunction{test-chain}\<%
\\
\>[3249I][@{}l@{\AgdaIndent{0}}]%
\>[13]\AgdaOperator{\AgdaFunction{++}}\AgdaSpace{}%
\AgdaString{"\textbackslash{}n"}\AgdaSpace{}%
\AgdaOperator{\AgdaFunction{++}}\AgdaSpace{}%
\AgdaFunction{chain-grad-sac}\AgdaSpace{}%
\AgdaFunction{test-chain}\AgdaSpace{}%
\AgdaSymbol{(}\AgdaFunction{chain-grad}\AgdaSpace{}%
\AgdaFunction{test-chain}\AgdaSpace{}%
\AgdaSymbol{(}\AgdaInductiveConstructor{one}\AgdaSymbol{))}\<%
\\
\>[0]\<%
\end{code}

We present the specification of our case study in \AF{E} using \AF{Chain}.  We start
with the context \AF{cnn-ctx} that contains the \texttt{target} digit that
is depicted on the image, the input image \texttt{inp} and the weights of the network.
The definition of the chain is a one-to-one copy of the definition found in
Section~\ref{sec:cnn}.  The only real difference is that we have to take care of
maintaining bindings between Agda variables and the variables in \AF{E}.  Fortunately,
let expressions in Agda make it possible to shadow the binding, which comes very
useful in this case.

{\small
\begin{code}%
\>[0][@{}l@{\AgdaIndent{1}}]%
\>[2]\AgdaFunction{cnn-ctx}\AgdaSpace{}%
\AgdaSymbol{:}\AgdaSpace{}%
\AgdaDatatype{ChainCtx}\<%
\\
\>[2]\AgdaFunction{cnn-ctx}%
\>[11]\AgdaSymbol{=}\AgdaSpace{}%
\AgdaInductiveConstructor{ε}\<%
\\
\>[11]\AgdaOperator{\AgdaInductiveConstructor{▹}}\AgdaSpace{}%
\AgdaSymbol{(}\AgdaString{"target"}%
\>[24]\AgdaOperator{\AgdaInductiveConstructor{,}}\AgdaSpace{}%
\AgdaInductiveConstructor{ι}\AgdaSpace{}%
\AgdaNumber{10}\AgdaSpace{}%
\AgdaOperator{\AgdaInductiveConstructor{⊗}}\AgdaSpace{}%
\AgdaSymbol{(}\AgdaInductiveConstructor{ι}\AgdaSpace{}%
\AgdaNumber{1}\AgdaSpace{}%
\AgdaOperator{\AgdaInductiveConstructor{⊗}}\AgdaSpace{}%
\AgdaSymbol{(}\AgdaInductiveConstructor{ι}\AgdaSpace{}%
\AgdaNumber{1}\AgdaSpace{}%
\AgdaOperator{\AgdaInductiveConstructor{⊗}}\AgdaSpace{}%
\AgdaSymbol{(}\AgdaInductiveConstructor{ι}\AgdaSpace{}%
\AgdaNumber{1}\AgdaSpace{}%
\AgdaOperator{\AgdaInductiveConstructor{⊗}}\AgdaSpace{}%
\AgdaInductiveConstructor{ι}\AgdaSpace{}%
\AgdaNumber{1}\AgdaSymbol{))))}%
\>[66]\AgdaComment{--\ 7}\<%
\\
\>[11]\AgdaOperator{\AgdaInductiveConstructor{▹}}\AgdaSpace{}%
\AgdaSymbol{(}\AgdaString{"inp"}%
\>[24]\AgdaOperator{\AgdaInductiveConstructor{,}}\AgdaSpace{}%
\AgdaInductiveConstructor{ι}\AgdaSpace{}%
\AgdaNumber{28}\AgdaSpace{}%
\AgdaOperator{\AgdaInductiveConstructor{⊗}}\AgdaSpace{}%
\AgdaInductiveConstructor{ι}\AgdaSpace{}%
\AgdaNumber{28}\AgdaSymbol{)}%
\>[66]\AgdaComment{--\ 6}\<%
\\
\>[11]\AgdaOperator{\AgdaInductiveConstructor{▹}}\AgdaSpace{}%
\AgdaSymbol{(}\AgdaString{"k₁"}%
\>[24]\AgdaOperator{\AgdaInductiveConstructor{,}}\AgdaSpace{}%
\AgdaInductiveConstructor{ι}\AgdaSpace{}%
\AgdaNumber{6}\AgdaSpace{}%
\AgdaOperator{\AgdaInductiveConstructor{⊗}}\AgdaSpace{}%
\AgdaSymbol{(}\AgdaInductiveConstructor{ι}\AgdaSpace{}%
\AgdaNumber{5}\AgdaSpace{}%
\AgdaOperator{\AgdaInductiveConstructor{⊗}}\AgdaSpace{}%
\AgdaInductiveConstructor{ι}\AgdaSpace{}%
\AgdaNumber{5}\AgdaSymbol{))}%
\>[66]\AgdaComment{--\ 5}\<%
\\
\>[11]\AgdaOperator{\AgdaInductiveConstructor{▹}}\AgdaSpace{}%
\AgdaSymbol{(}\AgdaString{"b₁"}%
\>[24]\AgdaOperator{\AgdaInductiveConstructor{,}}\AgdaSpace{}%
\AgdaInductiveConstructor{ι}\AgdaSpace{}%
\AgdaNumber{6}\AgdaSymbol{)}%
\>[66]\AgdaComment{--\ 4}\<%
\\
\>[11]\AgdaOperator{\AgdaInductiveConstructor{▹}}\AgdaSpace{}%
\AgdaSymbol{(}\AgdaString{"k₂"}%
\>[24]\AgdaOperator{\AgdaInductiveConstructor{,}}\AgdaSpace{}%
\AgdaInductiveConstructor{ι}\AgdaSpace{}%
\AgdaNumber{12}\AgdaSpace{}%
\AgdaOperator{\AgdaInductiveConstructor{⊗}}\AgdaSpace{}%
\AgdaSymbol{(}\AgdaInductiveConstructor{ι}\AgdaSpace{}%
\AgdaNumber{6}\AgdaSpace{}%
\AgdaOperator{\AgdaInductiveConstructor{⊗}}\AgdaSpace{}%
\AgdaSymbol{(}\AgdaInductiveConstructor{ι}\AgdaSpace{}%
\AgdaNumber{5}\AgdaSpace{}%
\AgdaOperator{\AgdaInductiveConstructor{⊗}}\AgdaSpace{}%
\AgdaInductiveConstructor{ι}\AgdaSpace{}%
\AgdaNumber{5}\AgdaSymbol{)))}%
\>[66]\AgdaComment{--\ 3}\<%
\\
\>[11]\AgdaOperator{\AgdaInductiveConstructor{▹}}\AgdaSpace{}%
\AgdaSymbol{(}\AgdaString{"b₂"}%
\>[24]\AgdaOperator{\AgdaInductiveConstructor{,}}\AgdaSpace{}%
\AgdaInductiveConstructor{ι}\AgdaSpace{}%
\AgdaNumber{12}\AgdaSymbol{)}%
\>[66]\AgdaComment{--\ 2}\<%
\\
\>[11]\AgdaOperator{\AgdaInductiveConstructor{▹}}\AgdaSpace{}%
\AgdaSymbol{(}\AgdaString{"fc"}%
\>[24]\AgdaOperator{\AgdaInductiveConstructor{,}}\AgdaSpace{}%
\AgdaInductiveConstructor{ι}\AgdaSpace{}%
\AgdaNumber{10}\AgdaSpace{}%
\AgdaOperator{\AgdaInductiveConstructor{⊗}}\AgdaSpace{}%
\AgdaSymbol{(}\AgdaInductiveConstructor{ι}\AgdaSpace{}%
\AgdaNumber{12}\AgdaSpace{}%
\AgdaOperator{\AgdaInductiveConstructor{⊗}}\AgdaSpace{}%
\AgdaSymbol{(}\AgdaInductiveConstructor{ι}\AgdaSpace{}%
\AgdaNumber{1}\AgdaSpace{}%
\AgdaOperator{\AgdaInductiveConstructor{⊗}}\AgdaSpace{}%
\AgdaSymbol{(}\AgdaInductiveConstructor{ι}\AgdaSpace{}%
\AgdaNumber{4}\AgdaSpace{}%
\AgdaOperator{\AgdaInductiveConstructor{⊗}}\AgdaSpace{}%
\AgdaInductiveConstructor{ι}\AgdaSpace{}%
\AgdaNumber{4}\AgdaSymbol{))))}%
\>[66]\AgdaComment{--\ 1}\<%
\\
\>[11]\AgdaOperator{\AgdaInductiveConstructor{▹}}\AgdaSpace{}%
\AgdaSymbol{(}\AgdaString{"b"}%
\>[24]\AgdaOperator{\AgdaInductiveConstructor{,}}\AgdaSpace{}%
\AgdaInductiveConstructor{ι}\AgdaSpace{}%
\AgdaNumber{10}\AgdaSymbol{)}%
\>[66]\AgdaComment{--\ 0}\<%
\\
\\[\AgdaEmptyExtraSkip]%
\>[2]\AgdaFunction{cnn-chain}\AgdaSpace{}%
\AgdaSymbol{:}\AgdaSpace{}%
\AgdaDatatype{Chain}\AgdaSpace{}%
\AgdaSymbol{\AgdaUnderscore{}}\<%
\\
\>[2]\AgdaFunction{cnn-chain}\AgdaSpace{}%
\AgdaSymbol{=}\AgdaSpace{}%
\AgdaKeyword{let}\<%
\\
\>[2][@{}l@{\AgdaIndent{0}}]%
\>[6]\AgdaBound{Γ}\AgdaSpace{}%
\AgdaOperator{\AgdaInductiveConstructor{,}}\AgdaSpace{}%
\AgdaBound{ρ}\AgdaSpace{}%
\AgdaSymbol{=}\AgdaSpace{}%
\AgdaFunction{ce-split}\AgdaSpace{}%
\AgdaFunction{cnn-ctx}\<%
\\
\>[6]\AgdaBound{inp}\AgdaSpace{}%
\AgdaSymbol{=}\AgdaSpace{}%
\AgdaInductiveConstructor{var}\AgdaSpace{}%
\AgdaInductiveConstructor{v₆}\AgdaSymbol{;}\AgdaSpace{}%
\AgdaBound{k₁}\AgdaSpace{}%
\AgdaSymbol{=}\AgdaSpace{}%
\AgdaInductiveConstructor{var}\AgdaSpace{}%
\AgdaInductiveConstructor{v₅}\AgdaSymbol{;}\AgdaSpace{}%
\AgdaBound{b₁}\AgdaSpace{}%
\AgdaSymbol{=}\AgdaSpace{}%
\AgdaInductiveConstructor{var}\AgdaSpace{}%
\AgdaInductiveConstructor{v₄}\AgdaSymbol{;}\AgdaSpace{}%
\AgdaBound{k₂}\AgdaSpace{}%
\AgdaSymbol{=}\AgdaSpace{}%
\AgdaInductiveConstructor{var}\AgdaSpace{}%
\AgdaInductiveConstructor{v₃}\AgdaSymbol{;}\AgdaSpace{}%
\AgdaBound{b₂}\AgdaSpace{}%
\AgdaSymbol{=}\AgdaSpace{}%
\AgdaInductiveConstructor{var}\AgdaSpace{}%
\AgdaInductiveConstructor{v₂}\AgdaSymbol{;}\AgdaSpace{}%
\AgdaBound{fc}\AgdaSpace{}%
\AgdaSymbol{=}\AgdaSpace{}%
\AgdaInductiveConstructor{var}\AgdaSpace{}%
\AgdaInductiveConstructor{v₁}\AgdaSymbol{;}\AgdaSpace{}%
\AgdaBound{b}\AgdaSpace{}%
\AgdaSymbol{=}\AgdaSpace{}%
\AgdaInductiveConstructor{var}\AgdaSpace{}%
\AgdaInductiveConstructor{v₀}\<%
\\
\>[6]\AgdaBound{C₁}\AgdaSpace{}%
\AgdaSymbol{=}\AgdaSpace{}%
\AgdaInductiveConstructor{ε}\AgdaSpace{}%
\AgdaSymbol{\{}\AgdaBound{Γ}\AgdaSymbol{\}}\AgdaSpace{}%
\AgdaBound{ρ}\AgdaSpace{}%
\AgdaOperator{\AgdaInductiveConstructor{▹}}\AgdaSpace{}%
\AgdaSymbol{(}\AgdaString{"c₁₁"}\AgdaSpace{}%
\AgdaOperator{\AgdaInductiveConstructor{,}}\AgdaSpace{}%
\AgdaFunction{mconv}\AgdaSpace{}%
\AgdaSymbol{(}\AgdaInductiveConstructor{ι}\AgdaSpace{}%
\AgdaOperator{\AgdaInductiveConstructor{⊗}}\AgdaSpace{}%
\AgdaInductiveConstructor{ι}\AgdaSymbol{)}\AgdaSpace{}%
\AgdaBound{inp}\AgdaSpace{}%
\AgdaBound{k₁}\AgdaSpace{}%
\AgdaBound{b₁}\AgdaSpace{}%
\AgdaSymbol{(}\AgdaInductiveConstructor{ι}\AgdaSpace{}%
\AgdaOperator{\AgdaInductiveConstructor{⊗}}\AgdaSpace{}%
\AgdaInductiveConstructor{ι}\AgdaSymbol{));}%
\>[71]\AgdaBound{k₂}\AgdaSpace{}%
\AgdaSymbol{=}\AgdaSpace{}%
\AgdaOperator{\AgdaFunction{↑↑}}\AgdaSpace{}%
\AgdaBound{k₂}\AgdaSymbol{;}\AgdaSpace{}%
\AgdaBound{b₂}\AgdaSpace{}%
\AgdaSymbol{=}\AgdaSpace{}%
\AgdaOperator{\AgdaFunction{↑↑}}\AgdaSpace{}%
\AgdaBound{b₂}\AgdaSymbol{;}%
\>[96]\AgdaBound{fc}\AgdaSpace{}%
\AgdaSymbol{=}\AgdaSpace{}%
\AgdaOperator{\AgdaFunction{↑↑}}\AgdaSpace{}%
\AgdaBound{fc}\AgdaSymbol{;}\AgdaSpace{}%
\AgdaBound{b}\AgdaSpace{}%
\AgdaSymbol{=}\AgdaSpace{}%
\AgdaOperator{\AgdaFunction{↑↑}}\AgdaSpace{}%
\AgdaBound{b}\AgdaSymbol{;}\AgdaSpace{}%
\AgdaBound{c₁₁}\AgdaSpace{}%
\AgdaSymbol{=}\AgdaSpace{}%
\AgdaInductiveConstructor{var}\AgdaSpace{}%
\AgdaInductiveConstructor{v₀}\<%
\\
\>[6]\AgdaBound{C₂}\AgdaSpace{}%
\AgdaSymbol{=}\AgdaSpace{}%
\AgdaBound{C₁}\AgdaSpace{}%
\AgdaOperator{\AgdaInductiveConstructor{▹}}\AgdaSpace{}%
\AgdaSymbol{(}\AgdaString{"c₁"}%
\>[23]\AgdaOperator{\AgdaInductiveConstructor{,}}\AgdaSpace{}%
\AgdaInductiveConstructor{logistic}\AgdaSpace{}%
\AgdaBound{c₁₁}\AgdaSymbol{);}%
\>[71]\AgdaBound{k₂}\AgdaSpace{}%
\AgdaSymbol{=}\AgdaSpace{}%
\AgdaOperator{\AgdaFunction{↑↑}}\AgdaSpace{}%
\AgdaBound{k₂}\AgdaSymbol{;}\AgdaSpace{}%
\AgdaBound{b₂}\AgdaSpace{}%
\AgdaSymbol{=}\AgdaSpace{}%
\AgdaOperator{\AgdaFunction{↑↑}}\AgdaSpace{}%
\AgdaBound{b₂}\AgdaSymbol{;}%
\>[96]\AgdaBound{fc}\AgdaSpace{}%
\AgdaSymbol{=}\AgdaSpace{}%
\AgdaOperator{\AgdaFunction{↑↑}}\AgdaSpace{}%
\AgdaBound{fc}\AgdaSymbol{;}\AgdaSpace{}%
\AgdaBound{b}\AgdaSpace{}%
\AgdaSymbol{=}\AgdaSpace{}%
\AgdaOperator{\AgdaFunction{↑↑}}\AgdaSpace{}%
\AgdaBound{b}\AgdaSymbol{;}\AgdaSpace{}%
\AgdaBound{c₁}\AgdaSpace{}%
\AgdaSymbol{=}\AgdaSpace{}%
\AgdaInductiveConstructor{var}\AgdaSpace{}%
\AgdaInductiveConstructor{v₀}\<%
\\
\>[6]\AgdaBound{C₃}\AgdaSpace{}%
\AgdaSymbol{=}\AgdaSpace{}%
\AgdaBound{C₂}\AgdaSpace{}%
\AgdaOperator{\AgdaInductiveConstructor{▹}}\AgdaSpace{}%
\AgdaSymbol{(}\AgdaString{"s₁"}%
\>[23]\AgdaOperator{\AgdaInductiveConstructor{,}}\AgdaSpace{}%
\AgdaFunction{Imap}\AgdaSpace{}%
\AgdaSymbol{λ}\AgdaSpace{}%
\AgdaBound{i}\AgdaSpace{}%
\AgdaSymbol{→}\AgdaSpace{}%
\AgdaFunction{avgp₂}\AgdaSpace{}%
\AgdaNumber{12}\AgdaSpace{}%
\AgdaNumber{12}\AgdaSpace{}%
\AgdaSymbol{(}\AgdaInductiveConstructor{sel}\AgdaSpace{}%
\AgdaSymbol{(}\AgdaOperator{\AgdaFunction{↑}}\AgdaSpace{}%
\AgdaBound{c₁}\AgdaSymbol{)}\AgdaSpace{}%
\AgdaBound{i}\AgdaSymbol{));}%
\>[71]\AgdaBound{k₂}\AgdaSpace{}%
\AgdaSymbol{=}\AgdaSpace{}%
\AgdaOperator{\AgdaFunction{↑↑}}\AgdaSpace{}%
\AgdaBound{k₂}\AgdaSymbol{;}\AgdaSpace{}%
\AgdaBound{b₂}\AgdaSpace{}%
\AgdaSymbol{=}\AgdaSpace{}%
\AgdaOperator{\AgdaFunction{↑↑}}\AgdaSpace{}%
\AgdaBound{b₂}\AgdaSymbol{;}%
\>[96]\AgdaBound{fc}\AgdaSpace{}%
\AgdaSymbol{=}\AgdaSpace{}%
\AgdaOperator{\AgdaFunction{↑↑}}\AgdaSpace{}%
\AgdaBound{fc}\AgdaSymbol{;}\AgdaSpace{}%
\AgdaBound{b}\AgdaSpace{}%
\AgdaSymbol{=}\AgdaSpace{}%
\AgdaOperator{\AgdaFunction{↑↑}}\AgdaSpace{}%
\AgdaBound{b}\AgdaSymbol{;}\AgdaSpace{}%
\AgdaBound{s₁}\AgdaSpace{}%
\AgdaSymbol{=}\AgdaSpace{}%
\AgdaInductiveConstructor{var}\AgdaSpace{}%
\AgdaInductiveConstructor{v₀}\<%
\\
\>[6]\AgdaBound{C₄}\AgdaSpace{}%
\AgdaSymbol{=}\AgdaSpace{}%
\AgdaBound{C₃}\AgdaSpace{}%
\AgdaOperator{\AgdaInductiveConstructor{▹}}\AgdaSpace{}%
\AgdaSymbol{(}\AgdaString{"c₂₁"}\AgdaSpace{}%
\AgdaOperator{\AgdaInductiveConstructor{,}}\AgdaSpace{}%
\AgdaFunction{mconv}\AgdaSpace{}%
\AgdaSymbol{(}\AgdaInductiveConstructor{ι}\AgdaSpace{}%
\AgdaOperator{\AgdaInductiveConstructor{⊗}}\AgdaSpace{}%
\AgdaSymbol{(}\AgdaInductiveConstructor{ι}\AgdaSpace{}%
\AgdaOperator{\AgdaInductiveConstructor{⊗}}\AgdaSpace{}%
\AgdaInductiveConstructor{ι}\AgdaSymbol{))}\AgdaSpace{}%
\AgdaBound{s₁}\AgdaSpace{}%
\AgdaBound{k₂}\AgdaSpace{}%
\AgdaBound{b₂}\AgdaSpace{}%
\AgdaSymbol{(}\AgdaInductiveConstructor{ι}\AgdaSpace{}%
\AgdaOperator{\AgdaInductiveConstructor{⊗}}\AgdaSpace{}%
\AgdaSymbol{(}\AgdaInductiveConstructor{ι}\AgdaSpace{}%
\AgdaOperator{\AgdaInductiveConstructor{⊗}}\AgdaSpace{}%
\AgdaInductiveConstructor{ι}\AgdaSymbol{)));}%
\>[96]\AgdaBound{fc}\AgdaSpace{}%
\AgdaSymbol{=}\AgdaSpace{}%
\AgdaOperator{\AgdaFunction{↑↑}}\AgdaSpace{}%
\AgdaBound{fc}\AgdaSymbol{;}\AgdaSpace{}%
\AgdaBound{b}\AgdaSpace{}%
\AgdaSymbol{=}\AgdaSpace{}%
\AgdaOperator{\AgdaFunction{↑↑}}\AgdaSpace{}%
\AgdaBound{b}\AgdaSymbol{;}\AgdaSpace{}%
\AgdaBound{c₂₁}\AgdaSpace{}%
\AgdaSymbol{=}\AgdaSpace{}%
\AgdaInductiveConstructor{var}\AgdaSpace{}%
\AgdaInductiveConstructor{v₀}\<%
\\
\>[6]\AgdaBound{C₅}\AgdaSpace{}%
\AgdaSymbol{=}\AgdaSpace{}%
\AgdaBound{C₄}\AgdaSpace{}%
\AgdaOperator{\AgdaInductiveConstructor{▹}}\AgdaSpace{}%
\AgdaSymbol{(}\AgdaString{"c₂"}%
\>[23]\AgdaOperator{\AgdaInductiveConstructor{,}}\AgdaSpace{}%
\AgdaInductiveConstructor{logistic}\AgdaSpace{}%
\AgdaBound{c₂₁}\AgdaSymbol{);}%
\>[96]\AgdaBound{fc}\AgdaSpace{}%
\AgdaSymbol{=}\AgdaSpace{}%
\AgdaOperator{\AgdaFunction{↑↑}}\AgdaSpace{}%
\AgdaBound{fc}\AgdaSymbol{;}\AgdaSpace{}%
\AgdaBound{b}\AgdaSpace{}%
\AgdaSymbol{=}\AgdaSpace{}%
\AgdaOperator{\AgdaFunction{↑↑}}\AgdaSpace{}%
\AgdaBound{b}\AgdaSymbol{;}\AgdaSpace{}%
\AgdaBound{c₂}\AgdaSpace{}%
\AgdaSymbol{=}\AgdaSpace{}%
\AgdaInductiveConstructor{var}\AgdaSpace{}%
\AgdaInductiveConstructor{v₀}\<%
\\
\>[6]\AgdaBound{C₆}\AgdaSpace{}%
\AgdaSymbol{=}\AgdaSpace{}%
\AgdaBound{C₅}\AgdaSpace{}%
\AgdaOperator{\AgdaInductiveConstructor{▹}}\AgdaSpace{}%
\AgdaSymbol{(}\AgdaString{"s₂"}%
\>[23]\AgdaOperator{\AgdaInductiveConstructor{,}}\AgdaSpace{}%
\AgdaFunction{Imap}\AgdaSpace{}%
\AgdaSymbol{λ}\AgdaSpace{}%
\AgdaBound{i}\AgdaSpace{}%
\AgdaSymbol{→}\AgdaSpace{}%
\AgdaFunction{Imap}\AgdaSpace{}%
\AgdaSymbol{λ}\AgdaSpace{}%
\AgdaBound{j}\AgdaSpace{}%
\AgdaSymbol{→}\AgdaSpace{}%
\AgdaFunction{avgp₂}\AgdaSpace{}%
\AgdaNumber{4}\AgdaSpace{}%
\AgdaNumber{4}\AgdaSpace{}%
\AgdaSymbol{(}\AgdaInductiveConstructor{sel}\AgdaSpace{}%
\AgdaSymbol{(}\AgdaInductiveConstructor{sel}\AgdaSpace{}%
\AgdaSymbol{(}\AgdaOperator{\AgdaFunction{↑↑}}\AgdaSpace{}%
\AgdaBound{c₂}\AgdaSymbol{)}\AgdaSpace{}%
\AgdaSymbol{(}\AgdaOperator{\AgdaFunction{↑}}\AgdaSpace{}%
\AgdaBound{i}\AgdaSymbol{))}\AgdaSpace{}%
\AgdaBound{j}\AgdaSymbol{));}%
\>[96]\AgdaBound{fc}\AgdaSpace{}%
\AgdaSymbol{=}\AgdaSpace{}%
\AgdaOperator{\AgdaFunction{↑↑}}\AgdaSpace{}%
\AgdaBound{fc}\AgdaSymbol{;}\AgdaSpace{}%
\AgdaBound{b}\AgdaSpace{}%
\AgdaSymbol{=}\AgdaSpace{}%
\AgdaOperator{\AgdaFunction{↑↑}}\AgdaSpace{}%
\AgdaBound{b}\AgdaSymbol{;}\AgdaSpace{}%
\AgdaBound{s₂}\AgdaSpace{}%
\AgdaSymbol{=}\AgdaSpace{}%
\AgdaInductiveConstructor{var}\AgdaSpace{}%
\AgdaInductiveConstructor{v₀}\<%
\\
\>[6]\AgdaBound{C₇}\AgdaSpace{}%
\AgdaSymbol{=}\AgdaSpace{}%
\AgdaBound{C₆}\AgdaSpace{}%
\AgdaOperator{\AgdaInductiveConstructor{▹}}\AgdaSpace{}%
\AgdaSymbol{(}\AgdaString{"r₁"}%
\>[23]\AgdaOperator{\AgdaInductiveConstructor{,}}\AgdaSpace{}%
\AgdaFunction{mconv}\AgdaSpace{}%
\AgdaSymbol{(}\AgdaInductiveConstructor{ι}\AgdaSpace{}%
\AgdaOperator{\AgdaInductiveConstructor{⊗}}\AgdaSpace{}%
\AgdaSymbol{(}\AgdaInductiveConstructor{ι}\AgdaSpace{}%
\AgdaOperator{\AgdaInductiveConstructor{⊗}}\AgdaSpace{}%
\AgdaSymbol{(}\AgdaInductiveConstructor{ι}\AgdaSpace{}%
\AgdaOperator{\AgdaInductiveConstructor{⊗}}\AgdaSpace{}%
\AgdaInductiveConstructor{ι}\AgdaSymbol{)))}\AgdaSpace{}%
\AgdaBound{s₂}\AgdaSpace{}%
\AgdaBound{fc}\AgdaSpace{}%
\AgdaBound{b}\AgdaSpace{}%
\AgdaSymbol{(}\AgdaInductiveConstructor{ι}\AgdaSpace{}%
\AgdaOperator{\AgdaInductiveConstructor{⊗}}\AgdaSpace{}%
\AgdaSymbol{(}\AgdaInductiveConstructor{ι}\AgdaSpace{}%
\AgdaOperator{\AgdaInductiveConstructor{⊗}}\AgdaSpace{}%
\AgdaSymbol{(}\AgdaInductiveConstructor{ι}\AgdaSpace{}%
\AgdaOperator{\AgdaInductiveConstructor{⊗}}\AgdaSpace{}%
\AgdaInductiveConstructor{ι}\AgdaSymbol{))));}%
\>[96]\AgdaBound{r₁}\AgdaSpace{}%
\AgdaSymbol{=}\AgdaSpace{}%
\AgdaInductiveConstructor{var}\AgdaSpace{}%
\AgdaInductiveConstructor{v₀}\<%
\\
\>[6]\AgdaBound{C₈}\AgdaSpace{}%
\AgdaSymbol{=}\AgdaSpace{}%
\AgdaBound{C₇}\AgdaSpace{}%
\AgdaOperator{\AgdaInductiveConstructor{▹}}\AgdaSpace{}%
\AgdaSymbol{(}\AgdaString{"r"}%
\>[23]\AgdaOperator{\AgdaInductiveConstructor{,}}\AgdaSpace{}%
\AgdaInductiveConstructor{logistic}\AgdaSpace{}%
\AgdaBound{r₁}\AgdaSymbol{)}\<%
\\
\>[6]\AgdaKeyword{in}\AgdaSpace{}%
\AgdaBound{C₈}\<%
\end{code}

\begin{code}[hide]%
\>[2]\AgdaFunction{test-cnn}\AgdaSpace{}%
\AgdaSymbol{:}\AgdaSpace{}%
\AgdaPostulate{String}\<%
\\
\>[2]\AgdaFunction{test-cnn}\<%
\\
\>[2][@{}l@{\AgdaIndent{0}}]%
\>[4]\AgdaSymbol{=}%
\>[3570I]\AgdaKeyword{let}\<%
\\
\>[3570I][@{}l@{\AgdaIndent{0}}]%
\>[8]\AgdaComment{--\ 2*8\ +\ 7\ =\ 23}\<%
\\
\>[8]\AgdaBound{target}\AgdaSpace{}%
\AgdaSymbol{=}\AgdaSpace{}%
\AgdaOperator{\AgdaFunction{↑↑}}\AgdaSpace{}%
\AgdaOperator{\AgdaFunction{↑↑}}\AgdaSpace{}%
\AgdaOperator{\AgdaFunction{↑↑}}\AgdaSpace{}%
\AgdaOperator{\AgdaFunction{↑↑}}\AgdaSpace{}%
\AgdaOperator{\AgdaFunction{↑↑}}%
\>[33]\AgdaOperator{\AgdaFunction{↑↑}}\AgdaSpace{}%
\AgdaOperator{\AgdaFunction{↑↑}}\AgdaSpace{}%
\AgdaOperator{\AgdaFunction{↑↑}}\AgdaSpace{}%
\AgdaOperator{\AgdaFunction{↑↑}}\AgdaSpace{}%
\AgdaOperator{\AgdaFunction{↑↑}}%
\>[49]\AgdaOperator{\AgdaFunction{↑↑}}\AgdaSpace{}%
\AgdaOperator{\AgdaFunction{↑}}\AgdaSpace{}%
\AgdaSymbol{(}\AgdaInductiveConstructor{var}\AgdaSpace{}%
\AgdaInductiveConstructor{v₀}\AgdaSymbol{)}\<%
\\
\>[.][@{}l@{}]\<[3570I]%
\>[6]\AgdaKeyword{in}%
\>[3584I]\AgdaFunction{chain-sac}\AgdaSpace{}%
\AgdaFunction{cnn-chain}\<%
\\
\>[3584I][@{}l@{\AgdaIndent{0}}]%
\>[13]\AgdaOperator{\AgdaFunction{++}}\AgdaSpace{}%
\AgdaString{"\textbackslash{}n"}\AgdaSpace{}%
\AgdaOperator{\AgdaFunction{++}}\AgdaSpace{}%
\AgdaFunction{chain-grad-sac}\AgdaSpace{}%
\AgdaFunction{cnn-chain}\AgdaSpace{}%
\AgdaSymbol{(}\AgdaFunction{chain-grad}\AgdaSpace{}%
\AgdaFunction{cnn-chain}\AgdaSpace{}%
\AgdaSymbol{(}\AgdaInductiveConstructor{var}\AgdaSpace{}%
\AgdaInductiveConstructor{v₀}\AgdaSpace{}%
\AgdaOperator{\AgdaInductiveConstructor{⊞}}\AgdaSpace{}%
\AgdaInductiveConstructor{minus}\AgdaSpace{}%
\AgdaBound{target}\AgdaSymbol{))}\<%
\end{code}
}

\section{Performance\label{sec:performance}}

One of the goals of this work is to demonstrate that it is possible to formulate
the problem in a proof assistant and then pass it on to the other system that can
run the algorithm efficiently.  In order to substantiate this claim, we compare
the running times of the code that we generate from the specification at the end of the
Section~\ref{sec:ad} and the hand-written SaC code from~\cite{cnn-array}.
We are not interested in an exhaustive performance study similar to what is provided
in \cite{cnn-array}. Instead, we take the version from that paper as reference point 
and we aim to find out whether we are in the same ballpark.

We take the code from~\cite{cnn-array}, make sure that it runs, and we replace
the hand-written CNN training with the Agda-generated one. 
Our first observation is that both versions 
produce the same results, and none of the shape constraints
that we defined in Section~\ref{sec:sac-primitives} fired.  This means that our
code generation is working.  Unfortunately, the runtime comparison revealed that
our version is about 10$\times$ slower than the hand-written SaC version.

We got in touch with the SaC team who provided a lot of support in identifying
the causes of the disappointing difference in performance.  It turns out that the main culprit has
to do with the inability to optimise away selections into tensor comprehensions in a few situations.
With-Loop-Folding~\cite{wlf}, SaC's mechanism for fusing tensor comprehensions fails to fold
tensor comprehensions that are nested and cannot be flattened statically.
In simple terms, the expression \texttt{\{iv -> e(iv)\}[jv]} in some situation does not reduce to
\texttt{e(jv)} which, in our generated code, is essential to match the hand-written performance.
As a result, several arrays were created
simply to make a single selection into them.  The original code never ran into this
problem as the hand-written code avoided such patterns.  Our \AF{E} optimiser
from Section~\ref{sec:opt} could not help either, because the problem was occurring after
the SaC primitives such as slide and block were inlined.

After numerous attempts on altering \AF{E} to fit SaC requirements and the SaC
team trying to implement some of the missing optimisations, on the February 23rd
(6 days before the deadline) we realised that the best runtime we can
obtain is still 6$\times$ slower than the hand-written code.  The compiler is too
sensitive to the flavour of the code that we pass to it, and when certain patterns
are not recognised, there is very little one can do other than trying to fit
those patterns.  However, this is not always possible with the generated code.
Performance \emph{is} frustrating!

\subsection{Generating C}

After overcoming the natural instinct to give up, we realised that the real
power of the proposed approach lies in the ability to modify any part of the
code generation pipeline.  This includes swapping the back-end language of choice
to something else.  Therefore, we decided to try generating C code instead of
SaC code.

While C is a canonical low-level language, it has excellent support for
multi-dimensional arrays, given that the ranks are statically known.
At runtime these arrays are flattened vectors, they do not have to live
on the stack, and the language takes care of multi-dimensional indexing.

However, the key difference between the C and SaC is memory management.
SaC is a functional language that uses reference counting to automate
operations on allocating and freeing memory.  In C these decisions are
manual, and as we have seen before, excessive memory allocation is detrimental
for the runtime.  For our use case we avoid memory management problem
entirely, by assuming that all the variables in the \AF{Chain} have
to be preallocated, and if we need any temporary array variables when
extracting array values, we fail extraction.  This way we guarantee
that no memory allocation is ever needed.

Meeting such a requirement means that we need to optimise away operations
like \AC{slide}/\AC{backslide} as they require conceptual array allocation.
The same goes for \AC{imap}s appearing in some of the argument positions.
Putting these considerations together, we extended \AF{E} with the following
explicit operations on indices:
\begin{code}[hide]%
\>[0]\AgdaKeyword{open}\AgdaSpace{}%
\AgdaKeyword{import}\AgdaSpace{}%
\AgdaModule{Data.Nat}\AgdaSpace{}%
\AgdaSymbol{as}\AgdaSpace{}%
\AgdaModule{ℕ}\AgdaSpace{}%
\AgdaKeyword{using}\AgdaSpace{}%
\AgdaSymbol{(}\AgdaDatatype{ℕ}\AgdaSymbol{;}\AgdaSpace{}%
\AgdaInductiveConstructor{zero}\AgdaSymbol{;}\AgdaSpace{}%
\AgdaInductiveConstructor{suc}\AgdaSymbol{)}\<%
\\
\>[0]\AgdaKeyword{open}\AgdaSpace{}%
\AgdaKeyword{import}\AgdaSpace{}%
\AgdaModule{Data.Unit}\<%
\\
\>[0]\AgdaKeyword{open}\AgdaSpace{}%
\AgdaKeyword{import}\AgdaSpace{}%
\AgdaModule{Data.Empty}\<%
\\
\>[0]\AgdaKeyword{open}\AgdaSpace{}%
\AgdaKeyword{import}\AgdaSpace{}%
\AgdaModule{Data.Product}\AgdaSpace{}%
\AgdaSymbol{as}\AgdaSpace{}%
\AgdaModule{Prod}\AgdaSpace{}%
\AgdaKeyword{using}\AgdaSpace{}%
\AgdaSymbol{(}\AgdaRecord{Σ}\AgdaSymbol{;}\AgdaSpace{}%
\AgdaFunction{∃}\AgdaSymbol{;}\AgdaSpace{}%
\AgdaOperator{\AgdaInductiveConstructor{\AgdaUnderscore{},\AgdaUnderscore{}}}\AgdaSymbol{;}\AgdaSpace{}%
\AgdaOperator{\AgdaFunction{\AgdaUnderscore{}×\AgdaUnderscore{}}}\AgdaSymbol{;}\AgdaSpace{}%
\AgdaField{proj₁}\AgdaSymbol{;}\AgdaSpace{}%
\AgdaField{proj₂}\AgdaSymbol{)}\<%
\\
\>[0]\AgdaKeyword{open}\AgdaSpace{}%
\AgdaKeyword{import}\AgdaSpace{}%
\AgdaModule{Relation.Nullary}\<%
\\
\>[0]\AgdaKeyword{open}\AgdaSpace{}%
\AgdaKeyword{import}\AgdaSpace{}%
\AgdaModule{Relation.Binary.PropositionalEquality}\AgdaSpace{}%
\AgdaKeyword{hiding}\AgdaSpace{}%
\AgdaSymbol{(}\AgdaOperator{\AgdaInductiveConstructor{[\AgdaUnderscore{}]}}\AgdaSymbol{)}\<%
\\
\>[0]\AgdaKeyword{open}\AgdaSpace{}%
\AgdaKeyword{import}\AgdaSpace{}%
\AgdaModule{Data.List}\AgdaSpace{}%
\AgdaSymbol{as}\AgdaSpace{}%
\AgdaModule{L}\AgdaSpace{}%
\AgdaKeyword{using}\AgdaSpace{}%
\AgdaSymbol{(}\AgdaDatatype{List}\AgdaSymbol{;}\AgdaSpace{}%
\AgdaInductiveConstructor{[]}\AgdaSymbol{;}\AgdaSpace{}%
\AgdaOperator{\AgdaInductiveConstructor{\AgdaUnderscore{}∷\AgdaUnderscore{}}}\AgdaSymbol{)}\<%
\\
\>[0]\AgdaKeyword{open}\AgdaSpace{}%
\AgdaKeyword{import}\AgdaSpace{}%
\AgdaModule{Function}\<%
\\
\>[0]\AgdaKeyword{open}\AgdaSpace{}%
\AgdaKeyword{import}\AgdaSpace{}%
\AgdaModule{arrays}\<%
\\
\>[0]\AgdaKeyword{open}\AgdaSpace{}%
\AgdaModule{Array}\AgdaSpace{}%
\AgdaKeyword{hiding}\AgdaSpace{}%
\AgdaSymbol{(}\AgdaFunction{sum}\AgdaSymbol{;}\AgdaSpace{}%
\AgdaFunction{slide}\AgdaSymbol{;}\AgdaSpace{}%
\AgdaFunction{backslide}\AgdaSymbol{)}\<%
\\
\\[\AgdaEmptyExtraSkip]%
\>[0]\AgdaKeyword{data}\AgdaSpace{}%
\AgdaDatatype{IS}\AgdaSpace{}%
\AgdaSymbol{:}\AgdaSpace{}%
\AgdaPrimitive{Set}\AgdaSpace{}%
\AgdaKeyword{where}\<%
\\
\>[0][@{}l@{\AgdaIndent{0}}]%
\>[2]\AgdaInductiveConstructor{ix}\AgdaSpace{}%
\AgdaSymbol{:}\AgdaSpace{}%
\AgdaDatatype{S}\AgdaSpace{}%
\AgdaSymbol{→}\AgdaSpace{}%
\AgdaDatatype{IS}\<%
\\
\>[2]\AgdaInductiveConstructor{ar}\AgdaSpace{}%
\AgdaSymbol{:}\AgdaSpace{}%
\AgdaDatatype{S}\AgdaSpace{}%
\AgdaSymbol{→}\AgdaSpace{}%
\AgdaDatatype{IS}\<%
\\
\\[\AgdaEmptyExtraSkip]%
\>[0]\AgdaKeyword{infixl}\AgdaSpace{}%
\AgdaNumber{15}\AgdaSpace{}%
\AgdaOperator{\AgdaInductiveConstructor{\AgdaUnderscore{}▹\AgdaUnderscore{}}}\<%
\\
\>[0]\AgdaKeyword{data}\AgdaSpace{}%
\AgdaDatatype{Ctx}\AgdaSpace{}%
\AgdaSymbol{:}\AgdaSpace{}%
\AgdaPrimitive{Set}\AgdaSpace{}%
\AgdaKeyword{where}\<%
\\
\>[0][@{}l@{\AgdaIndent{0}}]%
\>[2]\AgdaInductiveConstructor{ε}\AgdaSpace{}%
\AgdaSymbol{:}\AgdaSpace{}%
\AgdaDatatype{Ctx}\<%
\\
\>[2]\AgdaOperator{\AgdaInductiveConstructor{\AgdaUnderscore{}▹\AgdaUnderscore{}}}\AgdaSpace{}%
\AgdaSymbol{:}\AgdaSpace{}%
\AgdaDatatype{Ctx}\AgdaSpace{}%
\AgdaSymbol{→}\AgdaSpace{}%
\AgdaDatatype{IS}\AgdaSpace{}%
\AgdaSymbol{→}\AgdaSpace{}%
\AgdaDatatype{Ctx}\<%
\\
\\[\AgdaEmptyExtraSkip]%
\>[0]\AgdaKeyword{variable}\<%
\\
\>[0][@{}l@{\AgdaIndent{0}}]%
\>[2]\AgdaGeneralizable{Γ}\AgdaSpace{}%
\AgdaGeneralizable{Δ}\AgdaSpace{}%
\AgdaGeneralizable{Ξ}\AgdaSpace{}%
\AgdaGeneralizable{Ψ}\AgdaSpace{}%
\AgdaSymbol{:}\AgdaSpace{}%
\AgdaDatatype{Ctx}\<%
\\
\>[2]\AgdaGeneralizable{is}\AgdaSpace{}%
\AgdaGeneralizable{ip}\AgdaSpace{}%
\AgdaGeneralizable{iq}\AgdaSpace{}%
\AgdaGeneralizable{ir}\AgdaSpace{}%
\AgdaSymbol{:}\AgdaSpace{}%
\AgdaDatatype{IS}\<%
\\
\\[\AgdaEmptyExtraSkip]%
\>[0]\AgdaKeyword{data}\AgdaSpace{}%
\AgdaOperator{\AgdaDatatype{\AgdaUnderscore{}∈\AgdaUnderscore{}}}\AgdaSpace{}%
\AgdaSymbol{:}\AgdaSpace{}%
\AgdaDatatype{IS}\AgdaSpace{}%
\AgdaSymbol{→}\AgdaSpace{}%
\AgdaDatatype{Ctx}\AgdaSpace{}%
\AgdaSymbol{→}\AgdaSpace{}%
\AgdaPrimitive{Set}\AgdaSpace{}%
\AgdaKeyword{where}\<%
\\
\>[0][@{}l@{\AgdaIndent{0}}]%
\>[2]\AgdaInductiveConstructor{here}\AgdaSpace{}%
\AgdaSymbol{:}\AgdaSpace{}%
\AgdaGeneralizable{is}\AgdaSpace{}%
\AgdaOperator{\AgdaDatatype{∈}}\AgdaSpace{}%
\AgdaSymbol{(}\AgdaGeneralizable{Γ}\AgdaSpace{}%
\AgdaOperator{\AgdaInductiveConstructor{▹}}\AgdaSpace{}%
\AgdaGeneralizable{is}\AgdaSymbol{)}\<%
\\
\>[2]\AgdaInductiveConstructor{there}\AgdaSpace{}%
\AgdaSymbol{:}\AgdaSpace{}%
\AgdaGeneralizable{is}\AgdaSpace{}%
\AgdaOperator{\AgdaDatatype{∈}}\AgdaSpace{}%
\AgdaGeneralizable{Γ}\AgdaSpace{}%
\AgdaSymbol{→}\AgdaSpace{}%
\AgdaGeneralizable{is}\AgdaSpace{}%
\AgdaOperator{\AgdaDatatype{∈}}\AgdaSpace{}%
\AgdaSymbol{(}\AgdaGeneralizable{Γ}\AgdaSpace{}%
\AgdaOperator{\AgdaInductiveConstructor{▹}}\AgdaSpace{}%
\AgdaGeneralizable{ip}\AgdaSymbol{)}\<%
\\
\\[\AgdaEmptyExtraSkip]%
\>[0]\AgdaKeyword{pattern}\AgdaSpace{}%
\AgdaInductiveConstructor{v₀}\AgdaSpace{}%
\AgdaSymbol{=}\AgdaSpace{}%
\AgdaInductiveConstructor{here}\<%
\\
\>[0]\AgdaKeyword{pattern}\AgdaSpace{}%
\AgdaInductiveConstructor{v₁}\AgdaSpace{}%
\AgdaSymbol{=}\AgdaSpace{}%
\AgdaInductiveConstructor{there}\AgdaSpace{}%
\AgdaInductiveConstructor{v₀}\<%
\\
\>[0]\AgdaKeyword{pattern}\AgdaSpace{}%
\AgdaInductiveConstructor{v₂}\AgdaSpace{}%
\AgdaSymbol{=}\AgdaSpace{}%
\AgdaInductiveConstructor{there}\AgdaSpace{}%
\AgdaInductiveConstructor{v₁}\<%
\\
\>[0]\AgdaKeyword{pattern}\AgdaSpace{}%
\AgdaInductiveConstructor{v₃}\AgdaSpace{}%
\AgdaSymbol{=}\AgdaSpace{}%
\AgdaInductiveConstructor{there}\AgdaSpace{}%
\AgdaInductiveConstructor{v₂}\<%
\\
\>[0]\AgdaKeyword{pattern}\AgdaSpace{}%
\AgdaInductiveConstructor{v₄}\AgdaSpace{}%
\AgdaSymbol{=}\AgdaSpace{}%
\AgdaInductiveConstructor{there}\AgdaSpace{}%
\AgdaInductiveConstructor{v₃}\<%
\\
\>[0]\AgdaKeyword{pattern}\AgdaSpace{}%
\AgdaInductiveConstructor{v₅}\AgdaSpace{}%
\AgdaSymbol{=}\AgdaSpace{}%
\AgdaInductiveConstructor{there}\AgdaSpace{}%
\AgdaInductiveConstructor{v₄}\<%
\\
\>[0]\AgdaKeyword{pattern}\AgdaSpace{}%
\AgdaInductiveConstructor{v₆}\AgdaSpace{}%
\AgdaSymbol{=}\AgdaSpace{}%
\AgdaInductiveConstructor{there}\AgdaSpace{}%
\AgdaInductiveConstructor{v₅}\<%
\\
\>[0]\AgdaKeyword{pattern}\AgdaSpace{}%
\AgdaInductiveConstructor{v₇}\AgdaSpace{}%
\AgdaSymbol{=}\AgdaSpace{}%
\AgdaInductiveConstructor{there}\AgdaSpace{}%
\AgdaInductiveConstructor{v₆}\<%
\\
\>[0]\AgdaKeyword{pattern}\AgdaSpace{}%
\AgdaInductiveConstructor{v₈}\AgdaSpace{}%
\AgdaSymbol{=}\AgdaSpace{}%
\AgdaInductiveConstructor{there}\AgdaSpace{}%
\AgdaInductiveConstructor{v₇}\<%
\\
\>[0]\AgdaKeyword{pattern}\AgdaSpace{}%
\AgdaInductiveConstructor{v₉}\AgdaSpace{}%
\AgdaSymbol{=}\AgdaSpace{}%
\AgdaInductiveConstructor{there}\AgdaSpace{}%
\AgdaInductiveConstructor{v₈}\<%
\\
\\[\AgdaEmptyExtraSkip]%
\>[0]\AgdaFunction{unit}\AgdaSpace{}%
\AgdaSymbol{:}\AgdaSpace{}%
\AgdaDatatype{S}\<%
\\
\>[0]\AgdaFunction{unit}\AgdaSpace{}%
\AgdaSymbol{=}\AgdaSpace{}%
\AgdaInductiveConstructor{ι}\AgdaSpace{}%
\AgdaNumber{1}\<%
\\
\\[\AgdaEmptyExtraSkip]%
\>[0]\AgdaKeyword{data}\AgdaSpace{}%
\AgdaDatatype{Bop}\AgdaSpace{}%
\AgdaSymbol{:}\AgdaSpace{}%
\AgdaPrimitive{Set}\AgdaSpace{}%
\AgdaKeyword{where}\<%
\\
\>[0][@{}l@{\AgdaIndent{0}}]%
\>[2]\AgdaInductiveConstructor{plus}\AgdaSpace{}%
\AgdaInductiveConstructor{mul}\AgdaSpace{}%
\AgdaSymbol{:}\AgdaSpace{}%
\AgdaDatatype{Bop}\<%
\end{code}
\begin{code}%
\>[0]\AgdaKeyword{data}\AgdaSpace{}%
\AgdaDatatype{E}\AgdaSpace{}%
\AgdaSymbol{:}\AgdaSpace{}%
\AgdaDatatype{Ctx}\AgdaSpace{}%
\AgdaSymbol{→}\AgdaSpace{}%
\AgdaDatatype{IS}\AgdaSpace{}%
\AgdaSymbol{→}\AgdaSpace{}%
\AgdaPrimitive{Set}\AgdaSpace{}%
\AgdaKeyword{where}\<%
\\
\>[0][@{}l@{\AgdaIndent{0}}]%
\>[2]\AgdaInductiveConstructor{div}%
\>[13]\AgdaSymbol{:}\AgdaSpace{}%
\AgdaGeneralizable{s}\AgdaSpace{}%
\AgdaOperator{\AgdaDatatype{*}}\AgdaSpace{}%
\AgdaGeneralizable{p}\AgdaSpace{}%
\AgdaOperator{\AgdaDatatype{≈}}\AgdaSpace{}%
\AgdaGeneralizable{q}\AgdaSpace{}%
\AgdaSymbol{→}\AgdaSpace{}%
\AgdaSymbol{(}\AgdaBound{i}\AgdaSpace{}%
\AgdaSymbol{:}\AgdaSpace{}%
\AgdaDatatype{E}\AgdaSpace{}%
\AgdaGeneralizable{Γ}\AgdaSpace{}%
\AgdaSymbol{(}\AgdaInductiveConstructor{ix}\AgdaSpace{}%
\AgdaGeneralizable{q}\AgdaSymbol{))}\AgdaSpace{}%
\AgdaSymbol{→}\AgdaSpace{}%
\AgdaDatatype{E}\AgdaSpace{}%
\AgdaGeneralizable{Γ}\AgdaSpace{}%
\AgdaSymbol{(}\AgdaInductiveConstructor{ix}\AgdaSpace{}%
\AgdaGeneralizable{s}\AgdaSymbol{)}\<%
\\
\>[2]\AgdaInductiveConstructor{mod}%
\>[13]\AgdaSymbol{:}\AgdaSpace{}%
\AgdaGeneralizable{s}\AgdaSpace{}%
\AgdaOperator{\AgdaDatatype{*}}\AgdaSpace{}%
\AgdaGeneralizable{p}\AgdaSpace{}%
\AgdaOperator{\AgdaDatatype{≈}}\AgdaSpace{}%
\AgdaGeneralizable{q}\AgdaSpace{}%
\AgdaSymbol{→}\AgdaSpace{}%
\AgdaSymbol{(}\AgdaBound{i}\AgdaSpace{}%
\AgdaSymbol{:}\AgdaSpace{}%
\AgdaDatatype{E}\AgdaSpace{}%
\AgdaGeneralizable{Γ}\AgdaSpace{}%
\AgdaSymbol{(}\AgdaInductiveConstructor{ix}\AgdaSpace{}%
\AgdaGeneralizable{q}\AgdaSymbol{))}\AgdaSpace{}%
\AgdaSymbol{→}\AgdaSpace{}%
\AgdaDatatype{E}\AgdaSpace{}%
\AgdaGeneralizable{Γ}\AgdaSpace{}%
\AgdaSymbol{(}\AgdaInductiveConstructor{ix}\AgdaSpace{}%
\AgdaGeneralizable{p}\AgdaSymbol{)}\<%
\\
\>[2]\AgdaInductiveConstructor{ix-plus}%
\>[13]\AgdaSymbol{:}\AgdaSpace{}%
\AgdaSymbol{(}\AgdaBound{i}\AgdaSpace{}%
\AgdaSymbol{:}\AgdaSpace{}%
\AgdaDatatype{E}\AgdaSpace{}%
\AgdaGeneralizable{Γ}\AgdaSpace{}%
\AgdaSymbol{(}\AgdaInductiveConstructor{ix}\AgdaSpace{}%
\AgdaGeneralizable{s}\AgdaSymbol{))}\AgdaSpace{}%
\AgdaSymbol{→}\AgdaSpace{}%
\AgdaSymbol{(}\AgdaBound{j}\AgdaSpace{}%
\AgdaSymbol{:}\AgdaSpace{}%
\AgdaDatatype{E}\AgdaSpace{}%
\AgdaGeneralizable{Γ}\AgdaSpace{}%
\AgdaSymbol{(}\AgdaInductiveConstructor{ix}\AgdaSpace{}%
\AgdaGeneralizable{u}\AgdaSymbol{))}\AgdaSpace{}%
\AgdaSymbol{→}\AgdaSpace{}%
\AgdaOperator{\AgdaDatatype{suc}}\AgdaSpace{}%
\AgdaGeneralizable{p}\AgdaSpace{}%
\AgdaOperator{\AgdaDatatype{≈}}\AgdaSpace{}%
\AgdaGeneralizable{u}\AgdaSpace{}%
\AgdaSymbol{→}\AgdaSpace{}%
\AgdaGeneralizable{s}\AgdaSpace{}%
\AgdaOperator{\AgdaDatatype{+}}\AgdaSpace{}%
\AgdaGeneralizable{p}\AgdaSpace{}%
\AgdaOperator{\AgdaDatatype{≈}}\AgdaSpace{}%
\AgdaGeneralizable{r}\AgdaSpace{}%
\AgdaSymbol{→}\AgdaSpace{}%
\AgdaDatatype{E}\AgdaSpace{}%
\AgdaGeneralizable{Γ}\AgdaSpace{}%
\AgdaSymbol{(}\AgdaInductiveConstructor{ix}\AgdaSpace{}%
\AgdaGeneralizable{r}\AgdaSymbol{)}\<%
\\
\>[2]\AgdaInductiveConstructor{ix-minus}%
\>[13]\AgdaSymbol{:}\AgdaSpace{}%
\AgdaSymbol{(}\AgdaBound{i}\AgdaSpace{}%
\AgdaSymbol{:}\AgdaSpace{}%
\AgdaDatatype{E}\AgdaSpace{}%
\AgdaGeneralizable{Γ}\AgdaSpace{}%
\AgdaSymbol{(}\AgdaInductiveConstructor{ix}\AgdaSpace{}%
\AgdaGeneralizable{r}\AgdaSymbol{))}\AgdaSpace{}%
\AgdaSymbol{→}\AgdaSpace{}%
\AgdaSymbol{(}\AgdaBound{j}\AgdaSpace{}%
\AgdaSymbol{:}\AgdaSpace{}%
\AgdaDatatype{E}\AgdaSpace{}%
\AgdaGeneralizable{Γ}\AgdaSpace{}%
\AgdaSymbol{(}\AgdaInductiveConstructor{ix}\AgdaSpace{}%
\AgdaGeneralizable{s}\AgdaSymbol{))}\AgdaSpace{}%
\AgdaSymbol{→}\AgdaSpace{}%
\AgdaGeneralizable{s}\AgdaSpace{}%
\AgdaOperator{\AgdaDatatype{+}}\AgdaSpace{}%
\AgdaGeneralizable{p}\AgdaSpace{}%
\AgdaOperator{\AgdaDatatype{≈}}\AgdaSpace{}%
\AgdaGeneralizable{r}\AgdaSpace{}%
\AgdaSymbol{→}\AgdaSpace{}%
\AgdaOperator{\AgdaDatatype{suc}}\AgdaSpace{}%
\AgdaGeneralizable{p}\AgdaSpace{}%
\AgdaOperator{\AgdaDatatype{≈}}\AgdaSpace{}%
\AgdaGeneralizable{u}\<%
\\
\>[13]\AgdaSymbol{→}\AgdaSpace{}%
\AgdaSymbol{(}\AgdaBound{e}\AgdaSpace{}%
\AgdaSymbol{:}\AgdaSpace{}%
\AgdaDatatype{E}\AgdaSpace{}%
\AgdaSymbol{(}\AgdaGeneralizable{Γ}\AgdaSpace{}%
\AgdaOperator{\AgdaInductiveConstructor{▹}}\AgdaSpace{}%
\AgdaInductiveConstructor{ix}\AgdaSpace{}%
\AgdaGeneralizable{u}\AgdaSymbol{)}\AgdaSpace{}%
\AgdaSymbol{(}\AgdaInductiveConstructor{ar}\AgdaSpace{}%
\AgdaGeneralizable{q}\AgdaSymbol{))}\AgdaSpace{}%
\AgdaSymbol{→}\AgdaSpace{}%
\AgdaDatatype{E}\AgdaSpace{}%
\AgdaGeneralizable{Γ}\AgdaSpace{}%
\AgdaSymbol{(}\AgdaInductiveConstructor{ar}\AgdaSpace{}%
\AgdaGeneralizable{q}\AgdaSymbol{)}\<%
\\
\>[2]\AgdaInductiveConstructor{ix-minusᵣ}%
\>[13]\AgdaSymbol{:}\AgdaSpace{}%
\AgdaSymbol{(}\AgdaBound{i}\AgdaSpace{}%
\AgdaSymbol{:}\AgdaSpace{}%
\AgdaDatatype{E}\AgdaSpace{}%
\AgdaGeneralizable{Γ}\AgdaSpace{}%
\AgdaSymbol{(}\AgdaInductiveConstructor{ix}\AgdaSpace{}%
\AgdaGeneralizable{r}\AgdaSymbol{))}\AgdaSpace{}%
\AgdaSymbol{→}\AgdaSpace{}%
\AgdaSymbol{(}\AgdaBound{j}\AgdaSpace{}%
\AgdaSymbol{:}\AgdaSpace{}%
\AgdaDatatype{E}\AgdaSpace{}%
\AgdaGeneralizable{Γ}\AgdaSpace{}%
\AgdaSymbol{(}\AgdaInductiveConstructor{ix}\AgdaSpace{}%
\AgdaGeneralizable{u}\AgdaSymbol{))}\AgdaSpace{}%
\AgdaSymbol{→}\AgdaSpace{}%
\AgdaGeneralizable{s}\AgdaSpace{}%
\AgdaOperator{\AgdaDatatype{+}}\AgdaSpace{}%
\AgdaGeneralizable{p}\AgdaSpace{}%
\AgdaOperator{\AgdaDatatype{≈}}\AgdaSpace{}%
\AgdaGeneralizable{r}\AgdaSpace{}%
\AgdaSymbol{→}\AgdaSpace{}%
\AgdaOperator{\AgdaDatatype{suc}}\AgdaSpace{}%
\AgdaGeneralizable{p}\AgdaSpace{}%
\AgdaOperator{\AgdaDatatype{≈}}\AgdaSpace{}%
\AgdaGeneralizable{u}\<%
\\
\>[13]\AgdaSymbol{→}\AgdaSpace{}%
\AgdaSymbol{(}\AgdaBound{e}\AgdaSpace{}%
\AgdaSymbol{:}\AgdaSpace{}%
\AgdaDatatype{E}\AgdaSpace{}%
\AgdaSymbol{(}\AgdaGeneralizable{Γ}\AgdaSpace{}%
\AgdaOperator{\AgdaInductiveConstructor{▹}}\AgdaSpace{}%
\AgdaInductiveConstructor{ix}\AgdaSpace{}%
\AgdaGeneralizable{s}\AgdaSymbol{)}\AgdaSpace{}%
\AgdaSymbol{(}\AgdaInductiveConstructor{ar}\AgdaSpace{}%
\AgdaGeneralizable{q}\AgdaSymbol{))}\AgdaSpace{}%
\AgdaSymbol{→}\AgdaSpace{}%
\AgdaDatatype{E}\AgdaSpace{}%
\AgdaGeneralizable{Γ}\AgdaSpace{}%
\AgdaSymbol{(}\AgdaInductiveConstructor{ar}\AgdaSpace{}%
\AgdaGeneralizable{q}\AgdaSymbol{)}\<%
\\
\>[2]\AgdaComment{--\ ...}\<%
\end{code}
\begin{code}[hide]%
\>[2]\AgdaInductiveConstructor{zero}\AgdaSpace{}%
\AgdaInductiveConstructor{one}\AgdaSpace{}%
\AgdaSymbol{:}\AgdaSpace{}%
\AgdaDatatype{E}\AgdaSpace{}%
\AgdaGeneralizable{Γ}\AgdaSpace{}%
\AgdaSymbol{(}\AgdaInductiveConstructor{ar}\AgdaSpace{}%
\AgdaGeneralizable{s}\AgdaSymbol{)}\<%
\\
\>[2]\AgdaInductiveConstructor{var}\AgdaSpace{}%
\AgdaSymbol{:}\AgdaSpace{}%
\AgdaGeneralizable{is}\AgdaSpace{}%
\AgdaOperator{\AgdaDatatype{∈}}\AgdaSpace{}%
\AgdaGeneralizable{Γ}\AgdaSpace{}%
\AgdaSymbol{→}\AgdaSpace{}%
\AgdaDatatype{E}\AgdaSpace{}%
\AgdaGeneralizable{Γ}\AgdaSpace{}%
\AgdaGeneralizable{is}\<%
\\
\\[\AgdaEmptyExtraSkip]%
\>[2]\AgdaInductiveConstructor{imapₛ}\AgdaSpace{}%
\AgdaSymbol{:}\AgdaSpace{}%
\AgdaDatatype{E}\AgdaSpace{}%
\AgdaSymbol{(}\AgdaGeneralizable{Γ}\AgdaSpace{}%
\AgdaOperator{\AgdaInductiveConstructor{▹}}\AgdaSpace{}%
\AgdaInductiveConstructor{ix}\AgdaSpace{}%
\AgdaGeneralizable{s}\AgdaSymbol{)}\AgdaSpace{}%
\AgdaSymbol{(}\AgdaInductiveConstructor{ar}\AgdaSpace{}%
\AgdaFunction{unit}\AgdaSymbol{)}\AgdaSpace{}%
\AgdaSymbol{→}\AgdaSpace{}%
\AgdaDatatype{E}\AgdaSpace{}%
\AgdaGeneralizable{Γ}\AgdaSpace{}%
\AgdaSymbol{(}\AgdaInductiveConstructor{ar}\AgdaSpace{}%
\AgdaGeneralizable{s}\AgdaSymbol{)}\<%
\\
\>[2]\AgdaInductiveConstructor{selₛ}\AgdaSpace{}%
\AgdaSymbol{:}\AgdaSpace{}%
\AgdaDatatype{E}\AgdaSpace{}%
\AgdaGeneralizable{Γ}\AgdaSpace{}%
\AgdaSymbol{(}\AgdaInductiveConstructor{ar}\AgdaSpace{}%
\AgdaGeneralizable{s}\AgdaSymbol{)}\AgdaSpace{}%
\AgdaSymbol{→}\AgdaSpace{}%
\AgdaDatatype{E}\AgdaSpace{}%
\AgdaGeneralizable{Γ}\AgdaSpace{}%
\AgdaSymbol{(}\AgdaInductiveConstructor{ix}\AgdaSpace{}%
\AgdaGeneralizable{s}\AgdaSymbol{)}\AgdaSpace{}%
\AgdaSymbol{→}\AgdaSpace{}%
\AgdaDatatype{E}\AgdaSpace{}%
\AgdaGeneralizable{Γ}\AgdaSpace{}%
\AgdaSymbol{(}\AgdaInductiveConstructor{ar}\AgdaSpace{}%
\AgdaFunction{unit}\AgdaSymbol{)}\<%
\\
\\[\AgdaEmptyExtraSkip]%
\>[2]\AgdaInductiveConstructor{imap}\AgdaSpace{}%
\AgdaSymbol{:}\AgdaSpace{}%
\AgdaDatatype{E}\AgdaSpace{}%
\AgdaSymbol{(}\AgdaGeneralizable{Γ}\AgdaSpace{}%
\AgdaOperator{\AgdaInductiveConstructor{▹}}\AgdaSpace{}%
\AgdaInductiveConstructor{ix}\AgdaSpace{}%
\AgdaGeneralizable{s}\AgdaSymbol{)}\AgdaSpace{}%
\AgdaSymbol{(}\AgdaInductiveConstructor{ar}\AgdaSpace{}%
\AgdaGeneralizable{p}\AgdaSymbol{)}\AgdaSpace{}%
\AgdaSymbol{→}\AgdaSpace{}%
\AgdaDatatype{E}\AgdaSpace{}%
\AgdaGeneralizable{Γ}\AgdaSpace{}%
\AgdaSymbol{(}\AgdaInductiveConstructor{ar}\AgdaSpace{}%
\AgdaSymbol{(}\AgdaGeneralizable{s}\AgdaSpace{}%
\AgdaOperator{\AgdaInductiveConstructor{⊗}}\AgdaSpace{}%
\AgdaGeneralizable{p}\AgdaSymbol{))}\<%
\\
\>[2]\AgdaInductiveConstructor{sel}\AgdaSpace{}%
\AgdaSymbol{:}\AgdaSpace{}%
\AgdaDatatype{E}\AgdaSpace{}%
\AgdaGeneralizable{Γ}\AgdaSpace{}%
\AgdaSymbol{(}\AgdaInductiveConstructor{ar}\AgdaSpace{}%
\AgdaSymbol{(}\AgdaGeneralizable{s}\AgdaSpace{}%
\AgdaOperator{\AgdaInductiveConstructor{⊗}}\AgdaSpace{}%
\AgdaGeneralizable{p}\AgdaSymbol{))}\AgdaSpace{}%
\AgdaSymbol{→}\AgdaSpace{}%
\AgdaDatatype{E}\AgdaSpace{}%
\AgdaGeneralizable{Γ}\AgdaSpace{}%
\AgdaSymbol{(}\AgdaInductiveConstructor{ix}\AgdaSpace{}%
\AgdaGeneralizable{s}\AgdaSymbol{)}\AgdaSpace{}%
\AgdaSymbol{→}\AgdaSpace{}%
\AgdaDatatype{E}\AgdaSpace{}%
\AgdaGeneralizable{Γ}\AgdaSpace{}%
\AgdaSymbol{(}\AgdaInductiveConstructor{ar}\AgdaSpace{}%
\AgdaGeneralizable{p}\AgdaSymbol{)}\<%
\\
\\[\AgdaEmptyExtraSkip]%
\>[2]\AgdaComment{--\ Blocked\ operations\ for\ avgpool\ }\<%
\\
\>[2]\AgdaInductiveConstructor{imapb}\AgdaSpace{}%
\AgdaSymbol{:}\AgdaSpace{}%
\AgdaGeneralizable{s}\AgdaSpace{}%
\AgdaOperator{\AgdaDatatype{*}}\AgdaSpace{}%
\AgdaGeneralizable{p}\AgdaSpace{}%
\AgdaOperator{\AgdaDatatype{≈}}\AgdaSpace{}%
\AgdaGeneralizable{q}\AgdaSpace{}%
\AgdaSymbol{→}\AgdaSpace{}%
\AgdaDatatype{E}\AgdaSpace{}%
\AgdaSymbol{(}\AgdaGeneralizable{Γ}\AgdaSpace{}%
\AgdaOperator{\AgdaInductiveConstructor{▹}}\AgdaSpace{}%
\AgdaInductiveConstructor{ix}\AgdaSpace{}%
\AgdaGeneralizable{s}\AgdaSymbol{)}\AgdaSpace{}%
\AgdaSymbol{(}\AgdaInductiveConstructor{ar}\AgdaSpace{}%
\AgdaGeneralizable{p}\AgdaSymbol{)}\AgdaSpace{}%
\AgdaSymbol{→}\AgdaSpace{}%
\AgdaDatatype{E}\AgdaSpace{}%
\AgdaGeneralizable{Γ}\AgdaSpace{}%
\AgdaSymbol{(}\AgdaInductiveConstructor{ar}\AgdaSpace{}%
\AgdaGeneralizable{q}\AgdaSymbol{)}\<%
\\
\>[2]\AgdaInductiveConstructor{selb}\AgdaSpace{}%
\AgdaSymbol{:}\AgdaSpace{}%
\AgdaGeneralizable{s}\AgdaSpace{}%
\AgdaOperator{\AgdaDatatype{*}}\AgdaSpace{}%
\AgdaGeneralizable{p}\AgdaSpace{}%
\AgdaOperator{\AgdaDatatype{≈}}\AgdaSpace{}%
\AgdaGeneralizable{q}\AgdaSpace{}%
\AgdaSymbol{→}\AgdaSpace{}%
\AgdaDatatype{E}\AgdaSpace{}%
\AgdaGeneralizable{Γ}\AgdaSpace{}%
\AgdaSymbol{(}\AgdaInductiveConstructor{ar}\AgdaSpace{}%
\AgdaGeneralizable{q}\AgdaSymbol{)}\AgdaSpace{}%
\AgdaSymbol{→}\AgdaSpace{}%
\AgdaDatatype{E}\AgdaSpace{}%
\AgdaGeneralizable{Γ}\AgdaSpace{}%
\AgdaSymbol{(}\AgdaInductiveConstructor{ix}\AgdaSpace{}%
\AgdaGeneralizable{s}\AgdaSymbol{)}\AgdaSpace{}%
\AgdaSymbol{→}\AgdaSpace{}%
\AgdaDatatype{E}\AgdaSpace{}%
\AgdaGeneralizable{Γ}\AgdaSpace{}%
\AgdaSymbol{(}\AgdaInductiveConstructor{ar}\AgdaSpace{}%
\AgdaGeneralizable{p}\AgdaSymbol{)}\<%
\\
\\[\AgdaEmptyExtraSkip]%
\>[2]\AgdaComment{--\ zero-but\ i\ j\ e\ =\ i\ ==\ j\ ?\ e\ :\ 0}\<%
\\
\>[2]\AgdaInductiveConstructor{zero-but}\AgdaSpace{}%
\AgdaSymbol{:}\AgdaSpace{}%
\AgdaDatatype{E}\AgdaSpace{}%
\AgdaGeneralizable{Γ}\AgdaSpace{}%
\AgdaSymbol{(}\AgdaInductiveConstructor{ix}\AgdaSpace{}%
\AgdaGeneralizable{s}\AgdaSymbol{)}\AgdaSpace{}%
\AgdaSymbol{→}\AgdaSpace{}%
\AgdaDatatype{E}\AgdaSpace{}%
\AgdaGeneralizable{Γ}\AgdaSpace{}%
\AgdaSymbol{(}\AgdaInductiveConstructor{ix}\AgdaSpace{}%
\AgdaGeneralizable{s}\AgdaSymbol{)}\AgdaSpace{}%
\AgdaSymbol{→}\AgdaSpace{}%
\AgdaDatatype{E}\AgdaSpace{}%
\AgdaGeneralizable{Γ}\AgdaSpace{}%
\AgdaSymbol{(}\AgdaInductiveConstructor{ar}\AgdaSpace{}%
\AgdaGeneralizable{p}\AgdaSymbol{)}\AgdaSpace{}%
\AgdaSymbol{→}\AgdaSpace{}%
\AgdaDatatype{E}\AgdaSpace{}%
\AgdaGeneralizable{Γ}\AgdaSpace{}%
\AgdaSymbol{(}\AgdaInductiveConstructor{ar}\AgdaSpace{}%
\AgdaGeneralizable{p}\AgdaSymbol{)}\<%
\\
\>[2]\AgdaInductiveConstructor{sum}\AgdaSpace{}%
\AgdaSymbol{:}\AgdaSpace{}%
\AgdaDatatype{E}\AgdaSpace{}%
\AgdaSymbol{(}\AgdaGeneralizable{Γ}\AgdaSpace{}%
\AgdaOperator{\AgdaInductiveConstructor{▹}}\AgdaSpace{}%
\AgdaInductiveConstructor{ix}\AgdaSpace{}%
\AgdaGeneralizable{s}\AgdaSymbol{)}\AgdaSpace{}%
\AgdaSymbol{(}\AgdaInductiveConstructor{ar}\AgdaSpace{}%
\AgdaGeneralizable{p}\AgdaSymbol{)}\AgdaSpace{}%
\AgdaSymbol{→}\AgdaSpace{}%
\AgdaDatatype{E}\AgdaSpace{}%
\AgdaGeneralizable{Γ}\AgdaSpace{}%
\AgdaSymbol{(}\AgdaInductiveConstructor{ar}\AgdaSpace{}%
\AgdaGeneralizable{p}\AgdaSymbol{)}\<%
\\
\>[2]\AgdaInductiveConstructor{bin}\AgdaSpace{}%
\AgdaSymbol{:}\AgdaSpace{}%
\AgdaDatatype{Bop}\AgdaSpace{}%
\AgdaSymbol{→}\AgdaSpace{}%
\AgdaDatatype{E}\AgdaSpace{}%
\AgdaGeneralizable{Γ}\AgdaSpace{}%
\AgdaSymbol{(}\AgdaInductiveConstructor{ar}\AgdaSpace{}%
\AgdaGeneralizable{s}\AgdaSymbol{)}\AgdaSpace{}%
\AgdaSymbol{→}\AgdaSpace{}%
\AgdaDatatype{E}\AgdaSpace{}%
\AgdaGeneralizable{Γ}\AgdaSpace{}%
\AgdaSymbol{(}\AgdaInductiveConstructor{ar}\AgdaSpace{}%
\AgdaGeneralizable{s}\AgdaSymbol{)}\AgdaSpace{}%
\AgdaSymbol{→}\AgdaSpace{}%
\AgdaDatatype{E}\AgdaSpace{}%
\AgdaGeneralizable{Γ}\AgdaSpace{}%
\AgdaSymbol{(}\AgdaInductiveConstructor{ar}\AgdaSpace{}%
\AgdaGeneralizable{s}\AgdaSymbol{)}\<%
\\
\\[\AgdaEmptyExtraSkip]%
\>[2]\AgdaInductiveConstructor{slide}%
\>[468I]\AgdaSymbol{:}\AgdaSpace{}%
\AgdaDatatype{E}\AgdaSpace{}%
\AgdaGeneralizable{Γ}\AgdaSpace{}%
\AgdaSymbol{(}\AgdaInductiveConstructor{ix}\AgdaSpace{}%
\AgdaGeneralizable{s}\AgdaSymbol{)}\AgdaSpace{}%
\AgdaSymbol{→}\AgdaSpace{}%
\AgdaGeneralizable{s}\AgdaSpace{}%
\AgdaOperator{\AgdaDatatype{+}}\AgdaSpace{}%
\AgdaGeneralizable{p}\AgdaSpace{}%
\AgdaOperator{\AgdaDatatype{≈}}\AgdaSpace{}%
\AgdaGeneralizable{r}\AgdaSpace{}%
\AgdaSymbol{→}\AgdaSpace{}%
\AgdaDatatype{E}\AgdaSpace{}%
\AgdaGeneralizable{Γ}\AgdaSpace{}%
\AgdaSymbol{(}\AgdaInductiveConstructor{ar}\AgdaSpace{}%
\AgdaGeneralizable{r}\AgdaSymbol{)}\<%
\\
\>[.][@{}l@{}]\<[468I]%
\>[8]\AgdaSymbol{→}\AgdaSpace{}%
\AgdaOperator{\AgdaDatatype{suc}}\AgdaSpace{}%
\AgdaGeneralizable{p}\AgdaSpace{}%
\AgdaOperator{\AgdaDatatype{≈}}\AgdaSpace{}%
\AgdaGeneralizable{u}\AgdaSpace{}%
\AgdaSymbol{→}\AgdaSpace{}%
\AgdaDatatype{E}\AgdaSpace{}%
\AgdaGeneralizable{Γ}\AgdaSpace{}%
\AgdaSymbol{(}\AgdaInductiveConstructor{ar}\AgdaSpace{}%
\AgdaGeneralizable{u}\AgdaSymbol{)}\<%
\\
\>[2]\AgdaInductiveConstructor{backslide}%
\>[493I]\AgdaSymbol{:}\AgdaSpace{}%
\AgdaDatatype{E}\AgdaSpace{}%
\AgdaGeneralizable{Γ}\AgdaSpace{}%
\AgdaSymbol{(}\AgdaInductiveConstructor{ix}\AgdaSpace{}%
\AgdaGeneralizable{s}\AgdaSymbol{)}\AgdaSpace{}%
\AgdaSymbol{→}\AgdaSpace{}%
\AgdaDatatype{E}\AgdaSpace{}%
\AgdaGeneralizable{Γ}\AgdaSpace{}%
\AgdaSymbol{(}\AgdaInductiveConstructor{ar}\AgdaSpace{}%
\AgdaGeneralizable{u}\AgdaSymbol{)}\AgdaSpace{}%
\AgdaSymbol{→}\AgdaSpace{}%
\AgdaOperator{\AgdaDatatype{suc}}\AgdaSpace{}%
\AgdaGeneralizable{p}\AgdaSpace{}%
\AgdaOperator{\AgdaDatatype{≈}}\AgdaSpace{}%
\AgdaGeneralizable{u}\<%
\\
\>[.][@{}l@{}]\<[493I]%
\>[12]\AgdaSymbol{→}\AgdaSpace{}%
\AgdaGeneralizable{s}\AgdaSpace{}%
\AgdaOperator{\AgdaDatatype{+}}\AgdaSpace{}%
\AgdaGeneralizable{p}\AgdaSpace{}%
\AgdaOperator{\AgdaDatatype{≈}}\AgdaSpace{}%
\AgdaGeneralizable{r}\AgdaSpace{}%
\AgdaSymbol{→}\AgdaSpace{}%
\AgdaDatatype{E}\AgdaSpace{}%
\AgdaGeneralizable{Γ}\AgdaSpace{}%
\AgdaSymbol{(}\AgdaInductiveConstructor{ar}\AgdaSpace{}%
\AgdaGeneralizable{r}\AgdaSymbol{)}\<%
\\
\>[0]\<%
\\
\>[2]\AgdaInductiveConstructor{scaledown}\AgdaSpace{}%
\AgdaSymbol{:}\AgdaSpace{}%
\AgdaDatatype{ℕ}\AgdaSpace{}%
\AgdaSymbol{→}\AgdaSpace{}%
\AgdaDatatype{E}\AgdaSpace{}%
\AgdaGeneralizable{Γ}\AgdaSpace{}%
\AgdaSymbol{(}\AgdaInductiveConstructor{ar}\AgdaSpace{}%
\AgdaGeneralizable{s}\AgdaSymbol{)}\AgdaSpace{}%
\AgdaSymbol{→}\AgdaSpace{}%
\AgdaDatatype{E}\AgdaSpace{}%
\AgdaGeneralizable{Γ}\AgdaSpace{}%
\AgdaSymbol{(}\AgdaInductiveConstructor{ar}\AgdaSpace{}%
\AgdaGeneralizable{s}\AgdaSymbol{)}\<%
\\
\>[2]\AgdaInductiveConstructor{minus}\AgdaSpace{}%
\AgdaSymbol{:}\AgdaSpace{}%
\AgdaDatatype{E}\AgdaSpace{}%
\AgdaGeneralizable{Γ}\AgdaSpace{}%
\AgdaSymbol{(}\AgdaInductiveConstructor{ar}\AgdaSpace{}%
\AgdaGeneralizable{s}\AgdaSymbol{)}\AgdaSpace{}%
\AgdaSymbol{→}\AgdaSpace{}%
\AgdaDatatype{E}\AgdaSpace{}%
\AgdaGeneralizable{Γ}\AgdaSpace{}%
\AgdaSymbol{(}\AgdaInductiveConstructor{ar}\AgdaSpace{}%
\AgdaGeneralizable{s}\AgdaSymbol{)}\<%
\\
\\[\AgdaEmptyExtraSkip]%
\>[2]\AgdaInductiveConstructor{logistic}\AgdaSpace{}%
\AgdaSymbol{:}\AgdaSpace{}%
\AgdaDatatype{E}\AgdaSpace{}%
\AgdaGeneralizable{Γ}\AgdaSpace{}%
\AgdaSymbol{(}\AgdaInductiveConstructor{ar}\AgdaSpace{}%
\AgdaGeneralizable{s}\AgdaSymbol{)}\AgdaSpace{}%
\AgdaSymbol{→}\AgdaSpace{}%
\AgdaDatatype{E}\AgdaSpace{}%
\AgdaGeneralizable{Γ}\AgdaSpace{}%
\AgdaSymbol{(}\AgdaInductiveConstructor{ar}\AgdaSpace{}%
\AgdaGeneralizable{s}\AgdaSymbol{)}\<%
\end{code}
The \AC{div} and \AC{mod} constructors perform point-wise division or modulo
operation on the index $i$ and the shape $p$.  This is needed to express selections
into blocked arrays as we have seen in Section~\ref{sec:sac-primitives}.
The \AC{ix-plus} is a point-wise addition of $i$ and $j$.  The \AC{ix-minus} and
\AC{ix-minusᵣ} correspond to left and right subtraction from the Section~\ref{sec:cnn}.
The meaning of these constructors is follows: if $j$ can be subtracted from $i$
(in the sense of existence of inverse to \AF{⊕ₚ} exists) then we evaluate $e$ at that index,
otherwise we return zero.

\subsubsection{Optimisations}
We add the following optimisations to facilitate removal of temporary arrays in
the generated code.  We show the only ones that we added, all the optimisations
we defined before are still valid.
\begin{code}[hide]%
\>[0]\AgdaOperator{\AgdaFunction{\AgdaUnderscore{}/\AgdaUnderscore{}}}\AgdaSpace{}%
\AgdaSymbol{:}\AgdaSpace{}%
\AgdaSymbol{(}\AgdaBound{Γ}\AgdaSpace{}%
\AgdaSymbol{:}\AgdaSpace{}%
\AgdaDatatype{Ctx}\AgdaSymbol{)}\AgdaSpace{}%
\AgdaSymbol{→}\AgdaSpace{}%
\AgdaGeneralizable{is}\AgdaSpace{}%
\AgdaOperator{\AgdaDatatype{∈}}\AgdaSpace{}%
\AgdaBound{Γ}\AgdaSpace{}%
\AgdaSymbol{→}\AgdaSpace{}%
\AgdaDatatype{Ctx}\<%
\\
\>[0]\AgdaSymbol{(}\AgdaBound{Γ}\AgdaSpace{}%
\AgdaOperator{\AgdaInductiveConstructor{▹}}\AgdaSpace{}%
\AgdaBound{x}\AgdaSymbol{)}\AgdaSpace{}%
\AgdaOperator{\AgdaFunction{/}}\AgdaSpace{}%
\AgdaInductiveConstructor{here}\AgdaSpace{}%
\AgdaSymbol{=}\AgdaSpace{}%
\AgdaBound{Γ}\<%
\\
\>[0]\AgdaSymbol{(}\AgdaBound{Γ}\AgdaSpace{}%
\AgdaOperator{\AgdaInductiveConstructor{▹}}\AgdaSpace{}%
\AgdaBound{x}\AgdaSymbol{)}\AgdaSpace{}%
\AgdaOperator{\AgdaFunction{/}}\AgdaSpace{}%
\AgdaInductiveConstructor{there}\AgdaSpace{}%
\AgdaBound{v}\AgdaSpace{}%
\AgdaSymbol{=}\AgdaSpace{}%
\AgdaSymbol{(}\AgdaBound{Γ}\AgdaSpace{}%
\AgdaOperator{\AgdaFunction{/}}\AgdaSpace{}%
\AgdaBound{v}\AgdaSymbol{)}\AgdaSpace{}%
\AgdaOperator{\AgdaInductiveConstructor{▹}}\AgdaSpace{}%
\AgdaBound{x}\<%
\\
\\[\AgdaEmptyExtraSkip]%
\>[0]\AgdaComment{--\ See\ the\ actual\ definition\ in\ the\ ./code\ directory\ in\ the}\<%
\\
\>[0]\AgdaComment{--\ root\ of\ the\ repo,\ here\ we\ just\ make\ a\ stub\ to\ explain\ the}\<%
\\
\>[0]\AgdaComment{--\ code\ below.}\<%
\\
\>[0]\AgdaKeyword{postulate}\<%
\\
\>[0][@{}l@{\AgdaIndent{0}}]%
\>[2]\AgdaPostulate{wkv}\AgdaSpace{}%
\AgdaSymbol{:}\AgdaSpace{}%
\AgdaSymbol{(}\AgdaBound{v}\AgdaSpace{}%
\AgdaSymbol{:}\AgdaSpace{}%
\AgdaGeneralizable{is}\AgdaSpace{}%
\AgdaOperator{\AgdaDatatype{∈}}\AgdaSpace{}%
\AgdaGeneralizable{Γ}\AgdaSymbol{)}\AgdaSpace{}%
\AgdaSymbol{→}\AgdaSpace{}%
\AgdaGeneralizable{ip}\AgdaSpace{}%
\AgdaOperator{\AgdaDatatype{∈}}\AgdaSpace{}%
\AgdaSymbol{(}\AgdaGeneralizable{Γ}\AgdaSpace{}%
\AgdaOperator{\AgdaFunction{/}}\AgdaSpace{}%
\AgdaBound{v}\AgdaSymbol{)}\AgdaSpace{}%
\AgdaSymbol{→}\AgdaSpace{}%
\AgdaGeneralizable{ip}\AgdaSpace{}%
\AgdaOperator{\AgdaDatatype{∈}}\AgdaSpace{}%
\AgdaGeneralizable{Γ}\<%
\\
\>[2]\AgdaPostulate{wk}\AgdaSpace{}%
\AgdaSymbol{:}\AgdaSpace{}%
\AgdaSymbol{(}\AgdaBound{v}\AgdaSpace{}%
\AgdaSymbol{:}\AgdaSpace{}%
\AgdaGeneralizable{is}\AgdaSpace{}%
\AgdaOperator{\AgdaDatatype{∈}}\AgdaSpace{}%
\AgdaGeneralizable{Γ}\AgdaSymbol{)}\AgdaSpace{}%
\AgdaSymbol{→}\AgdaSpace{}%
\AgdaDatatype{E}\AgdaSpace{}%
\AgdaSymbol{(}\AgdaGeneralizable{Γ}\AgdaSpace{}%
\AgdaOperator{\AgdaFunction{/}}\AgdaSpace{}%
\AgdaBound{v}\AgdaSymbol{)}\AgdaSpace{}%
\AgdaGeneralizable{ip}\AgdaSpace{}%
\AgdaSymbol{→}\AgdaSpace{}%
\AgdaDatatype{E}\AgdaSpace{}%
\AgdaGeneralizable{Γ}\AgdaSpace{}%
\AgdaGeneralizable{ip}\<%
\\
\\[\AgdaEmptyExtraSkip]%
\>[0]\AgdaComment{--\ Nicer\ syntax\ for\ common\ case:}\<%
\\
\>[0]\AgdaKeyword{infixr}\AgdaSpace{}%
\AgdaNumber{18}\AgdaSpace{}%
\AgdaOperator{\AgdaFunction{↑\AgdaUnderscore{}}}\<%
\\
\>[0]\AgdaOperator{\AgdaFunction{↑\AgdaUnderscore{}}}\AgdaSpace{}%
\AgdaSymbol{:}\AgdaSpace{}%
\AgdaDatatype{E}\AgdaSpace{}%
\AgdaGeneralizable{Γ}\AgdaSpace{}%
\AgdaGeneralizable{is}\AgdaSpace{}%
\AgdaSymbol{→}\AgdaSpace{}%
\AgdaDatatype{E}\AgdaSpace{}%
\AgdaSymbol{(}\AgdaGeneralizable{Γ}\AgdaSpace{}%
\AgdaOperator{\AgdaInductiveConstructor{▹}}\AgdaSpace{}%
\AgdaGeneralizable{ip}\AgdaSymbol{)}\AgdaSpace{}%
\AgdaGeneralizable{is}\<%
\\
\>[0]\AgdaOperator{\AgdaFunction{↑\AgdaUnderscore{}}}\AgdaSpace{}%
\AgdaSymbol{=}\AgdaSpace{}%
\AgdaPostulate{wk}\AgdaSpace{}%
\AgdaInductiveConstructor{here}\<%
\\
\\[\AgdaEmptyExtraSkip]%
\>[0]\AgdaKeyword{infixr}\AgdaSpace{}%
\AgdaNumber{18}\AgdaSpace{}%
\AgdaOperator{\AgdaFunction{↑↑\AgdaUnderscore{}}}\<%
\\
\>[0]\AgdaOperator{\AgdaFunction{↑↑\AgdaUnderscore{}}}\AgdaSpace{}%
\AgdaSymbol{:}\AgdaSpace{}%
\AgdaDatatype{E}\AgdaSpace{}%
\AgdaGeneralizable{Γ}\AgdaSpace{}%
\AgdaGeneralizable{is}\AgdaSpace{}%
\AgdaSymbol{→}\AgdaSpace{}%
\AgdaDatatype{E}\AgdaSpace{}%
\AgdaSymbol{(}\AgdaGeneralizable{Γ}\AgdaSpace{}%
\AgdaOperator{\AgdaInductiveConstructor{▹}}\AgdaSpace{}%
\AgdaGeneralizable{ip}\AgdaSpace{}%
\AgdaOperator{\AgdaInductiveConstructor{▹}}\AgdaSpace{}%
\AgdaGeneralizable{iq}\AgdaSymbol{)}\AgdaSpace{}%
\AgdaGeneralizable{is}\<%
\\
\>[0]\AgdaOperator{\AgdaFunction{↑↑\AgdaUnderscore{}}}\AgdaSpace{}%
\AgdaSymbol{=}\AgdaSpace{}%
\AgdaOperator{\AgdaFunction{↑\AgdaUnderscore{}}}\AgdaSpace{}%
\AgdaOperator{\AgdaFunction{∘}}\AgdaSpace{}%
\AgdaOperator{\AgdaFunction{↑\AgdaUnderscore{}}}\<%
\\
\\[\AgdaEmptyExtraSkip]%
\>[0]\AgdaKeyword{data}\AgdaSpace{}%
\AgdaDatatype{Eq}\AgdaSpace{}%
\AgdaSymbol{:}\AgdaSpace{}%
\AgdaGeneralizable{is}\AgdaSpace{}%
\AgdaOperator{\AgdaDatatype{∈}}\AgdaSpace{}%
\AgdaGeneralizable{Γ}\AgdaSpace{}%
\AgdaSymbol{→}\AgdaSpace{}%
\AgdaGeneralizable{ip}\AgdaSpace{}%
\AgdaOperator{\AgdaDatatype{∈}}\AgdaSpace{}%
\AgdaGeneralizable{Γ}\AgdaSpace{}%
\AgdaSymbol{→}\AgdaSpace{}%
\AgdaPrimitive{Set}\AgdaSpace{}%
\AgdaKeyword{where}\<%
\\
\>[0][@{}l@{\AgdaIndent{0}}]%
\>[2]\AgdaInductiveConstructor{eq}\AgdaSpace{}%
\AgdaSymbol{:}\AgdaSpace{}%
\AgdaSymbol{\{}\AgdaBound{x}\AgdaSpace{}%
\AgdaSymbol{:}\AgdaSpace{}%
\AgdaGeneralizable{is}\AgdaSpace{}%
\AgdaOperator{\AgdaDatatype{∈}}\AgdaSpace{}%
\AgdaGeneralizable{Γ}\AgdaSymbol{\}}\AgdaSpace{}%
\AgdaSymbol{→}\AgdaSpace{}%
\AgdaDatatype{Eq}\AgdaSpace{}%
\AgdaBound{x}\AgdaSpace{}%
\AgdaBound{x}\<%
\\
\>[2]\AgdaInductiveConstructor{neq}\AgdaSpace{}%
\AgdaSymbol{:}\AgdaSpace{}%
\AgdaSymbol{(}\AgdaBound{x}\AgdaSpace{}%
\AgdaSymbol{:}\AgdaSpace{}%
\AgdaGeneralizable{is}\AgdaSpace{}%
\AgdaOperator{\AgdaDatatype{∈}}\AgdaSpace{}%
\AgdaGeneralizable{Γ}\AgdaSymbol{)}\AgdaSpace{}%
\AgdaSymbol{→}\AgdaSpace{}%
\AgdaSymbol{(}\AgdaBound{y}\AgdaSpace{}%
\AgdaSymbol{:}\AgdaSpace{}%
\AgdaGeneralizable{ip}\AgdaSpace{}%
\AgdaOperator{\AgdaDatatype{∈}}\AgdaSpace{}%
\AgdaSymbol{(}\AgdaGeneralizable{Γ}\AgdaSpace{}%
\AgdaOperator{\AgdaFunction{/}}\AgdaSpace{}%
\AgdaBound{x}\AgdaSymbol{))}\AgdaSpace{}%
\AgdaSymbol{→}\AgdaSpace{}%
\AgdaDatatype{Eq}\AgdaSpace{}%
\AgdaBound{x}\AgdaSpace{}%
\AgdaSymbol{(}\AgdaPostulate{wkv}\AgdaSpace{}%
\AgdaBound{x}\AgdaSpace{}%
\AgdaBound{y}\AgdaSymbol{)}\<%
\\
\\[\AgdaEmptyExtraSkip]%
\>[0]\AgdaKeyword{postulate}\<%
\\
\>[0][@{}l@{\AgdaIndent{0}}]%
\>[2]\AgdaPostulate{eq?}\AgdaSpace{}%
\AgdaSymbol{:}\AgdaSpace{}%
\AgdaSymbol{(}\AgdaBound{x}\AgdaSpace{}%
\AgdaSymbol{:}\AgdaSpace{}%
\AgdaGeneralizable{is}\AgdaSpace{}%
\AgdaOperator{\AgdaDatatype{∈}}\AgdaSpace{}%
\AgdaGeneralizable{Γ}\AgdaSymbol{)}\AgdaSpace{}%
\AgdaSymbol{→}\AgdaSpace{}%
\AgdaSymbol{(}\AgdaBound{y}\AgdaSpace{}%
\AgdaSymbol{:}\AgdaSpace{}%
\AgdaGeneralizable{ip}\AgdaSpace{}%
\AgdaOperator{\AgdaDatatype{∈}}\AgdaSpace{}%
\AgdaGeneralizable{Γ}\AgdaSymbol{)}\AgdaSpace{}%
\AgdaSymbol{→}\AgdaSpace{}%
\AgdaDatatype{Eq}\AgdaSpace{}%
\AgdaBound{x}\AgdaSpace{}%
\AgdaBound{y}\<%
\\
\>[2]\AgdaPostulate{sub}\AgdaSpace{}%
\AgdaSymbol{:}\AgdaSpace{}%
\AgdaSymbol{(}\AgdaBound{v}\AgdaSpace{}%
\AgdaSymbol{:}\AgdaSpace{}%
\AgdaGeneralizable{is}\AgdaSpace{}%
\AgdaOperator{\AgdaDatatype{∈}}\AgdaSpace{}%
\AgdaGeneralizable{Γ}\AgdaSymbol{)}\AgdaSpace{}%
\AgdaSymbol{→}\AgdaSpace{}%
\AgdaDatatype{E}\AgdaSpace{}%
\AgdaGeneralizable{Γ}\AgdaSpace{}%
\AgdaGeneralizable{ip}\AgdaSpace{}%
\AgdaSymbol{→}\AgdaSpace{}%
\AgdaDatatype{E}\AgdaSpace{}%
\AgdaSymbol{(}\AgdaGeneralizable{Γ}\AgdaSpace{}%
\AgdaOperator{\AgdaFunction{/}}\AgdaSpace{}%
\AgdaBound{v}\AgdaSymbol{)}\AgdaSpace{}%
\AgdaGeneralizable{is}\AgdaSpace{}%
\AgdaSymbol{→}\AgdaSpace{}%
\AgdaDatatype{E}\AgdaSpace{}%
\AgdaSymbol{(}\AgdaGeneralizable{Γ}\AgdaSpace{}%
\AgdaOperator{\AgdaFunction{/}}\AgdaSpace{}%
\AgdaBound{v}\AgdaSymbol{)}\AgdaSpace{}%
\AgdaGeneralizable{ip}\<%
\\
\>[0]\<%
\end{code}
\begin{code}%
\>[0]\AgdaFunction{opt}\AgdaSpace{}%
\AgdaSymbol{:}\AgdaSpace{}%
\AgdaDatatype{E}\AgdaSpace{}%
\AgdaGeneralizable{Γ}\AgdaSpace{}%
\AgdaGeneralizable{is}\AgdaSpace{}%
\AgdaSymbol{→}\AgdaSpace{}%
\AgdaDatatype{E}\AgdaSpace{}%
\AgdaGeneralizable{Γ}\AgdaSpace{}%
\AgdaGeneralizable{is}\<%
\\
\>[0]\AgdaFunction{opt}\AgdaSpace{}%
\AgdaSymbol{(}\AgdaInductiveConstructor{selₛ}\AgdaSpace{}%
\AgdaBound{e}\AgdaSpace{}%
\AgdaBound{e₁}\AgdaSymbol{)}\AgdaSpace{}%
\AgdaKeyword{with}\AgdaSpace{}%
\AgdaFunction{opt}\AgdaSpace{}%
\AgdaBound{e}\AgdaSpace{}%
\AgdaSymbol{|}\AgdaSpace{}%
\AgdaFunction{opt}\AgdaSpace{}%
\AgdaBound{e₁}\<%
\\
\>[0]\AgdaSymbol{...}\AgdaSpace{}%
\AgdaSymbol{|}\AgdaSpace{}%
\AgdaInductiveConstructor{imapb}\AgdaSpace{}%
\AgdaBound{m}\AgdaSpace{}%
\AgdaBound{e}%
\>[24]\AgdaSymbol{|}\AgdaSpace{}%
\AgdaBound{i}\AgdaSpace{}%
\AgdaSymbol{=}\AgdaSpace{}%
\AgdaInductiveConstructor{selₛ}\AgdaSpace{}%
\AgdaSymbol{(}\AgdaPostulate{sub}\AgdaSpace{}%
\AgdaInductiveConstructor{v₀}\AgdaSpace{}%
\AgdaBound{e}\AgdaSpace{}%
\AgdaSymbol{(}\AgdaInductiveConstructor{div}\AgdaSpace{}%
\AgdaBound{m}\AgdaSpace{}%
\AgdaBound{i}\AgdaSymbol{))}\AgdaSpace{}%
\AgdaSymbol{(}\AgdaInductiveConstructor{mod}\AgdaSpace{}%
\AgdaBound{m}\AgdaSpace{}%
\AgdaBound{i}\AgdaSymbol{)}\<%
\\
\>[0]\AgdaSymbol{...}\AgdaSpace{}%
\AgdaSymbol{|}\AgdaSpace{}%
\AgdaInductiveConstructor{slide}\AgdaSpace{}%
\AgdaBound{i}\AgdaSpace{}%
\AgdaBound{pl}\AgdaSpace{}%
\AgdaBound{a}\AgdaSpace{}%
\AgdaBound{su}%
\>[24]\AgdaSymbol{|}\AgdaSpace{}%
\AgdaBound{k}\AgdaSpace{}%
\AgdaSymbol{=}\AgdaSpace{}%
\AgdaInductiveConstructor{selₛ}\AgdaSpace{}%
\AgdaBound{a}\AgdaSpace{}%
\AgdaSymbol{(}\AgdaInductiveConstructor{ix-plus}\AgdaSpace{}%
\AgdaBound{i}\AgdaSpace{}%
\AgdaBound{k}\AgdaSpace{}%
\AgdaBound{su}\AgdaSpace{}%
\AgdaBound{pl}\AgdaSymbol{)}\<%
\\
\>[0]\AgdaComment{---\ |\ ...\ as\ before\ ...}\<%
\end{code}
Here we optimise away scalar selections into blocked imaps.  Recall that $m$ tells us
that we have an array of shape $s * p$, and $e$ computes blocks of shape $p$.  If we
are selecting into such a blocked array at the index $i$, we know that we are selecting
$(i / p)$-th block, and from that block we are selecting $(i \% p)$ element.  Existence
of explicit \AC{div} and \AC{mod} operations on indices makes it possible to implement
this rewrite rule that is again very similar to $\beta$-reduction.
\begin{code}[hide]%
\>[0]\AgdaCatchallClause{\AgdaSymbol{...}}\AgdaSpace{}%
\AgdaCatchallClause{\AgdaSymbol{|}}\AgdaSpace{}%
\AgdaCatchallClause{\AgdaBound{a}}%
\>[24]\AgdaCatchallClause{\AgdaSymbol{|}}\AgdaSpace{}%
\AgdaCatchallClause{\AgdaBound{i}}\AgdaSpace{}%
\AgdaSymbol{=}\AgdaSpace{}%
\AgdaInductiveConstructor{selₛ}\AgdaSpace{}%
\AgdaBound{a}\AgdaSpace{}%
\AgdaBound{i}\<%
\end{code}
\begin{code}%
\>[0]\AgdaFunction{opt}\AgdaSpace{}%
\AgdaSymbol{(}\AgdaInductiveConstructor{sum}\AgdaSpace{}%
\AgdaBound{e}\AgdaSymbol{)}\AgdaSpace{}%
\AgdaKeyword{with}\AgdaSpace{}%
\AgdaFunction{opt}\AgdaSpace{}%
\AgdaBound{e}\<%
\\
\>[0]\AgdaSymbol{...}\AgdaSpace{}%
\AgdaSymbol{|}\AgdaSpace{}%
\AgdaInductiveConstructor{zero-but}\AgdaSpace{}%
\AgdaSymbol{(}\AgdaInductiveConstructor{var}\AgdaSpace{}%
\AgdaBound{i}\AgdaSymbol{)}\AgdaSpace{}%
\AgdaSymbol{(}\AgdaInductiveConstructor{ix-plus}\AgdaSpace{}%
\AgdaSymbol{(}\AgdaInductiveConstructor{var}\AgdaSpace{}%
\AgdaBound{j}\AgdaSymbol{)}\AgdaSpace{}%
\AgdaSymbol{(}\AgdaInductiveConstructor{var}\AgdaSpace{}%
\AgdaBound{k}\AgdaSymbol{)}\AgdaSpace{}%
\AgdaBound{su}\AgdaSpace{}%
\AgdaBound{pl}\AgdaSymbol{)}\AgdaSpace{}%
\AgdaBound{a}%
\>[58]\AgdaKeyword{with}\AgdaSpace{}%
\AgdaPostulate{eq?}\AgdaSpace{}%
\AgdaInductiveConstructor{v₀}\AgdaSpace{}%
\AgdaBound{i}\AgdaSpace{}%
\AgdaSymbol{|}\AgdaSpace{}%
\AgdaPostulate{eq?}\AgdaSpace{}%
\AgdaInductiveConstructor{v₀}\AgdaSpace{}%
\AgdaBound{j}\AgdaSpace{}%
\AgdaSymbol{|}\AgdaSpace{}%
\AgdaPostulate{eq?}\AgdaSpace{}%
\AgdaInductiveConstructor{v₀}\AgdaSpace{}%
\AgdaBound{k}\<%
\\
\>[0]\AgdaSymbol{...}\AgdaSpace{}%
\AgdaSymbol{|}\AgdaSpace{}%
\AgdaInductiveConstructor{neq}\AgdaSpace{}%
\AgdaSymbol{\AgdaUnderscore{}}\AgdaSpace{}%
\AgdaBound{i′}%
\>[16]\AgdaSymbol{|}\AgdaSpace{}%
\AgdaInductiveConstructor{neq}\AgdaSpace{}%
\AgdaSymbol{\AgdaUnderscore{}}\AgdaSpace{}%
\AgdaBound{j′}%
\>[28]\AgdaSymbol{|}\AgdaSpace{}%
\AgdaInductiveConstructor{eq}%
\>[40]\AgdaSymbol{=}\AgdaSpace{}%
\AgdaInductiveConstructor{ix-minus}%
\>[53]\AgdaSymbol{(}\AgdaInductiveConstructor{var}\AgdaSpace{}%
\AgdaBound{i′}\AgdaSymbol{)}\AgdaSpace{}%
\AgdaSymbol{(}\AgdaInductiveConstructor{var}\AgdaSpace{}%
\AgdaBound{j′}\AgdaSymbol{)}\AgdaSpace{}%
\AgdaBound{pl}\AgdaSpace{}%
\AgdaBound{su}\AgdaSpace{}%
\AgdaBound{a}\<%
\\
\>[0]\AgdaSymbol{...}\AgdaSpace{}%
\AgdaSymbol{|}\AgdaSpace{}%
\AgdaInductiveConstructor{neq}\AgdaSpace{}%
\AgdaSymbol{\AgdaUnderscore{}}\AgdaSpace{}%
\AgdaBound{i′}%
\>[16]\AgdaSymbol{|}\AgdaSpace{}%
\AgdaInductiveConstructor{eq}%
\>[28]\AgdaSymbol{|}\AgdaSpace{}%
\AgdaInductiveConstructor{neq}\AgdaSpace{}%
\AgdaSymbol{\AgdaUnderscore{}}\AgdaSpace{}%
\AgdaBound{k′}%
\>[40]\AgdaSymbol{=}\AgdaSpace{}%
\AgdaInductiveConstructor{ix-minusᵣ}%
\>[53]\AgdaSymbol{(}\AgdaInductiveConstructor{var}\AgdaSpace{}%
\AgdaBound{i′}\AgdaSymbol{)}\AgdaSpace{}%
\AgdaSymbol{(}\AgdaInductiveConstructor{var}\AgdaSpace{}%
\AgdaBound{k′}\AgdaSymbol{)}\AgdaSpace{}%
\AgdaBound{pl}\AgdaSpace{}%
\AgdaBound{su}\AgdaSpace{}%
\AgdaBound{a}\<%
\\
\>[0]\AgdaCatchallClause{\AgdaSymbol{...}}\AgdaSpace{}%
\AgdaCatchallClause{\AgdaSymbol{|}}\AgdaSpace{}%
\AgdaCatchallClause{\AgdaSymbol{\AgdaUnderscore{}}}%
\>[16]\AgdaCatchallClause{\AgdaSymbol{|}}\AgdaSpace{}%
\AgdaCatchallClause{\AgdaSymbol{\AgdaUnderscore{}}}%
\>[28]\AgdaCatchallClause{\AgdaSymbol{|}}\AgdaSpace{}%
\AgdaCatchallClause{\AgdaSymbol{\AgdaUnderscore{}}}%
\>[40]\AgdaSymbol{=}\AgdaSpace{}%
\AgdaInductiveConstructor{sum}\AgdaSpace{}%
\AgdaSymbol{(}\AgdaInductiveConstructor{zero-but}\AgdaSpace{}%
\AgdaSymbol{(}\AgdaInductiveConstructor{var}\AgdaSpace{}%
\AgdaBound{i}\AgdaSymbol{)}\AgdaSpace{}%
\AgdaSymbol{(}\AgdaInductiveConstructor{ix-plus}\AgdaSpace{}%
\AgdaSymbol{(}\AgdaInductiveConstructor{var}\AgdaSpace{}%
\AgdaBound{j}\AgdaSymbol{)}\AgdaSpace{}%
\AgdaSymbol{(}\AgdaInductiveConstructor{var}\AgdaSpace{}%
\AgdaBound{k}\AgdaSymbol{)}\AgdaSpace{}%
\AgdaBound{su}\AgdaSpace{}%
\AgdaBound{pl}\AgdaSymbol{)}\AgdaSpace{}%
\AgdaBound{a}\AgdaSymbol{)}\<%
\\
\>[0]\AgdaComment{---\ |\ ...\ as\ before\ ...}\<%
\end{code}
\begin{code}[hide]%
\>[0]\AgdaCatchallClause{\AgdaFunction{opt}}\AgdaSpace{}%
\AgdaCatchallClause{\AgdaSymbol{(}}\AgdaCatchallClause{\AgdaInductiveConstructor{sum}}\AgdaSpace{}%
\AgdaCatchallClause{\AgdaBound{e}}\AgdaCatchallClause{\AgdaSymbol{)}}\AgdaSpace{}%
\AgdaCatchallClause{\AgdaSymbol{|}}\AgdaSpace{}%
\AgdaCatchallClause{\AgdaBound{a}}\AgdaSpace{}%
\AgdaSymbol{=}\AgdaSpace{}%
\AgdaInductiveConstructor{sum}\AgdaSpace{}%
\AgdaBound{a}\<%
\end{code}
Here we are dealing with the sum over summation index $t$ where the inner expression is
a conditional on indices of the form \texttt{i == j + k ? e : 0}.  Here we apply the
same comparison of index variables as before.  If $k$ happens to be the variable $t$,
then overall sum will only add one non-zero element at $(i-j)$-th index, given that this
(left) subtraction is possible in the sense of existence of the inverse to \AF{\_⊕ₚ\_} operation
defined in Section~\ref{sec:general-ix-ops}.  The same happens when the summation index
$t$ is equal to $j$, we only need to consider $(i-k)$-th element given that this (right)
subtraction is possible.  One could cover other cases where $t$ is equal to $i$, or
when $i$ and $j+k$ are swapped, but these are not occurring in our running example.

\begin{code}%
\>[0]\AgdaFunction{opt}\AgdaSpace{}%
\AgdaSymbol{(}\AgdaInductiveConstructor{scaledown}\AgdaSpace{}%
\AgdaBound{x}\AgdaSpace{}%
\AgdaBound{e}\AgdaSymbol{)}\AgdaSpace{}%
\AgdaKeyword{with}\AgdaSpace{}%
\AgdaFunction{opt}\AgdaSpace{}%
\AgdaBound{e}\<%
\\
\>[0]\AgdaSymbol{...}\AgdaSpace{}%
\AgdaSymbol{|}\AgdaSpace{}%
\AgdaInductiveConstructor{sum}\AgdaSpace{}%
\AgdaBound{a}\AgdaSpace{}%
\AgdaSymbol{=}\AgdaSpace{}%
\AgdaInductiveConstructor{sum}\AgdaSpace{}%
\AgdaSymbol{(}\AgdaInductiveConstructor{scaledown}\AgdaSpace{}%
\AgdaBound{x}\AgdaSpace{}%
\AgdaBound{a}\AgdaSymbol{)}\<%
\\
\>[0]\AgdaComment{---\ |\ ...\ as\ before\ ...}\<%
\end{code}
Finally, here is a rule that looks very innocent in the high-level language, yet
becomes of importance in the low-level one.  The rule says that if we are summing
the array and then dividing it by a constant, we should move division inside the
summation.  The reason for this rewrite rule being important is when the result
of the sum is non-scalar, we need to create a temporary array, before scaling down
all its elements.  A language with first class arrays can obviously take care of
such minor details, but in C we have to be explicit about it.
\begin{code}[hide]%
\>[0]\AgdaCatchallClause{\AgdaSymbol{...}}\AgdaSpace{}%
\AgdaCatchallClause{\AgdaSymbol{|}}\AgdaSpace{}%
\AgdaCatchallClause{\AgdaBound{a}}\AgdaSpace{}%
\AgdaSymbol{=}\AgdaSpace{}%
\AgdaInductiveConstructor{scaledown}\AgdaSpace{}%
\AgdaBound{x}\AgdaSpace{}%
\AgdaBound{a}\<%
\\
\>[0]\AgdaCatchallClause{\AgdaFunction{opt}}\AgdaSpace{}%
\AgdaCatchallClause{\AgdaBound{e}}\AgdaSpace{}%
\AgdaSymbol{=}\AgdaSpace{}%
\AgdaBound{e}\<%
\end{code}

\subsubsection{Code Generation}
Due to space limitations, we only consider the basic mechanisms we used in the
code generator, all the code is available in supplementary materials.  We use
heap-allocated multi-dimensional arrays that can be defined as follows:
\begin{lstlisting}[language=C]
  float(*k1)[6][5][5] = malloc(sizeof(*k1));
\end{lstlisting}
This ensures that \texttt{k1} is represented as a continuous region of memory
of size $6*5*5$ floats.  When such arrays are indexed (\eg{} \texttt{(*k1)[i][j][k]}),
the indices are translated into a single offset into the continuous memory.
Therefore, there is no pointer chasing which makes this approach efficient at
runtime.  As C uses row-major order to compute the offsets, we do obtain
partial array selections on the left, \eg \texttt{(*k1)[i]} is a $5\times 5$
array that can be further indexed or passed to \texttt{sizeof} that correctly
identifies the size of this subarray.  Surely, this is a pointer into the \texttt{k1}
array, so all the modifications to \texttt{(*k1)[i]} will modify \texttt{k1}.
As a great convenience feature, C compiler tracks the ranges of the indices
and produces warnings in cases when it figures out that ranges of indices
and the array we are indexing do not match.

Whenever we translate some $e$ in \AF{E} into C, we have to provide a storage
where $e$ has to be written to.  In case of compiling the \AF{Chain} every
local variable becomes such a storage for the bound expression.  Therefore,
our extractor always has a result variable as an argument.

For example, let us consider an expression $a ⊞ a$ of shape
(\AC{ι} 5 \AC{⊗} \AC{ι} 5), where $a$ is mapped to the C variable
\texttt{float (*a)[5][5]} that is written to the result variable
\texttt{float (*r)[5][5]}.  Here is the code that we generate:
\begin{lstlisting}[language=C]
  for (size_t x1_1 = 0; x1_1 < 5; x1_1++) { 
    for (size_t x1_2 = 0; x1_2 < 5; x1_2++) { 
      (*r)[x1_1][x1_2] = ((*a)[x1_1][x1_2] + (*a)[x1_1][x1_2]);
    }}
\end{lstlisting}
We started with checking that $a ⊞ a$ is a \emph{selectable} expression.
This means that we can always generate expression at the given index.
As we know that the shape of $a ⊞ a$ is (\AC{ι} 5 \AC{⊗} \AC{ι} 5),
we need to generate a loop nest of that shape that assigns where
we assign the expression at the given index to the result at the given
index.

We need to distinguish whether we are writing into the result or adding
into it as in cases when dealing with \AF{sum}.  Consider the code that
is generated for (\AC{sum} (\AC{selₛ} (\AB{a} (\AC{var v₀}))) where
we are adding all the elements of the array $a$ into result variable
\texttt{float (*r)[1]}:
\begin{lstlisting}[language=C]
  for (size_t x2_1 = 0; x2_1 < 5; x2_1++) {
    for (size_t x2_2 = 0; x2_2 < 5; x2_2++) {
      for (size_t x3_1 = 0; x3_1 < 1; x3_1++) { 
        (*r)[x3_1] += (&(*a)[x2_1][x2_2])[x3_1];
      }}}
\end{lstlisting}
Two things are happening here, first we generate \texttt{+=} assignment
and we make an implicit assumption that resulting variables are initialised
to zero.  In the extractor, additionally to the resulting variable we
track whether we need to do an assignment or assignment with addition.
Secondly, while $a$ is two-dimensional, we have three-dimensional loop
nest.  The latter comes from the representation of scalars as 1-element
vectors.  When we resolved the two-dimensional summation index \texttt{x2},
we know that we need to assign into the object of shape (\AC{ι} 1), but
the left-hand-side is a scalar (float).  The trick here is that in C we
can always turn scalars into 1-element vectors by simply taking the address
of the scalar.  This is why we have this 1-iteration for-loop over
\texttt{x3\_1} that will be immediately optimised away by the C compiler.

Finally, when we it comes to the operation on indices, such as addition,
subtraction, division or modulo, we generate the corresponding operation
on the individual loop indices.

Remaining details of the code generation take care of traversing through
the structure of \AF{E} with some plumbing that has to do with generating
loop-nests around expressions and checking that they are selectable.

\subsubsection{Running the Generated C Code}
In order to run the generated C code we translate the boilerplate code
from SaC to C.  While doing so, we made sure that our code can be run
in parallel.  While the  SaC compiler does this automatically, there is one
obvious loop that requires parallelisation which is computation of
the batch.  When we train the CNN, we take a batch of images and the
weights and we compute gradients for those weights per every image.
After that we average all the gradients in the batch, and we update
the weights, after all the batch is processed.  Clearly, all the
gradient computations in the batch can run in parallel.  We achieve
this by organising the batch loop such that all the gradients are
stored in a separate memory region, and we parallelise this loop
using OpenMP annotations.

We verify that the code that we generate compute the same results
as the hand-written SaC code.  Then we replicate the experiment from
the~\cite{cnn-array} using 40 epochs, 100 images in the batch, and
feeding 10000 training images.  We run the experiment on the 18-core
13th Gen Intel(R) Core(TM) i5-13600K machine using sac2c version
\texttt{1.3.3-MijasCosta-1161-gb543c} and the GCC compiler
version \texttt{12.2.0}.  The first thing that we learn is that
our generated C code is sensible (factor of 3 running time)
to the compilation flags that we enable.  We identified the set
of flags that when passed to both compilers\footnote{SaC compiler
generates C code, so we can control what flags it uses when
compiling it.}, the runtime at the
largest number of cores are 11s for the hand-written SaC implementation 
and 13.5s for the generated C code, with
very little variance.  This 20\% difference is orthogonal
to parallel execution, as it is also observed when running
the code on a single core.  The set of flags has to do with
floating point operations: \texttt{-fno-signed-zeros} ignores
the distinction between negative and positive zeroes that is given
by IEEE 754 standard, allowing to reduce (-0.0*x) to 0.0;
\texttt{-fno-math-errno} does not set errno after calling math functions;
\texttt{-fno-trapping-math} and \texttt{-fassociative-math} make
sure that we can assume associativity of floating point operations
which does not hold according to the IEEE 754.

The main performance difference comes from the fact that
compiled SaC code uses less intermediate arrays, significantly reducing the number
of memory writes.  There are numerous ways how to improve the performance
of the generated C code, but for the purposes of this paper we consider that getting within
20\% of the hand-written SaC code is sufficient evidence for our hypothesis
that the two-languages approach seems viable for achieving proved
correctness and performance.
We have automatic differentiation in the safe environment
that generates the C code that runs almost as fast as the hand-written
SaC code.

\section{Conclusions\label{sec:conclusions}}

The paper demonstrates a technique of developing high performance applications with
strong correctness guarantees.  The key insight lies in using a proof assistant
in cooperation with a high-performance language of choice.  This gives a clear
separation of concerns that is very difficult to achieve within a single
language.  The proof assistant is used to design a specification, prove
all the correctness invariants of interest and performs an extraction
into a high-performance language of choice.  This may take a non-trivial
effort, but correctness \emph{is} demanding!

Having a trusted specification as well as entire code-generation pipeline
within a single dependently-typed framework is incredibly powerful.
As we have demonstrated at the example of the neural network, we can introduce domain-specific
optimisations and even swap the backend in case its performance is unsatisfying.
For our example, the entire framework that includes array theory, DSL,
optimisations and extraction is about 2000 lines of Agda code.  We managed
to introduce a new backend in about two days and match performance of
the hand-written code.

A lot of pieces that we have developed in this
paper can be reused in other numerical applications.  We used dependent
types to guarantee the absence of out-of-bound indexing, certain function being
inverses as well as well-scopedness and well-typedness of our DSL.
However, there are many more opportunities that we did not explore.
For example, one can prove the correctness of optimisations, relating
evaluation of optimised an non-optimised expressions.  We can provide
more guarantees when we run extraction, \eg{} we can formalise some
aspects of the backend language and relate them to our DSL.
As for the DSL itself, we can try extending it with internal let
constructions which should improve our C code generation as well as
facilitate optimisations of the derivatives.

There are indeed plenty of opportunities, but the key point is this.
Correctness and performance are competing requirements when it
comes to application design.  Therefore, such a cooperation between
correctness-oriented and performance-oriented tools is likely to
persist.  With this work we demonstrate that such cooperation
is feasible in practice.

\bibliographystyle{ACM-Reference-Format}
\bibliography{paper}

\end{document}